# Assessing the CAD Capabilities of General-Purpose AI Models: A Case Study of GPT-6 Astra

Qiang Zou*, Jiajun Shen, Zhijie Yang, Yuhao Wang, Shitianyi Leng, Dongshuo Yin

State Key Laboratory of CAD&CG, Zhejiang University, China

**Abstract—** Recent advances in general-purpose AI models and agentic systems have enabled them to perform some CAD tasks. A systematic assessment is now needed to establish what these systems can accomplish and how well their outputs meet engineering requirements. This paper presents an empirical study of GPT-6 Astra, a SOTA general-purpose AI model, across four basic CAD tasks: model understanding, generation, editing, and evaluation. We design test cases spanning different levels of complexity and engineering requirements. The results show that GPT-6 Astra can support multiple stages of CAD workflows, with successful outcomes in cases involving nontrivial geometric complexity. This suggests that general-purpose AI is beginning to offer practical value in almost all steps in CAD workflow. However, gaps remain between generating visually plausible geometry and producing valid, functional, and editable CAD models. Recurring errors in freeform geometry, parameter dependencies, and mating relationships between parts limit engineering usefulness. Agentic iterative refinement can improve results but does not consistently resolve critical errors; long completion times further constrain practical use.

## I. Introduction

Recent advances in foundation models, code generation, and tool use have enabled general-purpose AI systems to perform a growing range of computer-aided design (CAD) tasks. For example, they can interpret design inputs, execute modeling commands, inspect geometry, and revise outputs using feedback [1–3]. Systematic assessment is therefore needed to establish which CAD tasks these systems can support, whether their outputs meet engineering requirements, and where specialized CAD methods remain necessary.

CAD models encode design intent and support subsequent editing, analysis, and manufacturing. Their quality therefore depends on geometric accuracy, solid validity, parametric editability, and conformance to functional and process requirements [4, 5]. A model may reproduce a part's visible appearance well while containing disconnected passages or incompatible assembly interfaces. It may also match the initial geometry yet lose required features when a dimension changes. These failures highlight the need to assess the accuracy, validity, usability, and behavior under modification of AI-generated CAD models.

Relevant evaluation studies exist. Picard et al. [6] investigated drawing interpretation, CAD generation, and machining feature recognition, although their tests used an oversimplified case: block with a blind hole. More recent benchmarks assess CAD generation, reconstruction and editing [7–11]. CADWorld evaluates agents operating FreeCAD through its graphical interface [12]. These studies address different subsets of CAD work, and most of their test cases are synthetically generated and differ substantially from those encountered in engineering practice.

This paper presents the first systematic, in-depth empirical assessment of general-purpose AI capabilities covering the very four basic CAD tasks of model understanding, generation, editing, and autonomous evaluation [13]. The full report can be found in the Appendix. Specifically, we use GPT-6 Astra as a case study, conducting six kinds of CAD tests: part classification, machining feature recognition, CAD generation from engineering drawings, text, and rendered images, and parametric reconstruction from boundary representation (B-rep) models. The cases span different levels of complexity and engineering requirements. All cases are drawn from real engineering practice.

The evaluation results suggest emerging practical value for AI in design intent understanding in conceptual design and initial geometric modeling for detailed design. AI can now effectively interpret design intent from text and images, construct geometry, build parts and assemblies, and develop designs from functional requirements. However, recurring errors in freeform geometry, parameter dependencies, and mating relationships between parts limit the usefulness of the outputs. Refinement corrects some defects but leaves others unresolved or exacerbates other geometric errors; moreover, long completion times impose a further practical constraint. Professional verification remains necessary in AI+CAD.

* Corresponding author. Qiangzou@cad.zju.edu.cn

The main contributions of this paper are as follows:

1. An assessment design that systematically covers four basic CAD tasks through six kinds of tests, with cases selected from real engineering practice, and spanning different levels of geometric complexity and engineering requirements.
2. An empirical study that systematically assesses geometric accuracy, solid validity, parametric editability, and compliance with functional requirements.
3. An analysis of recurring failures in AI+CAD, which identifies the system's practical capabilities and remaining engineering limitations.

## II. Assessment Design

This section provides a high-level overview of the assessment objectives, design rationale, and experimental settings. Detailed descriptions are available in the full evaluation report in the Appendix: Chapter 1 presents the overall framework and settings, while the first sections of Chapters 2–7 describes the corresponding task and evaluation criteria in each tested CAD tasks.

### *A. Assessment objectives and design rationale*

The objective of this study is to determine whether today's AI-generated CAD models are suitable for engineering use. This requires examining whether general-purpose AI can understand mechanical structures, develop designs autonomously from requirements, and repeatedly evaluate and revise its work as engineers do. These questions motivate a joint assessment of four basic CAD tasks: understanding, generation, editing, and evaluation. Understanding parts and features involves interpreting geometric structure and engineering meaning [14, 15], while generation expresses a design in a CAD representation [16]. Editing tests whether the model preserves intended relationships when design parameters change [5]. Evaluation supports these activities through checking and iterative refinement. Together, these tasks provide a basis for assessing both the creation of CAD models and their suitability for subsequent use. Also, they can be easily decomposed into the six kinds of CAD tests as shown in Table I, which can be connected to low-level detailed CAD workflow.

Engineering suitability depends on both the quality of the initial model and its behavior during subsequent modification. A CAD model must represent the intended geometry and function, maintain solid validity, and preserve required relationships as design parameters change. We therefore assess geometric accuracy, solid validity, functional adequacy, and parametric editability separately. We also examine the agent's checking and refinement process to determine whether successive revisions correct defects, preserve requirements already satisfied, and improve model quality. Together, these assessments establish what the system can accomplish and how reliably its outputs can support engineering design.

TABLE I Main test sets and task-specific sources of complexity.

| Task | No. of Cases | Input and requested output | Principal complexity factors |
|---|---|---|---|
| Part classification (model understanding) | 12 | STEP model → part class with geometric and semantic justification | Distinctiveness of overall geometry, compound structures, and potential confusion between local and whole-part functions |
| Feature recognition (model understanding) | 6 | B-rep model → machining feature instances, including their types, locations, and relationships | Feature intersections, nesting, interrupted boundaries, and ambiguity in feature grouping |
| Engineering drawings to CAD models (model generation) | 8 | Drawings → STEP models; five single-part and three multi-part cases | Correspondence between views, internal features, interpretation of dimensions, and component fit |
| Text to CAD models (model generation) | 12 | Six geometric and six functional descriptions → STEP models | Specified geometry, coupled requirements, component relationships, and design choices |
| Rendered images to CAD models (model generation) | 6 | Six orthographic views and one isometric view → STEP model | Shape composition, freeform surfaces, thin walls, ribs, interfaces, and scale ambiguity |

| Task | No. of Cases | Input and requested output | Principal complexity factors |
|---|---|---|---|
| B-rep to parametric feature-based models (model generation and editing) | 11 | STEP model → native parametric CAD model; editability tested through subsequent dimension changes | Modeling sequence, curved geometry, assembly structure, and parameter dependencies |
| Note: Model evaluation is assessed across all tests through the agent's autonomous checking and iterative refinement. | | | |

### *B. Case design principles*

Cases are drawn from engineering practice and selected to examine how geometric complexity and interactions among features, components, and parameters affect engineering usefulness. Case selection follows the capability being assessed, with distinct design considerations for understanding, generation, editing, and iteration/evaluation.

For understanding, cases test interpretation at both the part and feature levels. Overall geometry must support identification of the part, while local geometry must support recognition of machining features and their relationships. Intersecting features, nested structures, and ambiguous boundaries are included to probe interpretation beyond isolated, readily recognizable shapes.

For generation, cases vary both the input representation and the extent to which the design is specified. Drawings, rendered images, and text impose different interpretation demands, while functional requirements leave design decisions to the system. Internal structures and mating interfaces introduce requirements that cannot be assessed from external appearance alone.

For editing, cases exercise dependencies among features and components, testing whether reconstructed parametric models preserve required geometry and relationships when dimensions change. Initial geometric agreement and subsequent editability are therefore assessed separately.

For evaluation, it is examined throughout the tests by relating the agent's own checking and refinement to independently verified output quality. This establishes whether revisions correct defects, preserve requirements already satisfied, or introduce new errors.

Complexity is defined within each task according to the interpretation demands, feature interactions, and parameter dependencies involved. These levels support comparisons within tasks, while cross-task analysis examines recurring difficulties in meeting engineering requirements. Table I summarizes the six test sets and their principal complexity factors, and Figure 1 shows representative cases. For all the details of the tested cases, please refer to the first sections of Chapters 2–7 in the full report of the Appendix.

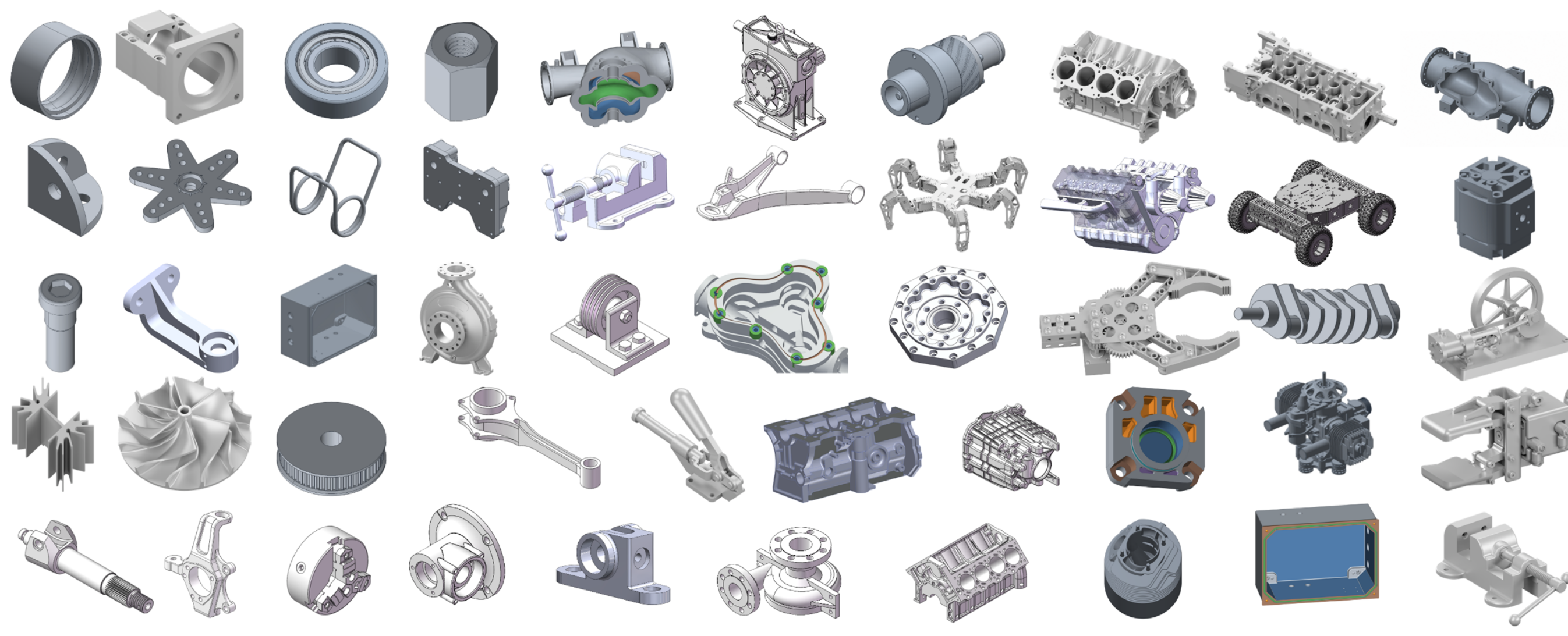

Figure 1. Representative cases from the six kinds of CAD tests.

### C. C. Experimental setting and verification

All tests use GPT-6 Astra at the MAX reasoning setting in the ChatGPT client, with Codex mode connected to a Linux server. Each run uses an independent conversation with the memoryless working mode. The agent can read the supplied files, execute code, configure its working environment, and use available tools subject to the task instructions. Web search is enabled in the main tests and disabled in selected supplementary runs. Within a single conversational turn, the agent can carry out multiple cycles of execution, checking, and revision. For all the details of the setup, please refer to Table 1-2 in Appendix.

Output quality is verified separately from the agent's own assessment, using geometric comparisons, solid-validity checks, and inspection against task-specific engineering requirements as appropriate. For parametric reconstructions that pass the initial solid-validity checks, prescribed manual dimension changes test whether the model rebuilds correctly and preserves the required relationships. These tests measure the editability of the generated model. Recorded completion times cover the full task execution, including environment setup, tool use, and autonomous refinement.

The full evaluation report in the Appendix provides the complete task instructions, case descriptions, task-specific procedures and verification criteria, and detailed results.

## III. Results

Table II summarizes the main tests. Successful recognition or generation often coexists with defects found by further checks. Supplementary runs are reported within each task and excluded from the main-test counts. Table III summarizes the working time for individual tests.

TABLE II Results for the main tests. Counts use the task-specific units stated in each row. Initial validity and subsequent editability are checked separately.

| Task | Principal outcome | Limitations observed |
|---|---|---|
| Part classification (model understanding) | 12/12 models assigned to the correct candidate class | Subtype and whole-part function can still be misidentified. |
| Feature recognition (model understanding) | 3/6 cases correct in feature grouping, type, and relationships | Two cases count base-shape regions as machining features; one also misgroups holes and assigns an incorrect type. |
| Engineering drawings to CAD models (model generation) | All components pass validity checks in 6/8 cases | Of three multi-part cases, one assembles, one has interference, and one cannot be assembled completely. |
| Text to CAD models (model generation) | 12/12 cases satisfy the listed requirements; 11/12 pass validity checks | Gaps and missing mechanical connections remain outside the listed requirements. |
| Rendered images to CAD models (model generation) | 6/6 main cases produce STEP files with individually valid solids | Mean silhouette IoU is 0.8790–0.9915; local geometry, scale, and connectivity remain imperfect. |
| B-rep to parametric feature-based models (model generation and editing) | 11/11 native CAD models with construction histories; 6/11 eligible for editability tests | 52/60 changes pass in the eligible subset; 4/11 initial models pass validity and all ten prescribed edits. |

TABLE III Completion times (minutes) for the main tests, computed from the per-case records. Means are in minutes:seconds and ranges in minutes. Complexity levels are task-specific; supplementary runs are excluded.

| Task | Simple mean | Moderate mean | Complex mean | Case range |
|---|---|---|---|---|
| Part classification | 2:30 | 2:00 | 2:30 | 2–4 |
| Feature recognition | 30:30 | 40:30 | 85:00 | 24–121 |

| Task | Simple mean | Moderate mean | Complex mean | Case range |
|---|---|---|---|---|
| Engineering drawings to CAD models | 26:30 | 43:40 | 31:20 | 25–68 |
| Text to CAD models | 42:00 | 60:15 | 97:30 | 23–197 |
| Rendered images to CAD models | 14:30 | 17:30 | 121:00 | 9–221 |
| B-rep to parametric feature-based models | 7:36 | 13:20 | 50:00 | 1–112 |

### *A. Part classification*

This task evaluates whether GPT-6 can correctly classify parts from the geometric and topological information in 3D CAD models. The test set contains 12 models spanning 10 part classes, with four simple, four moderately complex, and four complex cases. Figure 2 shows representative examples.

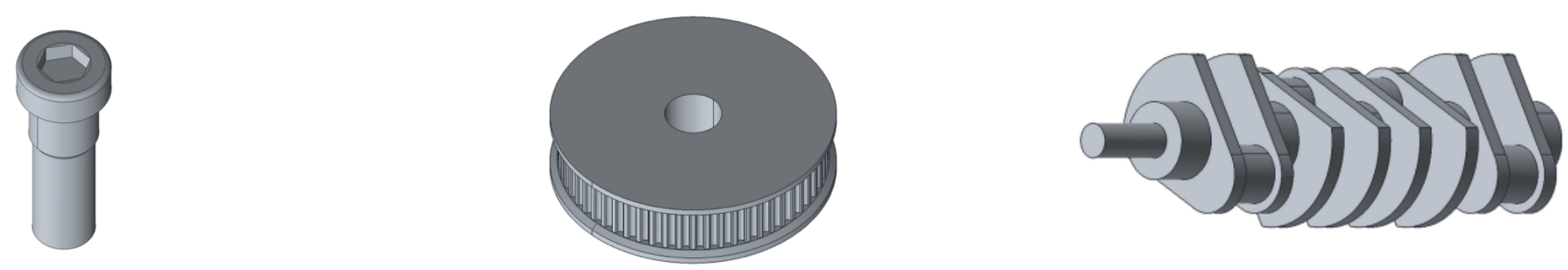

Simple: screw/bolt    Moderate: pulley    Complex: crankshaft

Figure 2  Part classification examples at different levels of complexity

With the names of 10 candidate part classes provided and web search enabled by default, GPT-6 correctly classified all 12 models, achieving 100% accuracy. Mean completion times for simple, moderately complex, and complex cases were 2 min 30 s, 2 min, and 2 min 30 s, respectively. The longest individual test took 4 min.

Two additional conditions were tested separately: withholding the candidate class names and disabling web search. Two representative complex cases were selected for each condition, and all were classified correctly.

A classification error arose when both conditions were applied together: with no candidate class names and web search disabled, GPT-6 misclassified a quick-disconnect coupling as a check valve (one-way valve), as shown in Figure 3. It identified the internal spring, valve stem, and valve disc, but treated the function of the valve mechanism as the purpose of the coupling as a whole.

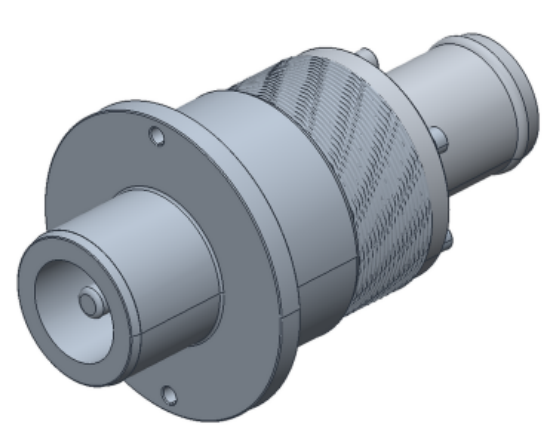

**Actual part: quick-disconnect coupling**

GPT-6 classification: check valve (one-way valve)

Figure 3  Misclassification of a quick-disconnect coupling

In summary, current capabilities and limitations are:

- For typical parts, GPT-6 can identify both the general part type and the specific type, as well as the part's intended use.
- For parts with atypical shapes or shapes resembling those of other part types, GPT-6 can identify the general type correctly but may still get the specific type wrong.
- For assemblies, GPT-6 may mistake the function of one structure within the assembly for the purpose of the whole assembly and assign the wrong part type.

### *B. Feature recognition*

This task tests whether GPT-6 can identify individual machining features in a B-rep model and recognize the relationships between them. It must identify both the general feature type and more specific forms, such as through holes and blind holes, without a list of candidate feature types. The test set contains two simple, two moderately complex, and two complex models. Figure 4 shows representative examples.

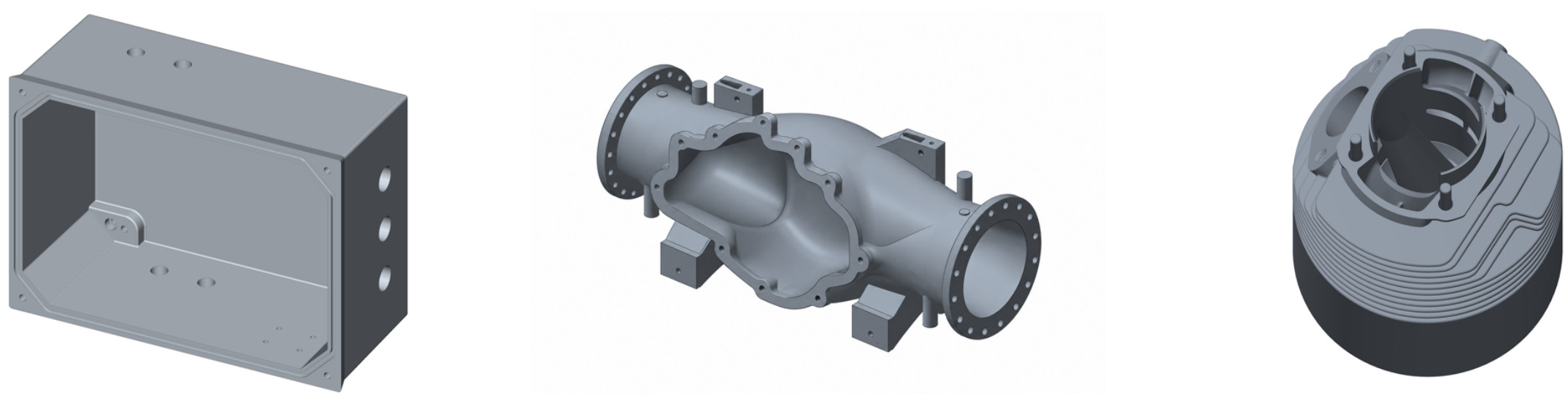

Simple Moderate Complex

Figure 4 Feature recognition examples at different levels of complexity

Feature recognition was correct in three of the six cases. In two others, GPT-6 identified the actual features correctly but also labeled some groups of faces as features when they were not. In the remaining case, most features were recognized correctly, with a small number of errors.

Mean completion times for simple, moderately complex, and complex cases were 30 min 30 s, 40 min 30 s, and 1 h 25 min, respectively. The longest individual test took 2 h 1 min. Results for the same model were similar with web search enabled and disabled. Each condition gave better results for some distinctions, such as whether holes were through or blind.

A recurring error was to label groups of faces that did not constitute machining features as features, producing false positives. In complex models, GPT-6 also treated two separate circular holes as a single feature and mistook an external cylindrical surface for a hole (Figure 5).

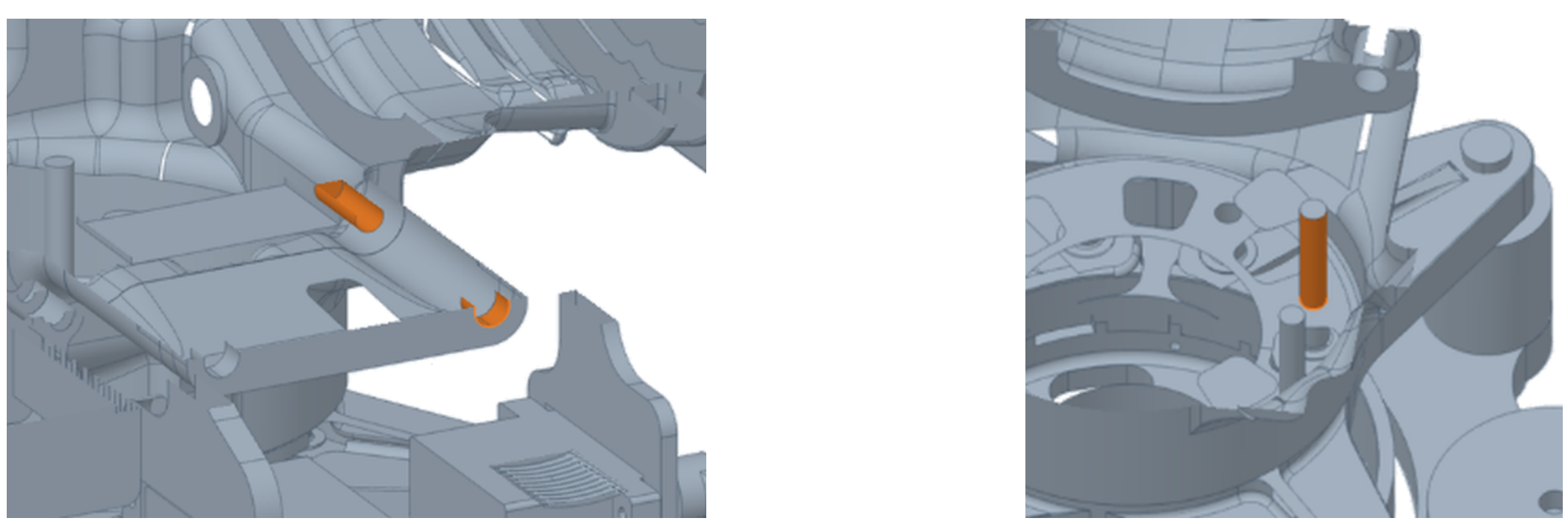

**Two circular holes identified as one feature**
The orange regions are spatially separate

**Cylindrical surface misidentified as a hole**
The orange region is on the outside of the solid

Figure 5 Errors in distinguishing feature instances and identifying feature types in complex models

In summary, current capabilities and limitations are:

- When features are independent or have only simple nesting relationships, GPT-6 can correctly recognize all machining features, including holes, slots, pockets, and fillets.
- When features are nested at multiple levels, intersect, or are truncated, GPT-6 can recognize most machining features but still makes some errors. In models with complex overall shapes, it may identify groups of faces as features when they are not. It also has difficulty recognizing internal features when the geometry is densely packed and internal structures are largely hidden from view.

### *C. CAD generation from engineering drawings*

This task evaluates whether GPT-6 can combine information from multiple views, section views, and dimension annotations to reconstruct the 3D geometry of parts, while ensuring that separately generated parts meet the assembly requirements. The test set contains five single-part cases and three cases involving multiple parts. Two cases are simple, three are moderately complex, and three are complex. Figure 6 shows representative reference models.

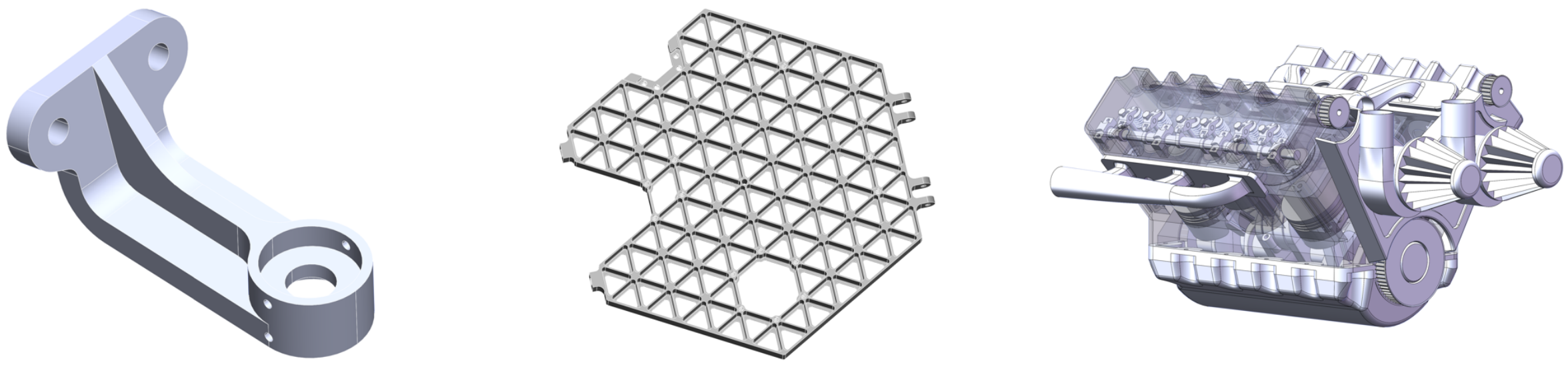

Simple Moderate Complex

Figure 6 Examples for CAD reconstruction from engineering drawings at different levels of complexity

All parts passed the solid validity checks in six of the eight cases. The other two cases failed because of defects such as self-intersections and non-manifold connections. Of the three cases involving multiple parts, one could be assembled successfully, one had interference between components, and one could not be assembled because of missing structures and incorrect dimensions.

Mean completion times for simple, moderately complex, and complex cases were 26 min 30 s, 43 min 40 s, and 31 min 20 s, respectively. The longest individual test took 1 h 8 min.

Typical errors involved inaccurate reconstruction of local structures and incorrect parameters at mating interfaces. In assemblies, the resulting mismatches in mating dimensions caused problems such as overlapping solid bodies (Figure 7).

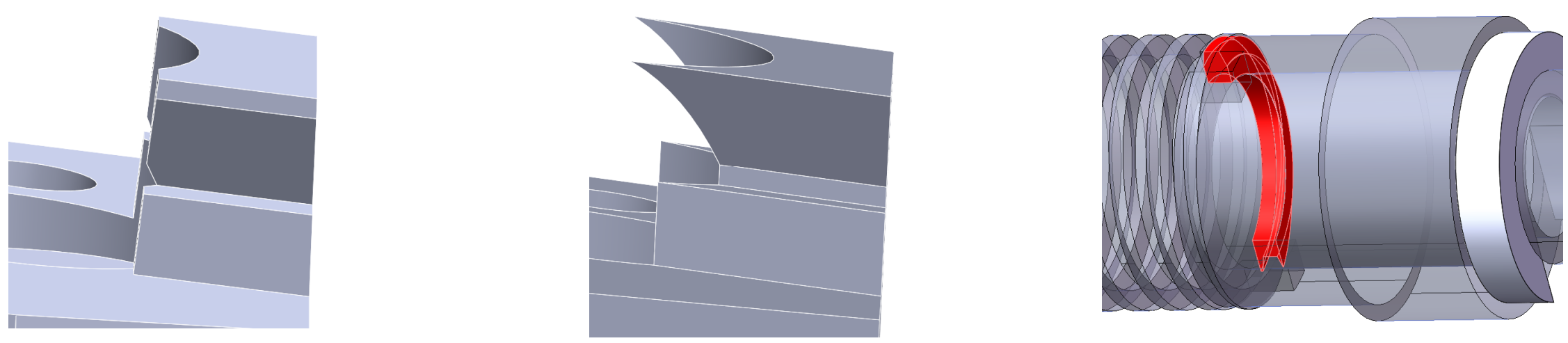

**Reference model**
Dovetail mating profile

**Generated model**
Deviation in the reconstructed dovetail profile

**Generated assembly**
Red regions show overlapping components

Figure 7 Dovetail profile reconstruction errors and assembly interference

In summary, current capabilities and limitations are:

- GPT-6 recovers the main geometry reasonably well when the relationships among drawing views are clear and the part primarily contains conventional holes, slots, and regular patterns. For parts with multiple steps, nested pockets, or complex transitions in thin walls, however, it remains prone to misinterpreting the structural arrangement and omitting local features. Invalid solids also occur in some cases.
- For assemblies with few parts and clearly specified mating relationships, GPT-6 can generate part models that fit together. When an assembly has more parts and more complex mating requirements, it struggles to reproduce critical mating dimensions accurately, leading to interference or preventing assembly altogether.
- In some complex cases, GPT-6 approximates geometry that would conventionally be modeled using draft, shell, or loft operations with basic extrusions and cuts. These approximations do not consistently preserve the required local geometry.

### *D. CAD generation from text*

This task tests whether GPT-6 can develop a design, determine its dimensions, and generate a CAD model from geometric or functional requirements. The test set contains four simple, four moderately complex, and four complex cases. Two additional complex assemblies were used for stress testing, and four complex cases were also tested with web search disabled. Figure 8 shows representative reference models.

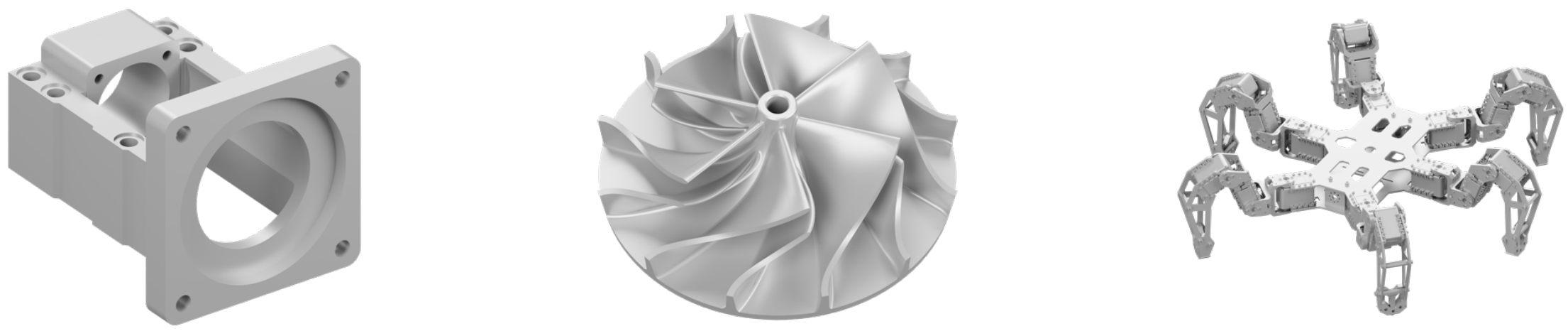

Simple (single part) | Moderate (single part) | Complex (assembly)

Figure 8 Examples for CAD generation from text at different levels of complexity

All 12 cases received full scores on the listed requirement checks, and 11 passed the solid validity checks. These are separate outcomes: a full requirement score does not establish complete engineering functionality. Mean completion times for simple, moderately complex, and complex cases were 42 min, 1 h 0 min 15 s, and 1 h 37 min 30 s, respectively. The longest test took 3 h 17 min.

Of the two stress-test cases, one received a full requirement score, but neither passed the solid validity checks. In the four cases tested with web search disabled, all four received full requirement scores, and two passed the validity checks.

Typical errors included leaving gaps where structures should connect continuously and adding unnecessary components that served no supporting, connecting, or motion-transmitting function (Figure 9).

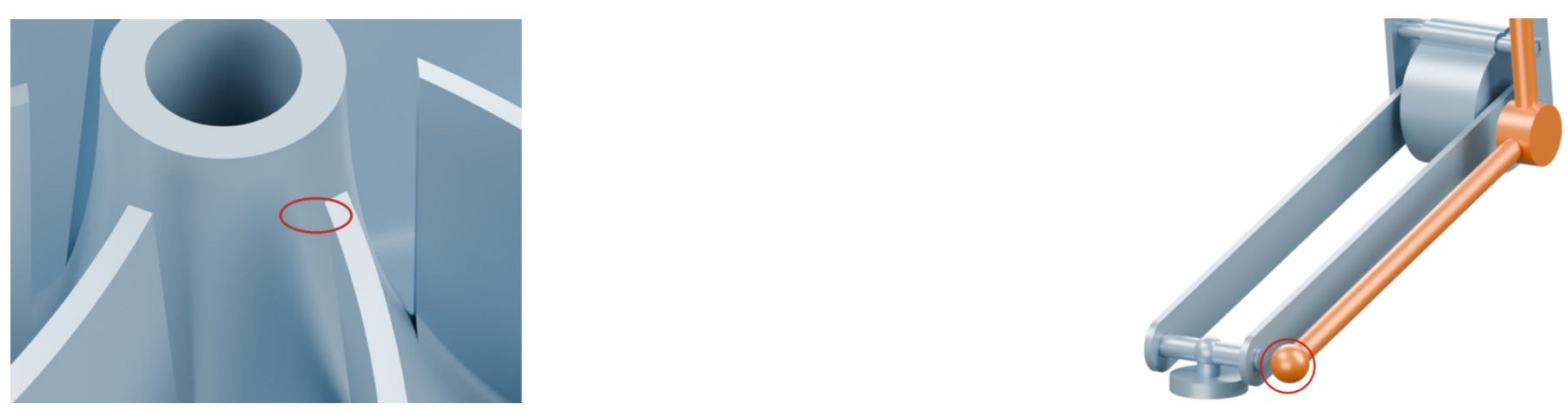

**Incomplete local connection**
The red circle marks a gap at the connection

**Unnecessary component**
The orange component serves no functional purpose

Figure 9 Typical errors in CAD generation from text

In summary, current capabilities and limitations are:

- For parts with clearly specified dimensions and feature positions, or mechanisms with clear load, motion, and mounting requirements, GPT-6 can develop a design from the text, determine its dimensions, and generate individual part models or assemblies. During iteration, it can add missing features and adjust component positions. The mean number of iterations across the 12 cases was four.
- For models requiring continuous surface transitions or coordinated motion between components, GPT-6 may leave gaps where parts should join, or leave drive components unsecured or disconnected so that motion cannot be transmitted. It also generates unnecessary components with no functional purpose. These errors indicate limitations in its understanding of how components fit together and what they are intended to do. Engineering requirements implicit in the text are therefore not reliably met.
- For parts with thin walls, intersecting internal holes and cavities, and hole walls close to the outer surface, GPT-6 cannot yet reliably handle surface geometry and the boundaries between adjacent faces. The generated models can contain self-intersections and non-manifold connections.

*E. CAD reconstruction from rendered images*

This task uses rendered views without dimension annotations to evaluate whether GPT-6 can recover 3D shape, feature positions, and connectivity. The test set contains two simple, two moderately complex, and two complex cases, with an additional complex housing used for stress testing. Tests also varied the combination of input views and included conditions with web search disabled. Figure 10 shows representative examples.

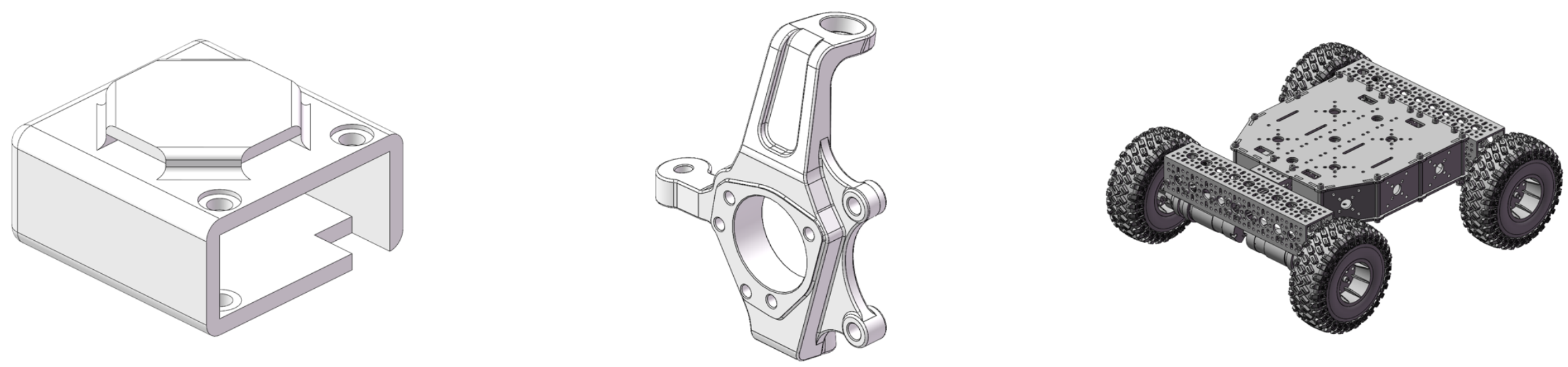

Simple Moderate Complex

Figure 10 Examples for CAD reconstruction from images at different levels of complexity

All 21 test runs produced STEP models, and every solid in the outputs passed the solid validity checks. Across all 21 test runs, mean completion times for simple, moderately complex, and complex cases were 12 min 30 s, 16 min 20 s, and 43 min, respectively.

For the six cases tested using six orthographic views and one axonometric view, the mean silhouette intersection over union (IoU) across views ranged from 0.8790 to 0.9915 (Table IV). GPT-6 recovered the main shape and component layout reasonably well, but local structures and absolute dimensions still showed substantial discrepancies.

For parts with freeform surfaces, thin walls, and intersecting ribs, GPT-6 often simplifies complex surfaces into planar faces or segmented surfaces and misinterprets feature orientations and connections. For example, the reconstructed control arm in Figure 11 retains the overall outline, but its smooth sidewall surfaces are replaced by segmented surfaces, the root fillets are missing, and local steps are introduced.

When the reference model consists of a single solid, GPT-6 sometimes produces multiple interpenetrating solids with overlapping volumes. Their connections do not match those of the reference part.

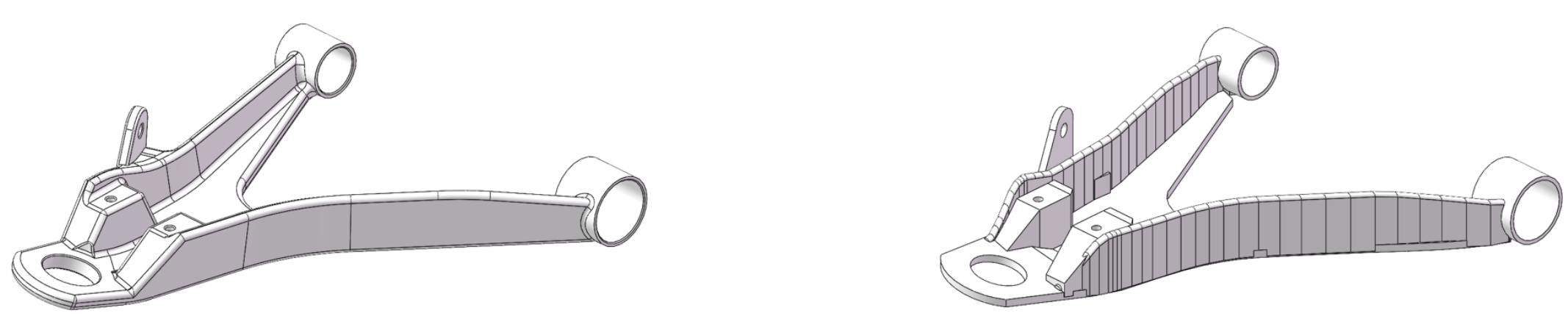

**Reference model** **Generated model**

Smooth curved surfaces and root fillets Segmented surfaces and missing root fillets

Figure 11 Reconstruction errors in the control arm surfaces and root transitions

In summary, current capabilities and limitations are:

- For plate-like parts and housings with regular geometry, clear outlines, and clearly visible main features, GPT-6 recovers the main shape and the layout of holes and slots reasonably well.
- For structures with freeform surfaces, thin walls, and closely spaced, intersecting ribs, GPT-6 tends to simplify the surfaces and the transitions where they join. Accurately reconstructing curved surfaces and fillets remains difficult.
- For complex assemblies, the generated models can have errors in connections between solids, holes and slots, and assembly interfaces.
- Without a dimensional reference, absolute scale cannot be determined uniquely from the images. GPT-6 also makes errors in relative proportions, which can be assessed independently of scale.

TABLE IV Image reconstruction with six orthographic views and one isometric view. Each output passes the checks on individual solids.

| Case | Mean IoU | HD (mm) |
|---|---|---|
| I2C-01 | 0.9788 | 14.758 |
| I2C-02 | 0.9915 | 64.207 |
| I2C-03 | 0.9483 | 62.401 |
| I2C-04 | 0.8790 | 45.211 |
| I2C-05 | 0.9468 | 9.959 |
| I2C-06 | 0.9391 | 80.297 |

### *F. Parametric reconstruction and editability*

This task tests whether GPT-6 can use the geometry and topology of an existing model to reconstruct sketches, modeling features, and parameter dependencies, and produce an editable CAD model that matches the input geometry. The test set contains five simple, three moderately complex, and three complex cases. The three complex cases were also tested with web search disabled. Figure 12 shows representative examples.

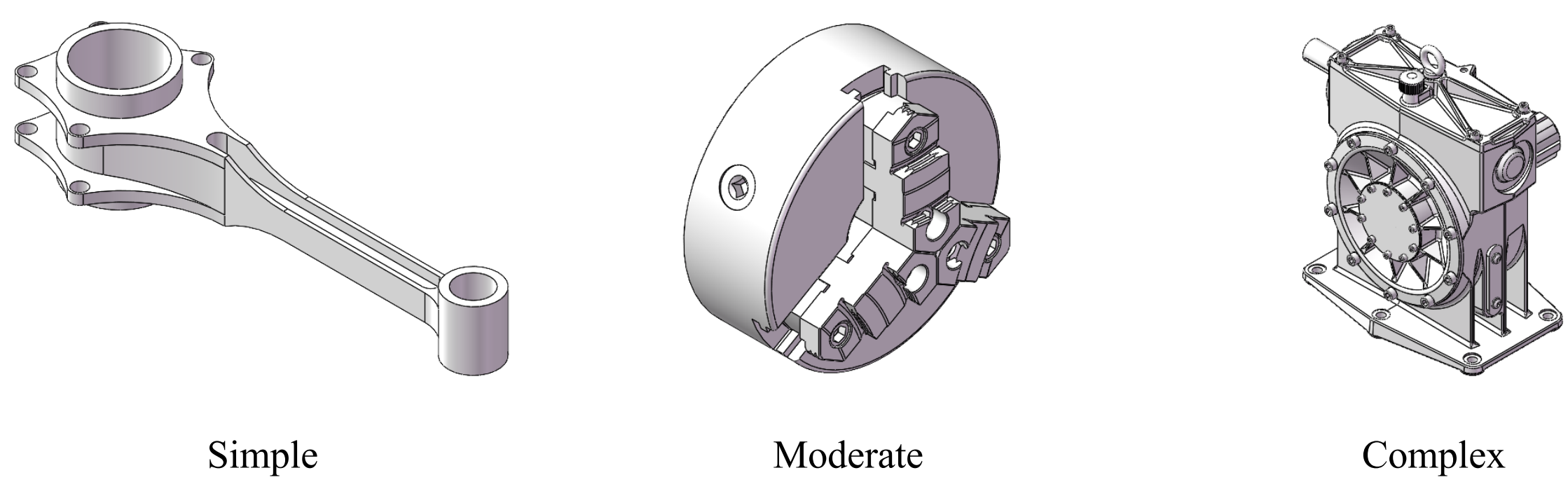

Figure 12 Examples for parametric reconstruction from B-rep models at different levels of complexity

All 11 cases produced parametric models in SOLIDWORKS format. Six models (54.5% of the total: five simple and one moderately complex) passed the solid validity checks and were therefore eligible for parameter-editing tests. The remaining two moderately complex models and three complex models contained invalid solids. They were excluded from parameter-editing tests because they failed the validity prerequisite; their behavior under editing was not tested.

Each of the six eligible models underwent 10 single-parameter edits, with only one dimensional parameter changed at a time. An edit passed if the parameter change took effect, dependent geometry updated correctly, the edited model passed the solid validity checks, and the model could be restored to its original state. Of the 60 edits, 52 passed (86.7%). Four models passed all 10 edits, accounting for 36.4% of all 11 cases: three simple models and one moderately complex model.

Mean completion times for simple, moderately complex, and complex cases were 7 min 36 s, 13 min 20 s, and 50 min, respectively. The longest individual test took 1 h 52 min. None of the three models tested with web search disabled passed the solid validity checks.

The reconstructed models contained sketches, features, and dimensional parameters and resembled the reference shapes, yet incomplete parameter dependencies limited their editability. In one case, increasing the housing depth did not correctly update the dependent hole features, and newly added material closed the existing hole openings.

For models with complex surfaces, GPT-6 often uses a series of extrusions to approximate the original surfaces with planar faces. Smooth surfaces become densely segmented and stepped, producing large geometric errors. The resulting feature structure is inconsistent with conventional modeling practice for such parts (Figure 13).

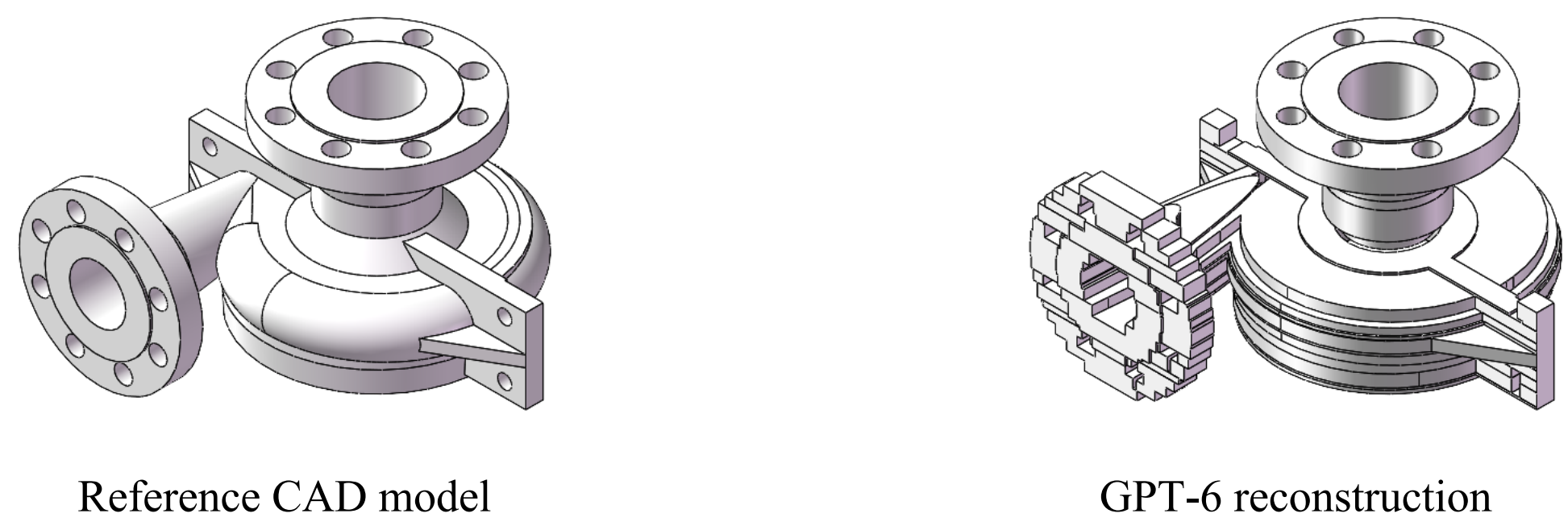

Figure 13  GPT-6 approximates curved surfaces with planar faces from multiple extrusions

In summary, current capabilities and limitations are:

- For models built mainly from conventional features such as extrusions, revolved features, holes, and slots, with relatively simple dimensional relationships, GPT-6 can reconstruct the main geometry with reasonable accuracy and support some dimensional changes.
- When hole and slot features depend on the main body dimensions, GPT-6 may fail to establish all the necessary feature dependencies. Changing the main body dimensions can then close hole openings or cause mounting holes to disappear.
- For models with complex surfaces, GPT-6 often uses a series of extrusions to approximate the original curved surfaces with planar faces. Surfaces that are smooth in the reference become segmented and stepped, resulting in large reconstruction errors.

## IV. Cross-Task Analysis

(1) GPT-6 demonstrates capabilities across several kinds of CAD tasks

GPT-6 can classify parts and recognize geometric and machining features in existing B-rep models, and generate 3D models from engineering drawings, images, or text descriptions. In some tasks, it can also select a design concept from functional requirements, determine dimensions, and perform engineering analyses and checks. The tests include successful cases of feature recognition in relatively complex parts, generation and reconstruction of moderately complex models, and construction of relatively complex assemblies.

(2) Generated models do not always form valid solids or meet functional requirements

Most generated CAD models resemble the required shape, but errors can remain in their internal structures, critical dimensions, and feature layouts. These errors occur in both individual parts and assemblies: internal passages may not connect or remain separate as required, and components may lack the connections needed to secure them or transmit motion. Some errors have little effect on appearance but directly compromise essential functions. Shape complexity and incomplete input information do not fully explain these failures. Errors also occur in simple models, and even complete information about the target geometry does not guarantee a correct reconstruction.

(3) Creating a 3D model does not guarantee that parametric edits will work correctly

For models with simple shapes, few features, and straightforward dependencies, GPT-6 can already create editable parametric models that support dimensional changes and the corresponding updates to dependent features. Even some geometrically simple models have incomplete feature dependencies, while complex reconstructions frequently contain invalid solids. Changing a dimension can close a hole opening, remove a local feature, or displace a component. GPT-6 cannot yet consistently produce parametric models that behave correctly when edited.

(4) GPT-6 can check and revise its own models, but cannot yet correct errors consistently

During modeling, GPT-6 can inspect its results, identify some problems, and revise the models repeatedly. These revisions contribute to successful outcomes in some complex tasks. Yet the tests show that error often fluctuates over successive iterations, and some changes increase it. GPT-6 is therefore still unreliable at identifying the source of an error, deciding what to modify, and selecting the appropriate operation. Dependencies among geometric features and parameters complicate this process: a single change may affect several parts of the geometry and multiple design requirements. GPT-6 does not yet fully identify these dependencies or reliably anticipate and control the consequences of an edit. As a result, correcting one problem can introduce new errors.

(5) Long modeling times limit practical use

GPT-6 often cycles repeatedly through model generation, inspection, and correction, which takes considerable time. Mean completion times for several kinds of modeling and reconstruction tasks were in the tens of minutes. Some complex cases took more than an hour, with the longest taking 3 h 41 min. These times make it difficult to provide the rapid feedback and frequent revisions needed in design work. Despite progress in modeling performance, practical use remains limited.

Table 1 GPT-6's current capabilities and main limitations in CAD tasks

| Task | Current capabilities | Main limitations |
|---|---|---|
| **Part classification** | GPT-6 uses geometric information and engineering knowledge to identify what kind of part a model represents. | GPT-6 may recognize the general type of part but still misidentify the specific type. It may also mistake the function of one structure within a part for the purpose of the whole part. |
| **Feature recognition** | GPT-6 identifies individual holes, slots, pockets, and other machining features in a model. | GPT-6 does not always distinguish individual features or identify their types correctly, especially where features are nested or intersect. |
| **CAD reconstruction from engineering drawings** | GPT-6 combines multiple views, section views, and dimensions to recover the main shape and internal structure. In some cases, the generated parts can be assembled as intended. | GPT-6 can misread how complex structures are arranged and get critical dimensions wrong, leaving out local features or creating interference between assembled components. |
| **CAD generation from text** | GPT-6 builds models from geometric descriptions, selects designs to meet functional requirements, and carries out modeling and engineering calculations. | GPT-6 still makes frequent errors in complex features, passage connectivity, local transitions, and connections that secure parts or transmit motion. The models do not consistently meet functional requirements or pass solid validity checks. |
| **CAD reconstruction from images** | GPT-6 recovers the main shape, main features, and component layout from images showing multiple views. | GPT-6 often simplifies complex surfaces and local transitions. Connections between solids and assembly interfaces may also be incorrect. Without a dimensional reference, absolute scale cannot be determined uniquely from the images. |
| **Parametric reconstruction from B-rep models** | GPT-6 recovers the main geometry of simple and moderately complex models, creates parametric models, and supports some dimensional changes. | Reconstructed complex surfaces and internal passages can differ substantially from the originals. GPT-6 often fails to establish the dependencies needed for related features to update correctly after an edit. |
| **Autonomous iteration across tasks** | GPT-6 uses feedback to add missing features, adjust dimensions and positions, reduce geometric error, and meet more of the requirements. | GPT-6 still struggles to decide what to change and how far those changes should extend. Even after several iterations, models may contain invalid solids, lack required functions, or fail when a parameter is edited. Long iterations limit practical use. |

## V. Conclusion

This study assessed GPT-6 Astra across CAD model understanding, generation, editing, and autonomous evaluation. The system interpreted geometry, constructed parts and assemblies, and developed designs from functional requirements. These successes indicate emerging practical value in conceptual design and initial modeling for detailed design.

Engineering reliability remains limited by errors in connectivity, part mating relationships, and parameter dependencies. Geometric similarity and valid individual solids do not establish functional adequacy or reliable editability. Autonomous refinement corrects some defects but leaves others unresolved and can increase geometric error. Long completion times further constrain practical use.

Reliable use requires professional verification of the initial model and its response to modification. Preserving the geometric and functional relationships needed for engineering use, while correcting errors without disrupting those relationships, remains a central challenge for CAD agents.

# Appendix

AI and CAD Evaluation
Technical White Paper Series

# Assessing the CAD Capabilities of General-Purpose AI Models:

## A Case Study of GPT-6 Astra

Qiang Zou, Jiajun Shen, Zhijie Yang
Yuhao Wang, Shitianyi Leng, Dongshuo Yin

Design and Manufacturing Automation Lab
Zhejiang University

# About this report

This report was prepared by the Design and Manufacturing Automation Lab (DMA Lab) at Zhejiang University.

## About the lab

The DMA Lab is part of the State Key Laboratory of CAD&CG at Zhejiang University. Its work focuses on geometric computing, intelligent CAD, and the development of CAD/CAM software.

Website (Chinese): https://person.zju.edu.cn/qiangzou

Website (English): https://qiang-zou.github.io/

## Author contributions

Qiang Zou: Designed the evaluation project and report outline, analyzed and synthesized the findings, and wrote and reviewed the report.

Jiajun Shen: Conducted the experiments and contributed to writing and reviewing the report.

Zhijie Yang: Contributed to the report outline, conducted the experiments, and contributed to writing the report.

Yuhao Wang, Shitianyi Leng, and Dongshuo Yin: Conducted the experiments and contributed to writing the report.

# Table of Contents

# Chapter 1 Evaluation objectives and design

## 1.1 Evaluation objectives

This report systematically evaluates how well OpenAI's GPT-6 can understand and generate CAD models. It examines how performance varies with task complexity, the strategies evident in its execution records, and what it can and cannot do. The findings help assess which CAD tasks GPT-6 is close to handling in practice, where its capabilities still fall short, and which research directions deserve priority.

- Model understanding: Assess how well GPT-6 understands a CAD model's geometry and topology, and what the model represents in engineering terms.
- Model generation: Assess whether GPT-6 can generate CAD models from engineering drawings, text descriptions, and rendered images, and construct a plausible parametric modeling history from an existing CAD model.
- Analyze the observable steps GPT-6 takes to complete each task and establish what it can and cannot do.
- Synthesize the test results to assess GPT-6's overall CAD capabilities and identify directions for future research in intelligent CAD.

## 1.2 CAD tasks evaluated

The evaluation covers six kinds of CAD tasks, listed in Table 1-1. For each task, the report presents the test design and results, analyzes GPT-6's approach, and discusses its capabilities and limitations. The complexity criteria, test data, input prompts, and evaluation metrics are chosen to suit each task.

Table 1-1 Overview of the CAD tasks

| Task | Input | Expected output | Capabilities assessed |
|---|---|---|---|
| **Part classification** | B-rep model (STEP file) | Part type, reasons for the classification, and typical functions or uses | Understand the part's overall geometry and topology, and identify what kind of part it is. |
| **Feature recognition** | B-rep model (STEP file) | Machining features and the faces that belong to each feature | Understand local geometry and topology and recognize the machining features they define. |
| **CAD reconstruction from engineering drawings** | 2D engineering drawings and related assembly information | CAD model of a single part, or CAD models of every part in an assembly | Read engineering drawings, infer geometry and dimensions, determine how parts fit together, and reconstruct CAD models. |

| Task | Input | Expected output | Capabilities assessed |
|---|---|---|---|
| **CAD generation from text** | Text describing geometric or functional requirements | CAD model that meets the stated geometric or functional requirements | Understand and implement geometric requirements, and translate functional requirements into CAD features. |
| **CAD reconstruction from images** | Rendered views of a product from multiple angles | CAD model that reproduces the visible shape and main structures | Interpret multiple views, infer features, dimensions, and proportions, reconstruct local features, and handle uncertainty. |
| **Parametric reconstruction from B-rep models** | B-rep model (STEP file) | Editable parametric CAD model with a modeling history and geometry that matches the input | Interpret geometry and topology, infer sketches and modeling features, plan a modeling sequence, and produce a model that can be edited parametrically. |

## 1.3 Test environment

Table 1-2 gives the testing setup. Task-specific prompts and exceptions are described in the corresponding chapters. Throughout this report, "GPT-6" refers to the model operating within this tool-enabled environment; the experiments do not isolate the foundation model from its execution tools.

Table 1-2 GPT-6 configuration

| Setting | Configuration |
|---|---|
| **Model** | GPT-6 Astra |
| **Reasoning effort** | MAX |
| **Interface and execution environment** | ChatGPT client in Codex mode, connected remotely to a Linux server. |
| **Conversations** | A separate conversation for each test task to avoid carrying over labels or solutions from other conversations. |
| **Memory** | Disabled |
| **File access** | GPT-6 could read STEP files, images, and any other attachments needed for the task. |
| **Code execution** | GPT-6 was allowed to set up its own execution environment. |
| **Web search** | Enabled. Additional tests were run with web search disabled. |
| **CAD and geometry tools** | No CAD backend was prescribed for the first five tasks. The parametric reconstruction prompt explicitly required the SOLIDWORKS API (Chapter 7). |
| **Iteration limit** | Task completion was limited to one conversational turn, within which GPT-6 could iterate autonomously. In Chapter 3, a subsequent visualization request was used only to inspect the completed recognition output. |

# Chapter 2 Part classification

## 2.1 Task objectives and evaluation criteria

The part classification task tests whether GPT-6 can use the geometry and topology in a B-rep model (provided as a STEP file) to assign a mechanical part to one of a predefined set of classes. This follows the conventional formulation of part classification, allowing results to be compared on the same basis.

Additional tests use open-ended prompts about the part and its engineering use. These go beyond selecting a label: they examine whether GPT-6 can identify the part, explain which geometric and structural features support its judgment, and infer its function and typical uses without a fixed list of candidate classes. Table 2-1 summarizes the task.

Table 2-1 Part classification task

| Test element | Description |
|---|---|
| Research question | Can GPT-6 correctly classify a mechanical part from a STEP model and, without a fixed label set, identify the part, explain the structural evidence, and infer its typical functions and uses? |
| Input | B-rep model of a mechanical part. |
| Expected output | Most likely part class or name, the main evidence for the judgment, typical functions and uses, and any necessary explanation of uncertainty. |
| Evaluation criteria | For the standard task, the primary criterion is classification correctness within the supplied label set. For the additional tests, the criteria are whether the proposed class or name is reasonable, whether the geometric and structural evidence is accurate, whether the inferred functions and uses are plausible, and whether uncertainty is recognized appropriately. |

## 2.2 Test design

### 2.2.1 Test case selection

The test cases were drawn from the TMCAD dataset. All represent real mechanical parts with known classes or names. The selection covers a range of part classes and identification difficulties, avoiding a test set made up only of parts with highly distinctive overall shapes or a single local feature that reveals their class.

Difficulty was assessed from how typical the part's overall shape is, how much of its geometry must be considered to identify it, how closely it resembles other part classes, and how directly its geometry and topology reveal its function. Table 2-2 gives the complexity criteria.

Table 2-2  Complexity criteria for part classification

| Complexity | Selection criteria | Typical characteristics |
|---|---|---|
| Simple | The overall shape and key functional structures are characteristic of the class. The part can be identified reliably from its overall shape or a few key structures. | Distinctive features clearly separate the part from related classes. |
| Moderate | Classification requires several local structures and their spatial relationships to be considered together. No single feature is decisive, and the geometry may resemble related classes. | The main body, mounting structures, hole layout, and supporting or connecting structures must be considered together. |
| Complex | The features that distinguish the class are not readily apparent, or the shape and structure closely resemble other classes. Local structures alone can be misleading. | Complex industrial parts, nonstandard parts, parts combining several functions, or parts for which more than one classification is plausible. |

Four cases were selected at each complexity level: simple, moderate, and complex. Table 2-3 lists all 12 cases.

Table 2-3  Part classification test cases

| Case ID | Rendered CAD model | Part class | Complexity | Basis for complexity rating |
|---|---|---|---|---|
| PC-01 | 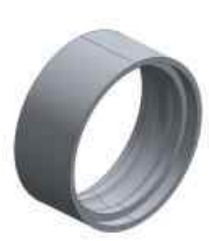 | Bearing | Simple | A single annular body with a continuous raceway on the inside is characteristic of a bearing ring. |
| PC-02 | 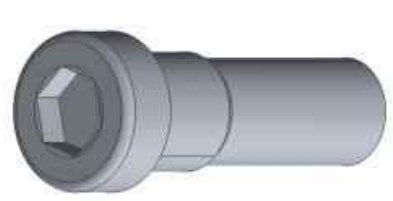 | Screw/bolt | Simple | A cylindrical shank, enlarged cylindrical head, and hexagonal socket form a typical bolt shape. |
| PC-03 | 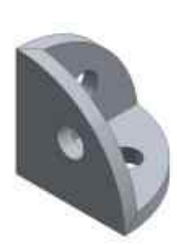 | Bracket | Simple | Three mutually perpendicular faces each have mounting holes, making the corner support and fastening arrangement clear. |
| PC-04 | 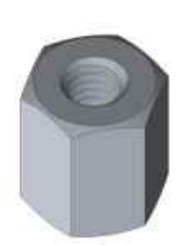 | Nut | Simple | The hexagonal exterior and threaded through hole are clearly visible. |
| PC-05 | 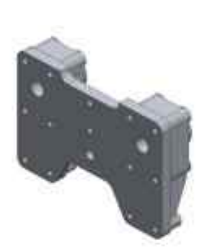 | Bracket | Moderate | Supports at both ends, a central connecting body, and mounting holes form a supporting structure. Multiple sets of mounting holes make classification more demanding. |

| Case ID | Rendered CAD model | Part class | Complexity | Basis for complexity rating |
|---|---|---|---|---|
| PC-06 | 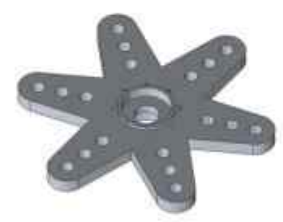 | Flange | Moderate | A central through hole and six radial mounting arms with holes at their ends form a connection interface. The arms alone could suggest a bracket. |
| PC-07 | 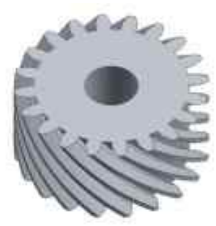 | Gear | Moderate | The helical teeth around the rim resemble blades when viewed locally. The circumferential tooth array and central shaft bore must be considered together to identify the gear. |
| PC-08 | 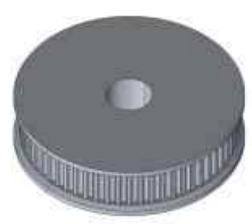 | Pulley | Moderate | Closely spaced teeth around the rim and flanges on both sides distinguish the part. The teeth alone could suggest a gear; the flanges, shaft bore, and overall pulley form are also needed. |
| PC-09 | 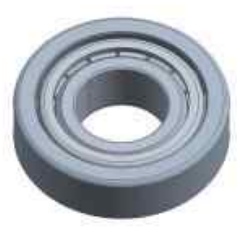 | Bearing | Complex | The inner and outer rings surround an array of rolling elements, some partly hidden by the rings. The external outline alone could suggest a plain ring or sleeve. |
| PC-10 | 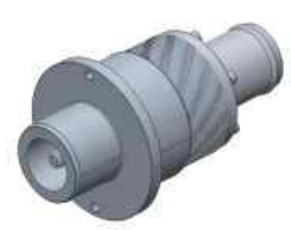 | Coupling | Complex | The coaxial cylindrical form resembles other shaft-related parts, but the internal connection structure is complex. |
| PC-11 | 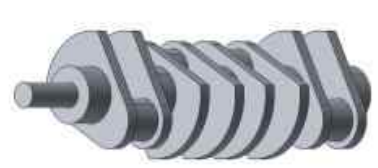 | Shaft | Complex | Several shaft sections are offset from one another, giving a shape unlike a conventional straight shaft. |
| PC-12 | 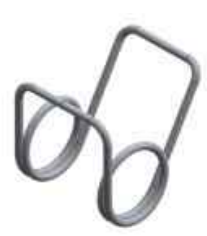 | Spring | Complex | Coils at both ends are linked by long bent arms. The part lacks the familiar close-wound helical shape. |

### 2.2.2 Test conditions

The prompts and model responses reproduced in this chapter are English translations of the Chinese records. Machine-readable class labels and case identifiers are retained; prose labels are formatted consistently for readability.

#### 2.2.2.1 Main experiments

The main experiments used the environment specified in Section 1.3 and the same prompt for every case. Following the conventional part classification formulation, the prompt supplied a predefined set of classes and asked GPT-6 to select the class of the input B-rep model. The candidate classes remained unchanged; only the input model varied. This allowed performance to be assessed across the three complexity levels. The prompt was: "Which of the following part classes does the given STEP model belong to: bearing bracket coupling flange gear nut pulley shaft screw_bolt spring?"

#### 2.2.2.2 Additional experiments

Additional experiments on representative complex cases examined the effects of prompt wording and access to web search. Two factors were varied. First, the standard classification prompt was replaced with an open-ended prompt that removed the candidate list and requested an explanation of the identification and typical functions or uses. Second, web search was disabled to test performance without that source of external information. The open-ended prompt was: "Identify the most likely type of part represented by this CAD model. Explain which geometric features support your judgment and what functions or uses this type of part typically has. If the part cannot be identified uniquely, give the most likely class and explain the uncertainty."

Combining these two factors gives the three additional test conditions in Table 2-4. Only the prompt and web search setting changed; all other settings matched the main experiments. Each test used a separate conversation to prevent information from one condition carrying over to another.

Table 2-4 Additional test conditions for part classification

| Condition | Prompt | Web search | Purpose |
|---|---|---|---|
| S1: Open-ended identification | P2: Open-ended identification | Enabled | Test identification without fixed labels and understanding of the part and its engineering function. |
| S2: Web search disabled | P1: Standard classification | Disabled | Test whether standard classification results depend on web search. |
| S3: Open-ended identification with web search disabled | P2: Open-ended identification | Disabled | Test identification and understanding of the part and its engineering function without fixed labels or web search. |

## 2.3 Test results

Results are reported separately for the main and additional experiments. The main experiments assess classification correctness with a fixed candidate list and compare performance across complexity levels. The additional experiments vary the prompt and web search setting on representative complex cases to examine whether GPT-6 can identify parts without candidate labels, support its judgments with structural evidence, and explain their engineering functions.

### 2.3.1 Main experiment results

The overall results are presented first, followed by each part model, GPT-6’s predicted class, and its explanation. The case analyses focus on the evidence used for classification.

#### 2.3.1.1 Overview

All cases were classified correctly in the main experiments (Table 2-5). Table 2-6 lists the completion time for each case.

Table 2-5 Main experiment classification results

| Case ID | Complexity | Reference class | Predicted class | Result |
|---|---|---|---|---|
| PC-01 | Simple | Bearing | Bearing | Correct |
| PC-02 | Simple | Screw/bolt | Screw/bolt | Correct |
| PC-03 | Simple | Bracket | Bracket | Correct |
| PC-04 | Simple | Nut | Nut | Correct |
| PC-05 | Moderate | Bracket | Bracket | Correct |
| PC-06 | Moderate | Flange | Flange | Correct |
| PC-07 | Moderate | Gear | Gear | Correct |
| PC-08 | Moderate | Pulley | Pulley | Correct |
| PC-09 | Complex | Bearing | Bearing | Correct |
| PC-10 | Complex | Coupling | Coupling | Correct |
| PC-11 | Complex | Shaft | Shaft | Correct |
| PC-12 | Complex | Spring | Spring | Correct |

Table 2-6 Completion times in the main experiments

| Case ID | PC-01 | PC-02 | PC-03 | PC-04 | PC-05 | PC-06 |
|---|---|---|---|---|---|---|
| Time | 3 min | 2 min | 2 min | 3 min | 2 min | 2 min |
| **Case ID** | **PC-07** | **PC-08** | **PC-09** | **PC-10** | **PC-11** | **PC-12** |
| Time | 2 min | 2 min | 2 min | 4 min | 2 min | 2 min |

#### 2.3.1.2 Case analyses

### PC-01 (bearing; simple)

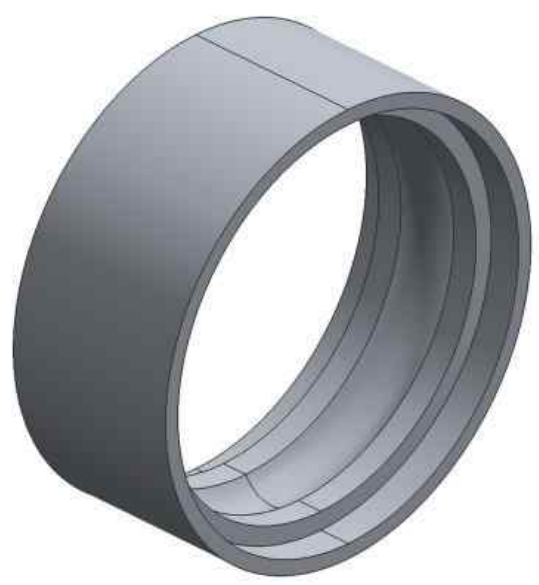

**GPT-6 response**

Completed easy_0: classified as "bearing", specifically a bearing outer ring, based on the recessed annular raceway on the inside and the steps at both ends.

**classification.json**

**case** easy_0

**category** bearing

**component** bearing outer ring

**confidence** high

**reason** The model is an annular solid with coaxial surfaces, an outside diameter of 6 mm, and an axial width of 2.5 mm. An annular recessed raceway with a circular-arc cross-section and steps at both ends is present on the inside, consistent with the geometry of a ball bearing outer ring.

**validation** STEP read successfully; valid model; 1 solid, 16 faces

GPT-6 correctly classified the model as a bearing and identified it more specifically as a bearing outer ring. Its explanation cited the annular groove on the inside and the steps at both ends, consistent with the model geometry.

### PC-02 (screw/bolt; simple)

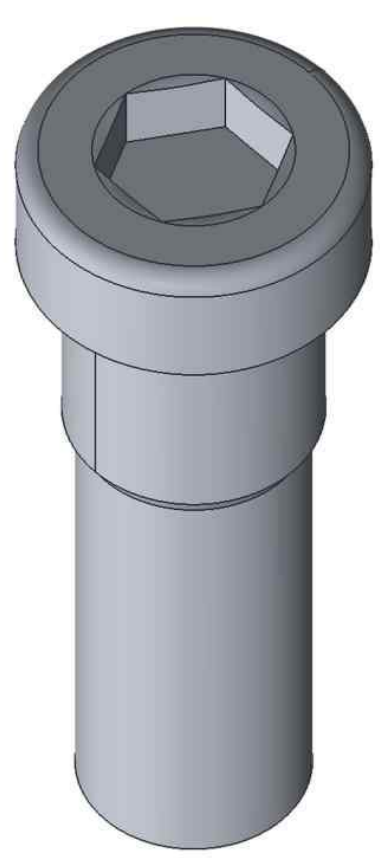

**GPT-6 response**

Completed easy_1: classified as "screw_bolt", based on the long cylindrical shank, enlarged head, and hexagonal drive socket.

**classification.json**

**case** easy_1

**category** screw_bolt

**component** socket head cap screw/bolt

**confidence** high

**reason** The model has a long cylindrical shank and an enlarged cylindrical head. Six internal planar faces at the end of the head form a hexagonal socket, characteristic of a headed fastener. The main shank diameter is 30 mm, the head diameter is 45 mm, and the total axial length is 117.5 mm. The helical thread profile is not explicitly modeled; classification is based on the overall form and hexagonal drive socket.

**validation** STEP read successfully; valid model; 1 solid, 22 faces

GPT-6 correctly classified the model as a screw/bolt, using the shank, head, and hexagonal socket as evidence. Although the thread was not explicitly modeled, these geometric features were sufficient for classification.

### PC-03 (bracket; simple)

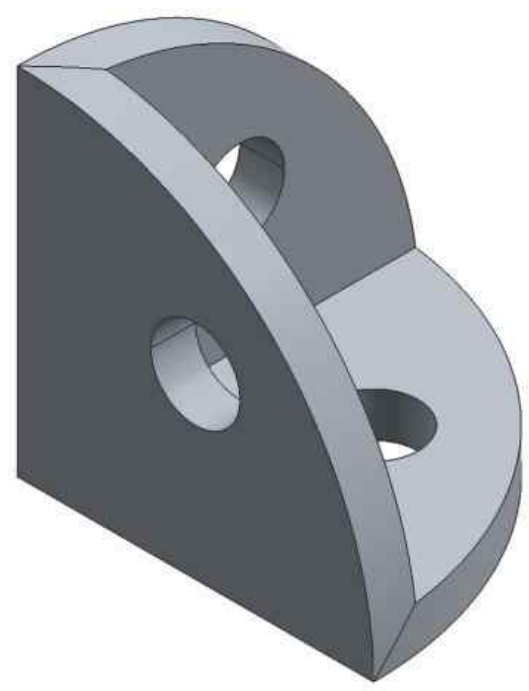

**GPT-6 response**

Completed easy_2: classified as "bracket", specifically a three-way right-angle bracket, based on the three orthogonal plates and their mounting holes.

**classification.json**

**case** easy_2

**category** bracket

**component** three-way right-angle bracket

**confidence** high

**reason** The model consists of three mutually perpendicular plates, each 5 mm thick and each containing a through mounting hole 10 mm in diameter. The three hole axes are mutually orthogonal, and the overall bounding dimensions are 40 × 40 × 40 mm. This three-way mounting arrangement is characteristic of a corner bracket.

**validation** STEP read successfully; valid model; 1 solid, 12 faces

GPT-6 correctly classified the model as a bracket. It identified the three perpendicular plates and their mounting holes, using the clearly defined overall structure to reach its judgment.

### PC-04 (nut; simple)

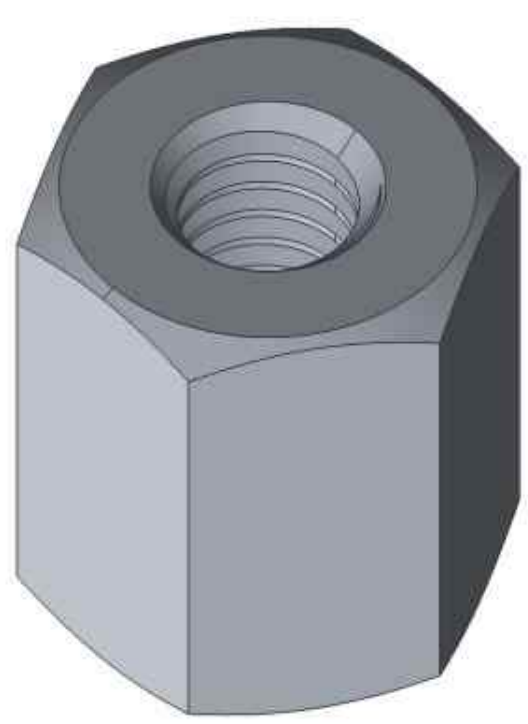

**GPT-6 response**

Completed easy_3: classified as "nut", based on the hexagonal prism shape and threaded through hole.

**classification.json**

**case** easy_3

**category** nut

**component** internally threaded hex standoff (elongated nut form)

**confidence** high

**reason** The model is a hexagonal prism measuring 5 mm across flats and 6 mm along its axis. It has a central threaded through hole and chamfers at both ends. The hexagonal wrenching surfaces and internal threaded connection are characteristic of a nut; among the supplied classes, it belongs to the "nut" class.

**validation** STEP read successfully; valid model; 1 solid, 44 faces

GPT-6 correctly classified the model as a nut, citing the hexagonal exterior and central threaded hole. It also gave a more specific name: an internally threaded hex standoff.

### PC-05 (bracket; moderate)

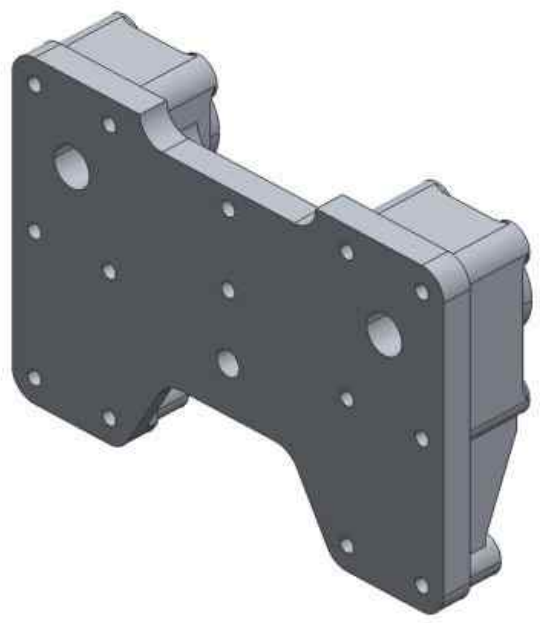

**GPT-6 response**

Completed medium_0: classified as "bracket", based on the reinforced plate body, which is not rotationally symmetric, and the multiple mounting holes and bosses.

**classification.json**

**case** medium_0

**category** bracket

**component** load-bearing mounting bracket with multiple holes

**confidence** high

**reason** The model has a plate body and reinforcing structures, rather than a rotationally symmetric form, with multiple mounting bosses and projecting connection features. Two parallel main holes, each 16 mm in diameter, are spaced 142 mm apart. A further 14 holes, each 6 mm in diameter, are distributed around the perimeter and center, with recessed openings on some holes. The overall geometry is consistent with a bracket for supporting loads, locating parts, and making mounting connections.

**validation** STEP read successfully; valid model; 1 solid, 179 faces

GPT-6 correctly classified the model as a bracket. The many holes did not distract it from the overall structure: it considered the plate body, reinforcement, and multiple mounting bosses together.

### PC-06 (flange; moderate)

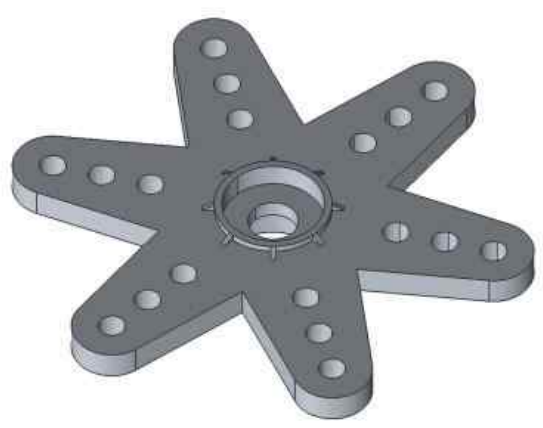

**GPT-6 response**

Completed medium_1: classified as "flange". The specific component is a six-arm servo horn. The central hub and radial arrangement of mounting holes support its assignment to the "flange" class among the supplied options.

**classification.json**

**case** medium_1

**category** flange

**component** six-arm servo horn (shaft-end connection plate with an irregular outline)

**confidence** medium_high

**reason** A short cylindrical hub with a stepped bore is surrounded by six thin radial arms at equal angular intervals. Each arm has three axial mounting holes, giving 18 holes in total. The arms are 1.8 mm thick and the mounting holes are 1.6 mm in diameter. The central hub connects to a shaft end, while the radial holes connect to external components. The overall structure most closely resembles a connecting flange with an irregular outline.

**validation** STEP read successfully; valid model; 1 solid, 72 faces

GPT-6 correctly classified the model as a flange, considering both the central connection and the mounting holes arranged around it. It also identified the actual component as a six-arm servo horn.

### PC-07 (gear; moderate)

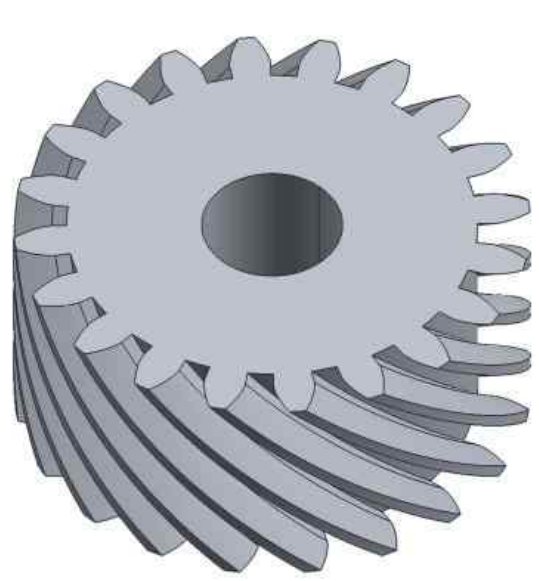

**GPT-6 response**

Completed medium_2: classified as "gear", specifically a 20-tooth helical gear.

**classification.json**

**case** medium_2

**category** gear

**component** 20-tooth helical gear

**confidence** high

**reason** The model has 20 equally spaced tooth profiles around its circumference. The tooth surfaces follow a clear helix along the axial direction, forming a helical gear. The root circle radius is 17.5 mm, the outside diameter is approximately 44 mm, the axial width is 20 mm, and the central through bore is 12 mm in diameter. The geometry, topology, and model views all support the gear classification.

**validation** STEP read successfully; valid model; 1 solid, 143 faces

GPT-6 correctly classified the model as a gear. It recognized the continuous array of teeth around the circumference and identified their helical form.

**PC-08 (pulley; moderate)**

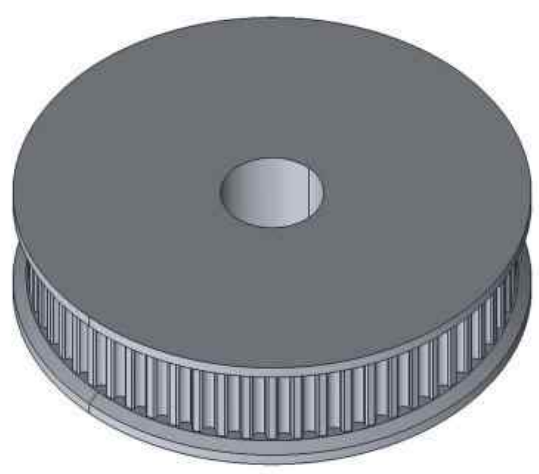

**GPT-6 response**

Completed medium_3: classified as "pulley", specifically a 60-tooth timing pulley, based on the fine teeth on the belt-contact surface and the flanges on both sides.

**classification.json**

**case** medium_3

**category** pulley

**component** 60-tooth timing pulley

**confidence** high

**reason** The model has 60 equally spaced fine tooth spaces around its circumference. The toothed belt-contact surface is 7 mm wide, with flanges 40 mm in diameter on both sides. The central through bore is 8 mm in diameter and the hub is 20 mm in diameter. Two radial set-screw holes, each 3 mm in diameter, are perpendicular to one another. The fine teeth and flanges on both sides are characteristic of a timing pulley.

**validation** STEP read successfully; valid model; 1 solid, 326 faces

GPT-6 correctly classified the model as a pulley and identified it more specifically as a timing pulley. Although the teeth around the rim could suggest a gear, the fine teeth and flanges on both sides supported the more specific identification.

**PC-09 (bearing; complex)**

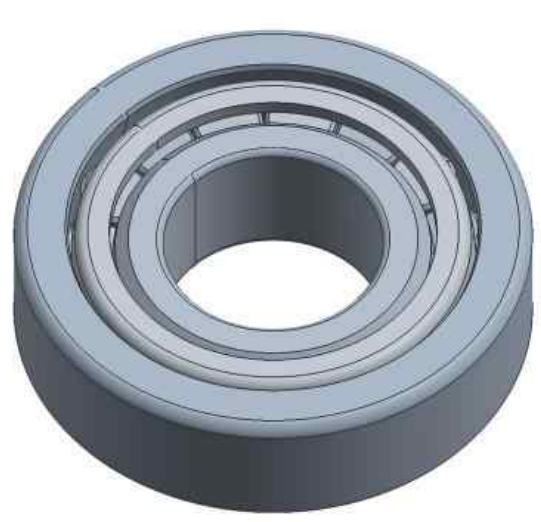

**GPT-6 response**

Completed hard_0: classified as "bearing", based on the inner and outer rings, cage, and 15 evenly spaced sets of rollers.

**classification.json**

**case** hard_0

**category** bearing

**component** roller bearing assembly

**confidence** high

**reason** The model consists of 33 solids: coaxial inner and outer rings, an annular cage with windows around its circumference, and 15 evenly spaced sets of rollers and pins. The inner ring bore is 17 mm in diameter and the outer ring has an outside diameter of 40 mm. The rolling elements sit between the rings and are positioned by the cage, clearly matching the structure of a rolling bearing.

**validation** STEP read successfully; valid model; 33 solids, 298 faces

GPT-6 correctly classified the model as a bearing. The file records show multiple components, including inner and outer rings, a cage, and rollers. GPT-6 used these components in its explanation, although the rendered appearance alone also makes the bearing class fairly apparent.

### PC-10 (coupling; complex)

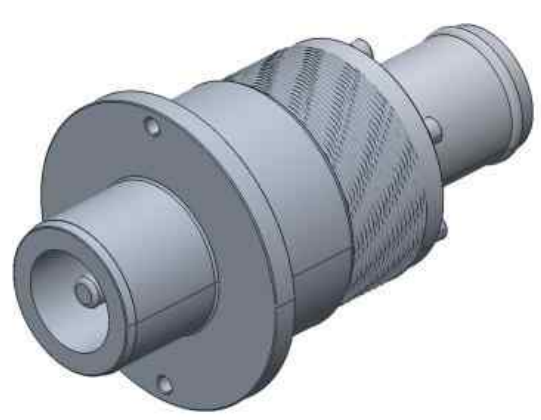

**GPT-6 response**

Completed hard_1: classified as "coupling" (connector), specifically the male end of a quick-disconnect fluid coupling with a spring-loaded valve element.

**classification.json**

**case** hard_1

**category** coupling (connector)

**component** male quick-disconnect fluid coupling with a spring-loaded valve element

**confidence** high

**reason** The model is a coaxial coupling assembly made up of 11 solids. It includes a cylindrical body, a connection end with an axial flow passage, six evenly spaced fasteners around the circumference, and an internal valve element, guide, and spring. The stepped insertion end, internal passage, and valve mechanism together form a quick-disconnect fluid coupling, so the complete assembly belongs to the "coupling" class. The original model name, Quick coupler (male), also agrees with the geometric identification.

**validation** STEP read successfully; valid model; 11 solids, 266 faces

GPT-6 assigned the model to the "coupling" class and used the Chinese equivalent of connector; this wording did not affect the correctness of the classification. It also identified the specific part as a quick-disconnect coupling for a fluid line, rather than a shaft coupling that joins two rotating shafts. Its explanation considered the insertion end, flow passage, and valve mechanism together. The response also explicitly cited the embedded model name, “Quick coupler (male),” so this case does not isolate identification from geometry alone.

### PC-11 (shaft; complex)

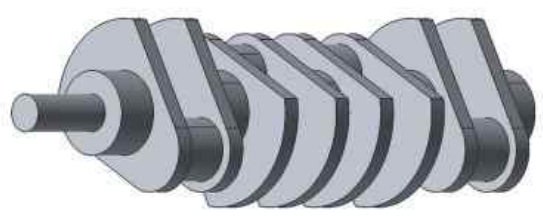

**GPT-6 response**

Completed hard_2: classified as "shaft", specifically a four-throw crankshaft, based on the coaxial main journals, eccentric crankpin journals, and crank structure.

**classification.json**

**case** hard_2

**category** shaft

**component** four-throw crankshaft

**confidence** high

**reason** The model has 5 coaxial main journals with a radius of 36 mm and 4 eccentric crankpin journals, each with a radius of 27 mm, an offset of 44 mm, and a length of 46 mm. Crank webs and counterweights connect both sides of each eccentric journal to the main shaft; the model also has a slender shaft end. This arrangement of multiple eccentric throws clearly identifies a crankshaft, which falls within the "shaft" class.

**validation** STEP read successfully; valid model; 1 solid, 61 faces

GPT-6 correctly classified the model as a shaft and identified it more specifically as a four-throw crankshaft. Despite the offsets between shaft sections, it recognized how they connect, using the eccentric journals and crank structure as evidence.

### PC-12 (spring; complex)

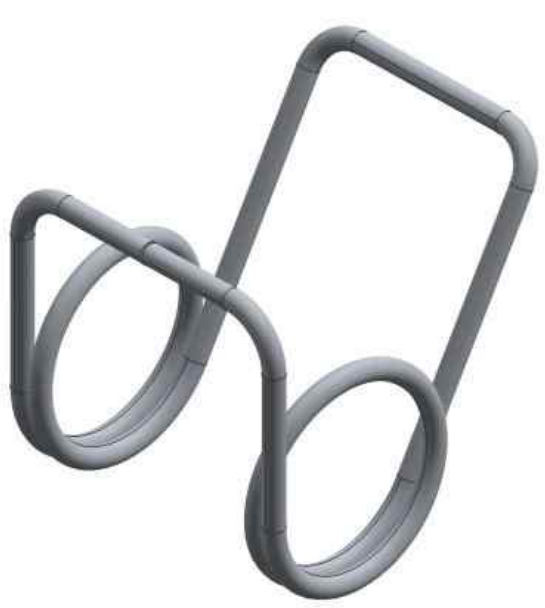

**GPT-6 response**

Completed hard_3: classified as "spring", specifically a double torsion spring.

**classification.json**

**case** hard_3

**category** spring

**component** double torsion spring

**confidence** high

**reason** The model is formed from continuous round wire 2 mm in diameter. Coiled sections and long arms on the left and right are joined by a bent bridging segment, and the ends have circular cross-sections. The symmetric pair of coils and arms clearly identifies a double torsion spring. The geometric analysis agrees with the model views.

**validation** STEP read successfully; valid model; 1 solid, 15 faces

GPT-6 correctly classified the model as a spring and identified it as a double torsion spring. Although its shape differs from a conventional helical spring, the coils on both sides and the projecting arms supported the correct judgment.

### 2.3.2 Additional experiment results

The following results are organized by conditions S1, S2, and S3, defined in Section 2.2.2. For each condition, representative complex cases were selected to illustrate the effect of the change, using the output for the same case in the main experiment as the baseline. Table 2-7 lists the completion times.

Table 2-7 Completion times in the additional experiments

| Case ID | S1_PC-09 | S1_PC-10 | S2_PC-10 | S2_PC-12 | S3_PC-09 | S3_PC-10 |
|---|---|---|---|---|---|---|
| Completion time | 11 min | 17 min | 1 min | 3 min | 7 min | 7 min |

#### 2.3.2.1 S1: Open-ended identification

This condition retained access to web search but removed the candidate list and requested an explanation of the geometry, uses, and uncertainty. PC-09 was selected to examine how GPT-6 checks the geometry when identifying a specific bearing type; PC-10 was selected to examine whether it can identify a specific product beyond the fixed labels.

**PC-09 (bearing)**

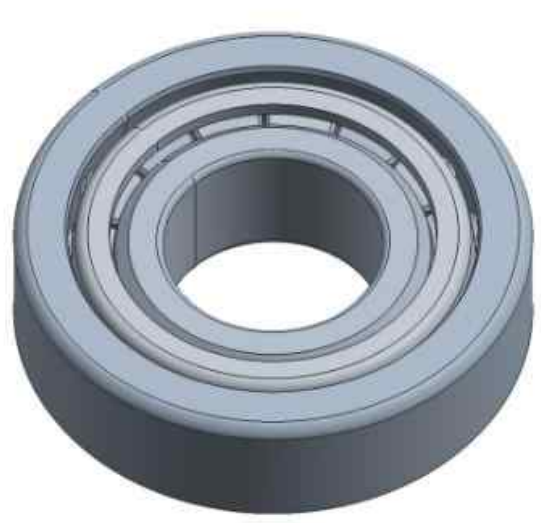

**Main experiment output**

Classification: bearing; specifically, a roller bearing assembly.

Evidence: The model consists of 33 solids: coaxial inner and outer rings, an annular cage with windows around its circumference, and 15 evenly spaced sets of rollers and pins. The inner ring bore is 17 mm in diameter and the outer ring has an outside diameter of 40 mm. The rolling elements sit between the rings and are positioned by the cage, clearly matching the structure of a rolling bearing.

**S1 output**

Identification: A roller bearing assembly, most likely a simplified or nonstandard model of a single-row tapered roller bearing.

Evidence: The inner and outer rings have conical raceways, and the 15 main rollers form a single inclined row. However, the main working surfaces of the rollers have no taper: all are cylindrical surfaces with a radius of 2.75 mm.

Uses and qualifications: Roller bearings support rotating shafts and transmit support loads through rolling contact. Confidence is high for roller bearing assembly, but medium for the more specific identification as a standard single-row tapered roller bearing.

The main experiment returned the label "bearing" and identified the model as a roller bearing assembly. S1 retained that general identification but proposed a more specific interpretation: a simplified or nonstandard model of a single-row tapered roller bearing. It also identified geometric evidence against a standard tapered roller bearing: the rollers have cylindrical working surfaces. It therefore assigned only medium confidence to that more specific identification. S1 added detail about the bearing type, checked the relevant geometry, and explained the uncertainty.

**PC-10 (coupling)**

<table>
<tr>
<td>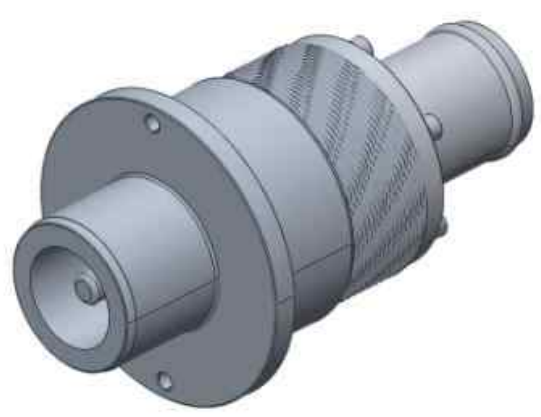</td>
<td>
Main experiment output<br>
Classification: coupling (connector); specifically, a male quick-disconnect fluid coupling with a spring-loaded valve element.<br>
Evidence: The model is a coaxial coupling assembly made up of 11 solids. It includes a cylindrical body, a connection end with an axial flow passage, six evenly spaced fasteners around the circumference, and an internal valve element, guide, and spring. The stepped insertion end, internal passage, and valve mechanism together form a quick-disconnect fluid coupling, so the complete assembly belongs to the "coupling" class.<br>
S1 output<br>
Identification: A water-line connector assembly with a spring-loaded self-closing valve; more specifically, a male quick-disconnect coupling for water lines.<br>
Evidence: The coaxial housing and flow passage are represented by 11 valid solids, including a valve stem, conical valve disc, guide, and return spring.<br>
Uses and qualifications: Typically used in water or cooling-water lines that need to be disconnected. Without the mating connector, the precise locking mechanism, opening mechanism, and interface standard cannot be determined. Sealing performance cannot be confirmed from the static shape alone.
</td>
</tr>
</table>

The main experiment selected "coupling" from the candidate list and had already identified the part as the male end of a quick-disconnect fluid coupling with a spring-loaded valve element. S1 likewise identified the complete assembly as a male quick-disconnect coupling for water lines. It added the likely water-line application and explained that the locking and opening mechanisms could not be confirmed without the mating connector.

#### 2.3.2.2 S2: Web search disabled

This condition retained the standard classification prompt and candidate list from the main experiments, disabling only web search. PC-10 tested classification of an assembly without web search; PC-12 tested whether GPT-6 could still recognize a spring with an atypical shape.

### PC-10 (coupling)

| | |
|---|---|
| 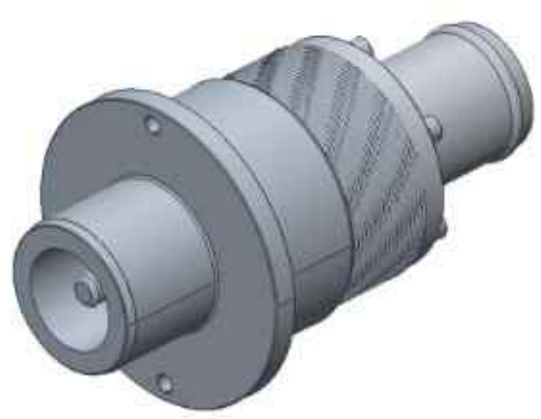 | **Main experiment output**<br>Classification: coupling (connector); specifically, a male quick-disconnect fluid coupling with a spring-loaded valve element.<br>Evidence: The model is a coaxial coupling assembly made up of 11 solids. It includes a cylindrical body, a connection end with an axial flow passage, six evenly spaced fasteners around the circumference, and an internal valve element, guide, and spring. The stepped insertion end, internal passage, and valve mechanism together form a quick-disconnect fluid coupling, so the complete assembly belongs to the "coupling" class.<br>**S2 output**<br>Classification: coupling; the Chinese name given means connector (quick-disconnect coupling).<br>Evidence: The model consists of a stepped cylindrical body, internal components fitted along the same axis, and fasteners distributed around the circumference, consistent with a coupling assembly. |

Both the main experiment and S2 assigned the model to the "coupling" class and identified it as a quick-disconnect coupling for a fluid line. PC-10 was therefore classified correctly in this run without web search. The main experiment nevertheless gave a more detailed explanation of the evidence.

### PC-12 (spring)

| | |
|---|---|
| 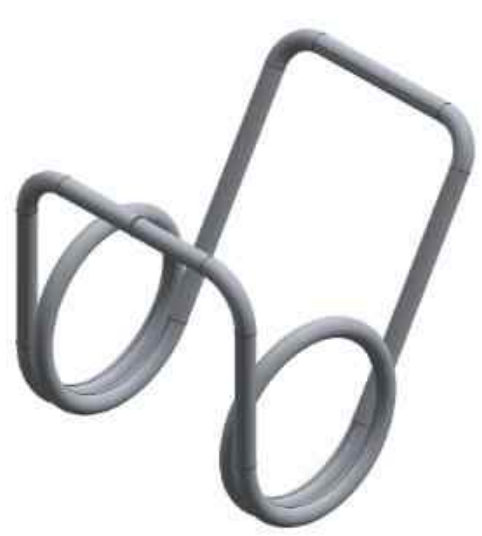 | **Main experiment output**<br>Classification: spring; specifically, a double torsion spring.<br>Evidence: The model is formed from continuous round wire 2 mm in diameter. Coiled sections and long arms on the left and right are joined by a bent bridging segment, and the ends have circular cross-sections. The symmetric pair of coils and arms clearly identifies a double torsion spring. The geometric analysis agrees with the model views.<br>**S2 output**<br>Classification: spring; the Chinese name given means spring (double torsion spring).<br>Evidence: The model is formed from continuous wire with a circular cross-section, with symmetric torsion coils on both sides and projecting arms. This is the characteristic structure of a double torsion spring, placing it in the "spring" class. |

The main experiment returned the label "spring" and identified a double torsion spring. S2 gave the same class and specific type, citing the continuous wire with a circular cross-section, the two torsion coils, and the projecting arms. PC-12 was therefore classified correctly in this run without web search. The explanation without web search was simpler, however; for example, it omitted dimensions.

#### 2.3.2.3 S3: Open-ended identification with web search disabled

This condition removed both the fixed candidate labels and access to web search. PC-09 tested the identification of a specific bearing type and the qualifications attached to it; PC-10 tested how well GPT-6 could identify the assembly and its engineering function under these conditions.

**PC-09 (bearing)**

| | |
|---|---|
| 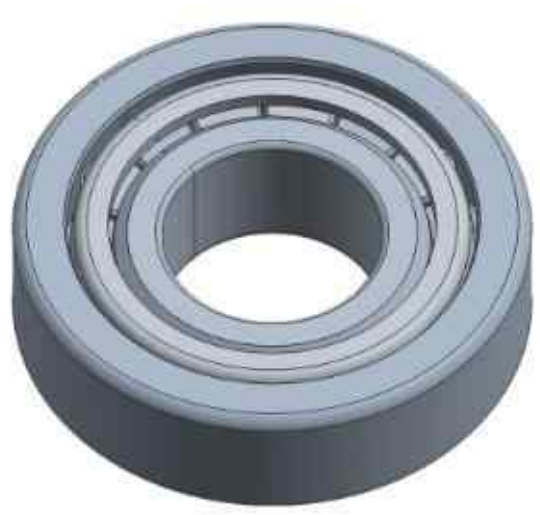 | **Main experiment output**<br>Classification: bearing; specifically, a roller bearing assembly.<br>Evidence: The model consists of 33 solids: coaxial inner and outer rings, an annular cage with windows around its circumference, and 15 evenly spaced sets of rollers and pins. The inner ring bore is 17 mm in diameter and the outer ring has an outside diameter of 40 mm. The rolling elements sit between the rings and are positioned by the cage, clearly matching the structure of a rolling bearing.<br>**S3 output**<br>Identification: Most likely a single-row tapered roller bearing assembly. The identification that can be made with relatively high confidence is a single-row roller bearing with a cage.<br>Evidence: The inner ring, outer ring, cage, and one row of rollers form a bearing assembly. The file contains 33 valid solids, including 15 roller bodies. The roller bodies have cylindrical surfaces of constant diameter, rather than conical surfaces.<br>Uses and qualifications: Supports a rotating shaft and transmits loads. If interpreted as the most likely type, a tapered roller bearing, it can carry radial loads and axial loads in one direction. Confidence is high for roller bearing assembly, but medium for the strict identification as a standard single-row tapered roller bearing. |

In the main experiment, GPT-6 selected "bearing" and named the part as a roller bearing assembly; the predefined labels may have influenced how specific its answer was. Without web search, it still gave a more detailed identification: it first established that the model was a bearing assembly, then proposed a single-row tapered roller bearing, and finally limited its confidence to medium because the main roller surfaces were cylindrical.

**PC-10 (coupling)**

| | |
|---|---|
| 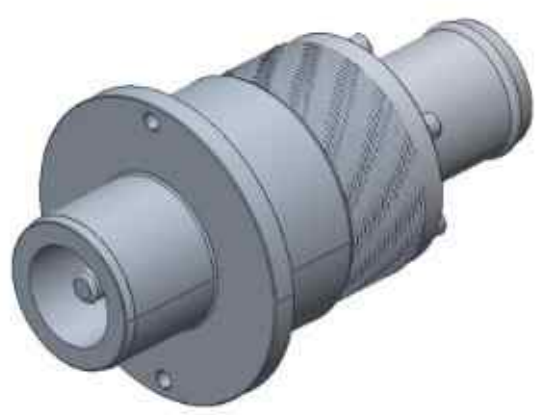 | **Main experiment output**<br>Classification: coupling (connector); specifically, a male quick-disconnect fluid coupling with a spring-loaded valve element.<br>Evidence: The model is a coaxial coupling assembly made up of 11 solids. It includes a cylindrical body, a connection end with an axial flow passage, six evenly spaced fasteners around the circumference, and an internal valve element, guide, and spring. The stepped insertion end, internal passage, and valve mechanism together form a quick-disconnect fluid coupling, so the complete assembly belongs to the "coupling" class.<br>**S3 output**<br>Identification: Most likely a spring-loaded lift check valve (one-way valve) assembly with a conical disc that moves axially on a guided valve stem.<br>Evidence: The model contains 11 solids. Two coaxial ports, a conical valve disc, a guide support, and a return spring support the interpretation as a spring-loaded valve mechanism.<br>Uses and qualifications: This type of check valve allows flow in one direction and closes automatically when the pressure difference is insufficient or flow tends to reverse. It could also be the check valve mechanism within a self-sealing fluid coupling. Without the mating connector and assembly context, the complete product type cannot be determined uniquely. |

The main experiment correctly selected "coupling" and identified a quick-disconnect coupling for a fluid line. S3 instead called it a lift check valve assembly with axial valve motion. With neither candidate labels nor web search, GPT-6 still recognized the internal spring, guided valve stem, and valve disc, but used the internal valve mechanism as its final identification. It did not correctly identify the complete part or infer its overall purpose.

## 2.4 Analysis of GPT-6’s approach

The recorded outputs and tool operations suggest the following workflow for part classification:

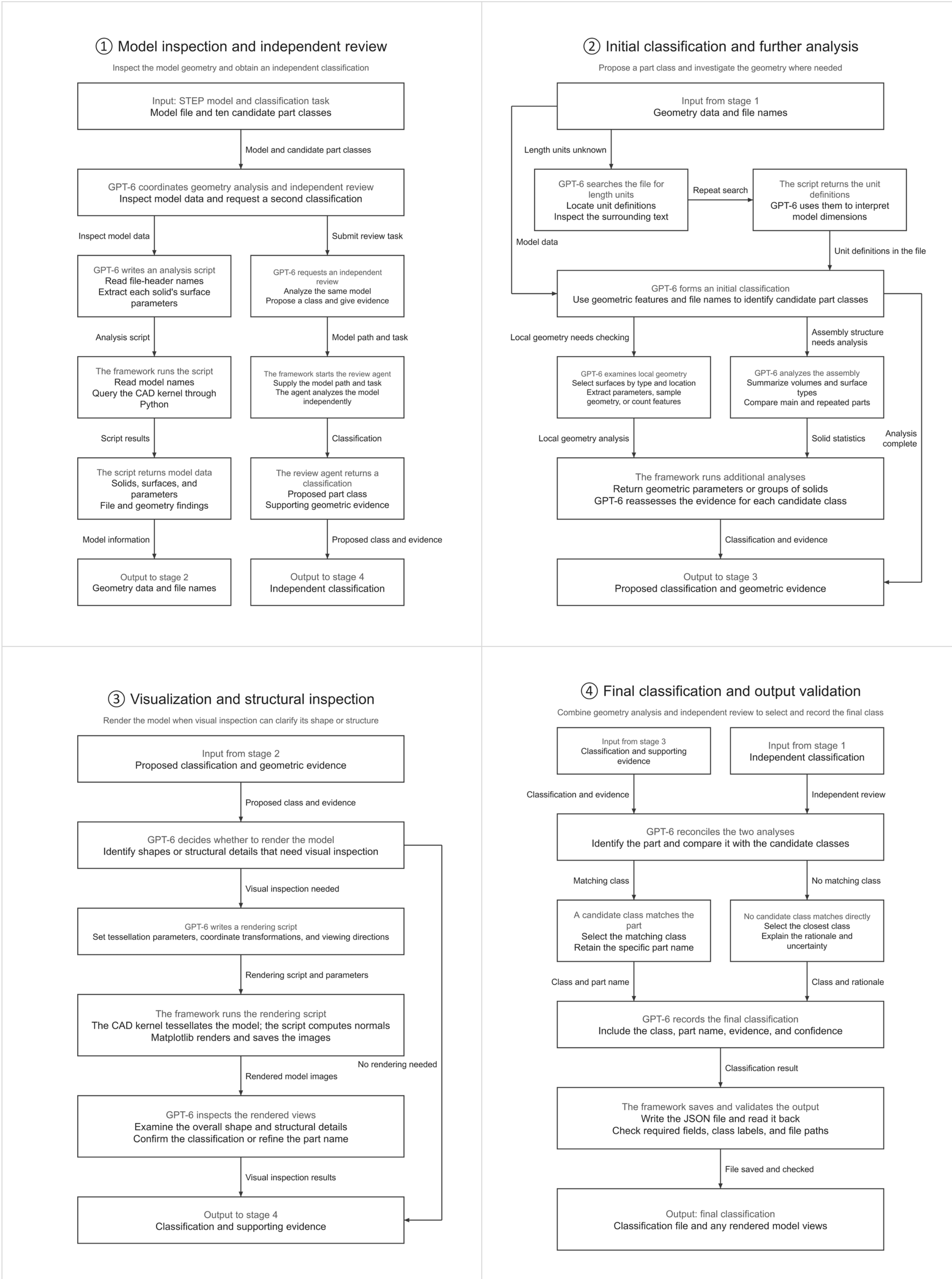

Figure 2-1  Inferred workflow for part classification

## 2.5 Capabilities and limitations

For part classification, GPT-6 can query the CAD kernel and inspect model views as needed, combining evidence from the overall shape, local geometric features, and assembly components. It classified every case correctly in the main experiments with predefined classes. Its ability is not limited to parts with typical shapes: when a part has an unusual form or resembles another class, GPT-6 can consider several structural features together and provide a more specific name and explanation of its function. The two additional tests that retained the candidate classes also returned correct classifications with web search disabled.

The main difficulty is determining exactly which type of part a model represents. When an assembly lacks information about mating components or its installation, GPT-6 sometimes assigns the whole part a class based on the structure and function of just one component. When the overall shape and local geometric features do not fully support the same interpretation, it often proposes a specific type without being able to confirm that identification reliably.

Overall, GPT-6 has been successful at identifying the general class of a part, but is less reliable when asked to determine its specific type. The remaining challenge is to connect the geometric evidence accurately to what the part does and which type it belongs to.

These results apply to a small, selected set of parts. The 100% result measures classification within the supplied candidate list, not unrestricted identification. In PC-10, the output also cites an embedded STEP model name. The contribution of such metadata and the variability between repeated runs were not isolated in these comparisons.

# Chapter 3 Feature recognition

## 3.1 Task objectives and evaluation criteria

This task tests whether GPT-6 can identify machining features from the geometry and topology of a B-rep model, moving from a general feature type to a more specific form without a predefined list of feature names. It also examines whether GPT-6 can interpret complex interactions between features, including intersections, nesting, interruptions, and shared boundaries, and associate each recognized feature with the correct geometric region. Table 3-1 defines the task.

Table 3-1 Feature recognition task

| Test element | Description |
|---|---|
| Evaluation question | Can GPT-6 accurately recognize machining features using only the geometry and topology of a B-rep model, identify both their general types and more specific forms without a fixed label list, and correctly interpret intersections, nesting, and other relationships between complex features? |
| Input | A B-rep model containing one or more machining features. |
| Output | The recognized features, including their general types, more specific forms, main supporting evidence, relationships to other features, and any necessary statements of uncertainty. |
| Evaluation criteria | Correctness of the general and specific classifications, and handling of uncertainty. For complex cases, particular attention is given to missed or misidentified regions where features intersect or are nested. |

## 3.2 Test design

### 3.2.1 Test case selection

The test cases use B-rep models with explicitly defined geometry and machining features that can be verified manually. They cover different feature types and counts, including intersecting and nested features.

Complexity is determined primarily by the number of features, the completeness of their boundaries, the difficulty of identifying their specific forms, and interactions such as intersections and nesting. It is not determined by the number of faces or features alone. Table 3-2 sets out the criteria. Two cases were selected at each of the simple, moderate, and complex levels; Table 3-3 lists the cases.

Table 3-2  Complexity criteria for feature recognition

| Complexity | Selection criteria | Typical characteristics |
|---|---|---|
| Simple | A small number of basic machining features with largely complete, separate boundaries. Their general and specific types can be determined directly from local geometry. | One or a few independent features; no substantial interactions; complete feature boundaries. |
| Moderate | Multiple machining features that require several geometric regions to be considered together. Some features interact, or their specific forms can be established only from the surrounding structure. | Several feature types coexist; limited intersections, local interruptions, or nesting; some boundaries affected by other features. |
| Complex | Multiple levels of nesting, extensive intersections, or compound features. Other features interrupt, obscure, or share some original boundaries, making detection, separation into instances, and detailed classification difficult. | Several pairwise intersections between features, or multiple levels of nesting. |

Table 3-3  Feature recognition test cases

| Case ID | Rendered CAD model | Complexity | Basis for complexity rating |
|---|---|---|---|
| FR-01 | 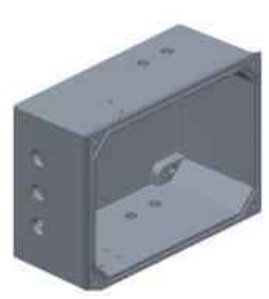 | Simple | Few features. Holes and recesses are relatively independent and have clear boundaries. |
| FR-02 | 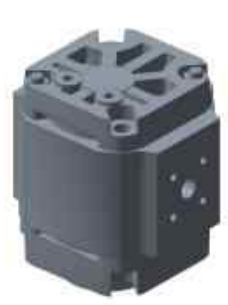 | Simple | More holes and slots, but clear outlines and largely independent features allow them to be recognized individually. |
| FR-03 | 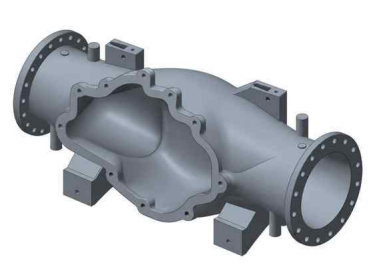 | Moderate | Side ports meet the main cavity, leaving incomplete boundaries at the junctions. The geometry must be considered from several directions. |
| FR-04 | 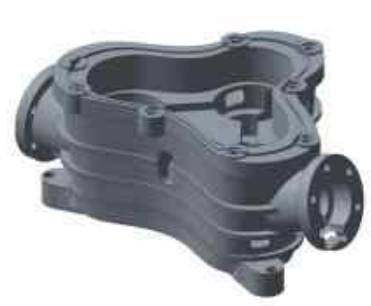 | Moderate | The internal cavity connects to side openings, interrupting some boundaries and creating numerous feature intersections. |

| Case ID | Rendered CAD model | Complexity | Basis for complexity rating |
|---|---|---|---|
| FR-05 | 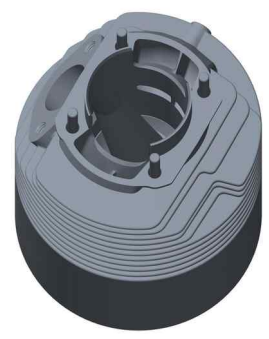 | Complex | Several annular structures are closely stacked. Internal features are occluded, making the boundaries of each layer difficult to distinguish. |
| FR-06 | 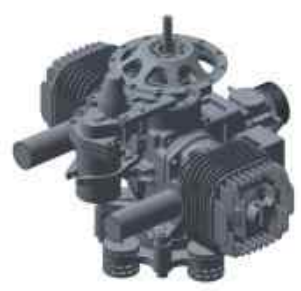 | Complex | Multiple components occlude one another, with densely arranged holes and slots, many feature intersections, and multiple levels of nesting. Feature boundaries and the assignment of regions to features are difficult to establish. |

### 3.2.2 Experimental conditions

#### 3.2.2.1 Main experiment

*The prompt reproduced below is an English translation of the Chinese experimental record.*

The main experiment uses the same prompt for every case. It provides no predefined list of feature types and asks GPT-6 to identify machining features in the B-rep model, give a general type and a more specific form for each feature, and explain the geometric evidence and relationships between features. Only the input model changes, allowing performance to be compared across complexity levels. The prompt reads: "Identify the machining features in the supplied B-rep model. For each individual feature instance: 1. First give its general feature type (for example, a hole; this is only an example), then identify its more specific form where possible (for example, a through hole; this is only an example). 2. Locate the feature in the model and explain the main geometric or topological evidence supporting your judgment. 3. Report any intersections, nesting, containment, interruptions, repetition, or other apparent relationships between features. Examine the entire part systematically; do not identify only the most obvious features. If a feature cannot be classified more specifically with confidence, retain the level of classification that can be established and explain the uncertainty."

#### 3.2.2.2 Additional experiment

An additional experiment uses a representative complex case to examine the effect of access to web search. It focuses on feature detection, the level of detail in classification, and the recognition of relationships between features when that source of external information is unavailable.

Both the main and additional experiments use separate conversations. After feature recognition is complete, the same follow-up visualization prompt is submitted to produce HTML inspection material that highlights each feature individually. This step does not modify or replace the original recognition results.

## 3.3 Results

The main and additional experiments are presented and evaluated separately. The main experiment examines whether GPT-6 can find machining features, classify them at different levels of detail, locate them geometrically, and identify their interactions. The additional experiment examines whether disabling web search produces clear changes in the workflow or results.

### 3.3.1 Main experiment results

The results are first summarized across complexity levels. Each case is then illustrated with its CAD model, representative machining features recognized by GPT-6, and the corresponding B-rep geometry. The analysis examines whether features are found and separated into appropriate instances, whether their general and specific types are identified correctly, and whether their relationships are recognized. Particular attention is given to intersections, nesting, and interrupted boundaries.

#### 3.3.1.1 Overview

Table 3-4 summarizes the main experiment along three dimensions: finding features and distinguishing individual instances; identifying their general and specific types; and recognizing relationships between them. Table 3-5 reports completion times.

Table 3-4 Feature recognition results in the main experiment

| Case ID | Complexity | Finding features and distinguishing instances | Identifying general and specific feature types | Recognizing relationships between features |
|---|---|---|---|---|
| FR-01 | Simple | Correct | Correct | Correct |
| FR-02 | Simple | Correct | Correct | Correct |
| FR-03 | Moderate | Mostly correct; some face groups that are not features were counted as features | Correct | Correct |
| FR-04 | Moderate | Correct | Correct | Correct |
| FR-05 | Complex | Mostly correct; some face groups that are not features were counted as features | Correct | Correct |
| FR-06 | Complex | Mostly correct, with a few recognition errors | Mostly correct, with a few errors in feature type | Correct |

Table 3-5 Completion times for feature recognition in the main experiment

| Case ID | FR-01 | FR-02 | FR-03 | FR-04 | FR-05 | FR-06 |
|---|---|---|---|---|---|---|
| Completion time | 24 min | 37 min | 38 min | 43 min | 49 min | 2 h 1 min |

#### 3.3.1.2 Case analyses

**FR-01 (simple)**

Figure 3-1 shows representative results for FR-01. In Figure 3-1(b), the deep blind cavity, closed groove around its opening, and flared rim are identified as separate instances. A band of solid material separates the groove from the cavity, so the two are adjacent rather than directly intersecting. In Figure 3-1(c), GPT-6 distinguishes the corner boss, stepped through hole, and blind hole with a conical bottom. The larger- and smaller-diameter sections of the stepped hole are grouped into one compound hole instance, while the interrupted portions of a local fillet remain part of the same transition feature. In Figure 3-1(d), the upper, middle, and lower through holes in the side wall form separate groups and connect to the main internal cavity.

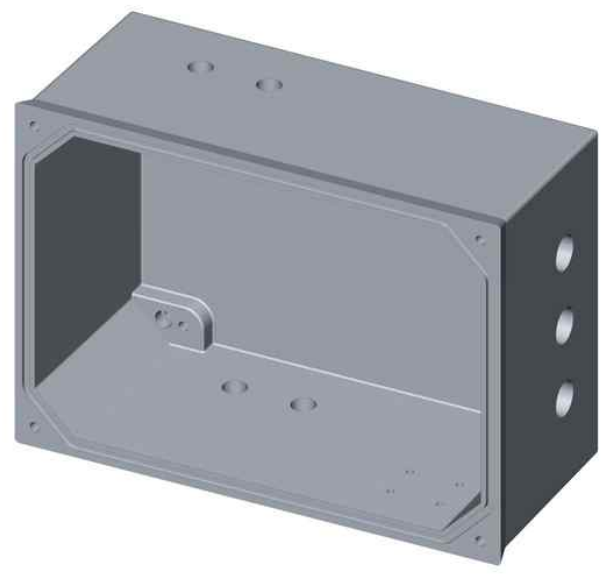

**(a) Input CAD model**

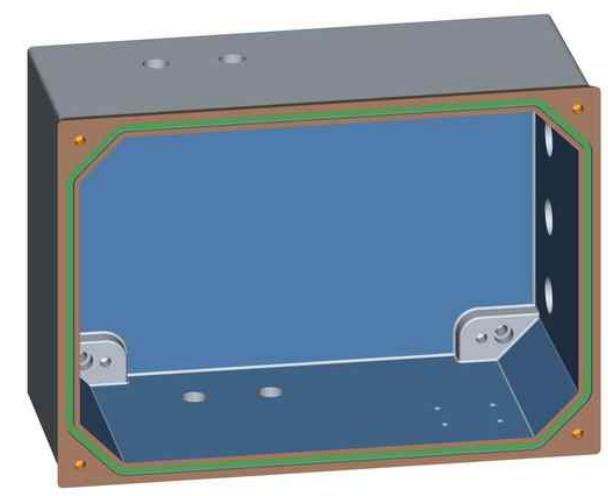

**(b) Cavity opening and surrounding groove**

Blue: deep blind cavity; green: closed groove around the opening; brown: flared rim; orange: corner blind hole; cyan: rim fillet; purple: groove-bottom fillet

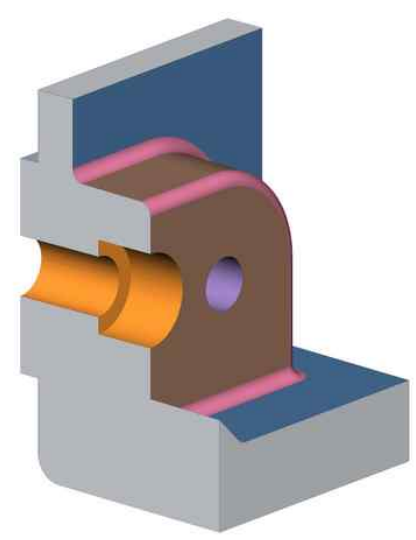

**(c) Compound hole at a corner (sectioned solid)**

Blue: deep blind cavity; brown: boss; light blue: shallow pad on the outer bottom; orange: stepped through hole; purple: blind hole with a conical bottom; magenta: platform fillet; cyan: bottom-wall fillet

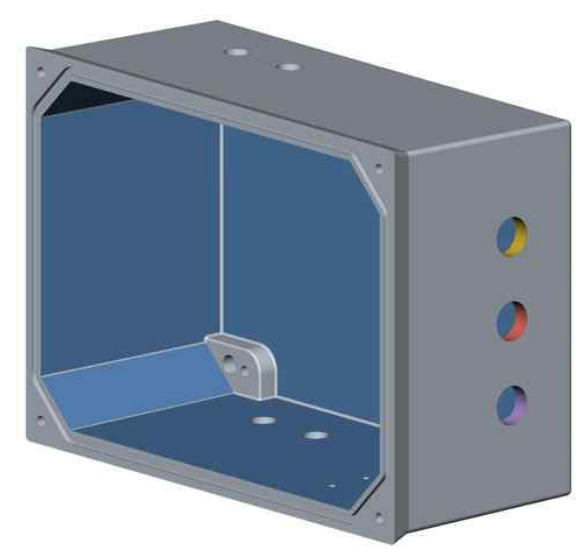

**(d) Side-wall through holes and main cavity**

Blue: deep blind cavity; purple/red/gold: upper/middle/lower through holes

Figure 3-1 Representative feature recognition results for FR-01

### FR-02 (simple)

Figure 3-2 shows representative results for FR-02. In Figure 3-2(b), six blue pockets with flat bottoms and irregular outlines surround the brown central boss. Although they share the side walls of the boss, each has its own bottom face, so they are identified as six instances. The narrow green slot with rounded ends, orange through hole in the boss, and purple recesses open at the corners are also distinguished. In Figure 3-2(c), the blue and green blind holes have conical bottoms and extend toward one another along the same axis. Gray solid material separates their tips, confirming that they are two independent blind holes; the small orange blind hole forms another instance. In Figure 3-2(d), a deeper green annular recess is nested in the bottom of the blue central blind cavity. Each of the four orange pockets has two levels: its shallow shoulder and deeper bottom are grouped as one compound instance. The shallow purple pocket and brown recesses open at the corners are grouped separately.

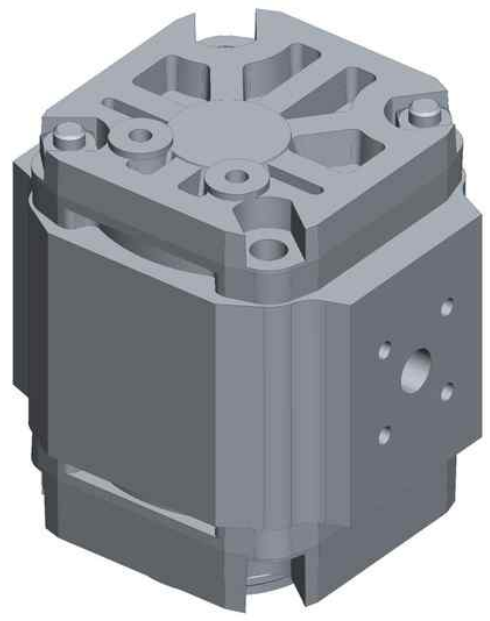

**(a) Input CAD model**

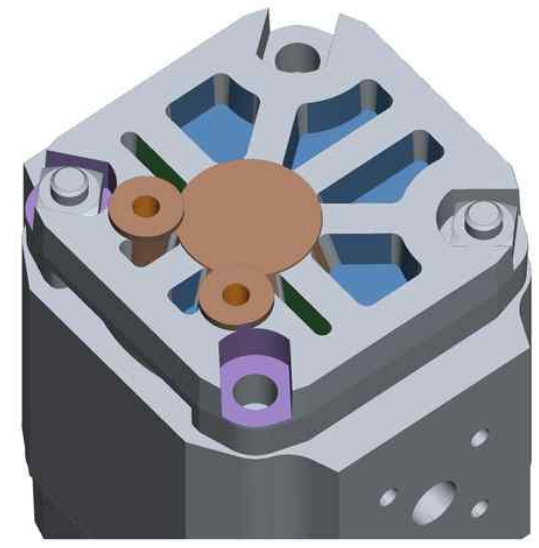

**(b) Pockets, slot, and bosses in the top cover**

Blue: pockets with flat bottoms and irregular outlines; green: narrow slot with rounded ends; brown: central boss and bosses with holes; orange: slightly tapered through hole; purple: recesses open at the corners

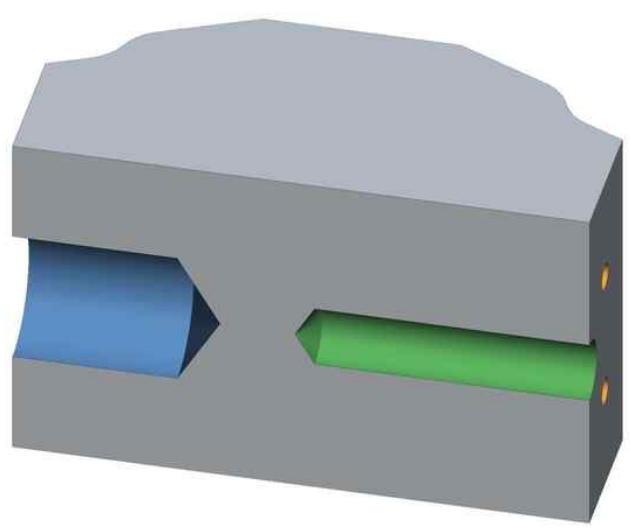

**(c) Opposing blind holes in the main block (sectioned solid)**

Blue/green: large and small opposing blind holes with conical bottoms; orange: small blind hole on the side

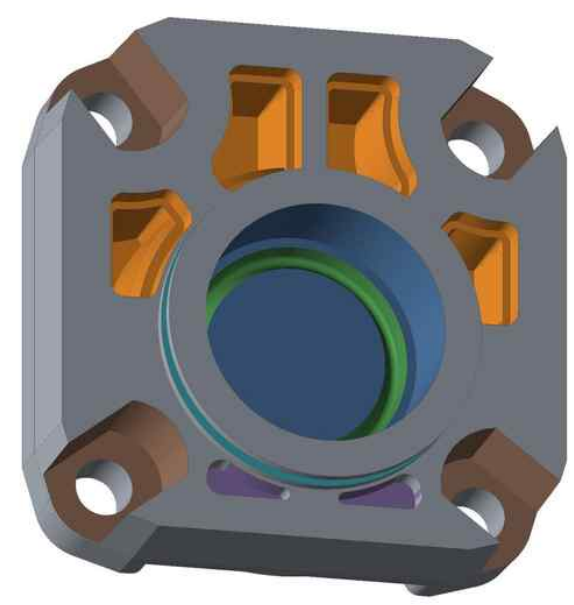

**(d) Compound features on the underside of the lower cover**

Blue: central blind cavity; green: deeper annular recess inside the cavity; orange: two-level pockets; purple: shallow pocket; brown: recesses open at the corners; cyan: outer circumferential groove

Figure 3-2 Representative feature recognition results for FR-02

### FR-03 (moderate)

Figure 3-3 shows representative results for FR-03. In Figure 3-3(b), the magenta region is reported as a main feature described as the "overall hollow housing." This describes the base form of the part, rather than a machining feature that can be counted independently, and illustrates the false feature detections reported in Table 3-4. The brown region is the mating flange at the split plane. In Figure 3-3(c), a curved brown partition partly separates the blue cavity following the outer housing from the green central curved passage; the two connect inside the semicircular opening. In Figure 3-3(d), the blue semicylindrical seat along the axis contains an orange groove of larger diameter with a sloping shoulder on each side. The green stepped semicircular seat contains a purple groove with a rectangular cross-section. Both structures have open boundaries at the split plane.

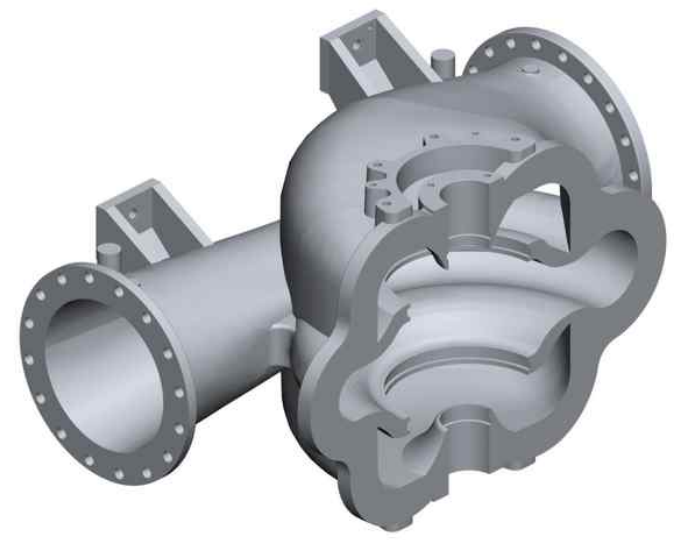

**(a) Input CAD model**

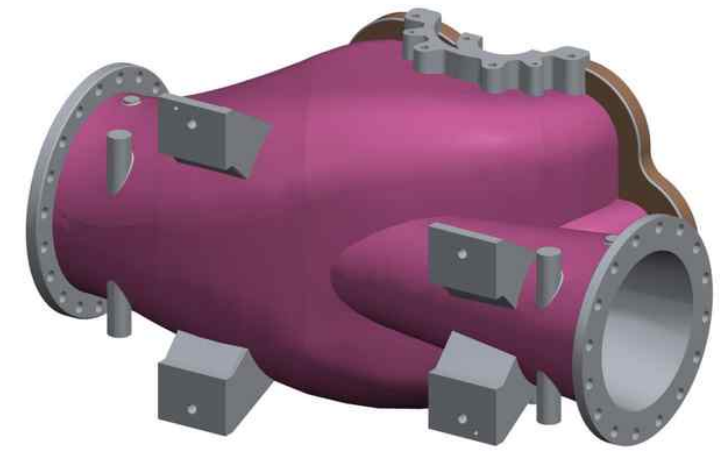

**(b) Overall housing surface**

Magenta: housing surface incorrectly identified as a machining feature; brown: mating flange at the split plane

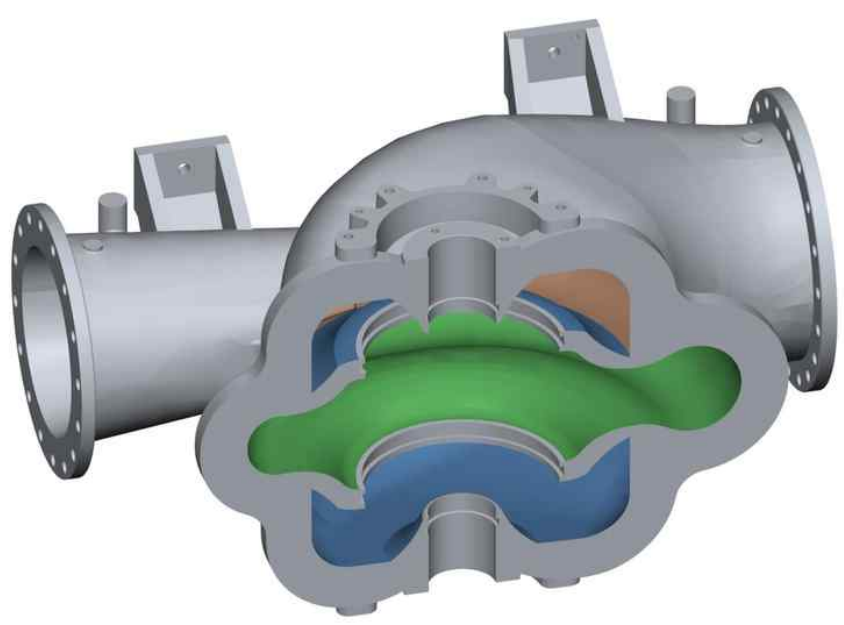

**(c) Two passages and a partial partition**

Blue: cavity following the outer housing; green: central curved passage; brown: curved partition

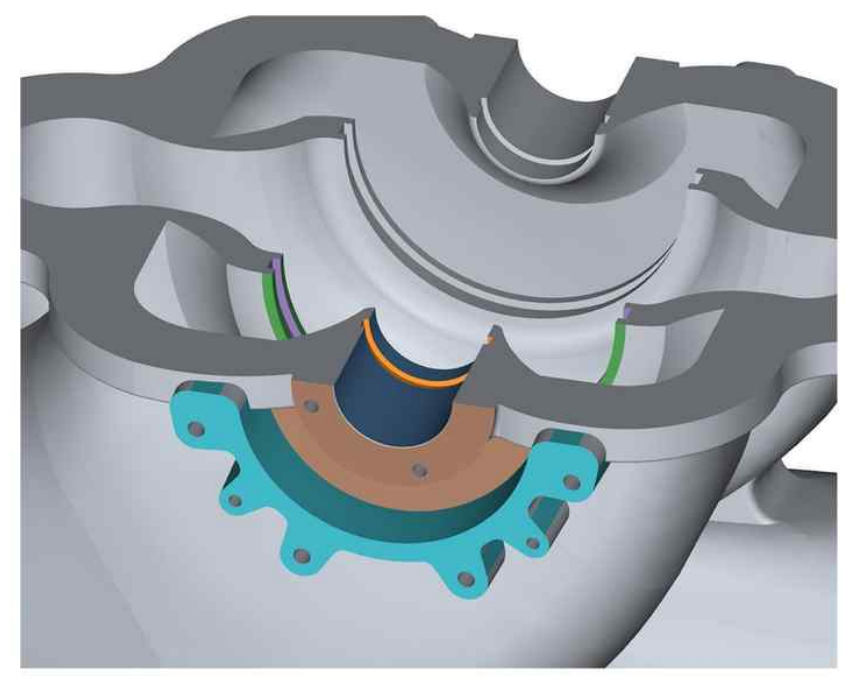

**(d) Open seats and semicircular grooves**

Brown: semicylindrical boss; cyan: lobed semicircular boss; blue: semicylindrical seat along the axis; orange: groove of larger diameter with two sloping shoulders; green: stepped semicircular seat; purple: semicircular groove with a rectangular cross-section

Figure 3-3 Representative feature recognition results for FR-03

### FR-04 (moderate)

Figure 3-4 shows representative results for FR-04. Figure 3-4(b) shows three repeated groups of nested cavities and slots inside the main cavity. Below the light blue main cavity, the blue, green, orange, and purple features become progressively narrower with depth. Their successive levels are clearly visible, as are the actual boundaries where surrounding passages interrupt them. Figure 3-4(c) highlights the circular holes, corresponding bosses, and groove with an irregular outline at the uppermost level. Small blue holes run around the top rim, with corresponding green bosses. The brown groove follows the rim continuously and intersects the hole openings and bosses in places, revealing substantial interaction among these features. In Figure 3-4(d), the blue stepped port intersects the cyan transverse passage at the side connection. Four purple mounting through holes in the flange retain complete circular boundaries, whereas two orange mounting

holes are interrupted by the transverse passage. These results illustrate hole recognition under complex intersections.

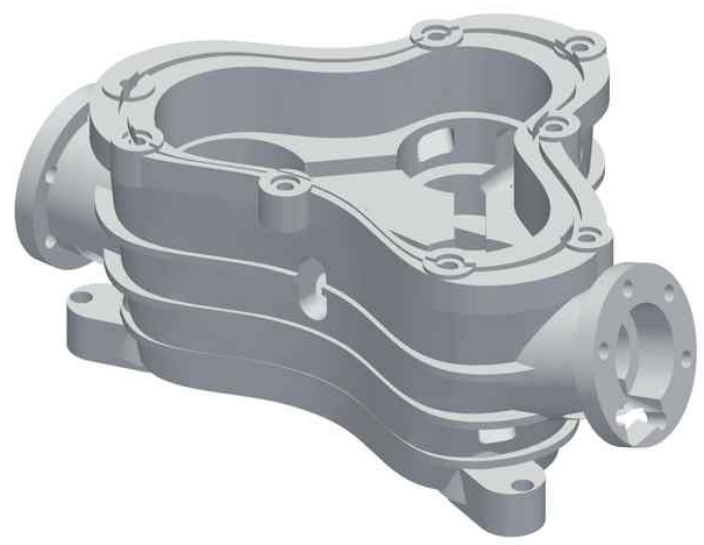

**(a) Input CAD model**

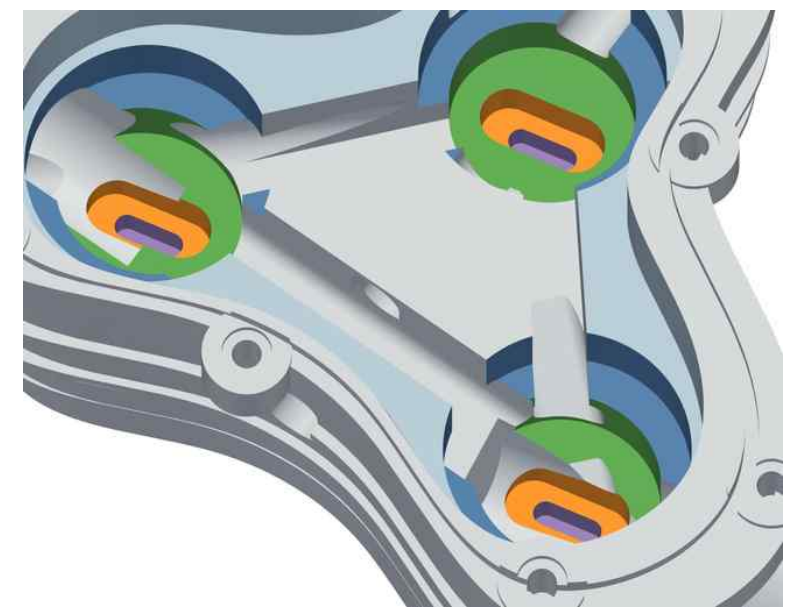

**(b) Four levels of nested cavities and slots**

Light blue: main cavity; blue: first-level circular recess; green: second-level elliptical recess with a flat bottom; orange: third-level slot with rounded ends; purple: fourth-level blind slot with rounded ends and a flat bottom

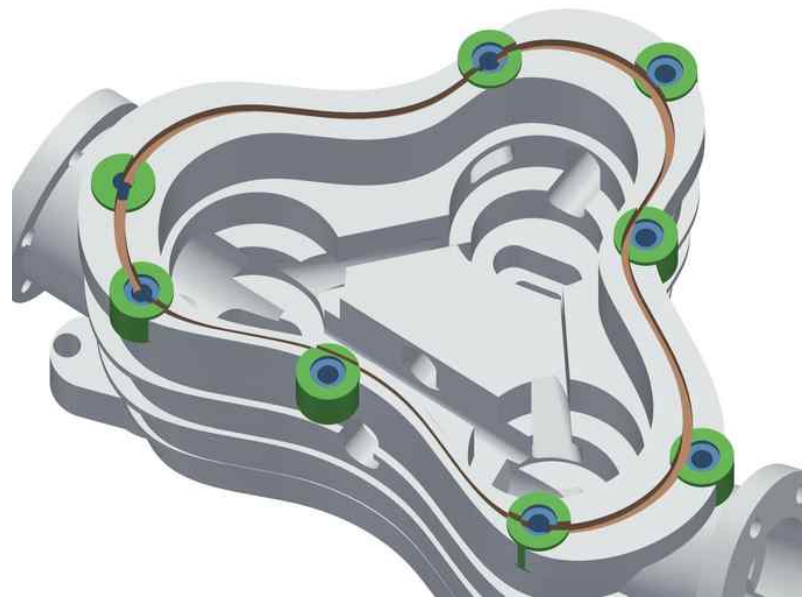

**(c) Circular holes and an irregular groove around the top rim**

Blue: small holes around the top rim; green: corresponding bosses; brown: irregular groove at the top

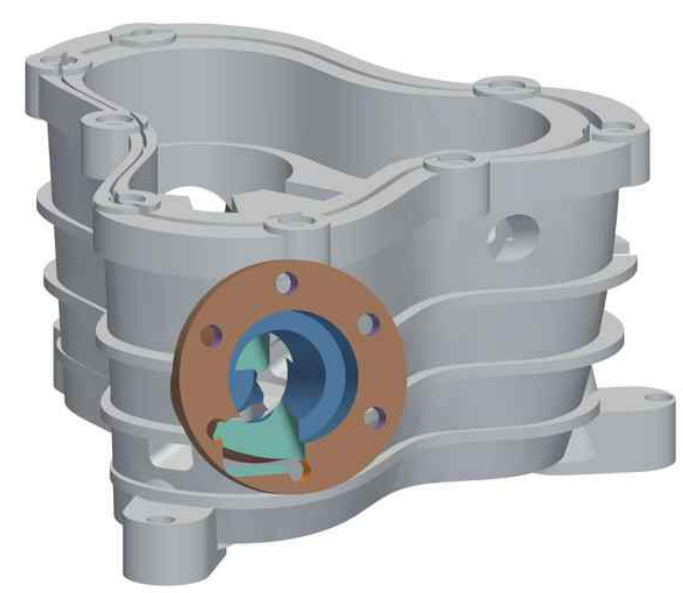

**(d) Side port and interrupted mounting holes**

Brown: side-port flange; blue: stepped port; cyan: transverse passage;

Purple: complete mounting through holes; orange: two mounting holes interrupted by the transverse passage

Figure 3-4 Representative feature recognition results for FR-04

### FR-05 (complex)

Figure 3-5 shows representative results for FR-05. In Figure 3-5(b), the magenta regions represent the overall form of the upper cover, including its spherical dome and disc-shaped base. These belong to the cover body and are not independently countable machining features, illustrating the false feature detections in FR-05. In Figure 3-5(c), a secondary green spherical recess is nested within the blue main spherical cavity and joined to it by a purple fillet. An inclined orange through hole connects the secondary recess to the exterior. Outside the cyan annular locating step, the lower mounting flange has a rose-colored, wavy outline and surrounds four brown mounting through holes in the cover. The view shows both the nested cavity and the surrounding mounting features. In Figure 3-5(d), blue circumferential cooling fins are arranged in layers. Brown bosses surround the orange radial blind holes with flat bottoms, and the upper mating flange with a wavy outline is shown in green. The bosses intersect adjacent fins and interrupt parts of their edges, illustrating how complex external features interact with local hole features.

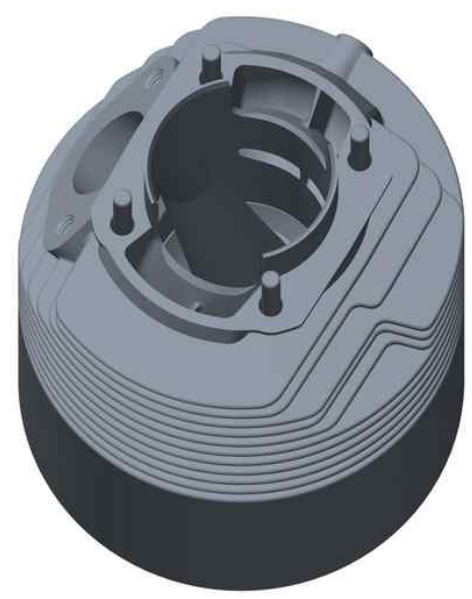

**(a) Input CAD model**

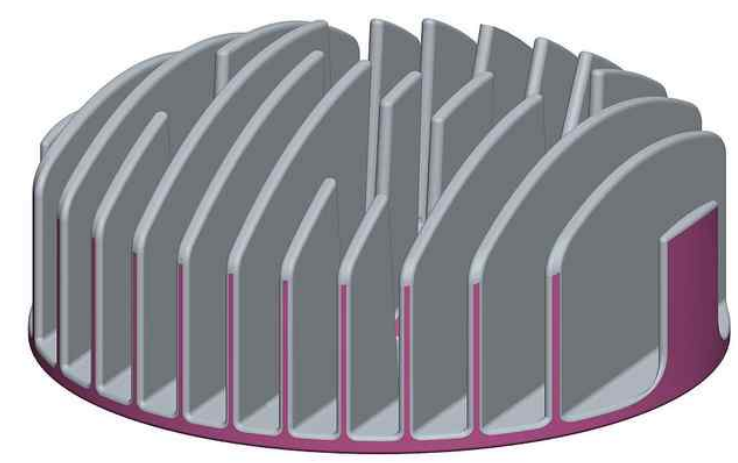

**(b) Upper cover surface**

Magenta: main cover surface incorrectly identified as a machining feature

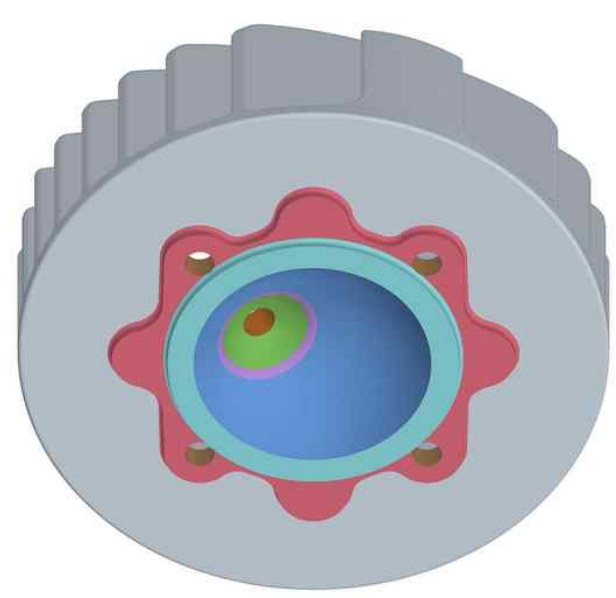

**(c) Nested spherical recesses and an inclined through hole**

Blue: main spherical cavity; green: secondary spherical recess; purple: transition fillet;

Orange: inclined through hole; cyan: annular locating step; brown: mounting through holes in the upper cover;

Rose: lower mounting flange with a wavy outline

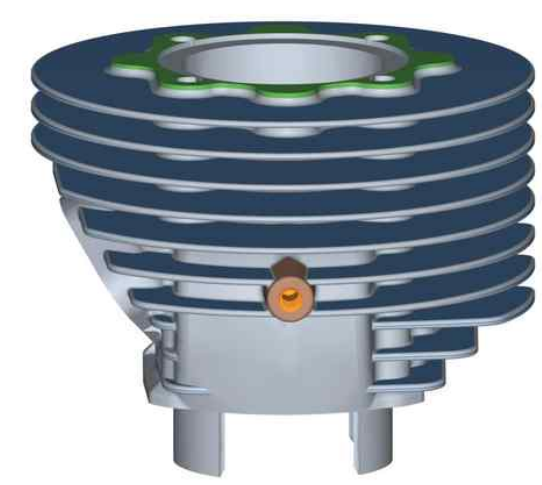

**(d) Cooling fins, radial blind holes, and upper flange**

Blue: circumferential cooling fins; brown: bosses around radial holes; orange: blind holes with flat bottoms;

Green: upper mating flange with a wavy outline

Figure 3-5 Representative feature recognition results for FR-05

### FR-06 (complex)

Figure 3-6 shows representative results for FR-06. Figure 3-6(b) uses the same local cutaway as the HTML view to expose an error in grouping feature instances: two spatially separate circular holes are identified as a single hole feature. Figure 3-6(c) uses another local cutaway to show a classification error. The highlighted cylindrical surface lies on the outside of the solid but is misidentified as a hole feature. Figure 3-6(d) shows a complex region containing cooling fins,

cooling ribs on an end face, end recesses and openings between ribs, and interrupted walls of transverse holes.

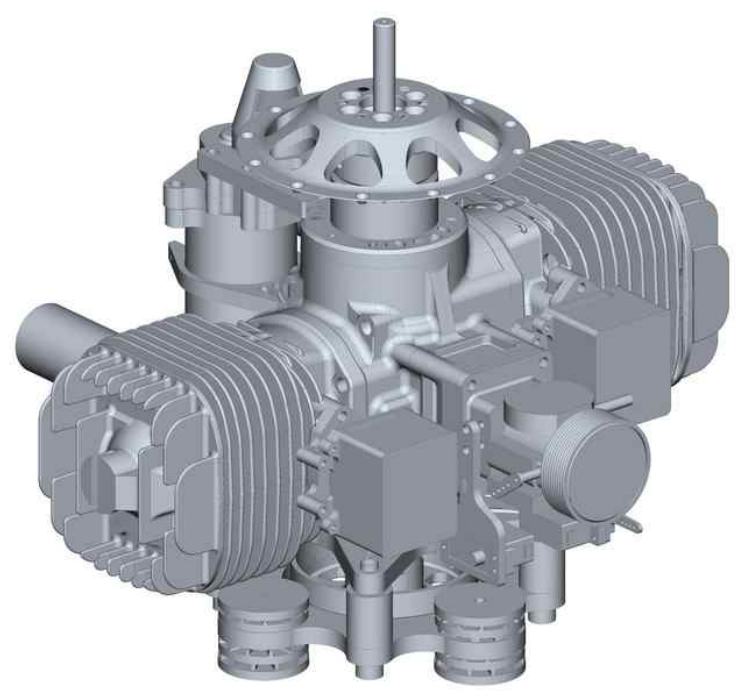

**(a) Input CAD model**

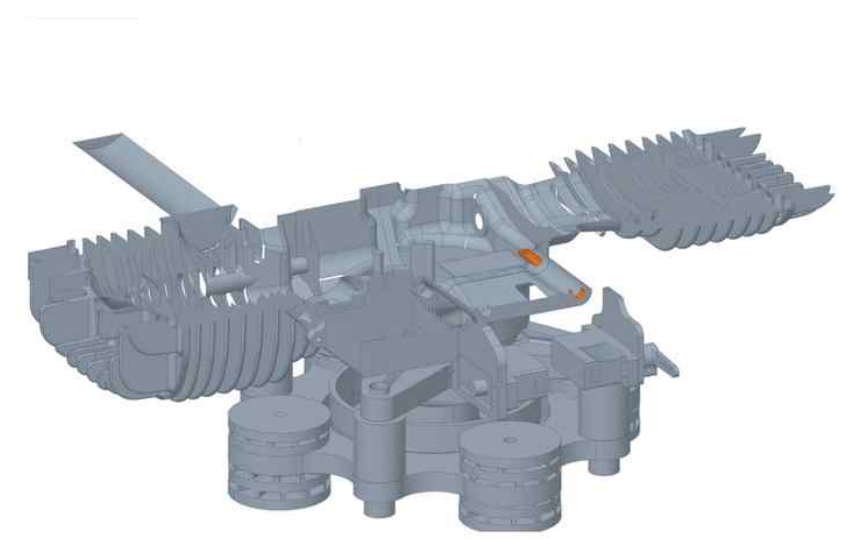

**(b) Two circular holes (local cutaway)**

Two circular holes incorrectly identified as one feature

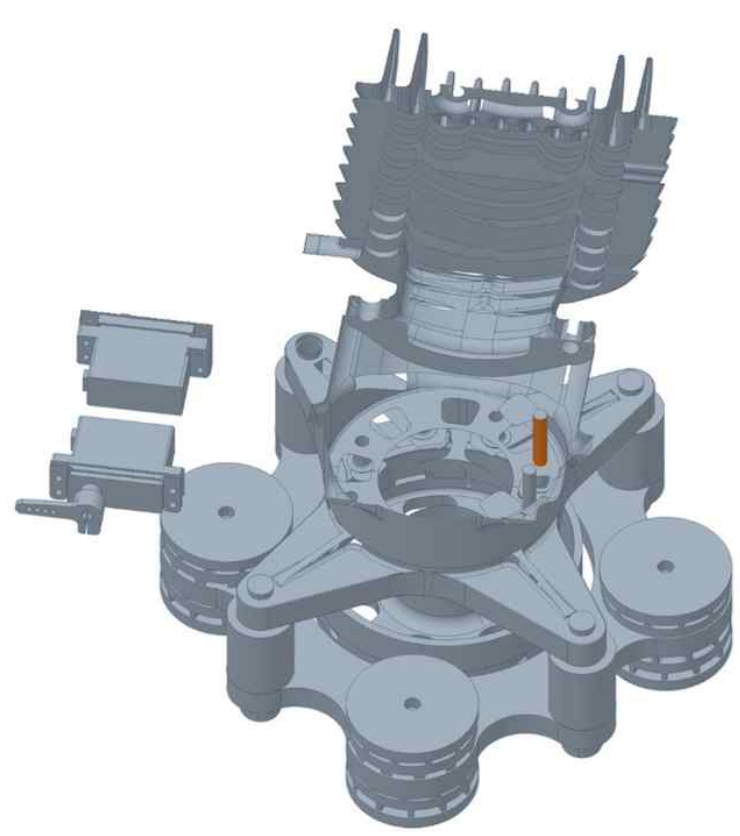

**(c) Cylindrical surface mistaken for a hole (local cutaway)**

External cylindrical surface incorrectly identified as a hole feature

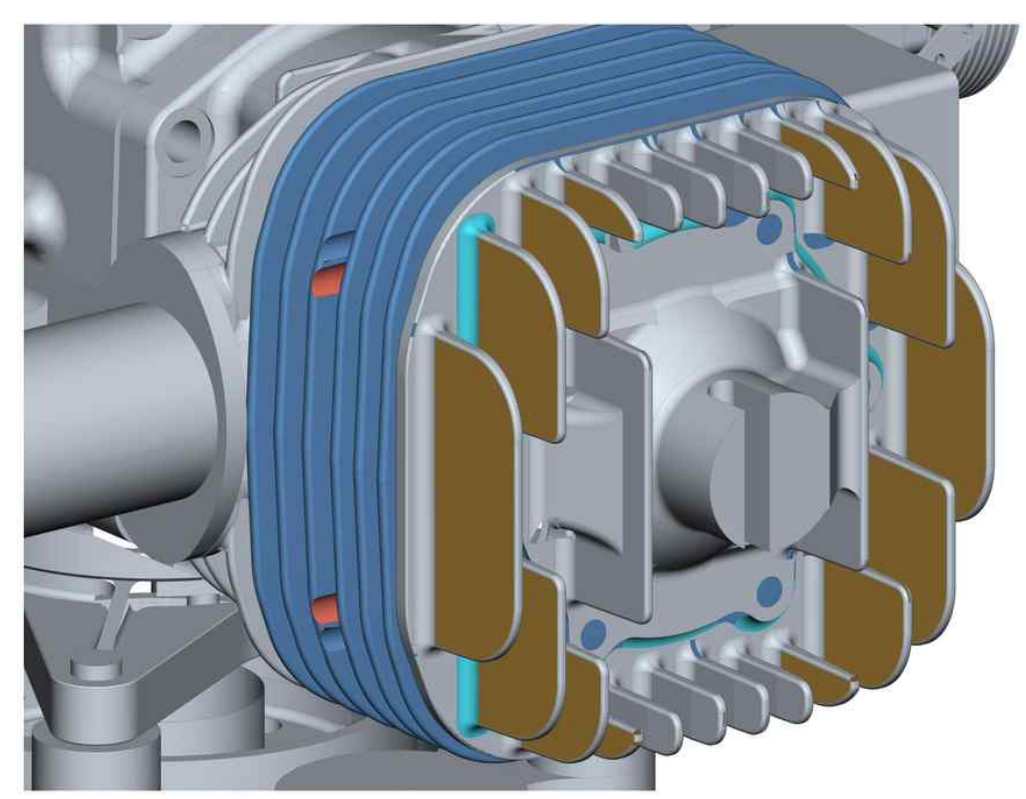

**(d) Region containing compound cooling features**

Blue: transverse cooling fins; gold: cooling ribs with draft on the end face;

Cyan: end recesses/openings between ribs; red: interrupted walls of transverse holes

Figure 3-6 Representative feature recognition results for FR-06

### 3.3.2 Additional experiment results

The additional experiment uses only the complex case FR-06 and compares recognition with web search enabled and disabled. Each comparison shows the same geometric region from the same viewpoint, with web search enabled on the left and disabled on the right. Colored regions indicate features recognized in the corresponding result. Gray surfaces provide context and do not by themselves indicate that no feature is present. The comparisons focus on feature types, separation into individual instances, and the geometric regions assigned to those instances. Table 3-6 gives the completion time with web search disabled.

Table 3-6 Completion time for feature recognition with web search disabled

| Case ID | FR-06 (web search disabled) |
| --- | --- |
| Completion time | 47 min |

#### 3.3.2.1 Upper disc: similar recognition of windows and hole openings

Figure 3-7 compares the results for the upper disc of FR-06 with web search enabled and disabled. Its sector-shaped windows, two rings of conical hole openings, and central stepped hub lie close together, requiring their geometric boundaries to be distinguished.

**(a) Web search enabled**

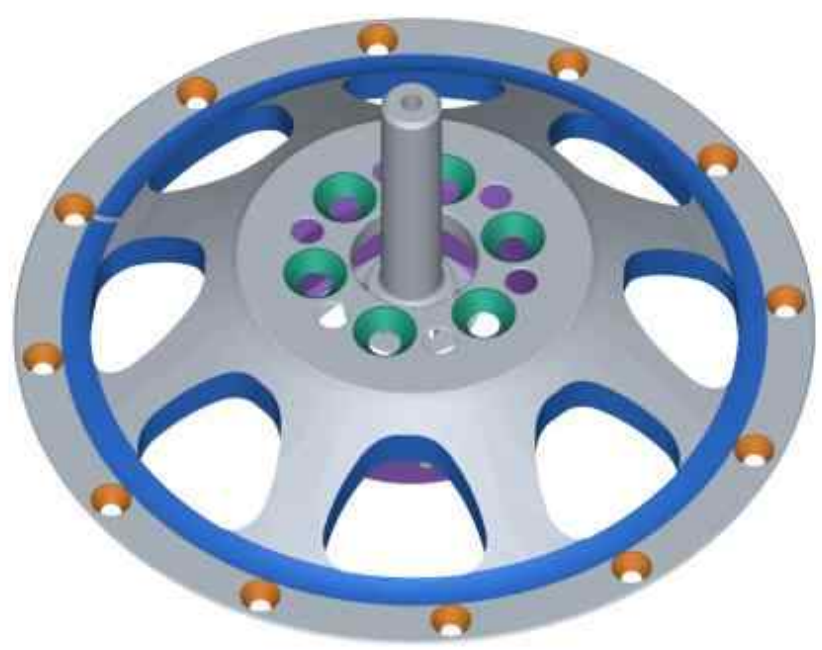

Blue: 8 sector-shaped windows with rounded corners; orange: 12 conical recesses in the outer ring;

Green: 6 conical hole openings in the inner ring; purple: central stepped hub.

**(b) Web search disabled**

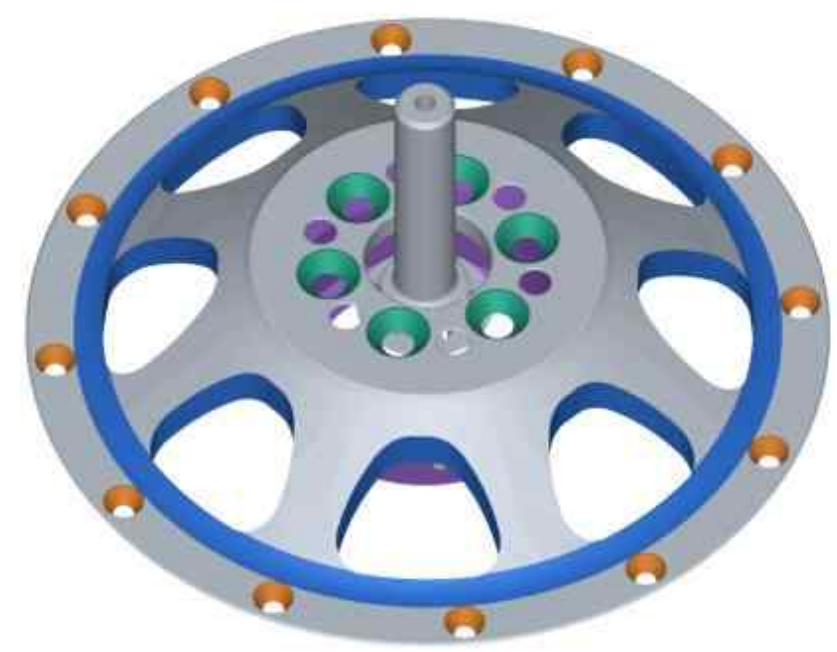

Blue: 8 sector-shaped windows with rounded corners; orange: 12 conical recesses in the outer ring;

Green: 6 conical hole openings in the inner ring; purple: central stepped hub.

Figure 3-7 FR-06 upper disc with web search enabled and disabled

In both Figure 3-7(a) and Figure 3-7(b), the blue sector-shaped windows are grouped individually around the disc. The orange outer recesses and green inner hole openings are colored separately, and the purple hub is identified as a separate feature. The feature types, counts, and corresponding regions are broadly consistent between the two conditions. There is no evident merging of windows, confusion between hole openings, or shift in their locations.

#### 3.3.2.2 Clamping link: through holes identified with web search enabled

Four small holes are evenly spaced along one side of the clamping link. Their local walls and openings at both ends are clearly visible, making this region suitable for comparing how specifically GPT-6 identifies holes under each condition. Figure 3-8 shows the same region under both conditions.

| **(a) Web search enabled** | **(b) Web search disabled** |
| --- | --- |
| 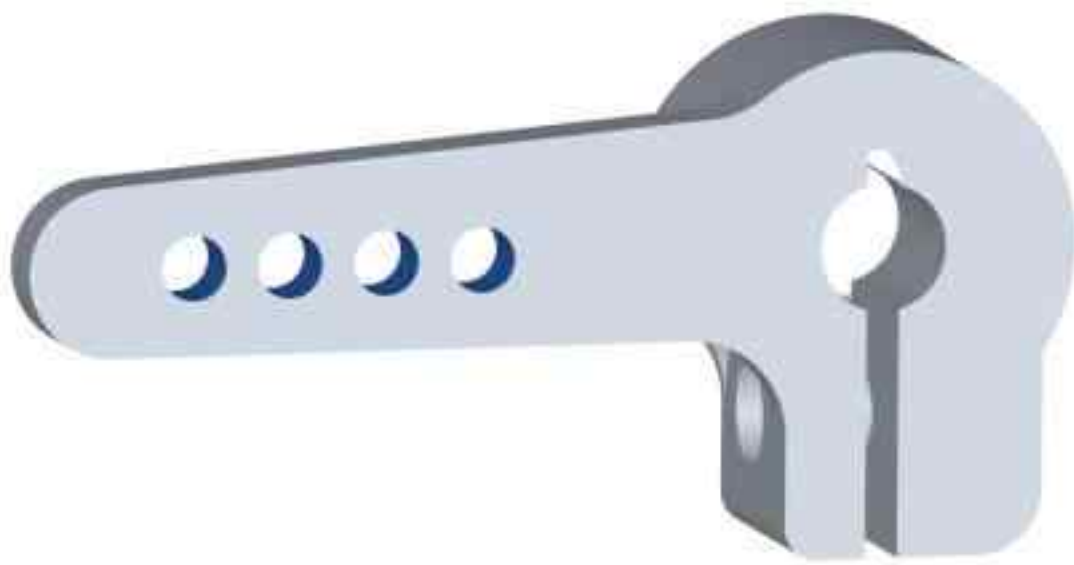 | 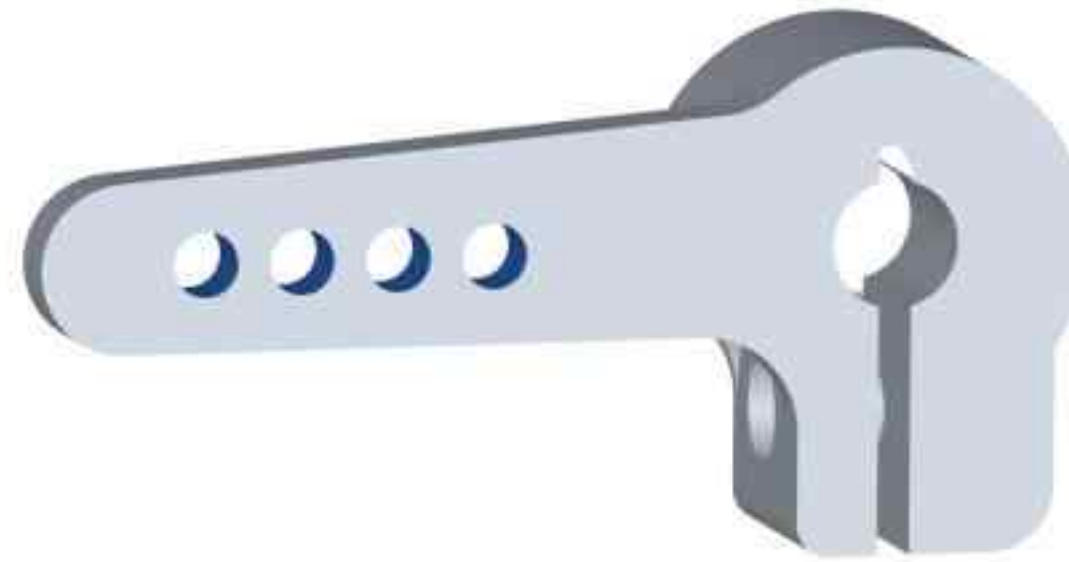 |
| Blue: 4 cylindrical through holes | Blue: 4 circular hole openings (through or blind unresolved). |

Figure 3-8 Four small holes in the FR-06 clamping link with web search enabled and disabled

GPT-6 locates the walls of the same four holes under both conditions, as shown in Figure 3-8(a) and Figure 3-8(b). Their geometric positions are consistent. With web search enabled, GPT-6 further identifies them as cylindrical through holes. With web search disabled, it finds the same holes but identifies them only as circular holes, without determining whether they are through or blind. Recognition is therefore more specific in this region when web search is enabled.

### 3.3.2.3 Stepped hub: a blind hole identified with web search disabled

Two radial holes lie close together on the side of the stepped hub. The inner end of one hole wall meets a flat bottom, whereas the other is represented only by a cylindrical section. Figure 3-9 enlarges the same side of the hub to compare whether GPT-6 uses the bottom face to distinguish through and blind holes under each condition.

| **(a) Web search enabled** | **(b) Web search disabled** |
| --- | --- |
| 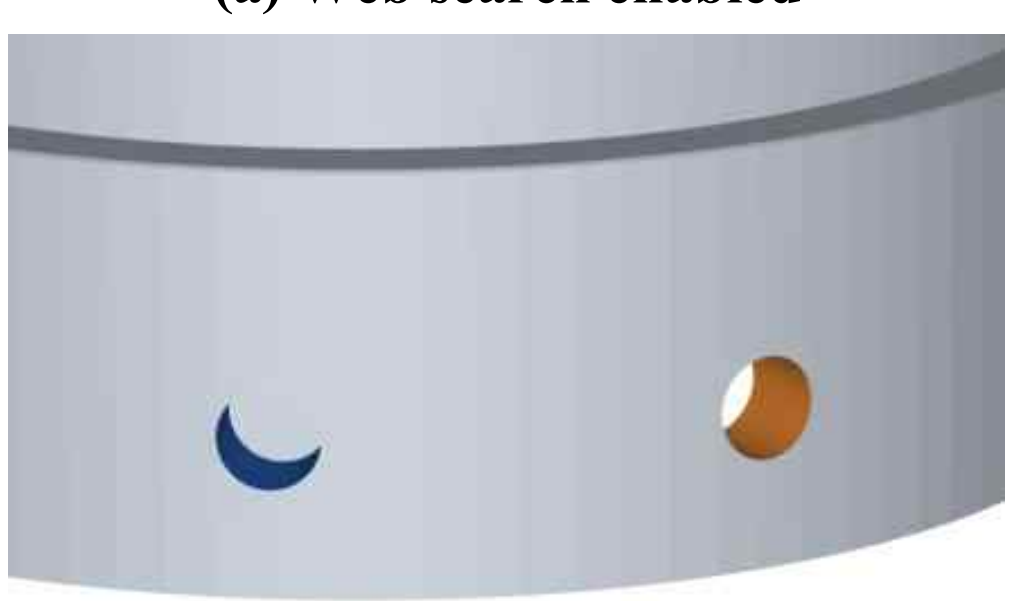 | 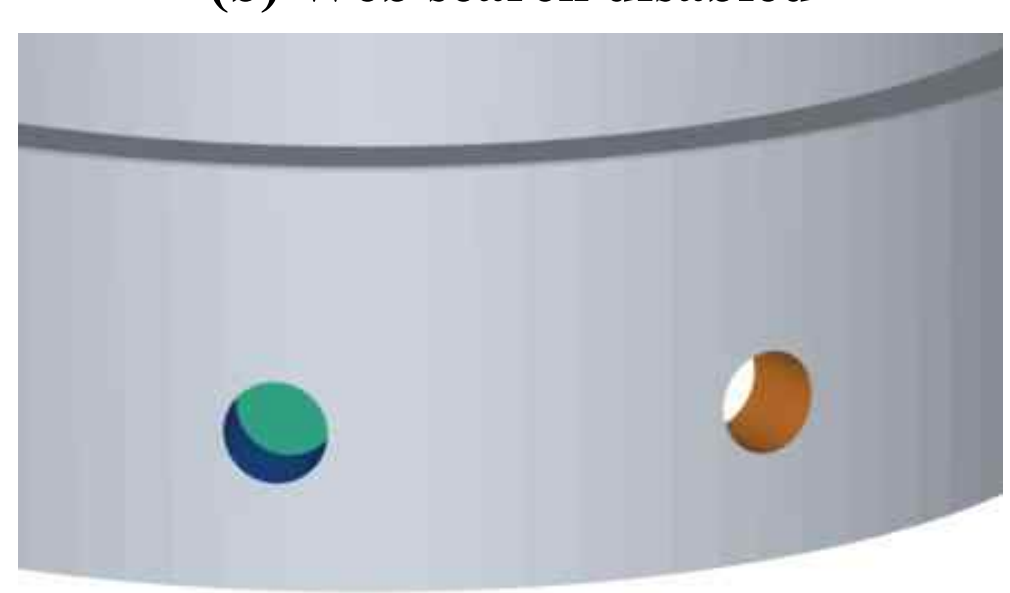 |
| Blue: left radial hole wall (through or blind unresolved)<br>Orange: right radial hole wall (through or blind unresolved) | Blue: left blind-hole wall; green: adjoining flat bottom;<br>Orange: right radial hole section (through or blind unresolved) |

Figure 3-9 Radial holes in the FR-06 hub with web search enabled and disabled

With web search enabled, GPT-6 locates the two radial hole walls but identifies both only as holes, as shown in Figure 3-9(a). With web search disabled, it combines the blue wall on the left with its green flat bottom and identifies a blind hole with a flat bottom in Figure 3-9(b). It retains uncertainty about whether the orange hole section on the right is through or blind. Thus, the result with web search disabled uses the bottom face to reach a more specific identification here, while the result with web search enabled finds the hole but does not identify it as blind.

#### 3.3.2.4 Analysis of the additional experiment

The three comparisons show mixed local differences. Recognition of the disc windows and hole openings is broadly unchanged. With web search enabled, GPT-6 identifies the four small holes in the clamping link as through holes; with web search disabled, it identifies a flat-bottomed blind hole in the stepped hub. This single-case comparison does not establish an overall benefit or disadvantage of web search. The execution analysis in Section 3.4 indicates that CAD-kernel queries, code execution, and interpretation of model views are relevant to the task. Recognition nevertheless remains incomplete in complex regions, as illustrated in Figures 3-8 and 3-9.

## 3.4 Analysis of GPT-6's approach

The recorded outputs and tool operations suggest the following workflow for feature recognition:

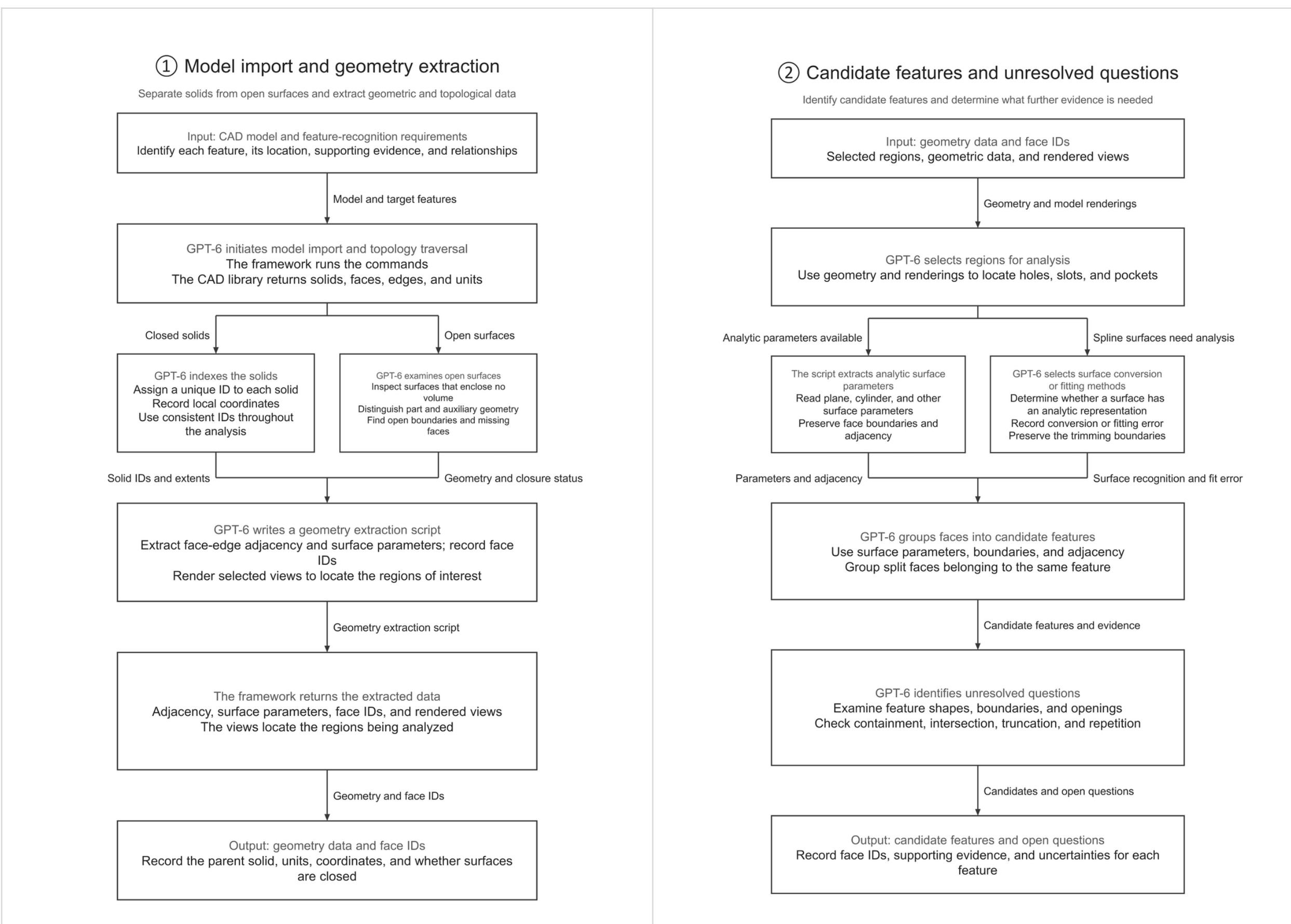

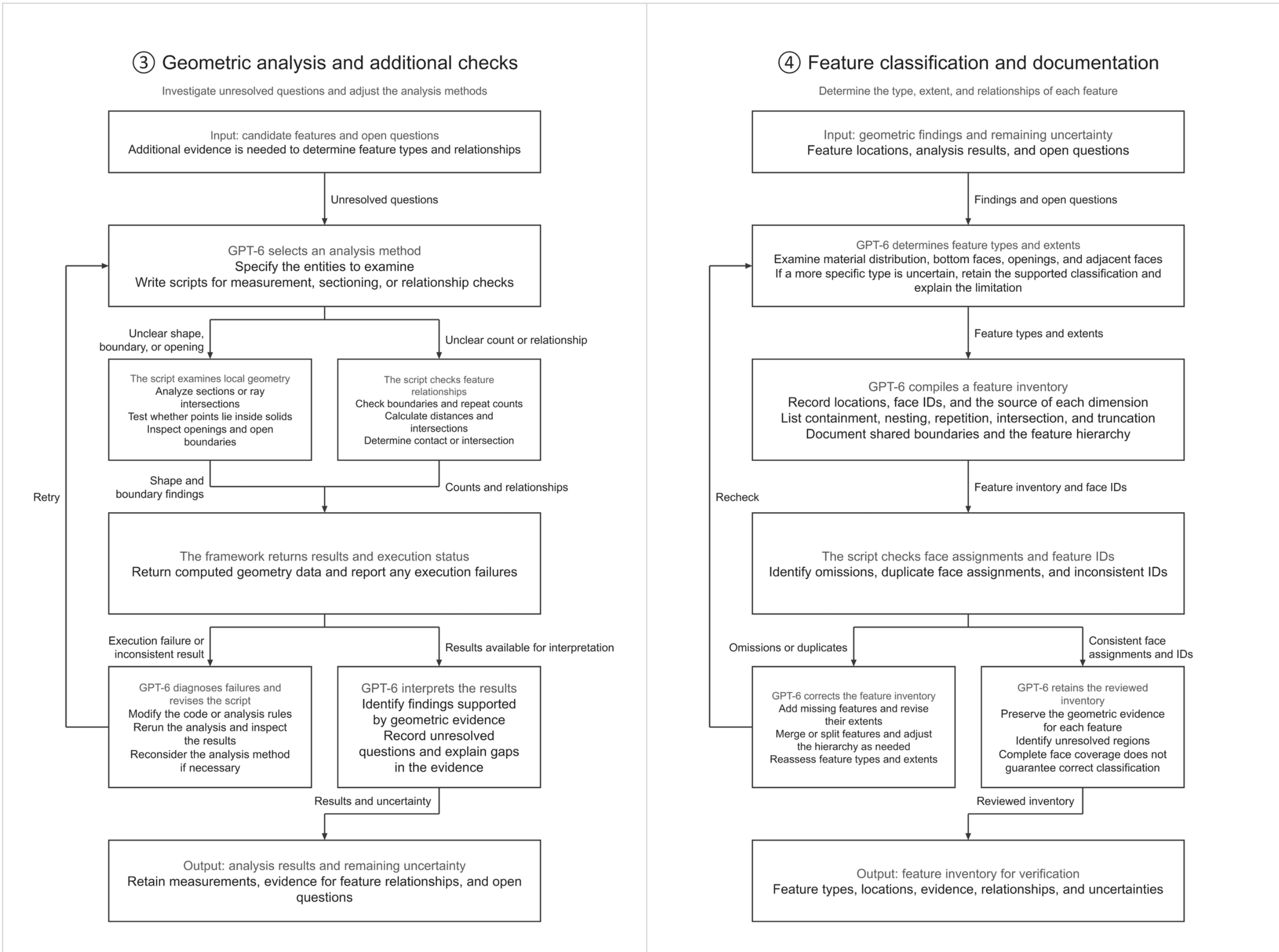
③ Geometric analysis and additional checks
Investigate unresolved questions and adjust the analysis methods
Input: candidate features and open questions
Additional evidence is needed to determine feature types and relationships
Unresolved questions
GPT-6 selects an analysis method
Specify the entities to examine
Write scripts for measurement, sectioning, or relationship checks
Unclear shape, boundary, or opening
Unclear count or relationship
The script examines local geometry
Analyze sections or ray intersections
Test whether points lie inside solids
Inspect openings and open boundaries
The script checks feature relationships
Check boundaries and repeat counts
Calculate distances and intersections
Determine contact or intersection
Shape and boundary findings
Counts and relationships
Retry
The framework returns results and execution status
Return computed geometry data and report any execution failures
Execution failure or inconsistent result
Results available for interpretation
GPT-6 diagnoses failures and revises the script
Modify the code or analysis rules
Rerun the analysis and inspect the results
Reconsider the analysis method if necessary
GPT-6 interprets the results
Identify findings supported by geometric evidence
Record unresolved questions and explain gaps in the evidence
Results and uncertainty
Output: analysis results and remaining uncertainty
Retain measurements, evidence for feature relationships, and open questions
④ Feature classification and documentation
Determine the type, extent, and relationships of each feature
Input: geometric findings and remaining uncertainty
Feature locations, analysis results, and open questions
Findings and open questions
GPT-6 determines feature types and extents
Examine material distribution, bottom faces, openings, and adjacent faces
If a more specific type is uncertain, retain the supported classification and explain the limitation
Feature types and extents
GPT-6 compiles a feature inventory
Record locations, face IDs, and the source of each dimension
List containment, nesting, repetition, intersection, and truncation
Document shared boundaries and the feature hierarchy
Feature inventory and face IDs
Recheck
The script checks face assignments and feature IDs
Identify omissions, duplicate face assignments, and inconsistent IDs
Omissions or duplicates
Consistent face assignments and IDs
GPT-6 corrects the feature inventory
Add missing features and revise their extents
Merge or split features and adjust the hierarchy as needed
Reassess feature types and extents
GPT-6 retains the reviewed inventory
Preserve the geometric evidence for each feature
Identify unresolved regions
Complete face coverage does not guarantee correct classification
Reviewed inventory
Output: feature inventory for verification
Feature types, locations, evidence, relationships, and uncertainties

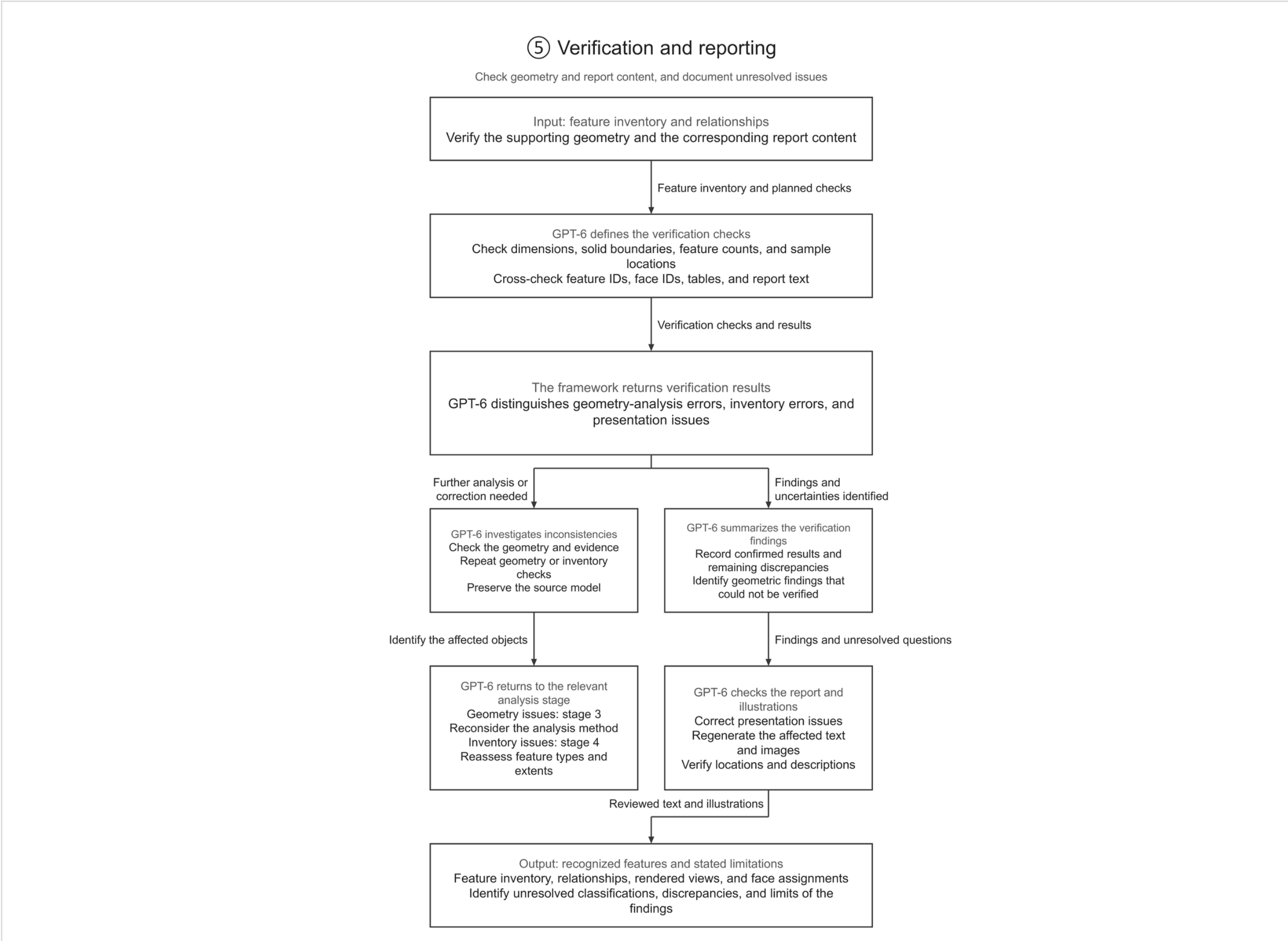

Figure 3-10  Inferred workflow for feature recognition

## 3.5 Capabilities and limitations

For feature recognition, GPT-6 can combine surface parameters, face adjacency, sectioning where needed, and checks of whether geometry lies inside or outside a solid to recognize holes, slots, pockets, bosses, and transition features. It can also associate those features with specific B-rep faces. When simple features have clearly separate boundaries, GPT-6 can recognize them correctly and identify their local structure and relationships at different levels. For holes, it can distinguish through holes, blind holes with conical bottoms, and stepped holes. For pockets, it can recognize successive levels and use bottom-face positions and connectivity to distinguish individual features.

The main limitations occur in complex regions with multiple levels of nesting, intersections, or interrupted boundaries. GPT-6 sometimes combines two spatially separate holes into one feature or mistakes an external cylindrical surface for a hole. Even after locating a hole, it may fail to use its bottom and openings fully to determine whether it is through or blind. The workflow analysis in Section 3.4 shows that GPT-6 sometimes uses rendered views to help interpret geometry. Where viewpoints are restricted, occlusion is substantial, or local structures are difficult to discern, reliance on image information may contribute to errors in the grouping of feature faces and the identification of feature types. This may be one cause of the errors in local detail.

Overall, GPT-6 can recognize basic features with reasonable reliability and handle some complex combinations. However, it does not yet consistently identify all machining features in complex models. When boundaries are unclear, features interact, or local structures are difficult to interpret from model views, it is more likely to merge several features into one or leave their specific types unresolved.

# Chapter 4 CAD reconstruction from engineering drawings

## 4.1 Task objectives and evaluation criteria

This task evaluates GPT-6's ability to interpret mechanical engineering drawings and reconstruct three-dimensional CAD models. It tests whether GPT-6 can correctly interpret projections across multiple views, dimensions, and local geometric features, then generate models with accurate geometry, topology, and key dimensions. The main tests cover both individual parts and the separate parts of an assembly. Additional tests vary the available dimensions and access to web search to examine performance under different input conditions.

Table 4-1 CAD reconstruction from engineering drawings

| Test element | Description |
|---|---|
| Evaluation question | Can GPT-6 correctly interpret the three-dimensional structure of a part from engineering drawings and generate the corresponding CAD model? |
| Input | Engineering drawings containing multiple views, dimensions, and necessary annotations. |
| Output | A three-dimensional CAD model. |
| Evaluation criteria | Check that the generated CAD model has valid and consistent geometry and topology (solid validity), that its structure matches the drawing, and that the stated dimensions are reproduced accurately. For cases with multiple parts, also verify that the parts can be assembled as intended. |

## 4.2 Test design

### 4.2.1 Test case selection

The test cases are drawn primarily from publicly available teaching materials on mechanical drawing and CAD. Selection favors drawings that provide complete information, represent characteristic part geometries, and have a corresponding 3D model or another reliable reference. To cover different levels of difficulty, the cases vary in structural complexity, the amount of information in the drawings, and the number of parts. Both single-part and multiple-part modeling tasks are included.

Complexity is assessed from the amount of information in the drawings, the complexity of the part geometry, and the relationships that must be resolved during modeling. For cases with multiple parts, the assessment also considers the number of parts and their assembly relationships. Table 4-2 defines the criteria.

Table 4-2 Complexity criteria for CAD reconstruction from engineering drawings

| Complexity | Selection criteria | Typical characteristics |
|---|---|---|
| Simple | Few geometric elements and views. The main structure can be determined directly from basic projection views and a small number of section views, with little need to relate information across views or infer internal structures. | The overall structure can be reconstructed from a few views, with few local features and clear relationships between them. |
| Moderate | More information must be related across views, or the part contains numerous local details, curved profiles, and repeated features while retaining a largely regular main structure. Cases with multiple parts have relatively few parts and clear assembly relationships, but require information to be related between parts. | Modeling requires information from several views or parts, but regular shapes, repeated features, or clear assembly relationships reduce the difficulty of interpretation. |
| Complex | Dense information across numerous projection, section, or detail views. Diverse structural details follow no single repetition pattern, so the complete geometry must be inferred from widely distributed information. Cases with multiple parts may involve many parts and local structures that can be understood only with reference to other parts. | A large amount of dispersed geometric and dimensional information must be combined, including complex relationships across views. Some structures may also require interpretation of the correspondence between parts. |

Eight cases were selected for the main experiment: five single-part cases and three cases with multiple parts. The single-part cases comprise two simple, two moderate, and one complex case; the cases with multiple parts comprise one moderate and two complex cases. Table 4-3 lists the cases.

Table 4-3 Test cases for CAD reconstruction from engineering drawings

| Case ID | Rendered CAD model | Case type | Complexity | Basis for complexity rating |
|---|---|---|---|---|
| D2C-P01 | 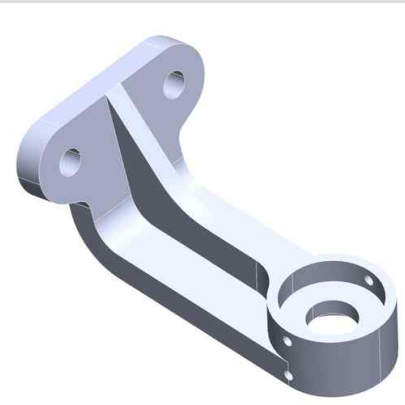 | Single part | Simple | One drawing with three principal views and 1 section view. The main geometry and internal relationships can be determined directly from basic projection views, with little need to combine information across views. |
| D2C-P02 | 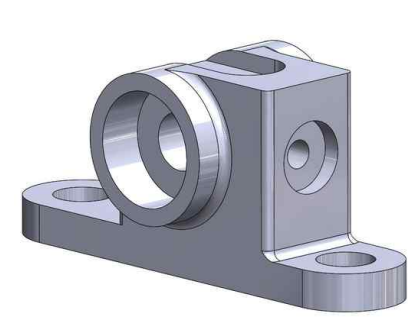 | Single part | Simple | One drawing with three principal views and 2 section views. The structure consists of basic features such as standard holes, steps, and bosses, allowing the main geometric relationships to be reconstructed from a few views. |

| Case ID | Rendered CAD model | Case type | Complexity | Basis for complexity rating |
|---|---|---|---|---|
| D2C-P03 | 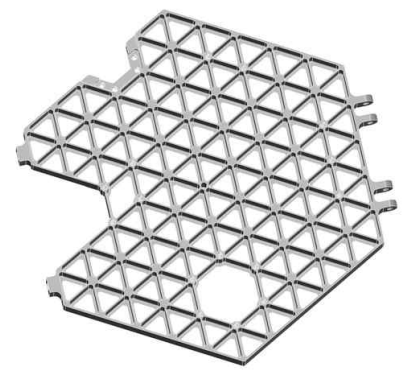 | Single part | Moderate | 3 drawings with 12 views provide a substantial amount of information, including local openings and other details that vary across the part. The main structure consists largely of regular patterns, however, so much of the geometry can be determined from repetition. The case is therefore rated moderate. |
| D2C-P04 | 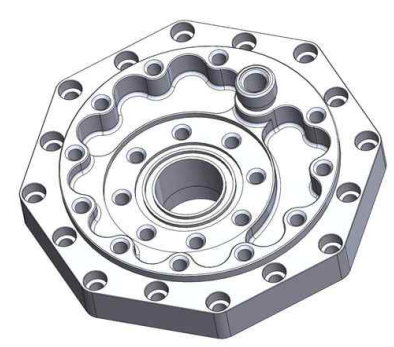 | Single part | Moderate | 4 drawings with 8 views show numerous curved profiles and intersections. Local boundaries and geometric correspondences must be established across views, but the amount of information and structural complexity remain well below those of the complex cases. |
| D2C-P05 | 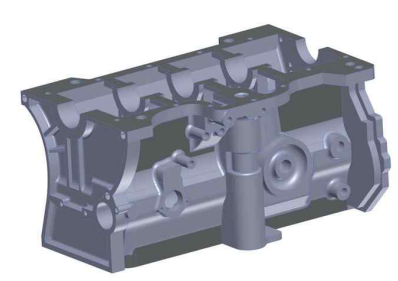 | Single part | Complex | 11 views show extensive internal and external structures, hole layouts, bosses, cavities, and local details. These are distributed across projection and section views without a single repetition pattern, requiring substantial information to be combined across views. |
| D2C-A01 | 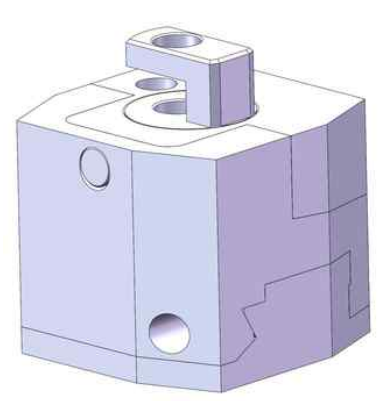 | Multiple parts | Moderate | 4 separate parts with relatively simple individual structures and assembly relationships. Each must be modeled separately, then assembled using dimensions and mating relationships across parts. The case is therefore rated moderate. |
| D2C-A02 | 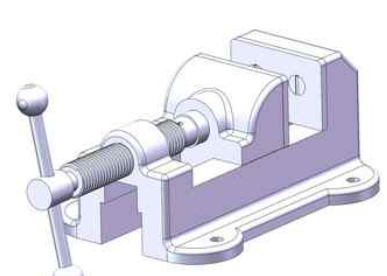 | Multiple parts | Complex | 11 parts, with several kinds of assembly relationship, including pin connections and threaded connections. Modeling must address the geometry of each part, correspondence between parts, and their different mating relationships. |
| D2C-A03 | 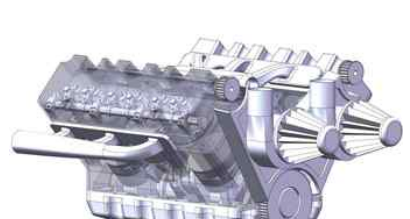 | Multiple parts | Complex | 28 part items of varied types, with complex internal structures. Individual parts require operations such as shelling and lofting, while the assembly contains numerous internal connections and relationships between parts. The amount of information and the modeling process are substantially more complex than in the other cases. |

### 4.2.2 Experimental conditions

#### 4.2.2.1 Main experiment

*The prompts reproduced below are English translations of the Chinese experimental records.*

All drawings undergo the same input preparation, and a consistent prompt is used for each kind of modeling task. Before testing, website names, logos, proprietary model names, product numbers, and other details that could reveal the source are masked, and files are given neutral names. In cases with multiple parts, identifiers linking the part drawings to the assembly drawing are retained so that the correspondence remains clear.

The same prompt is used for all cases of a given kind, with no additional hints about complexity or specific structures. GPT-6 is asked to construct a three-dimensional CAD model, specifically a B-rep model, from the drawings and provide the model file and a preview. For cases with multiple parts, assembly verification follows a consistent procedure after all parts have been modeled. Web search remains enabled in the main experiment to assess performance under normal operating conditions. For a single part, the prompt reads: “Using the supplied engineering drawing, generate a three-dimensional CAD model of the part shown and export a STEP file.” For multiple parts, it reads: “Using the supplied engineering drawings of the individual parts in the assembly, generate a three-dimensional CAD model of each part and export STEP files.”

#### 4.2.2.2 Additional experiments

Representative cases are selected for additional experiments that vary the information available in the drawings or access to web search. Two conditions are tested: removing local dimensions while retaining overall dimensions describing the part’s bounding box, and disabling web search. Each additional test uses the corresponding prompt from the main experiment, without extra hints for the case. All other test settings and input preparation remain unchanged. Each result is compared with the main experiment for the same case. Table 4-4 summarizes the conditions.

Table 4-4 Additional experimental conditions

| Condition | Input preparation | Web search | Purpose |
|---|---|---|---|
| S1 Local dimensions removed | Retain only overall dimensions such as total length, width, height, and maximum outside diameter; remove all other local dimensions. | Enabled | Test whether GPT-6 can infer part structures, local proportions, and mating relationships between parts when local dimensions are missing. |
| S2 Web search disabled | Use the same inputs as the main experiment, without further changes. | Disabled | Assess modeling performance without information obtained through web search. |

## 4.3 Results

The main and additional experiments are presented separately. The main experiment begins with an overview of solid validity, reconstruction of structural features, dimensional accuracy, and assembly results, followed by the results and analysis for each case. The additional experiments follow the conditions defined in Section 4.2.2 and compare each result with the corresponding case in the main experiment.

### 4.3.1 Main experiment results

#### 4.3.1.1 Overview

The main experiment includes five single-part cases and three cases with multiple parts. Table 4-5 summarizes the results, and Table 4-6 reports completion times.

Table 4-5 CAD reconstruction results in the main experiment

| Case ID | Complexity | Solid validity | Reconstruction of structural features | Dimensional accuracy | Assembly results |
|---|---|---|---|---|---|
| D2C-P01 | Simple | Pass | Two small holes missing | Pass elsewhere | N/A |
| D2C-P02 | Simple | Pass | Pass | Incorrect fillets at the rear | N/A |
| D2C-P03 | Moderate | Fail 0/1 passed | Pass | Incorrect chamfers at threaded holes | N/A |
| D2C-P04 | Moderate | Pass | Incorrect cavity side walls | Fail due to structural errors | N/A |
| D2C-P05 | Complex | Pass | Broadly correct | Fail due to structural errors | N/A |
| D2C-A01 | Moderate | Pass | Almost entirely correct | Minor issues that do not affect assembly | Pass |
| D2C-A02 | Complex | Pass | Almost entirely correct | Some dimensions do not match | Parts overlap when assembled |
| D2C-A03 | Complex | Fail 25/28 passed | Some structures missing | Some dimensions do not match | Cannot be assembled |

Solid validity is reported for each complete case. A fraction gives the number of valid part models divided by the number checked; “Pass” means that all part models passed.

Table 4-6  Completion times for CAD reconstruction in the main experiment

| **Case ID** | **D2C-P01** | **D2C-P02** | **D2C-P03** | **D2C-P04** |
|---|---|---|---|---|
| Completion time | 28 min | 25 min | 37 min | 1 h 8 min |
| **Case ID** | **D2C-P05** | **D2C-A01** | **D2C-A02** | **D2C-A03** |
| Completion time | 30 min | 26 min | 28 min | 36 min |

### 4.3.1.2 Case analyses

The following analysis presents each case in order of its identifier, describing the case, showing the generated models, and examining the results. For cases with multiple parts, it also presents the assembly verification performed by the evaluation team.

### D2C-P01 General-purpose bracket (simple)

**Reconstruction results**

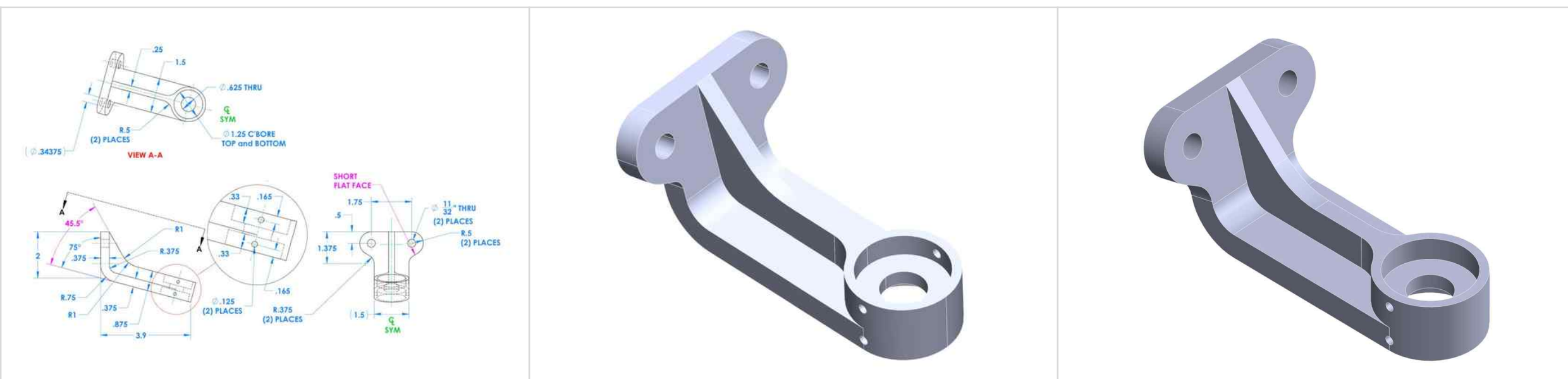

**(a) Example engineering drawing** | **(b) Rendered reference model** | **(c) Rendered generated model**

Figure 4-1  CAD reconstruction results for D2C-P01

The generated model is a valid solid, and its structure and dimensions largely match the drawing. The stepped hole in the cylindrical boss on the right has approximately the correct dimensions, but the radial holes at the corresponding symmetric positions are missing, as shown in Figure 4-2.

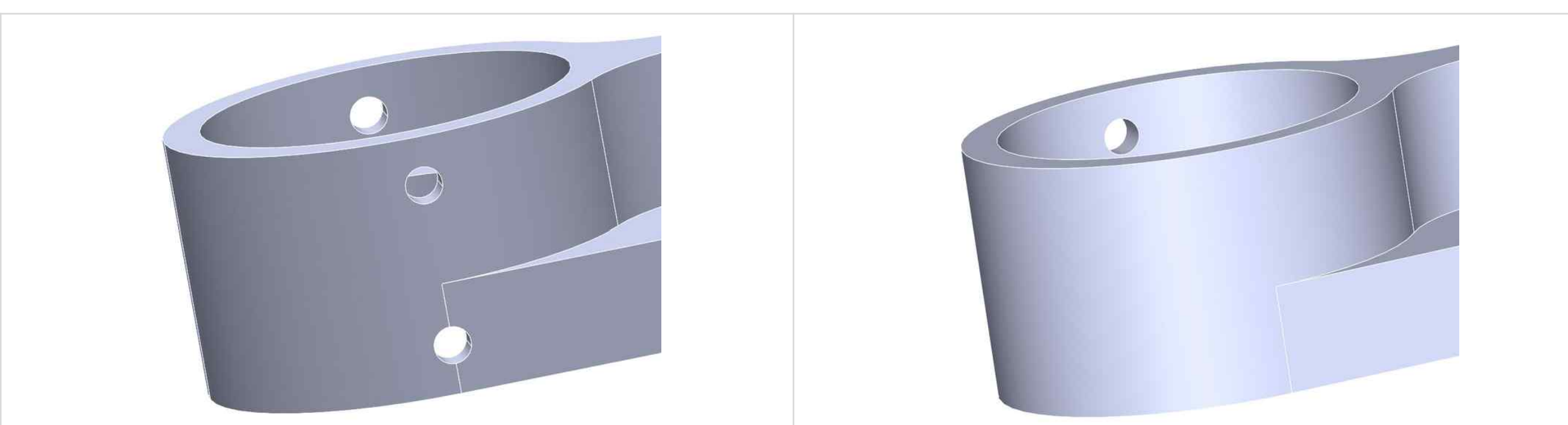

**(a) Rendered reference model** | **(b) Rendered generated model**

Figure 4-2  Details of CAD reconstruction for D2C-P01

## D2C-P02 Bearing support (simple)

### Reconstruction results

| 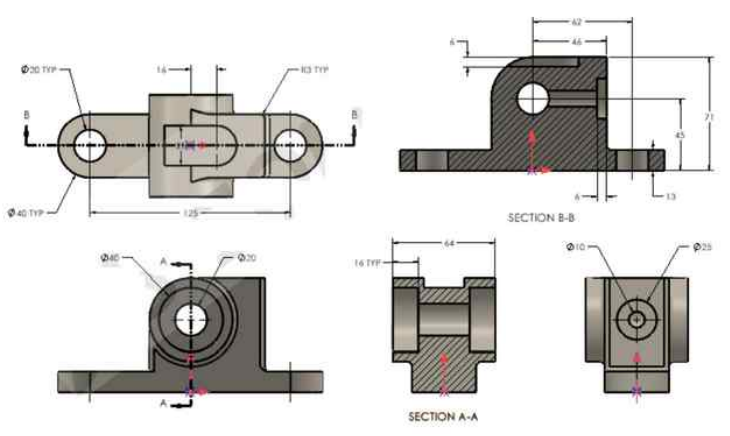 | 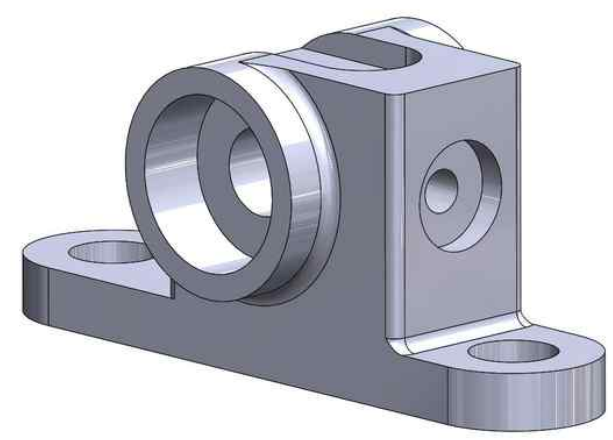 | 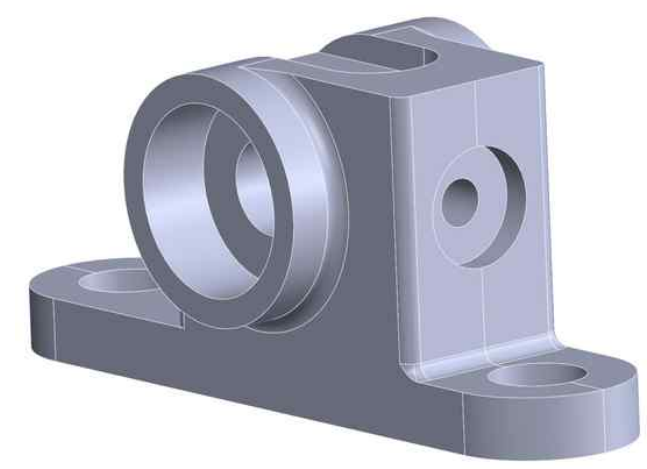 |
|---|---|---|
| **(a) Example engineering drawing** | **(b) Rendered reference model** | **(c) Rendered generated model** |

Figure 4-3 CAD reconstruction results for D2C-P02

The generated model is a valid solid, with structure and dimensions largely consistent with the drawing. However, the transitions between the two sides of the rear vertical support and the cylindrical structure have incorrect profiles, as shown in Figure 4-4.

| 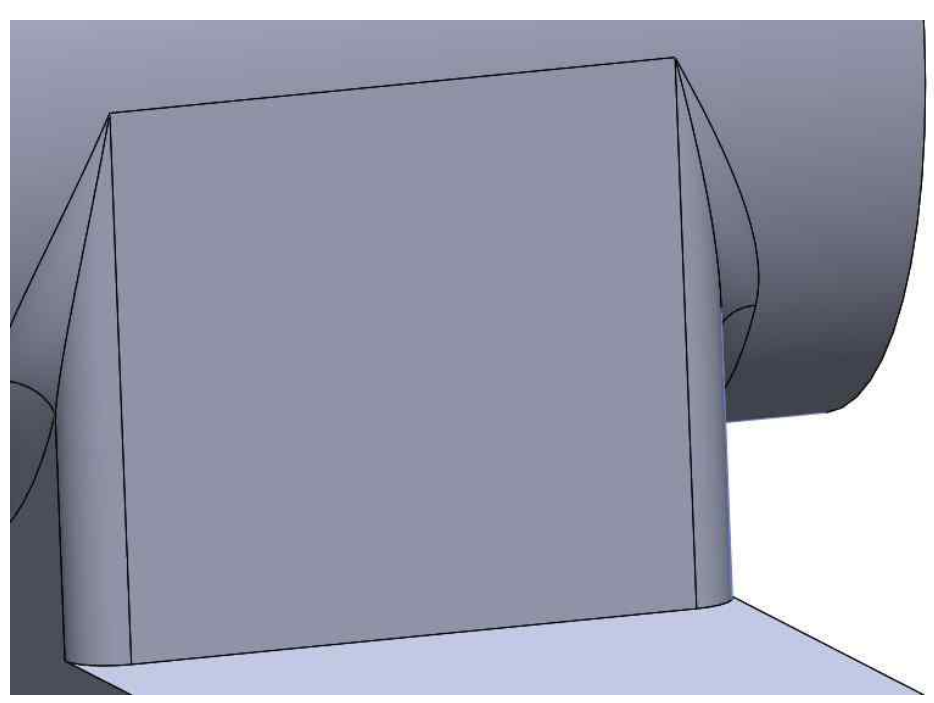 | 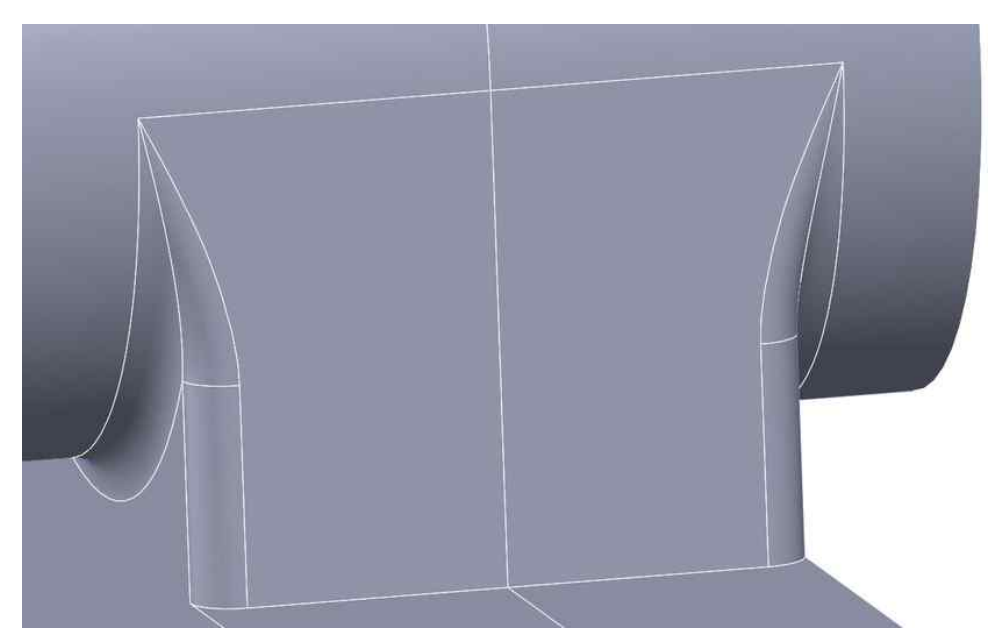 |
|---|---|
| **(a) Rendered reference model** | **(b) Rendered generated model** |

Figure 4-4 Details of CAD reconstruction for D2C-P02

## D2C-P03 TITAN-513M (moderate)

### Reconstruction results

| | | |
|---|---|---|
| 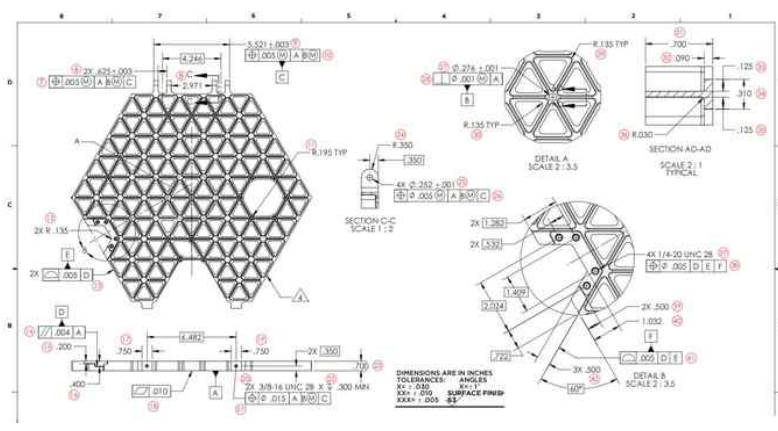 | 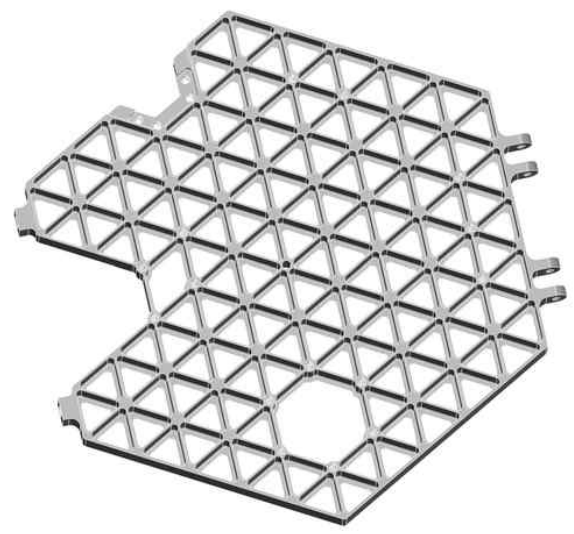 | 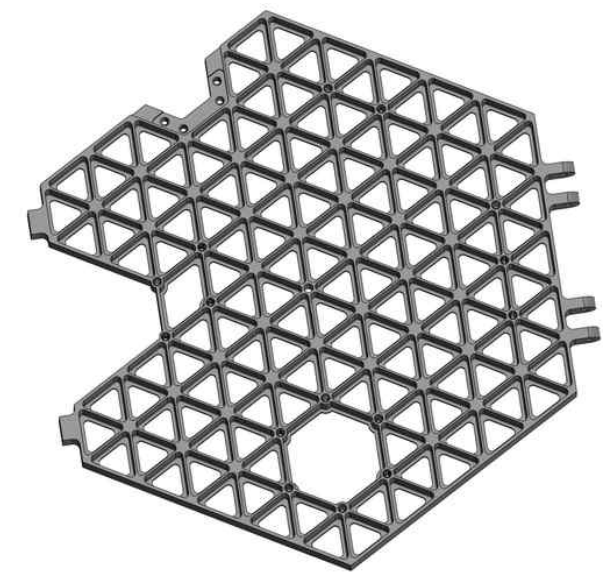 |
| **(a) Example engineering drawing** | **(b) Rendered reference model** | **(c) Rendered generated model** |

Figure 4-5 CAD reconstruction results for D2C-P03

The generated model contains invalid face intersections. Its overall structure nevertheless largely matches the reference model: the regular patterns, local openings, and main geometric relationships are reproduced correctly, so it passes the structural reconstruction check. The problem lies at the entrance to a threaded hole, where the chamfer differs from the reference model (Figure 4-6). The generated chamfer does not transition correctly to the thread's minor diameter, causing errors in the local geometry and dimensions at the thread entrance and resulting in invalid face intersections. The case therefore passes the structural reconstruction check but fails the dimensional accuracy check.

| | |
|---|---|
| 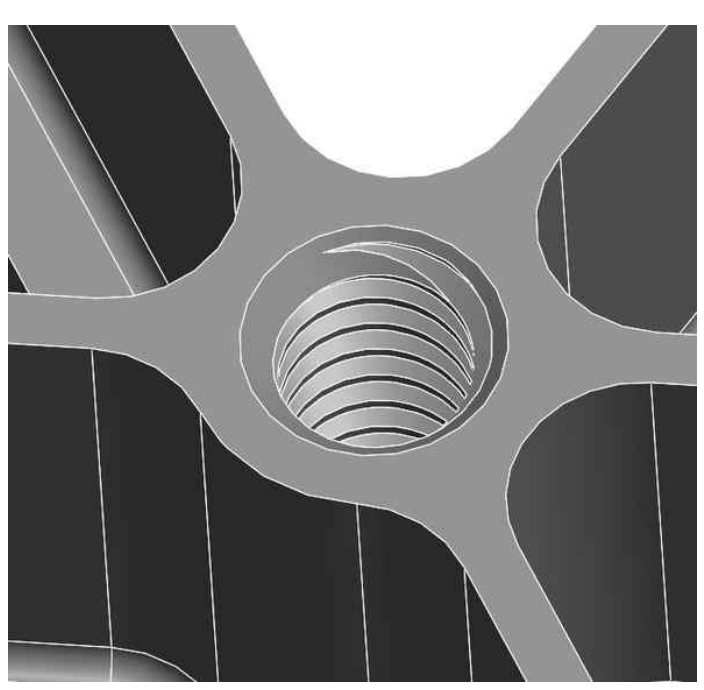 | 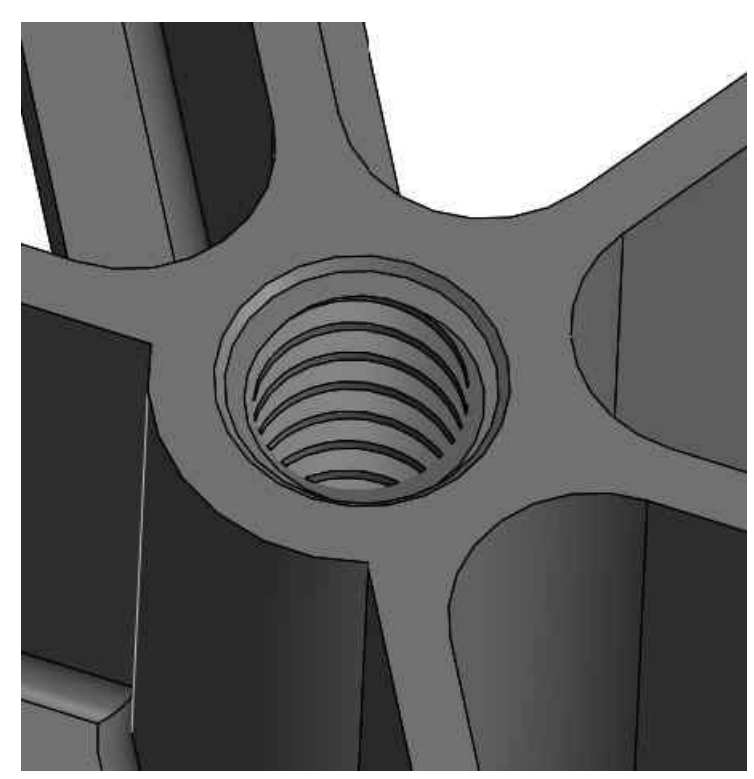 |
| **(a) Rendered reference model** | **(b) Rendered generated model** |

Figure 4-6 Details of CAD reconstruction for D2C-P03

## D2C-P04 TITAN-820 (moderate)

### Reconstruction results

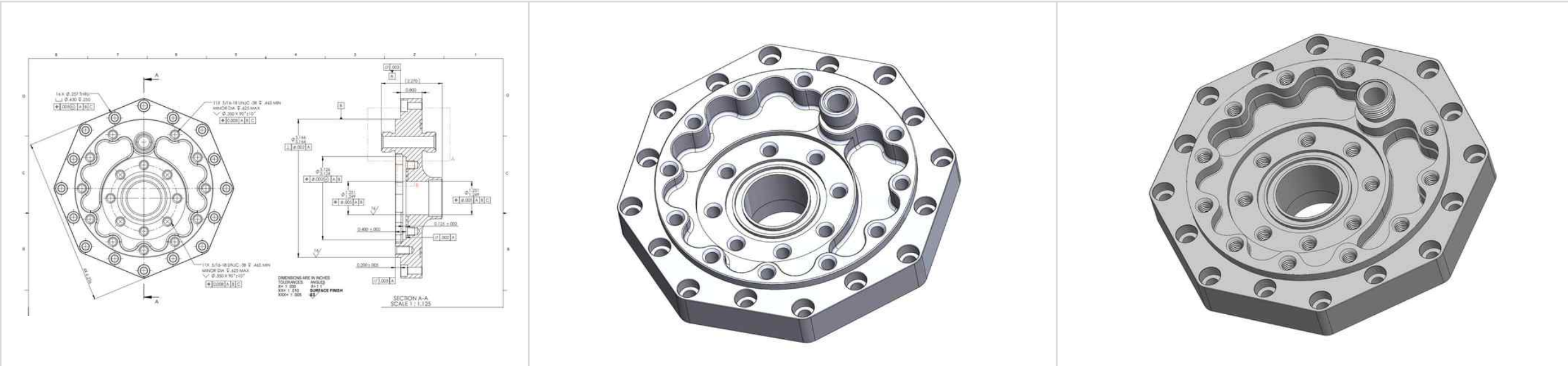

| (a) Example engineering drawing | (b) Rendered reference model | (c) Rendered generated model |
|---|---|---|

Figure 4-7 CAD reconstruction results for D2C-P04

The generated model is a valid solid and largely reproduces the reference model's overall structure. GPT-6 also constructs the threads correctly from the thread callouts in the drawing, indicating accurate recognition of these local machining features. However, it misinterprets the sequence of surfaces along an internal side wall shown in a section view. As Figure 4-8 illustrates, the reference geometry has a continuous transition from a chamfer to a side wall and then a fillet. The generated model instead uses two levels of side wall followed by a fillet, giving an incorrect internal profile and local dimensional relationships. The main structure is reproduced well, but the complex internal section is not interpreted accurately.

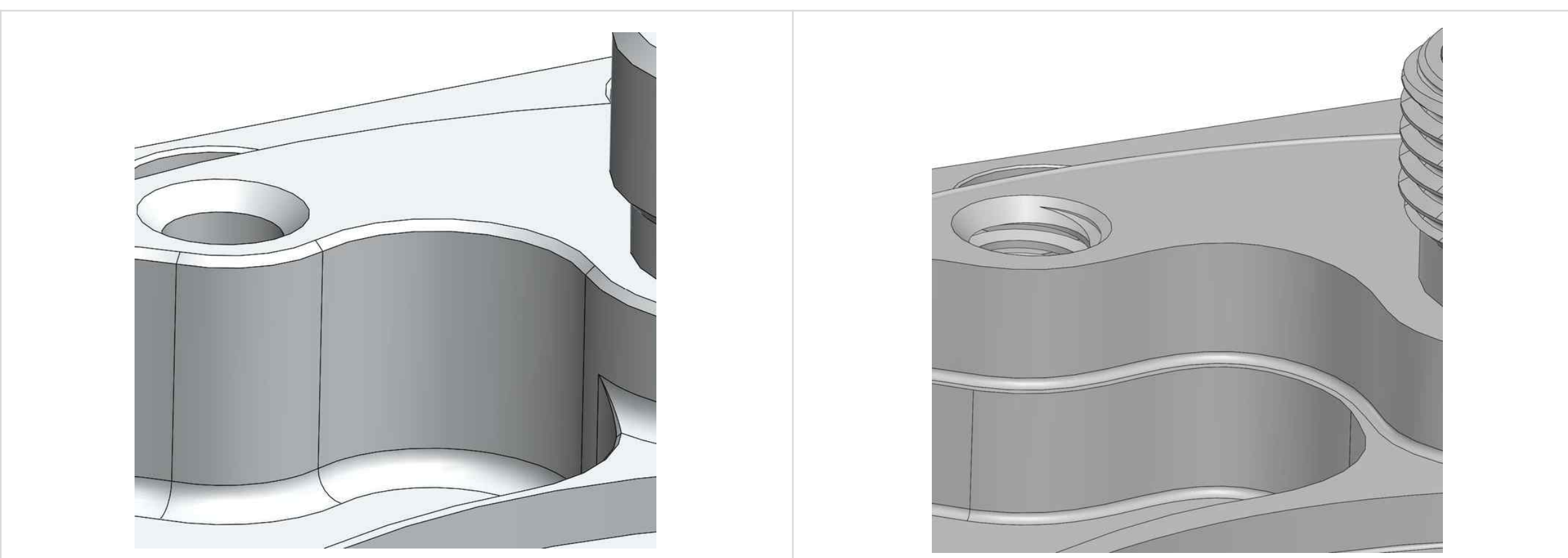

| (a) Rendered reference model | (b) Rendered generated model |
|---|---|

Figure 4-8 Details of CAD reconstruction for D2C-P04

## D2C-P05 Engine block (complex)

### Reconstruction results

| | | |
|---|---|---|
| 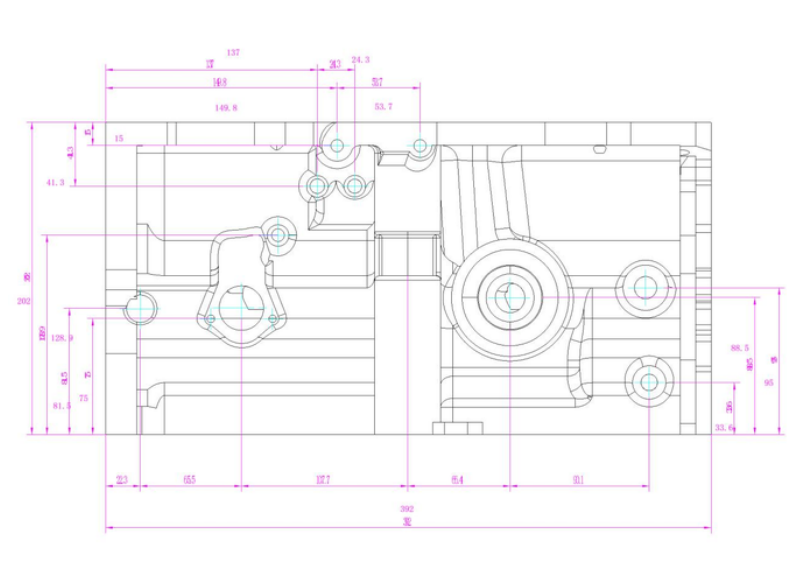 | 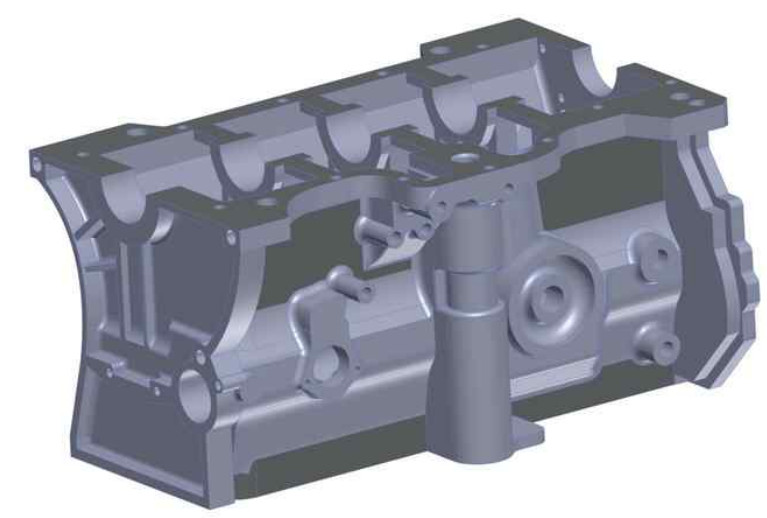 | 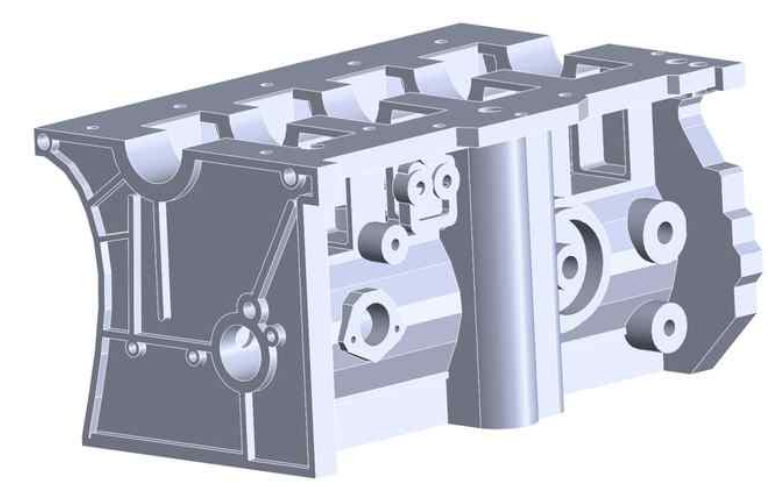 |
| **(a) Example engineering drawing** | **(b) Rendered reference model** | **(c) Rendered generated model** |

Figure 4-9 CAD reconstruction results for D2C-P05

The generated model is a valid solid and reproduces the engine block's main shape, hole layout, internal cavities, and reinforcing structures reasonably well. Its main structure largely matches the reference model, but complex local geometry and dimensional relationships show substantial errors. In some regions, GPT-6 fails to reproduce holes opening into a cavity, and the central cavity has an incorrect sharp-cornered profile (Figure 4-10). Some holes that should pass through the part are modeled as blind holes, changing the local topology. Step dimensions, cavity and hole depths, and local wall thicknesses also differ from the reference. Overall, the main structure is broadly correct, but the case fails the checks for completeness of local structures and dimensional accuracy.

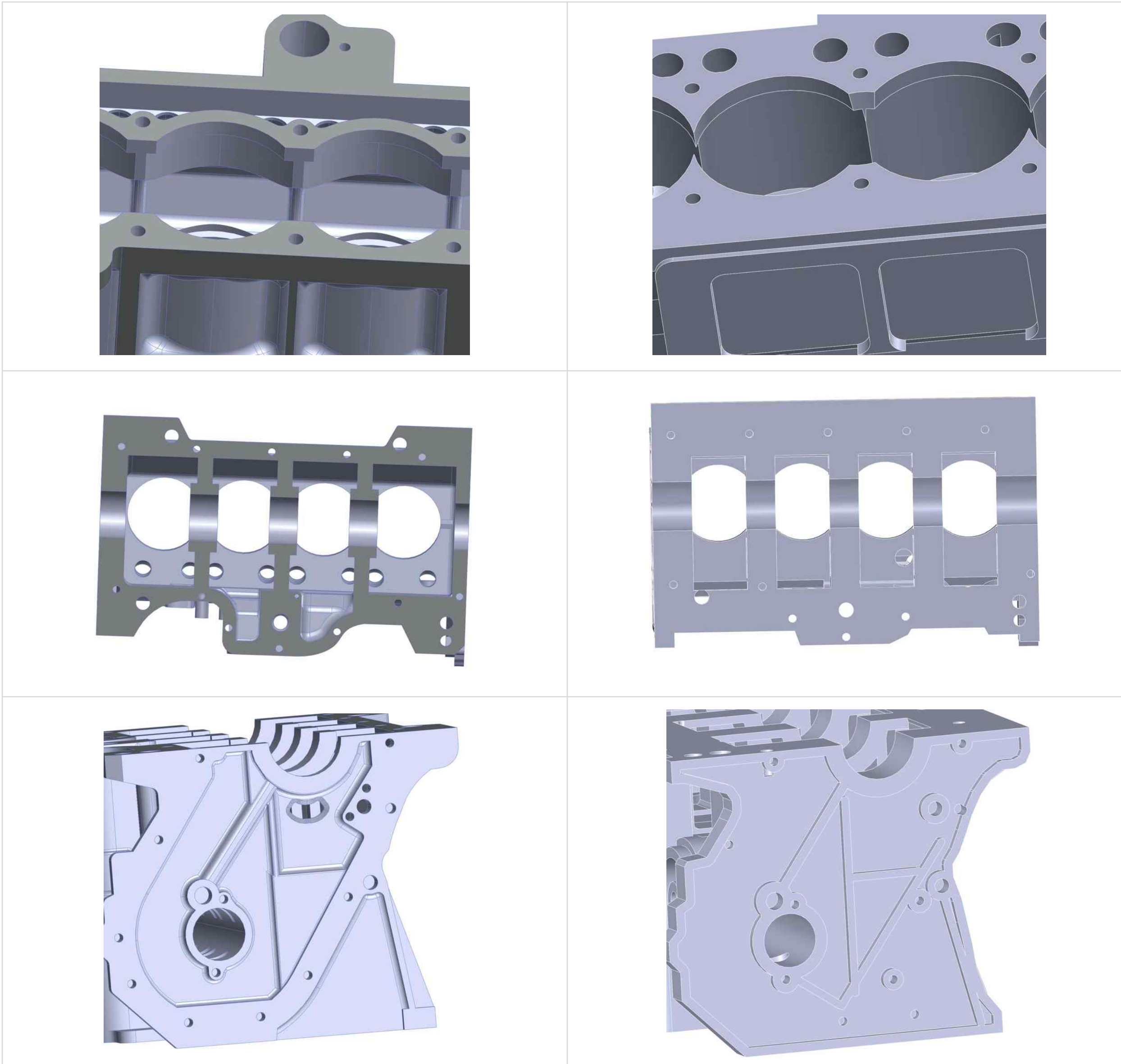

| **(a) Rendered reference model** | **(b) Rendered generated model** |
|---|---|

Figure 4-10  Details of CAD reconstruction for D2C-P05

## D2C-A01 Sliding assembly for a milling task (moderate)

### Reconstruction results

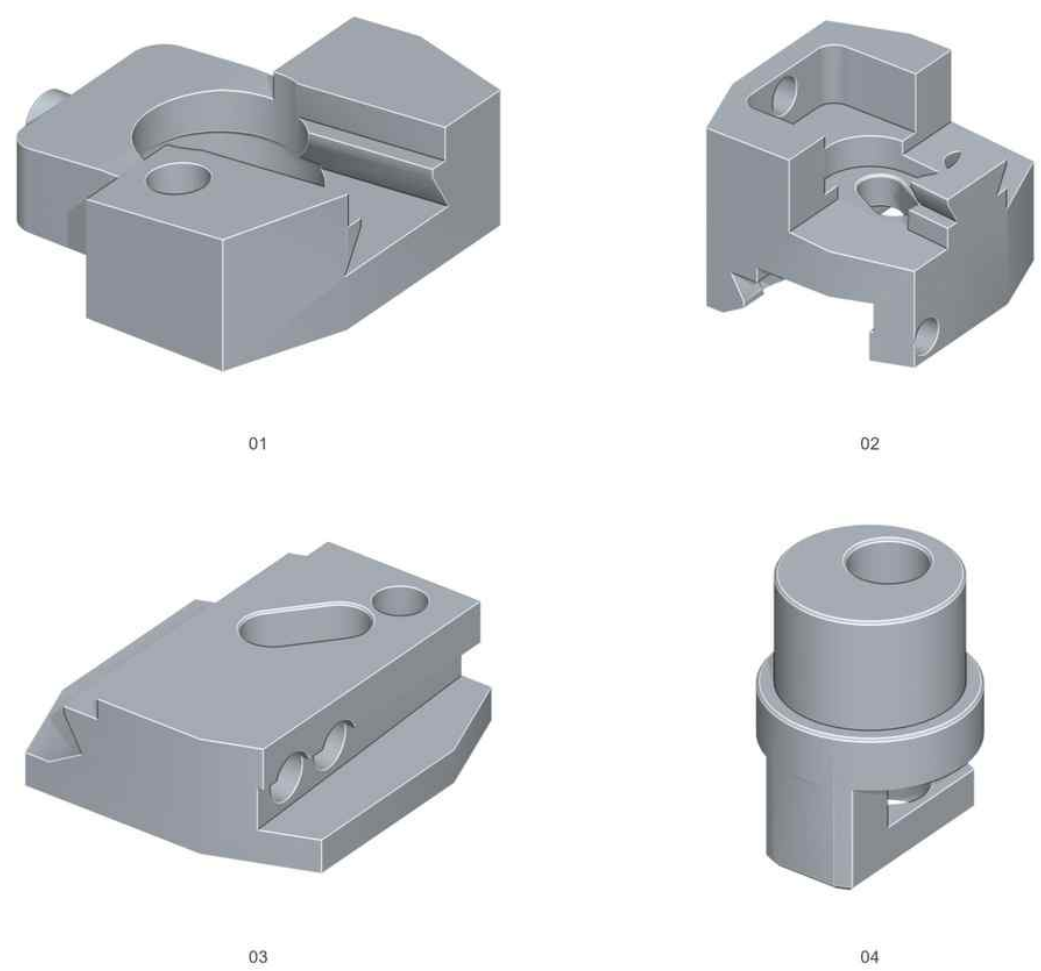

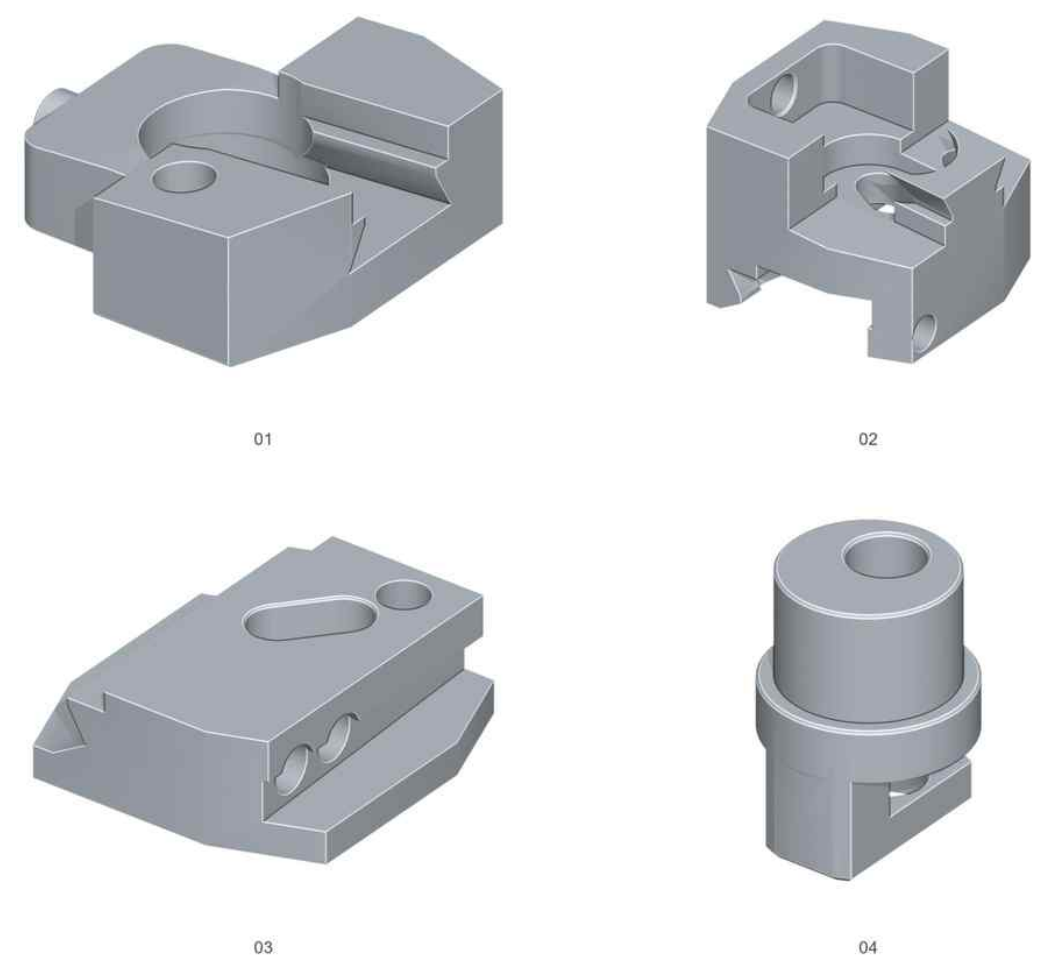

| (a) Rendered reference model | (b) Rendered generated model |
| --- | --- |

Figure 4-11 CAD reconstruction results for D2C-A01

### Assembly verification

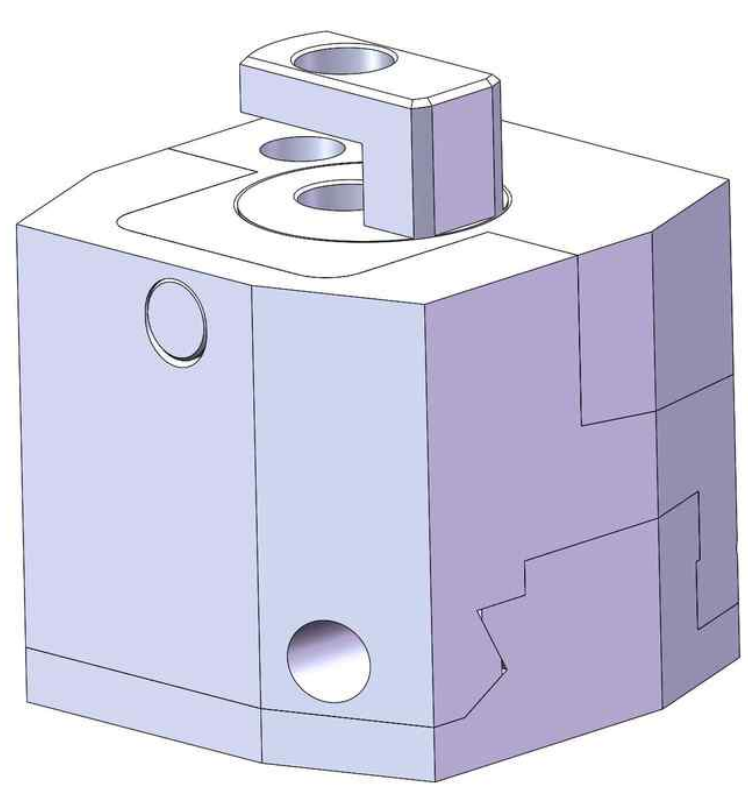
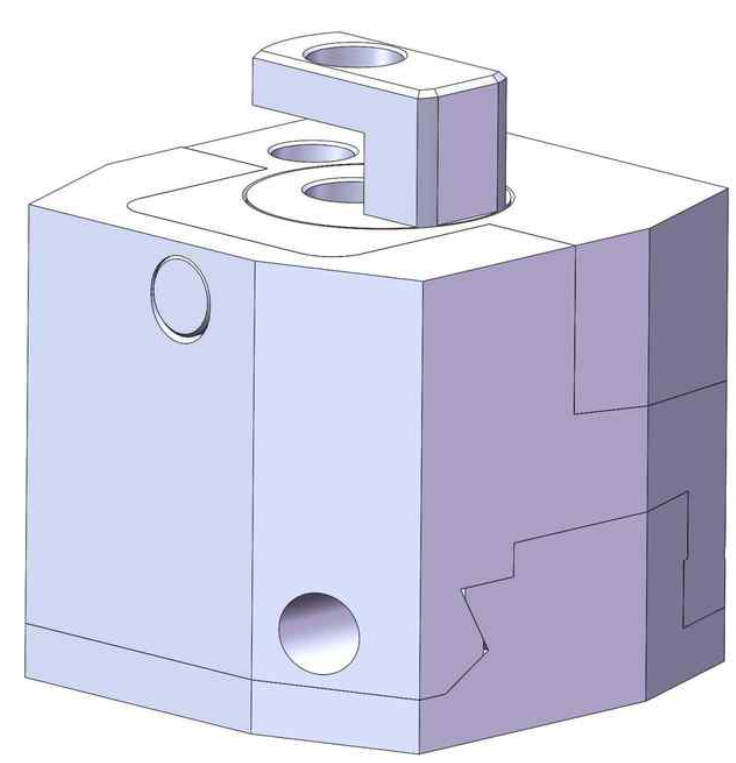

| (a) Rendered reference model | (b) Rendered generated model |
| --- | --- |

Figure 4-12 Assembly verification for D2C-A01

All generated parts are valid solids. Their main structures, principal holes and slots, and mating relationships are reproduced almost completely. The assembled geometry largely matches the reference, and the parts can be assembled as intended, so the overall structural reconstruction is correct. A few local dimensions and geometric parameters differ. In Figure 4-13, for example, an inclined cut should remove a sharp corner, but the generated model does not fully reproduce the inclined face and leaves a sharp remnant. This has no apparent effect on the overall assembly relationships, although the local angle and dimensions are inaccurate.

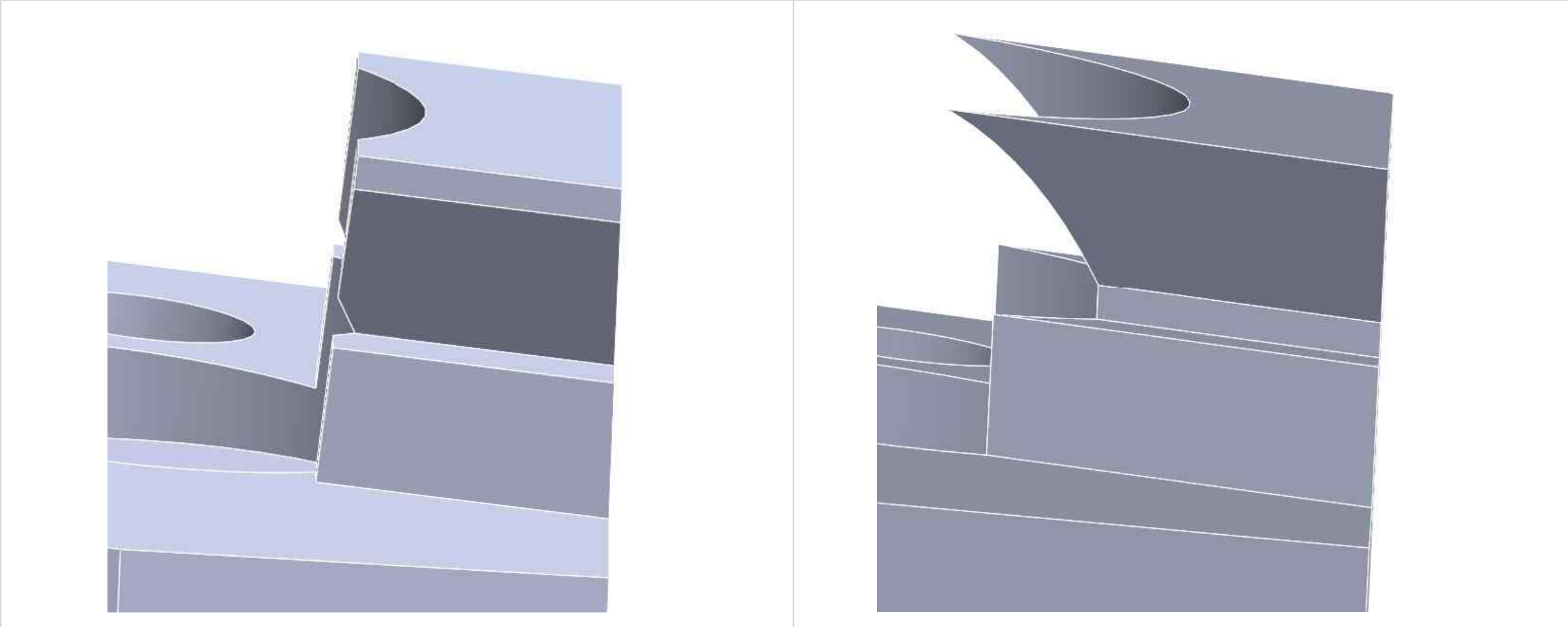

**(a) Rendered reference model** | **(b) Rendered generated model**

Figure 4-13  Details of CAD reconstruction for D2C-A01

## D2C-A02 Vise assembly (complex)

### Reconstruction results

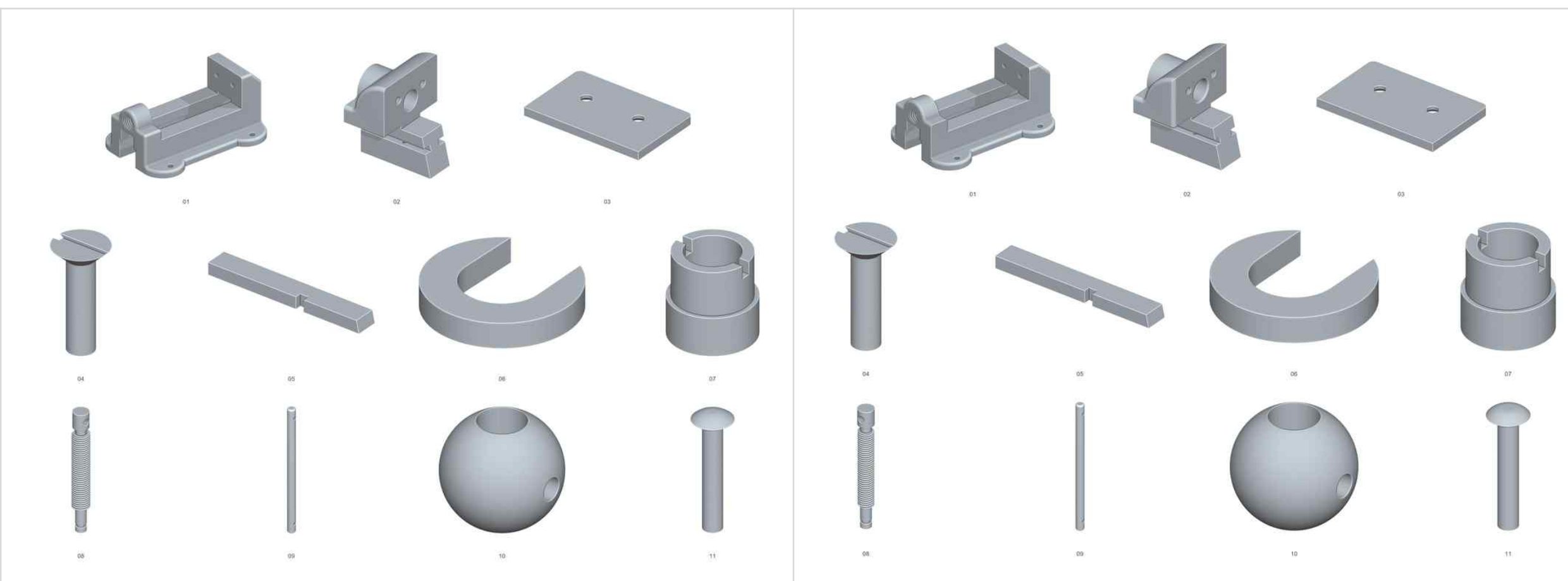

**(a) Rendered reference model** | **(b) Rendered generated model**

Figure 4-14  CAD reconstruction results for D2C-A02

**Assembly verification**

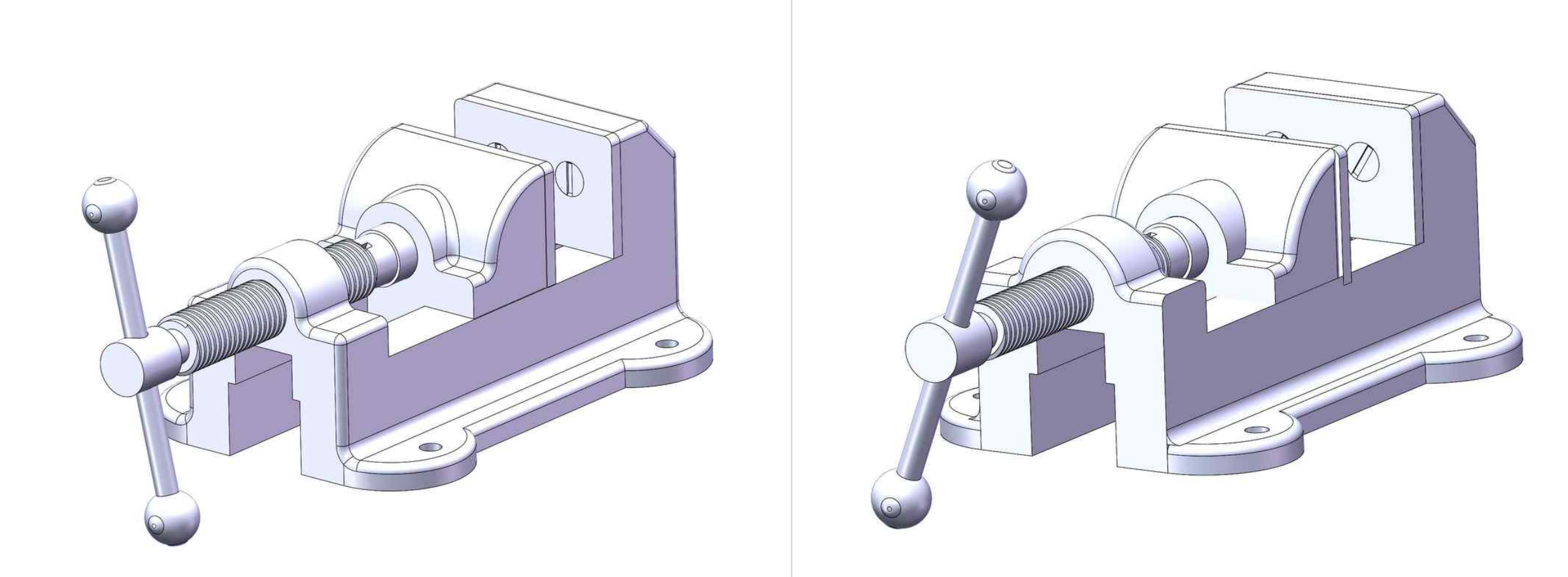

| (a) Rendered reference model | (b) Rendered generated model |
| --- | --- |

Figure 4-15 Assembly verification for D2C-A02

All generated parts are valid solids, and their main structures, principal features, and overall assembly relationships are reproduced almost completely. The overall shape largely matches the reference, indicating that GPT-6 can interpret how the structures of different parts correspond. However, dimensional errors in some parts cause local overlaps and interference after assembly. Figure 4-16 shows overlapping solids in the screw-drive region and at internal mating interfaces. Structural reconstruction and interpretation of the assembly relationships are therefore broadly correct, but dimensional accuracy is insufficient and affects the assembled result. The vise body also contains an abnormal configuration in which one fillet lies below another, as visible around its filleted and chamfered regions in the preceding assembly view. This error is consistent with difficulty interpreting the local roles of chamfers and fillets, although the output alone does not establish its cause.

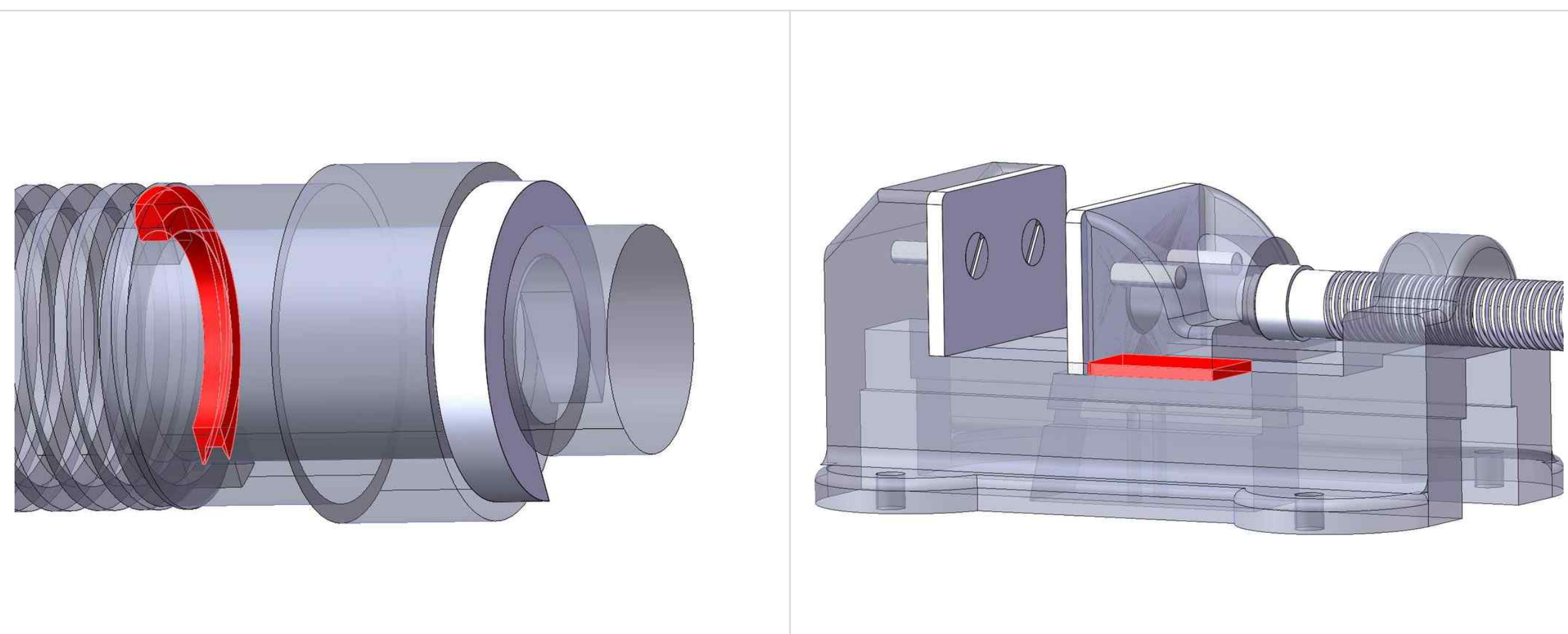

| (a) Overlapping parts, location 1 | (b) Overlapping parts, location 2 |
| --- | --- |

Figure 4-16 Details of CAD reconstruction for D2C-A02

## D2C-A03 Six-cylinder engine (complex)

### Reconstruction results

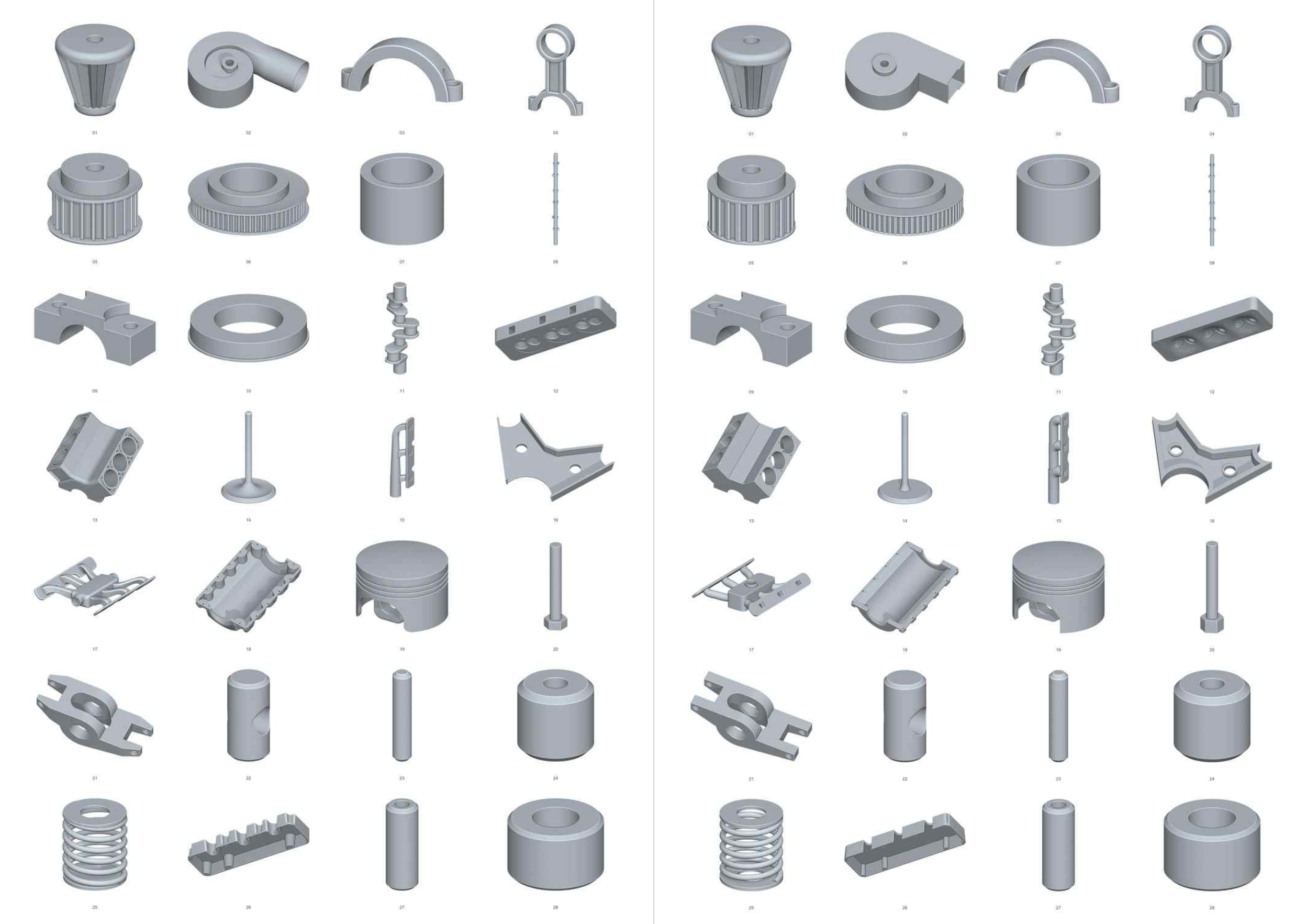

| (a) Rendered reference model | (b) Rendered generated model |
|---|---|

Figure 4-17 CAD reconstruction results for D2C-A03

Assembly verification could not be completed. Key holes needed for pipe connections and part alignment are not generated correctly in some parts, preventing the corresponding pipes and connectors from being assembled as intended.

Three part models fail the solid validity check. The defects include non-manifold edges and vertices, invalid face intersections, and a model consisting of only two surfaces with no solid. The main shapes and basic structures of most parts are reproduced, but some essential hole layouts and complex local structures are missing, and some dimensions are inaccurate. Figure 4-18 shows that the complex surfaces, thin walls, and transitions created by operations such as lofting and shelling in the reference model are not reproduced accurately. Some regions are approximated with more basic modeling operations. Missing holes for pipe connections and part location leave the affected parts without the interfaces needed for assembly, preventing complete assembly verification. Other defects include missing pipe openings, and some holes are placed incorrectly. The generated models thus capture the overall forms to some extent, but fail to meet the requirements for completeness of complex structures and dimensional accuracy.

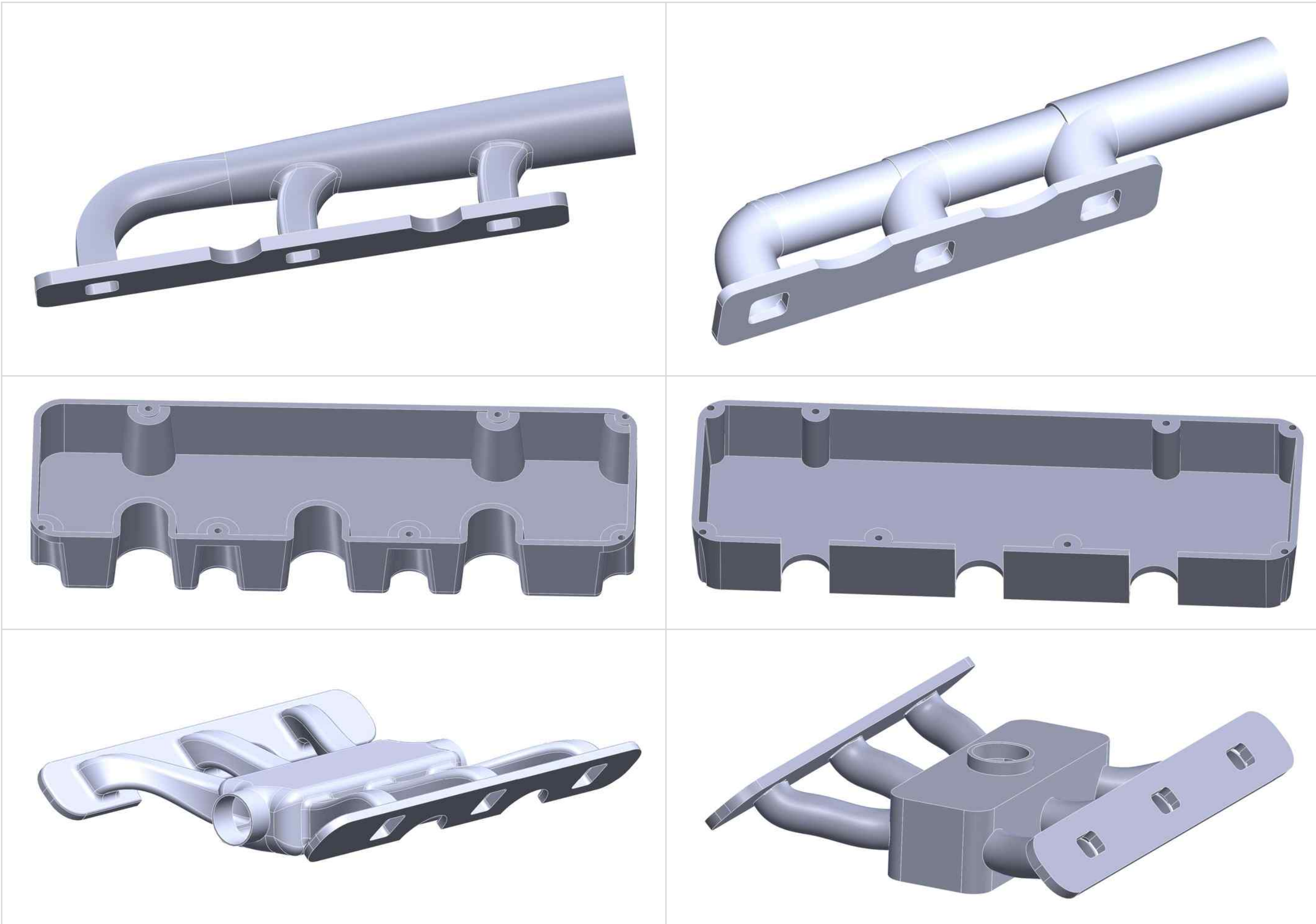

**(a) Rendered reference model** | **(b) Rendered generated model**

Figure 4-18 Details of CAD reconstruction for D2C-A03

Another observation concerns the drawing in Figure 4-19. Comparison with the shaft, whose radius is 8 mm, shows that the hexagonal head should be 6 mm thick. GPT-6 fails to account for the inverted numeral and generates a head thickness of 9 mm instead.

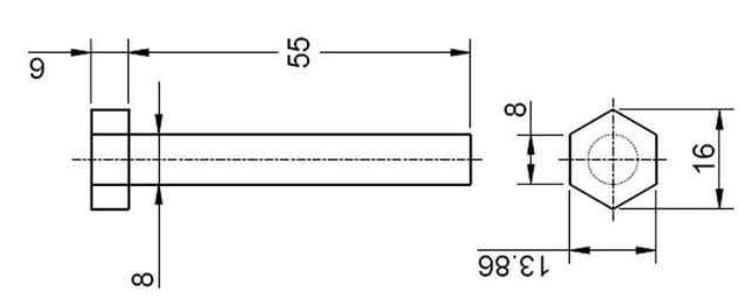

Figure 4-19 Engineering drawing misread by GPT-6 in D2C-A03

### 4.3.2 Additional experiment results

The additional experiments are presented in the order defined in Section 4.2.2: missing local dimensions, followed by disabled web search. Each test is compared with the main experiment for the same case. Only the specified condition changes; input preparation, the prompt, and all other test settings remain the same.

#### 4.3.2.1 Missing local dimensions

In this condition, only overall dimensions, such as total length, width, height, and maximum outside diameter, are retained; all other dimensions are removed. The following comparisons show the inputs and outputs for selected cases in the main experiment and under this condition, then examine the changes associated with missing local dimensions. Table 4-8 reports completion times.

Table 4-7 CAD reconstruction results with local dimensions removed

| Case ID | Complexity | Solid validity | Reconstruction of structural features | Dimensional accuracy | Assembly results |
|---|---|---|---|---|---|
| D2C-P01 | Simple | Pass | Pass | Pass | N/A |
| D2C-P03 | Moderate | Pass | Slight deviations | Slight deviations | N/A |
| D2C-A01 | Moderate | Pass | Structural errors in 3 part models | Some deviations | Cannot be assembled |
| D2C-A02 | Complex | Pass | Largely reconstructed | Some deviations | Parts overlap when assembled |

Solid validity is reported for each complete case. A fraction gives the number of valid part models divided by the number checked; “Pass” means that all part models passed.

Table 4-8 Completion times with local dimensions removed

| Case ID | S1_D2C-P01 | S1_D2C-P03 | S1_D2C-A01 | S1_D2C-A02 |
|---|---|---|---|---|
| Completion time | 18 min | 24 min | 19 min | 34 min |

## D2C-P01 General-purpose bracket (simple)

### Comparison of results

| | |
|---|---|
| 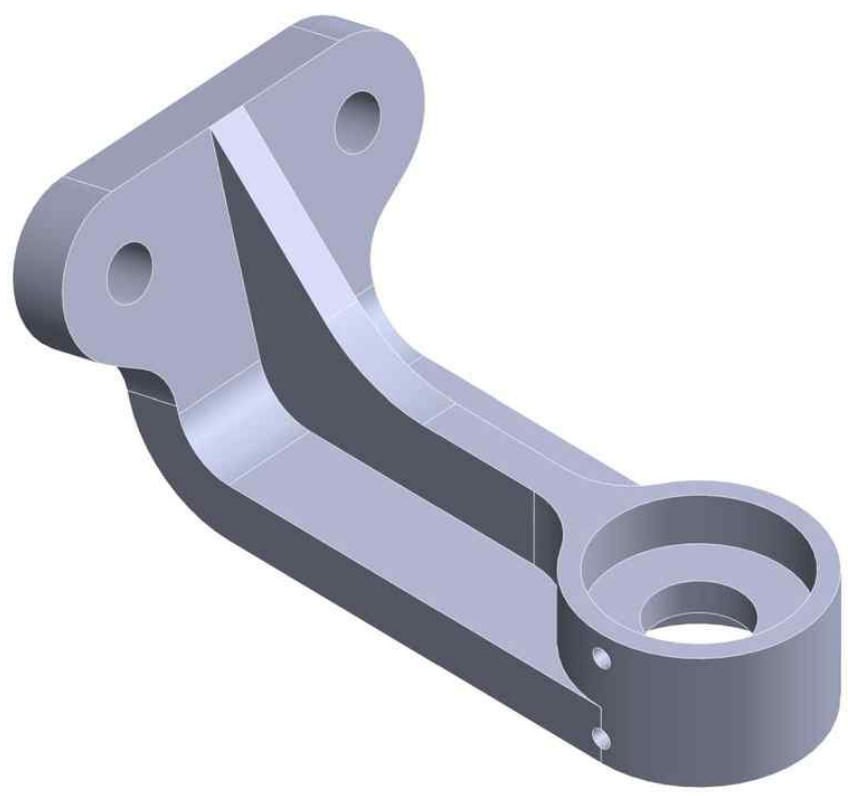 | 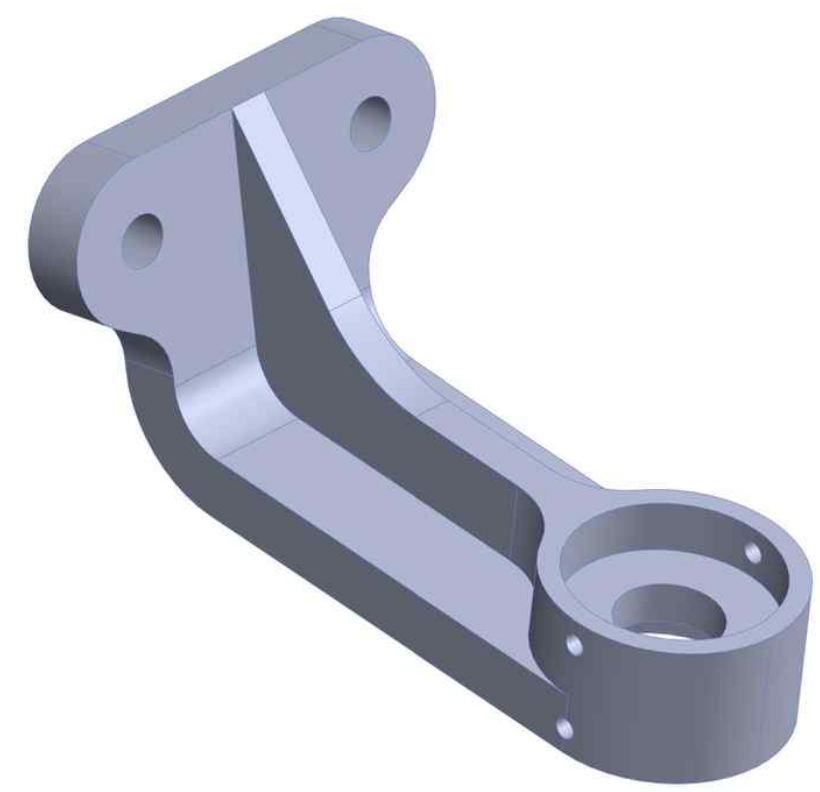 |
| **(a) Model generated with complete dimensions** | **(b) Model generated with only overall dimensions** |

Figure 4-20 D2C-P01 with complete and reduced dimensional information

With local dimensions removed, the generated model reproduces the part in the drawing well. It also includes all the holes that were missing from the model generated with complete dimensions.

## D2C-P03 TITAN-513M (moderate)

### Comparison of results

| | |
|---|---|
| 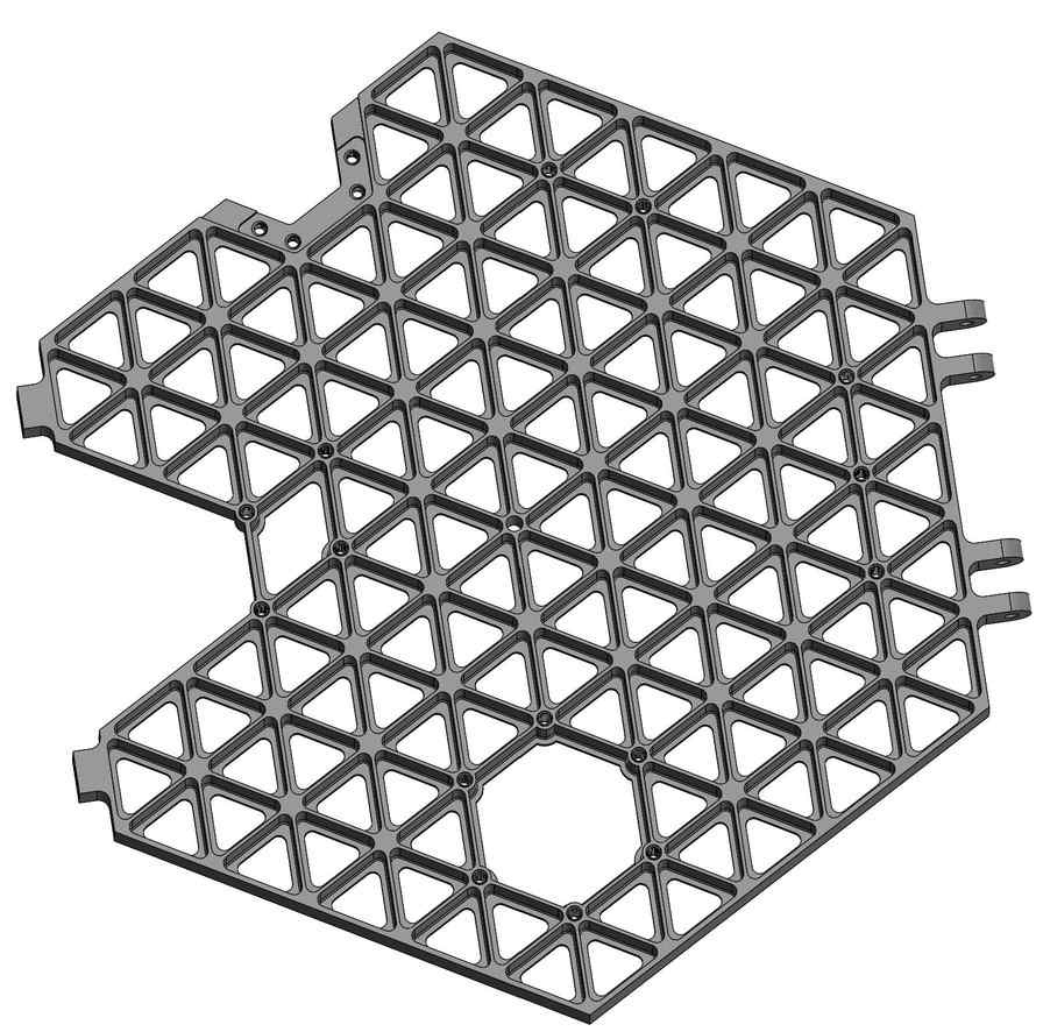 | 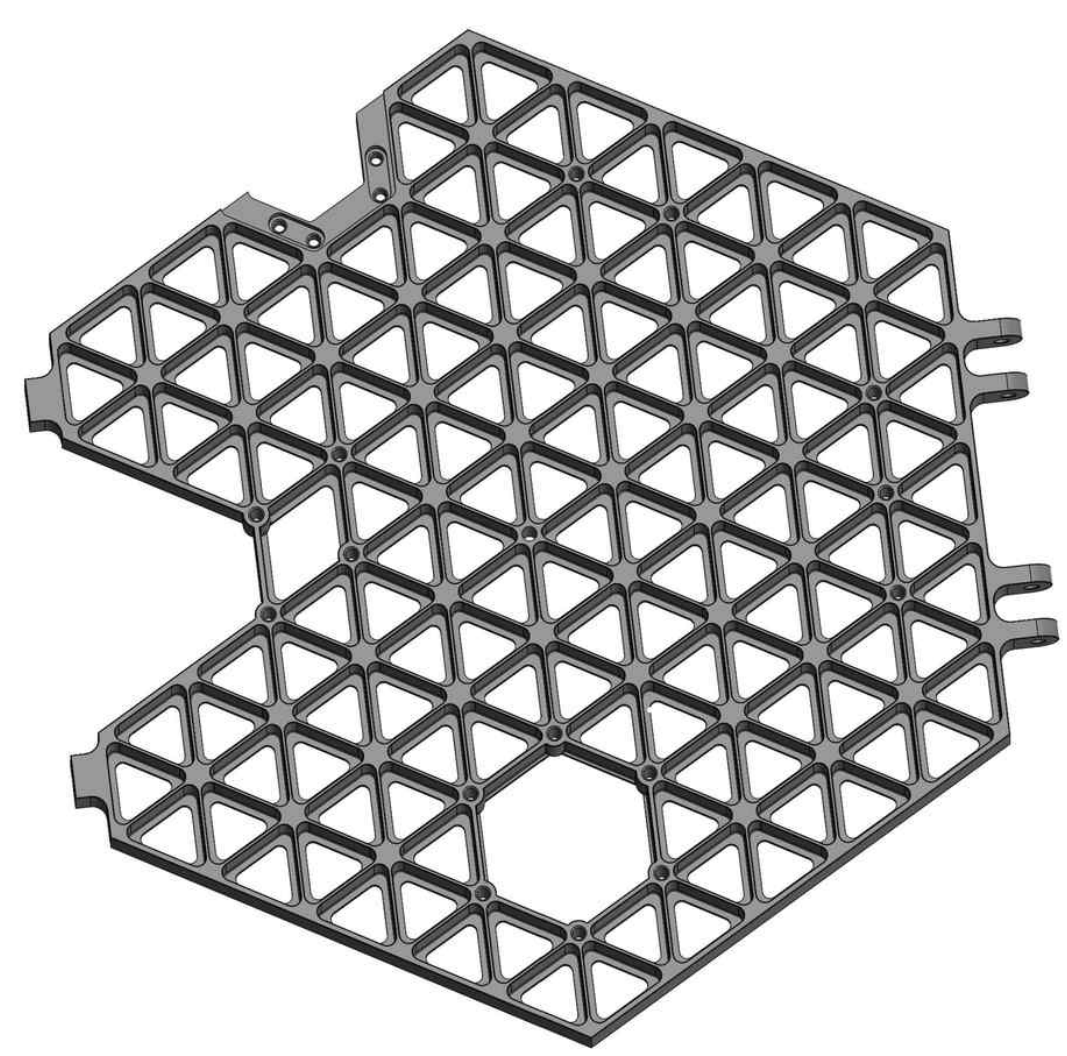 |
| **(a) Model generated with complete dimensions** | **(b) Model generated with only overall dimensions** |

Figure 4-21 D2C-P03 with complete and reduced dimensional information

Even with local dimensions removed, the generated model reproduces almost the entire part. Threads are not modeled because their callouts have been removed, while some thickness differences are reported as acceptable in the case evaluation. GPT-6 nevertheless adds chamfers to the holes that should contain threads, indicating that it has captured some of the detail in the drawing's top view.

## D2C-A01 Sliding assembly for a milling task (moderate)

### Comparison of results

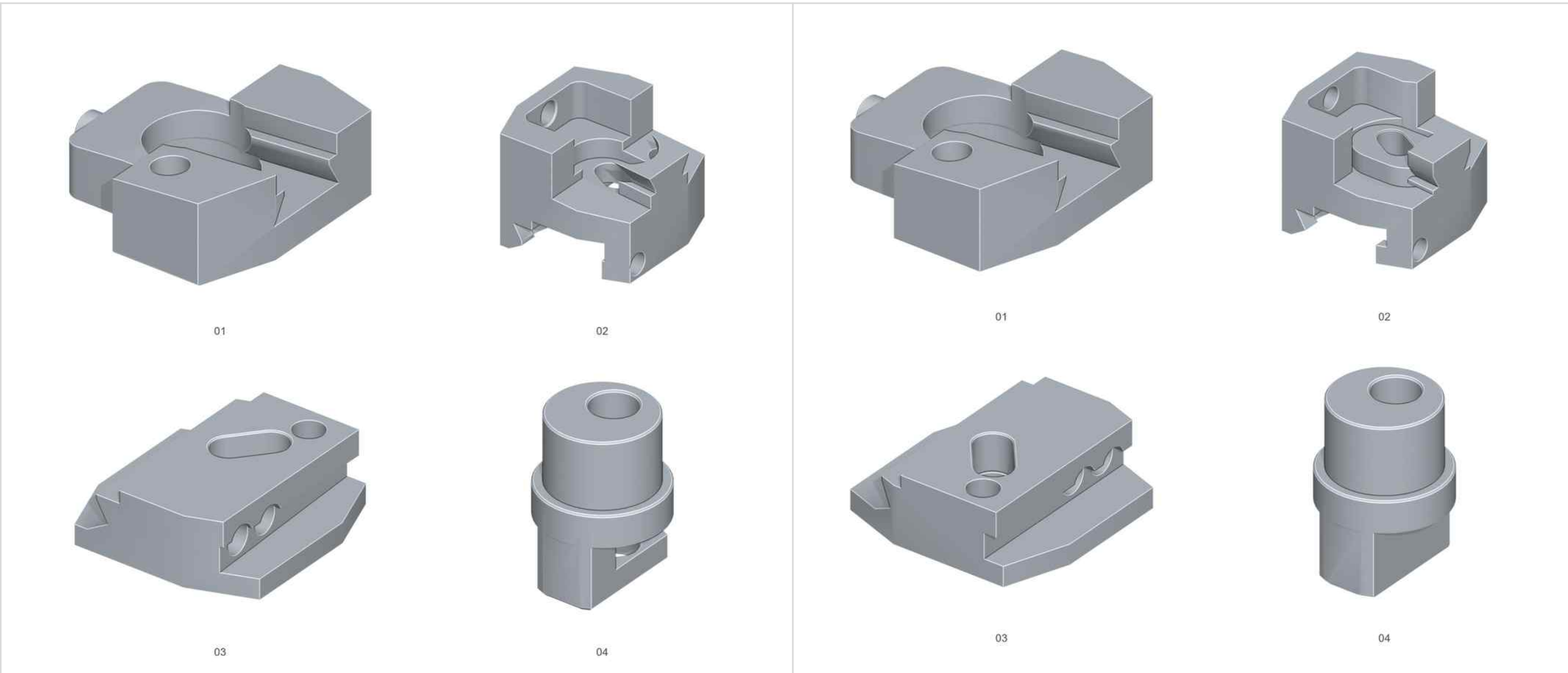

| (a) Model generated with complete dimensions | (b) Model generated with only overall dimensions |
|---|---|

Figure 4-22 D2C-A01 with complete and reduced dimensional information

With the full set of views still available, GPT-6 reconstructs the overall outlines and most main structures of the four parts despite the removal of local dimensions. The specific shapes and locations of features, and their dimensional relationships, show larger errors. The holes, slots, and local pockets in the figure differ substantially from those generated with complete dimensions.

## D2C-A02 Vise assembly (complex)

### Comparison of results

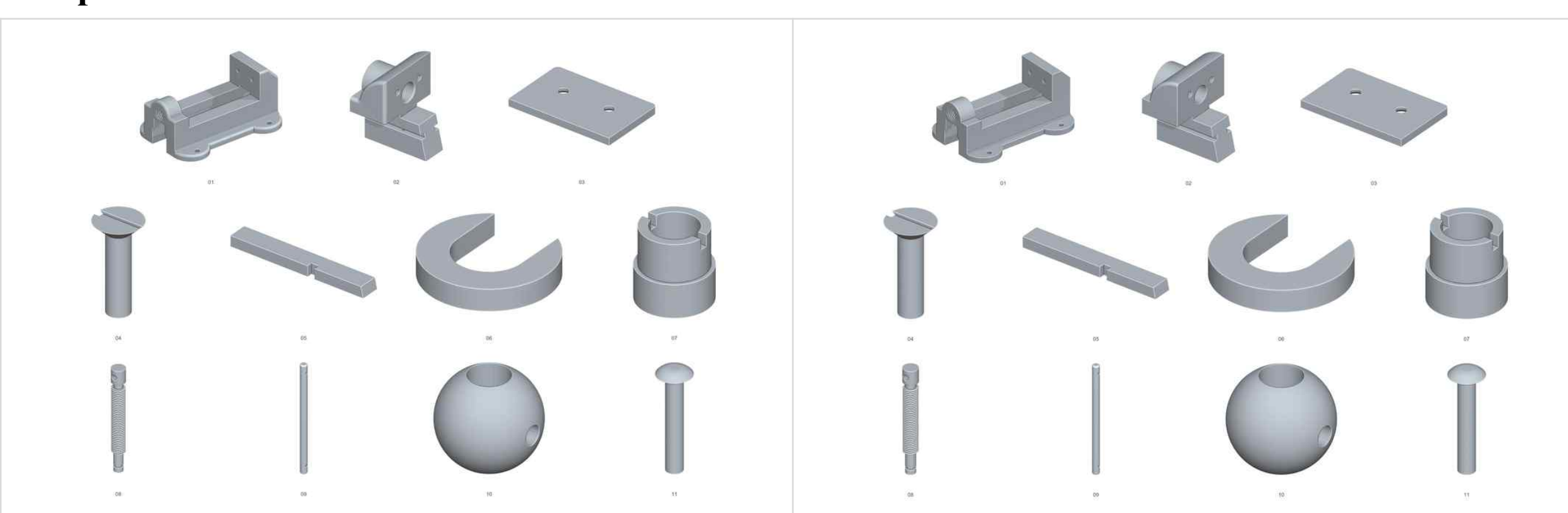

| (a) Model generated with complete dimensions | (b) Model generated with only overall dimensions |
|---|---|

Figure 4-23 D2C-A02 with complete and reduced dimensional information

With local dimensions removed, GPT-6 still largely reproduces the main structures and overall shapes of the vise components. However, local dimensions and mating dimensions differ more substantially than in the models generated with complete dimensions, and assembly quality deteriorates markedly. Figure 4-24 shows overlapping solids and contact surfaces that do not meet correctly in some mating regions. In this comparison, missing local dimensions are associated with poorer dimensional compatibility between parts and a less accurate assembled model.

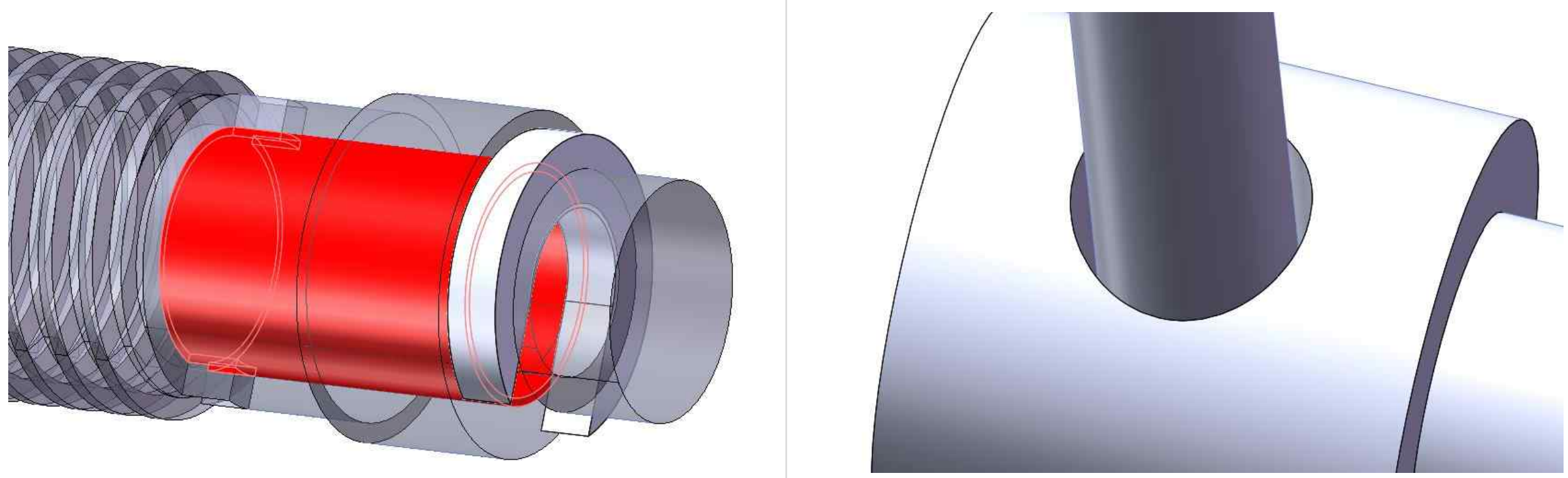

**(a) Overlapping parts** | **(b) Mating surfaces that do not meet correctly**

Figure 4-24  Details of D2C-A02 with local dimensions removed

#### 4.3.2.2 Web search disabled

Web search is disabled in this condition, while the input drawings and prompts remain the same as in the main experiment. The following comparisons show the models generated for the same cases with web search enabled and disabled. Table 4-10 reports completion times.

Table 4-9  CAD reconstruction results with web search disabled

| Case ID | Complexity | Solid validity | Reconstruction of structural features | Dimensional accuracy | Assembly results |
|---|---|---|---|---|---|
| D2C-P01 | Simple | Pass | Pass | Pass | N/A |
| D2C-P03 | Moderate | Pass | Slight deviations | Slight deviations | N/A |
| D2C-A01 | Moderate | Fail<br>3/4 passed | Largely reconstructed | Some deviations | Parts overlap when assembled |
| D2C-A02 | Complex | Pass | Some structures missing | Some deviations | Cannot be assembled |

Solid validity is reported for each complete case. A fraction gives the number of valid part models divided by the number checked; "Pass" means that all part models passed.

Table 4-10  Completion times with web search disabled

| Case ID | S2_D2C-P01 | S2_D2C-P03 | S2_D2C-A01 | S2_D2C-A02 |
|---|---|---|---|---|
| Completion time | 21 min | 46 min | 21 min | 45 min |

## D2C-P01 General-purpose bracket (simple)

### Comparison of results

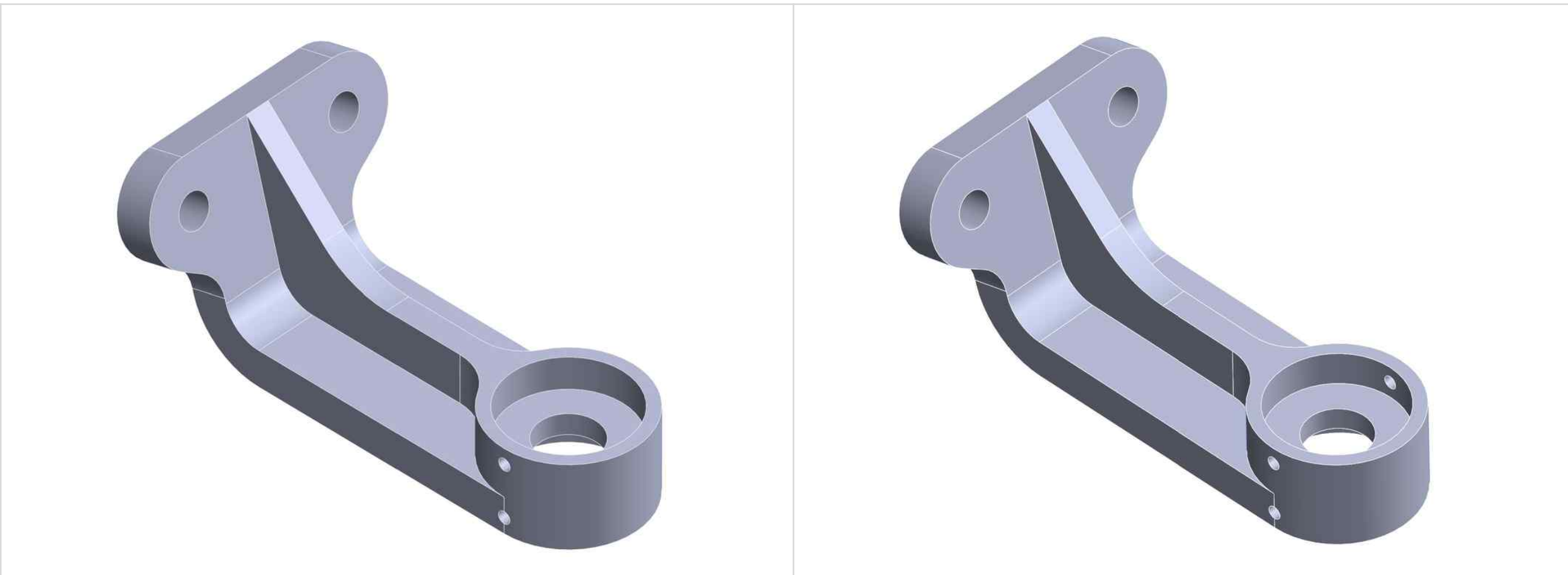

**(a) Model generated with web search enabled** | **(b) Model generated with web search disabled**

Figure 4-25  D2C-P01 with web search enabled and disabled

With web search disabled, the generated model reproduces the part in the drawing well. It includes all the holes that were missing from the model generated with web search enabled.

## D2C-P03 TITAN-513M (moderate)

### Comparison of results

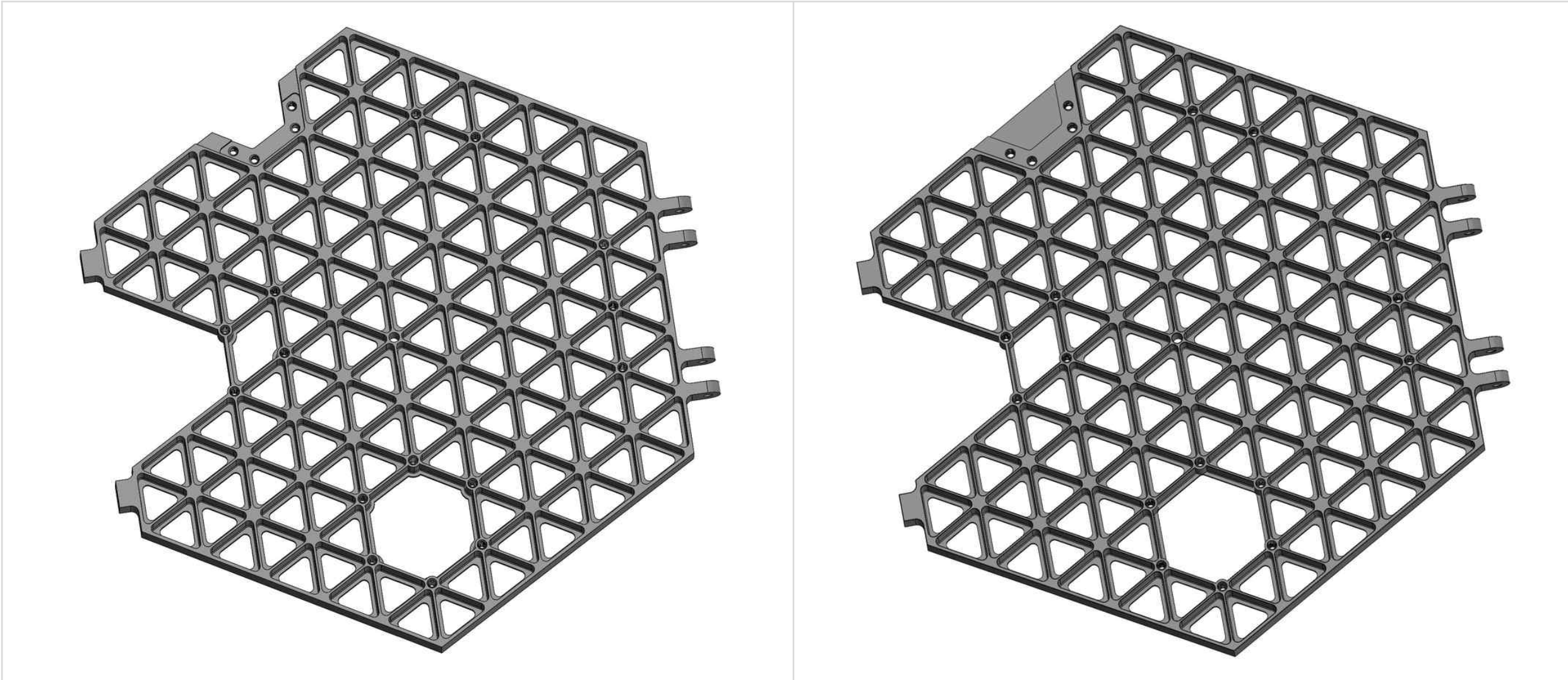

| (a) Model generated with web search enabled | (b) Model generated with web search disabled |
|---|---|

Figure 4-26 D2C-P03 with web search enabled and disabled

The generated model largely reproduces the part in the drawing, although some structural and local details differ. Notably, the earlier problem with chamfers at the threaded holes does not recur.

## D2C-A01 Sliding assembly for a milling task (moderate)

### Comparison of results

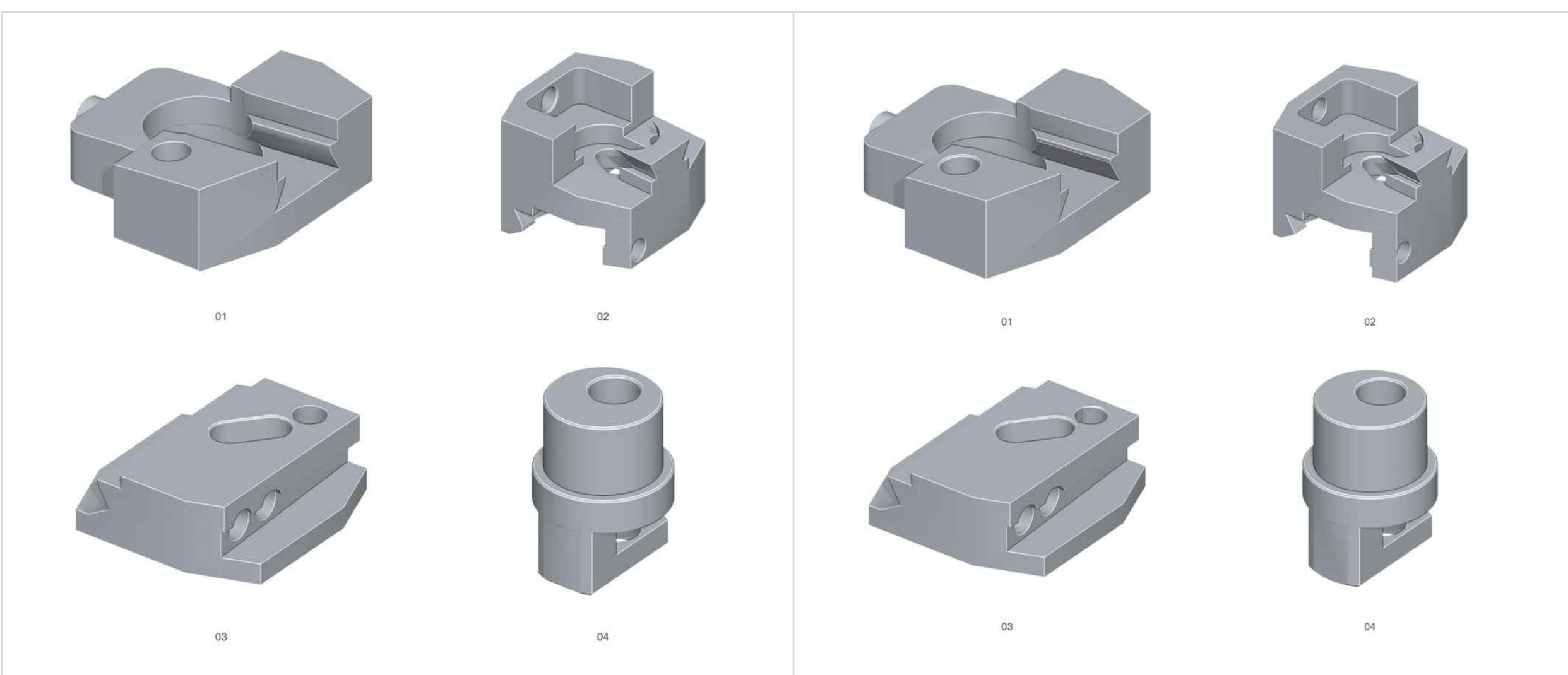

| (a) Model generated with web search enabled | (b) Model generated with web search disabled |
|---|---|

Figure 4-27 D2C-A01 with web search enabled and disabled

With web search disabled, GPT-6 still reproduces the main structures of the four parts reasonably well, but the assembly outcome is worse than in the main experiment. However, the part shown at the upper right contains extra solid material in its lower region. This changes its local shape relative to the reference and causes it to overlap an adjacent part during subsequent assembly.

Further defects include unintended hole geometry, as shown in Figure 4-28(a), and a non-manifold edge caused by line contact, as shown in Figure 4-28(b). These defects result from geometric parameters that do not match the drawing.

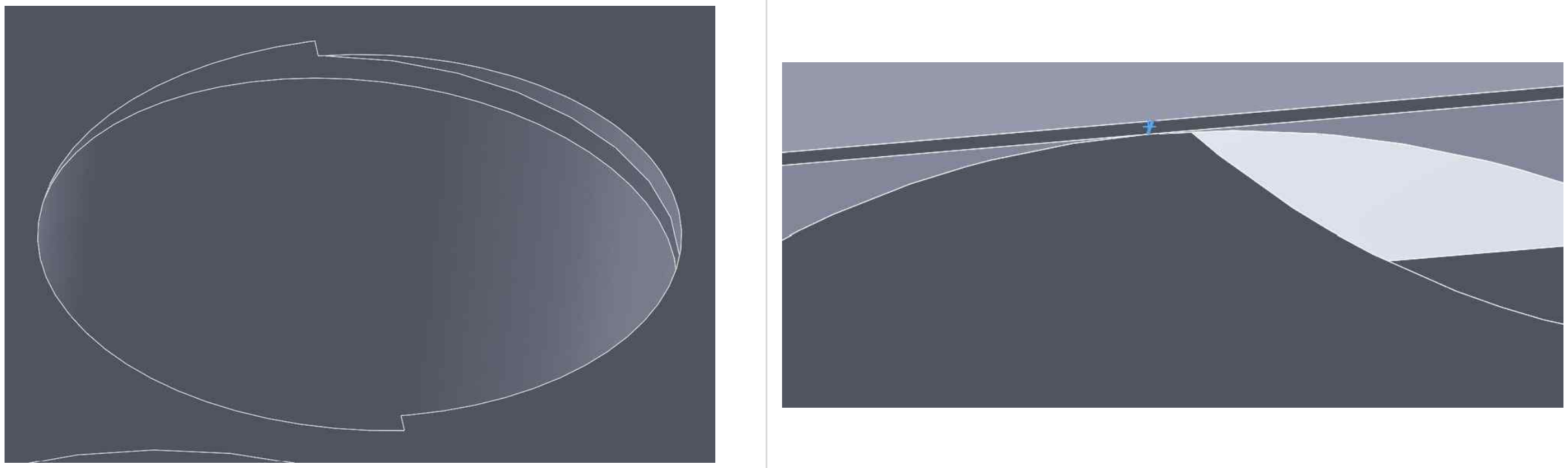

| **(a) Unnatural geometry** | **(b) Invalid solid with line contact** |
|---|---|

Figure 4-28  Details of D2C-A01 with web search disabled

### D2C-A02 Vise assembly (complex)

#### Comparison of results

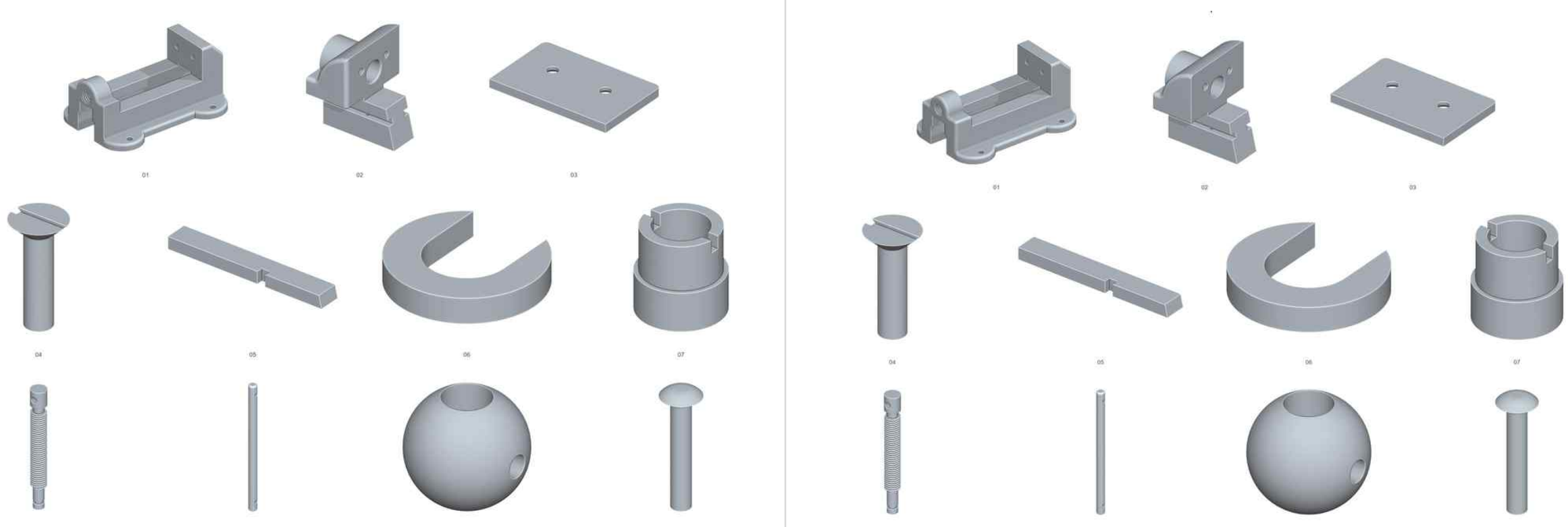

| **(a) Model generated with web search enabled** | **(b) Model generated with web search disabled** |
|---|---|

Figure 4-29  D2C-A02 with web search enabled and disabled

With web search disabled, GPT-6 retains the main component shapes and many principal features, and the previously incorrect fillet configuration is corrected. However, some internal threads are not reconstructed: the affected holes retain only their basic geometry. Table 4-9 records missing structures, dimensional deviations, and failure to assemble the parts. The overall result is therefore worse than in the main experiment, despite the improvement in the fillets.

## 4.4 Analysis of GPT-6's approach

The recorded outputs and tool operations suggest the following workflow for CAD reconstruction from engineering drawings:

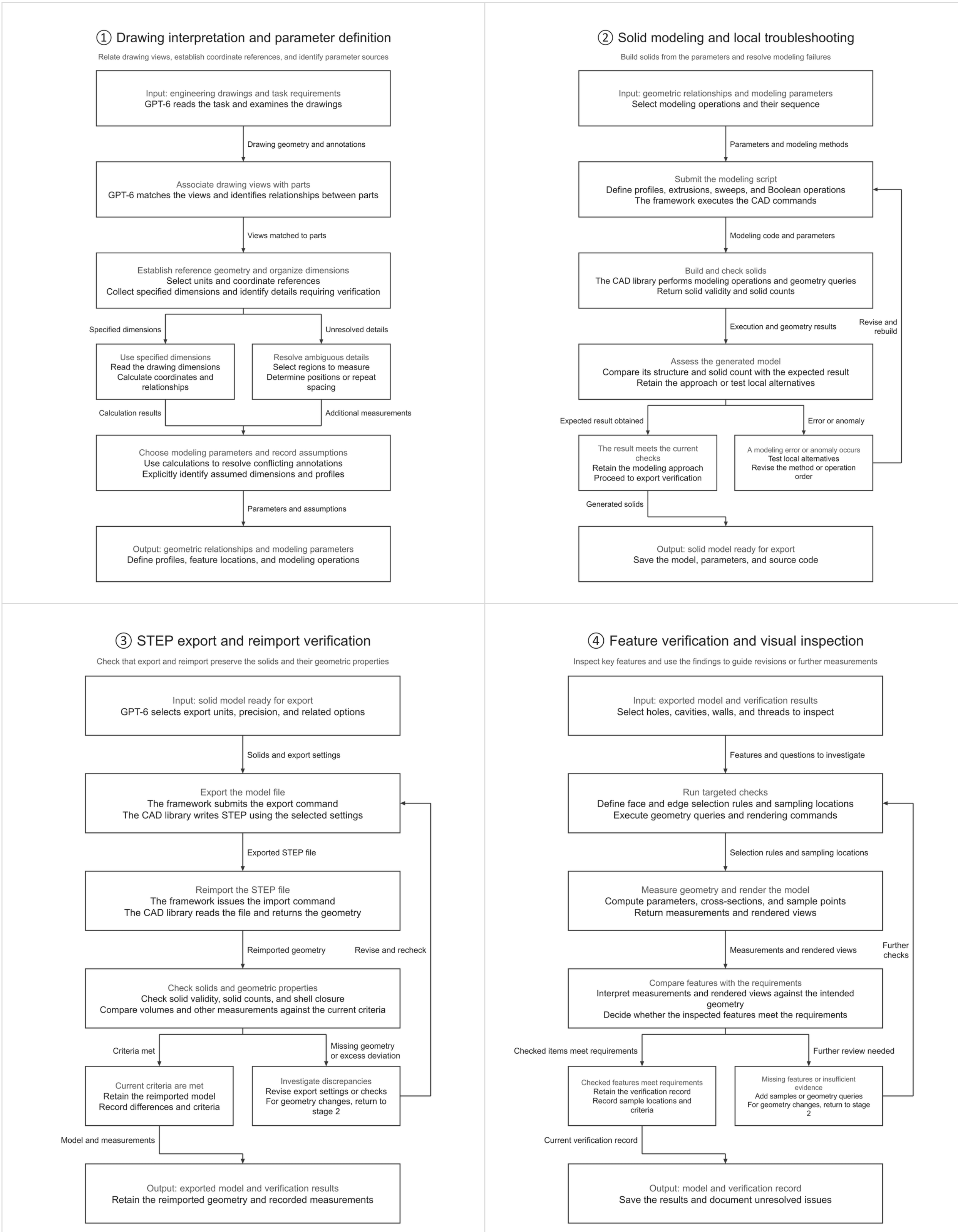

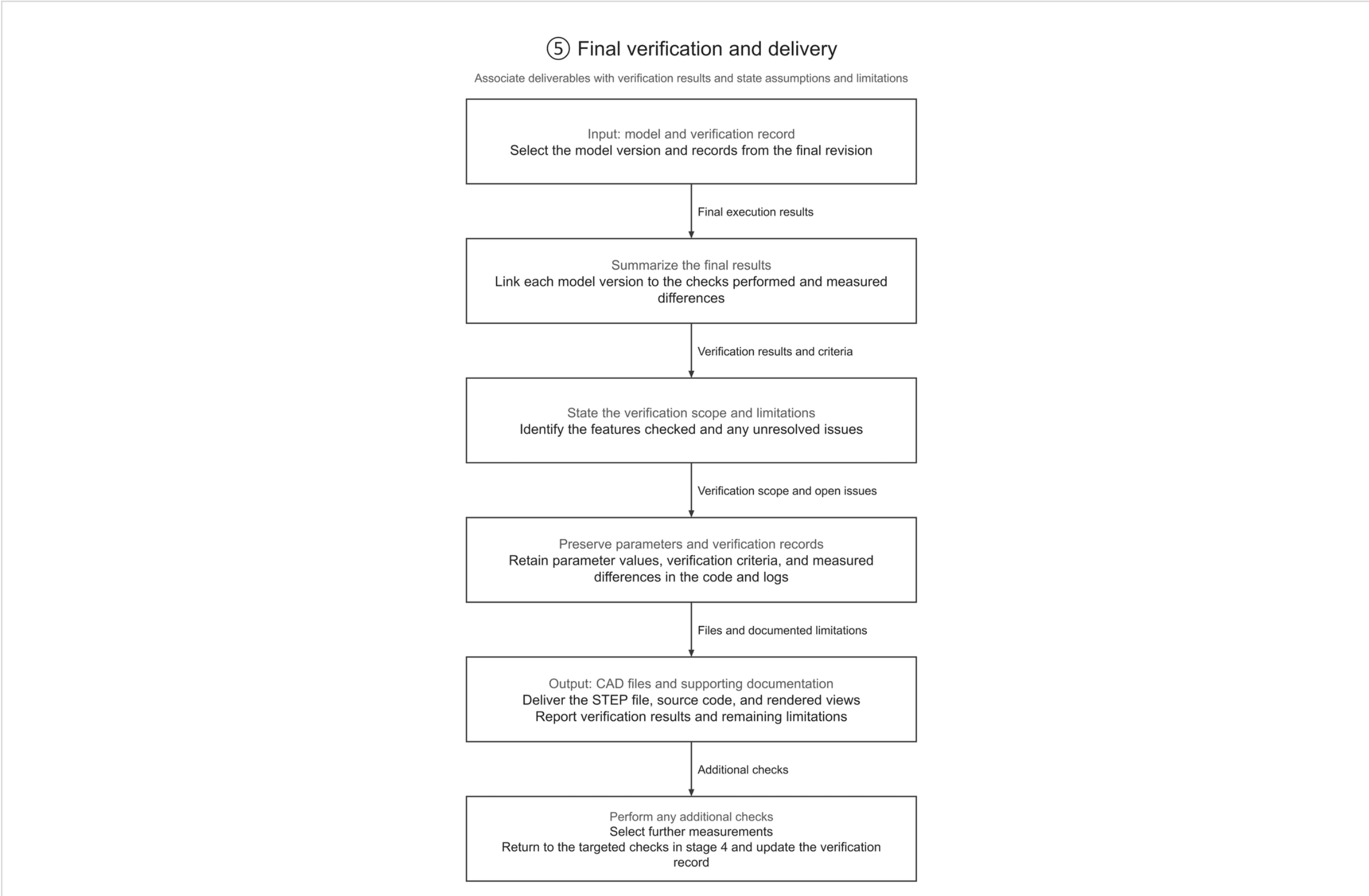

Figure 4-30 Inferred workflow for reconstruction from engineering drawings

## 4.5 Capabilities and limitations

GPT-6 can combine projections across multiple views, section views, dimensions, and relationships between local features to reconstruct the main three-dimensional structures of parts from engineering drawings. For simple and moderately complex cases, it generally reproduces the main outlines, holes and slots, repeated patterns, and most internal structures correctly. For complex individual parts such as the engine block, it can also identify the main hole layouts, cavities, reinforcing structures, and their spatial relationships. When several parts are involved, GPT-6 can model them individually and interpret their correspondence and assembly relationships. The moderately complex case can be assembled as intended, and the main components and overall structures of complex assemblies are also reproduced to a substantial extent.

The main limitations concern interpretation of complex local geometry, recovery of dimensions, and consistency at assembly interfaces. GPT-6 is prone to simplifying geometry, omitting features, or introducing local topological errors when interpreting multiple steps, nested cavities, holes opening into cavities, through and blind holes, and transitions such as chamfers and fillets. Some models have broadly correct overall shapes but inaccurate depths, wall thicknesses, hole positions, and mating dimensions. In an individual part, these appear mainly as dimensional errors. Across several parts, they can accumulate into interference, mating surfaces that fail to meet, or missing assembly interfaces where essential holes have been omitted. Complex cases also contain invalid face intersections, non-manifold geometry, and outputs with no valid solid. GPT-6 therefore cannot yet consistently ensure the validity of complex CAD geometry, or reliably detect and correct local errors and assembly interference.

GPT-6 also uses higher-level CAD operations such as shelling, applying draft, and lofting inconsistently. In some complex cases, thin walls, variable cross-sections, and complex transitions that should be formed with these operations are approximated using basic extrusions and cuts. Such approximations can recover the overall shape to some extent, but do not accurately preserve wall thickness, surface transitions, or local assembly interfaces, resulting in distorted or missing structures. This may reflect an incomplete understanding of higher-level CAD features and their engineering purpose, or may be related to the CAD modeling tools used.

Overall, GPT-6 shows a strong ability to interpret multiple drawing views and reconstruct three-dimensional CAD models, but remains well short of reliable engineering use. It can often establish the general shape of a part. The remaining challenge is to reconstruct complex local structures accurately, choose appropriate 3D modeling representations, satisfy dimensional and mating requirements, and produce geometrically valid models whose parts assemble consistently.

# Chapter 5 CAD generation from text

## 5.1 Task objectives and evaluation criteria

CAD generation from text comprises two tasks: modeling from geometric requirements and modeling from functional requirements. The first takes a textual description of the target shape, dimensions, quantities, and spatial relationships. It tests whether GPT-6 can interpret these requirements, plan a modeling sequence independently, and use CAD software APIs to generate the corresponding B-rep model. The second takes a description of the intended application, functional objectives, operating conditions, interfaces, and space constraints. It tests whether GPT-6 can select a structural design, determine its parameters, and produce a CAD model that meets those needs.

Web search is enabled in the main experiments. Additional experiments comprise stress tests for both tasks and tests with web search disabled. These examine GPT-6's ability to handle more complex tasks and to model from the supplied text without web search. Table 5-1 sets out the objectives and evaluation criteria.

Table 5-1 Objectives and evaluation criteria for CAD generation from text

| Test element | Description |
|---|---|
| Evaluation question | Can GPT-6 generate CAD models whose shapes, dimensions, and spatial relationships match geometric requirements? Can it use functional requirements to select a suitable structure and parameters, then generate a CAD model that meets the design requirements? |
| Input | English text describing geometric or functional requirements. Geometric requirements specify target shapes, dimensions, quantities, and spatial relationships; functional requirements specify the intended application, objectives, operating conditions, interfaces, and constraints. |
| Output | A B-rep model generated from the input text and delivered as a STEP file. A single-solid part should be a valid solid; an assembly should retain its individual parts and their assembly relationships. |
| Evaluation criteria | Evaluate solid validity and fulfillment of the input requirements. Geometric checks cover shapes, dimensions, quantities, and spatial relationships. Functional checks cover structures, interfaces, space for motion, and design parameters. |

## 5.2 Test design

### 5.2.1 Test case selection

Each task includes simple, moderate, and complex cases, with one single-solid part and one assembly at each level, giving 12 cases in the main experiments. Complexity is assessed from the structures specified in the input and the difficulty of designing and modeling them. For geometric requirements, the main challenges are constructing the shapes and resolving spatial relationships between components. For functional requirements, the challenges lie in selecting a structure and reconciling interdependent requirements.

#### 5.2.1.1 Cases with geometric requirements

For modeling from geometric requirements, complexity depends on the regularity of the geometry, the difficulty of constructing surfaces and internal passages, and the complexity of component positions and connections. Single-solid parts are assessed primarily by the connections and transitions between local features and the main body. Assemblies are assessed by how parts fit together, how mechanisms are connected, and how the assembly is arranged. Table 5-2 defines the criteria.

Table 5-2 Complexity criteria for modeling from geometric requirements

| Complexity | Selection criteria | Typical characteristics |
|---|---|---|
| Simple | Predominantly constant cross-sections, symmetry, or repeated structures, with little interaction between local features; assemblies consist mainly of one simple mechanism. | Regular main shape, few holes or slots, and symmetric or repeated structures; straightforward positioning and connections between parts. |
| Moderate | Twisted surfaces or continuous transitions, or a mechanism with several interconnected transmission elements. | Curved or twisted surfaces and local connections requiring smooth transitions; several parts transmit motion together through gears or linkages. |
| Complex | Closely packed internal cavities and passages, with interdependent wall and passage positions; assemblies contain several kinds of components and mechanisms whose spatial arrangements affect one another. | Passages in several directions, closely packed cavities, and complex local wall thicknesses and connections; several mechanisms share limited space and require coordinated positioning. |

The six cases are an electronics heat sink body, a push-pull toggle clamp, an impeller with curved blades, a VEX robot gripper, a Rover V8 engine block, and a horizontal single-cylinder steam engine. The reference models are publicly available mechanical models from GrabCAD. The inputs are English geometric requirements written from these reference models, specifying target dimensions, shapes, quantities, and relative positions. Table 5-3 lists the cases, with screenshots of excerpts from the input text, reference models, and the basis for each complexity rating.

Table 5-3 Test cases for modeling from geometric requirements

| Case ID | Case type | Excerpt from the input text | Reference model rendering | Complexity | Complexity rating rationale |
|---|---|---|---|---|---|
| TC-01 | Single-solid part | Create a single solid heat-sink body approximately 63.5 mm long, with a 42 × 25.4 mm overall cross-section. A thin, flat central web connects two mirrored groups of longitudinal fins. Each group contains eight fins, arranged as four above and four below the web. | 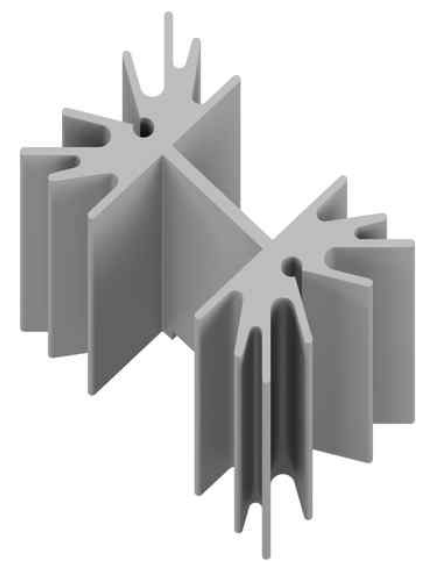 | Simple | Symmetrically repeated fins, a constant body cross-section along its length, and specified hole and slot positions. |
| TC-02 | Assembly | Create an assembled push-pull toggle clamp approximately 148 mm long, 40 mm wide and 101 mm high. A square mounting plate, approximately 40 × 40 × 5 mm, has rounded corners and four symmetrically arranged mounting holes. A raised longitudinal support carries a cylindrical guide at the front and a fixed hinge at the rear. | 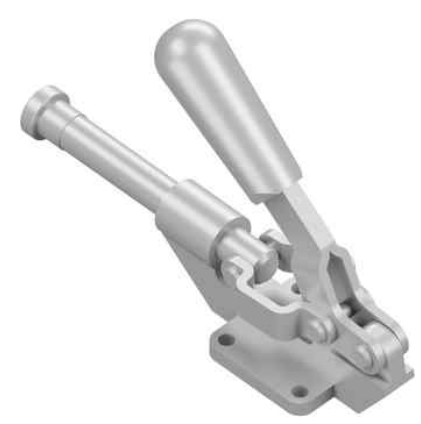 | Simple | One push-pull mechanism; the push rod and guide must be coaxial, and the pivot axes parallel. |
| TC-03 | Single-solid part | Create a single solid open impeller approximately 67.2 mm in overall diameter and 22.9 mm in axial depth. It has a circular backplate approximately 64.8 mm in diameter, a concentric 4.1 mm through-bore and a slender central nose. | 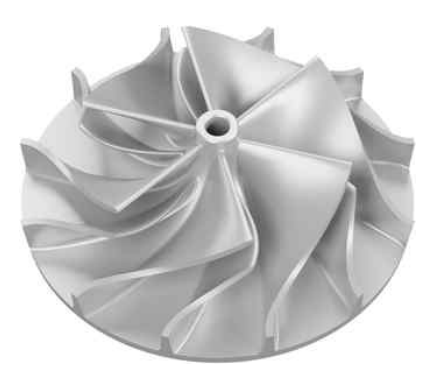 | Moderate | Two kinds of blades with different curvatures, heights, and radial extents, requiring smooth connections to the hub and backplate. |
| TC-04 | Assembly | Create an open two-jaw robotic gripper approximately 195 mm long, 120.5 mm wide and 67.3 mm thick. Two parallel perforated support plates carry a planar gear-and-linkage mechanism, with portions of the gear rims exposed. A perforated mounting tail extends rearward, and a rectangular motor housing sits outside one support plate. | 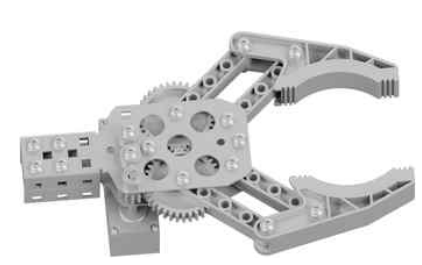 | Moderate | Gears and several linkages jointly control finger positions; gear engagement, pivots, and parallelism are interdependent. |
| TC-05 | Single-solid part | Create a single solid bare V8 cylinder block approximately 520 mm long, 475 mm wide and 290 mm high. Two cylinder banks form a 90° V, each containing four parallel bores approximately 92 mm in diameter and 110 mm deep. The cylinders repeat at approximately 107.5 mm longitudinal spacing, with the two banks slightly staggered. | 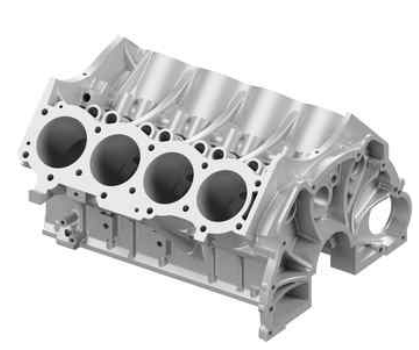 | Complex | Closely packed cylinder bores, water jackets, and other passages must connect or remain separate as specified, with continuous passage walls. |

| Case ID | Case type | Excerpt from the input text | Reference model rendering | Complexity | Complexity rating rationale |
|---|---|---|---|---|---|
| TC-06 | Assembly | Create a complete horizontal single-cylinder steam-engine assembly on a rectangular base approximately 170 × 80 × 14.5 mm, with rounded corners, mounting holes and a recessed underside. The cylinder lies along the long direction of the base and extends beyond one end. | 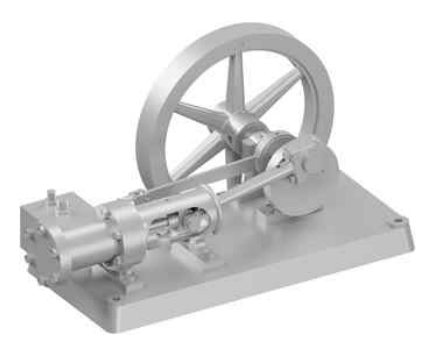 | Complex | The main drive and valve gear share a main shaft; cylinder, connecting-rod, and flywheel positions are interdependent. |

#### 5.2.1.2 Cases with functional requirements

For modeling from functional requirements, complexity is assessed primarily by the conditions that must be considered together and how they affect one another. Simple tasks typically focus on a main purpose such as support or clamping. Moderate tasks require one mechanism or component to satisfy related requirements for motion, loading, contact, or flow. Complex tasks involve several functions, so a structural change often affects the spatial arrangement, motion, and performance simultaneously. Table 5-4 defines the criteria.

Table 5-4 Complexity criteria for modeling from functional requirements

| Complexity | Selection criteria | Typical characteristics |
|---|---|---|
| Simple | A clearly defined main purpose, with installation, load-bearing, and maintenance requirements concentrated in one component or simple mechanism. | Mainly support, location, or clamping, with straightforward structures and simple functional relationships between components. |
| Moderate | One mechanism or component must meet several related requirements; local structural choices simultaneously affect motion, loading, contact, or flow. | Guiding, adjustment, and transmission elements used together; dimensions and mating relationships affect travel, loading, or contact simultaneously. |
| Complex | Several functions subject to common space, load, or operating constraints, requiring coordinated design of structures and parameters. | Several components or mechanisms work together; motion, load-bearing, and space requirements are interdependent, and local structural changes affect overall performance. |

The six cases are a coaxial support bracket for a motor and ball screw, a manual bench vise, a volute casing for an end-suction centrifugal pump, a pneumatic linkage gripper, a four-cylinder engine cylinder head, and a hexapod robot chassis. All reference models are from GrabCAD; the bracket and cylinder head use the HM15-86 and Nissan CR14DE models, respectively. The input text specifies functional objectives, operating conditions, interfaces, and space constraints. Table 5-5 lists the cases and the basis for their complexity ratings.

Table 5-5 Test cases for modeling from functional requirements

| Case ID | Case type | Excerpt from the input text | Reference model rendering | Complexity | Complexity rating rationale |
|---|---|---|---|---|---|
| TRC-01 | Single-solid part | Design a one-piece motor and ball-screw support for a compact feed mechanism. The motor has an 86 mm square mounting face, a 73 mm diameter locating pilot projecting 2 mm, and four 6.6 mm through-holes on a 69.6 mm square pattern for M6 mounting bolts. | 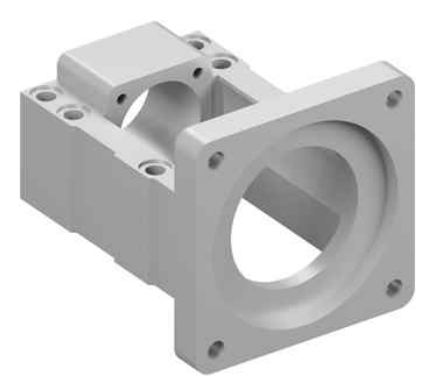 | Simple | Mounting interfaces and stiffness requirements center on coaxial support for the motor and ball screw. |
| TRC-02 | Assembly | Design a manually operated bench vise for holding small steel and aluminium workpieces during filing, sawing, and assembly. The workpieces have opposing flat gripping surfaces and a clamped thickness of 10–60 mm. | 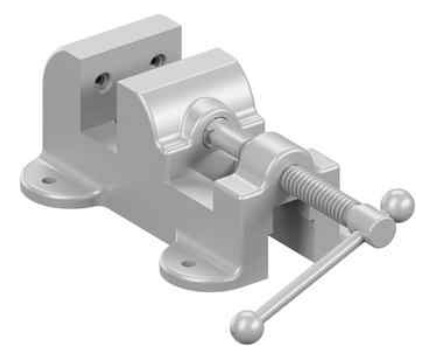 | Simple | Adjustment, clamping, and release mechanisms hold the workpiece, with straightforward functional relationships. |
| TRC-03 | Single-solid part | Design a one-piece cast-iron casing for a single-stage end-suction pump conveying clean water at 10–60 °C. The application has a nominal pump duty of 500 $m^3/h$ at 30 m total head. | 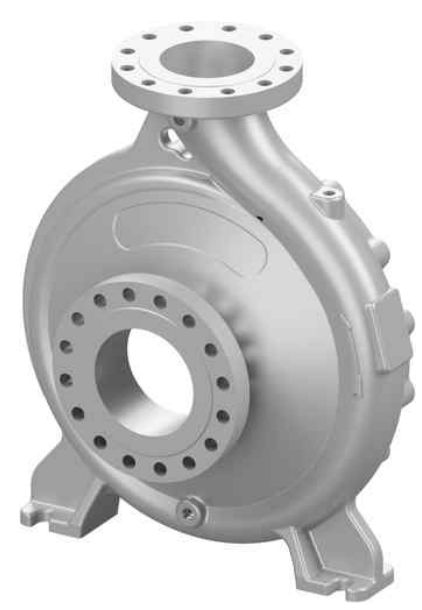 | Moderate | The casing must satisfy requirements for passage shape, pressure containment, and rotor installation and removal clearance. |
| TRC-04 | Assembly | Design a pneumatic two-finger gripper for transferring dry aluminium workpieces weighing up to 1 kg. Each workpiece has opposing flat gripping areas of at least 30 × 30 mm, separated by 20–60 mm, with its centre of mass within 20 mm of the midpoint between the contact areas. | 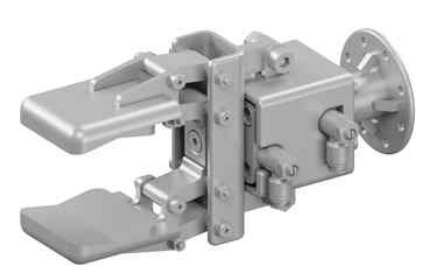 | Moderate | The transmission affects both opening range and gripping force; contact pads must resist slipping and protect the workpiece surface. |
| TRC-05 | Single-solid part | Design a one-piece aluminium cylinder-head body for a naturally aspirated, water-cooled, inline-four development engine with a 73 mm bore, 82.8 mm stroke, and 79 mm cylinder spacing. | 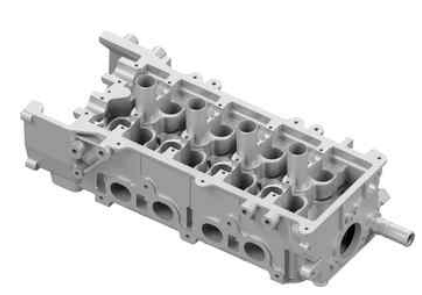 | Complex | The head must accommodate pressure containment, heat removal, and valve-train layout, with closely packed internal passages. |
| TRC-06 | Assembly | Design a six-legged, battery-powered chassis for indoor inspection on dry, hard floors, using 0.5 as the minimum friction coefficient between the feet and ground. | 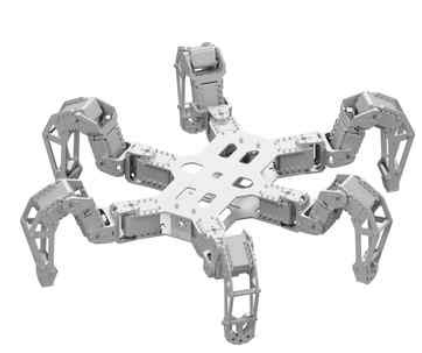 | Complex | Leg motion affects obstacle traversal, platform stability, and camera view; mass and operating-time requirements are also interdependent. |

### 5.2.2 Experimental conditions

#### 5.2.2.1 Main experiments

The main experiments use texts describing geometric requirements or functional requirements. A consistent prompt is used within each task. Each case is tested in a separate conversation with web search enabled.

*The prompts quoted below are English translations of the Chinese prompt records. The geometric and functional requirement texts supplied to GPT-6 were in English.*

The prompt for modeling from geometric requirements reads: "Generate the corresponding CAD model from the following geometric requirements, accurately representing the dimensions, shapes, and spatial relationships described. Save the model after each iteration and export the final model as a STEP file."

The prompt for modeling from functional requirements reads: "Generate a CAD model that satisfies the following functional requirements. Save the model after each iteration and export the final model as a STEP file."

#### 5.2.2.2 Additional experiments

The additional experiments comprise S1, a stress test for modeling from geometric requirements; S2, a stress test for modeling from functional requirements; and S3, tests with web search disabled. S1 and S2 increase task complexity, whereas S3 changes access to web search. All use the prompt for the corresponding main experiment. Each case is tested in a separate conversation, with results recorded separately. Table 5-6 gives the settings.

S1 uses a seven-cylinder radial engine. Its input describes the geometric requirements of the reference model, including dimensions, shapes, and spatial relationships between components. The test uses the geometric-requirements prompt from the main experiment, with web search enabled, to examine GPT-6's ability to handle several mechanisms and complex spatial relationships.

S2 uses an HK-450 electric helicopter mechanical assembly. Its input describes the functional requirements associated with the reference model, including the intended use, operating conditions, and performance requirements. The test uses the functional-requirements prompt from the main experiment, with web search enabled, to examine GPT-6's ability to select a structural design and complete the model when several functions impose interdependent requirements. Figure 5-1 shows excerpts from both inputs and renderings of the reference models.

S3 uses TC-05 and TC-06 from the geometric-requirements task and TRC-05 and TRC-06 from the functional-requirements task. The input texts and corresponding prompts remain the same as in the main experiments, but web search is disabled. The following instruction is added to the prompt: "Do not search the web or download external models or reference materials for this task. Use only the supplied text and local tools to complete the model."

Table 5-6 Additional experiments for CAD generation from text

| Condition | Prompt | Web search | Purpose |
|---|---|---|---|
| S1: Geometric requirements stress test | Use the prompt from the main geometric-requirements experiment. | Enabled | Test whether GPT-6 can generate the required geometric features and assembly relationships when several mechanisms and complex spatial relationships are involved. |
| S2: Functional requirements stress test | Use the prompt from the main functional-requirements experiment. | Enabled | Test whether GPT-6 can select a structural design and generate a CAD model when several functional requirements are interdependent. |
| S3: Web search disabled | Use the corresponding main-experiment prompt, adding an instruction prohibiting internet access, web search, and downloads of external materials. | Disabled | Test whether GPT-6 can complete complex models from the supplied text with web search disabled. |

| | |
|---|---|
| Create a complete single-row, seven-cylinder radial-engine assembly approximately 280 mm long, with a 440 mm propeller span. The cylinder assemblies, including rocker covers, have an overall envelope diameter of approximately 240 mm. Seven cylinders surround the main shaft at equal angles in a common transverse plane, with one upright. The central crankcase is approximately 100 mm in diameter and 67 mm long. Each cylinder has a 24 mm bore, a separate liner, closely spaced annular cooling fins up to approximately 46 mm in diameter, a finned head and a rounded rectangular rocker cover. | 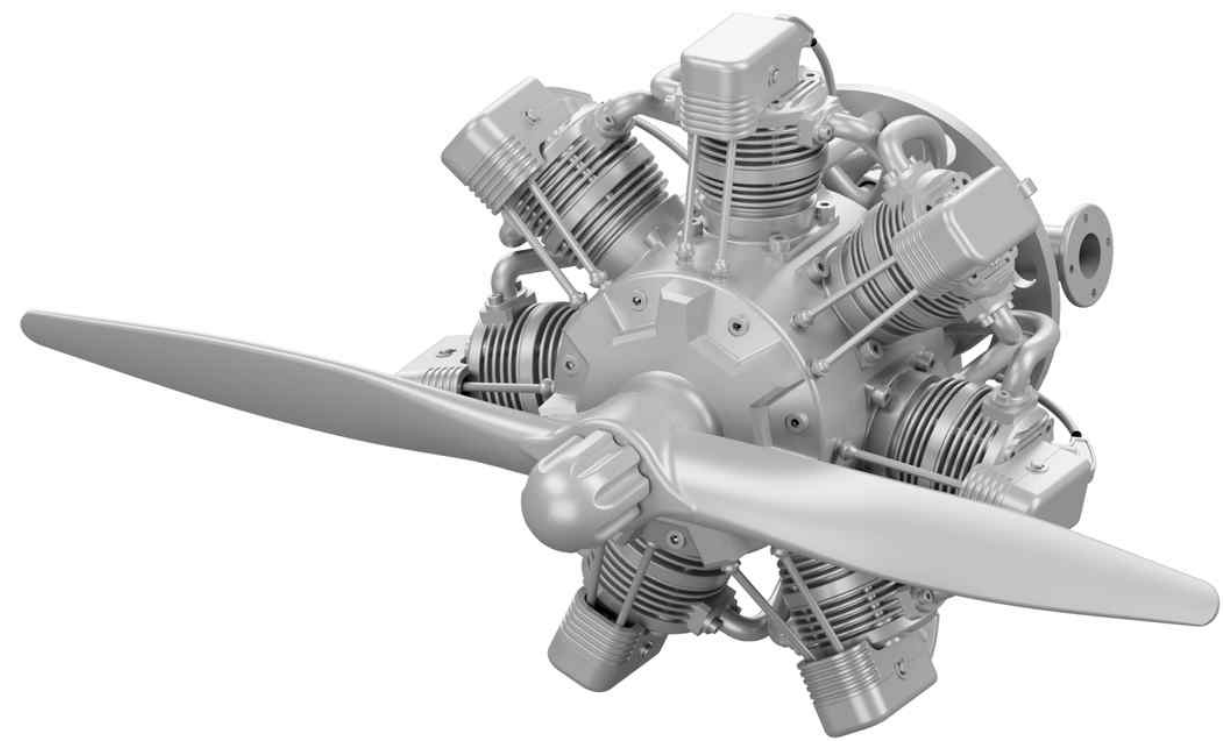 |
| S1: Excerpt from the input text | S1: Rendered reference model |
| Design a compact electric helicopter for visual inspection with a 150 g camera module measuring 60 × 50 × 40 mm. The module has four M3 attachment points on a 40 × 30 mm rectangular pattern and a centre of mass 20 mm above its mounting face. The complete aircraft, including the camera, battery, and electronics, shall weigh no more than 1.4 kg. Its operating envelope, including all rotating and moving components, shall remain within 1,100 mm in length, 850 mm in width, and 350 mm in height. | 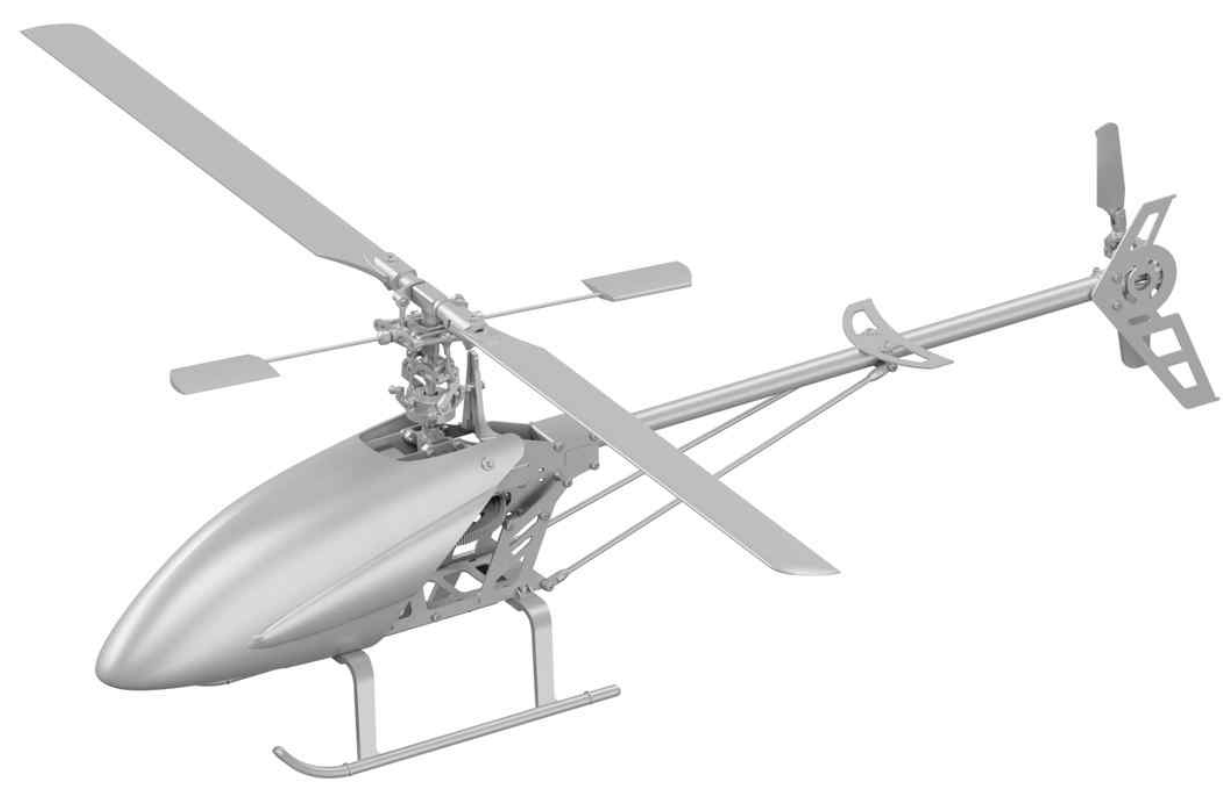 |
| S2: Excerpt from the input text | S2: Rendered reference model |

Figure 5-1 Input text excerpts and reference models for the two stress tests

## 5.3 Results

This section first presents the main results for modeling from geometric and functional requirements, then the additional stress tests and tests with web search disabled. All experiments assess whether the models meet the input requirements and pass the solid validity check.

The percentage of geometric requirements met is determined by checking shape, dimensions, quantities, and spatial relationships item by item. Functional requirements are checked against the structures, interfaces, available space for motion, and relevant design calculations. The resulting percentage is the proportion of listed requirements recorded as met under those checks. It is not a comprehensive measure of functionality or manufacturing readiness. Solid validity is assessed separately, and local inspections may reveal defects not captured by the scored items. Tables 5-7 and 5-8 define the criteria and pass conditions for the two tasks.

For example, TC-03 and TC-04 receive full scores on the geometric requirement checklist, yet local inspection reveals an incomplete blade-to-hub connection and an unsecured pinion, respectively. The score and these defects must be reported together. Functional results based on design calculations remain conditional on the loads, material properties, boundary conditions, and other assumptions used in those calculations.

Table 5-7 Criteria for evaluating geometric requirements

| Criterion | Pass condition |
|---|---|
| Geometric features and component counts | All specified holes, slots, blades, ribs, other features, and assembly components are present in the required numbers. |
| Dimensions, angles, and positions | Overall dimensions, hole diameters, spacings, heights, and angles meet the input requirements. Approximate dimensions are assessed using the tolerance applied in the corresponding case. |
| Symmetry, parallelism, and coaxiality | Features or components have the required arrangement, with axis positions and directions satisfying the specified coaxial, parallel, or symmetric relationships. |
| Surface shapes and local transitions | Changes in height, thickness, sweepback, and inclination in the inspected regions match the text. Local transitions are assessed using the checks actually performed on profiles, cross-sections, or angles between adjacent face normals. |
| Cross-sectional shapes and variation | For bodies with a constant cross-section, corresponding sections match after alignment within the threshold used. For structures with varying cross-sections, changes in profile and dimensions meet the input requirements. |
| Passage connectivity and separation | Passages that should connect remain connected; complete walls separate passages and cavities that should remain isolated. |
| Assembly components and layout | Parts are modeled separately as required. Mounting positions, pivot axes, connecting pins, and the specified assembly configuration match the text. |

Table 5-8 Criteria for evaluating functional requirements

| Criterion | Pass condition |
|---|---|
| Overall dimensions and mounting interfaces | Overall dimensions remain within the specified limits. The sizes and positions of locating holes, mounting holes, bearing seats, and other interfaces match the supplied components. |
| Component clearance and assembly connections | Specified components have no unintended volumetric interference at the installation positions checked. Connections that secure components or transmit motion are complete. |
| Range of motion and configuration | Opening, travel, rotation, calculated speed, and motion envelope meet the input requirements. No unintended interference occurs at the motion positions checked. |
| Support stability | At the support configurations and loads checked, the center-of-mass projection lies within the support region, and calculated joint torque demands do not exceed the selected actuators' capacity. |

| Criterion | Pass condition |
| --- | --- |
| Displacement under load, stress, and fastening force | Using the finite-element or simplified mechanical model actually adopted, calculated displacements and stresses in the inspected regions remain within their limits, and fastening forces can resist the specified loads. |
| Clamping force, resistance to slipping, and self-locking | At the specified hand force or supply pressure, calculated clamping force meets the requirement, and the safety factor against slipping meets the specified minimum. Where self-locking is required, reverse driving still requires an applied positive torque. |
| Contact pads and surface protection | Soft pads contact the workpiece. At the compression assumed in the design, metal backing plates and fasteners remain clear of the workpiece. |
| Flow-path connectivity and head loss | The flow path connects the specified inlet and outlet. At the specified flow rate and the resistance coefficients used in the calculation, total head loss in the suction passage does not exceed 0.5 m. |
| Combustion-chamber volume and compression ratio | The geometric compression ratio reaches 9.9:1, calculated from the model's chamber volume and the specified bore, stroke, and gasket dimensions. |
| Cooling, lubrication, and heat balance | Water jackets cover the specified heated regions. Cooling and lubrication passages remain connected, while gas, coolant, and oil passages remain separate. Supply interfaces meet the requirements, and mean temperature rise is checked using the specified flow rate, inlet temperature, and heat load. |
| Mass, power consumption, and operating time | The total mass does not exceed the specified limit. Operating time and remaining battery energy, calculated from the available energy and power consumption assumed in the design, meet the input requirements. |
| Camera field of view and rotation range | At the configurations checked, the model does not obstruct the required field of view. The camera mechanism covers the specified rotation angles, with complete mounting and drive connections. |
| Adjustment and removal clearance | Tools and components to be removed do not interfere with retained components at the access, operating, and removal positions checked. |

### 5.3.1 Main experiment results

#### 5.3.1.1 Overview

Each main experiment includes six cases: two simple, two moderate, and two complex. With web search enabled, GPT-6 delivered STEP models for all 12 cases. Tables 5-9 and 5-10 summarize the results. The evaluation team separated each assembly into individual part STEP files and checked the solids in those files. Fractions in the tables give the number of solids that passed divided by the total number of solids. Table 5-11 reports completion times.

In modeling from geometric requirements, all six cases receive full scores on the listed geometric requirement checks and pass the solid validity check. GPT-6 constructs the main shapes and assembly relationships of the heat sink, clamp, impeller, gripper, engine block, and steam engine.

Some local connections remain incomplete, however. In TC-03, a gap remains between the upper portions of the main blades' inner edges and the hub. In TC-04, the pinion lacks a fastening feature that would transmit torque from the motor output shaft.

In modeling from functional requirements, GPT-6 completes the structural and parameter design for all six cases, each ultimately receiving a full score on the listed functional requirement checks. The simple cases meet the requirements for coaxial support, installation and maintenance, workpiece clamping, and self-locking. The moderate cases provide the pump casing's flow passage and rotor removal clearance, and the gripper's transmission mechanism. The complex cases provide the cylinder head's valve, cooling, and lubrication structures, and the hexapod chassis design for motion, load capacity, and operating time. Five of the six cases pass the solid validity check. In TRC-06, three motor supports fail because their solid boundaries intersect themselves. The model also contains 36 separate open shells, so the complete chassis fails the solid validity check.

Table 5-9  Main results for modeling from geometric requirements

| **Case ID** | **Complexity** | **Solid validity** | **Listed requirements met** | **Results** |
|---|---|---|---|---|
| TC-01 | Simple | Pass | All 12 met | Fin shapes and arrangement, and hole and slot positions, meet the requirements. |
| TC-02 | Simple | Pass | All 10 met | The clamp structure is complete; push-rod, link, and handle positions meet the requirements. |
| TC-03 | Moderate | Pass | All 10 met | Blade counts and arrangement are correct; a gap remains between the upper inner edges of the main blades and the hub. |
| TC-04 | Moderate | Pass | All 12 met | Finger and gear layouts meet the requirements, but the motor shaft and pinion lack a fastening connection to transmit torque. |
| TC-05 | Complex | Pass | All 12 met | The block structure is complete; hole and internal-passage layouts meet the requirements. |
| TC-06 | Complex | Pass | All 12 met | Overall assembly relationships are correct, with complete main-drive and valve-gear connections. |

Table 5-10 Main results for modeling from functional requirements

| Case ID | Complexity | Solid validity | Listed requirements met | Results |
|---|---|---|---|---|
| TRC-01 | Simple | Pass | All 8 met | The bracket meets requirements for coaxial support of the motor and ball screw, installation, and maintenance. |
| TRC-02 | Simple | Pass | All 9 met | The vise meets opening-adjustment and workpiece-clamping requirements and can retain its clamping position. |
| TRC-03 | Moderate | Pass | All 11 met | The casing meets requirements for flow-path connectivity, pipe connections, and rotor removal from the rear. |
| TRC-04 | Moderate | Pass | All 9 met | The gripper meets requirements for synchronized opening and closing, gripping, surface protection, and mounting connections. |
| TRC-05 | Complex | Pass | All 12 met | Intake, exhaust, cooling, and lubrication structures are complete, with motion and maintenance clearance for valve-train components. |
| TRC-06 | Complex | Fail 904/907 valid | All 12 met | Motion relationships for walking, turning, and obstacle traversal are correct; 3 motor supports have self-intersecting solid boundaries, and 36 open shells are also present. |

Table 5-11 Completion times in the main experiments

| Case ID | TC-01 | TC-02 | TC-03 | TC-04 | TC-05 | TC-06 |
|---|---|---|---|---|---|---|
| Completion time | 23 min | 26 min | 35 min | 37 min | 46 min | 47 min |
| **Case ID** | **TRC-01** | **TRC-02** | **TRC-03** | **TRC-04** | **TRC-05** | **TRC-06** |
| Completion time | 39 min | 1 h 20 min | 1 h 9 min | 1 h 40 min | 1 h 40 min | 3 h 17 min |

#### 5.3.1.2 Case analyses

The following cases are grouped by task. Each presents an excerpt from the input text, the reference and generated models, the percentage of requirements met over successive iterations, and an analysis of the result.

#### 5.3.1.2.1 Modeling from geometric requirements

### TC-01 (simple)

Create a single solid heat-sink body approximately 63.5 mm long, with a 42 × 25.4 mm overall cross-section. A thin, flat central web connects two mirrored groups of longitudinal fins. Each group contains eight fins, arranged as four above and four below the web. The fins spread outward in a fan-shaped profile, with progressively changing inclinations.

**(a) Excerpt from the input text**

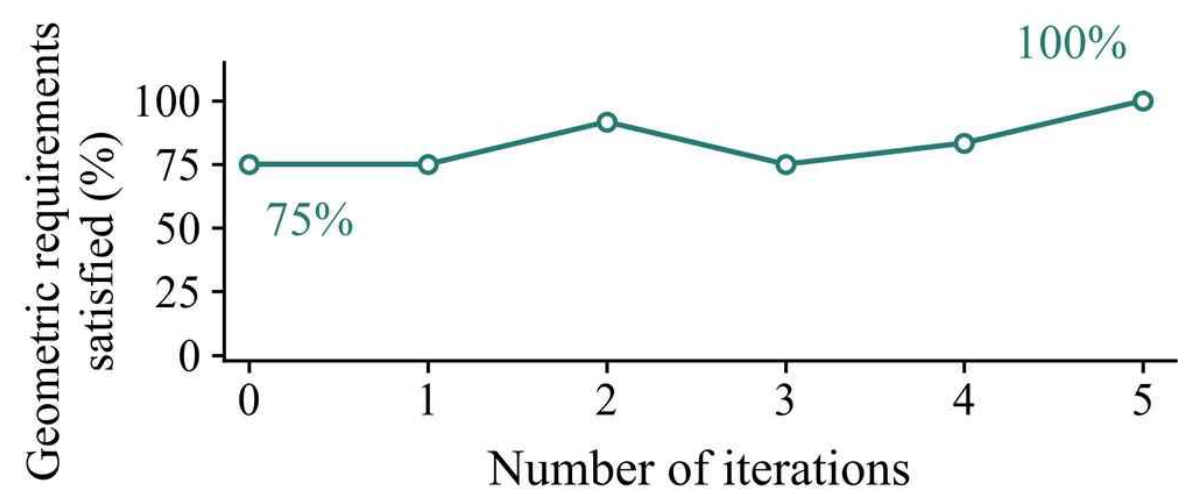

**(b) Percentage of geometric requirements met**

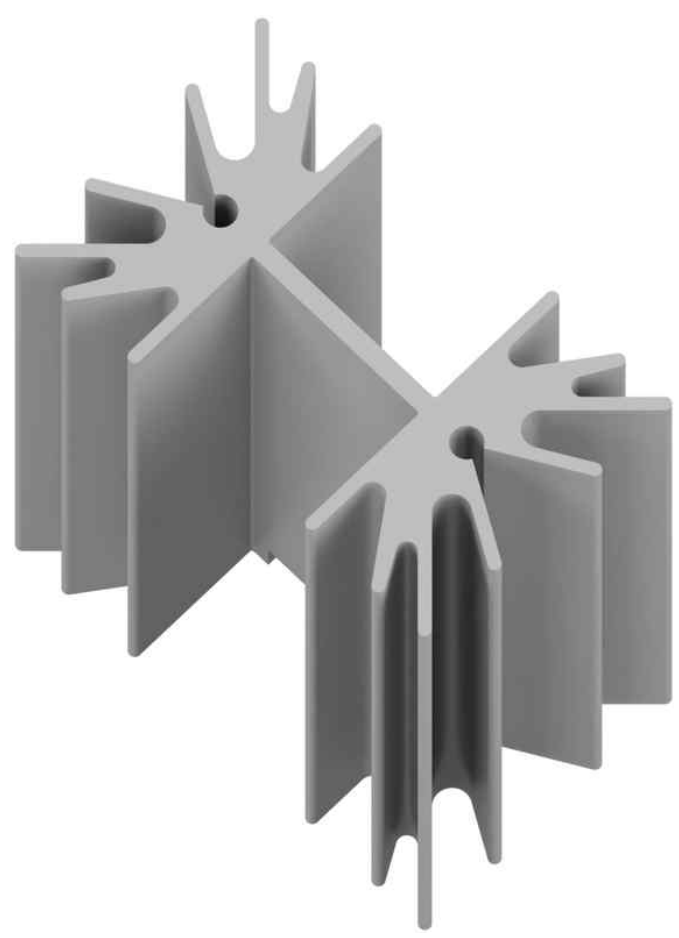

**(c) Rendered reference CAD model**

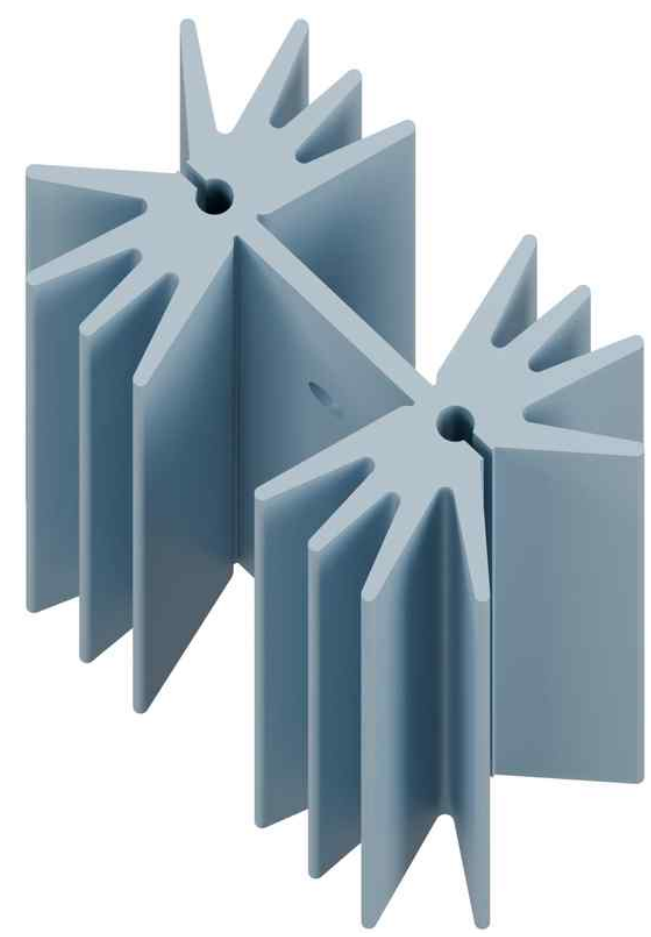

**(d) Rendered generated CAD model**

Figure 5-2 TC-01: modeling from geometric requirements

TC-01 asks GPT-6 to generate a single-solid heat sink with fan-shaped fins, through-slots, and mounting holes, following specified dimensions and symmetry relationships. Across iterations, GPT-6 adjusts the fin cross-sections, improves the transitions at their roots, and adds missing mounting holes and end notches. The final model has 16 regularly arranged fins, with smooth transitions from the thin outer portions to the wider roots. C-shaped slots on both sides extend through the body. The percentage of geometric requirements met rises from 75% to 100% (Figure 5-2). The final model passes the solid validity check.

### TC-02 (simple)

| | |
|---|---|
| Create an assembled push-pull toggle clamp approximately 148 mm long, 40 mm wide and 101 mm high. A square mounting plate, approximately 40 × 40 × 5 mm, has rounded corners and four symmetrically arranged mounting holes. A raised longitudinal support carries a cylindrical guide at the front and a fixed hinge at the rear. | 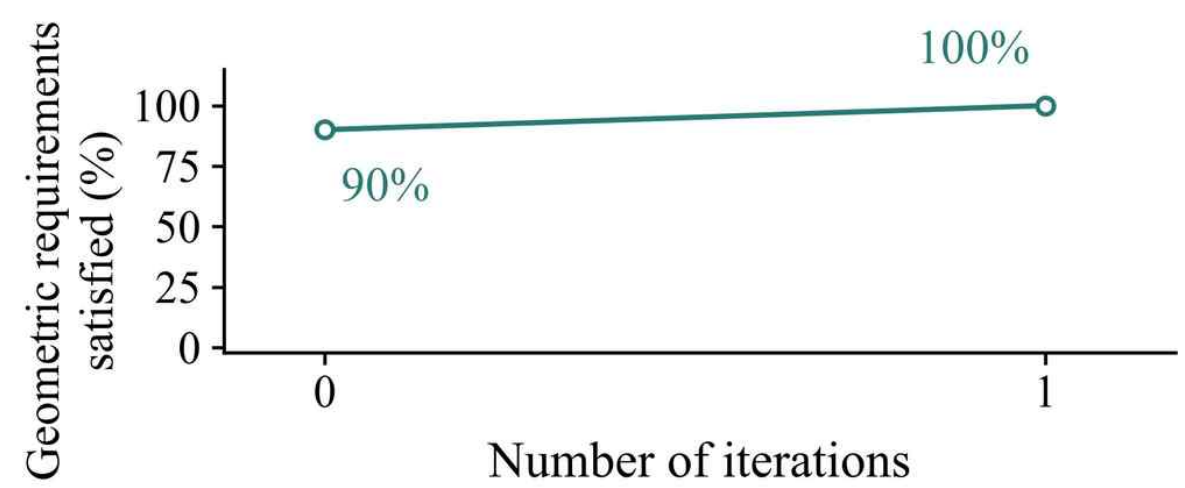  |
| **(a) Excerpt from the input text** | **(b) Percentage of geometric requirements met** |
| 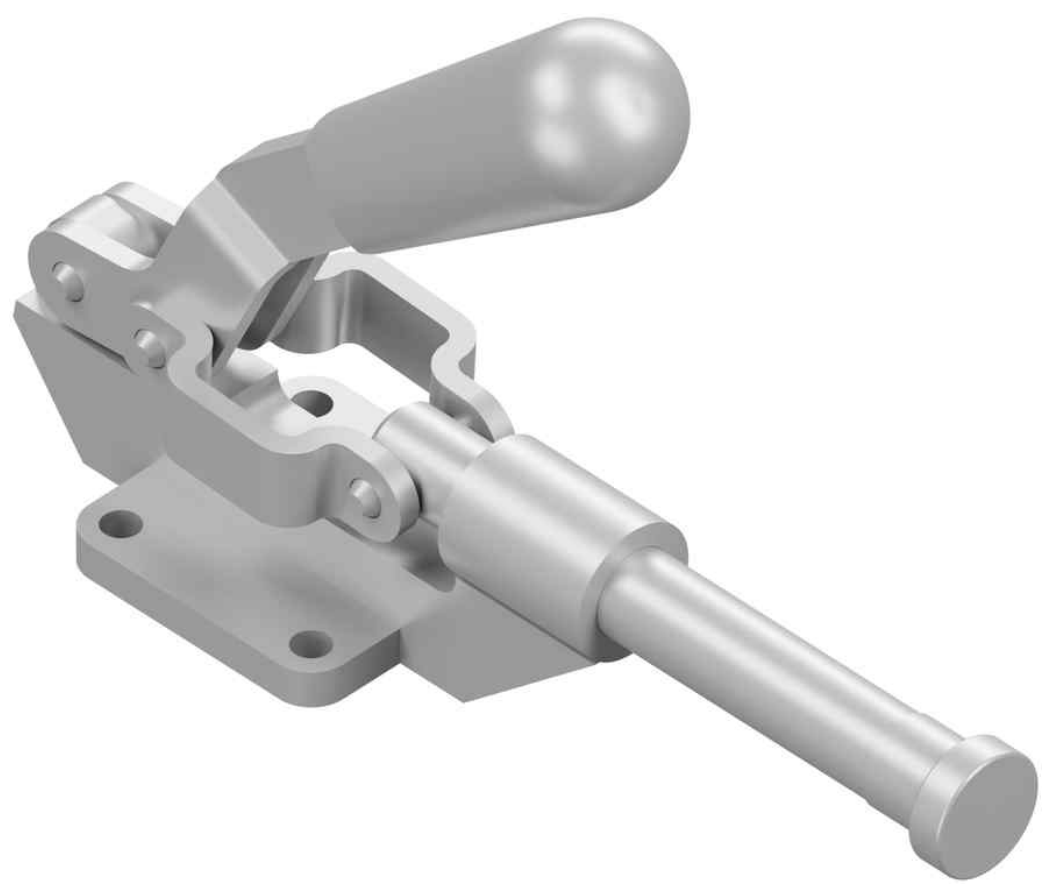 | 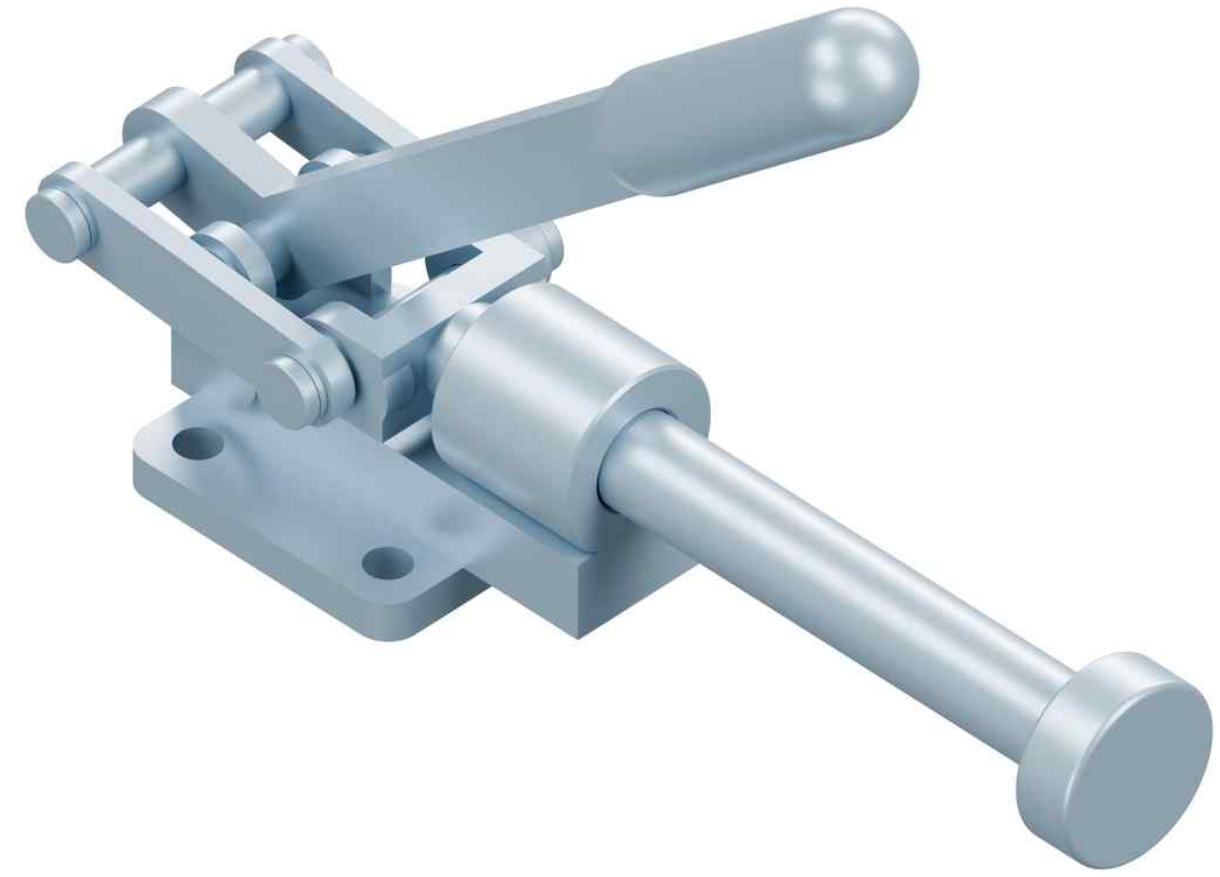 |
| **(c) Rendered reference CAD model** | **(d) Rendered generated CAD model** |

Figure 5-3  TC-02: modeling from geometric requirements

TC-02 asks GPT-6 to construct a push-pull toggle clamp with specified dimensions and an assembly configuration. GPT-6 positions the horizontal push rod within a coaxial guide and connects it to the raised handle through linkages on both sides. It adds missing pivot pins and fasteners over successive iterations. In the final model, the links are nearly horizontal, with the intermediate pivot slightly above the line joining the two end pivots, as required. The percentage of geometric requirements met rises from 90% to 100% (Figure 5-3). All 22 parts pass the solid validity check.

## TC-03 (moderate)

Create a single solid open impeller approximately 67.2 mm in overall diameter and 22.9 mm in axial depth. It has a circular backplate approximately 64.8 mm in diameter, a concentric 4.1 mm through-bore and a slender central nose. The hub rises steeply near the axis and expands smoothly into a broad, shallow profile towards the backplate rim. Six tall main blades alternate with six shorter splitter blades.

**(a) Excerpt from the input text**

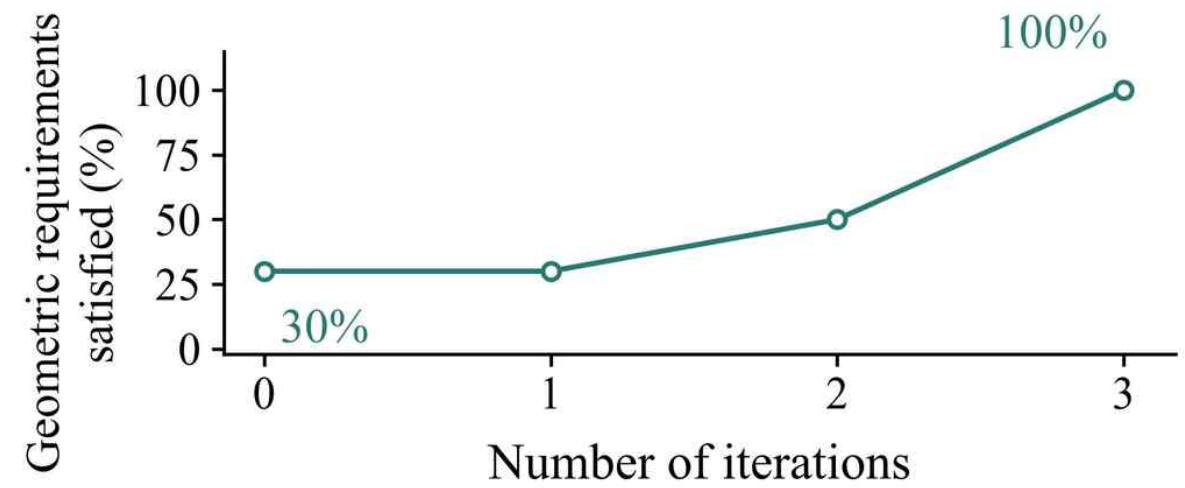

**(b) Percentage of geometric requirements met**

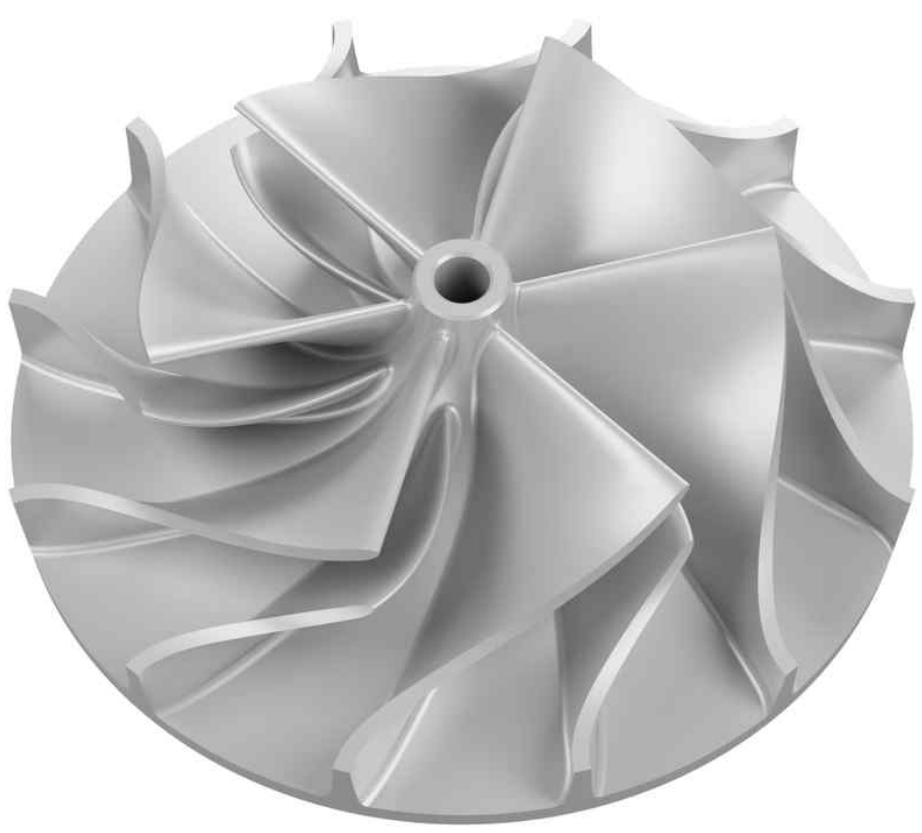

**(c) Rendered reference CAD model**

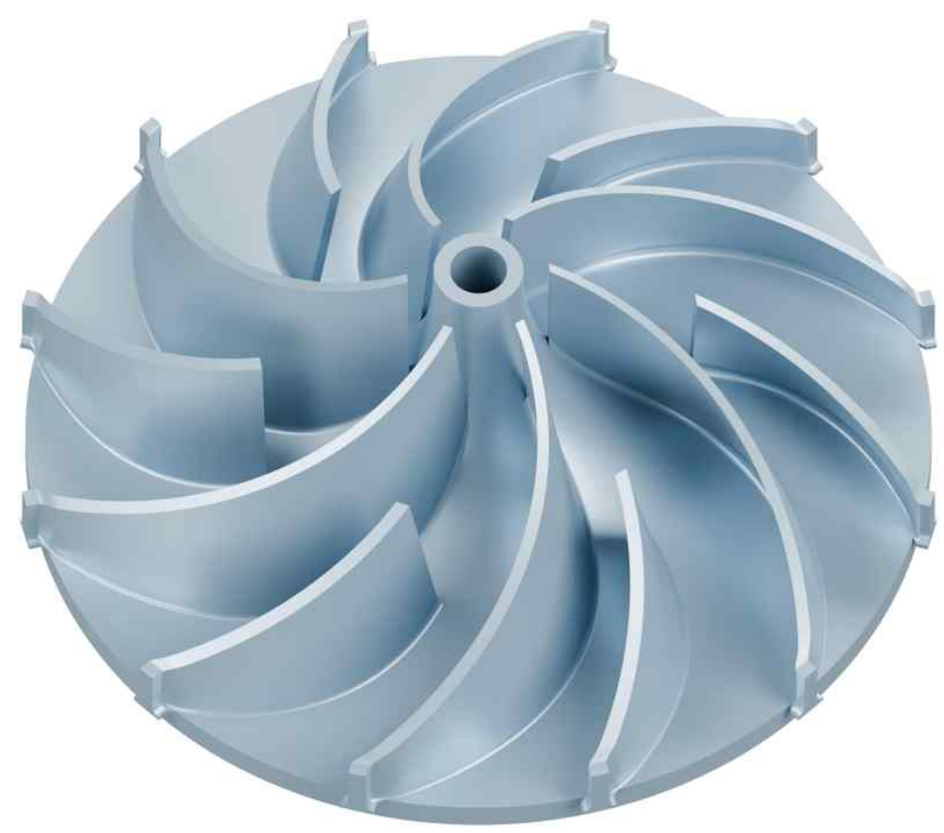

**(d) Rendered generated CAD model**

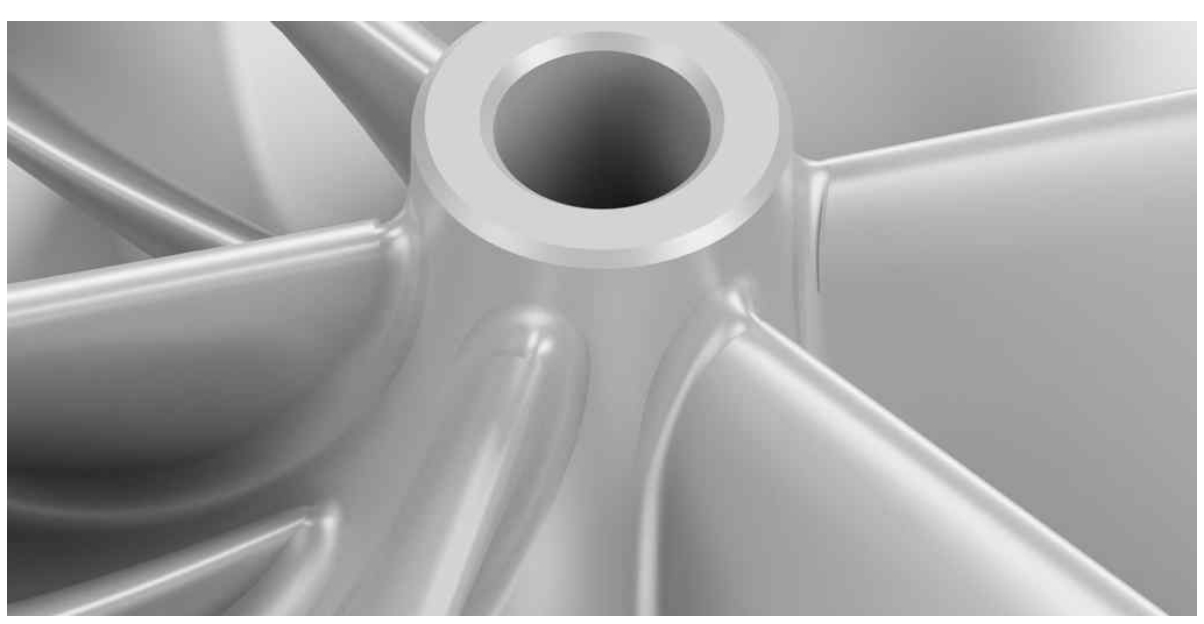

**(e) Reference model: continuous connection along the inner edge**

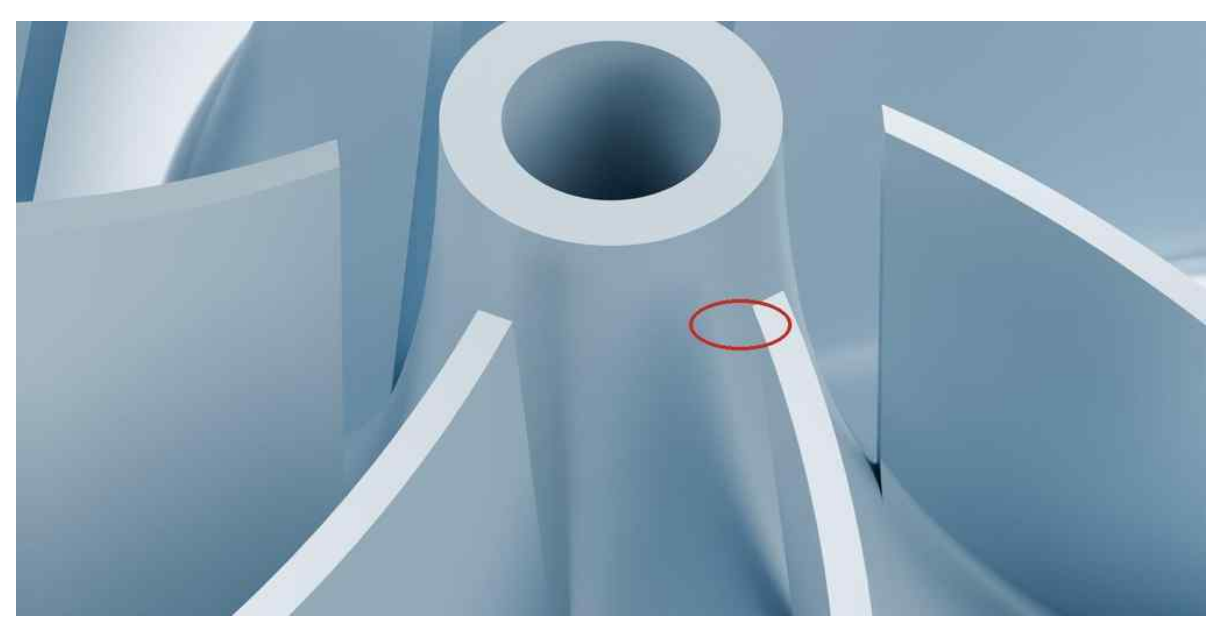

**(f) Generated model: gap along the upper inner edge**

Figure 5-4 TC-03: modeling from geometric requirements

TC-03 asks GPT-6 to generate a single-solid impeller with alternating main and splitter blades, following specified dimensions and twisted shapes. GPT-6 progressively adds both kinds of blades, raising the percentage of geometric requirements met from 30% to 100%. Local inspection nevertheless reveals a gap: although the lower portions of the blades connect to the hub and backplate, the upper portions of the main blades' inner edges remain separated from the central hub. The reference model has a continuous connection in this region (Figure 5-4). GPT-6 reproduces the main blade shapes but does not fully resolve the blade-to-hub transitions. The final impeller passes the solid validity check.

## TC-04 (moderate)

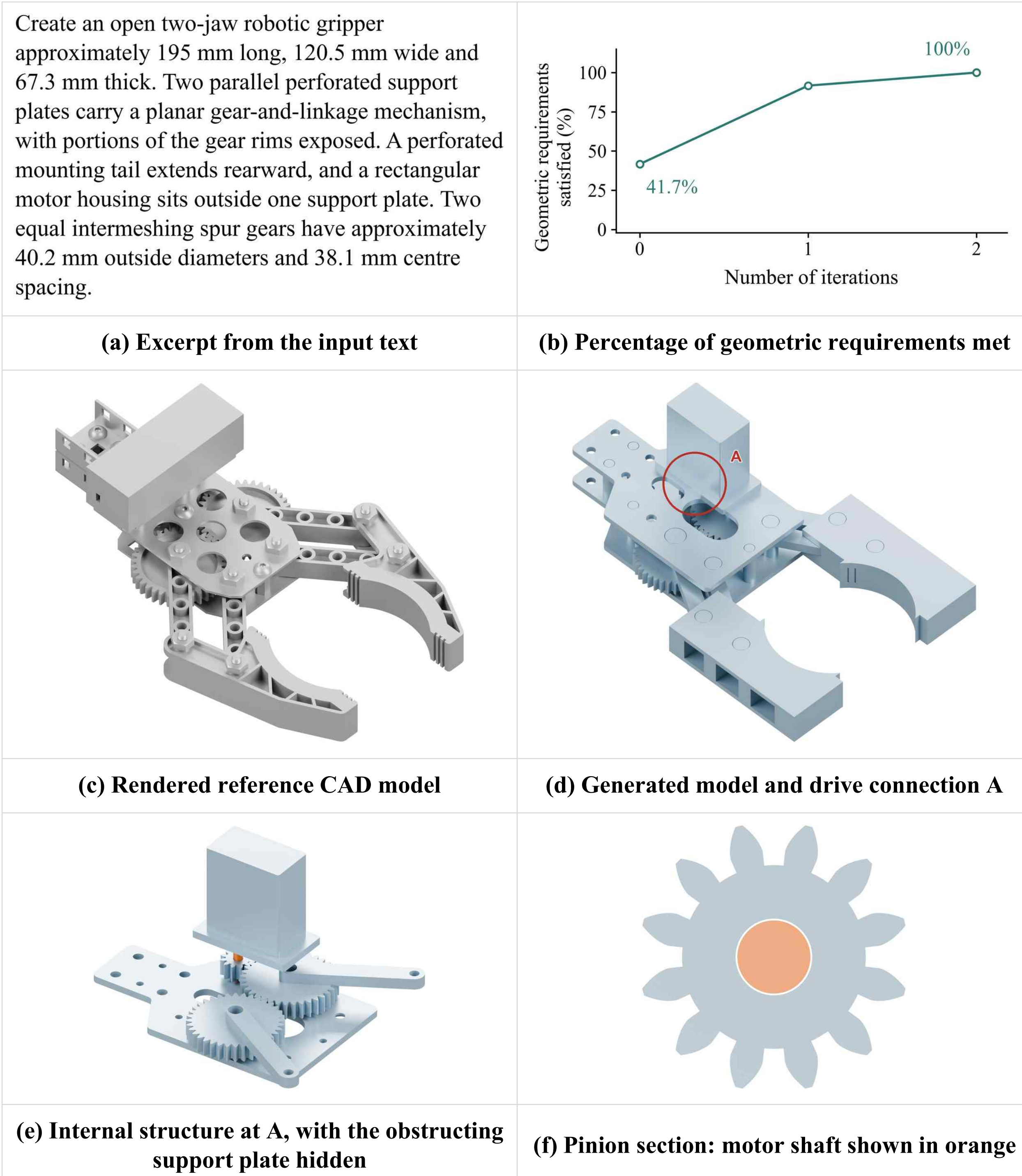

**(a) Excerpt from the input text**
**(b) Percentage of geometric requirements met**
**(c) Rendered reference CAD model**
**(d) Generated model and drive connection A**
**(e) Internal structure at A, with the obstructing support plate hidden**
**(f) Pinion section: motor shaft shown in orange**

Figure 5-5 TC-04: modeling from geometric requirements

TC-04 asks GPT-6 to construct a two-finger gripper driven by gears and linkages, with specified dimensions, gear engagement, and finger positions. Across iterations, GPT-6 adds missing fingers, links, and mounting components, raising the percentage of geometric requirements met from 41.7% to 100%. The two fingers in the final model remain parallel, but the drive connection between the motor and gear is incomplete. A radial clearance of 0.1 mm remains between the motor shaft and the pinion bore, and there is no set screw or other fastening feature. The pinion therefore cannot reliably rotate with the shaft. Location A in Figure 5-5 marks this region, and the detail views show the actual internal connection in the generated model. All 47 parts pass the solid validity check.

### TC-05 (complex)

Create a single solid bare V8 cylinder block approximately 520 mm long, 475 mm wide and 290 mm high. Two cylinder banks form a 90° V, each containing four parallel bores approximately 92 mm in diameter and 110 mm deep. The cylinders repeat at approximately 107.5 mm longitudinal spacing, with the two banks slightly staggered.

**(a) Excerpt from the input text**

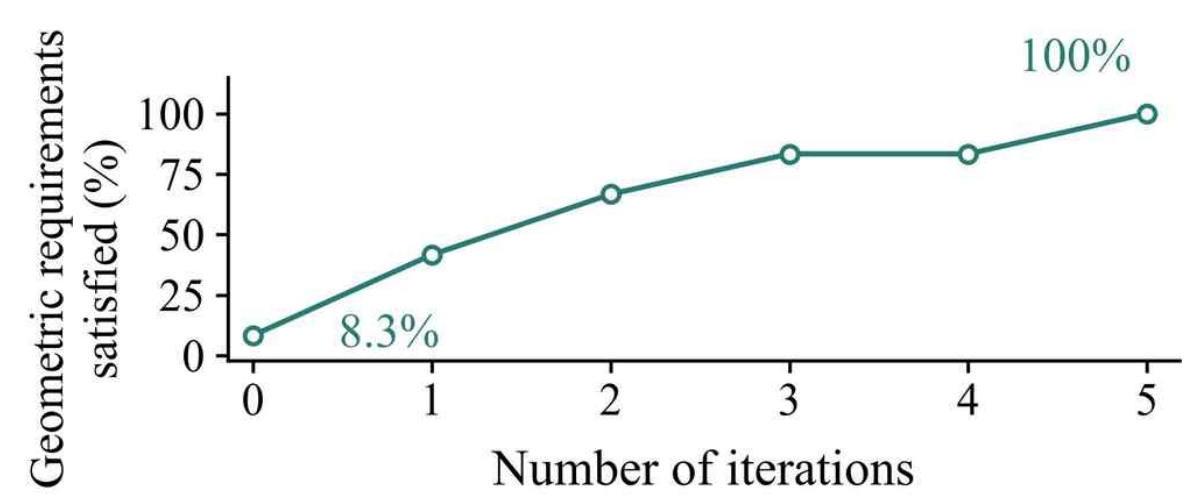

**(b) Percentage of geometric requirements met**

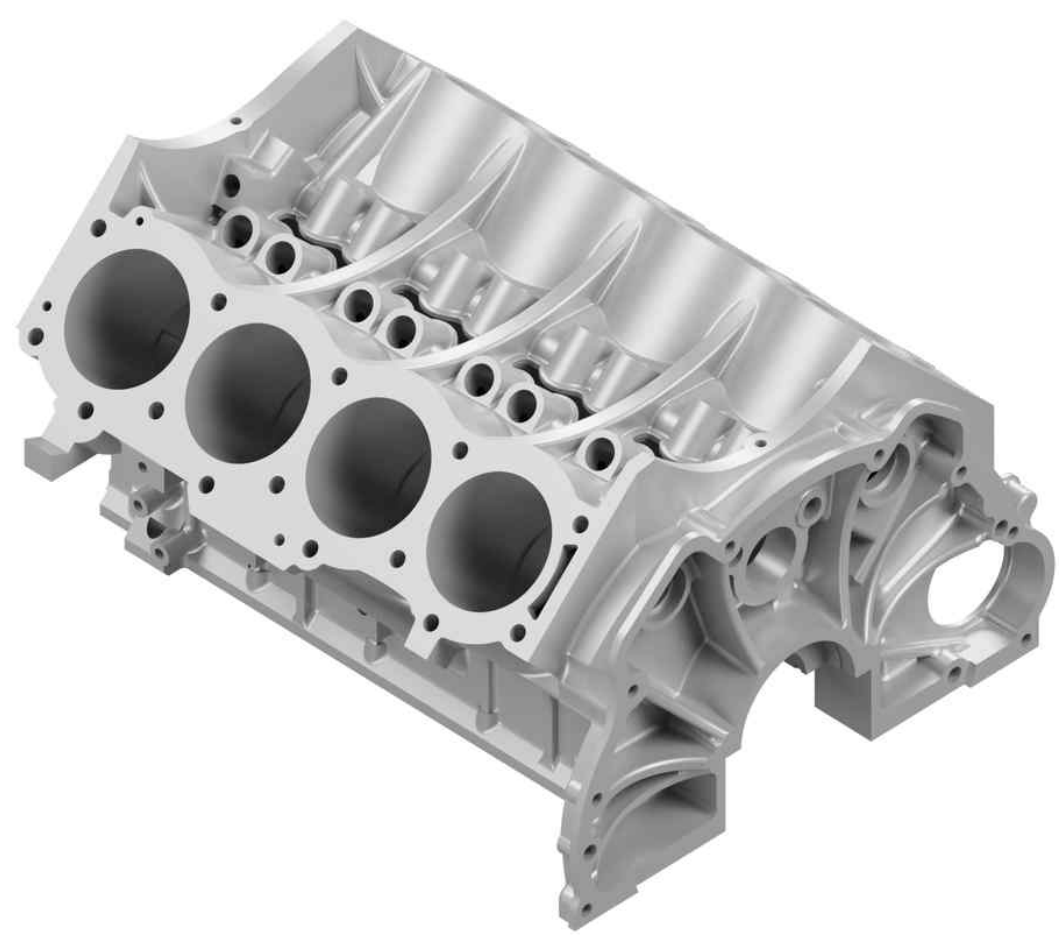

**(c) Rendered reference CAD model**

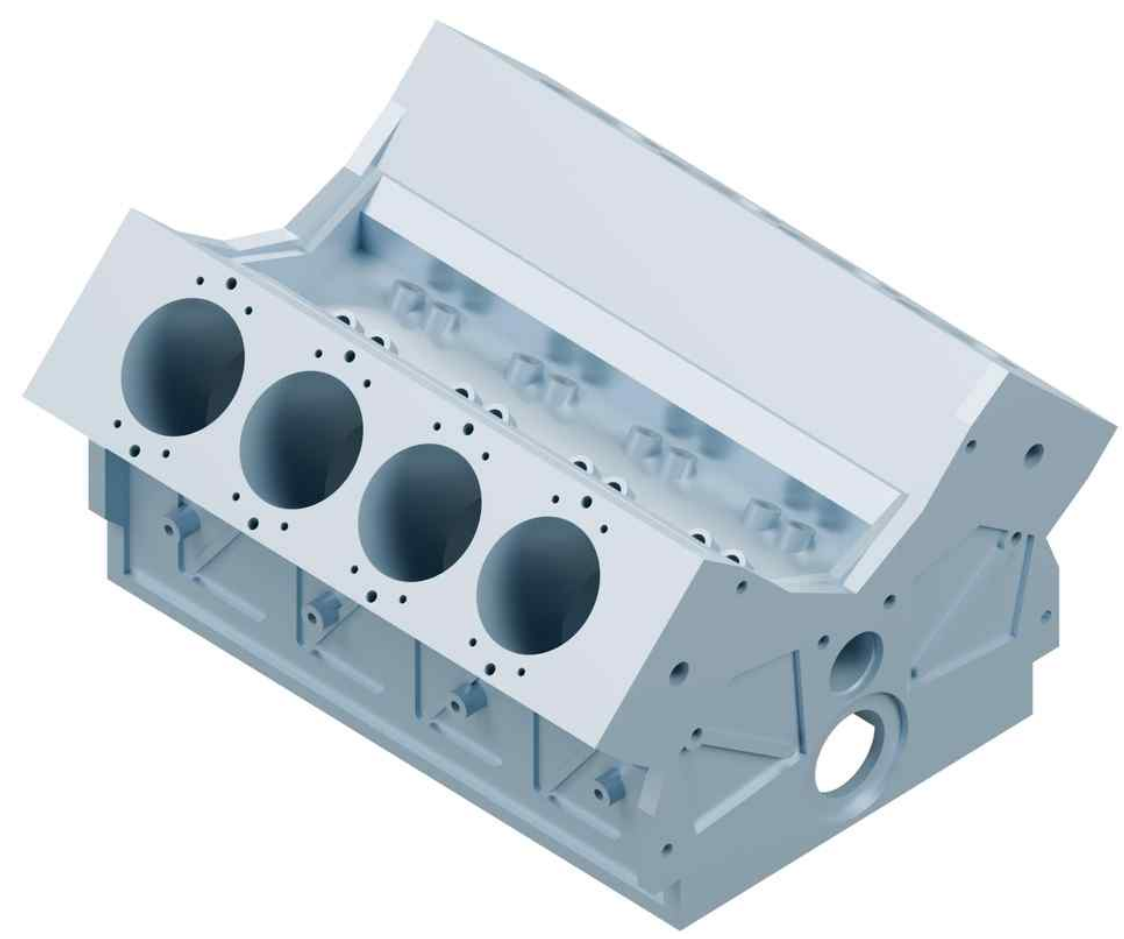

**(d) Rendered generated CAD model**

Figure 5-6 TC-05: modeling from geometric requirements

TC-05 asks GPT-6 to generate a single-solid V8 engine block with cylinders, water jackets, bearing seats, and internal passages. GPT-6 first constructs the two cylinder banks and crankcase, then the water jackets and longitudinal passages. It subsequently corrects water jackets that open into the central recess and longitudinal passages that are blocked. In the final model, the cylinder banks are arranged at 90°, the water jackets are connected but remain separate from the cylinder bores, and five webs along the bottom support coaxial crankshaft bearing seats. The percentage of geometric requirements met rises from 8.3% to 100% (Figure 5-6). The final block is a single solid and passes the validity check.

## TC-06 (complex)

Create a complete horizontal single-cylinder steam-engine assembly on a rectangular base approximately 170 × 80 × 14.5 mm, with rounded corners, mounting holes and a recessed underside. The cylinder lies along the long direction of the base and extends beyond one end. Its body is 35 mm long, with a 19 mm bore, a narrower barrel approximately 22 mm in diameter and wider end flanges approximately 32 mm in diameter.

**(a) Excerpt from the input text**

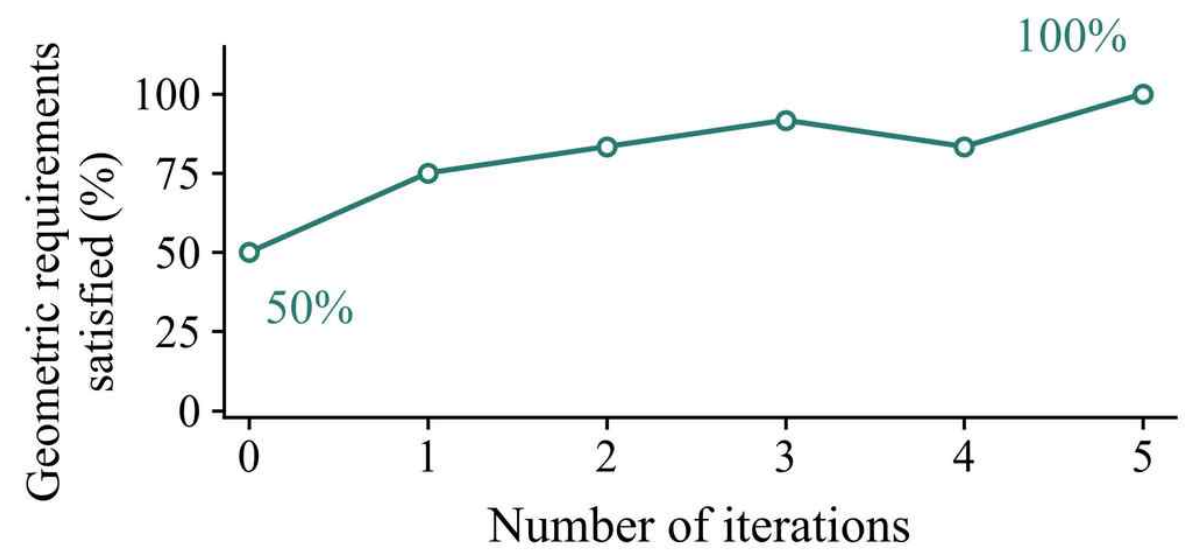

**(b) Percentage of geometric requirements met**

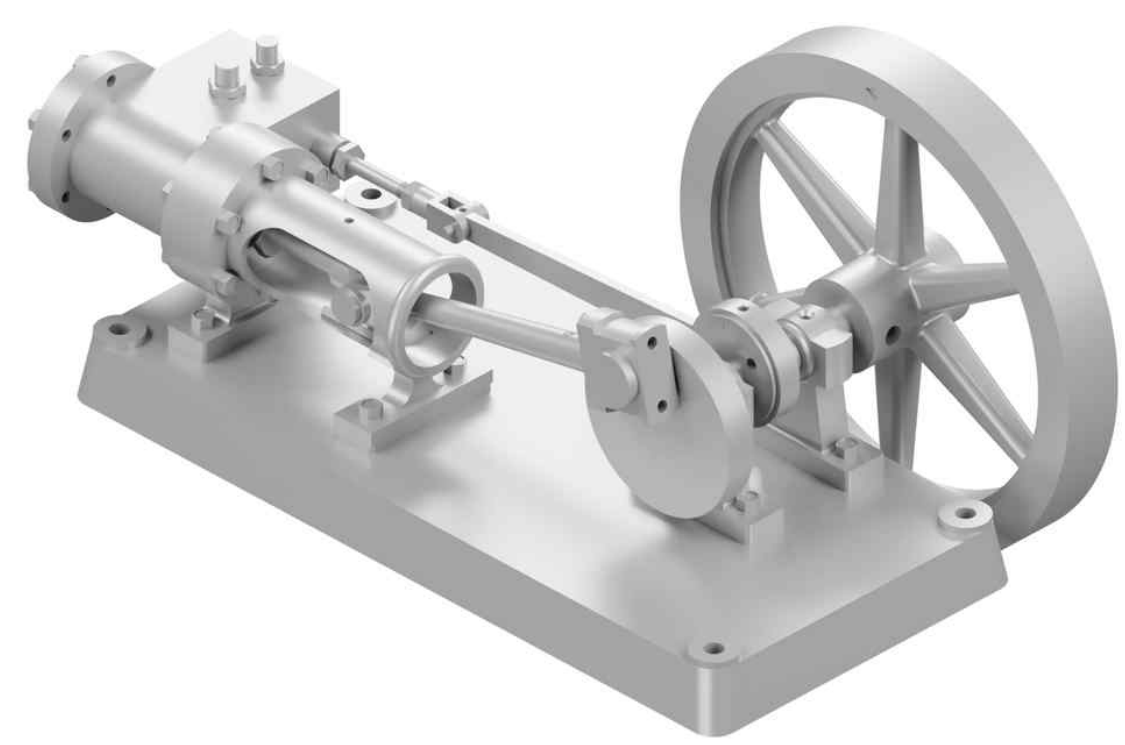

**(c) Rendered reference CAD model**

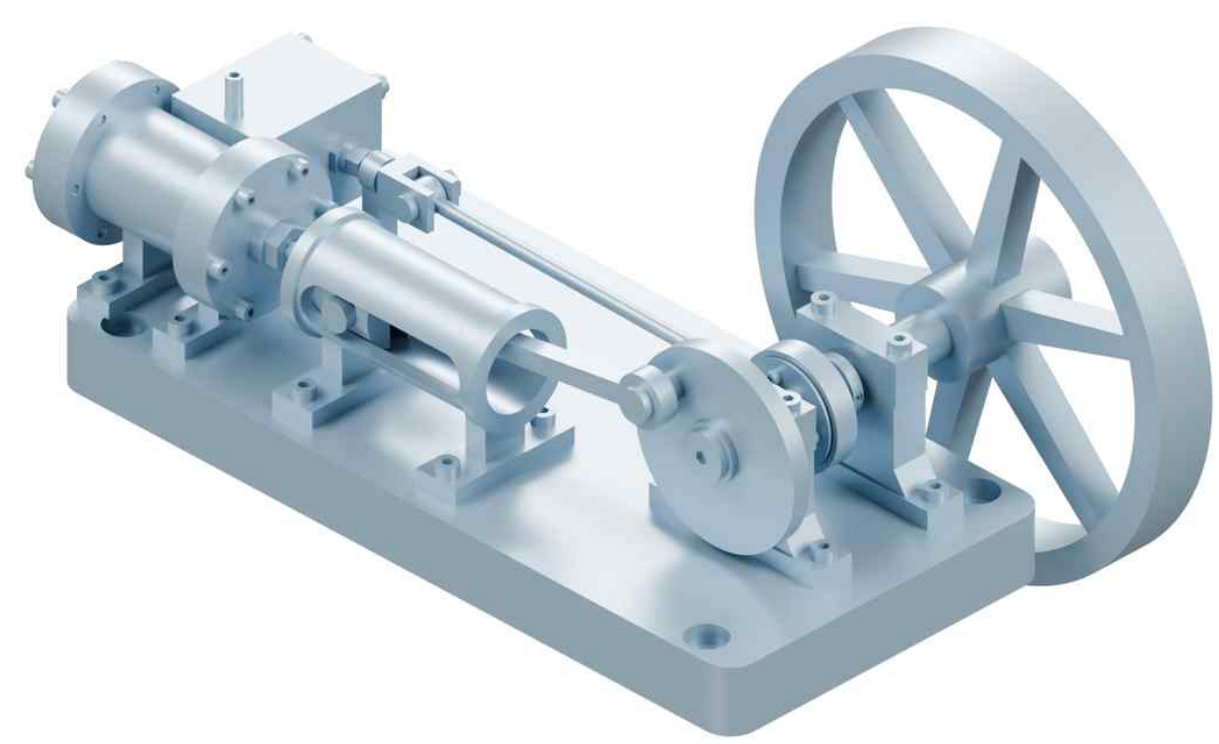

**(d) Rendered generated CAD model**

Figure 5-7 TC-06: modeling from geometric requirements

TC-06 asks GPT-6 to construct a horizontal single-cylinder steam engine to specified dimensions and assemble its main drive and valve gear. GPT-6 progressively constructs the cylinder, crank and connecting rod, flywheel, and eccentric-driven valve gear, then adds fasteners and adjusts component positions. In the final model, the piston rod and crosshead lie along the cylinder axis, the main connecting rod connects to the crank, and a separate eccentric rod drives the valve gear. The mechanisms have the required relative positions. The percentage of geometric requirements met rises from 50% to 100% (Figure 5-7). All 106 parts pass the solid validity check.

#### 5.3.1.2.2 Modeling from functional requirements

### TRC-01 (simple)

Design a one-piece motor and ball-screw support for a compact feed mechanism. The motor has an 86 mm square mounting face, a 73 mm diameter locating pilot projecting 2 mm, and four 6.6 mm through-holes on a 69.6 mm square pattern for M6 mounting bolts. The screw support uses a matched pair of angular-contact bearings, each with a 15 mm bore, 32 mm outside diameter, and 9 mm width.

**(a) Excerpt from the input text**

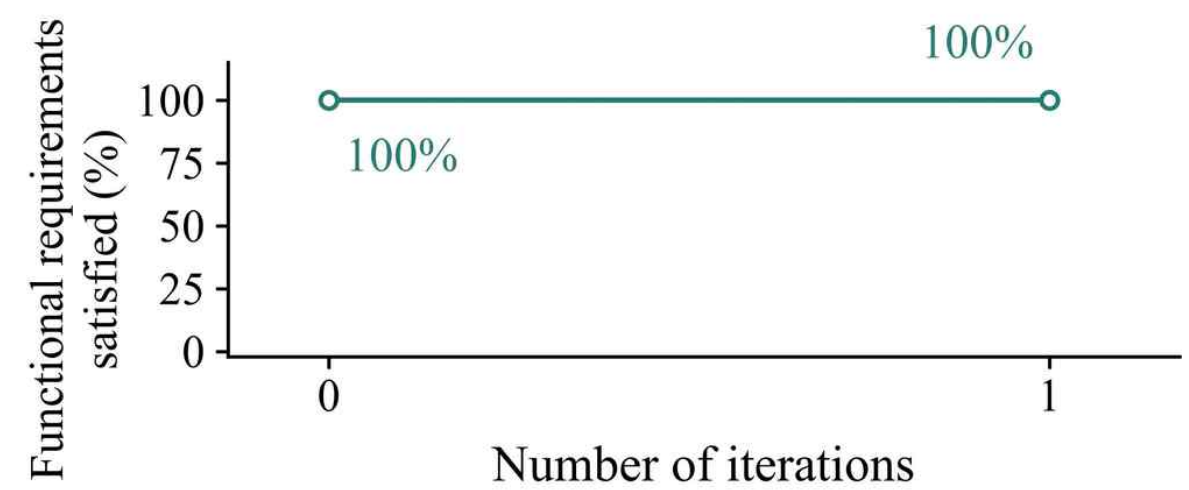

**(b) Percentage of functional requirements met**

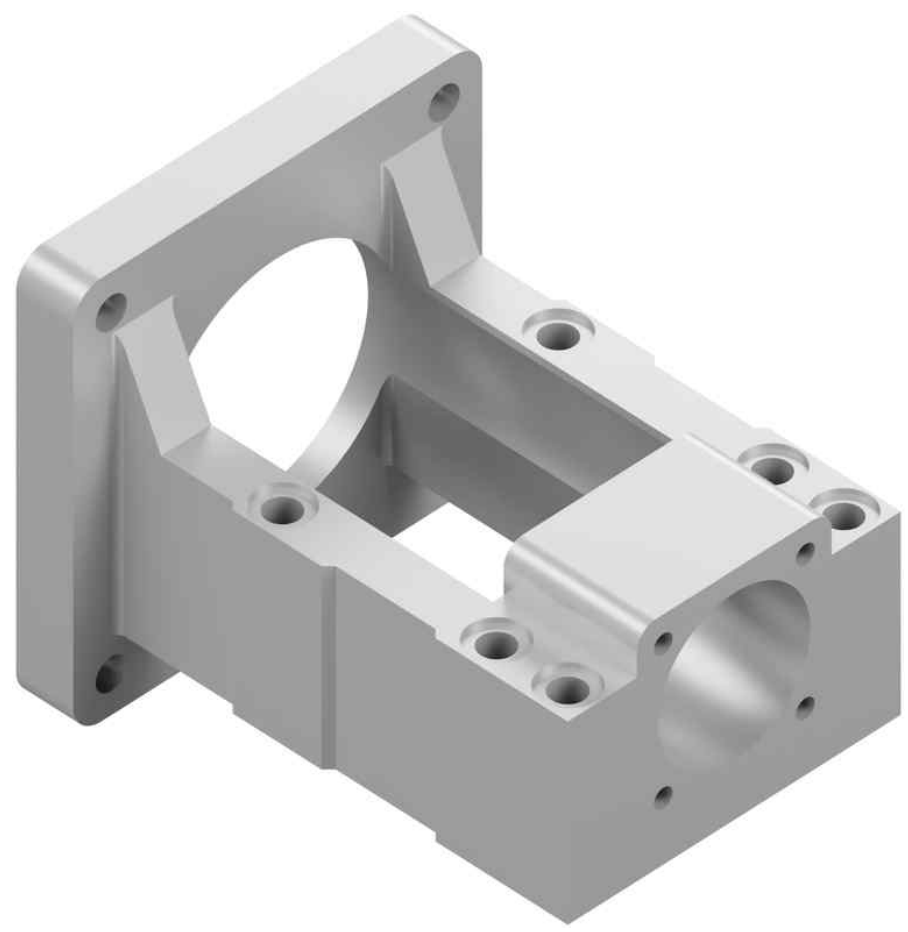

**(c) Rendered reference CAD model**

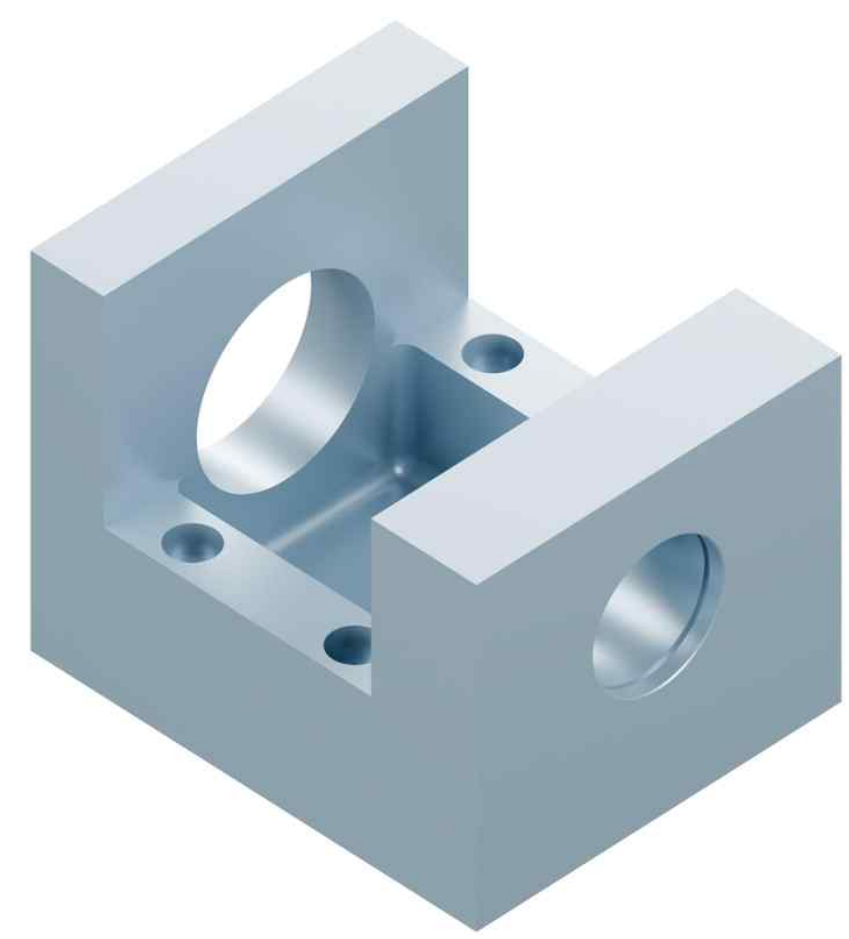

**(d) Rendered generated CAD model**

Figure 5-8 TRC-01: modeling from functional requirements

TRC-01 asks GPT-6 to design a one-piece motor and ball-screw support bracket that provides coaxial alignment, load-bearing stiffness, and space for assembly and adjustment. GPT-6 uses supports at both ends with an open central region, keeping the motor locating bore coaxial with the bearing seat while allowing access for coupling adjustment and bearing replacement. Under the specified load, the calculated relative radial displacement between the two shafts is approximately 0.0033 mm, below the 0.05 mm limit. The first saved model already meets all eight functional design requirements, and the percentage remains at 100% in subsequent iterations (Figure 5-8). The final bracket is a single solid and passes the validity check.

## TRC-02 (simple)

Design a manually operated bench vise for holding small steel and aluminium workpieces during filing, sawing, and assembly. The workpieces have opposing flat gripping surfaces and a clamped thickness of 10–60 mm. Provide at least 70 mm of usable jaw width and a clamping force of at least 1 kN with a hand force no greater than 100 N at the handle. The vise shall retain its clamping force after the handle is released.

**(a) Excerpt from the input text**

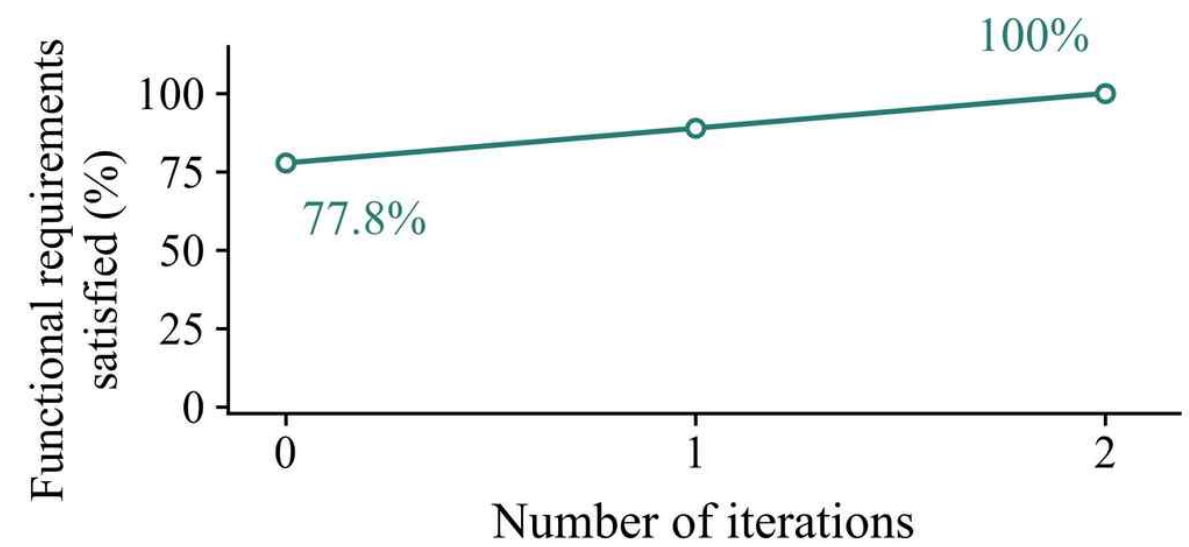

**(b) Percentage of functional requirements met**

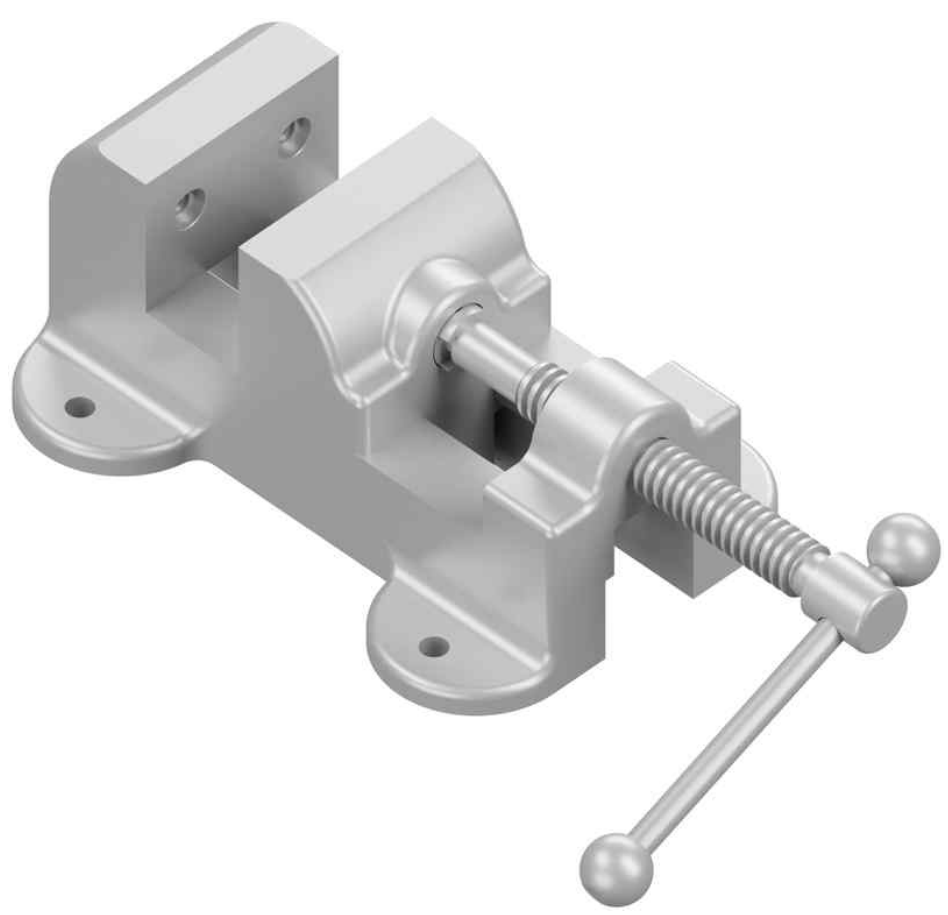

**(c) Rendered reference CAD model**

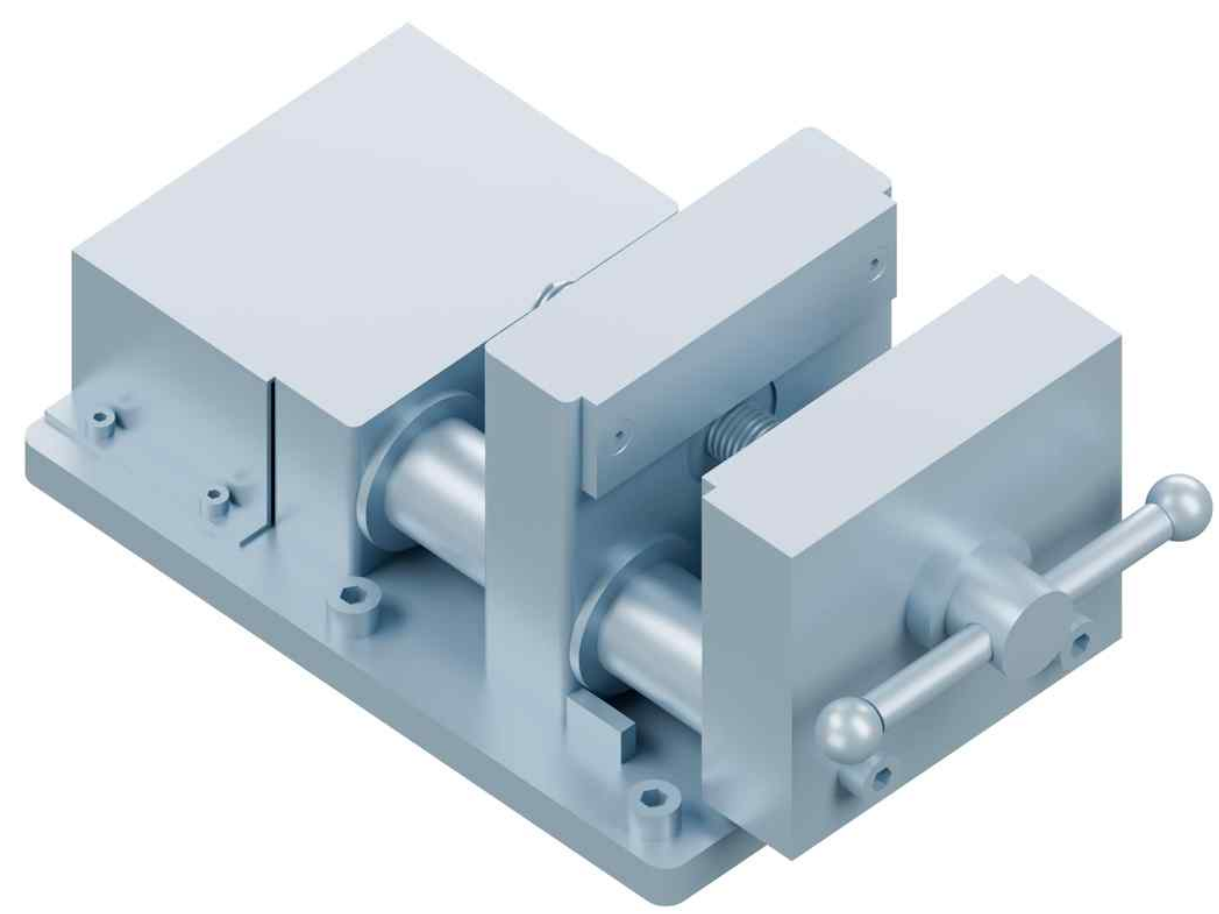

**(d) Rendered generated CAD model**

Figure 5-9 TRC-02: modeling from functional requirements

TRC-02 asks GPT-6 to design a manual bench vise that meets requirements for clamping range, clamping force, and self-locking, while allowing installation and maintenance. GPT-6 uses a screw drive and two guide rods to keep the moving jaw parallel. During iteration, it adjusts the thrust washer, stops, and jaw fastenings to eliminate local interference between moving components. The final design has a calculated clamping force of approximately 1.15 kN under a hand force of 100 N; the jaws and soft pads can each be replaced separately. The percentage of functional requirements met rises from 77.8% to 100% (Figure 5-9). All 36 parts pass the solid validity check.

## TRC-03 (moderate)

Design a one-piece cast-iron casing for a single-stage end-suction pump conveying clean water at 10–60 °C. The application has a nominal pump duty of 500 m³/h at 30 m total head. The matching rotor is fixed: its closed impeller has a 650 mm outside diameter, 100 mm overall axial width, 250 mm inlet opening, and 60 mm outlet width, and rotates clockwise at 730 rpm when viewed from the suction side.

**(a) Excerpt from the input text**

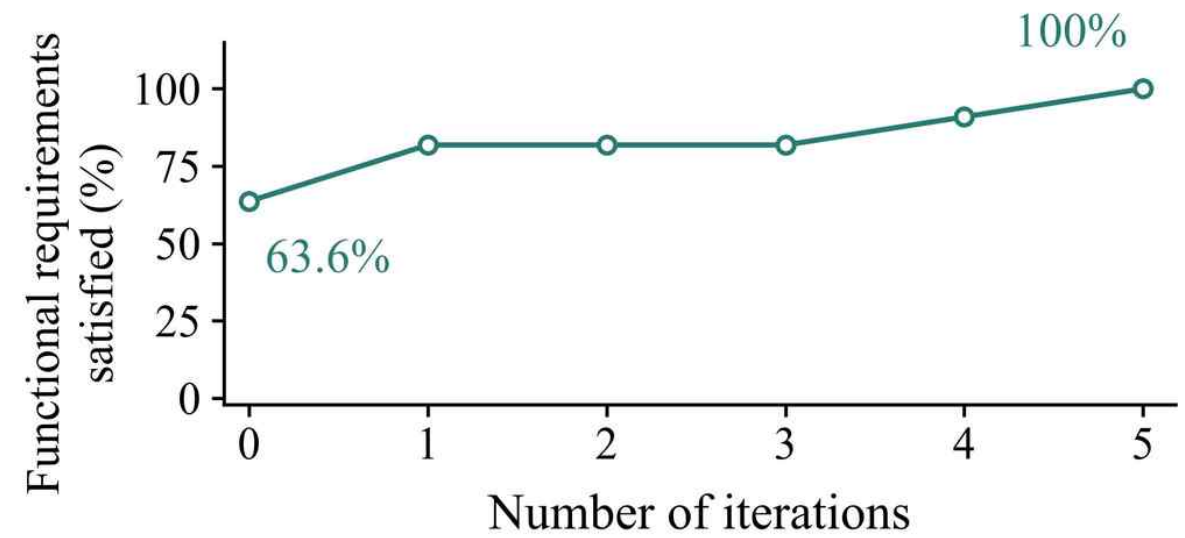

**(b) Percentage of functional requirements met**

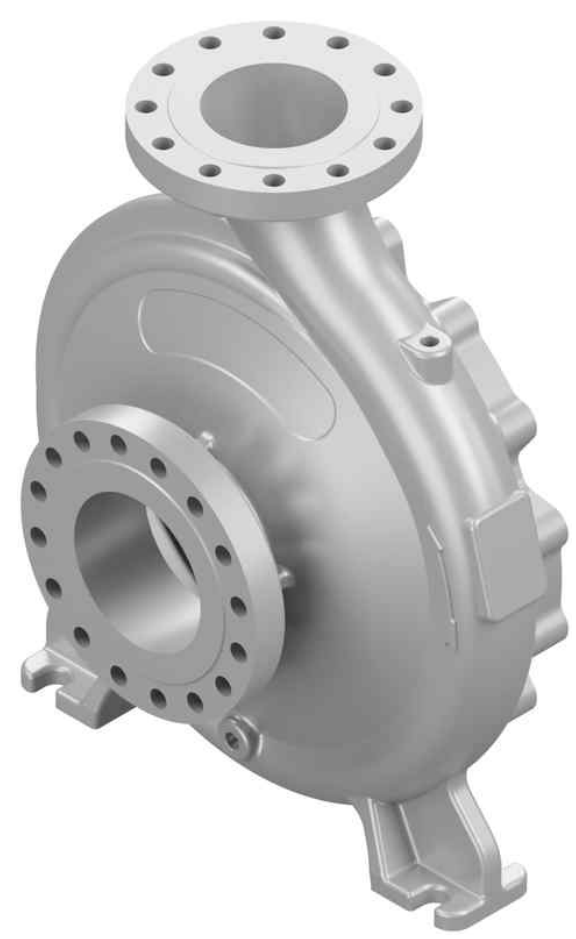

**(c) Rendered reference CAD model**

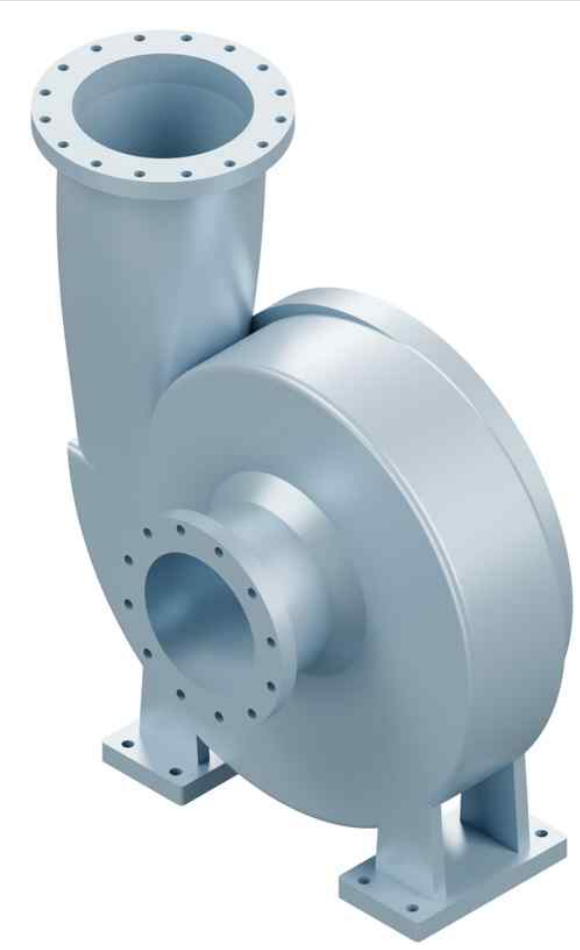

**(d) Rendered generated CAD model**

Figure 5-10 TRC-03: modeling from functional requirements

TRC-03 asks GPT-6 to design a one-piece pump casing around a specified impeller, meeting requirements for flow guidance, pressure containment, pipe connections, and rotor removal. GPT-6 constructs a connected volute passage and a rear removal opening, then adjusts the outlet shape, thickens local pressure-bearing regions, and moves the foot mounting holes outward. The final model provides access to flange bolts and clearance for tools at the mounting feet, and allows the rotor to be withdrawn from the rear. Calculations of head loss and pressure-bearing capacity satisfy the requirements for the specified operating conditions. The percentage of functional requirements met rises from 63.6% to 100% (Figure 5-10). The final casing is a single solid and passes the validity check.

### TRC-04 (moderate)

| | |
|---|---|
| Design a pneumatic two-finger gripper for transferring dry aluminium workpieces weighing up to 1 kg. Each workpiece has opposing flat gripping areas of at least 30 × 30 mm, separated by 20–60 mm, with its centre of mass within 20 mm of the midpoint between the contact areas. | 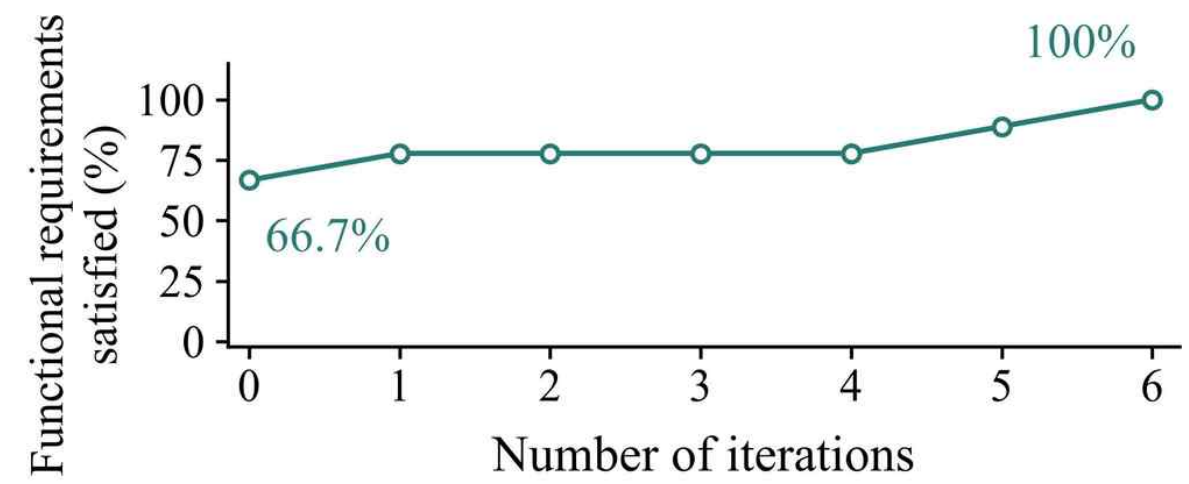 |
| **(a) Excerpt from the input text** | **(b) Percentage of functional requirements met** |
| 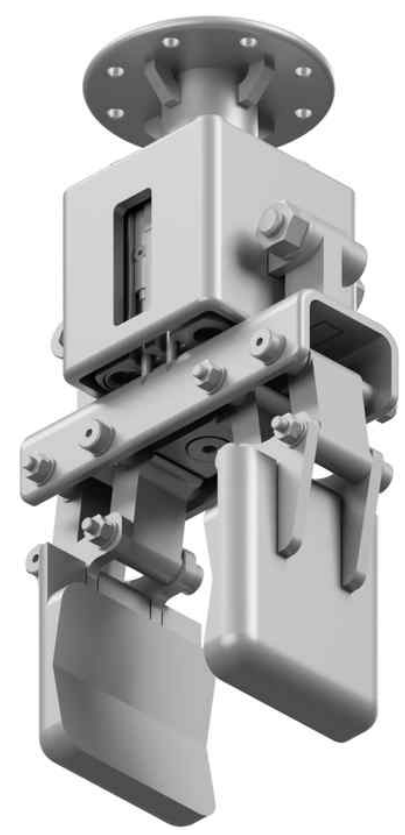 | 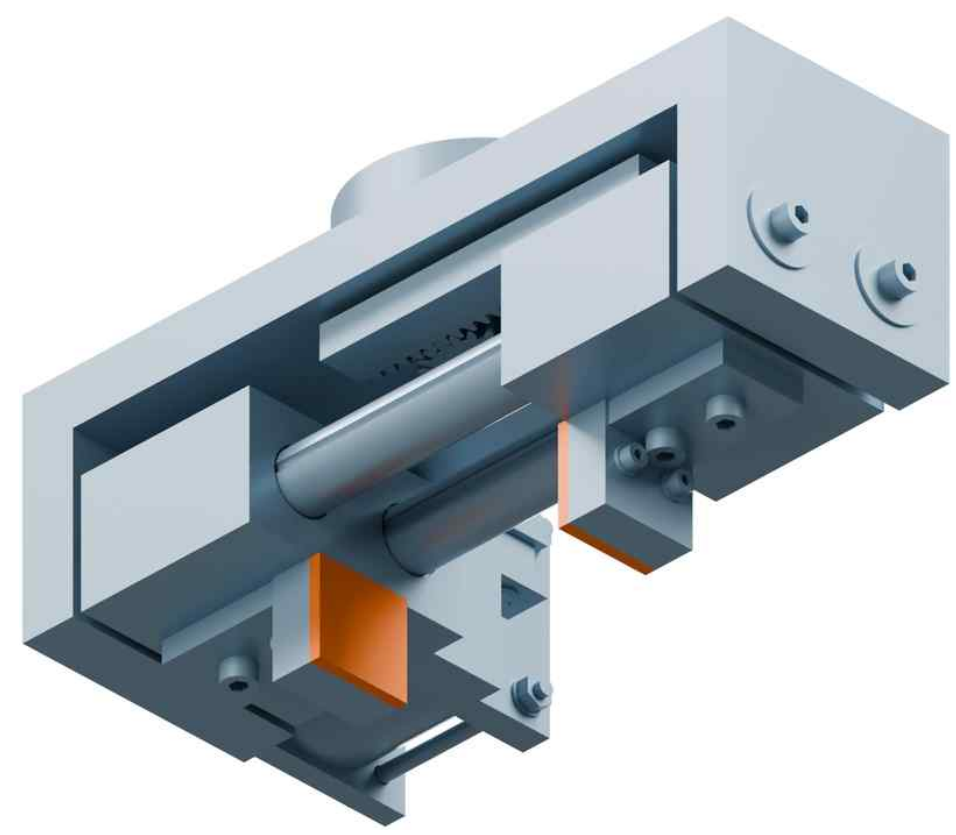 |
| **(c) Rendered reference CAD model** | **(d) Generated model: gripping pads shown in orange** |
| 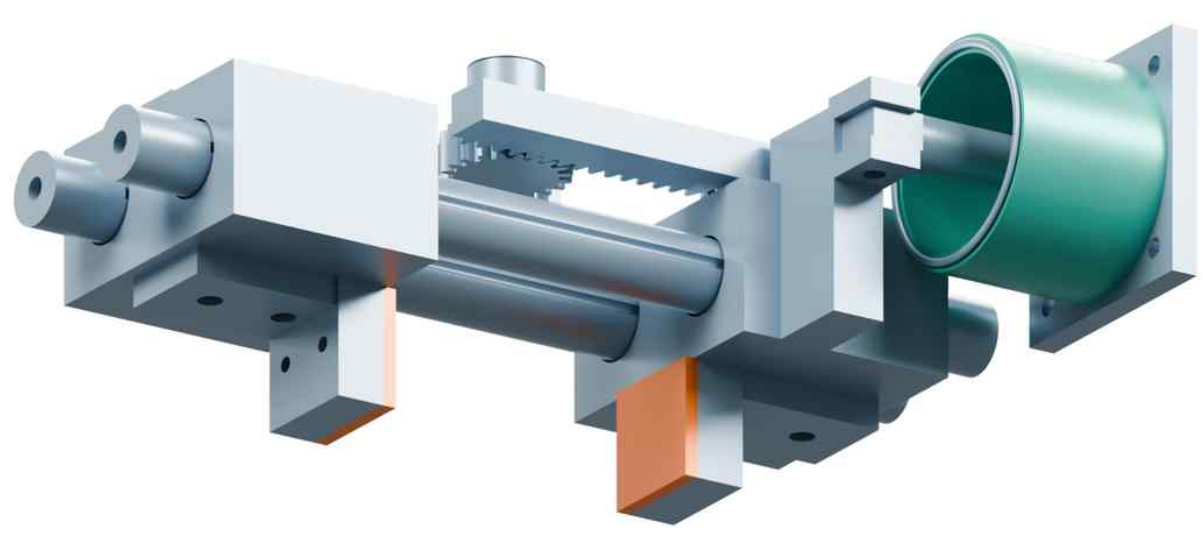 | 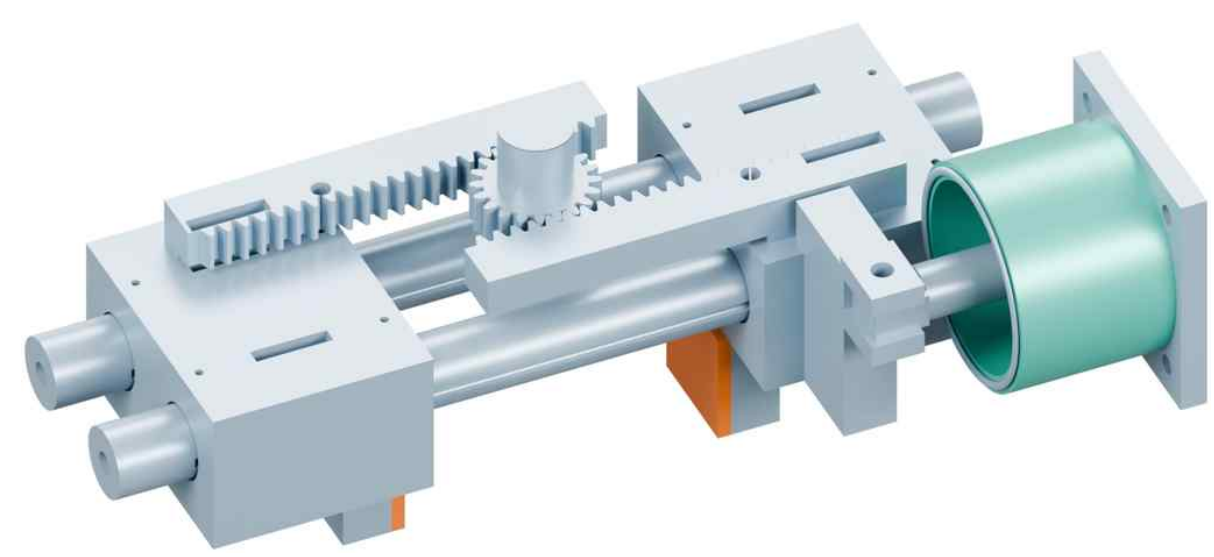 |
| **(e) Fingers and guide shafts with the outer frame removed** | **(f) Internal transmission: cylinder barrel shown in green** |

Figure 5-11 TRC-04: modeling from functional requirements

TRC-04 asks GPT-6 to design a pneumatic two-finger gripper that can handle a 1 kg workpiece securely while meeting opening-range and surface-protection requirements. GPT-6 selects a cylinder-driven parallel gripper with rack-and-pinion synchronization. Two guide shafts constrain finger motion, and an outer frame supports the shafts and transmission components. This produces a transverse frame different from the reference linkage mechanism. The fingers lie below the frame, with orange soft pads contacting the workpiece directly. The design has a clear opening of 71 mm and overall dimensions of 229 × 117 × 99 mm, close to the specified upper limits. GPT-6 improves the guiding and support structures over successive iterations, raising the percentage of functional requirements met from 66.7% to 100% (Figure 5-11). All 89 parts pass the solid validity check.

### TRC-05 (complex)

Design a one-piece aluminium cylinder-head body for a naturally aspirated, water-cooled, inline-four development engine with a 73 mm bore, 82.8 mm stroke, and 79 mm cylinder spacing. The fixed block interface has a flat deck and ten M10 attachment points arranged in two parallel rows 100 mm apart; each row contains five equally spaced points at 79 mm pitch, centred on the four-cylinder group.

**(a) Excerpt from the input text**

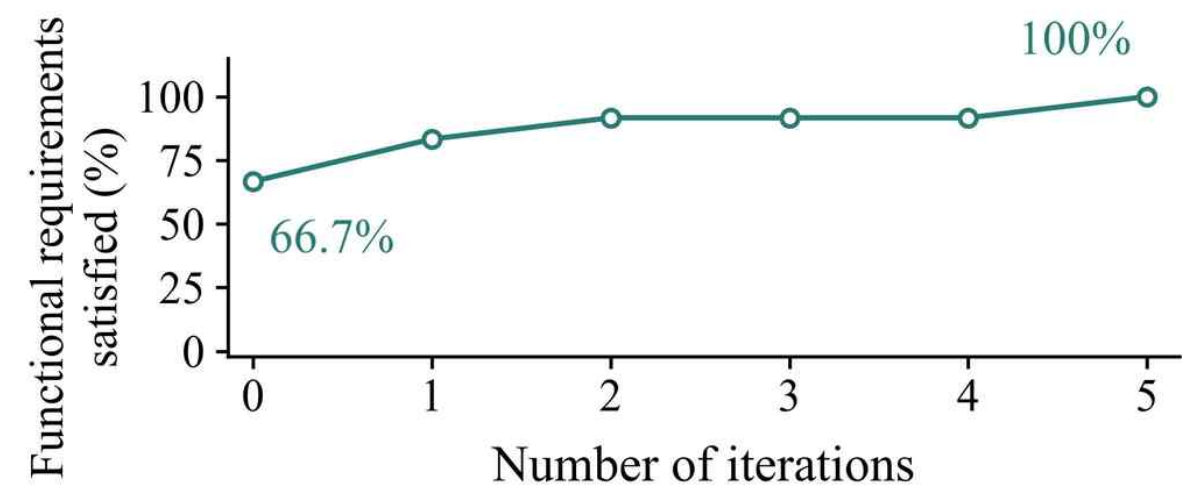

**(b) Percentage of functional requirements met**

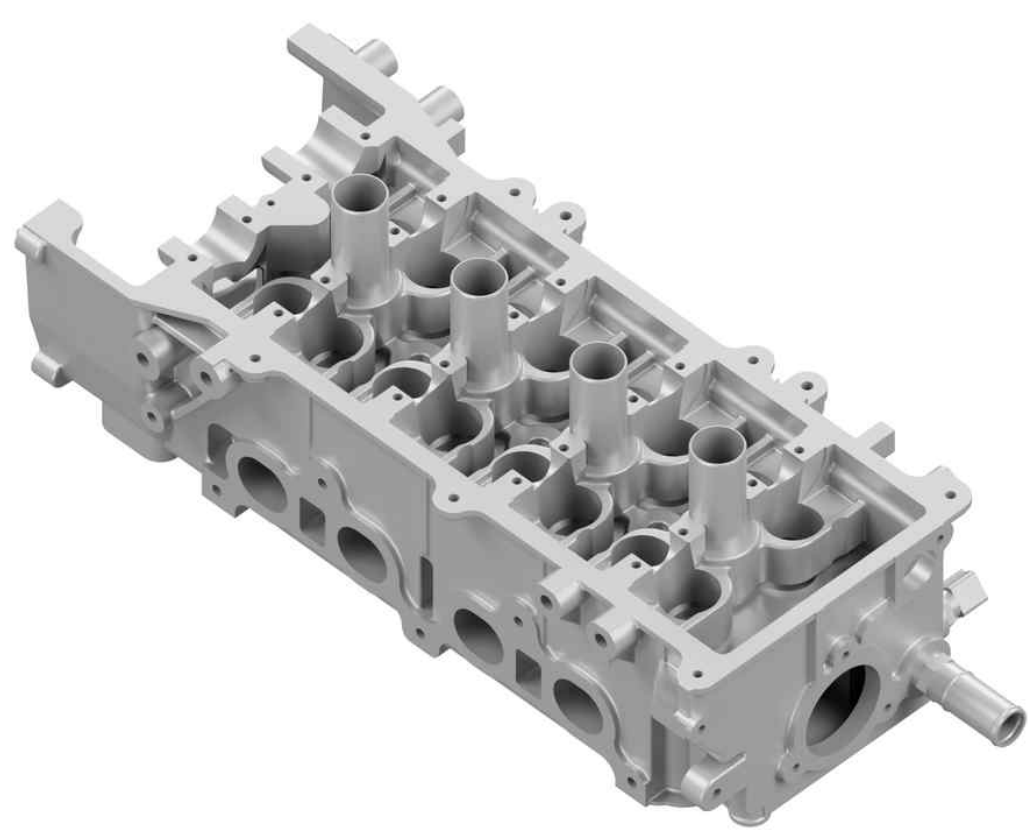

**(c) Rendered reference CAD model**

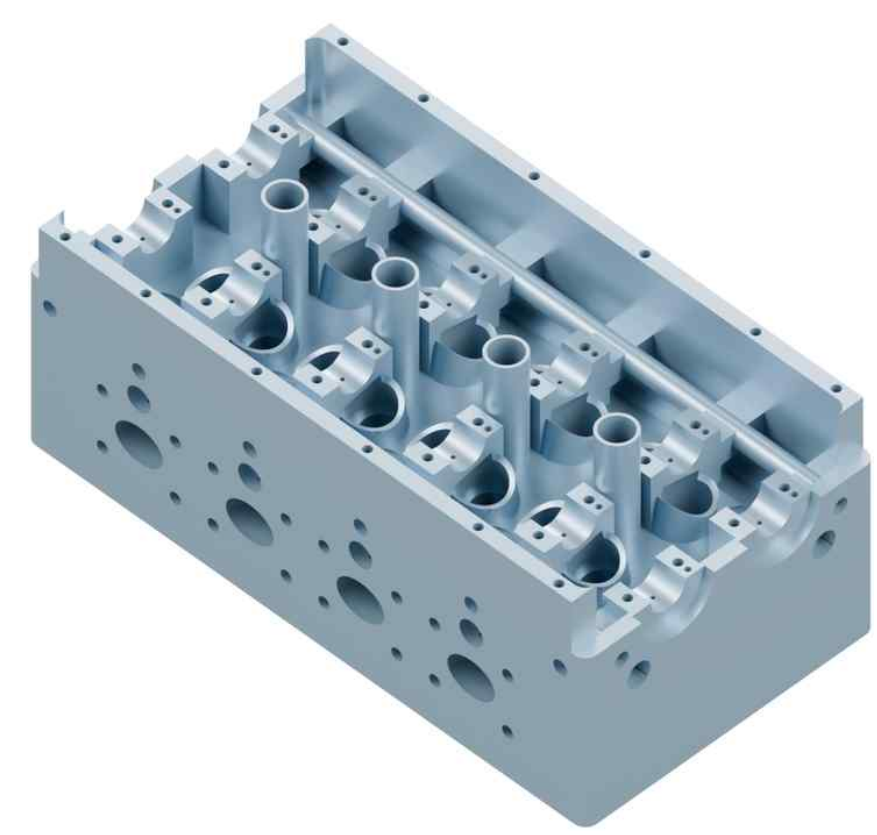

**(d) Rendered generated CAD model**

Figure 5-12 TRC-05: modeling from functional requirements

TRC-05 asks GPT-6 to design a one-piece cylinder head that provides valve operation, pressure containment, cooling, lubrication, and maintenance access. Across iterations, GPT-6 thickens the walls between passages, improves the cooling inlet, and opens a path for camshaft removal, raising the percentage of functional requirements met from 66.7% to 100%. The final model provides valve and spark-plug mounting holes for each of the four cylinders. Intake and exhaust ports, water jackets, and lubrication passages remain separate, and the valves and camshafts have the required space for motion, assembly, and removal (Figure 5-12). The final cylinder head passes the solid validity check.

### TRC-06 (complex)

Design a six-legged, battery-powered chassis for indoor inspection on dry, hard floors, using 0.5 as the minimum friction coefficient between the feet and ground. Carry a centrally mounted 1 kg inspection module measuring 100 × 100 × 60 mm, with its centre of mass 30 mm above the mounting face and four M4 attachment points on an 80 mm square pattern.

**(a) Excerpt from the input text**

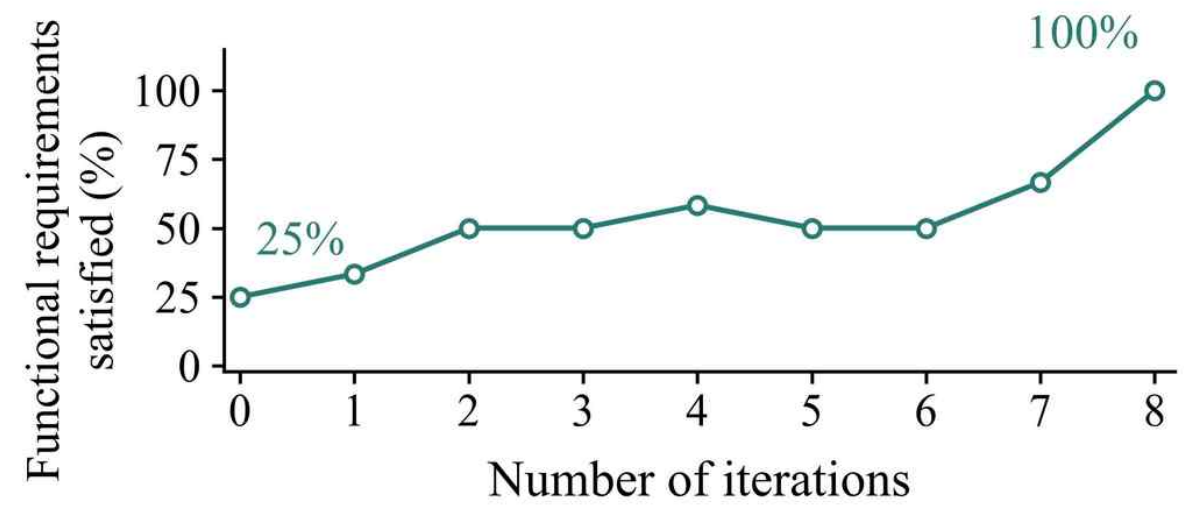

**(b) Percentage of functional requirements met**

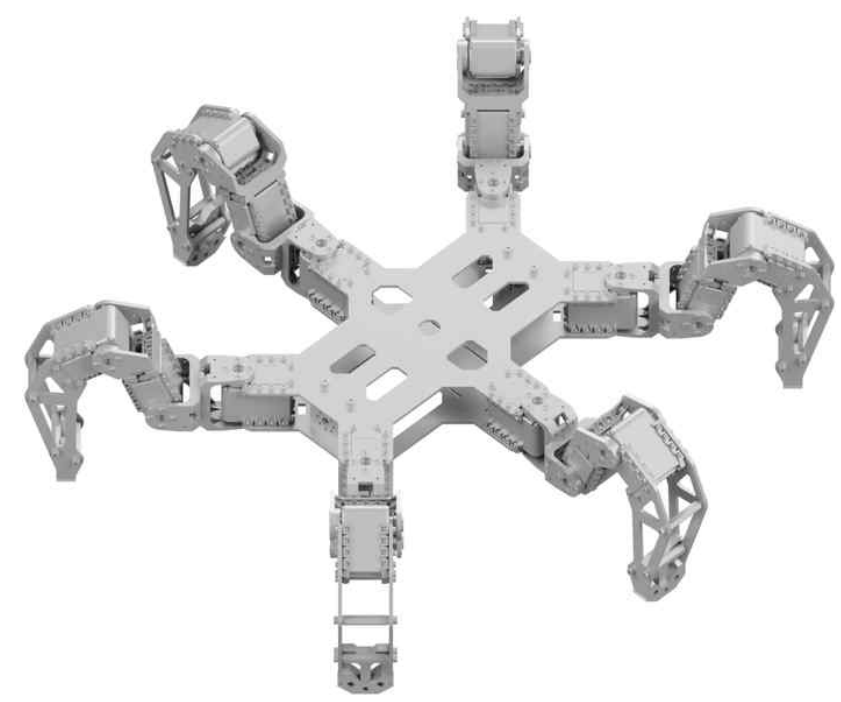

**(c) Rendered reference CAD model**

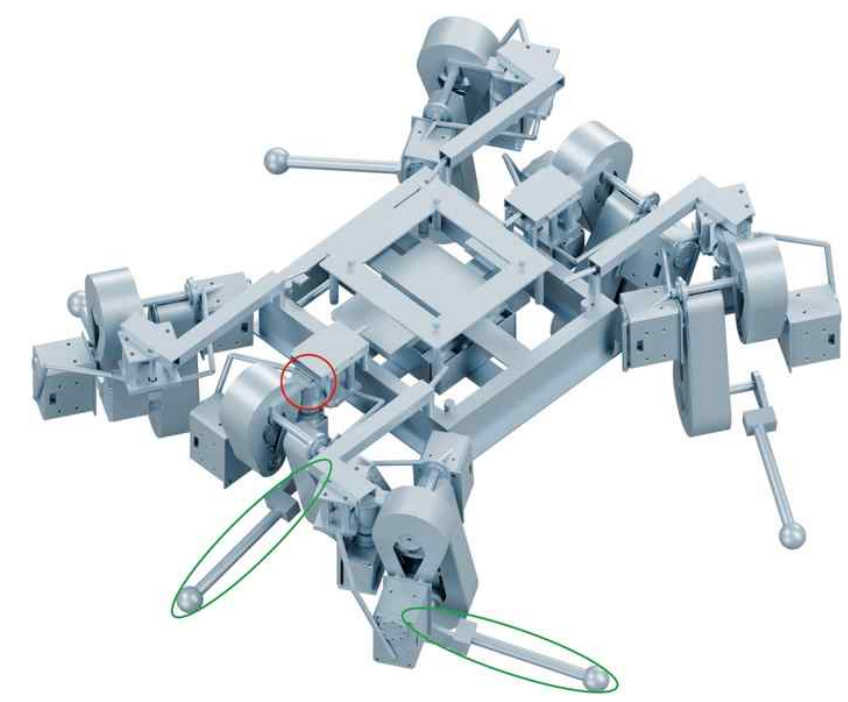

**(d) Generated model: red circle locates (e); green circles mark the legs**

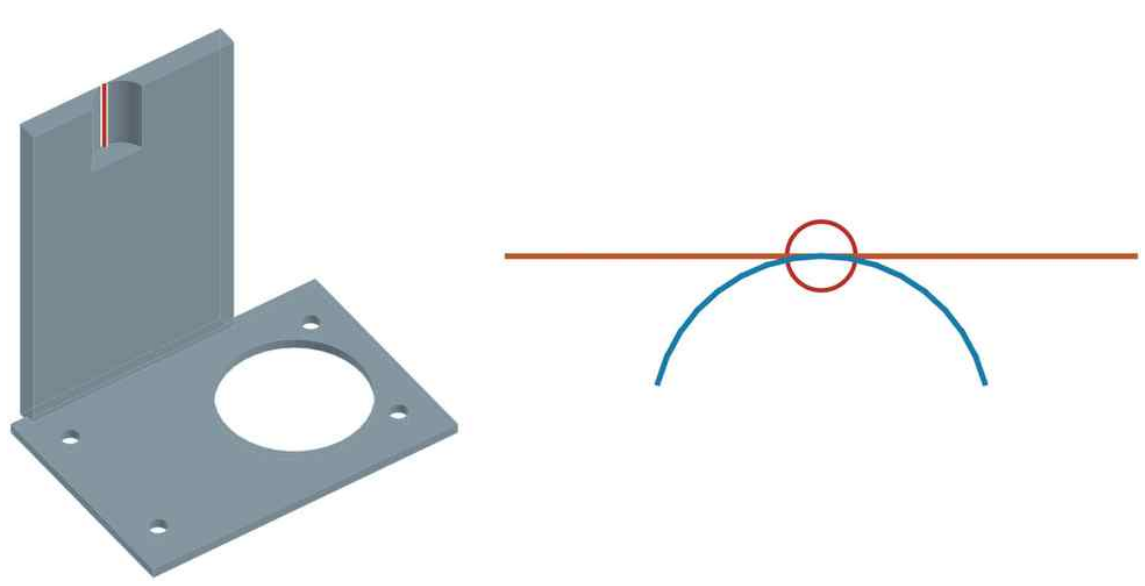

**(e) Motor support: self-intersecting solid boundary**

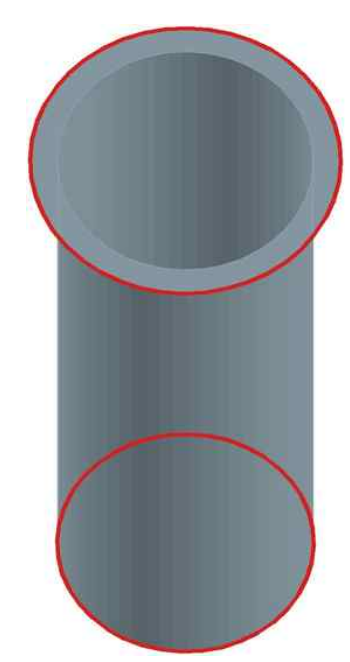

**(f) Open shell: red boundary edges have no adjacent faces**

Figure 5-13 TRC-06: modeling from functional requirements

TRC-06 asks GPT-6 to design a hexapod chassis that meets requirements for load capacity, walking, obstacle traversal, and operating time. By reducing weight and adjusting the space available for leg motion, GPT-6 raises the percentage of functional requirements met from 25% to 100%. Green circles in the figure mark the outward-extending lower legs and spherical feet. All six feet contact a level ground plane, and the center-of-mass projection lies within the support region. Static calculations indicate that this arrangement can support the chassis and a 1 kg inspection module. The chassis has a mass of approximately 2.94 kg and a calculated operating time of 23.25 min. Of its 907 solids, 904 pass the validity check. Three motor supports fail because their solid boundaries intersect themselves: a hole wall is tangent to an outer face along a line that is not a shared edge of the two faces. The red circle in Figure 5-13(d) locates one of these supports within the chassis, and Figure 5-13(e) shows its local self-intersection. The model also contains 36 separate open shells. The complete chassis therefore fails the solid validity check.

### 5.3.2 Additional experiment results

The additional experiments comprise six tests. S1 and S2 use a seven-cylinder radial engine and an electric helicopter mechanical assembly, respectively, with inputs describing the geometric or functional requirements of the reference model. Web search is enabled for both. S3 tests TC-05, TC-06, TRC-05, and TRC-06 with web search disabled. The evaluation team separates each assembly into individual part STEP files and checks the solids in those files. Fractions in the tables give the number of solids that pass divided by the total number of solids. Errors encountered during checking are counted as failures.

#### 5.3.2.1 Overview

Tables 5-12 and 5-13 summarize the complex assemblies produced in the two stress tests. In S1, GPT-6 models the seven-cylinder radial arrangement, master and articulated connecting rods, valve gear, and intake and exhaust pipes. The final model meets all 12 geometric requirements. In S2, GPT-6 completes the main component layout and addresses mass and operating time in its design. However, the rotor pitch links are disconnected from the swashplate, and the gimbal motor is not fastened to its support plate. The final design meets nine functional requirements but fails three: flight control, attitude adjustment in crosswinds, and camera pointing stability. Neither model passes the solid validity check. In S1, 385 of 394 solids pass and nine fail; in S2, 187 of 188 pass and one fails.

Tables 5-14 and 5-15 summarize S3, with web search disabled. For geometric requirements, the shapes, dimensions, and spatial relationships in TC-05 and TC-06 meet the input requirements, and both models pass the solid validity check. For functional requirements, GPT-6 designs the cylinder head's passages, space for valve gear, and pressure-bearing and cooling structures in TRC-05. In TRC-06, it develops a chassis design that accommodates the payload and meets the requirements for walking, obstacle traversal, and operating time. Both cases meet all 12 functional design requirements. However, the chassis still has protective tubing extending to feet with no electrical components, leaving the tube ends unsupported. Neither case passes the solid validity check. In TRC-05, some face boundary loops meet at multiple points. In TRC-06, one foot connection pin has an incorrectly oriented shell; the remaining 348 solids pass. Table 5-16 reports completion times.

Table 5-12 S1 stress test: modeling from geometric requirements

| Case ID | Complexity | Solid validity | Listed requirements met | Results |
|---|---|---|---|---|
| TC-07 | Complex | Fail 385/394 valid | All 12 met | The seven-cylinder layout and main mechanisms meet the requirements; the master rod has a non-manifold edge, and intake-pipe surfaces intersect themselves. |

Table 5-13 S2 stress test: modeling from functional requirements

| Case ID | Complexity | Solid validity | Listed requirements met | Results |
|---|---|---|---|---|
| TRC-07 | Complex | Fail 187/188 valid | 9 met; 3 unmet | Pitch links are disconnected from the swashplate, and the gimbal lacks a fastening connection; the lower fairing fails the solid validity check. |

Table 5-14 S3 geometric-requirements results with web search disabled

| Case ID | Complexity | Solid validity | Listed requirements met | Results |
|---|---|---|---|---|
| TC-05 | Complex | Pass | All 12 met | The block is a complete single solid; cylinder-bore, inclined-hole, and water-jacket layouts meet the requirements. |
| TC-06 | Complex | Pass | All 12 met | The cylinder, crank and connecting rod, and valve gear have the correct positions; the assembly is complete. |

Table 5-15 S3 functional-requirements results with web search disabled

| Case ID | Complexity | Solid validity | Listed requirements met | Results |
|---|---|---|---|---|
| TRC-05 | Complex | Fail 0/1 valid | All 12 met | Combustion chambers, gas passages, and water jackets remain separate; some face boundary loops meet at multiple points. |
| TRC-06 | Complex | Fail 348/349 valid | All 12 met | Protective tubing outside the legs forms unnecessary unsupported branches; 1 foot connection pin has an incorrectly oriented shell. |

Table 5-16  Completion times in the additional experiments

| Case ID | S1_TC-07 | S2_TRC-07 | S3_TC-05 | S3_TC-06 | S3_TRC-05 | S3_TRC-06 |
|---|---|---|---|---|---|---|
| Completion time | 40 min | 1 h 8 min | 48 min | 45 min | 44 min | 1 h 51 min |

#### 5.3.2.2 Case analyses

##### 5.3.2.2.1 S1: Geometric-requirements stress test

### TC-07 (complex)

Create a complete single-row, seven-cylinder radial-engine assembly approximately 280 mm long, with a 440 mm propeller span. The cylinder assemblies, including rocker covers, have an overall envelope diameter of approximately 240 mm. Seven cylinders surround the main shaft at equal angles in a common transverse plane, with one upright. The central crankcase is approximately 100 mm in diameter and 67 mm long.

**(a) Excerpt from the input text**

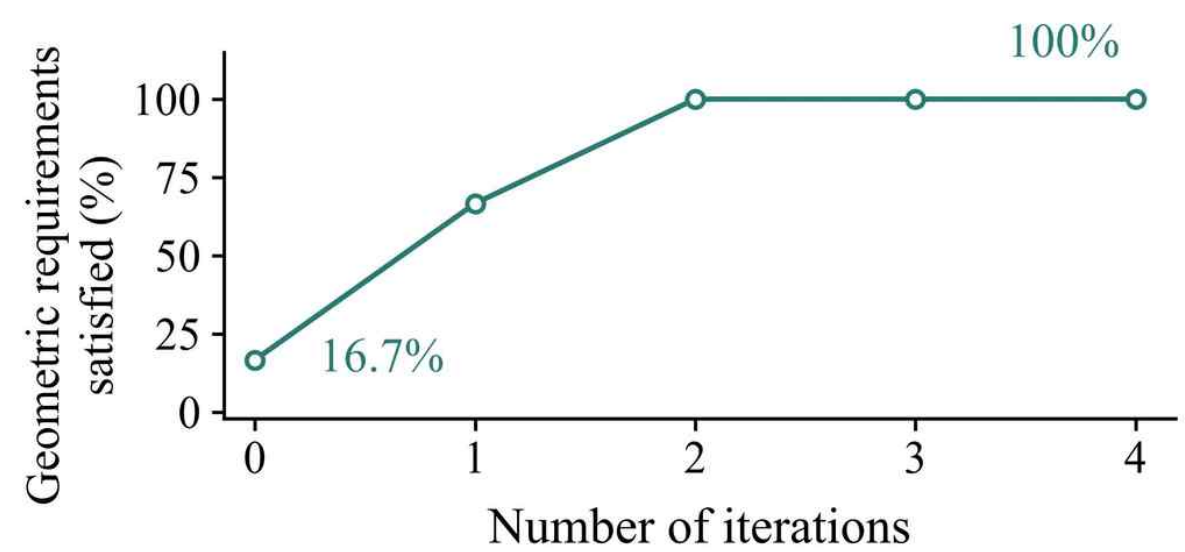

**(b) Percentage of geometric requirements met**

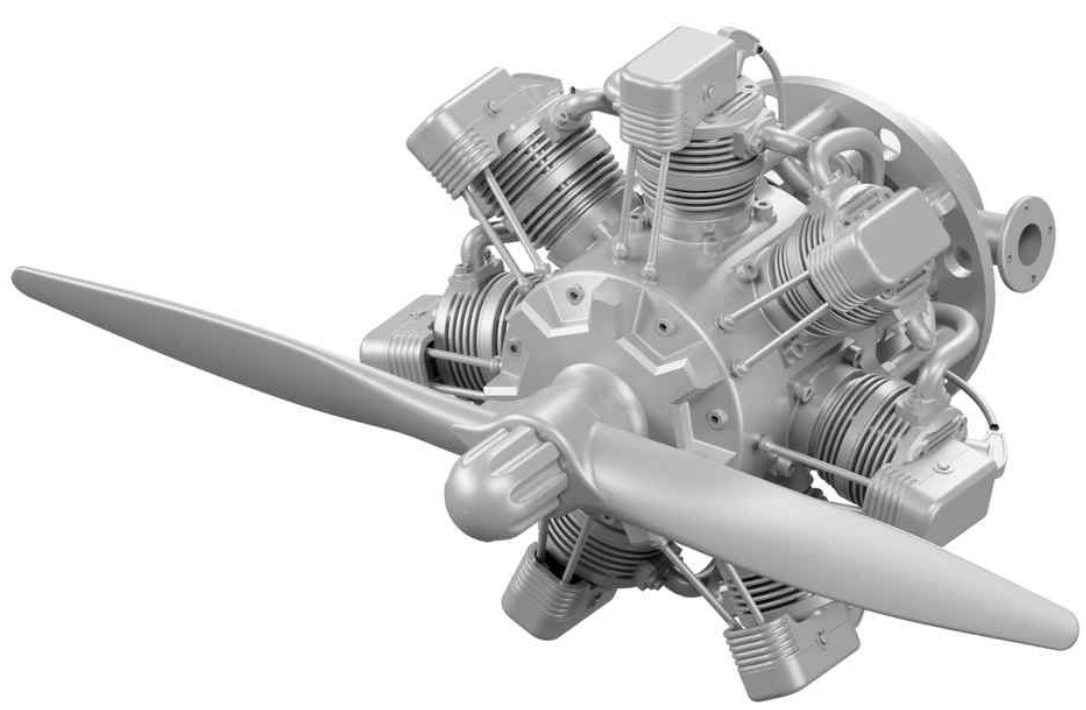

**(c) Rendered reference CAD model**

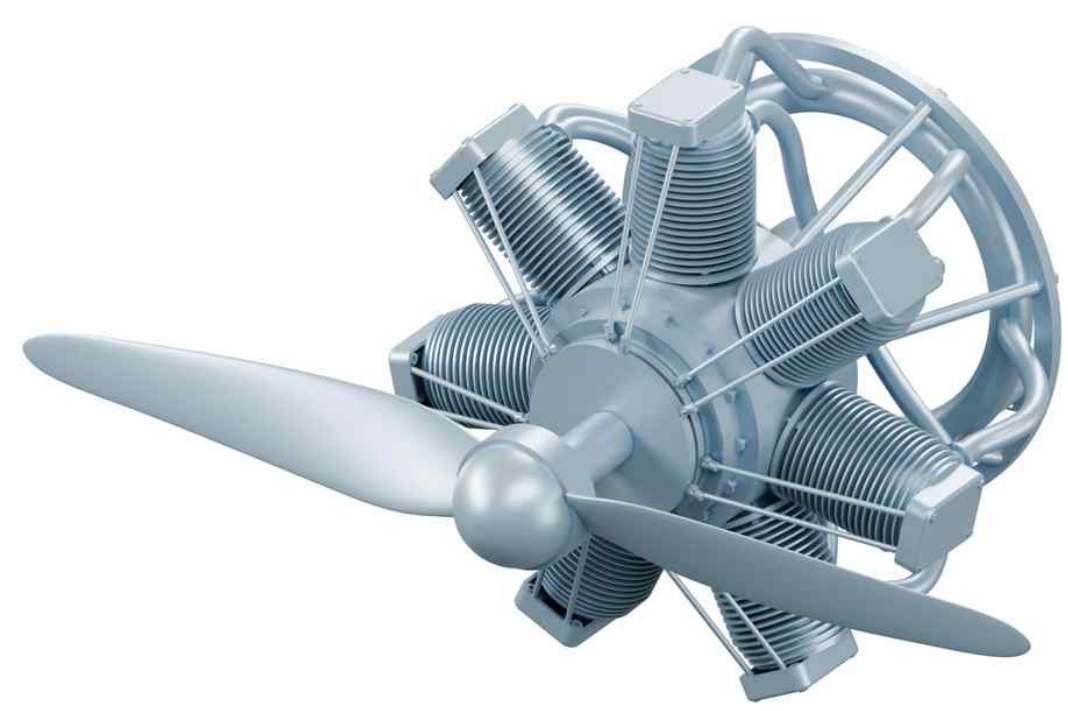

**(d) Rendered generated CAD model**

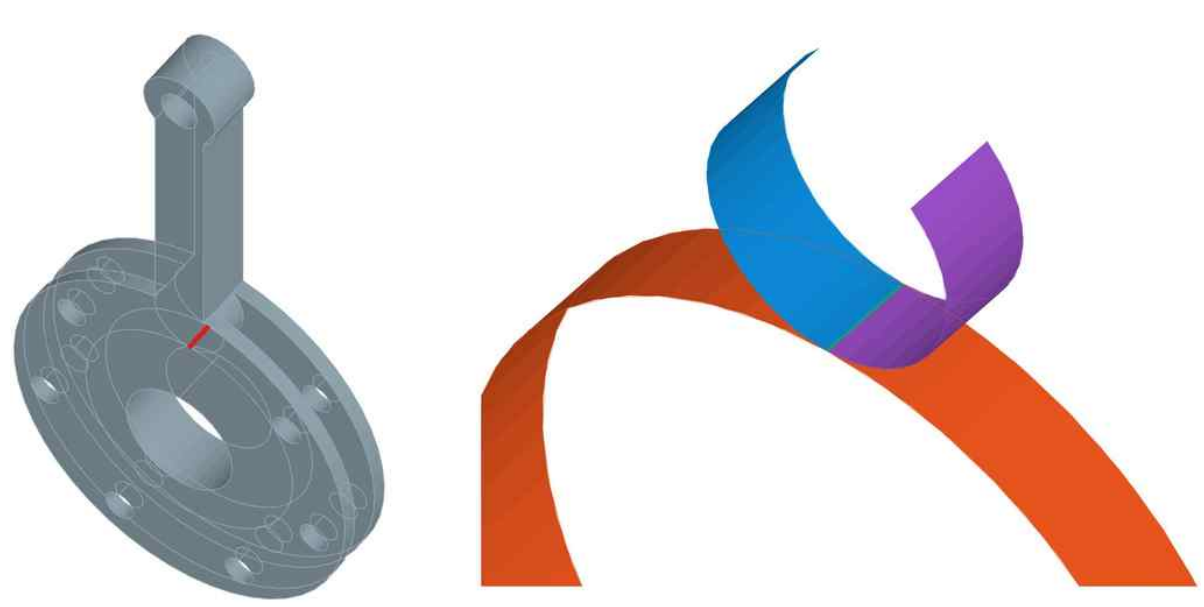

**(e) Master rod: red edge used repeatedly in face boundaries**

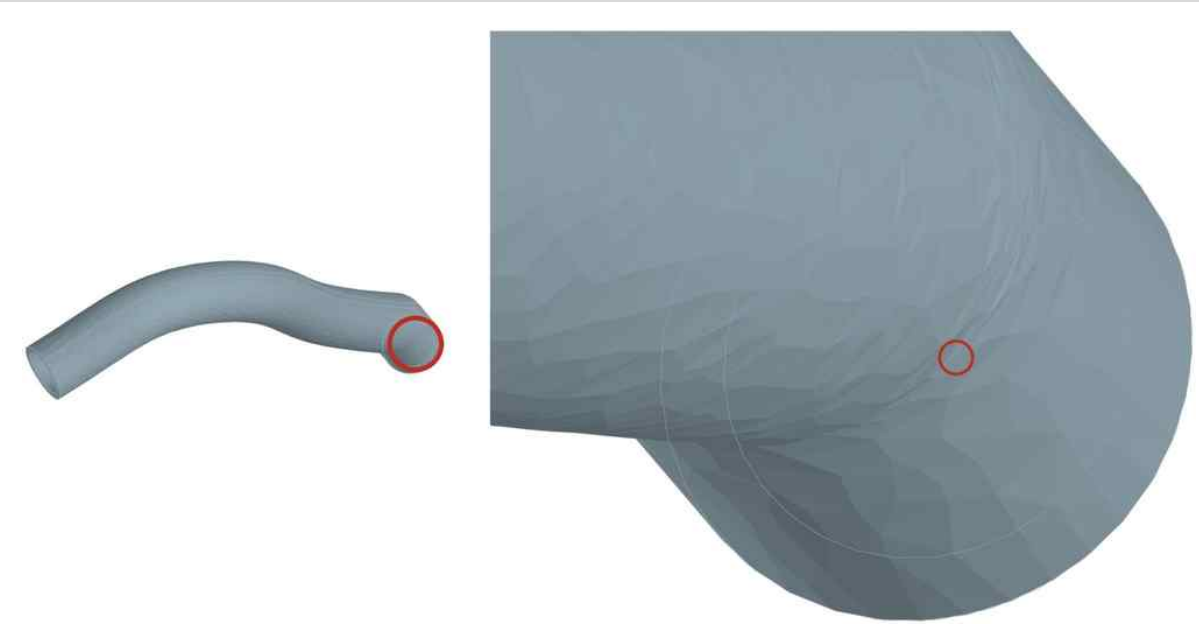

**(f) Intake pipe: red circle marks surface self-intersection**

Figure 5-14  TC-07: modeling from geometric requirements: stress test

TC-07 asks GPT-6 to model and assemble a seven-cylinder radial engine, including its main mechanisms and intake and exhaust pipes. GPT-6 progressively adds missing components and adjusts the mounting ring and propeller blade cross-sections, raising the percentage of geometric requirements met from 16.7% to 100%. The final model retains the seven-cylinder radial arrangement and the relationships between the master and articulated connecting rods. However, the master rod has a local non-manifold connection, with the same edge used four times in face boundaries. Local regions of the wall surfaces of all seven intake pipes fold back and intersect themselves (Figure 5-14). The rear intake housing also fails the parametric trimming-curve check. Of the 394 solids, 385 pass and nine fail. The complete engine fails the solid validity check.

#### 5.3.2.2.2 S2: Functional-requirements stress test

### TRC-07 (complex)

<table>
<tr><td>Design a compact electric helicopter for visual inspection with a 150 g camera module measuring 60 × 50 × 40 mm. The module has four M3 attachment points on a 40 × 30 mm rectangular pattern and a centre of mass 20 mm above its mounting face. The complete aircraft, including the camera, battery, and electronics, shall weigh no more than 1.4 kg.</td><td>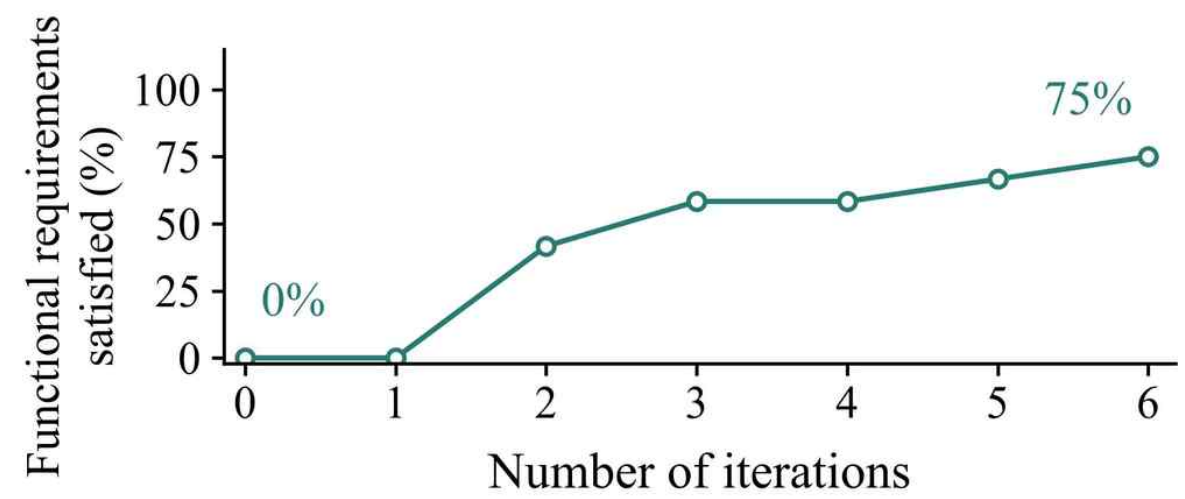
</td></tr>
<tr><td>(a) Excerpt from the input text</td><td>(b) Percentage of functional requirements met</td></tr>
<tr><td>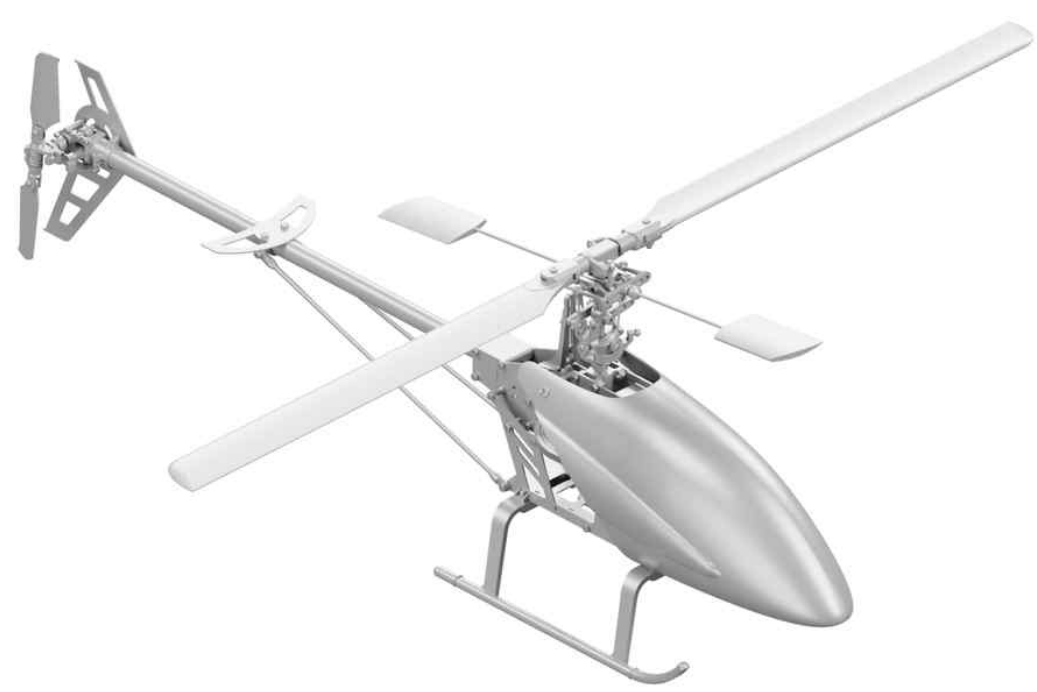</td><td>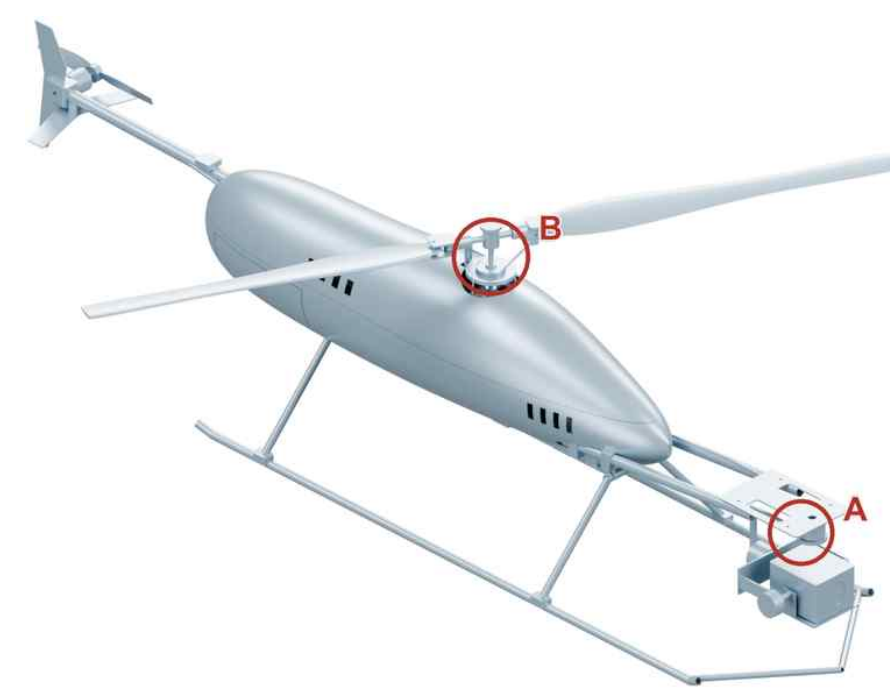
</td></tr>
<tr><td>(c) Rendered reference CAD model</td><td>(d) Generated model and faulty connections A and B</td></tr>
<tr><td>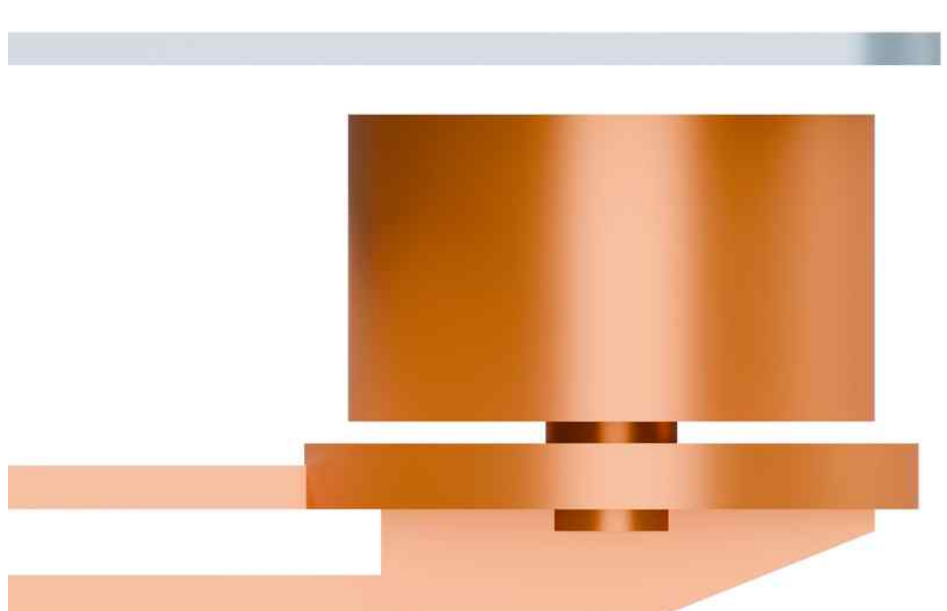</td><td>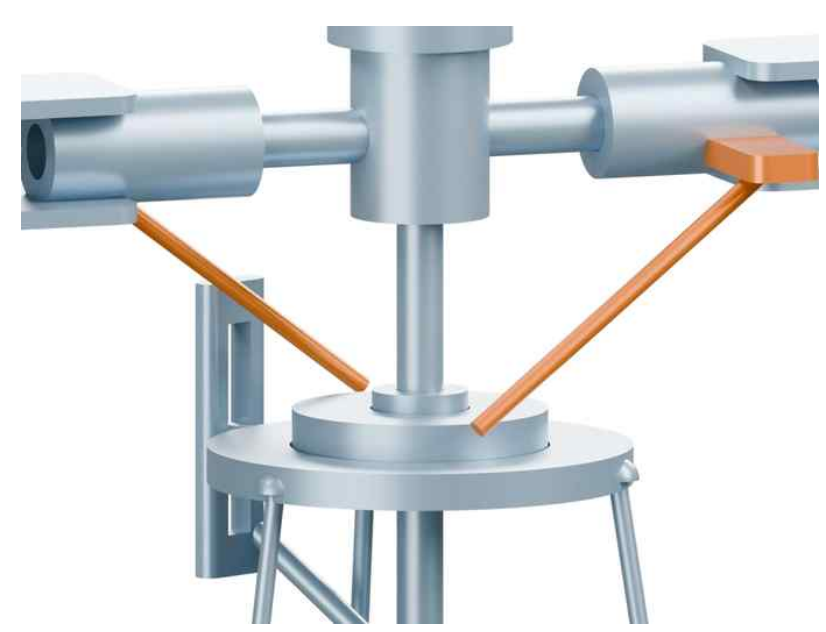</td></tr>
<tr><td>(e) A: gimbal motor not fastened to the support plate</td><td>(f) B: pitch link disconnected from the swashplate</td></tr>
</table>

Figure 5-15 TRC-07: modeling from functional requirements: stress test

TRC-07 asks GPT-6 to design an electric helicopter that meets requirements for flight control, hovering endurance, and camera pointing. GPT-6 places the camera and gimbal on a bracket projecting from the nose to provide an unobstructed view. By reducing weight and adjusting component positions over successive iterations, it raises the percentage of functional requirements met from 0% to 75%. In the final model, the gimbal motor at location A is separated from its support plate by 2.25 mm and has no fastening connection. At location B, the pitch link is approximately 3.15 mm from the swashplate and cannot change blade pitch (Figure 5-15). Three requirements remain unmet: flight control, attitude adjustment in crosswinds, and camera pointing stability. The lower fairing fails the parametric trimming-curve check. Of the 188 solids, 187 pass and one fails. The complete helicopter fails the solid validity check.

#### 5.3.2.2.3 S3: Geometric requirements with web search disabled

**TC-05 (complex)**

Create a single solid bare V8 cylinder block approximately 520 mm long, 475 mm wide and 290 mm high. Two cylinder banks form a 90° V, each containing four parallel bores approximately 92 mm in diameter and 110 mm deep. The cylinders repeat at approximately 107.5 mm longitudinal spacing, with the two banks slightly staggered.

**(a) Excerpt from the input text**

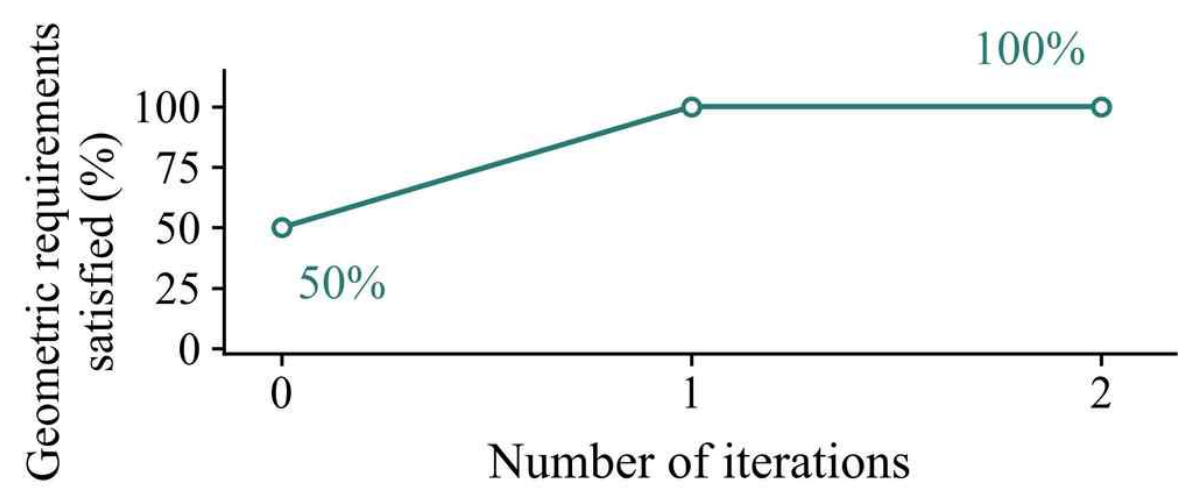

**(b) Percentage of geometric requirements met**

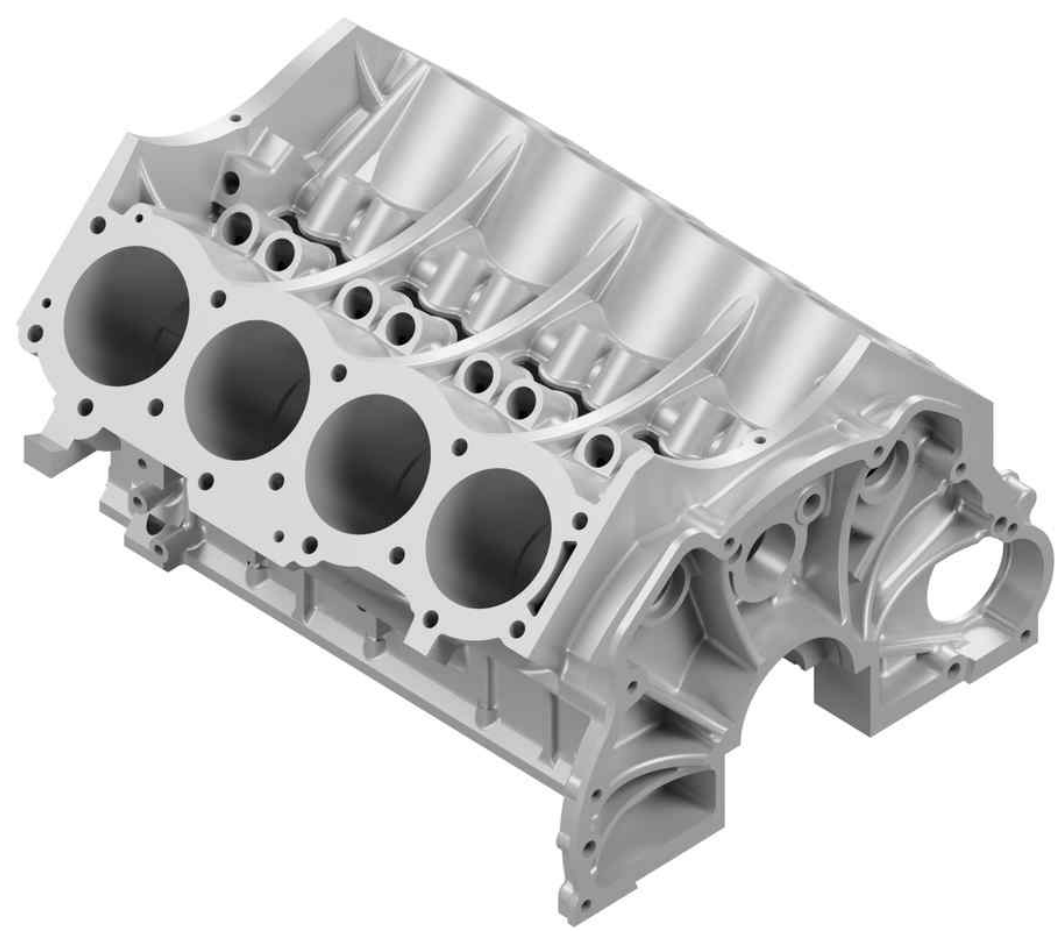

**(c) Rendered reference CAD model**

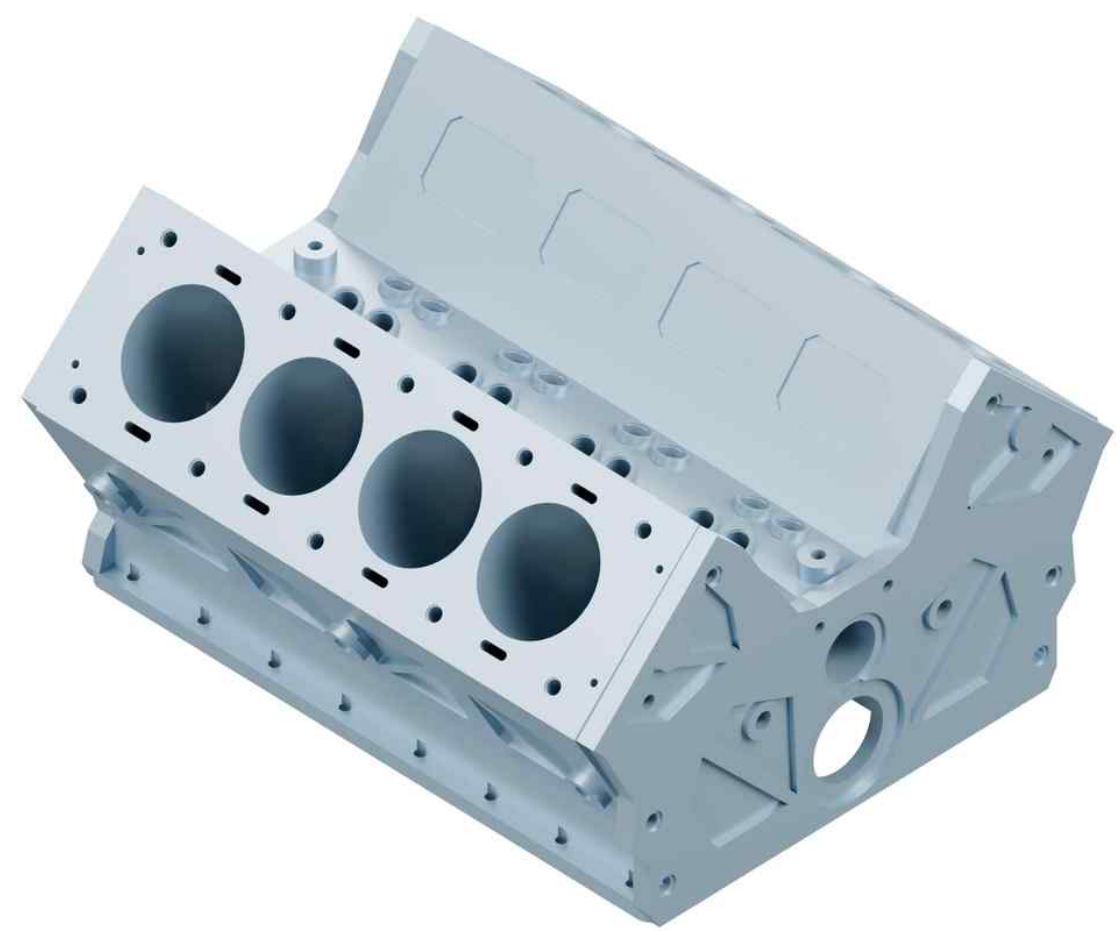

**(d) Rendered generated CAD model**

Figure 5-16 TC-05: modeling from geometric requirements, web search disabled

TC-05 asks GPT-6 to generate a V8 engine block with complete internal passages and mounting structures, with web search disabled. GPT-6 first constructs the two cylinder banks, water jackets, and longitudinal passages, then adds tappet bores, bearing seats, the bottom mating face, and external reinforcement. In the final model, the water jackets remain connected, complete walls separate the cavities, and the five webs and crankshaft bearing seats have the required positions. The percentage of geometric requirements met rises from 50% to 100% (Figure 5-16). The final block is a single solid and passes the validity check.

## TC-06 (complex)

Create a complete horizontal single-cylinder steam-engine assembly on a rectangular base approximately 170 × 80 × 14.5 mm, with rounded corners, mounting holes and a recessed underside. The cylinder lies along the long direction of the base and extends beyond one end. Its body is 35 mm long, with a 19 mm bore, a narrower barrel approximately 22 mm in diameter and wider end flanges approximately 32 mm in diameter.

**(a) Excerpt from the input text**

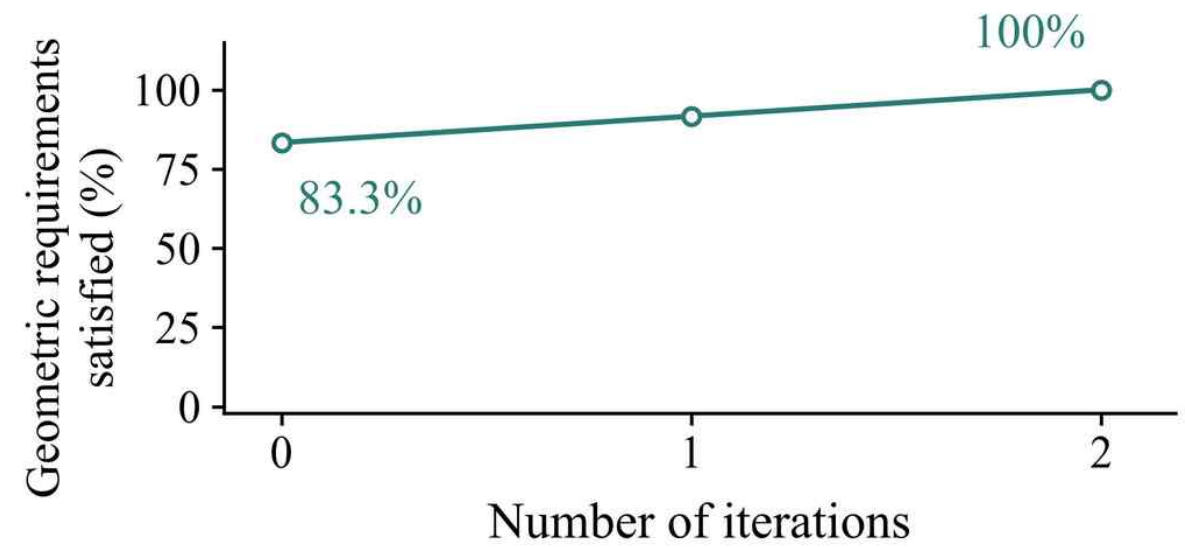

**(b) Percentage of geometric requirements met**

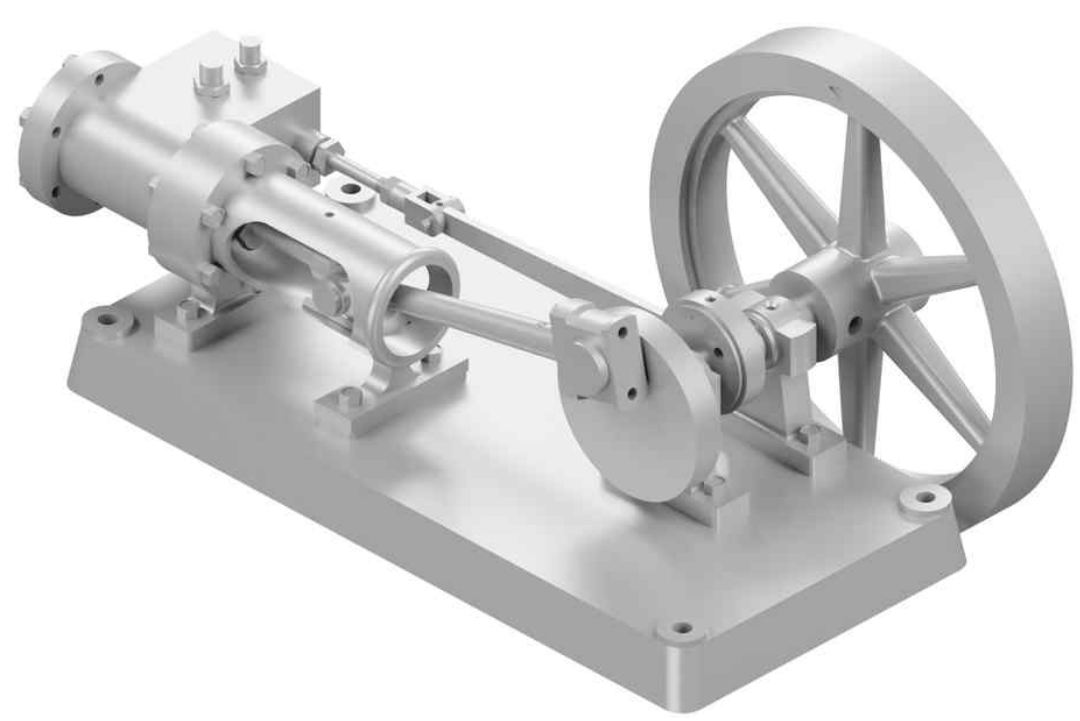

**(c) Rendered reference CAD model**

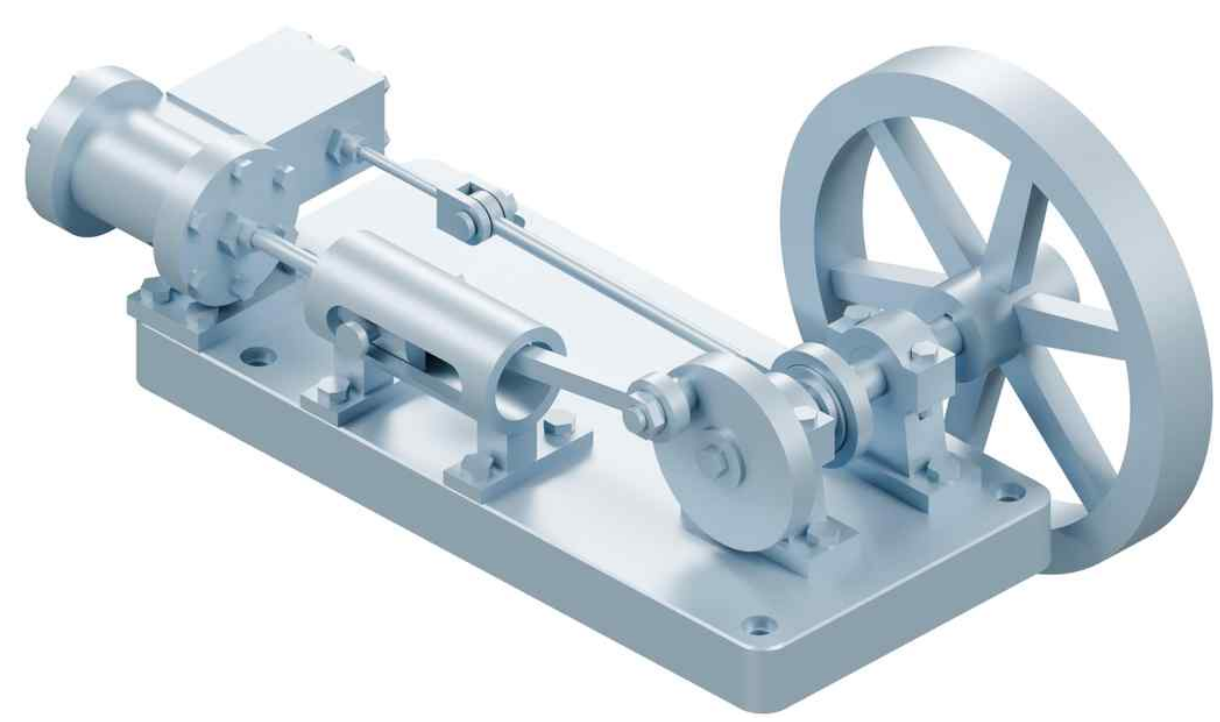

**(d) Rendered generated CAD model**

Figure 5-17 TC-06: modeling from geometric requirements, web search disabled

TC-06 asks GPT-6 to construct a horizontal single-cylinder steam engine and assemble its main drive and valve gear, with web search disabled. After constructing the main mechanisms, GPT-6 adds fasteners and adjusts the positions of the eccentric, collar, and mounting bolts to eliminate local assembly interference. In the final model, the cylinder, crank and connecting rod, and flywheel form a complete main drive, and the eccentric rod and valve stem have the required positions. The percentage of geometric requirements met rises from 83.3% to 100% (Figure 5-17). All 110 parts pass the solid validity check.

### 5.3.2.2.4 S3: Functional requirements with web search disabled

### TRC-05 (complex)

Design a one-piece aluminium cylinder-head body for a naturally aspirated, water-cooled, inline-four development engine with a 73 mm bore, 82.8 mm stroke, and 79 mm cylinder spacing. The fixed block interface has a flat deck and ten M10 attachment points arranged in two parallel rows 100 mm apart; each row contains five equally spaced points at 79 mm pitch, centred on the four-cylinder group.

**(a) Excerpt from the input text**

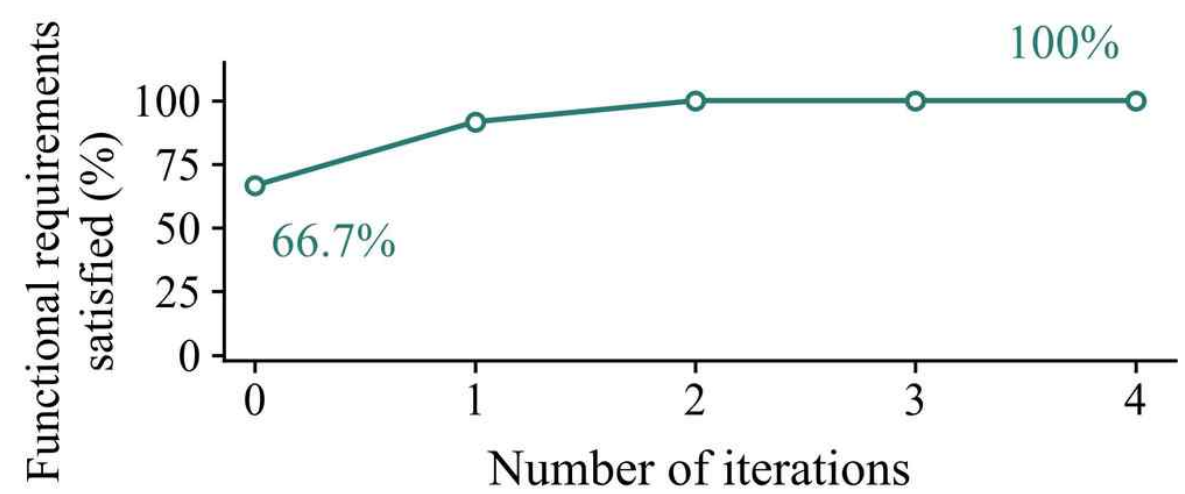

**(b) Percentage of functional requirements met**

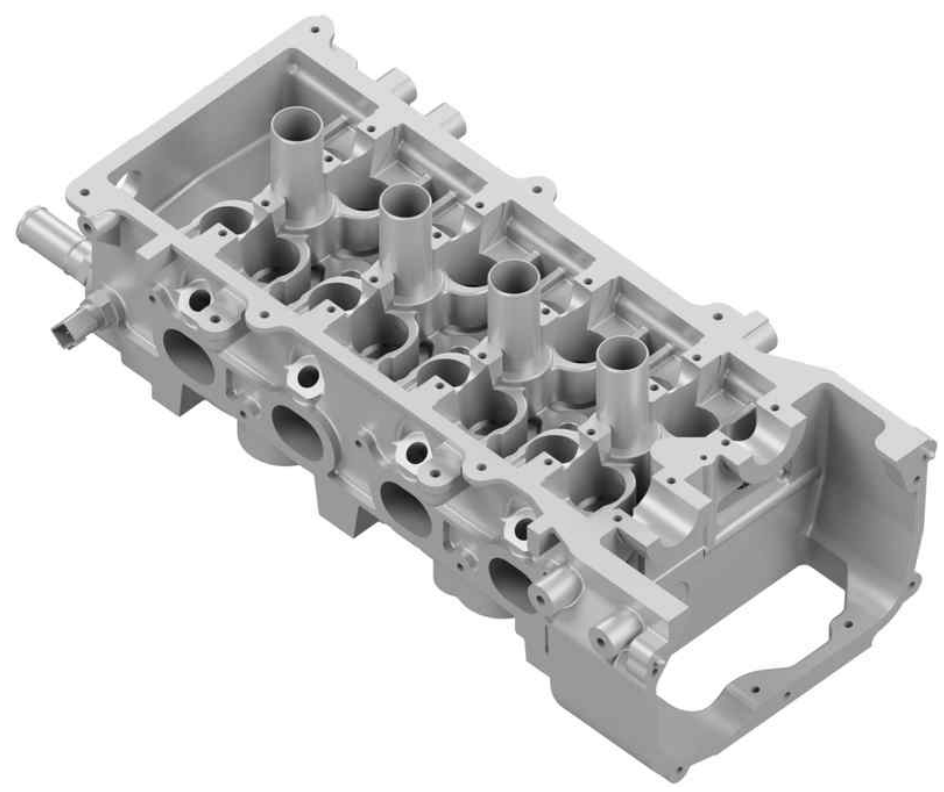

**(c) Rendered reference CAD model**

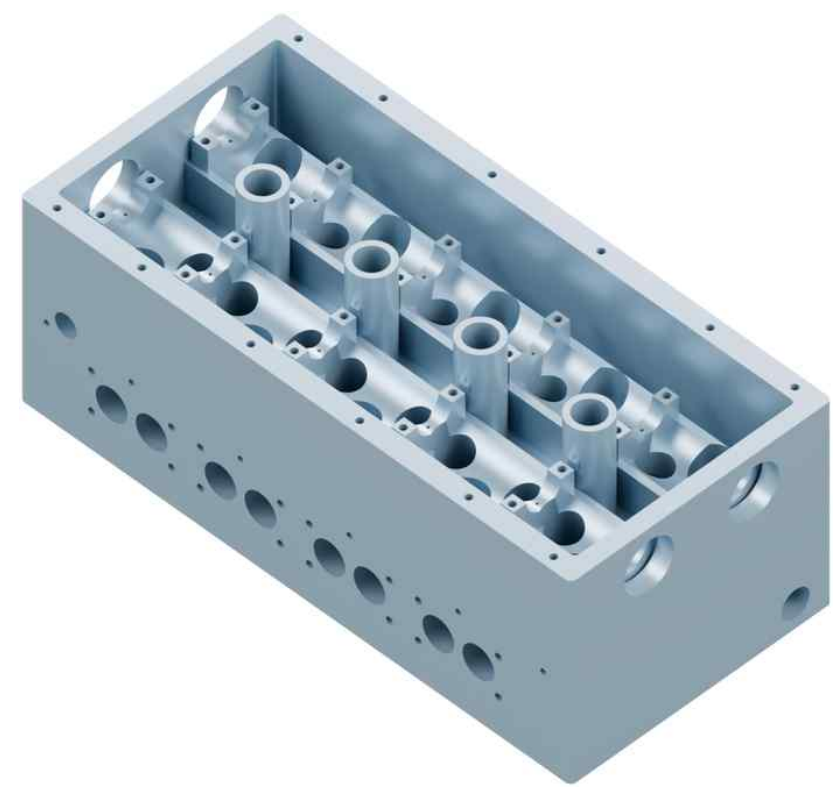

**(d) Rendered generated CAD model**

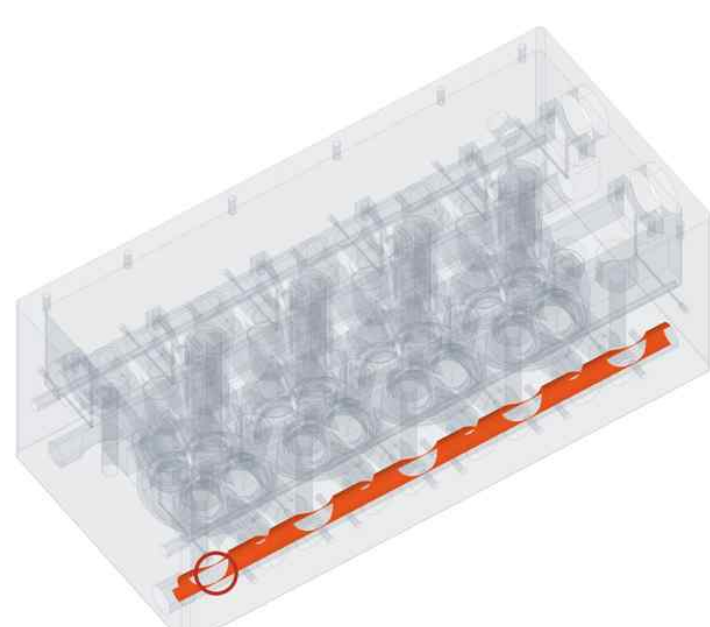

**(e) Intersection of internal passages in the cylinder head**

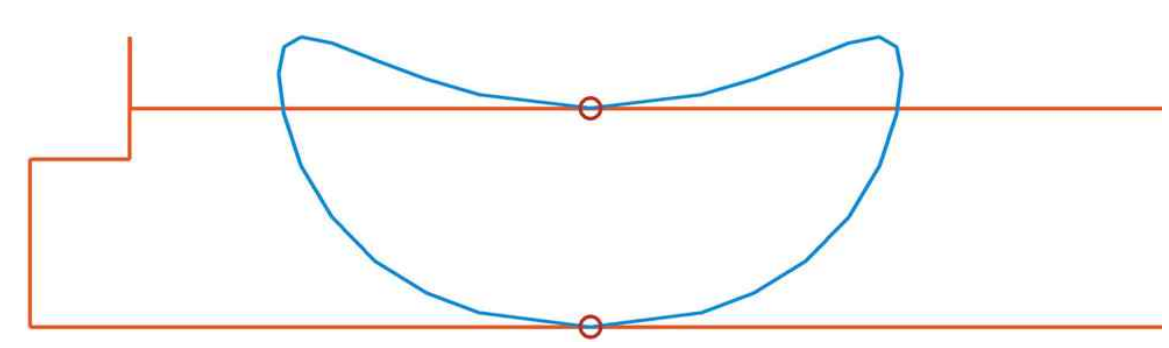

**(f) The same pair of boundary loops meets at both red circles**

Figure 5-18 TRC-05: modeling from functional requirements, web search disabled

TRC-05 asks GPT-6 to design the valve, pressure-bearing, cooling, and lubrication structures of a cylinder head, with web search disabled. GPT-6 adjusts the separating walls between water jackets and gas passages to correct tangential contact between the two in the early design. It also preserves space for valve motion and removal of valve-train components. The percentage of functional requirements met rises from 66.7% to 100%. However, invalid face boundaries remain at intersections between internal passages in the final model. Across two faces, ten pairs of boundary loops each meet at two distinct vertices. The detail view distinguishes one pair of loops by color and circles the two contact points (Figure 5-18). The final cylinder head fails the solid validity check.

### TRC-06 (complex)

Design a six-legged, battery-powered chassis for indoor inspection on dry, hard floors, using 0.5 as the minimum friction coefficient between the feet and ground. Carry a centrally mounted 1 kg inspection module measuring 100 × 100 × 60 mm, with its centre of mass 30 mm above the mounting face and four M4 attachment points on an 80 mm square pattern.

**(a) Excerpt from the input text**

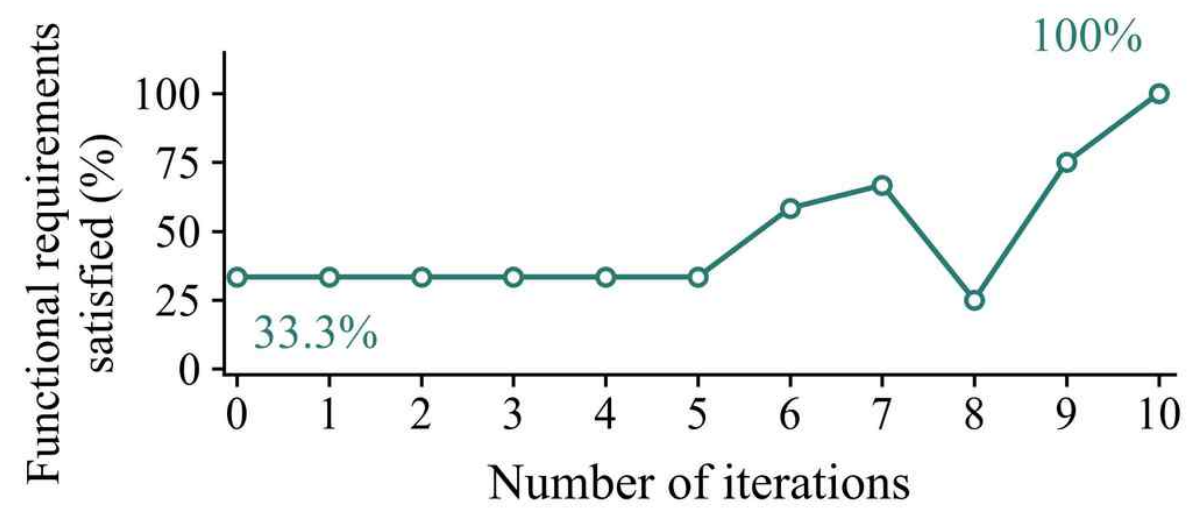

**(b) Percentage of functional requirements met**

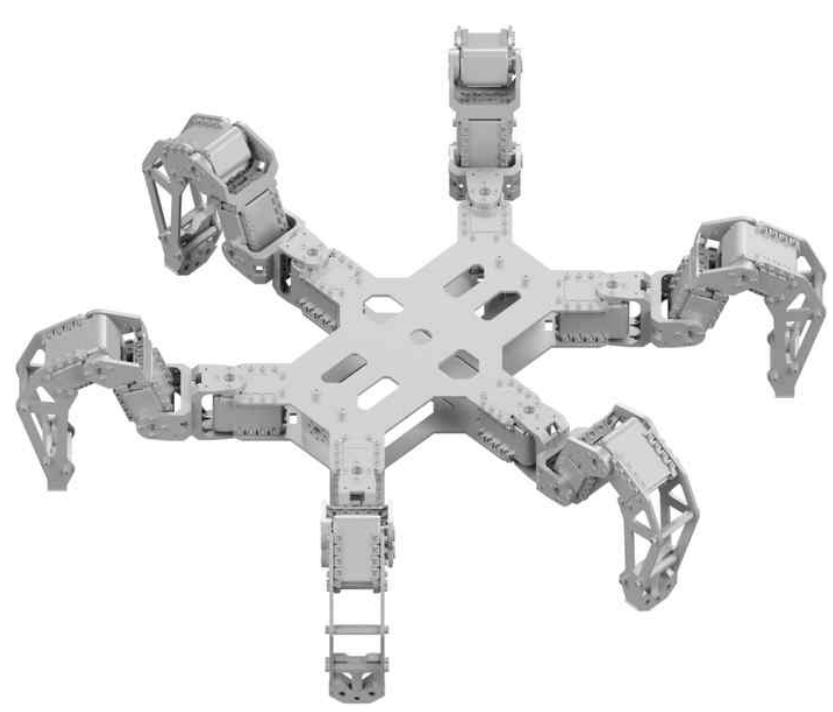

**(c) Rendered reference CAD model**

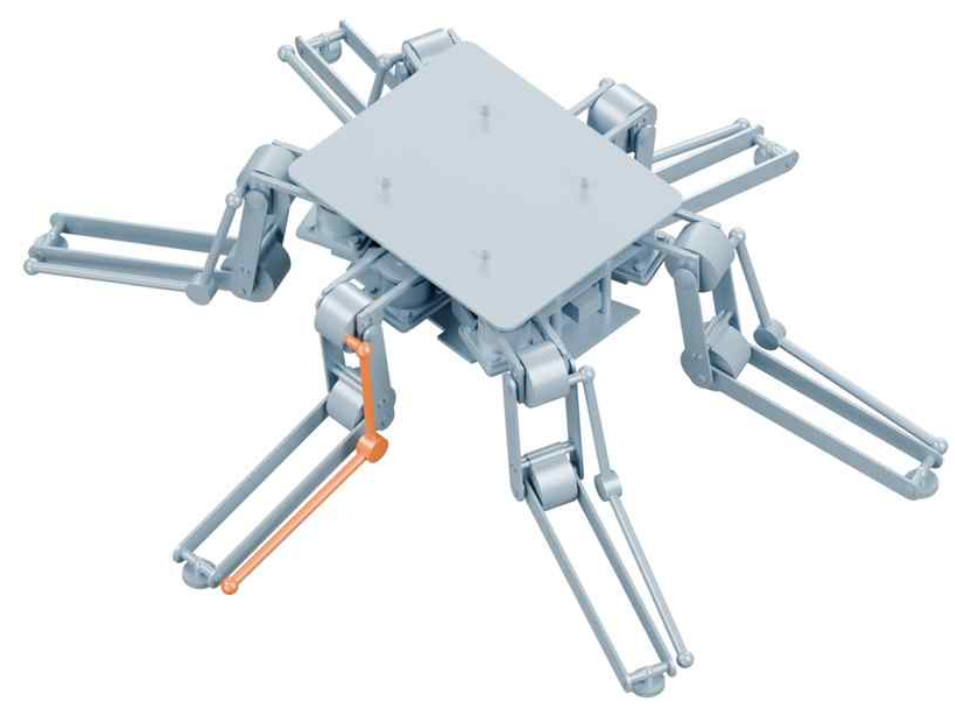

**(d) Generated model: protective tubing outside the legs shown in orange**

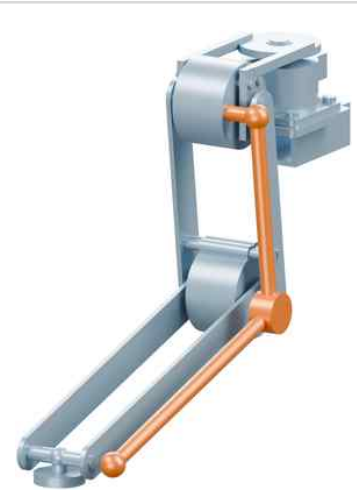

**(e) Corresponding leg: tubing extends to the foot**

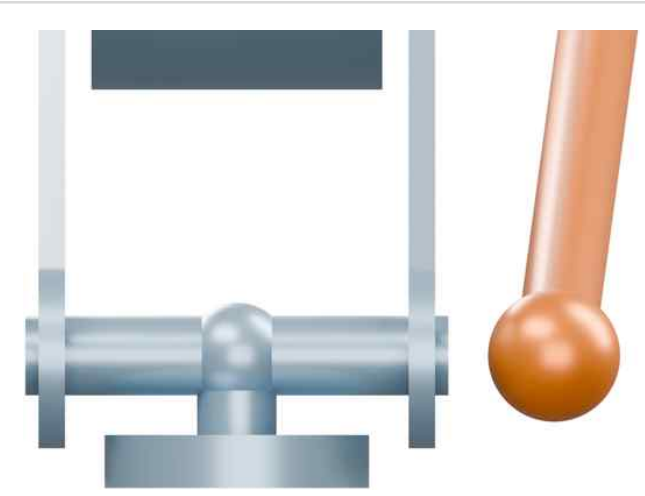

**(f) Tube end separated from the foot connection pin**

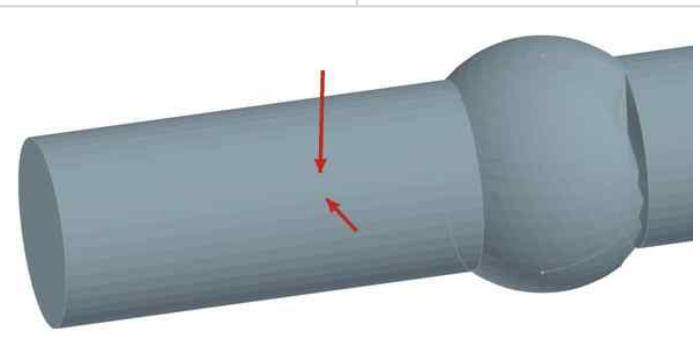

**(g) Incorrect shell orientation in the foot connection pin: red arrows point into the solid**

Figure 5-19 TRC-06: modeling from functional requirements, web search disabled

TRC-06 asks GPT-6 to design a hexapod chassis that meets requirements for load capacity, walking, and obstacle traversal. GPT-6 progressively improves leg motion, raising the percentage of functional requirements met from 33.3% to 100%. However, the orange protective tubing in Figure 5-19 extends all the way to feet that contain no electrical components. The tube ends are unsupported and lie 3 mm from the connection pins, creating unnecessary branches outside the legs. The solid validity check identifies an incorrectly oriented shell in one foot connection pin: the normal to a local exterior face points into the solid. Of the 349 solids, 348 pass and one fails. The chassis fails the solid validity check.

## 5.4 Analysis of GPT-6's approach

### 5.4.1 Modeling from geometric requirements

The recorded outputs and tool operations suggest the following workflow to generate CAD models from geometric requirements:

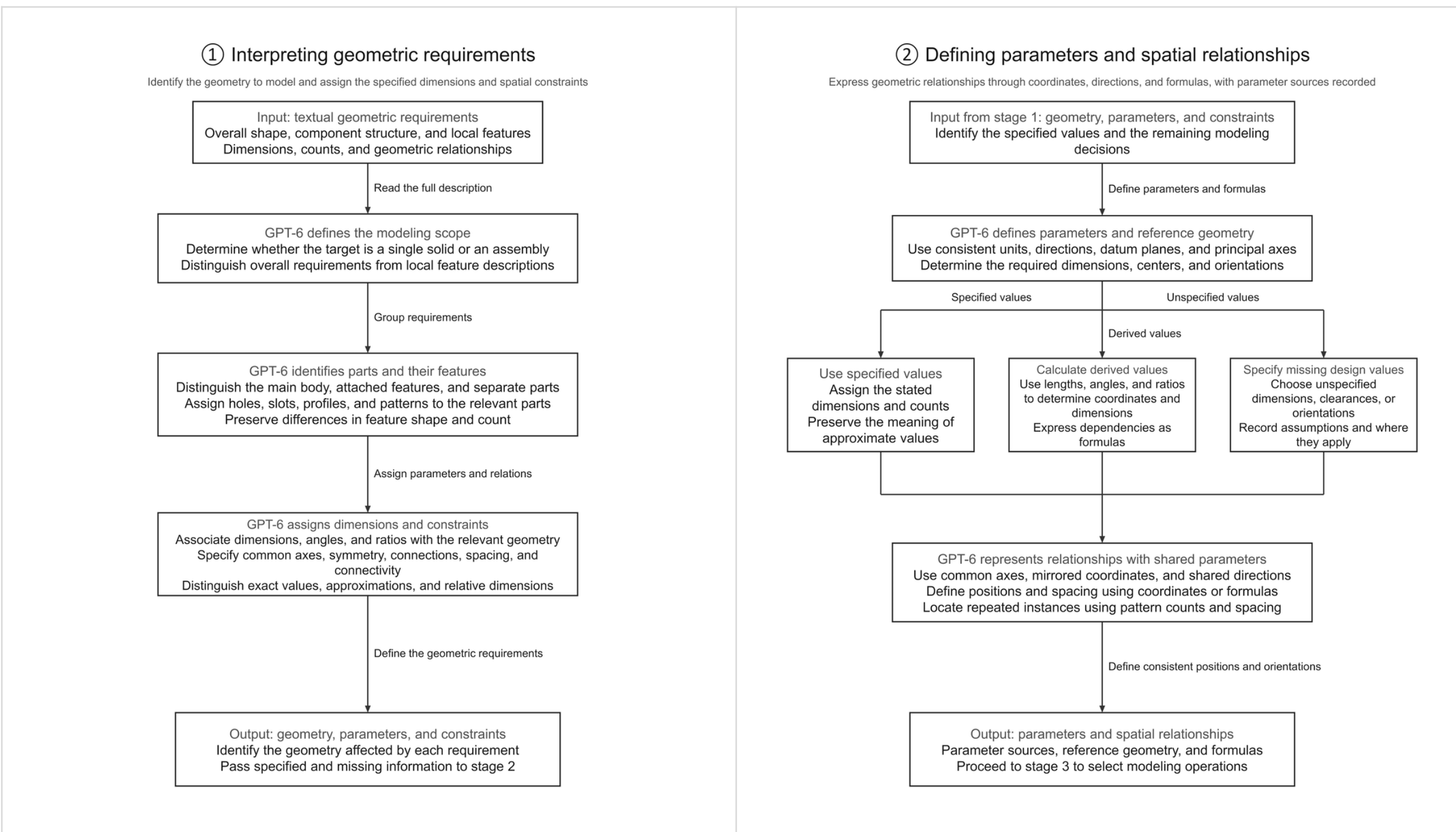

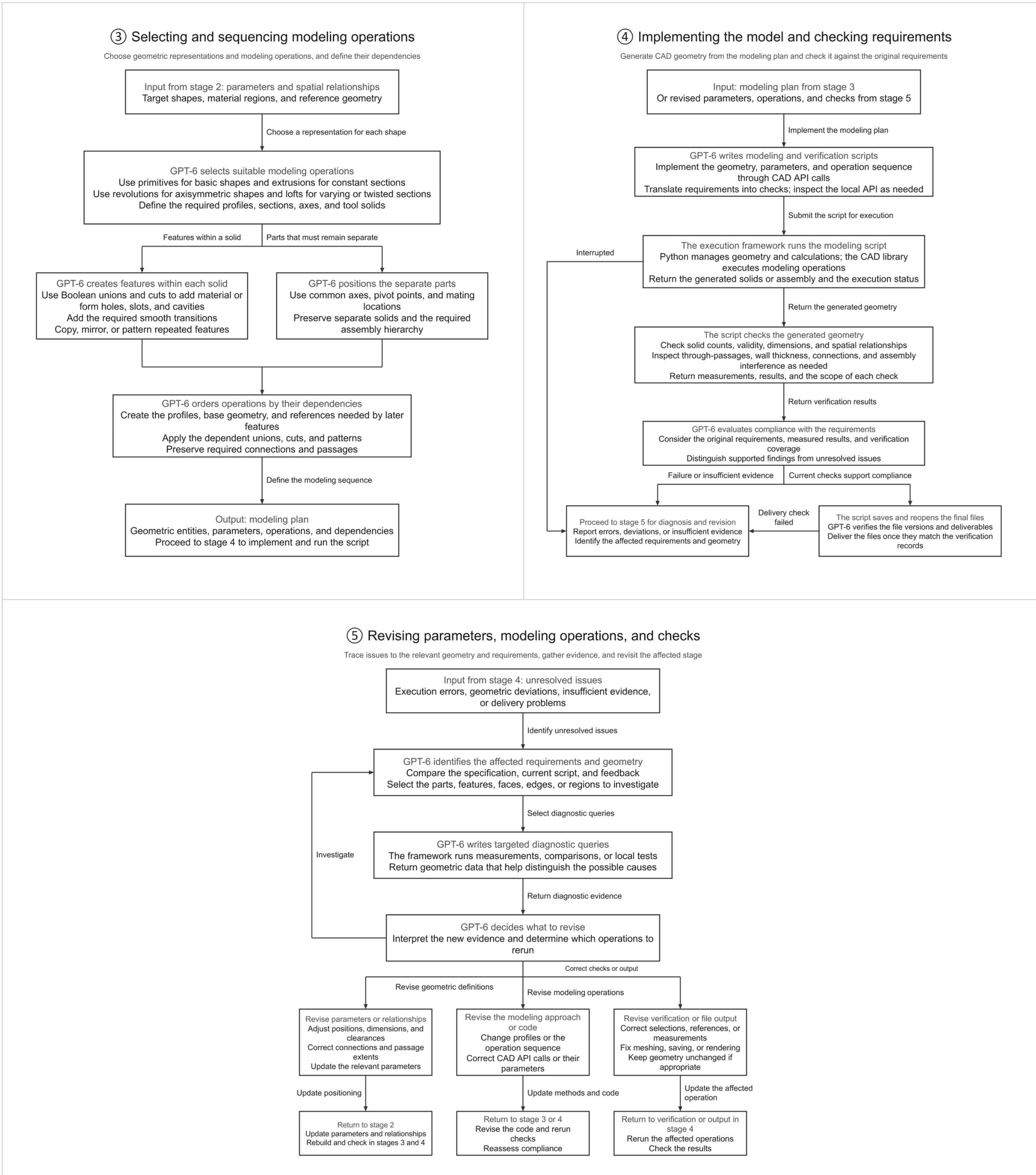

Figure 5-20  Inferred workflow for modeling from geometric requirements

### 5.4.2 Modeling from functional requirements

The recorded outputs and tool operations suggest the following workflow to generate CAD models from functional requirements:

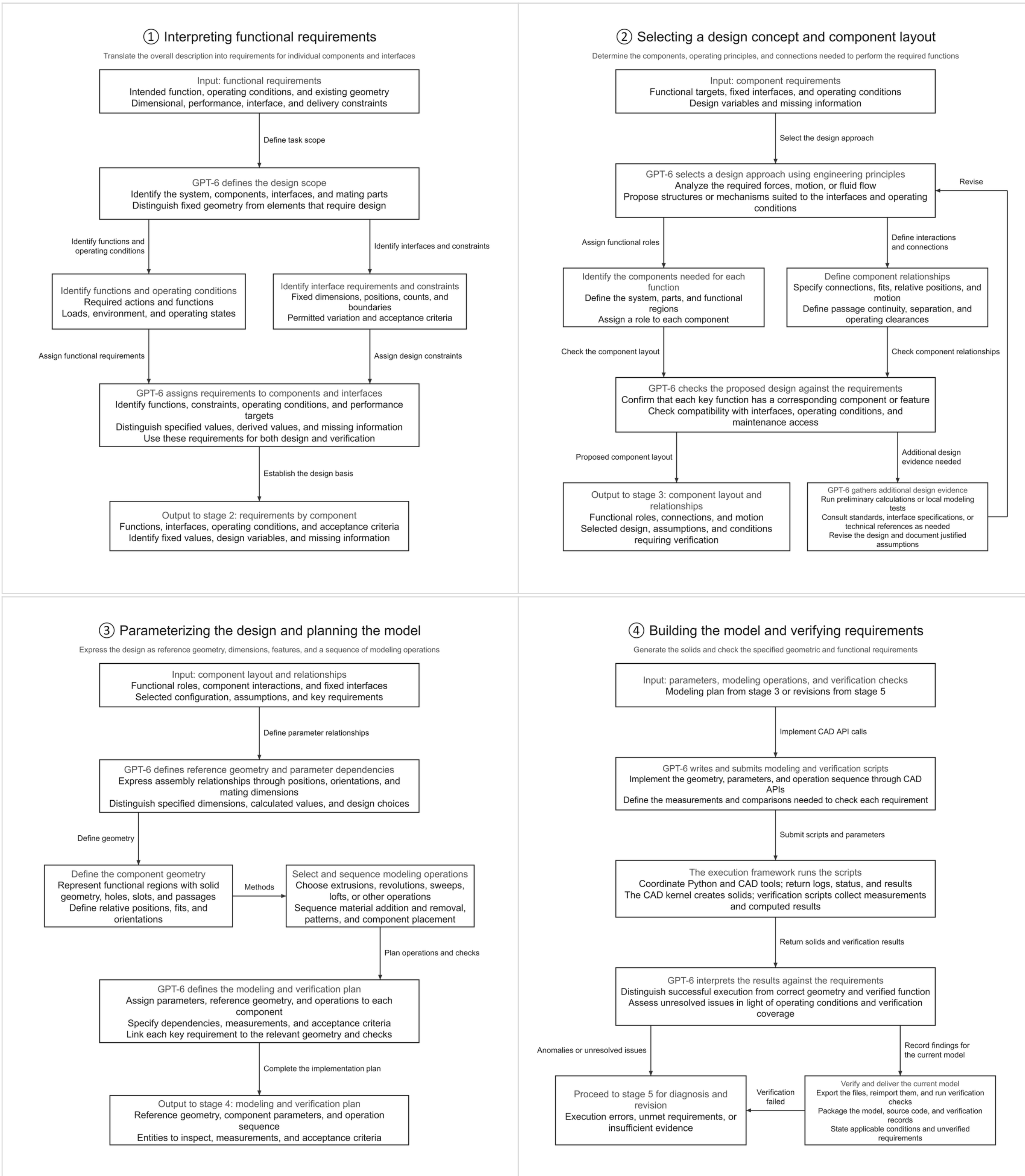

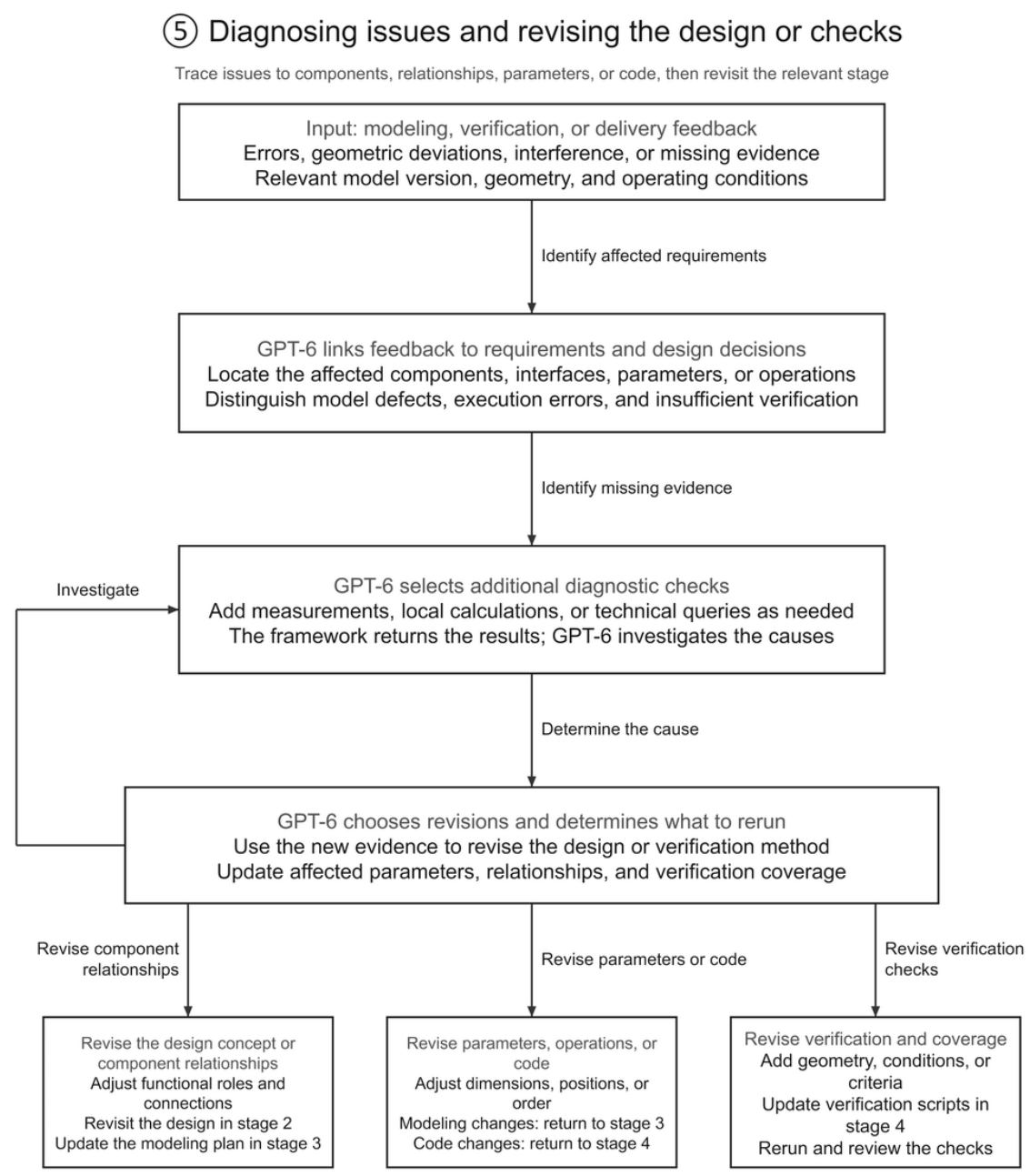

Figure 5-21  Inferred workflow for modeling from functional requirements

## 5.5 Capabilities and limitations

In this chapter's text-to-CAD tests, GPT-6 converts descriptions of geometry and intended use into structural designs. It independently determines dimensions and component relationships, plans the modeling sequence, and uses CAD tools to generate individual parts and assemblies. It can construct repeated features, complex surfaces, and internal passages, and arrange and assemble several mechanisms. During modeling, GPT-6 also uses load and motion calculations and geometric checks to revise structures and parameters, add omitted features, and correct some interference, progressively increasing the proportion of input requirements met.

Limitations remain in local detail and implicit engineering requirements. Some models satisfy the explicitly listed shape, dimensional, or layout requirements but have incomplete transitions, fastenings, or transmission connections, or contain unnecessary structures with no practical purpose. These defects affect how components fit together and function. In complex assemblies, some generated parts also contain geometric and topological defects, including self-intersections, non-manifold connections, and open shells, which cause the models to fail the solid validity check. GPT-6 cannot yet reliably detect and correct these problems on its own. It struggles to ensure sound structural details, working mechanisms, and valid solids simultaneously, leaving a substantial gap between its outputs and models suitable for engineering use.

# Chapter 6 CAD reconstruction from images

## 6.1 Task objectives and evaluation criteria

This task evaluates GPT-6's ability to interpret renderings from multiple viewpoints and reconstruct a three-dimensional CAD model. The main experiment uses a consistent set of views and a simple Chinese prompt so that reconstruction results can be compared.

The main experiment tests six cases using six orthographic views and one axonometric view. The same six cases are also tested with three orthographic views and with six orthographic views alone. Two complex cases are retested with web search disabled, and a seventh case is used for a stress test, giving 21 runs in total. These comparisons examine how view availability and search access relate to reconstruction quality.

Table 6-1 Objectives and evaluation criteria for CAD reconstruction from images

| Test element | Description |
| --- | --- |
| Evaluation question | Can GPT-6 reconstruct a CAD model consistent with renderings from multiple viewpoints? How does reconstruction quality change when fewer views are supplied? |
| Input | Renderings from multiple viewpoints |
| Output | A readable B-rep model |
| Evaluation criteria | Check solid validity and assess geometric agreement using silhouette IoU, HD between three-dimensional surfaces, and model comparisons. |

Geometric agreement is assessed in two ways: by comparing view silhouettes and by comparing three-dimensional surfaces. The reference STEP files are used only to check the generated results. Silhouette intersection over union (IoU) measures the overlap between projections of the generated model and the input renderings. Hausdorff distance (HD), in millimeters, estimates the maximum surface deviation between the generated and reference models; a smaller value indicates a smaller sampled deviation. To calculate HD, the nearest distance from each sampled reference point to the generated point cloud is found, and the calculation is then repeated in the reverse direction. The largest distance across both directions is reported as HD.

All models are declared to use millimeters. Points are sampled in proportion to surface area, and the models are rigidly aligned using multiple initial poses and bidirectional iterative closest point (ICP) registration. Alignment permits rotation and translation only, with no scaling or mirroring. Each model uses 12,000 points for registration and a separate set of 180,000 points for HD evaluation. All evaluation points are retained. The reported HD is therefore the symmetric Hausdorff distance between finite sampled point clouds, not an exact distance between continuous B-rep surfaces. Because alignment excludes scaling, it reflects differences in absolute scale as well as local shape. Without a dimensional cue in the input images, absolute scale is not uniquely recoverable; HD should be interpreted with that limitation in mind.

## 6.2 Test design

### 6.2.1 Test case selection

The cases are publicly available mechanical parts and assemblies. Selection favors representative structures with corresponding STEP reference files. To cover a range of modeling difficulties, the cases vary in structural complexity, local feature types, and number of parts. They include sheet-like parts, housings, supports with holes, brackets with curved surfaces, and a chassis assembly, covering both single-part and assembly modeling.

With the number of input views held constant, complexity is classified primarily by the geometry, spatial relationships between local features, and degree of occlusion. For assemblies, the number of parts and their assembly relationships are also considered. Table 6-2 defines the criteria.

Table 6-2 Complexity criteria for CAD reconstruction from images

| Complexity | Selection criteria | Typical characteristics |
|---|---|---|
| Simple | The main shape is regular, with few local features. Major outlines, holes, slots, and bosses can be matched clearly across views, with few complex surfaces or occluded structures. | Clear outlines and a few local features reveal the overall structure, requiring little cross-view matching or inference of local shapes. |
| Moderate | More local detail and spatial relationships are present, including holes in different directions, inclined connections, or varying cross-sections. Multiple views are needed to determine feature shapes, orientations, and connections, although the main structure remains regular. | Local shapes and positions must be inferred from several views. Clear geometric regularities, repeated features, or well-defined connections can still guide modeling. |
| Complex | The renderings contain dense structural detail, including complex surfaces, thin walls, and intersecting features, with some regions occluded. Overall outlines and repetition alone do not reveal local shapes. Assemblies also contain many components and connections that must be interpreted in relation to other parts. | Geometry distributed across views must be combined to resolve surface variations, structural layers, and connection transitions. In assemblies, local structures must also be interpreted through the positions and correspondence of parts; occlusion adds difficulty. |

The main experiment comprises six cases: five single parts and one assembly, with two cases at each of the simple, moderate, and complex levels. Table 6-3 lists the cases.

Table 6-3 Test cases for CAD reconstruction from images

| Case ID | Rendered CAD model | Case type | Complexity | Basis for complexity rating |
|---|---|---|---|---|
| I2C-01 | 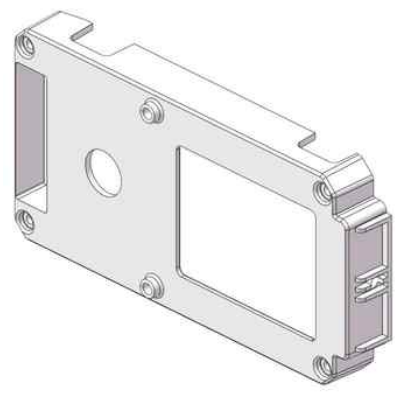 | Single part | Simple | A thin plate with perimeter flanges and regularly arranged windows, circular holes, and bosses. Its main boundaries can be matched across orthographic views. |
| I2C-02 | 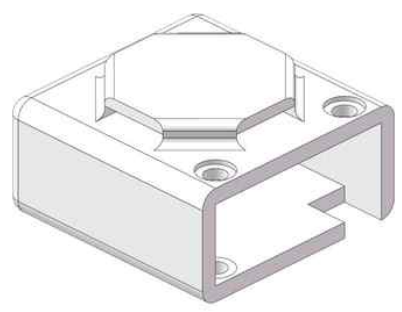 | Single part | Simple | A regular housing with a top platform and few openings or guides. Major outlines and feature locations are readily matched across views, with few complex surfaces and little cross-view inference required. |
| I2C-03 | 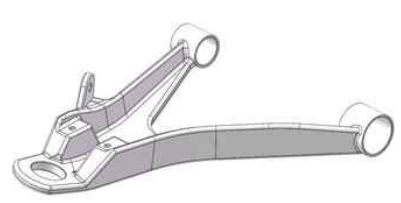 | Single part | Moderate | Transverse bushings and an inclined lug connect to reinforcing walls. Multiple views are needed to determine hole-axis directions, lug inclination, and local connections. The forked body and main support layout are clear, however, and can be decomposed into recognizable modeling features. |
| I2C-04 | 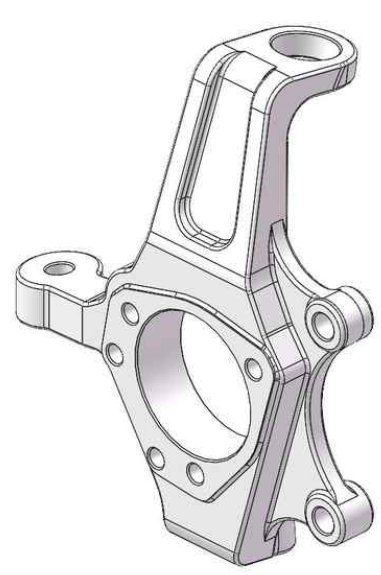 | Single part | Moderate | Central holes and mounting lugs connect to a curved arm with a varying cross-section. Multiple views establish the arm's spatial bends, section changes, and end-hole locations. There are relatively few main structures, with clear connections between them. |
| I2C-05 | 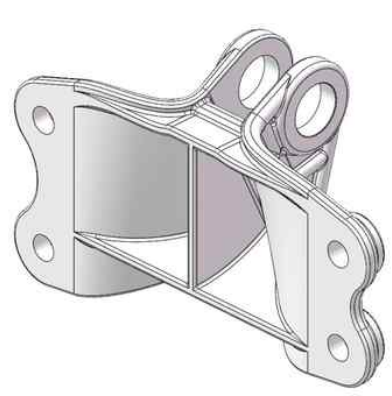 | Single part | Complex | Asymmetric mounting lugs connect through a thin curved wing to paired lugs with holes. Ribs lie on both sides of the wing. Multiple views are needed to determine curvature, local thickness, and transitions at rib intersections. Interacting features make local shapes difficult to infer from simple outlines or repetition. |
| I2C-06 | 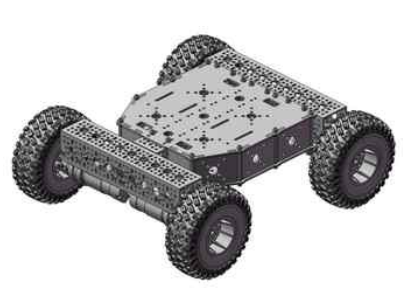 | Assembly | Complex | Many components, including plates, crossmembers, wheels, and motor supports, with dense hole patterns and connections and partial occlusion. Modeling requires both individual component geometry and their relative positions and connections to be determined from multiple views. |

### 6.2.2 Experimental conditions

#### 6.2.2.1 Main experiment

All six cases use six orthographic views plus an axonometric view and the same modeling prompt. No additional hints are provided about the case's complexity or specific structure. GPT-6 is required to construct a B-rep model from the renderings and export it as a STEP file.

Web search is enabled in the main experiment. The prompt reads: "Build the corresponding CAD model from these renderings and export it as a STEP file." Model validity and geometric agreement are then assessed as described in Section 6.1.

*The quoted prompt is an English translation of the Chinese prompt used in the experiment.*

#### 6.2.2.2 Additional experiments

Additional experiments use representative cases to examine the effects of changes to the rendering inputs and web-search conditions. Two main conditions are varied: the number of input views is reduced, or web search is disabled. Each experiment retains the corresponding main-experiment prompt, with no extra hints for the specific case. Results are compared with the main-experiment result for the same case. Table 6-4 lists the settings.

Table 6-4 Additional experiments for CAD reconstruction from images

| Condition | Input views | Web search | Purpose |
|---|---|---|---|
| S1: Reduced view count | Three orthographic views | Enabled | Assess GPT-6's modeling with substantially fewer views. |
| S2: Axonometric view removed | Six orthographic views | Enabled | Assess GPT-6's modeling without the axonometric view. |
| S3: Web search disabled | Six orthographic views plus an axonometric view | Disabled | Assess modeling when external information cannot be retrieved through web search. |
| S4: Stress test | Six orthographic views plus an axonometric view | Enabled | Assess GPT-6's modeling of a reference with very high complexity. |

## 6.3 Results

This section first summarizes the main and additional experiments, then presents the input renderings, reference and generated models, internal iteration curves, and analysis for each case.

### 6.3.1 Main experiment results

#### 6.3.1.1 Overview

The main experiment includes five single parts and one assembly. All six cases produced STEP files whose individual solids passed checks for topological validity, shell closure, and positive volume. Mean IoU across the six orthographic views and axonometric view ranges from 0.8790 to 0.9915; HD ranges from 9.959 to 80.297 mm (Table 6-5). Most generated models preserve the overall shape, but local structures and actual dimensions still differ. For example, I2C-02 achieves an IoU of 0.9915, yet its overall dimensions are approximately 3.9 times those of the reference; HD is 64.207 mm. Silhouette overlap and dimensional accuracy must therefore be assessed separately. Passing basic solid checks also does not establish that freeform surfaces, connections between solids, and assembly interfaces have been reconstructed correctly.

Table 6-5 Main experiment results for CAD reconstruction from images

| Case ID | Complexity | Validity of each solid | Mean IoU (seven views) | HD (mm) | IoU by iteration |
|---|---|---|---|---|---|
| I2C-01 | Simple | Pass | 0.9788 | 14.758 | 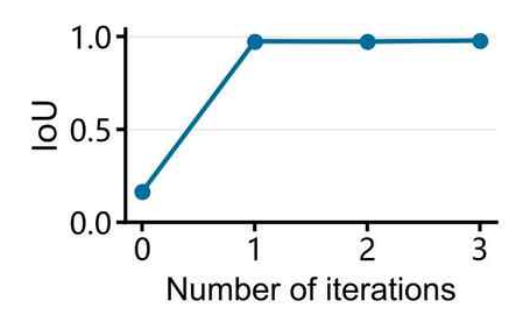 |
| I2C-02 | Simple | Pass | 0.9915 | 64.207 | 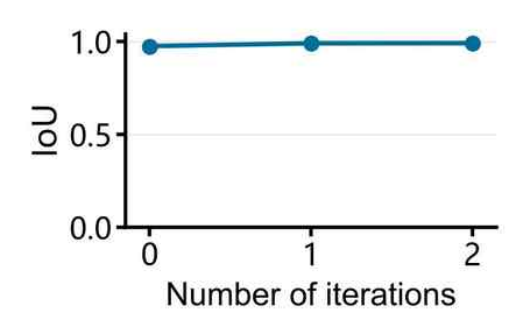 |
| I2C-03 | Moderate | Pass | 0.9483 | 62.401 | 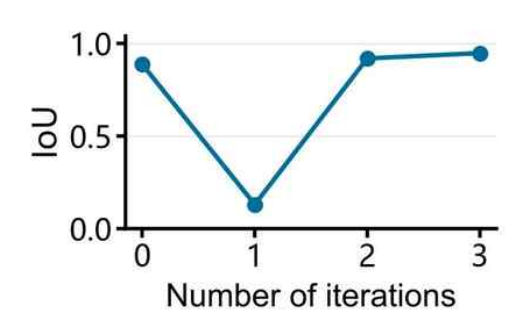 |
| I2C-04 | Moderate | Pass | 0.8790 | 45.211 | 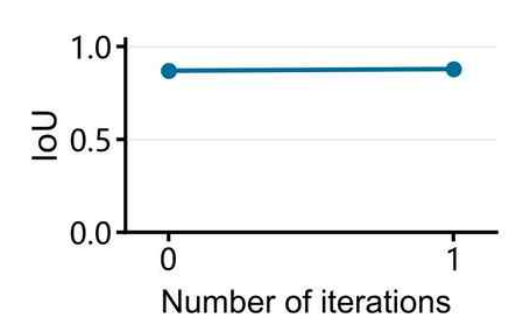 |
| I2C-05 | Complex | Pass | 0.9468 | 9.959 | 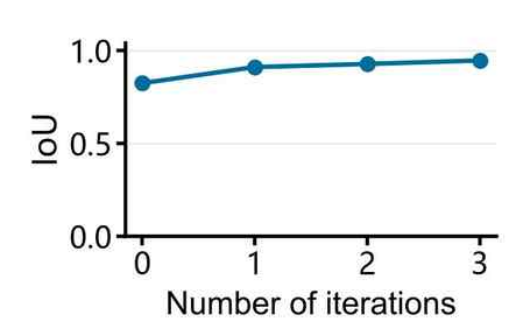 |
| I2C-06 | Complex | Pass | 0.9391 | 80.297 | 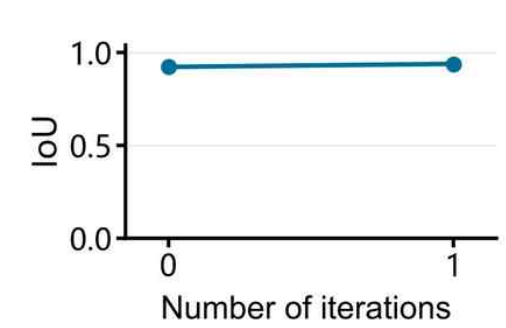 |

Table 6-6 Time taken for each case in the main experiment

| Case ID | I2C-01 | I2C-02 | I2C-03 | I2C-04 | I2C-05 | I2C-06 |
|---|---|---|---|---|---|---|
| Time | 16 min | 13 min | 26 min | 9 min | 21 min | 3 h 41 min |

#### 6.3.1.2 Case analyses

The following cases are presented in ID order. Each includes a description, input renderings, results, and analysis, focusing on reconstruction of the main structure and local differences.

## I2C-01 TrackerOne backplate (simple)

### Input renderings

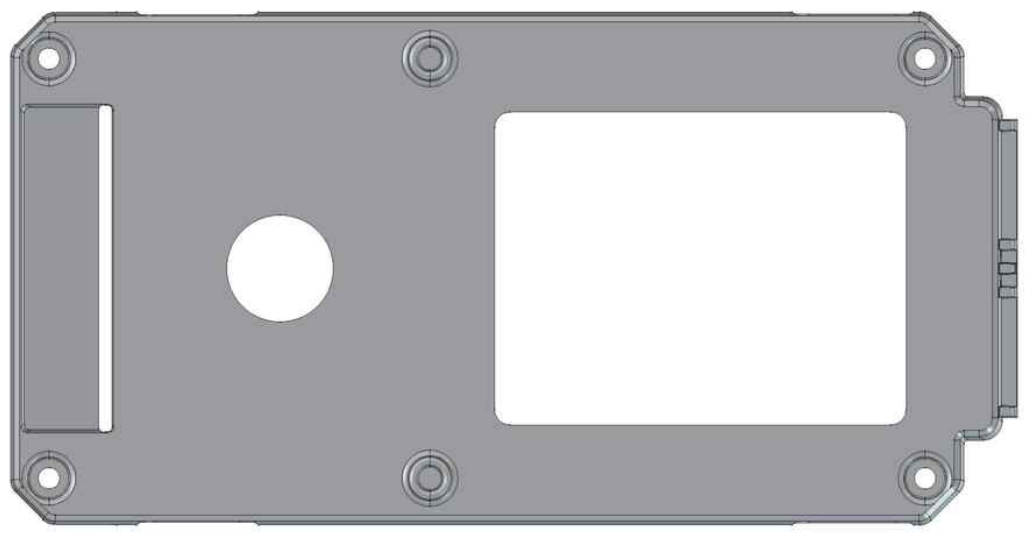

**Front view**

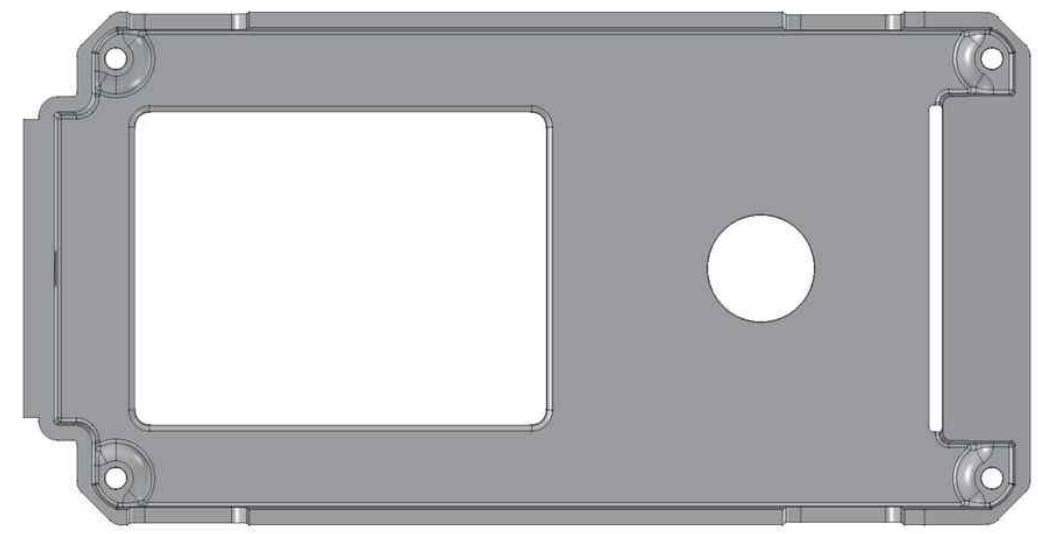

**Rear view**

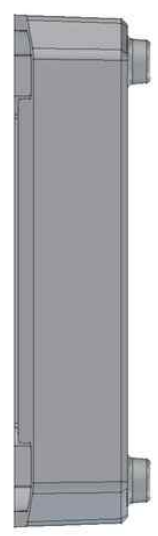

**Left view**

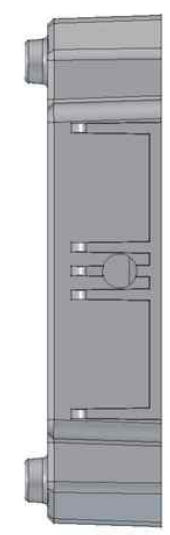

**Right view**

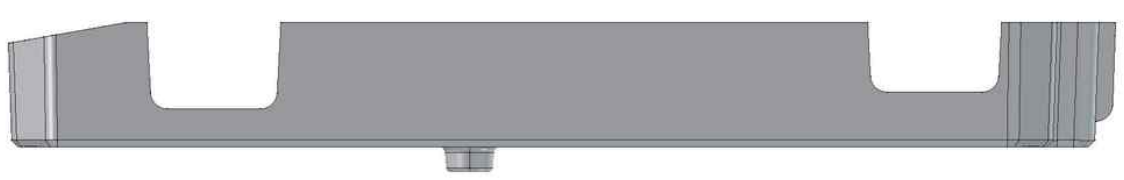

**Top view**

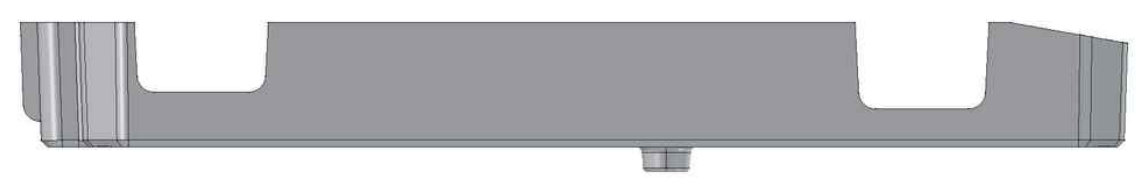

**Bottom view**

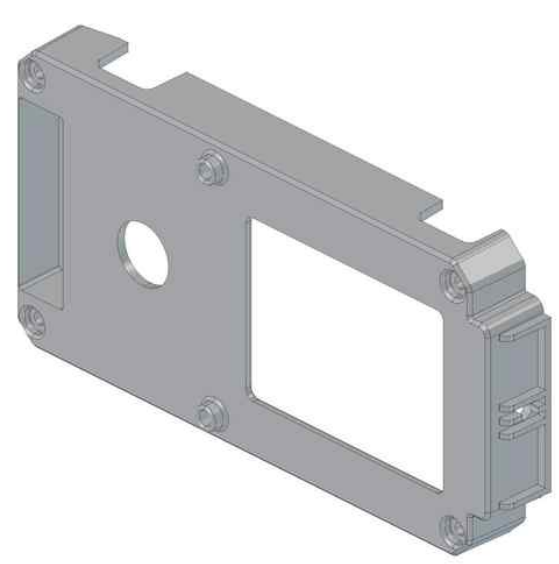

**Axonometric view**

Figure 6-1 Seven input renderings for I2C-01

## Results

| | |
|---|---|
| 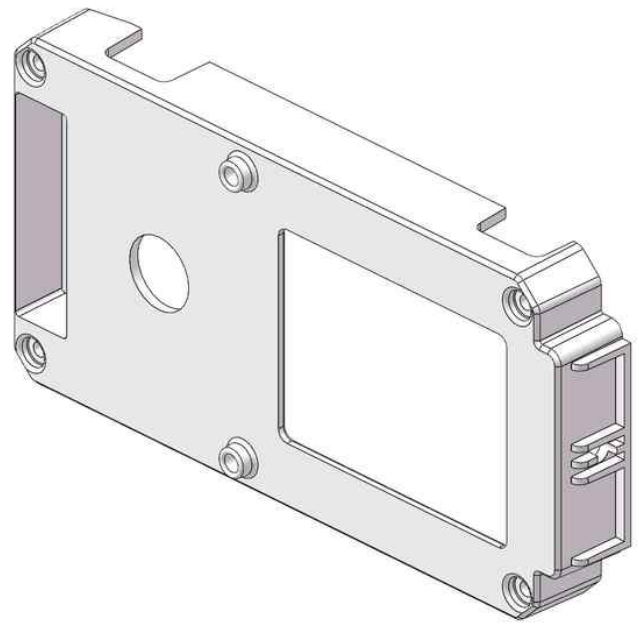 | 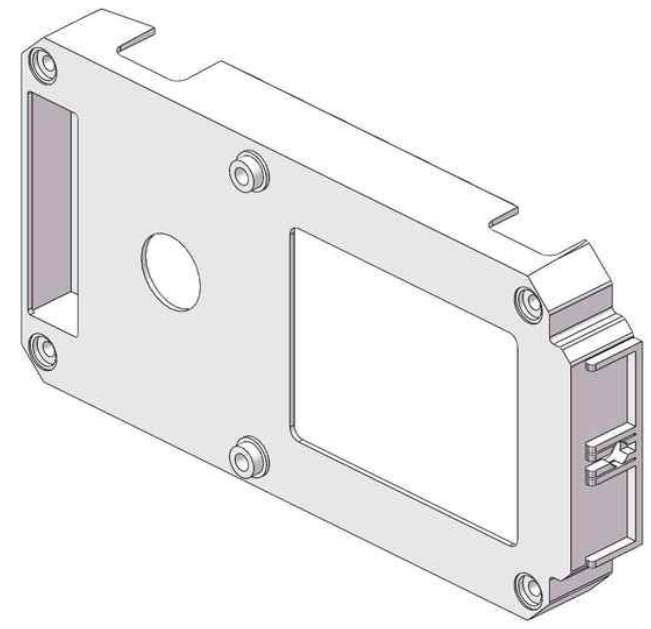 |
| **Reference model (axonometric view)** | **Generated model (axonometric view)** |
| 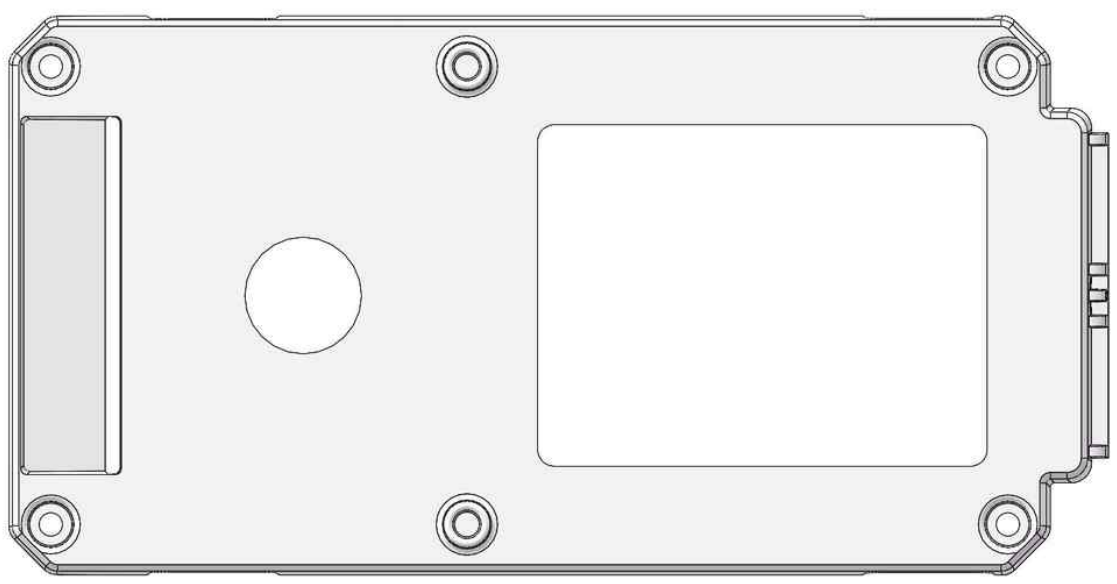 | 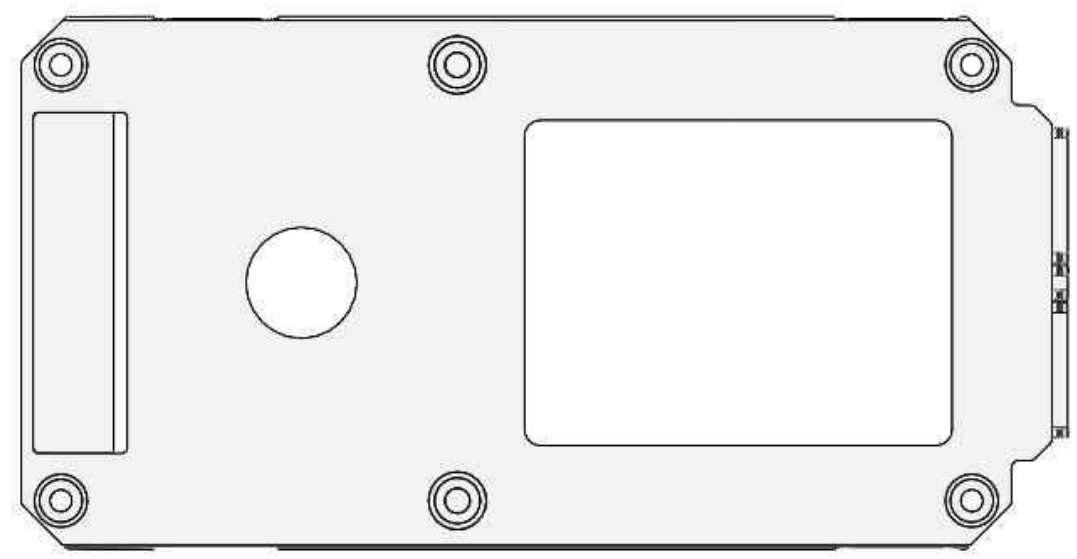 |
| **Reference model (front view)** | **Generated model (front view)** |
| 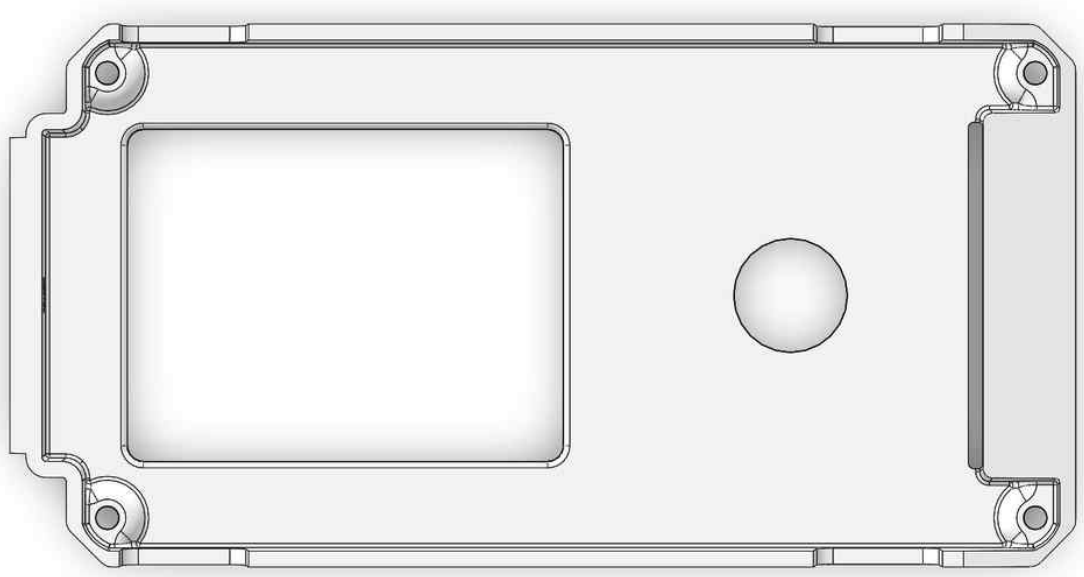 | 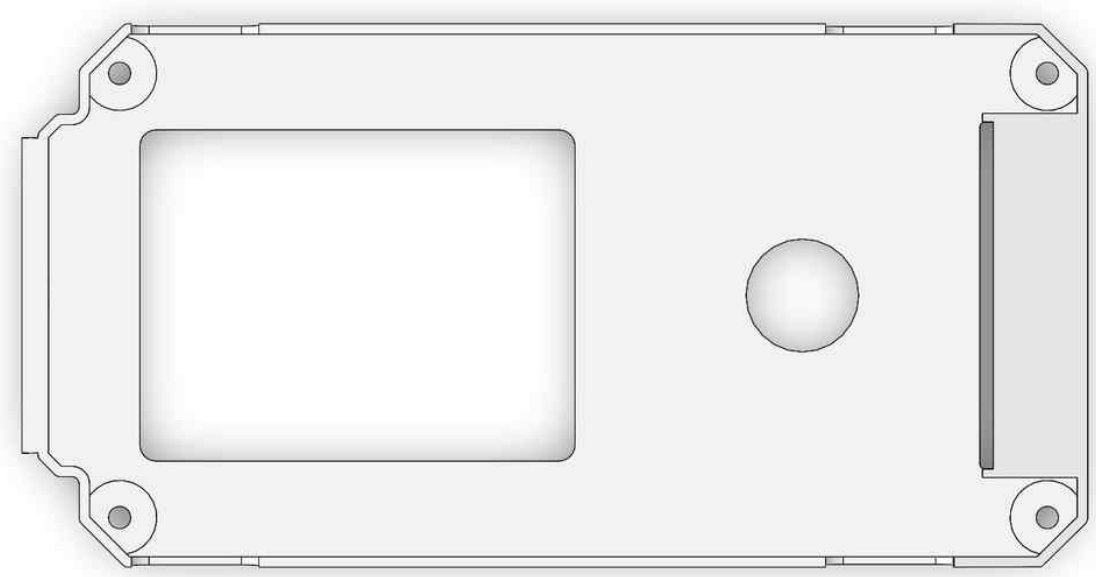 |
| **Reference model (rear view)** | **Generated model (rear view)** |
| 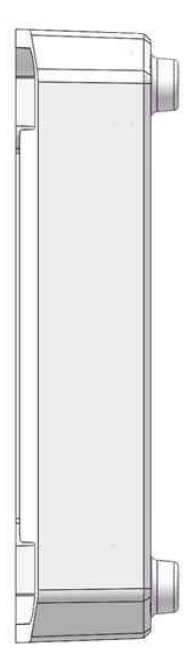 | 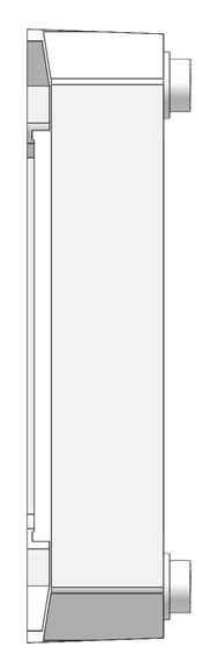 |
| **Reference model (left view)** | **Generated model (left view)** |
| 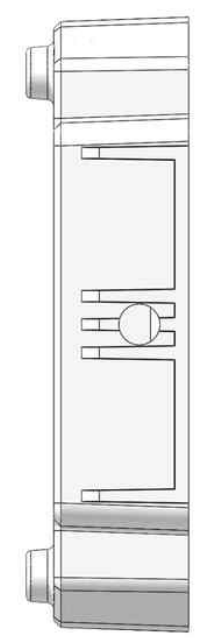 | 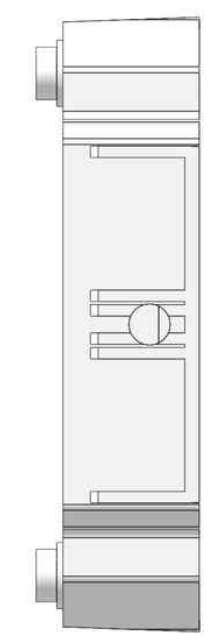 |
| **Reference model (right view)** | **Generated model (right view)** |
| 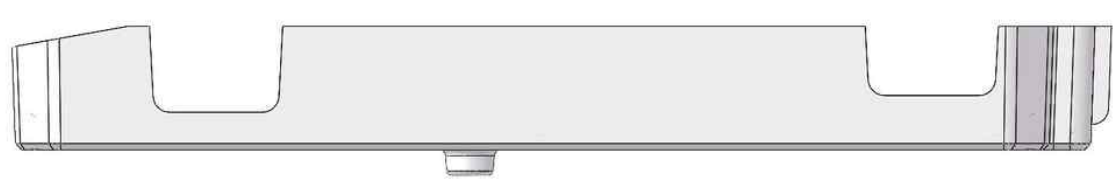 | 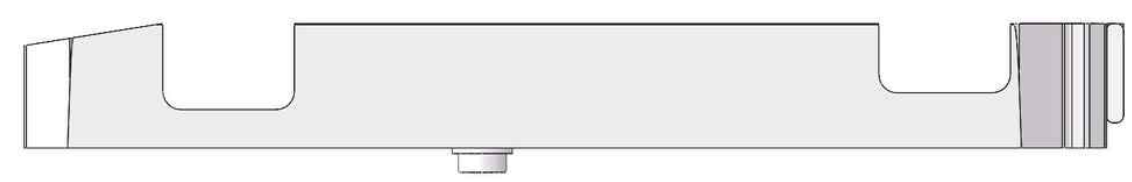 |
| **Reference model (top view)** | **Generated model (top view)** |
| 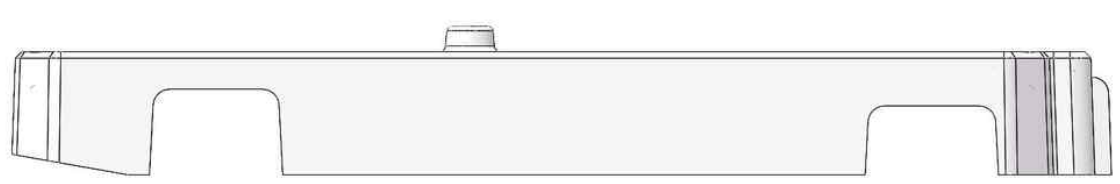 | 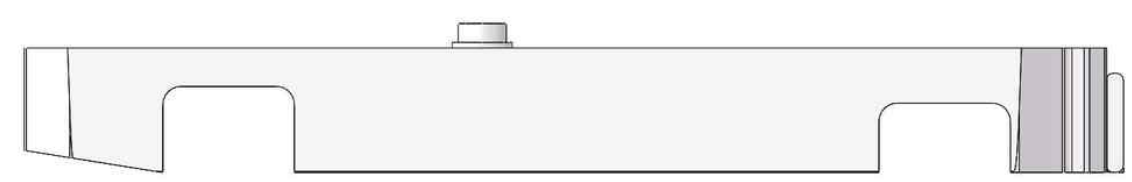 |
| **Reference model (bottom view)** | **Generated model (bottom view)** |

Figure 6-2  Renderings of the reference and generated models for I2C-01

The generated I2C-01 model passes the basic solid checks. The overall arrangement of the window, circular holes, and perimeter flanges is close to the reference. Figure 6-2 shows differences in the side clips, corner hole bosses, and boss roots, with some small transitions simplified. The generated model is approximately 158.00 mm long, compared with 132.11 mm for the reference, so its overall scale is too large; HD is 14.758 mm. The main structure has been reconstructed, but the shapes of the clips and hole bosses matter for snap-fit positioning and fastening.

Figure 6-3 plots silhouette IoU for each view against iteration number.

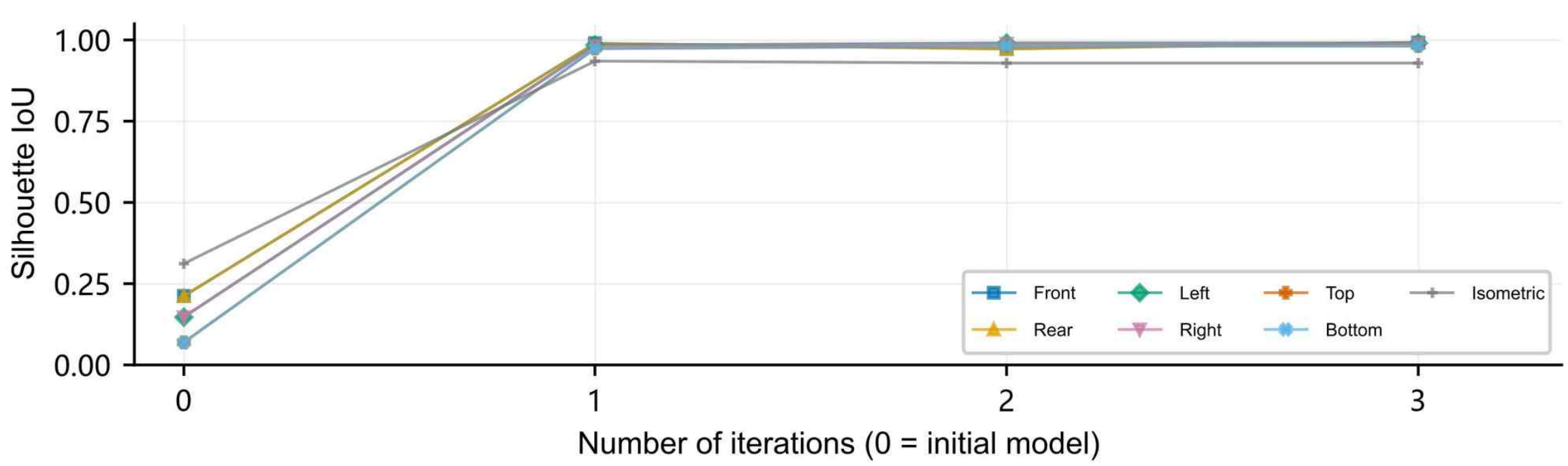

Figure 6-3  Silhouette IoU by iteration for I2C-01

## I2C-02 KnowCO2 sensor housing (simple)

### Input renderings

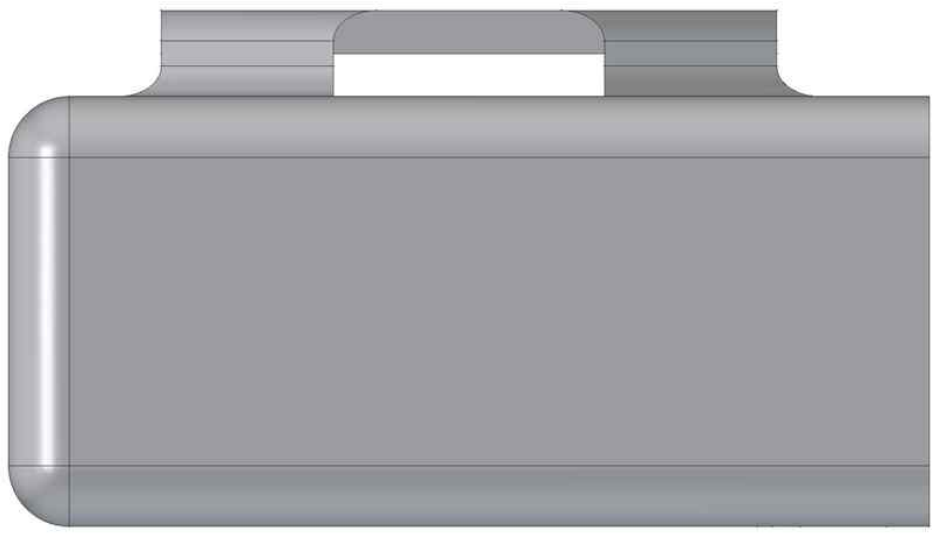

**Front view**

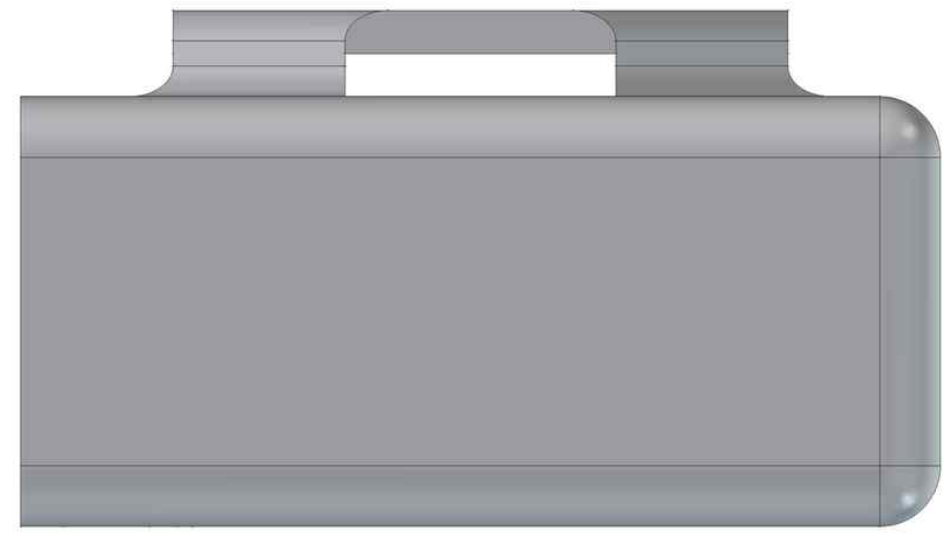

**Rear view**

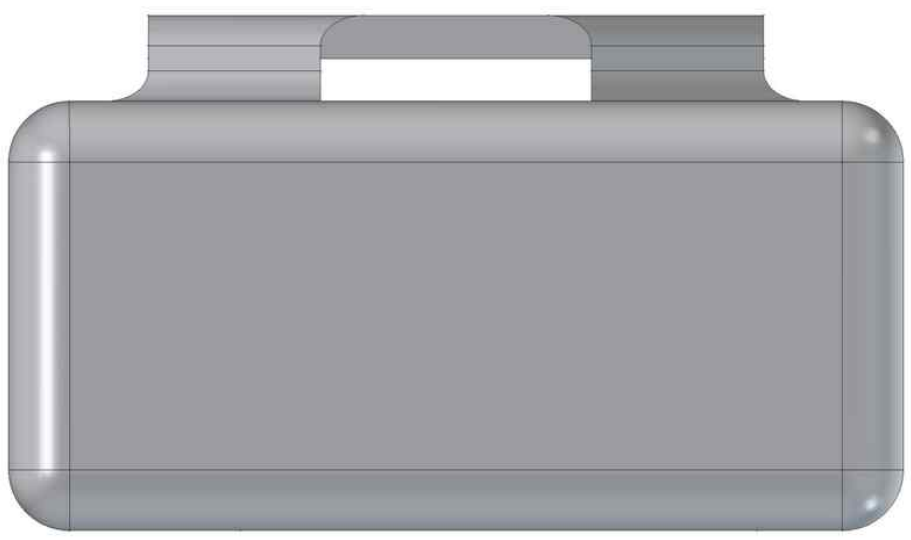

**Left view**

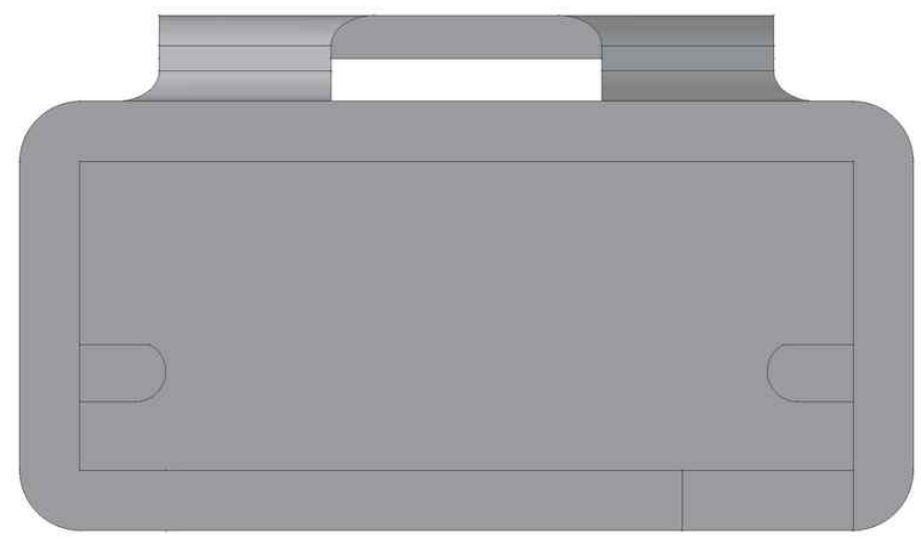

**Right view**

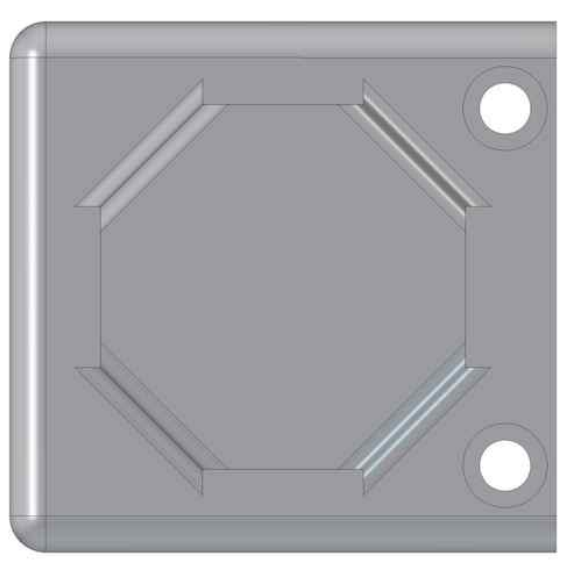

**Top view**

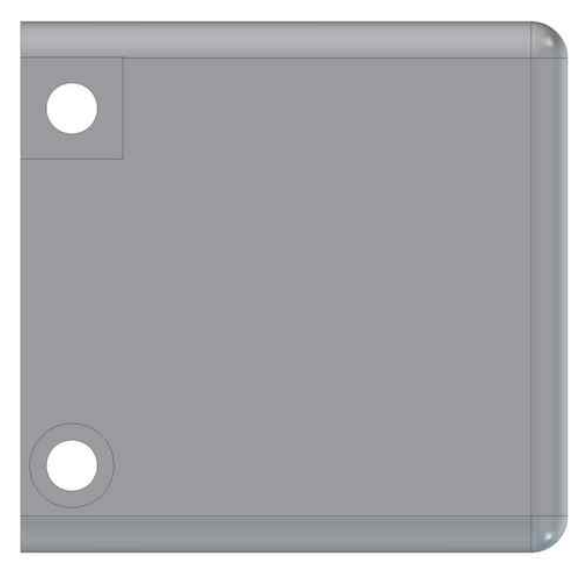

**Bottom view**

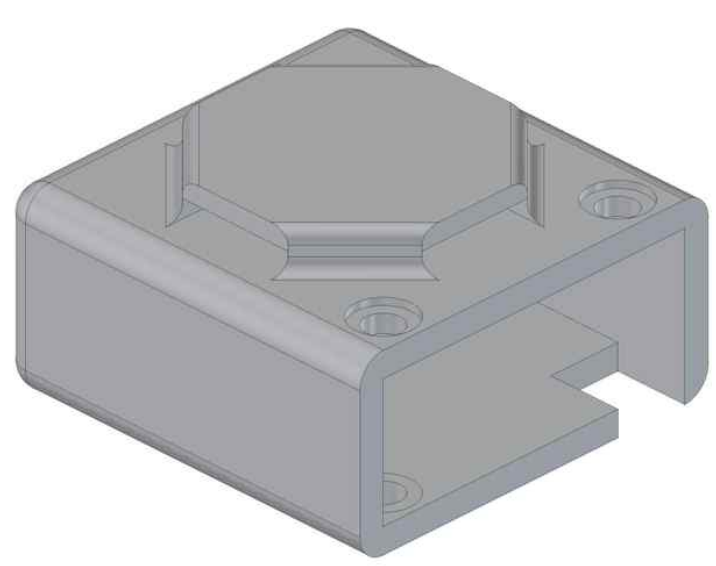

**Axonometric view**

Figure 6-4 Seven input renderings for I2C-02

## Results

| | |
|---|---|
| 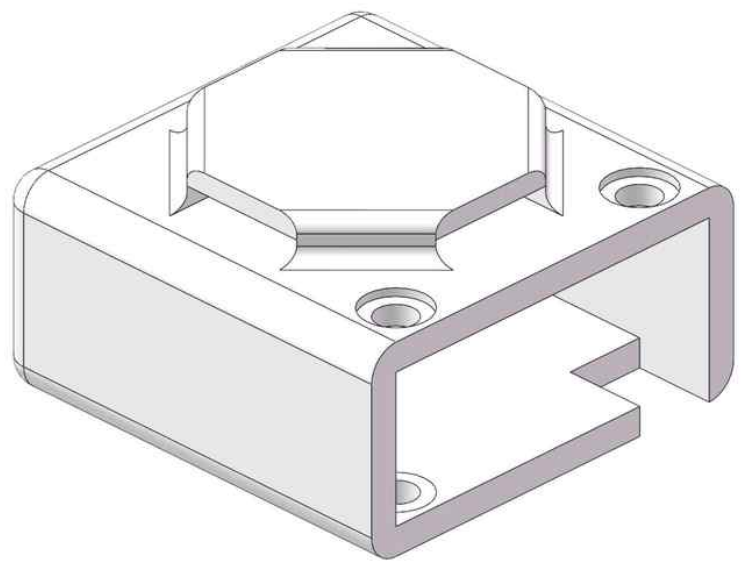 | 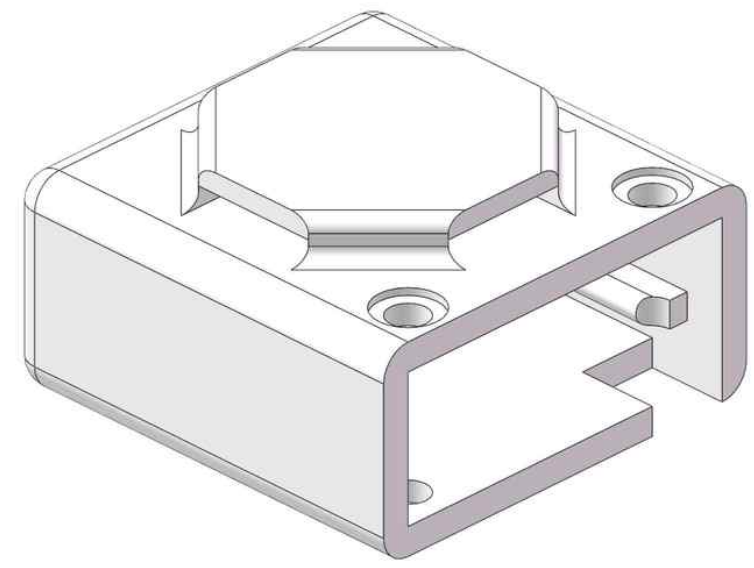 |
| **Reference model (axonometric view)** | **Generated model (axonometric view)** |
| 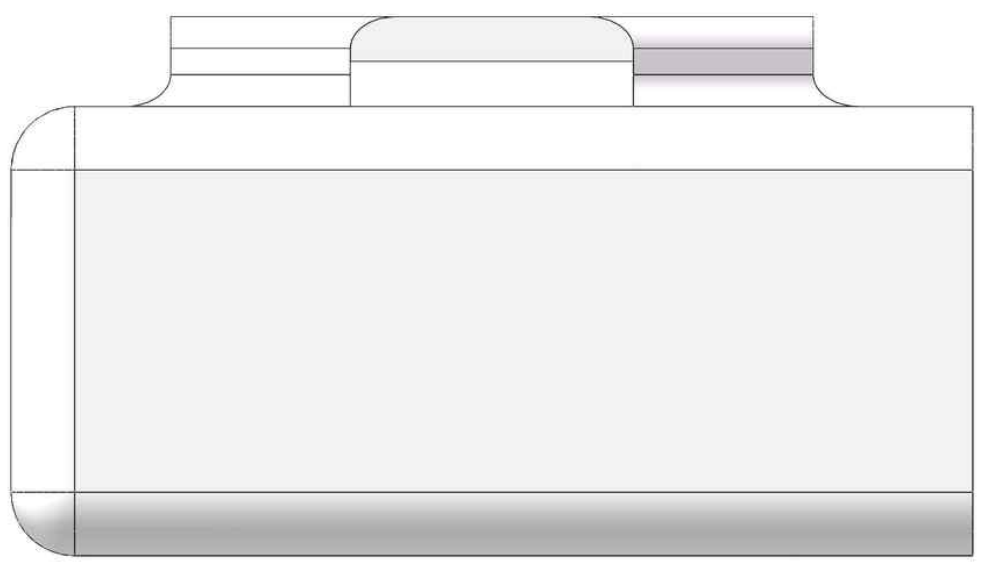 | 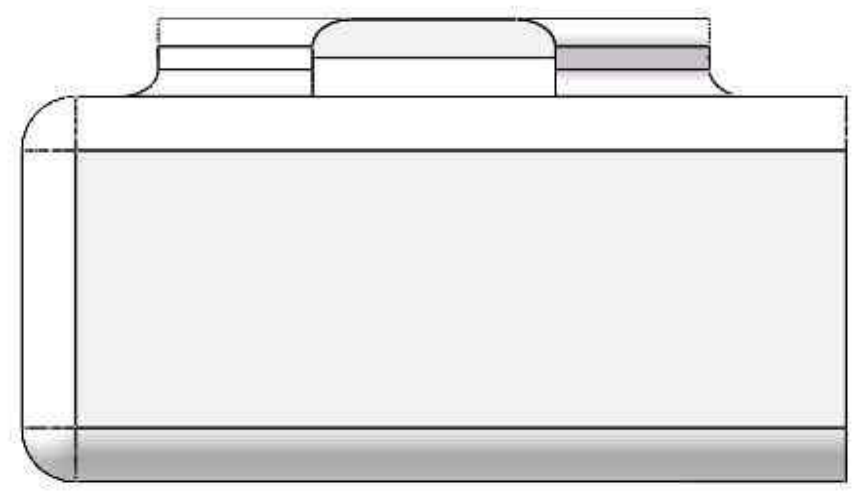 |
| **Reference model (front view)** | **Generated model (front view)** |
| 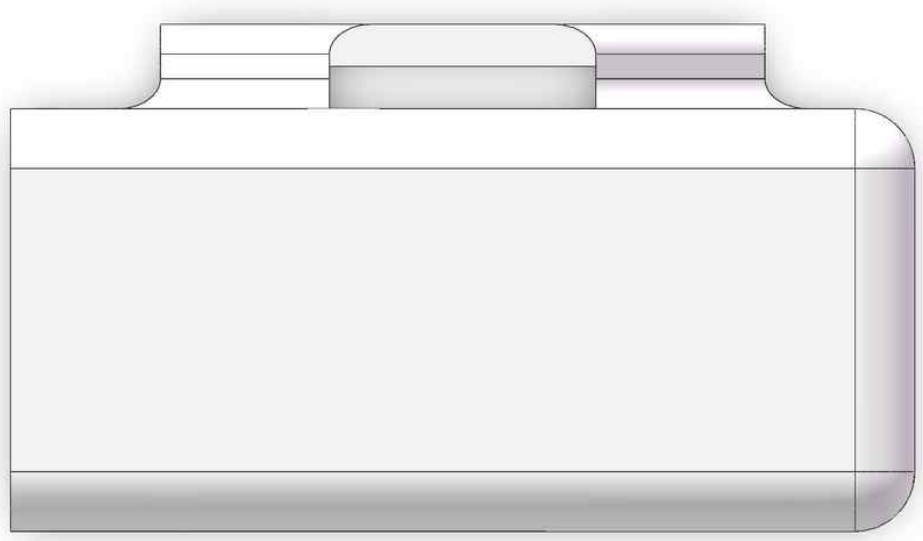 | 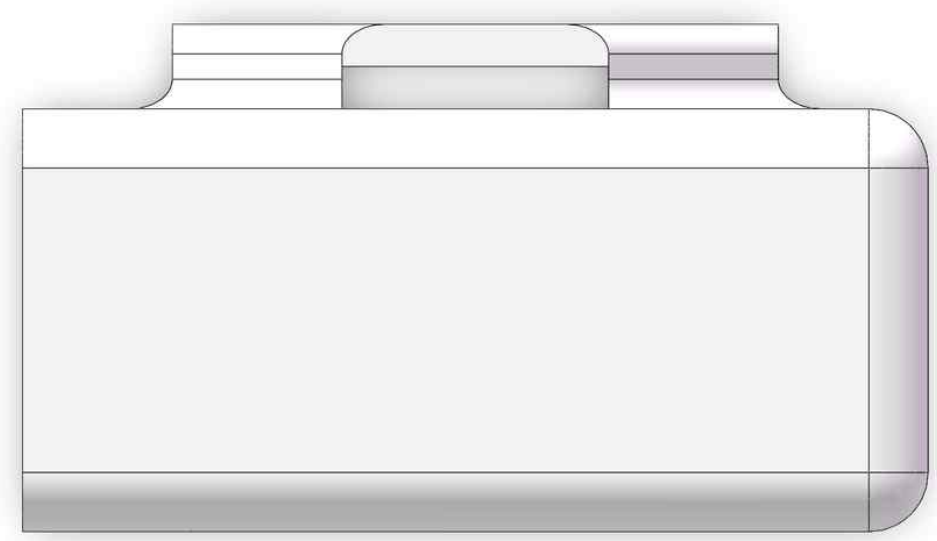 |
| **Reference model (rear view)** | **Generated model (rear view)** |
| 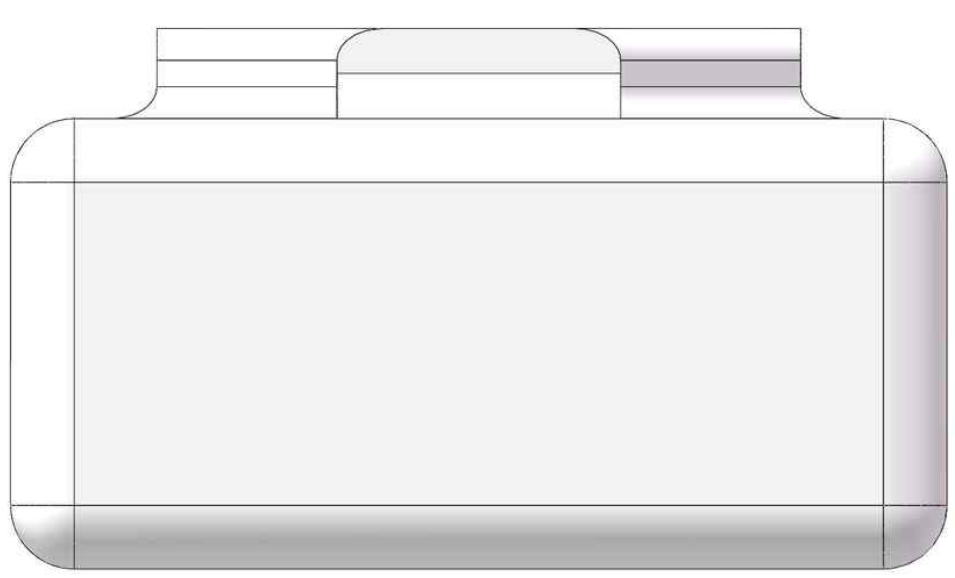 | 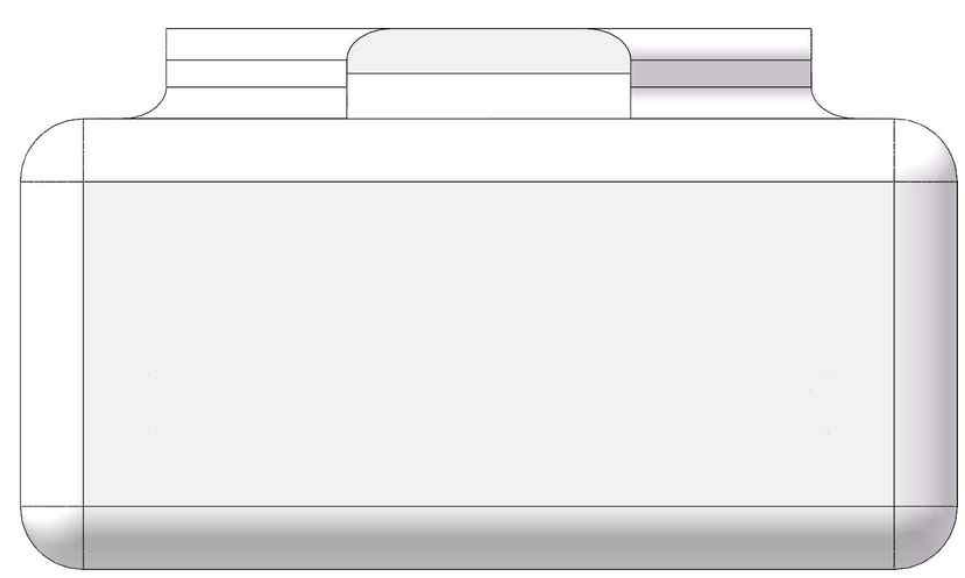 |
| **Reference model (left view)** | **Generated model (left view)** |
| 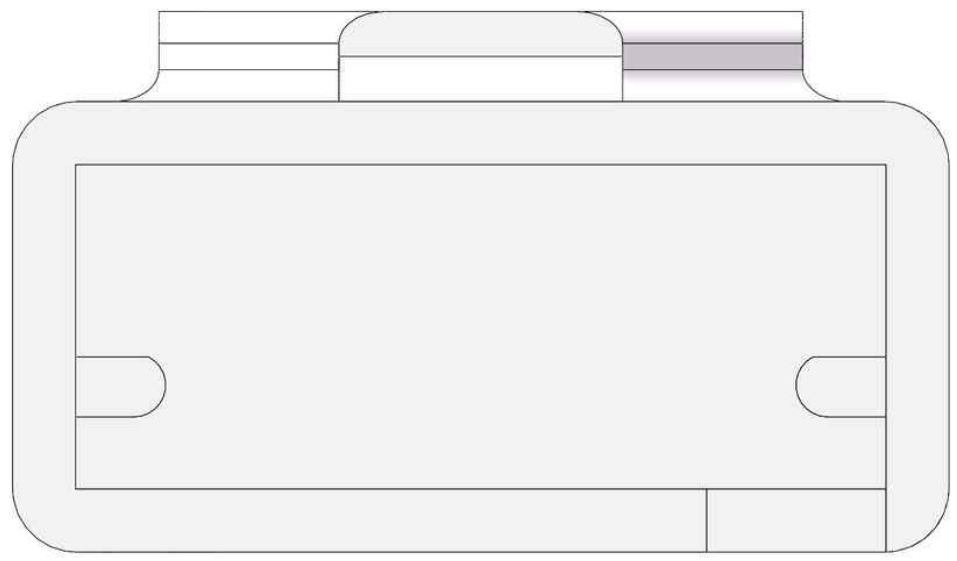 | 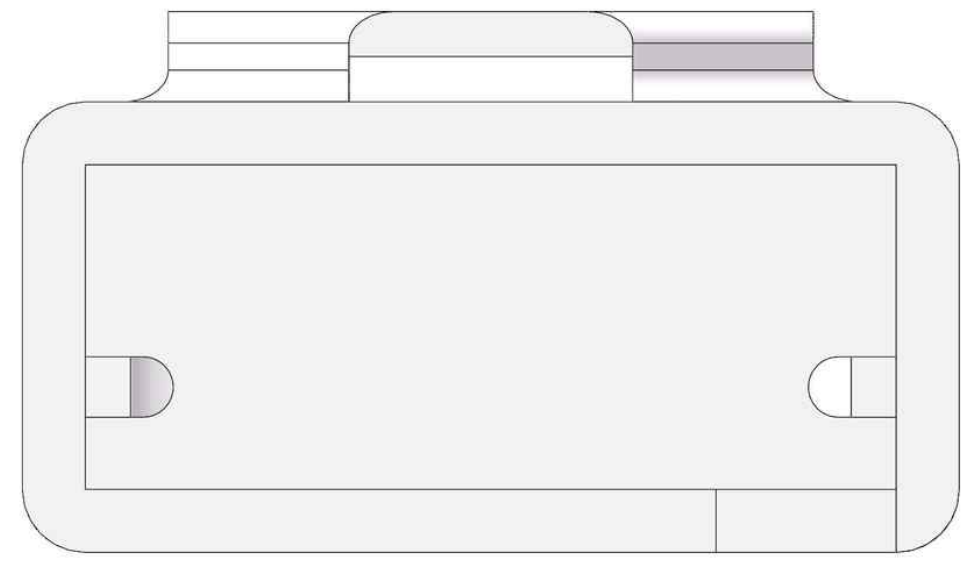 |
| **Reference model (right view)** | **Generated model (right view)** |
| 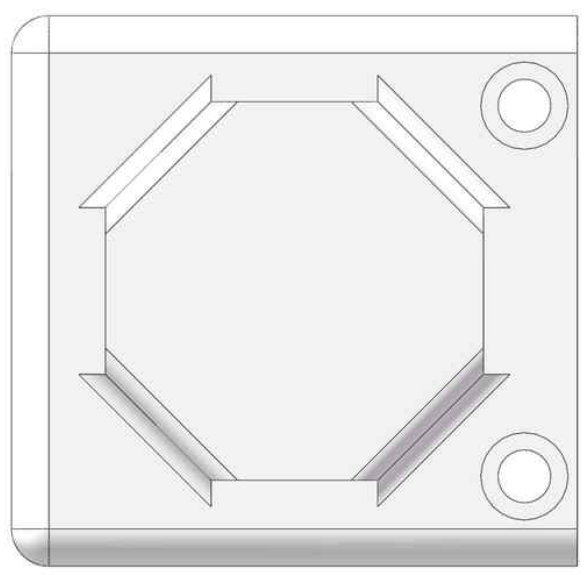 | 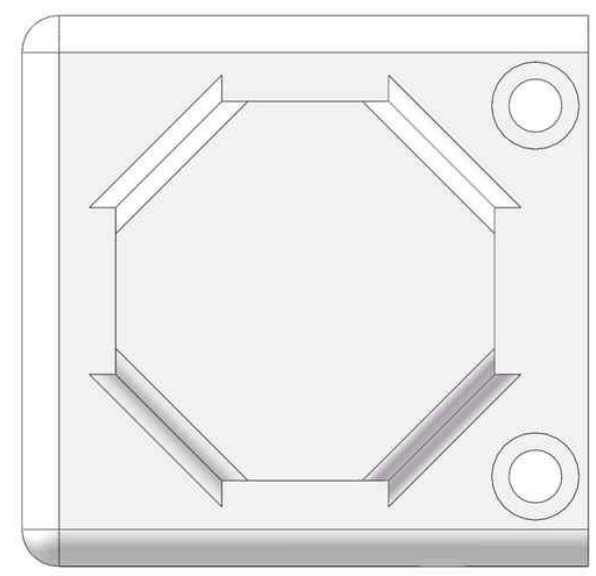 |
| **Reference model (top view)** | **Generated model (top view)** |
| 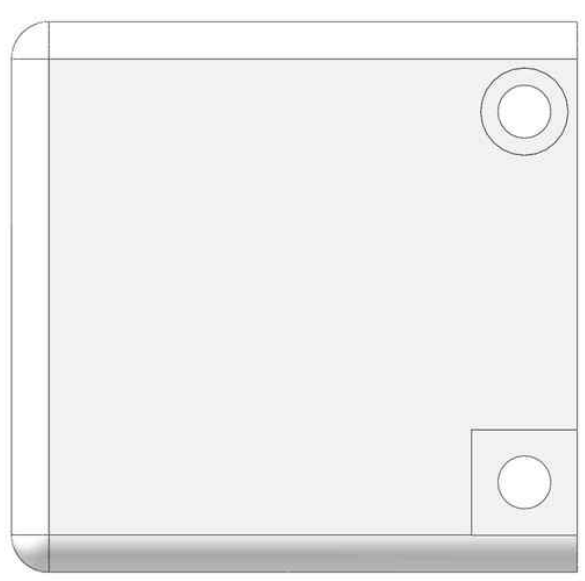 | 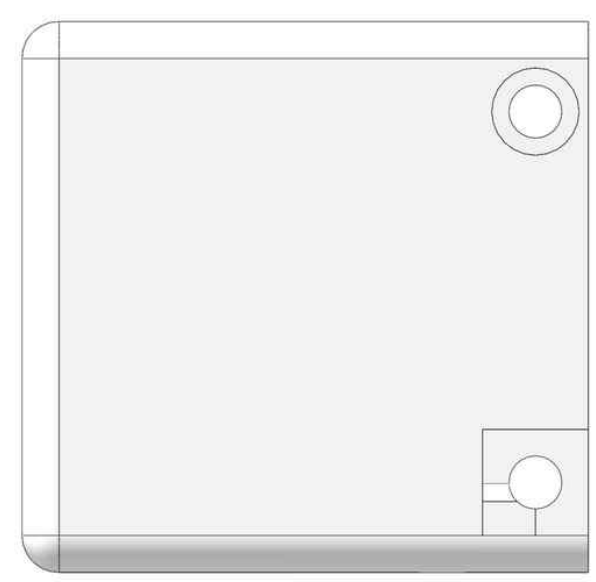 |
| **Reference model (bottom view)** | **Generated model (bottom view)** |

Figure 6-5 Renderings of the reference and generated models for I2C-02

The generated I2C-02 model passes the basic solid checks. The housing opening, top platform, and hole bosses on both sides are arranged similarly to the reference, although details at the platform root and in the internal guides differ. Mean IoU across the seven views is 0.9915. However, the reference measures approximately 27.18 × 26.42 × 15.24 mm, while the generated model measures 107.00 × 104.00 × 60.00 mm. The generated model is substantially larger, with an HD of 64.207 mm. Its overall proportions are close to the reference, but its absolute dimensions have not been recovered.

Figure 6-6 plots silhouette IoU for each view against iteration number.

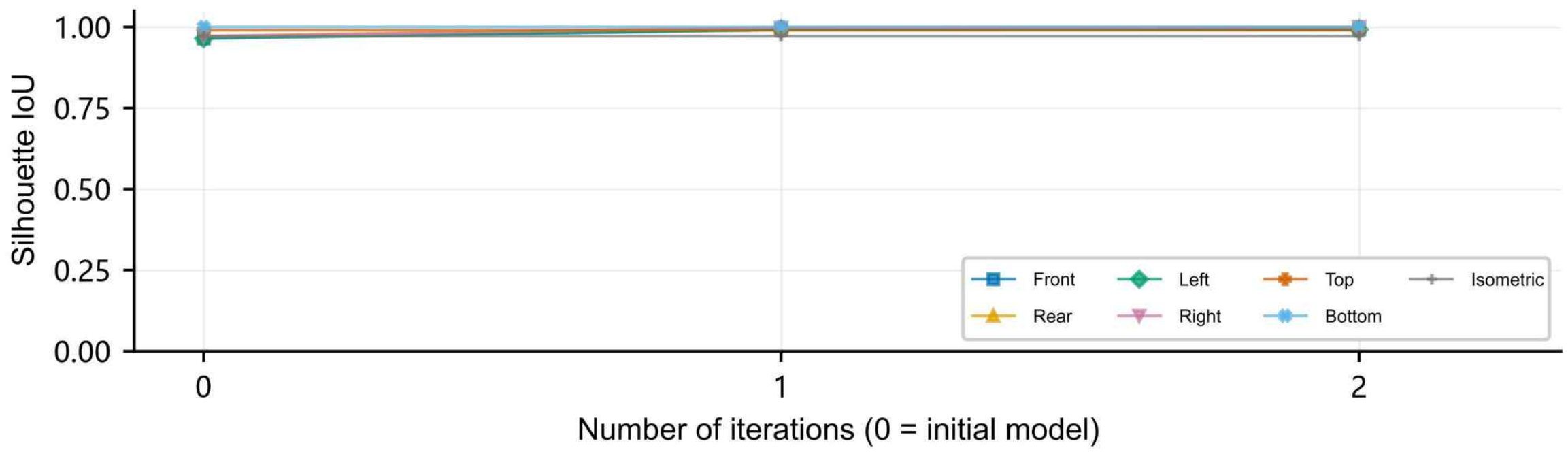

Figure 6-6 Silhouette IoU by iteration for I2C-02

## I2C-03 C5 Corvette lower control arm (moderate)

### Input renderings

| | |
|---|---|
| 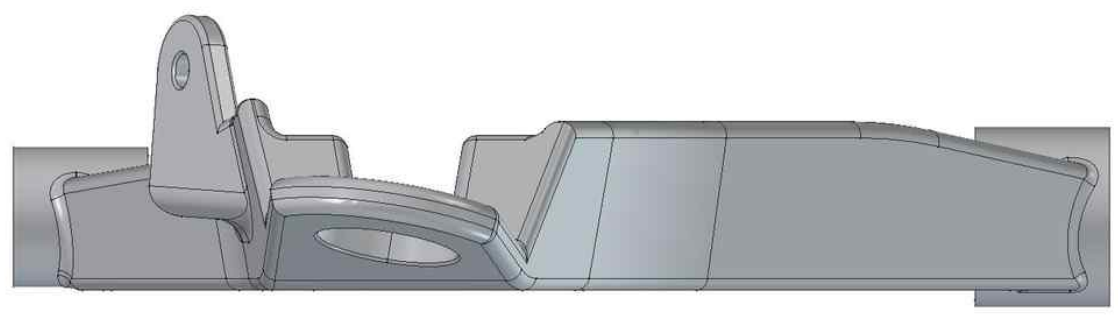 | 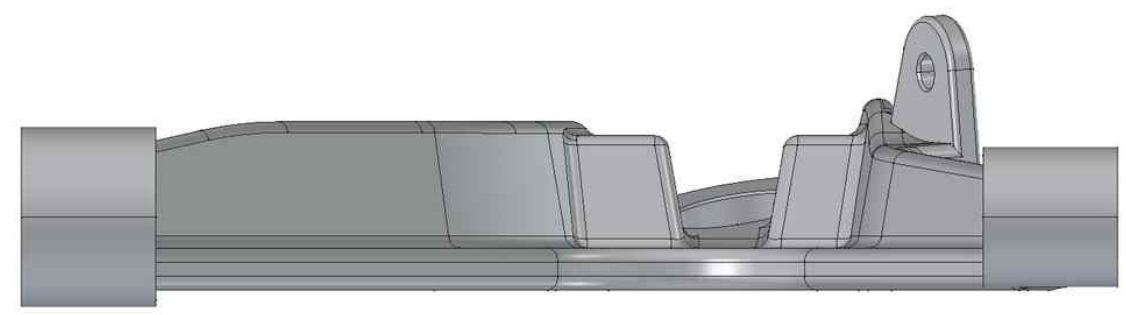 |
| **Front view** | **Rear view** |
| 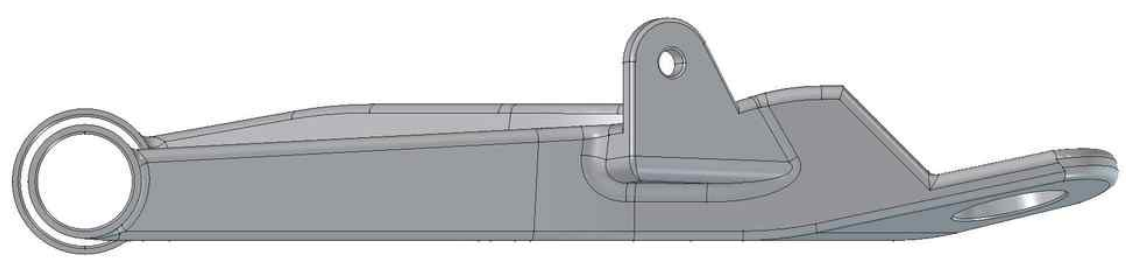 | 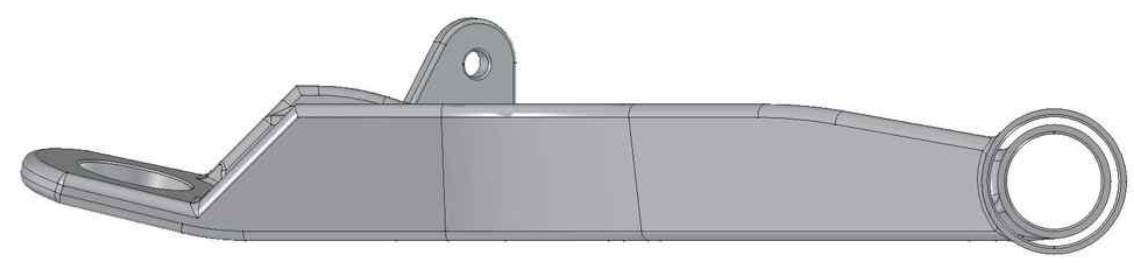 |
| **Left view** | **Right view** |
| 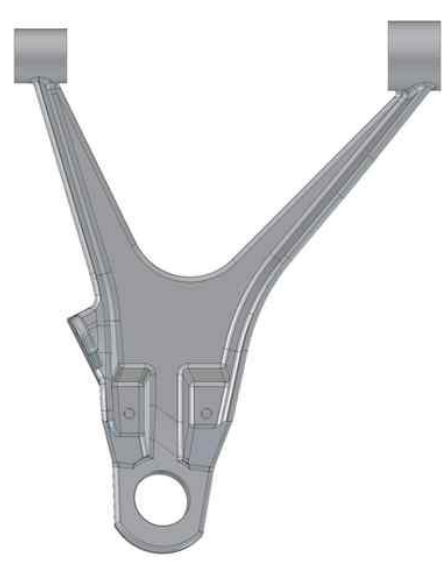 | 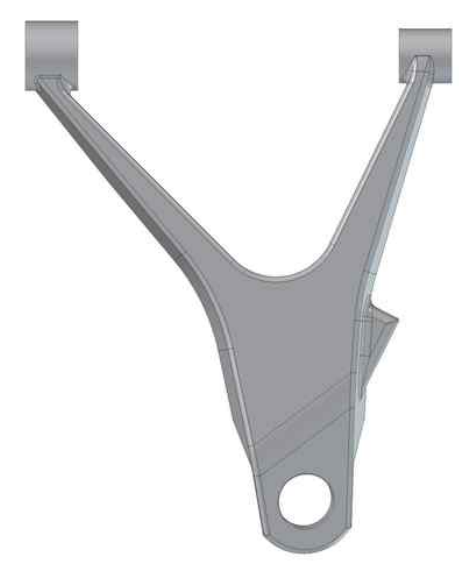 |
| **Top view** | **Bottom view** |
| 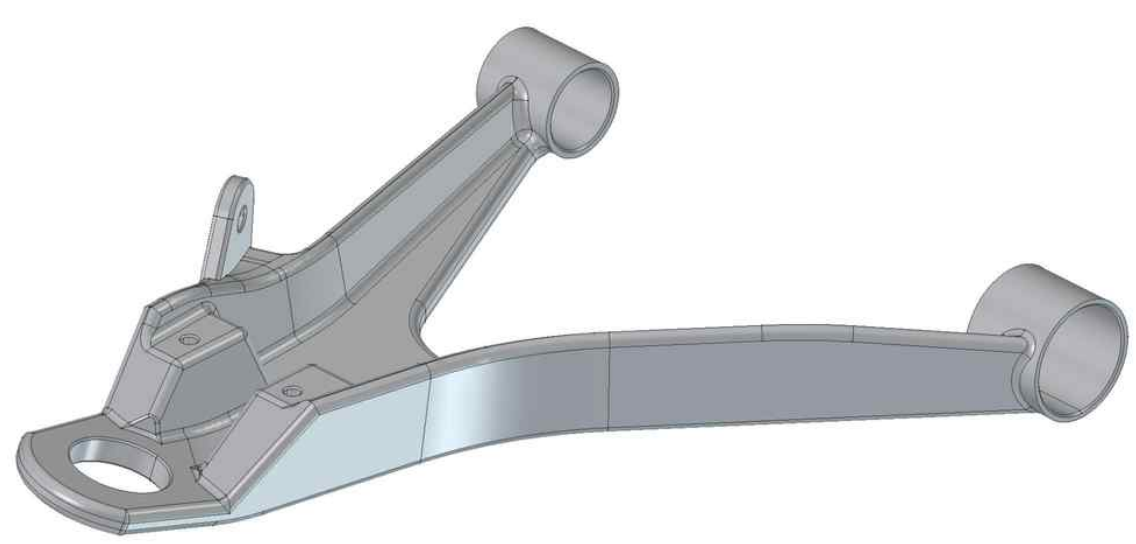 | |
| **Axonometric view** | |

Figure 6-7  Seven input renderings for I2C-03

**Results**

| | |
|---|---|
| 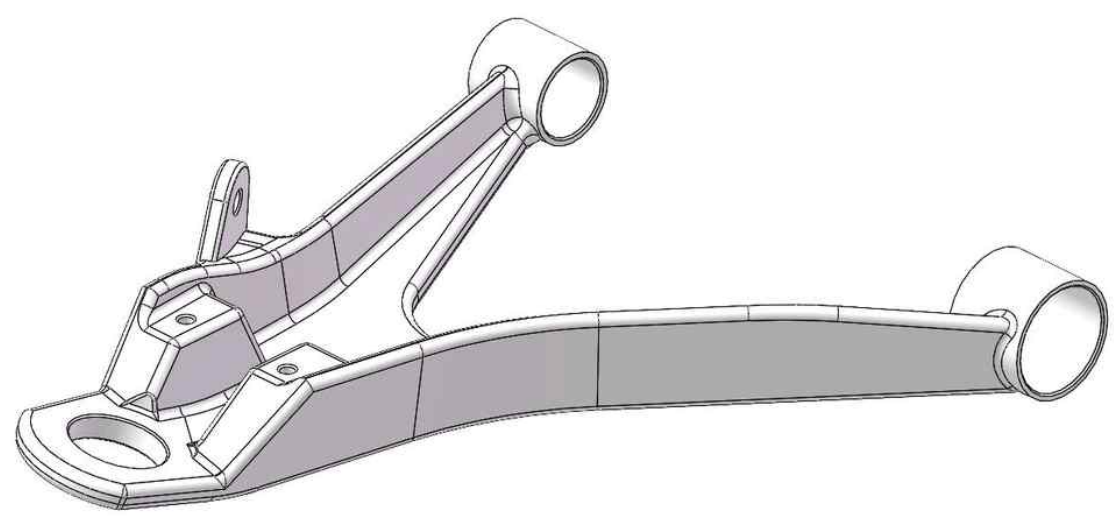 | 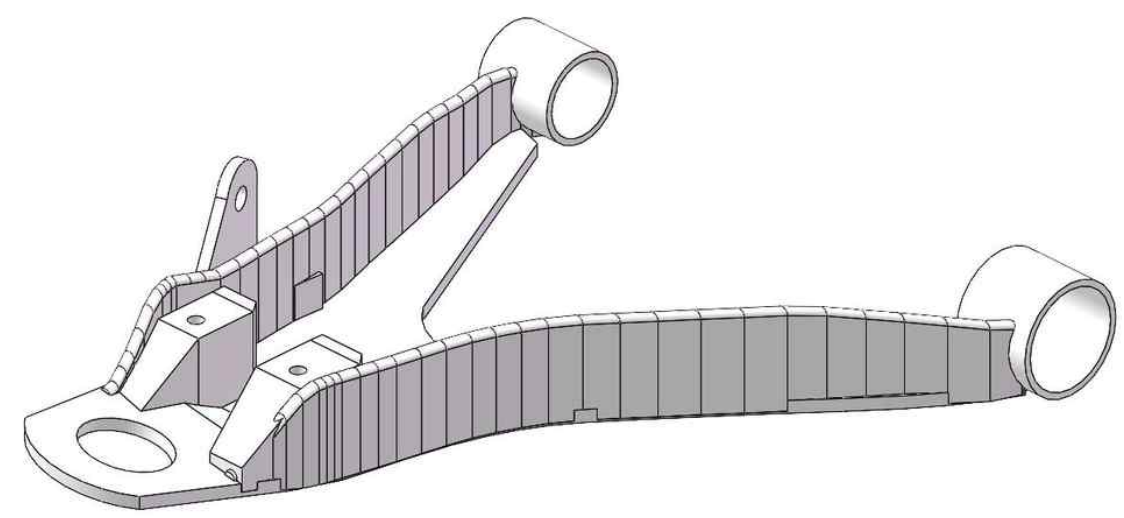 |
| **Reference model (axonometric view)** | **Generated model (axonometric view)** |
| 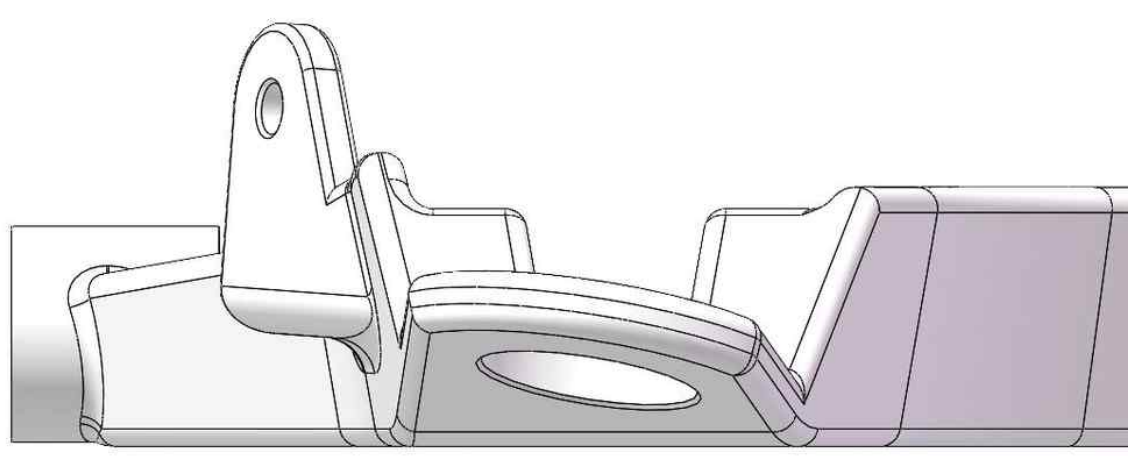 | 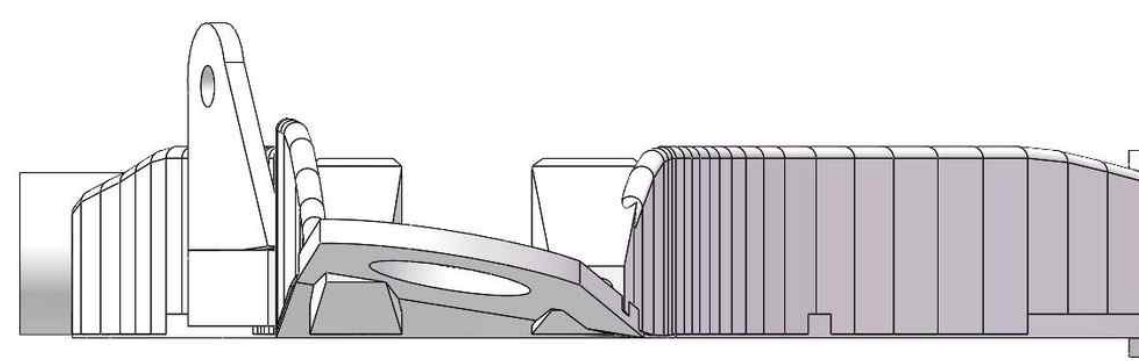 |
| **Reference model (front view)** | **Generated model (front view)** |
| 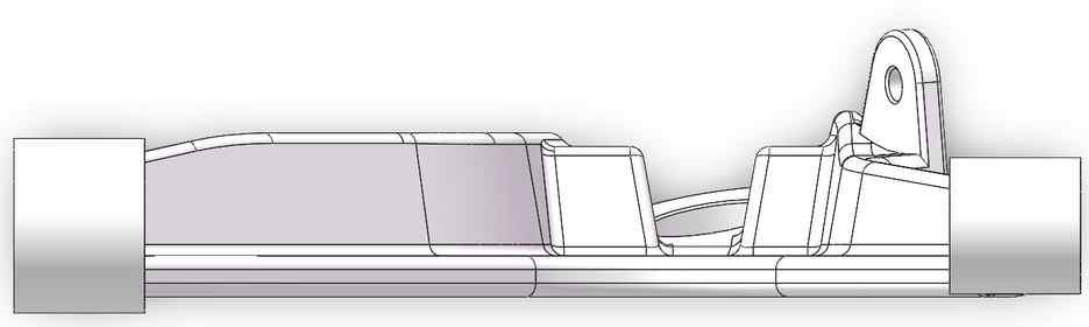 | 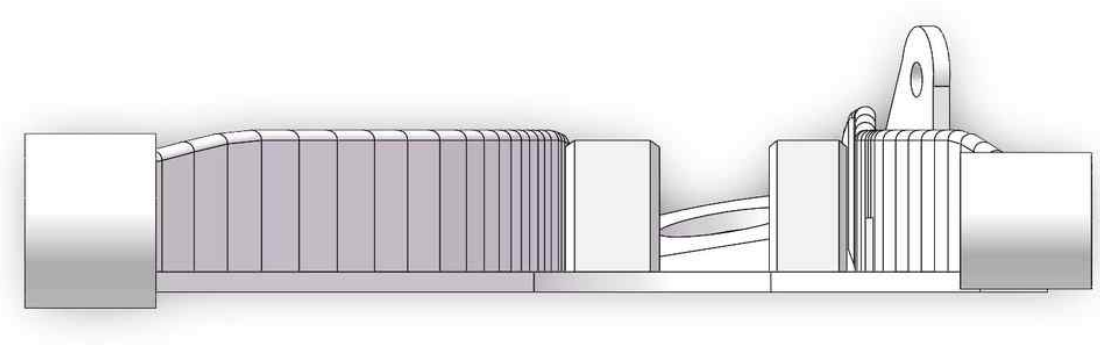 |
| **Reference model (rear view)** | **Generated model (rear view)** |
| 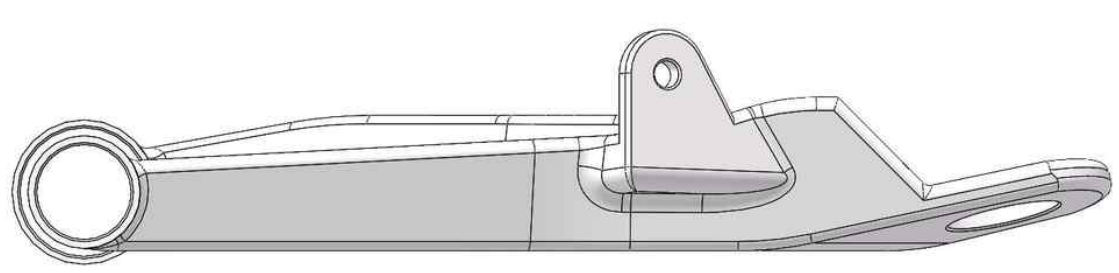 | 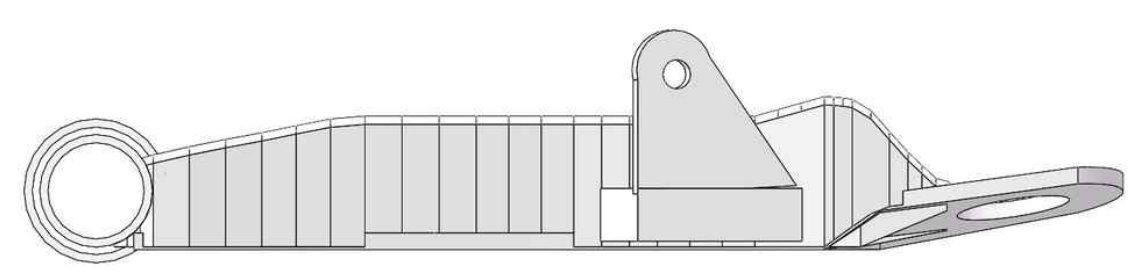 |
| **Reference model (left view)** | **Generated model (left view)** |
| 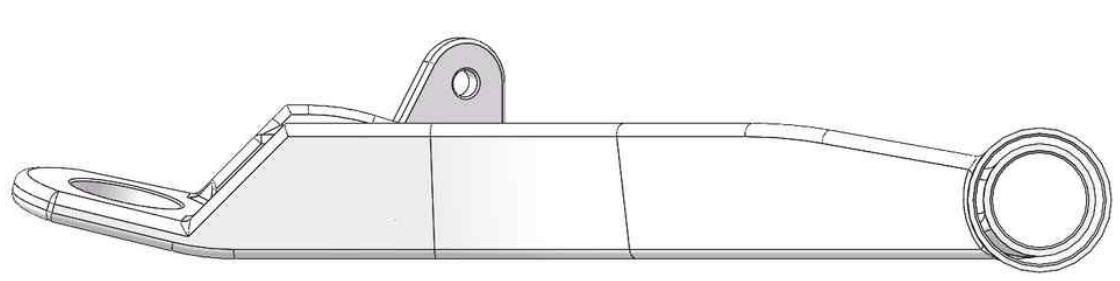 | 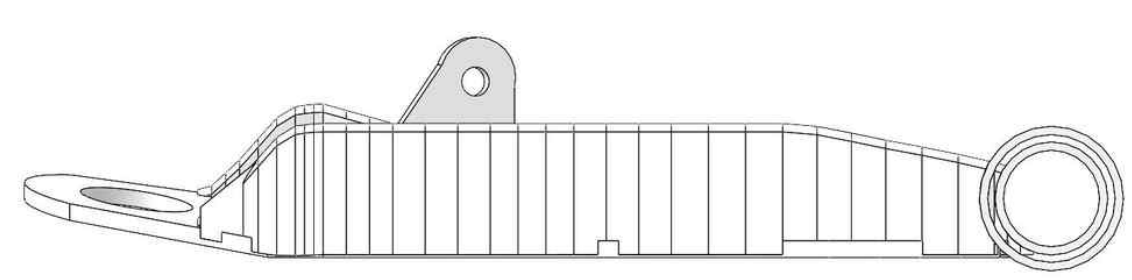 |
| **Reference model (right view)** | **Generated model (right view)** |

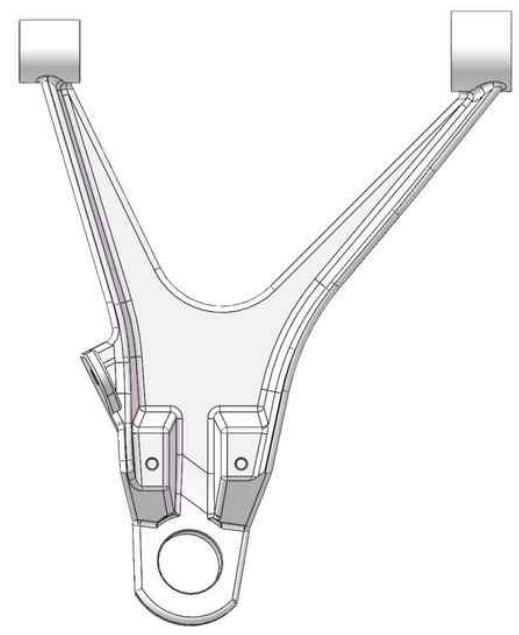

**Reference model (top view)**

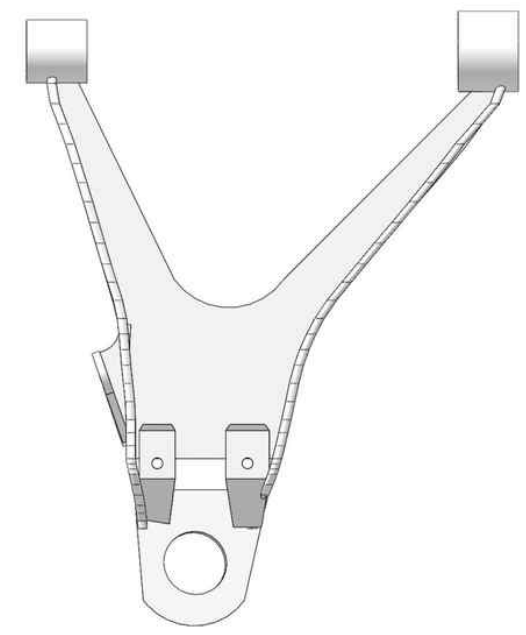

**Generated model (top view)**

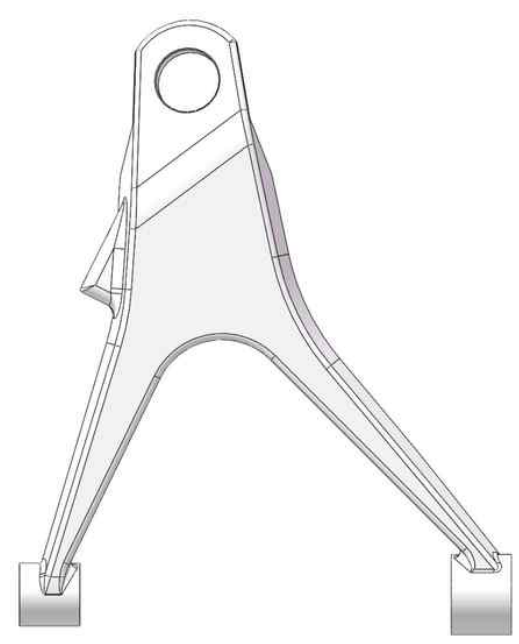

**Reference model (bottom view)**

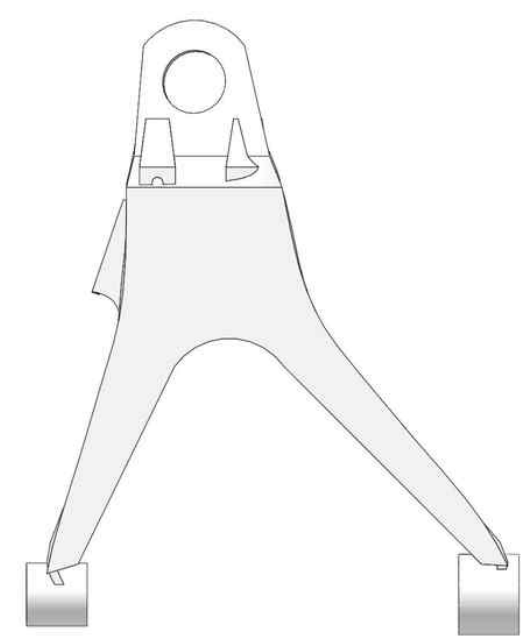

**Generated model (bottom view)**

Figure 6-8 Renderings of the reference and generated models for I2C-03

The generated I2C-03 model is a valid solid. The forked arms, bushings at the two ends, and front lug are in approximately the correct relative positions. However, the arm sidewalls are approximated by ruled lofts through multiple sections. Segmented surfaces replace the original smooth surfaces, root fillets are missing, and local steps have been introduced, as shown below. The longest side of the generated model's bounding box is approximately 394.76 mm, compared with 488.73 mm for the reference. The model is undersized, with an HD of 62.401 mm. The main errors are in surface transitions and connections between sections.

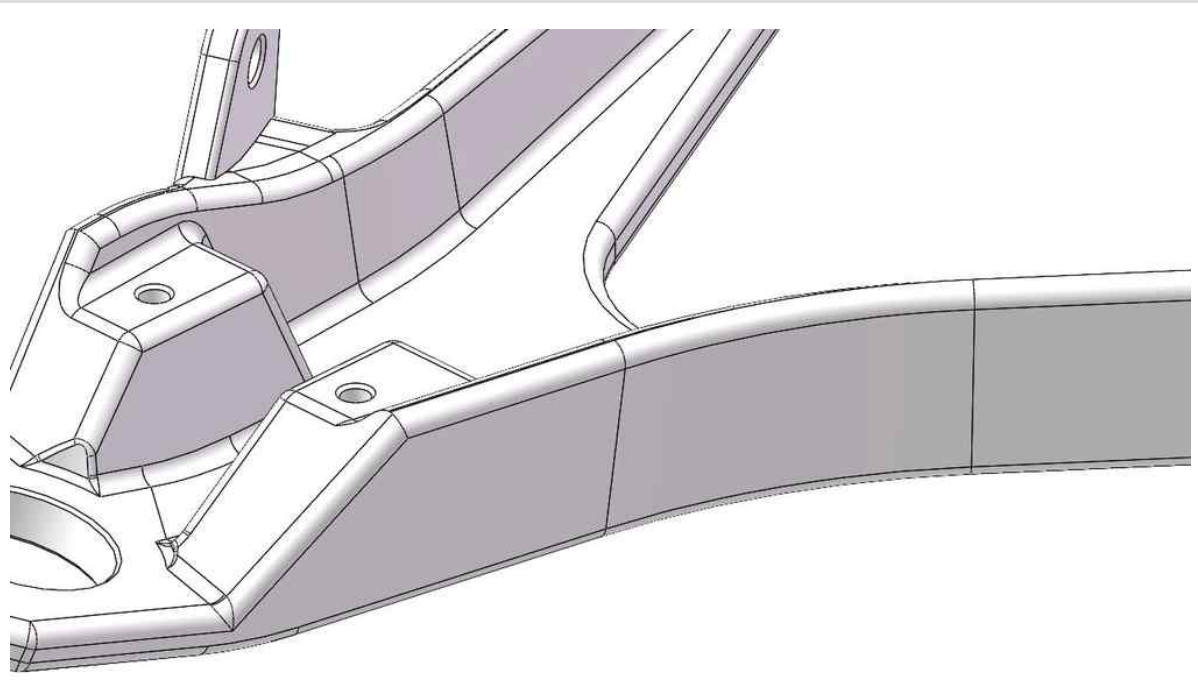

**Reference model: close-up**

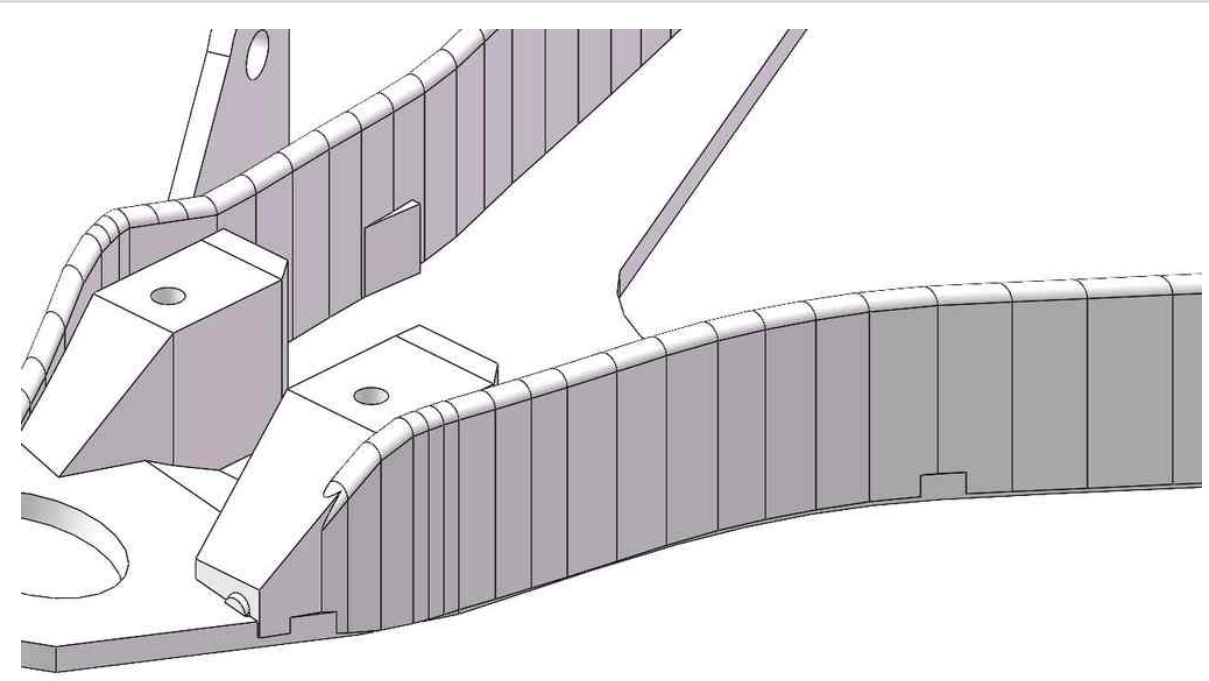

**Generated model: close-up**

Figure 6-9 Arm sidewalls and root transitions in I2C-03

Figure 6-10 plots silhouette IoU for each view against iteration number.

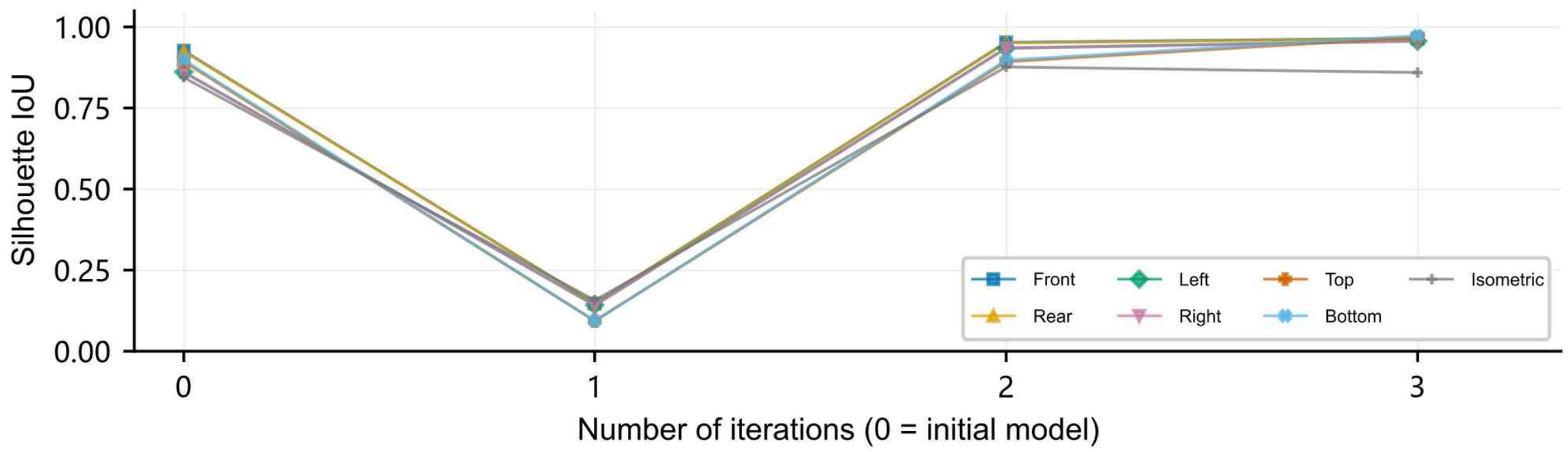

Figure 6-10 Silhouette IoU by iteration for I2C-03

## I2C-04 C5 Corvette steering knuckle (moderate)

### Input renderings

<table>
<tr><td>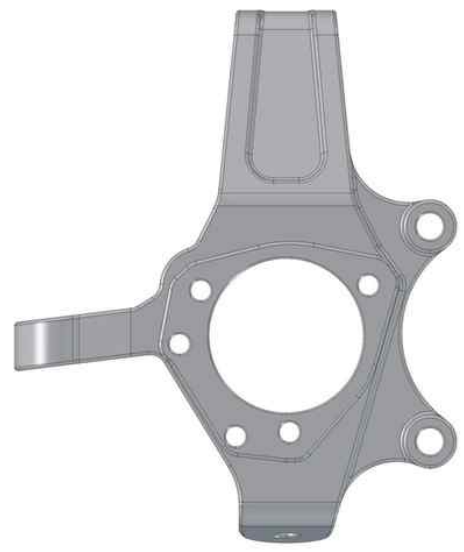</td><td>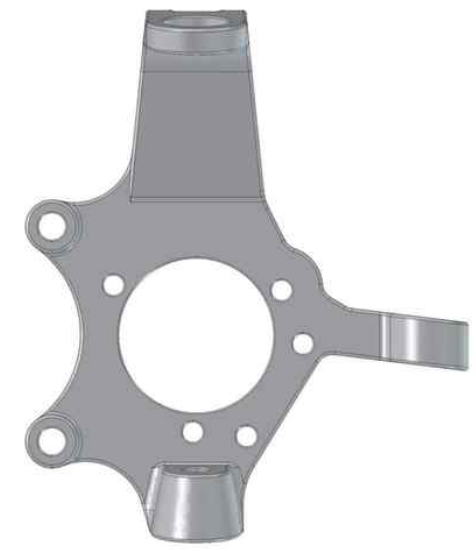</td></tr>
<tr><td>Front view</td><td>Rear view</td></tr>
<tr><td>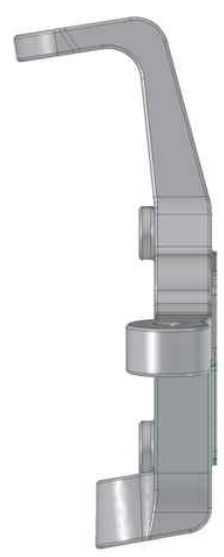</td><td>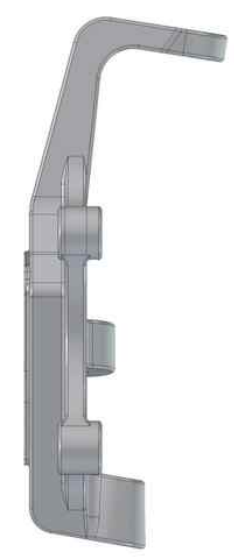</td></tr>
<tr><td>Left view</td><td>Right view</td></tr>
<tr><td>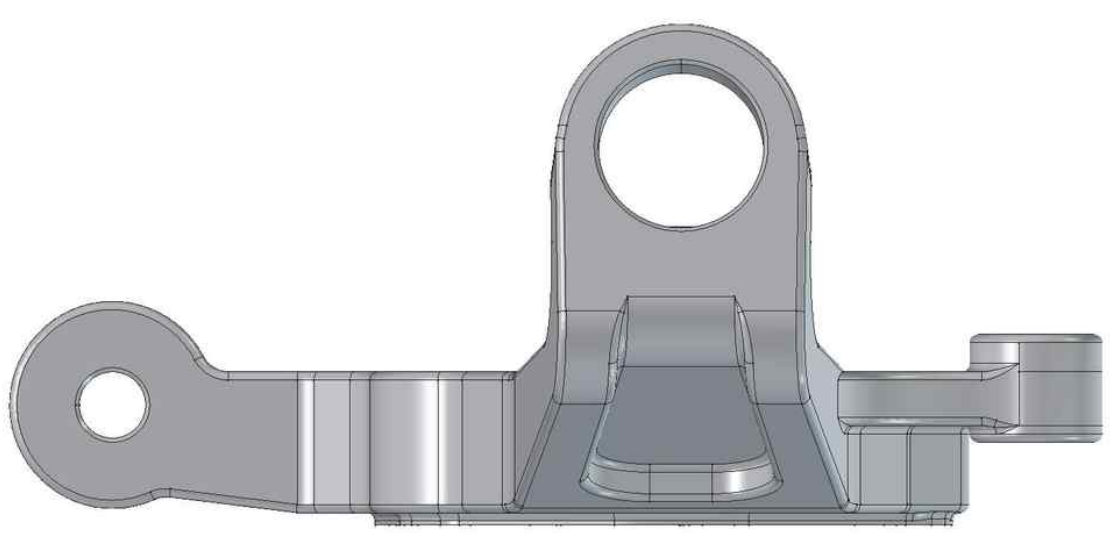</td><td>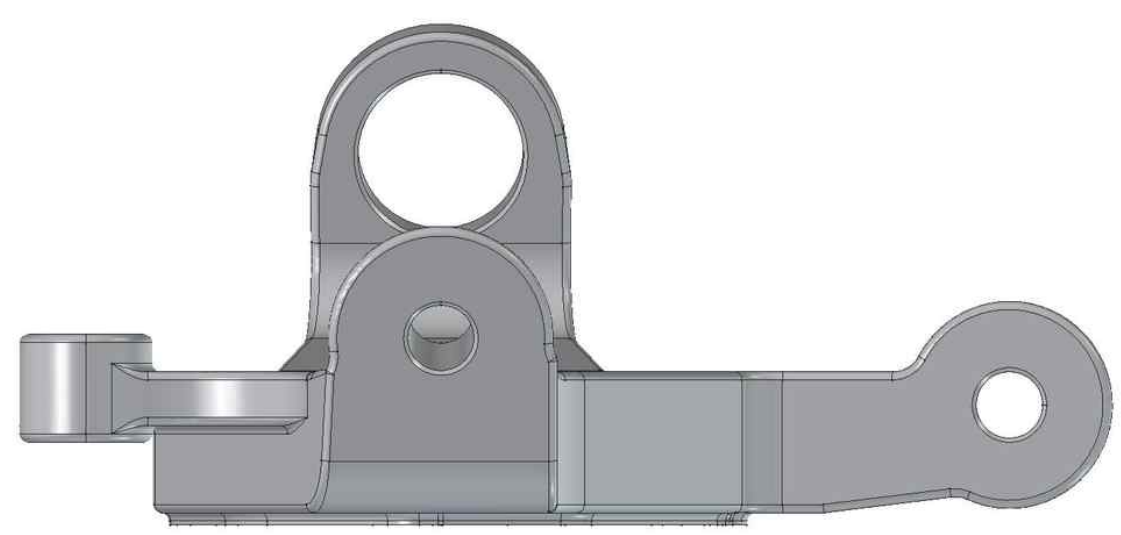</td></tr>
<tr><td>Top view</td><td>Bottom view</td></tr>
<tr><td colspan="2">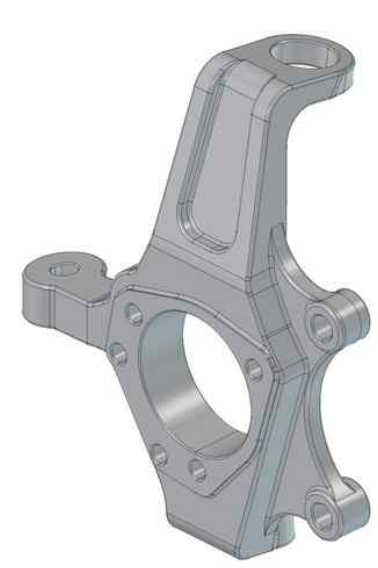</td></tr>
<tr><td colspan="2">Axonometric view</td></tr>
</table>

Figure 6-11 Seven input renderings for I2C-04

**Results**

| | |
|---|---|
| 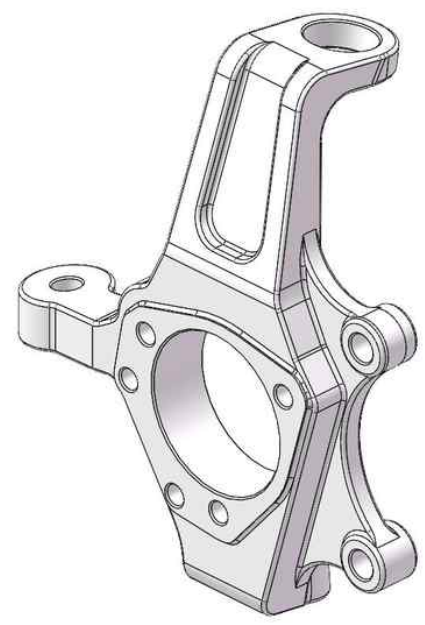 | 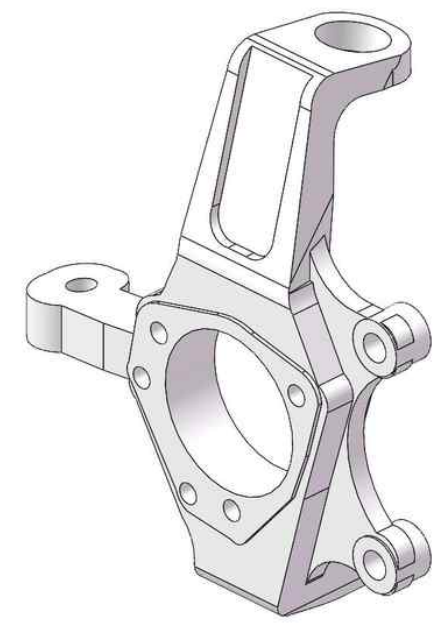 |
| **Reference model (axonometric view)** | **Generated model (axonometric view)** |
| 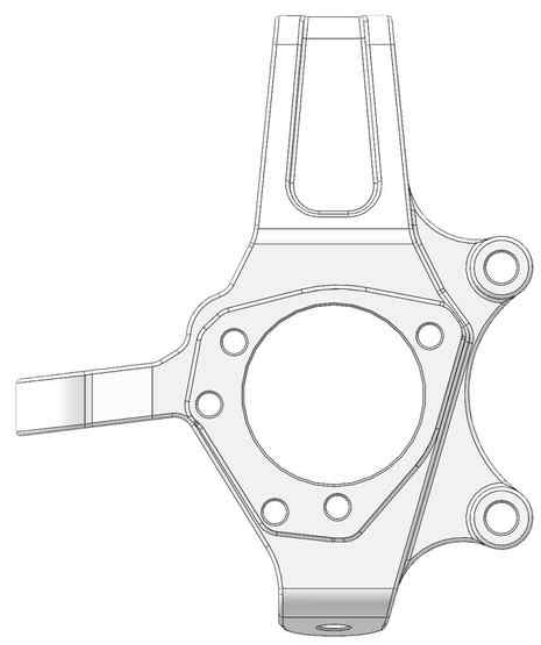 | 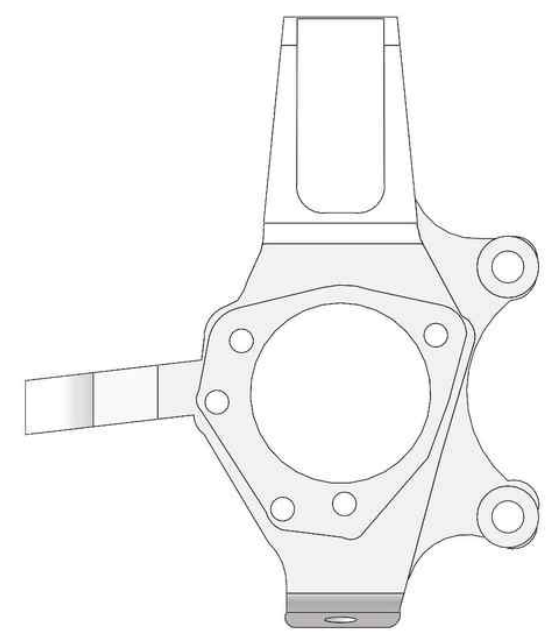 |
| **Reference model (front view)** | **Generated model (front view)** |
| 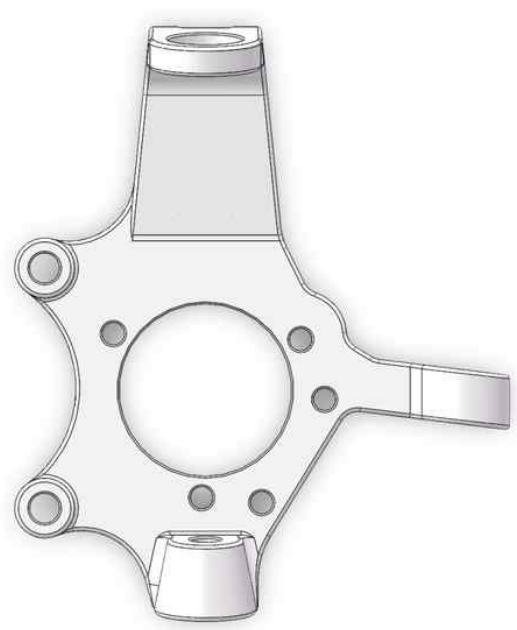 | 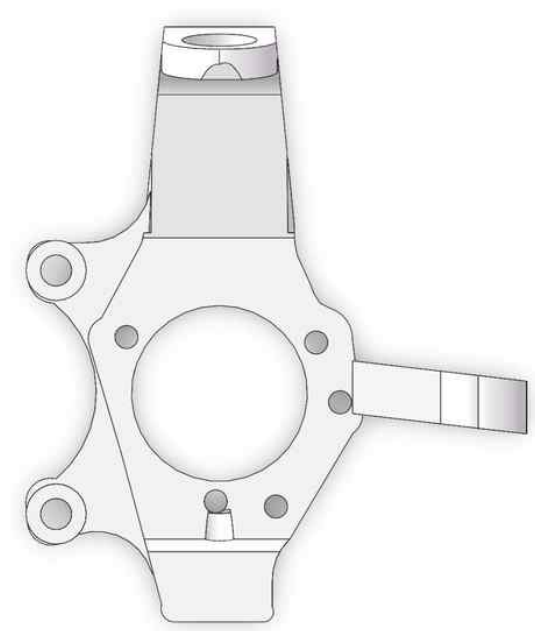 |
| **Reference model (rear view)** | **Generated model (rear view)** |
| 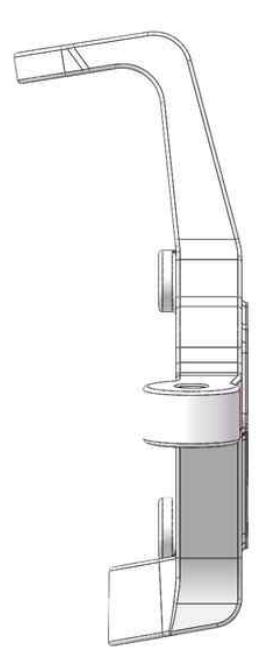 | 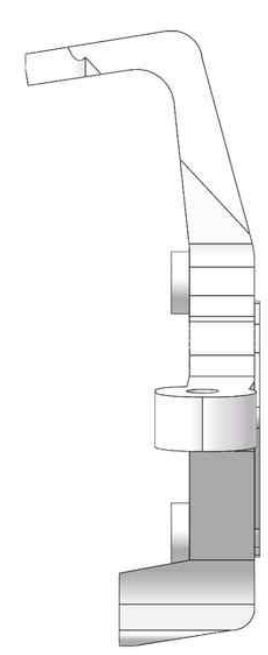 |
| **Reference model (left view)** | **Generated model (left view)** |

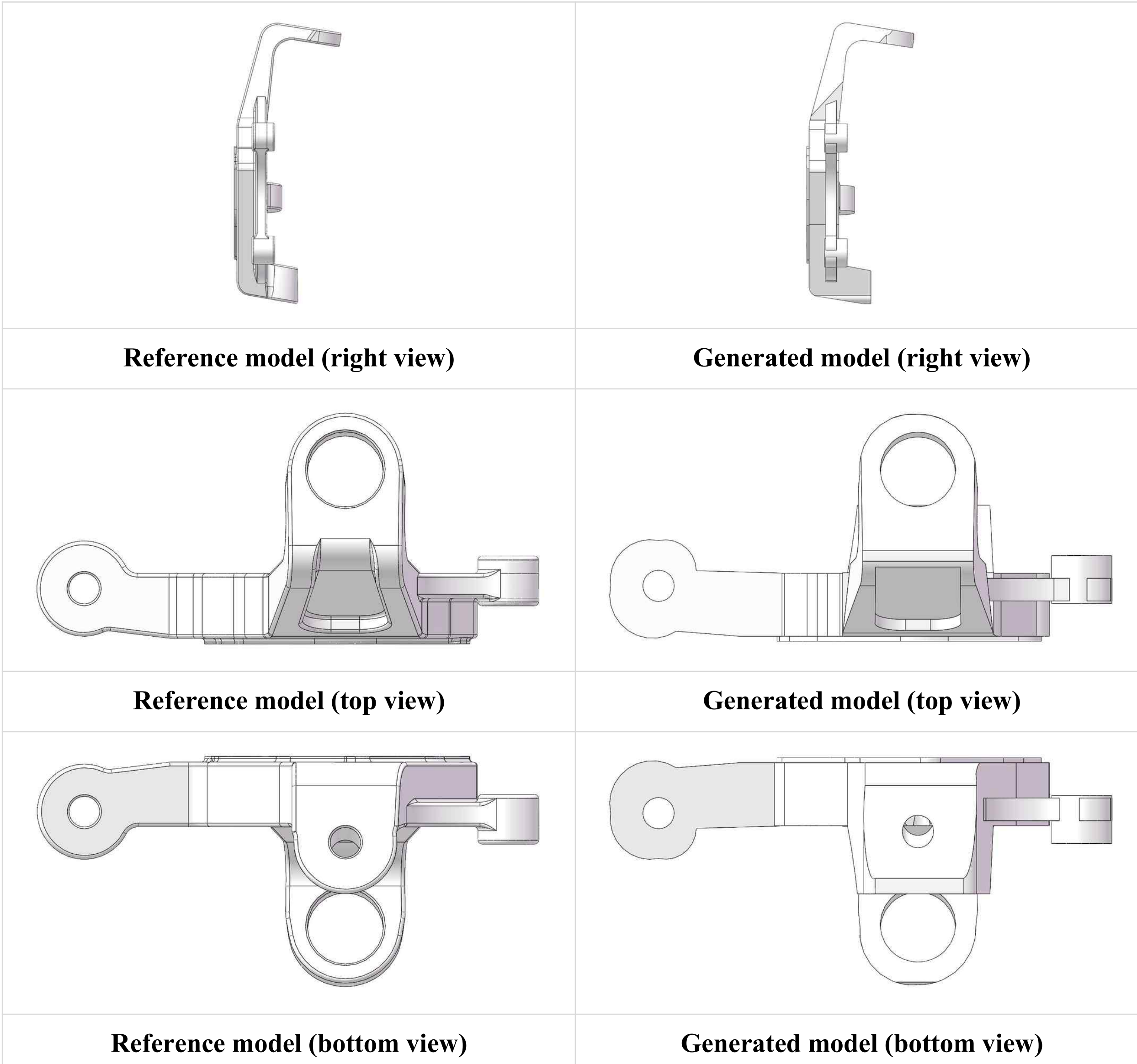

Figure 6-12 Renderings of the reference and generated models for I2C-04

The generated I2C-04 model is a valid solid. It includes the central holes, side mounting lugs, and upper curved arm, with the main features and connections broadly matching the reference. Steps around the mounting holes, the bottom profile, and edge fillets are simplified. The longest side of the generated model's bounding box is 237.03 mm, compared with 317.22 mm for the reference, so the model is undersized. HD is 45.211 mm, and hole locations, wall thicknesses, and mounting dimensions still differ.

Figure 6-13 plots silhouette IoU for each view against iteration number.

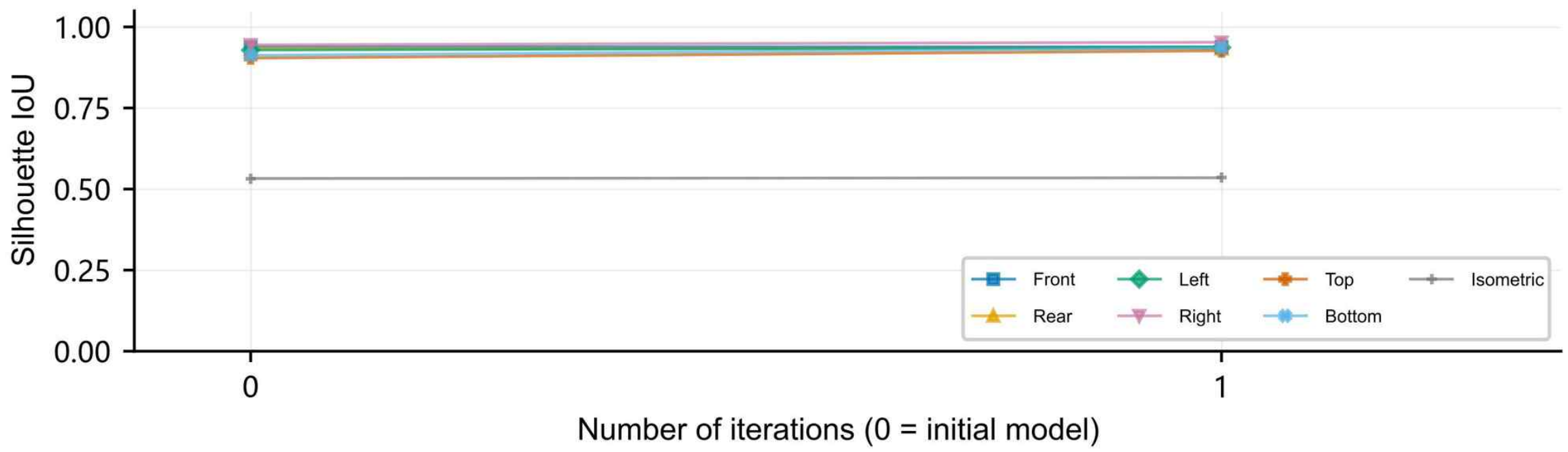

Figure 6-13 Silhouette IoU by iteration for I2C-04

## I2C-05 GE jet engine bracket IVO (complex)

### Input renderings

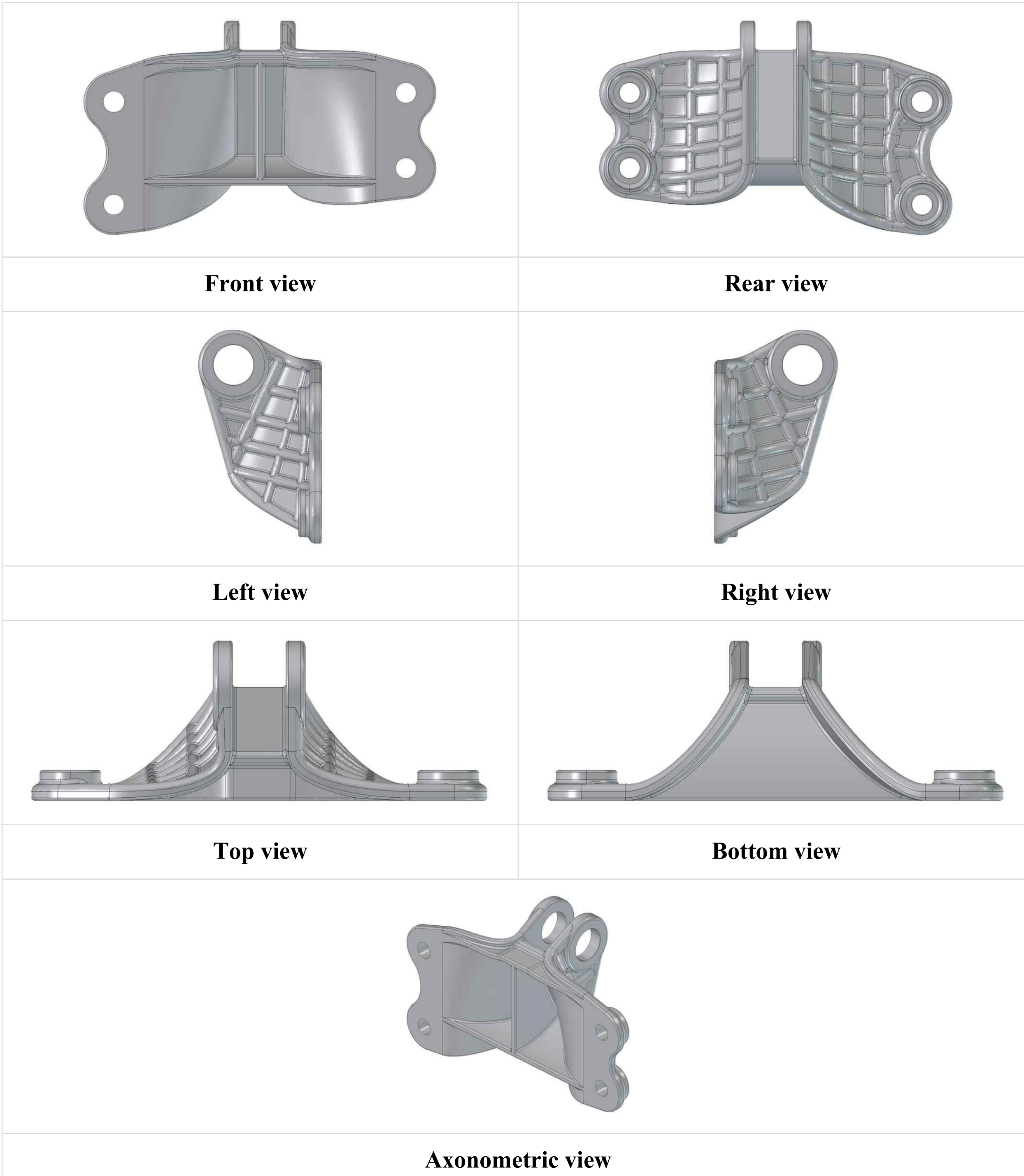

Figure 6-14 Seven input renderings for I2C-05

## Results

| | |
|---|---|
| 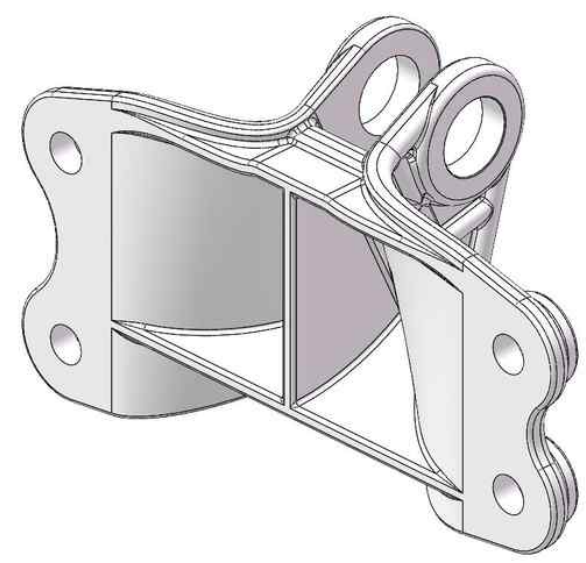 | 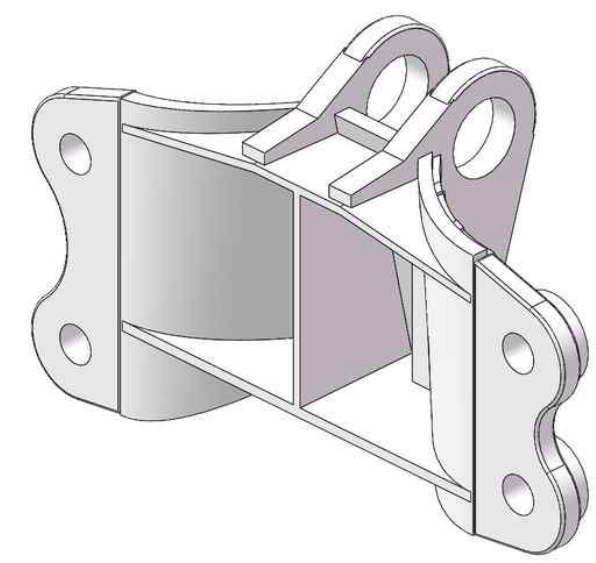 |
| **Reference model (axonometric view)** | **Generated model (axonometric view)** |
| 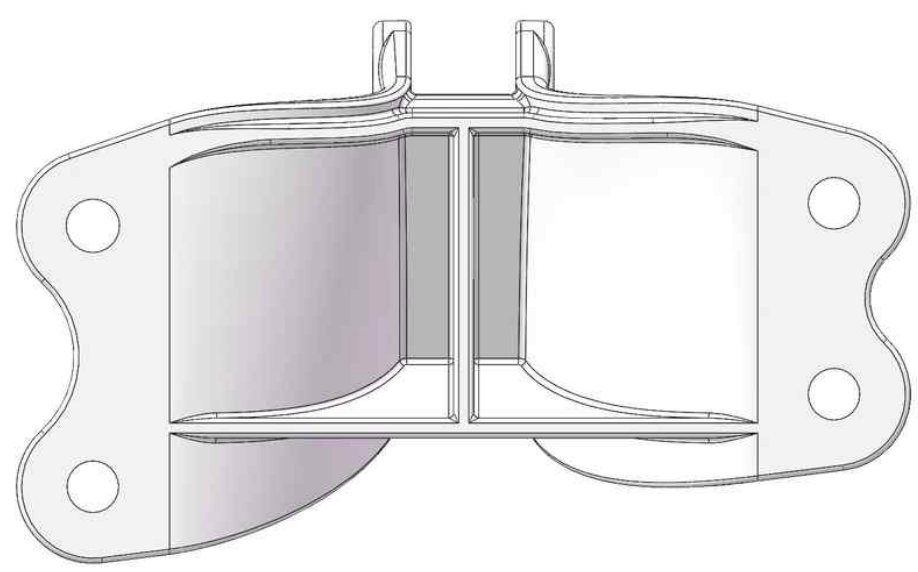 | 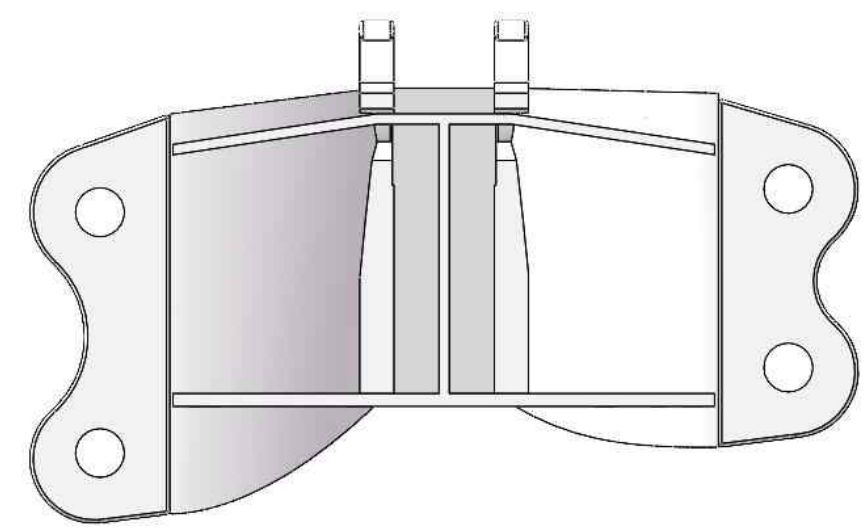 |
| **Reference model (front view)** | **Generated model (front view)** |
| 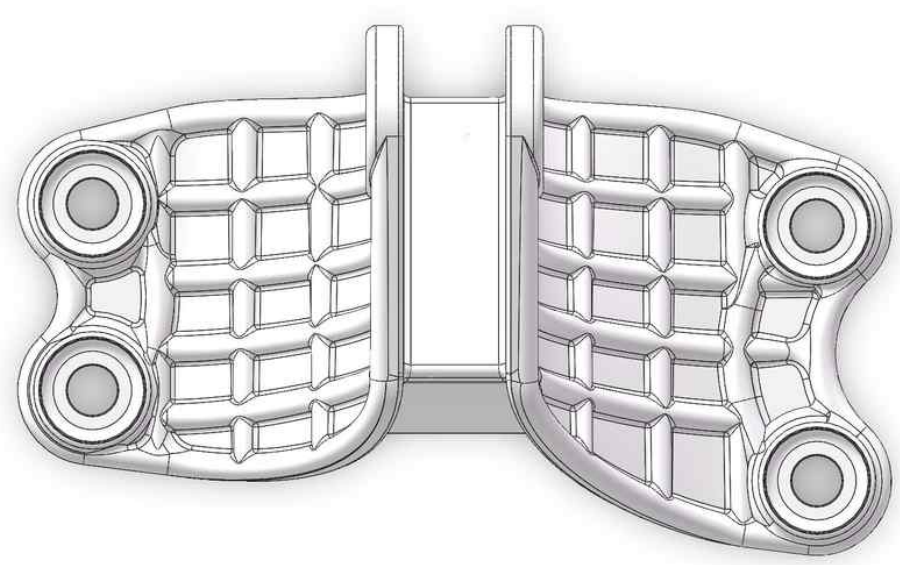 | 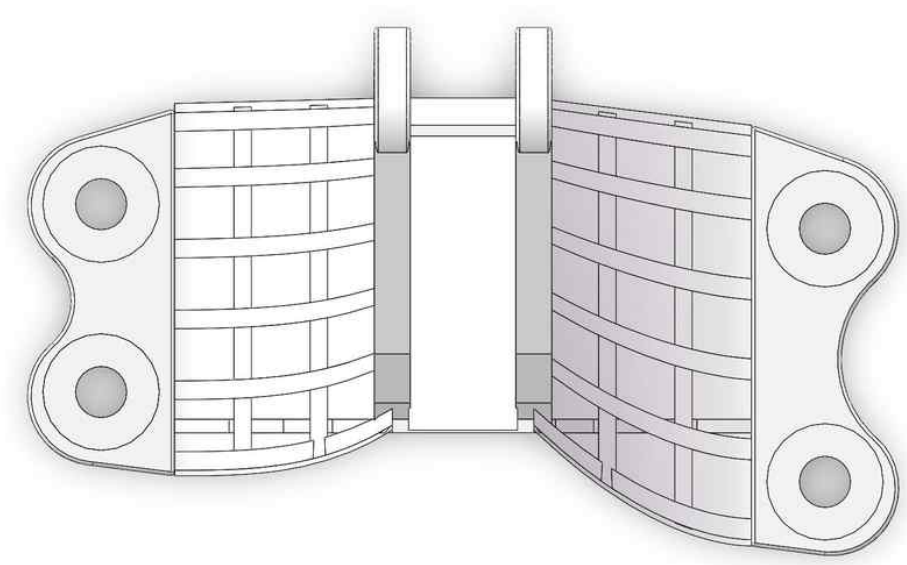 |
| **Reference model (rear view)** | **Generated model (rear view)** |
| 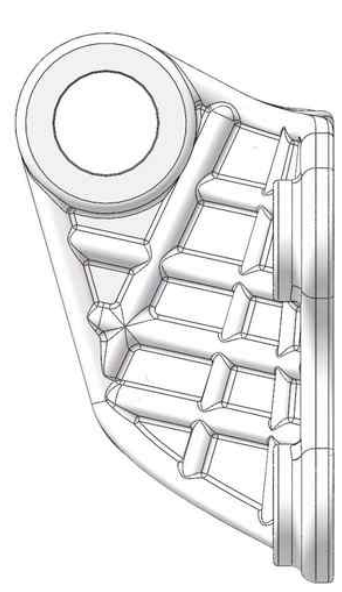 | 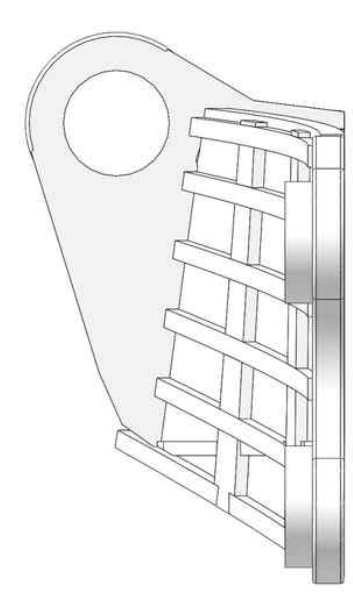 |
| **Reference model (left view)** | **Generated model (left view)** |
| 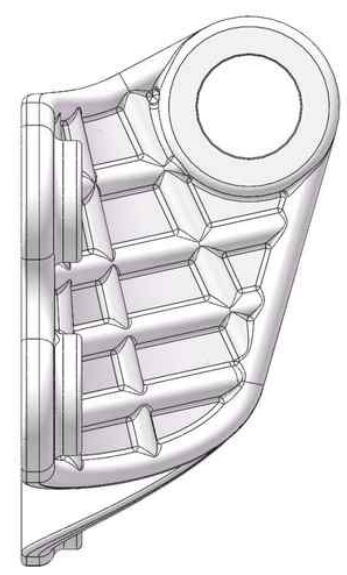 | 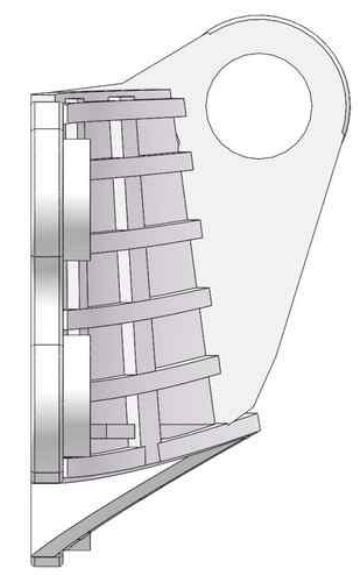 |
| **Reference model (right view)** | **Generated model (right view)** |
| 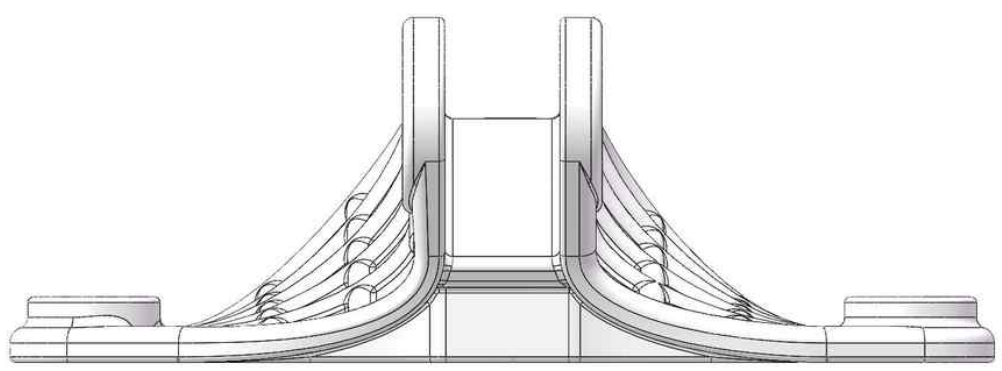 | 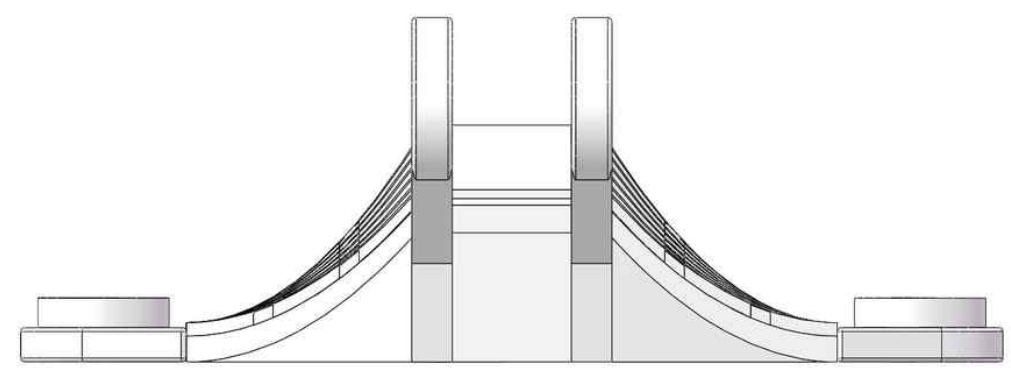 |
| **Reference model (top view)** | **Generated model (top view)** |
| 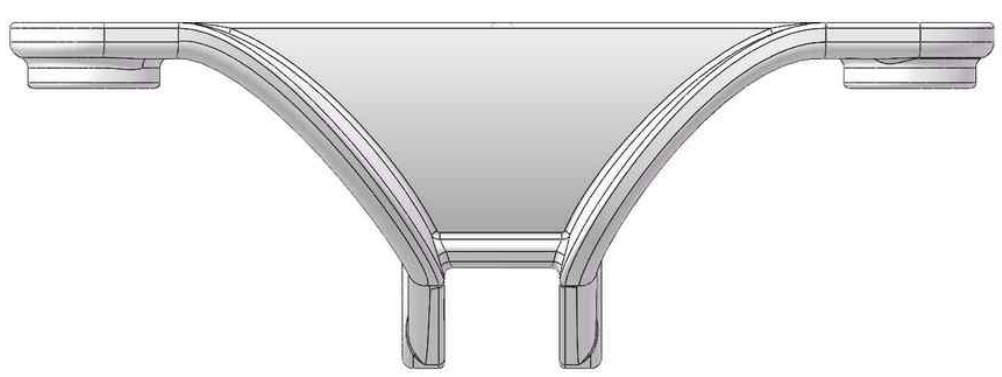 | 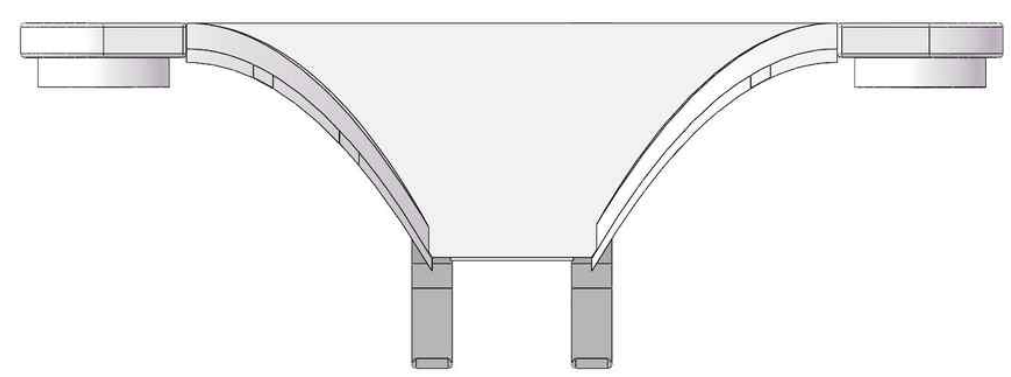 |
| **Reference model (bottom view)** | **Generated model (bottom view)** |

Figure 6-15 Renderings of the reference and generated models for I2C-05

The generated I2C-05 model is a valid solid. It includes the asymmetric mounting lugs, holes through the paired lugs, and main supporting structure. Transitions between the wing surfaces and lug bases are not sufficiently smooth. Rear reinforcing ribs are simplified to strips with rectangular cross-sections, and fillets at rib intersections and roots are incomplete, as shown below. HD is 9.959 mm. The longest side of the generated model's bounding box is approximately 160.07 mm, compared with 178.58 mm for the reference. Both local surfaces and overall dimensions differ.

| | |
|---|---|
| 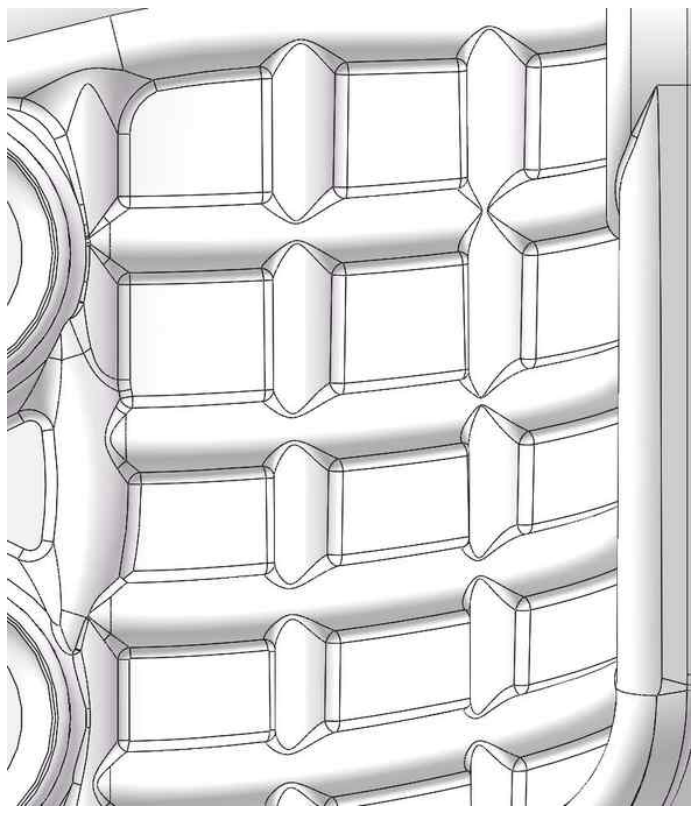 | 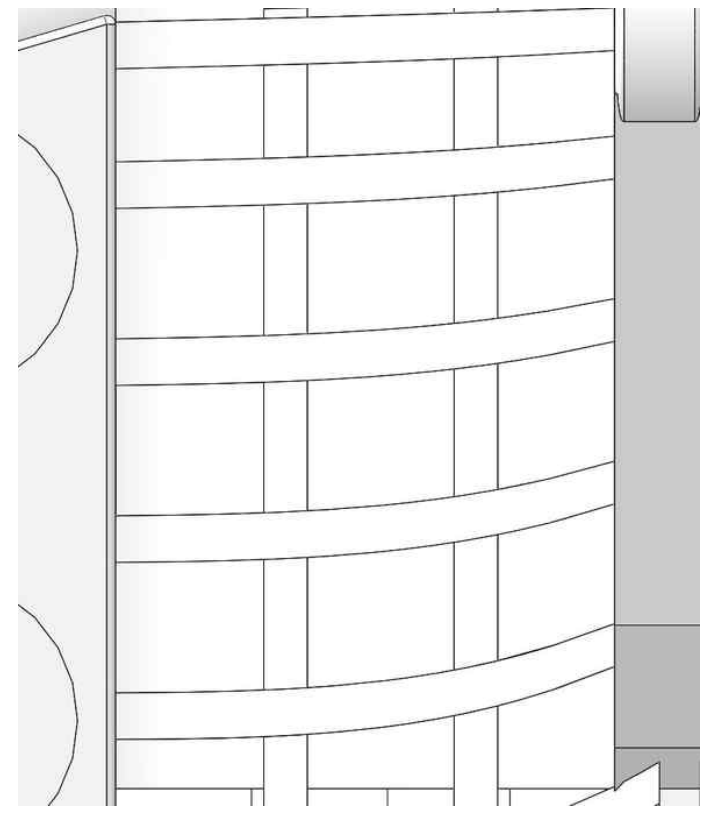 |
| **Reference model: close-up** | **Generated model: close-up** |

Figure 6-16  Rear reinforcing ribs and fillets in I2C-05

Figure 6-17 plots silhouette IoU for each view against iteration number.

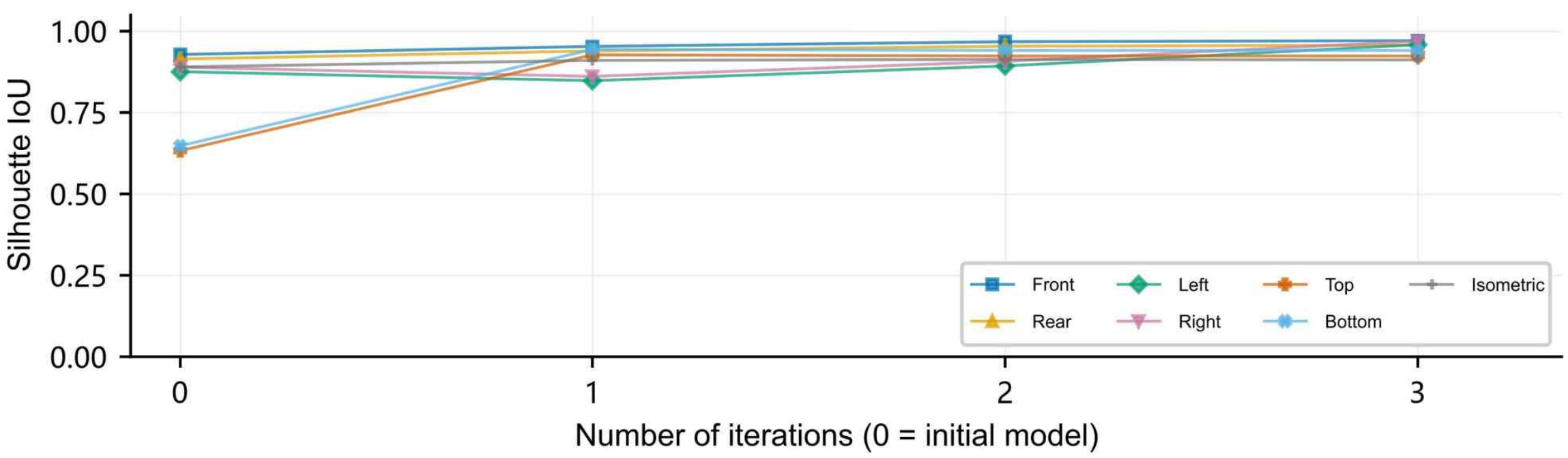

Figure 6-17  Silhouette IoU by iteration for I2C-05

## I2C-06 Hammerhead robot chassis (complex)

### Input renderings

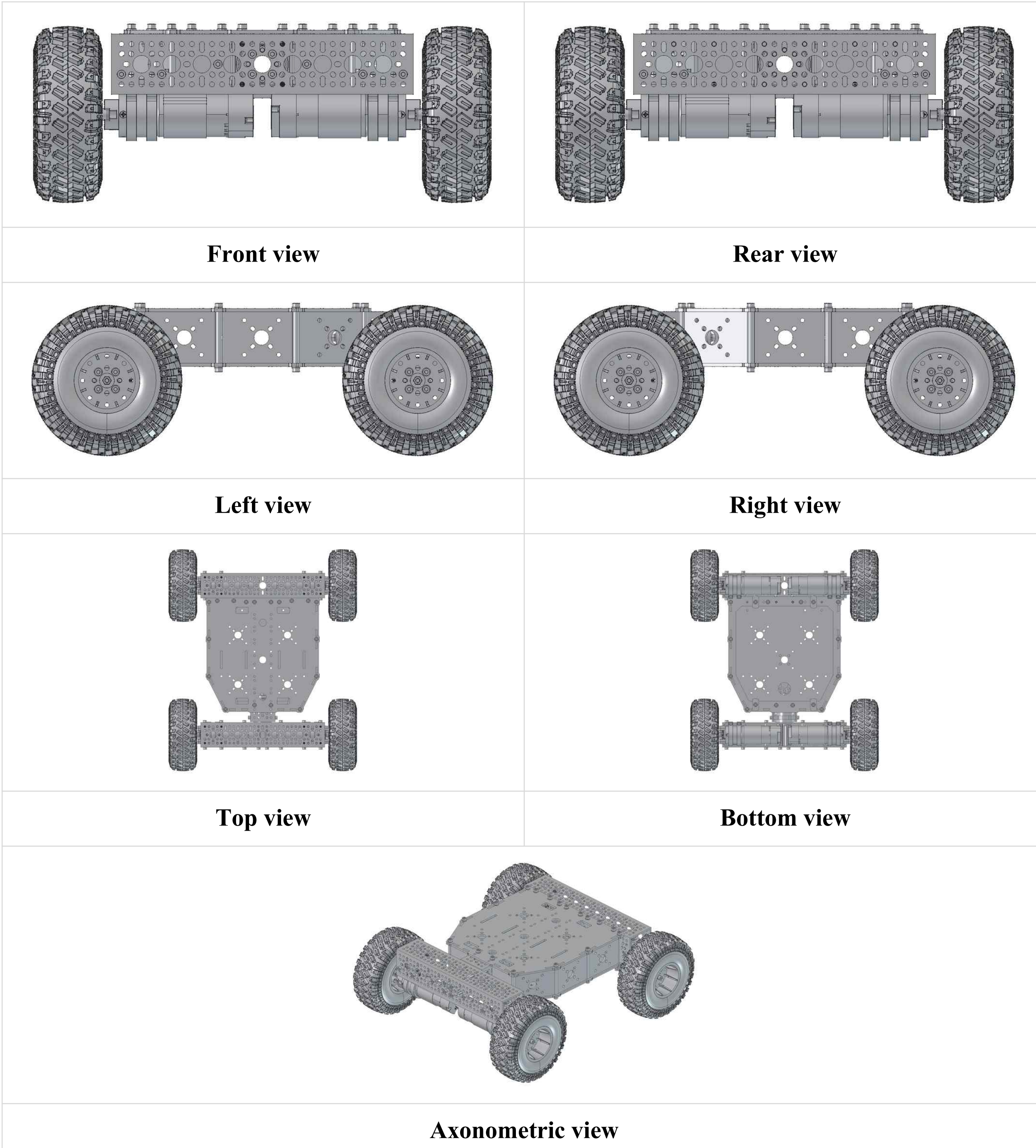

Figure 6-18 Seven input renderings for I2C-06

## Results

| | |
|---|---|
| 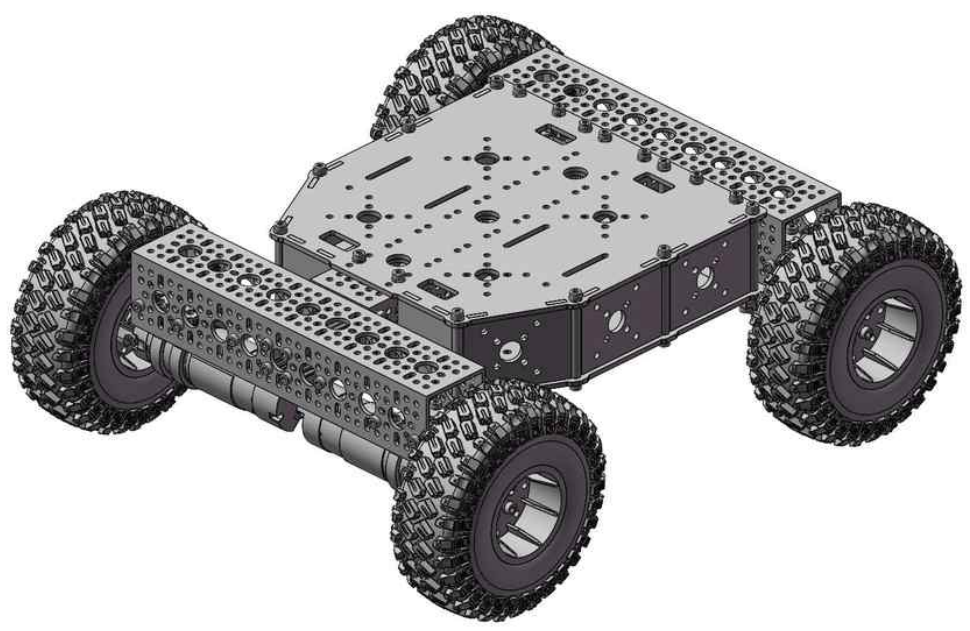 | 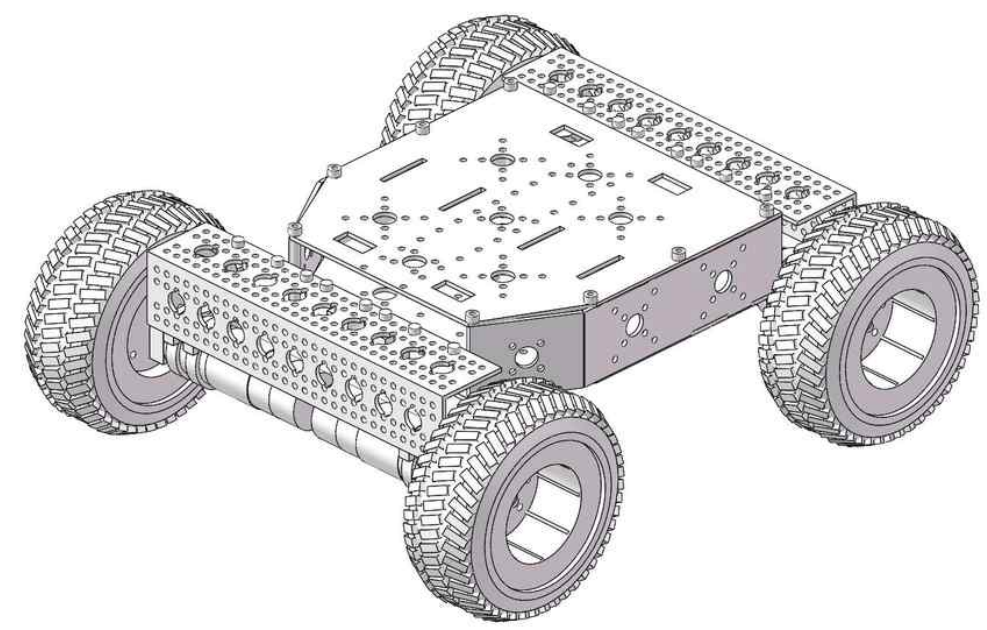 |
| **Reference model (axonometric view)** | **Generated model (axonometric view)** |
| 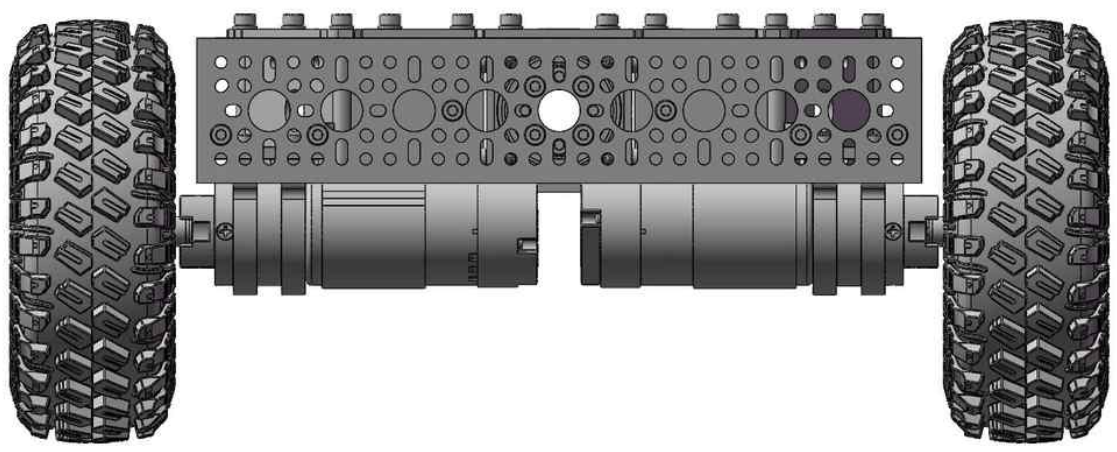 | 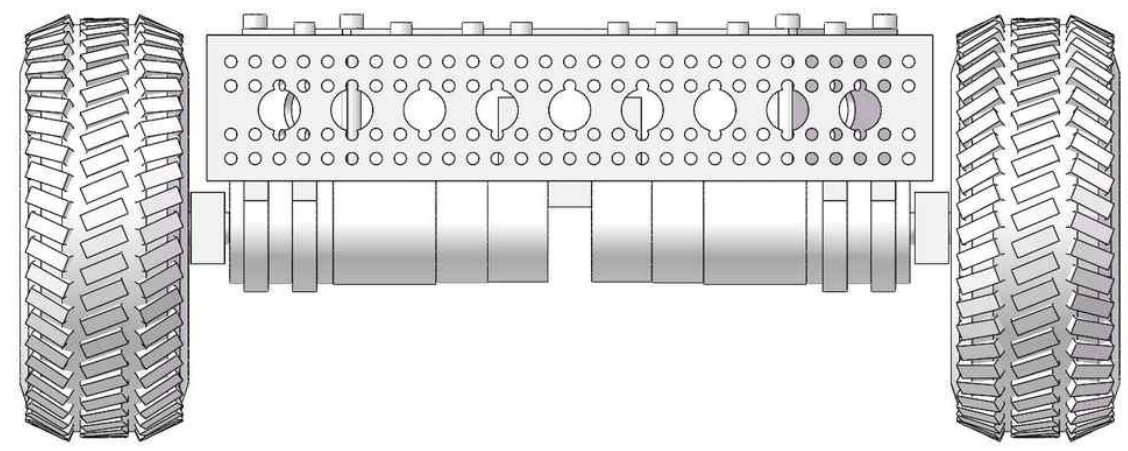 |
| **Reference model (front view)** | **Generated model (front view)** |
| 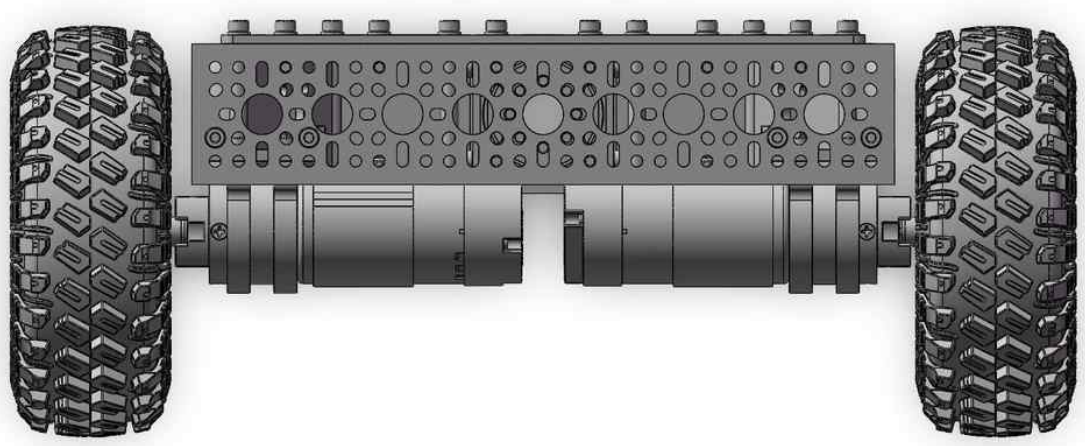 | 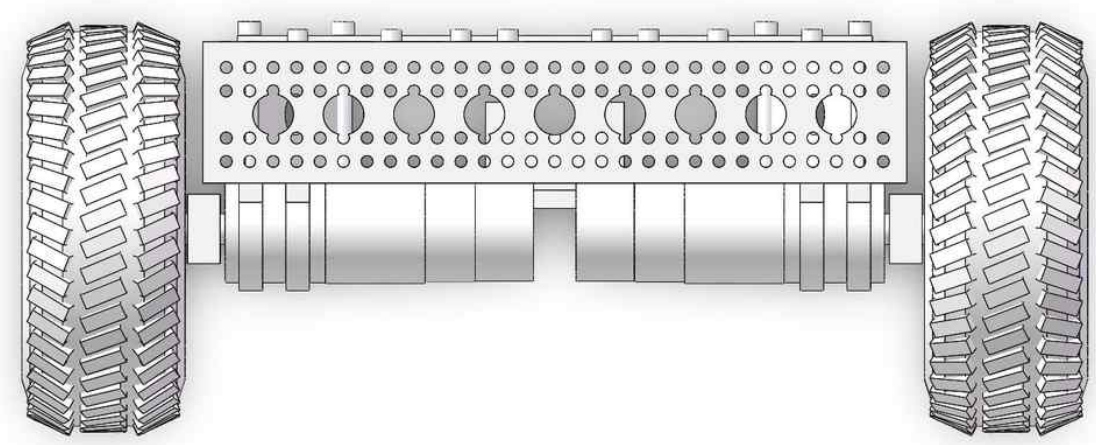 |
| **Reference model (rear view)** | **Generated model (rear view)** |
| 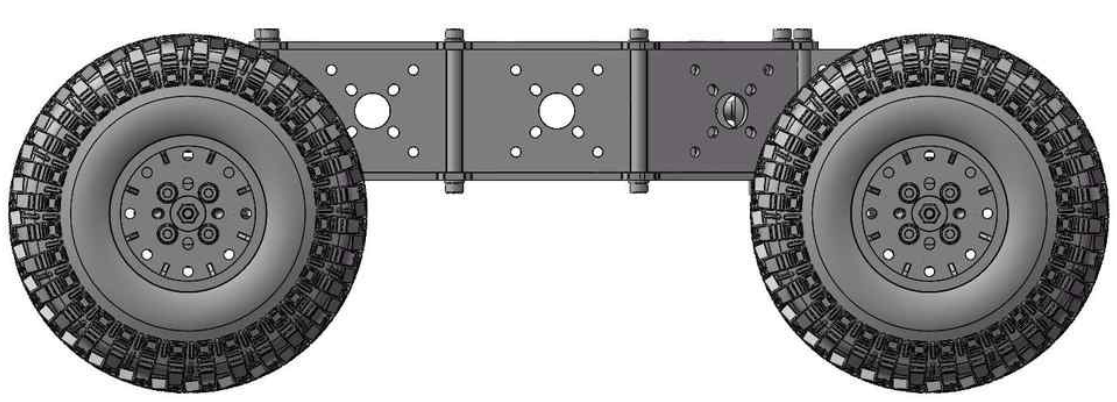 | 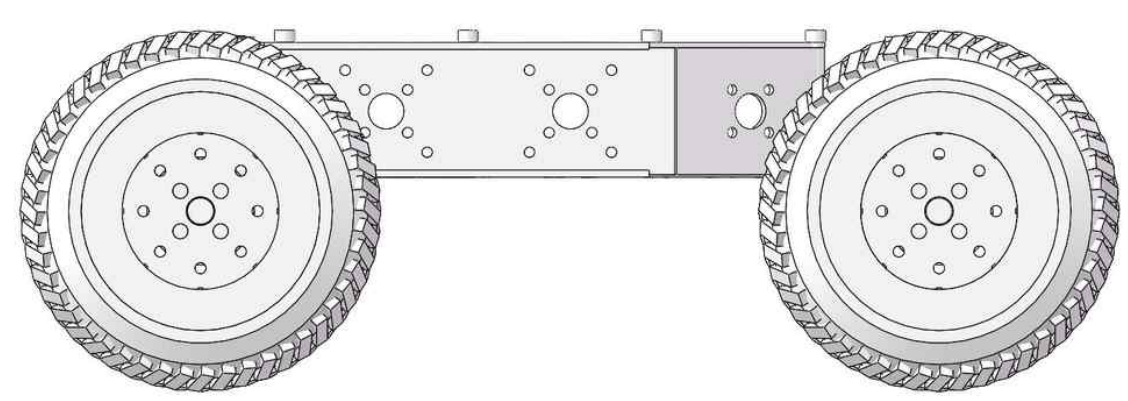 |
| **Reference model (left view)** | **Generated model (left view)** |
| 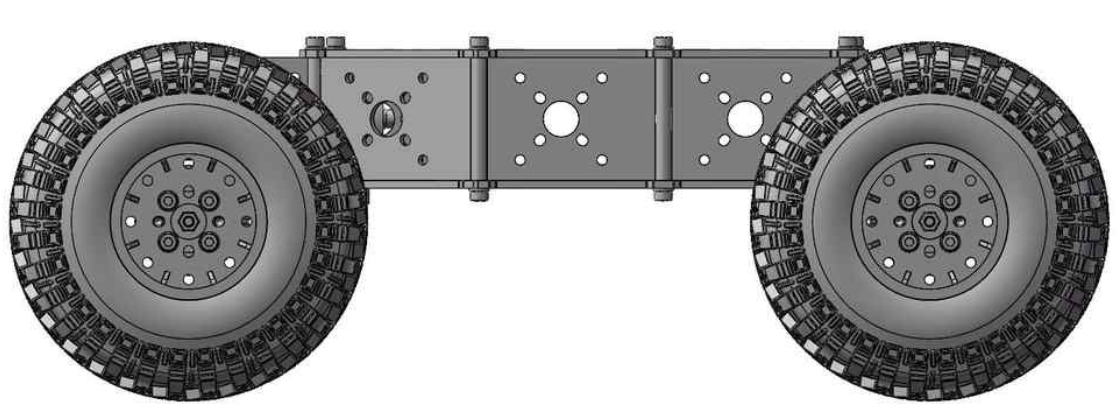 | 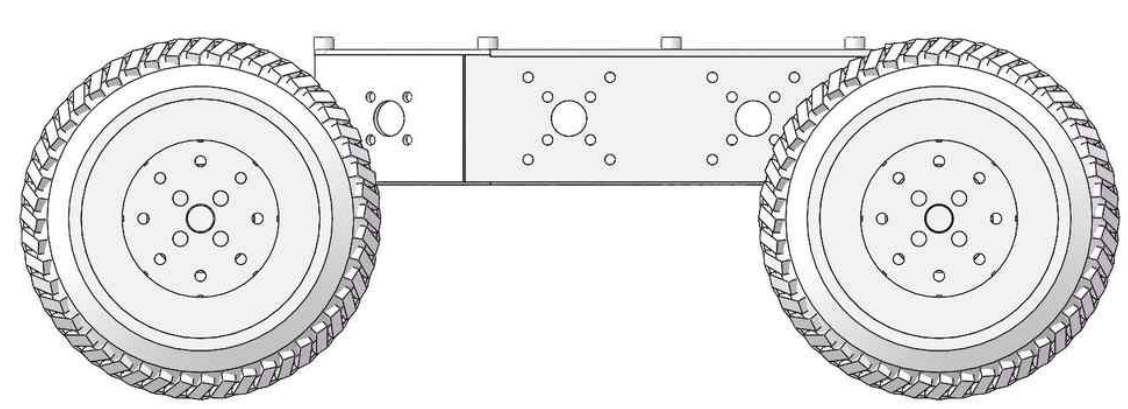 |
| **Reference model (right view)** | **Generated model (right view)** |
| 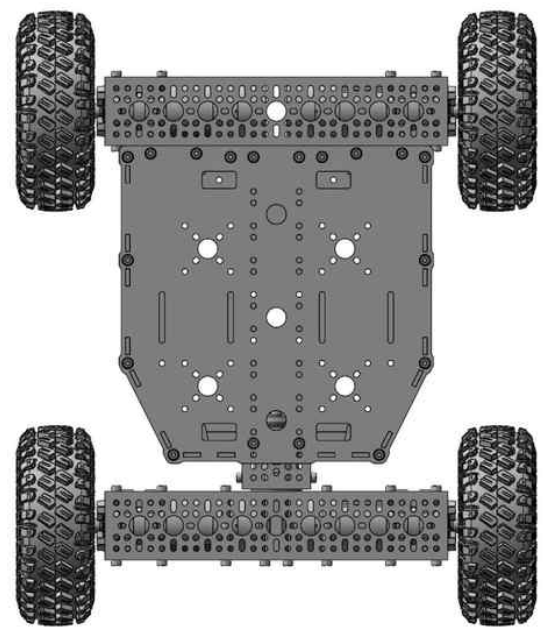 | 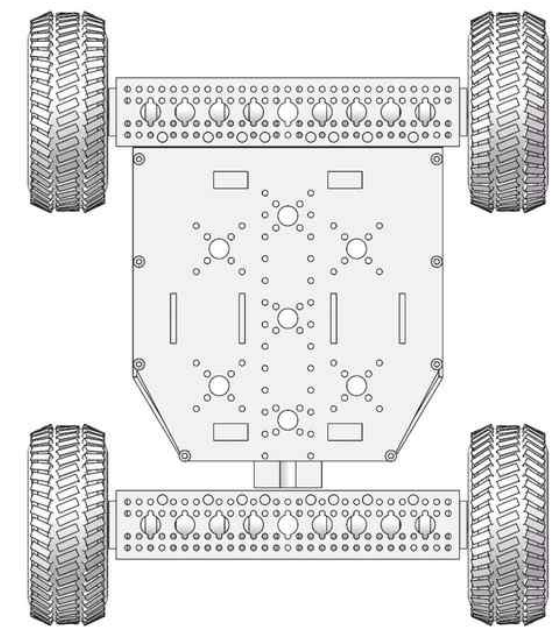 |
| **Reference model (top view)** | **Generated model (top view)** |
| 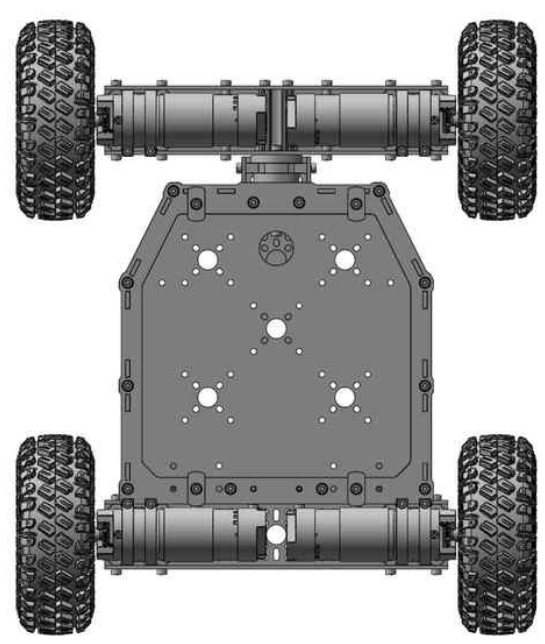 | 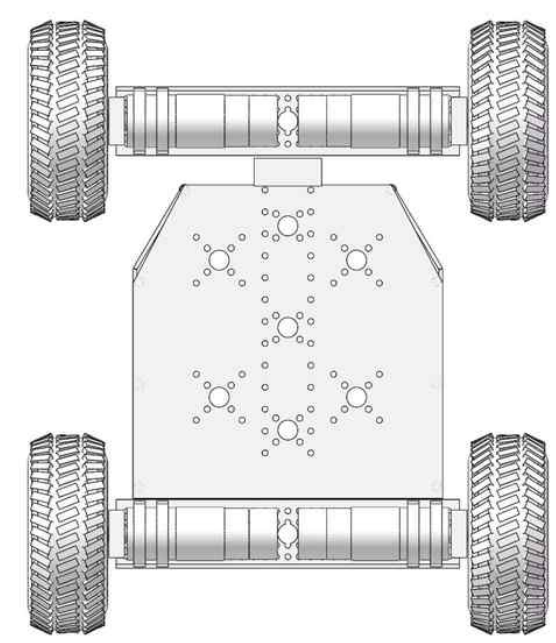 |
| **Reference model (bottom view)** | **Generated model (bottom view)** |

Figure 6-19 Renderings of the reference and generated models for I2C-06

The generated I2C-06 model contains 721 solids, all of which pass the basic validity checks. The overall arrangement of the four wheels, plates, and front and rear crossmembers is close to the reference. Wheel hubs, hole patterns in the plates, support locations, and connection details still differ; correspondence between the densely packed components has not been checked individually. The generated chassis has a maximum extent of approximately 308.47 mm, compared with 429.45 mm for the reference. It is undersized, with an HD of 80.297 mm. The main chassis components have been reconstructed, but the number of solids cannot be treated directly as the number of assembly parts.

Figure 6-20 plots silhouette IoU for each view against iteration number.

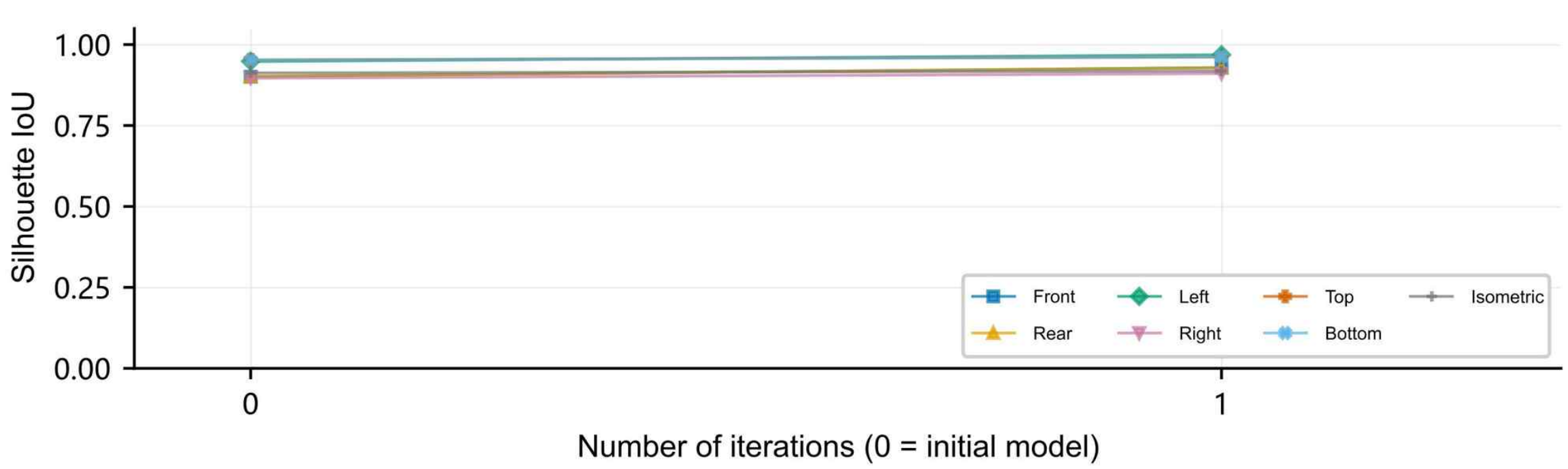

Figure 6-20 Silhouette IoU by iteration for I2C-06

### 6.3.2 Additional experiment results

The following results cover three-view inputs, six-view inputs, disabled web search, and a stress test using a complex housing. S1 and S2 are compared with the seven-view main experiment for the same cases, focusing on changes to the main structure, local features, and model dimensions when fewer views are supplied. HD is calculated against the same reference STEP file using the same method. Each case includes an axonometric view of the reference and one of the generated model.

Mean IoU is calculated over the views available in each condition: seven in the main experiment, three in S1, and six in S2. These means summarize different view sets and should not be treated as directly comparable scores on a common set of views. The paired case analyses and HD values provide complementary evidence, subject to the scale limitation described in Section 6.1.

#### 6.3.2.1 S1: Reduced view count

This group uses only the front, right, and top orthographic views. Web search remains enabled, and the prompt is unchanged from the main experiment. Table 6-7 summarizes the results.

Table 6-7 Results with three input views (S1)

| Case ID | Validity of each solid | Mean IoU (three views) | HD (mm) | IoU by iteration | Main differences from the reference |
|---|---|---|---|---|---|
| I2C-01 | Pass | 0.9912 | 15.147 | 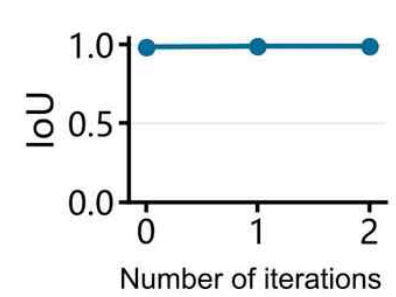 | The centers of the clips and corner hole bosses differ; overall scale is too large. |
| I2C-02 | Pass | 0.9910 | 27.924 | 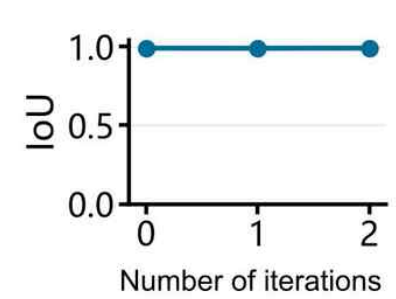 | The lower edge of the opening is incorrectly closed; overall dimensions exceed the reference. |
| I2C-03 | Pass | 0.9123 | 112.981 | 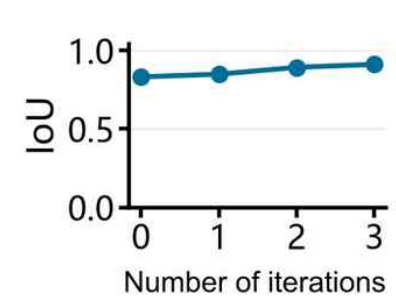 | The bushings and main body form 3 separate solids; connections are simplified, and overall scale is substantially too small. |
| I2C-04 | Pass | 0.9821 | 21.494 | 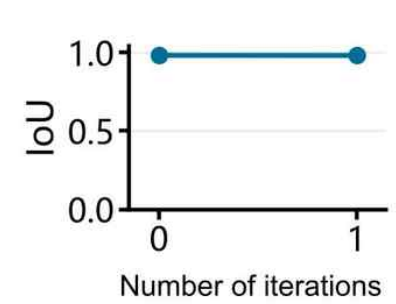 | Connecting walls are faceted; overall dimensions are close, but local transitions remain simplified. |
| I2C-05 | Pass | 0.9514 | 14.098 | 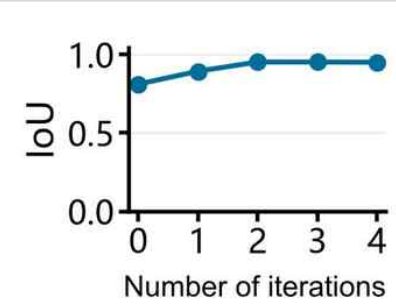 | Front ribs run in a different direction; wing-surface transitions are simplified. |

| Case ID | Validity of each solid | Mean IoU (three views) | HD (mm) | IoU by iteration | Main differences from the reference |
|---|---|---|---|---|---|
| I2C-06 | Pass | 0.9378 | 78.811 | 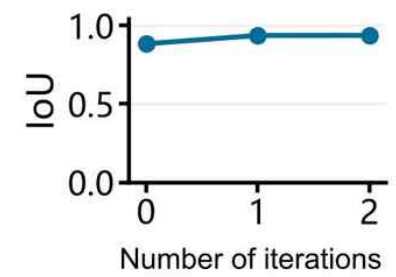 | Chassis layout is retained; scale is too small, and hole patterns and connection details differ. |

Table 6-8 Time taken for each case with three input views (S1)

| Case ID | I2C-01 | I2C-02 | I2C-03 | I2C-04 | I2C-05 | I2C-06 |
|---|---|---|---|---|---|---|
| Time | 13 min | 13 min | 19 min | 10 min | 24 min | 20 min |

All six cases in the three-view group pass the validity checks for individual solids and largely retain their overall shapes. However, I2C-03 consists of separate solids: its bushings overlap the main body, and the connections have not been combined into a single solid as in the reference part.

Differences are concentrated in local structures such as side clips, connections at hole bosses, and curved ribs. Compared with the main experiment, HD for I2C-03 increases from 62.401 to 112.981 mm, while HD for I2C-02 and I2C-04 decreases to 27.924 and 21.494 mm, respectively. The latter reductions are mainly associated with overall dimensions closer to the reference; local surfaces and connection features still differ.

### I2C-01 TrackerOne backplate (simple)

| 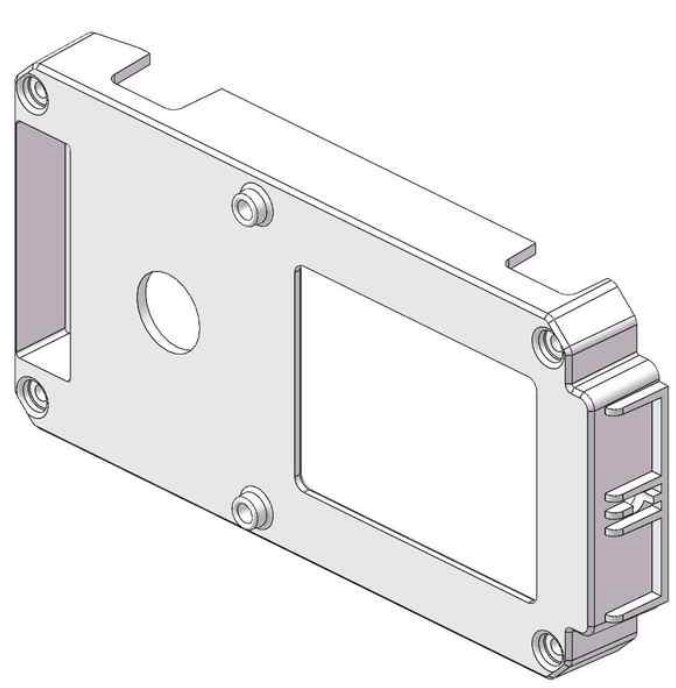 | 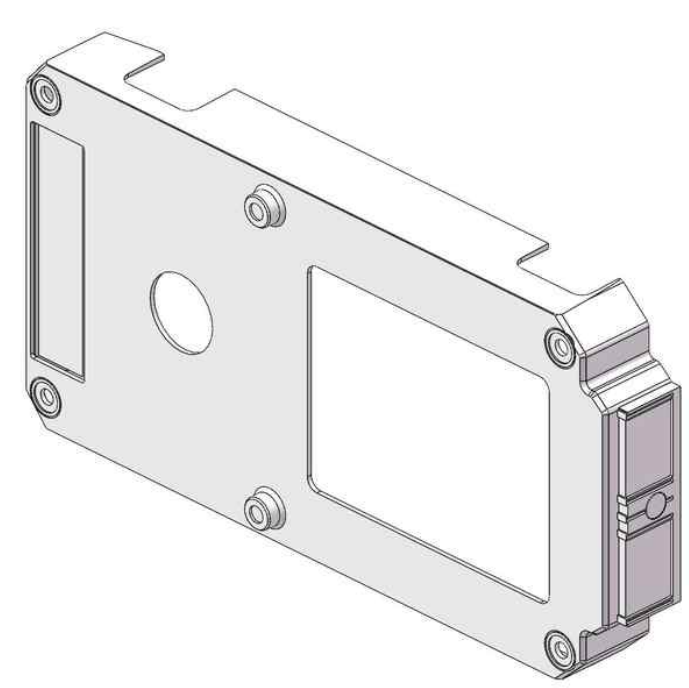 |
|---|---|
| **Reference CAD model** | **S1 generated model** |

Figure 6-21 Reference and generated models for I2C-01 with three input views

With three input views, the backplate’s window, circular holes, and perimeter flanges are reconstructed. The side clips become segmented structures with circular recesses, while the corner hole bosses and small transitions differ from the reference. HD is 15.147 mm, close to the main experiment’s 14.758 mm. Both generated models are approximately 158 mm long, compared with 132.11 mm for the reference. Reducing the number of views has little effect on the overall layout, but differences in local structures such as the clips are more pronounced.

### I2C-02 KnowCO2 sensor housing (simple)

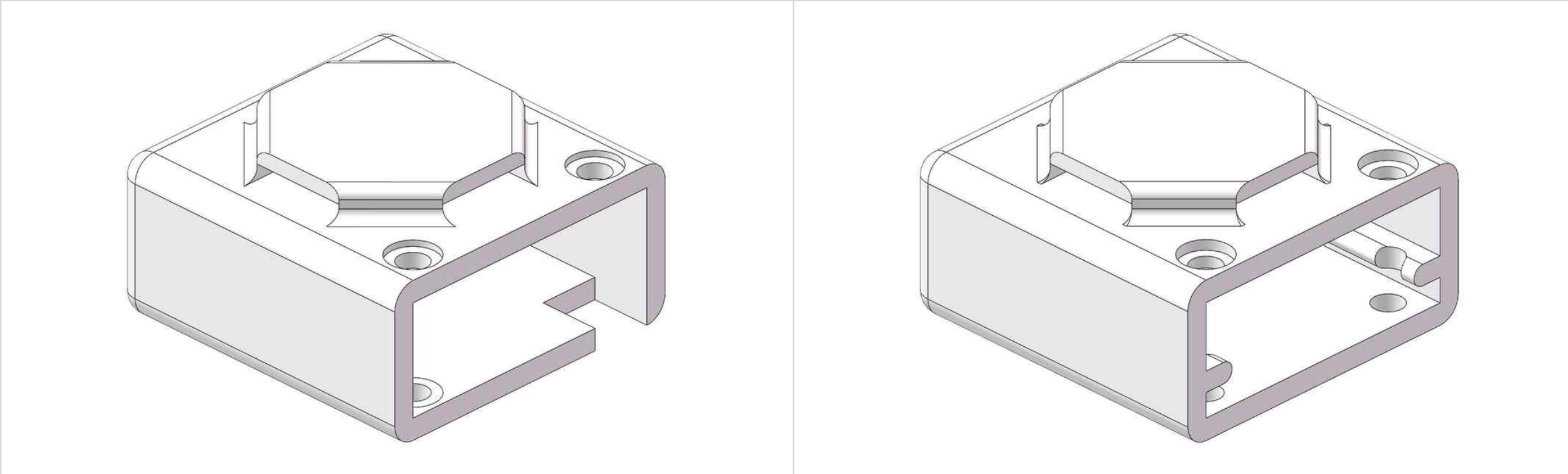

| Reference CAD model | S1 generated model |
| --- | --- |

Figure 6-22 Reference and generated models for I2C-02 with three input views

The three-view result retains the overall shapes of the housing and top platform. However, a continuous lower frame closes the bottom of the side opening, where the reference has an open notch, and the internal guides are not accurately matched. The generated model measures 64.00 × 62.25 × 36.00 mm. It remains larger than the reference but is closer in size than the main-experiment result; HD falls to 27.924 mm. This decrease largely accompanies the reduction in dimensional error. The continuous lower frame occupies space left open in the reference and may obstruct insertion and installation of a mating part.

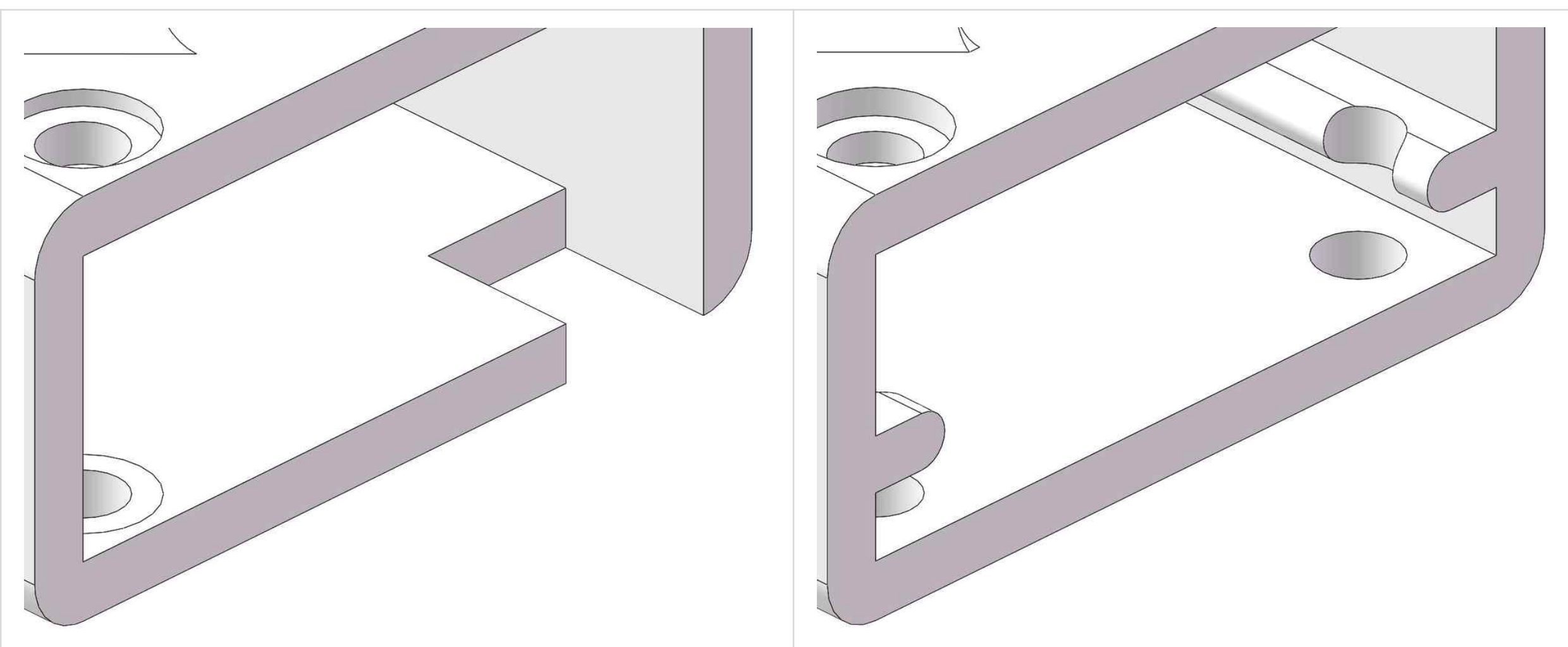

| Reference model: close-up | S1 generated model: close-up |
| --- | --- |

Figure 6-23 Opening boundaries in the three-view result for I2C-02

### I2C-03 C5 Corvette lower control arm (moderate)

| | |
|---|---|
| 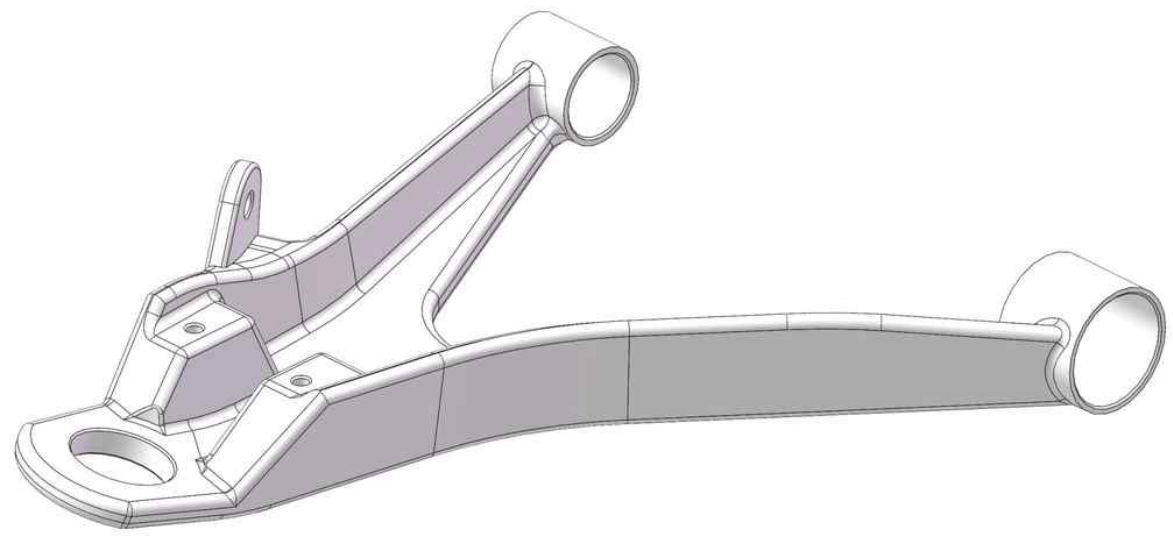 | 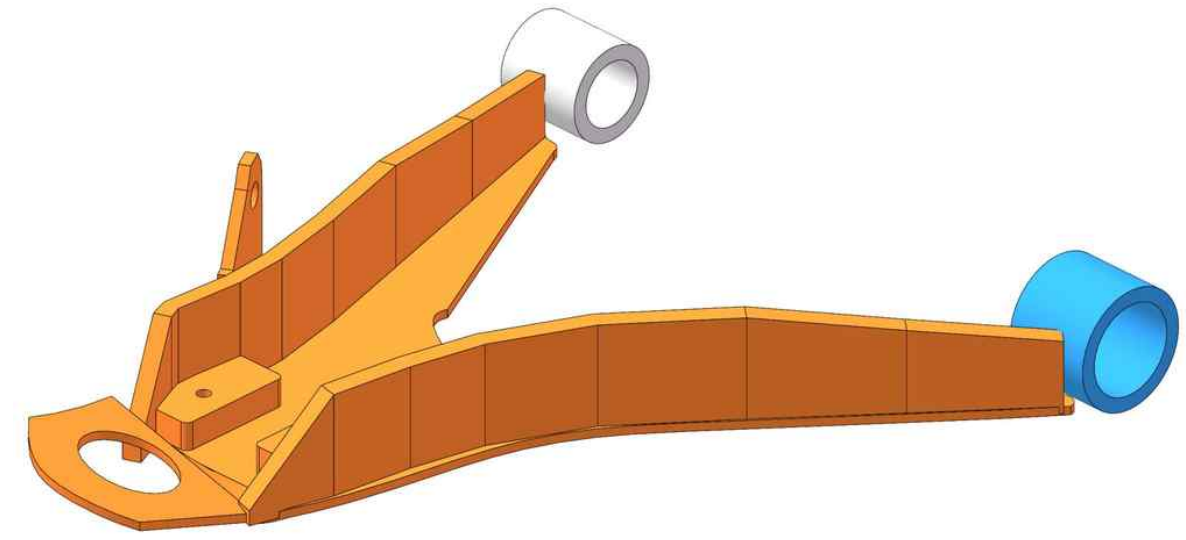 |
| **Reference model (1 solid)** | **S1 generated model (3 solids, shown in different colors)** |

Figure 6-24 Reference and generated models for I2C-03 with three input views

The three-view result includes the forked arms, bushings, and front lug, but the reference's single solid is reconstructed as three separate solids, shown in different colors. Pairwise Boolean intersections show that the two bushings overlap the main body by approximately 0.454 mm³ and 135.024 mm³, respectively. The solids overlap at the connections and have not been combined by a Boolean union. The close-up also shows missing fillets at the bushing roots. Although the individual solids are valid, the topology of the part as a whole differs from the reference. The longest side of the generated model's bounding box is 315.52 mm, below the reference's 488.73 mm, and HD is 112.981 mm.

| | |
|---|---|
| 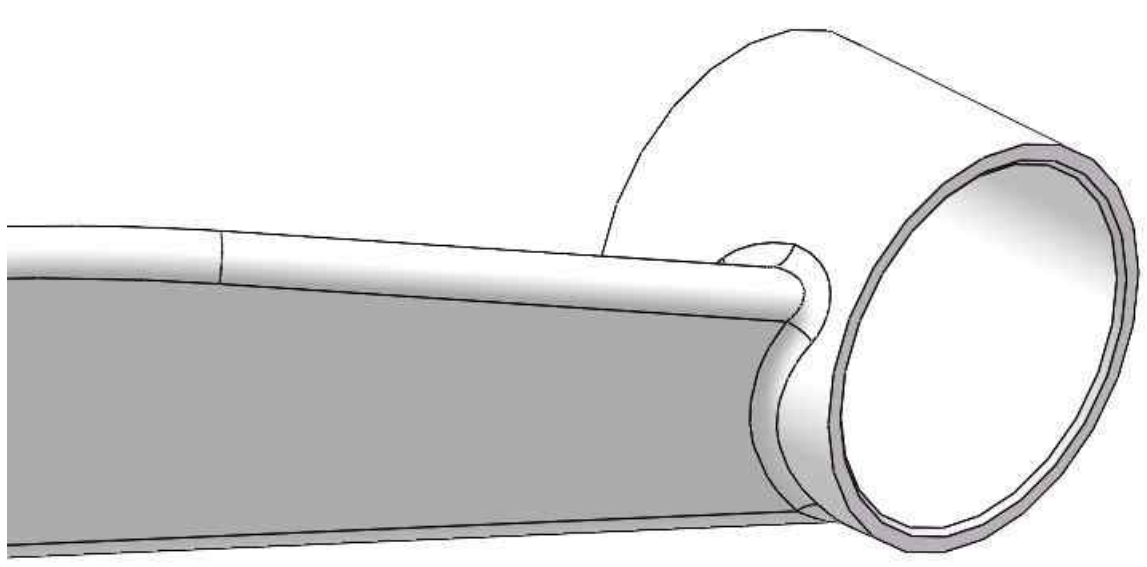 | 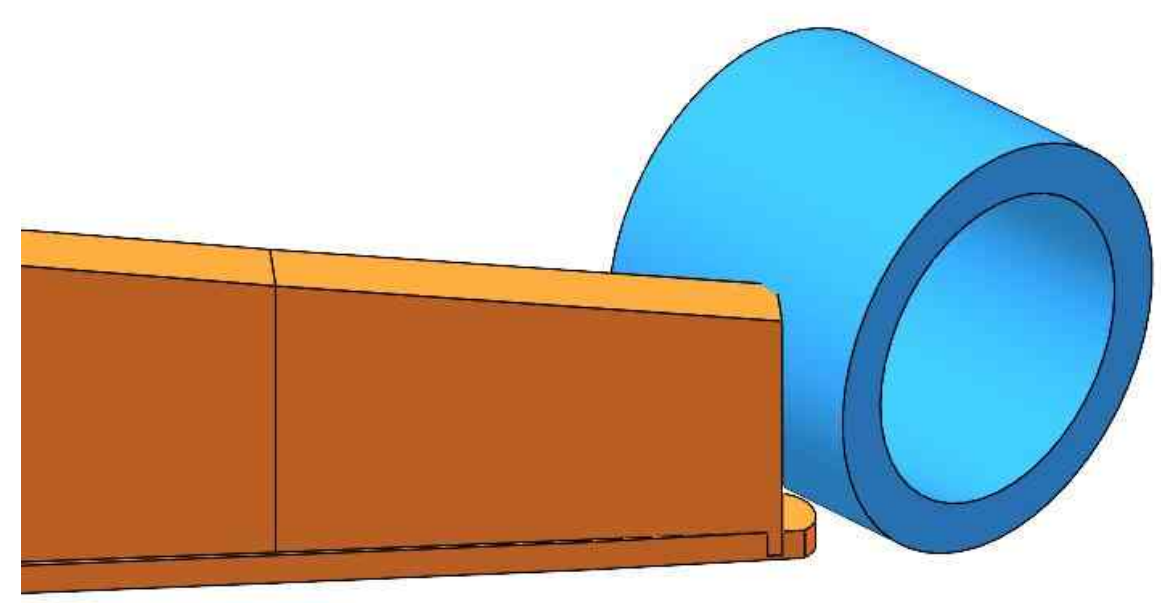 |
| **Reference model: transition within one solid** | **S1 generated model: solids not united** |

Figure 6-25 Bushing connections in the three-view result for I2C-03

### I2C-04 C5 Corvette steering knuckle (moderate)

| 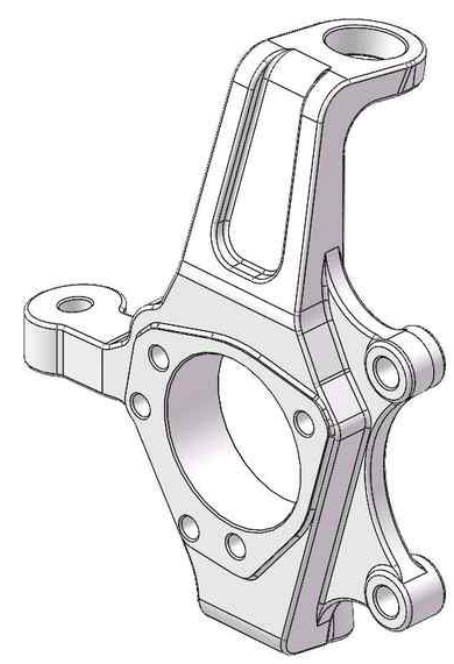 | 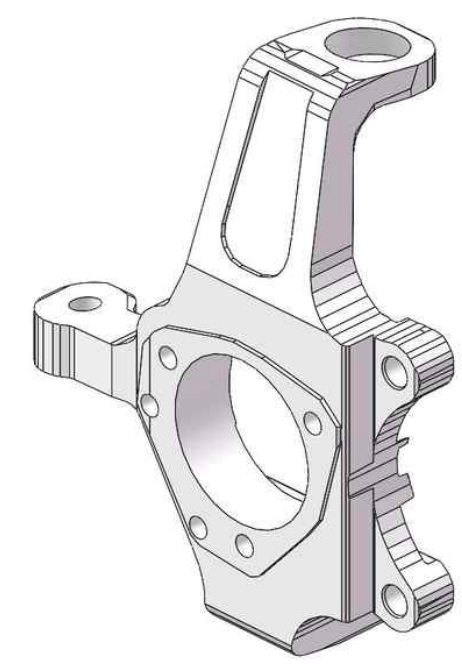 |
|---|---|
| **Reference CAD model** | **S1 generated model** |

Figure 6-26 Reference and generated models for I2C-04 with three input views

The three-view result includes the central holes, upper curved arm, and side mounting lugs. Connecting walls and mounting lugs have faceted surfaces, while steps at the hole openings and casting fillets are incomplete. HD is 21.494 mm, below the main experiment's 45.211 mm. The maximum extent is 315.83 mm, close to the reference's 317.22 mm. Although overall dimensions are closer, errors remain at hole steps, lug connections, and fillet transitions, so the associated contact surfaces still differ from the reference part.

### I2C-05 GE jet engine bracket IVO (complex)

| 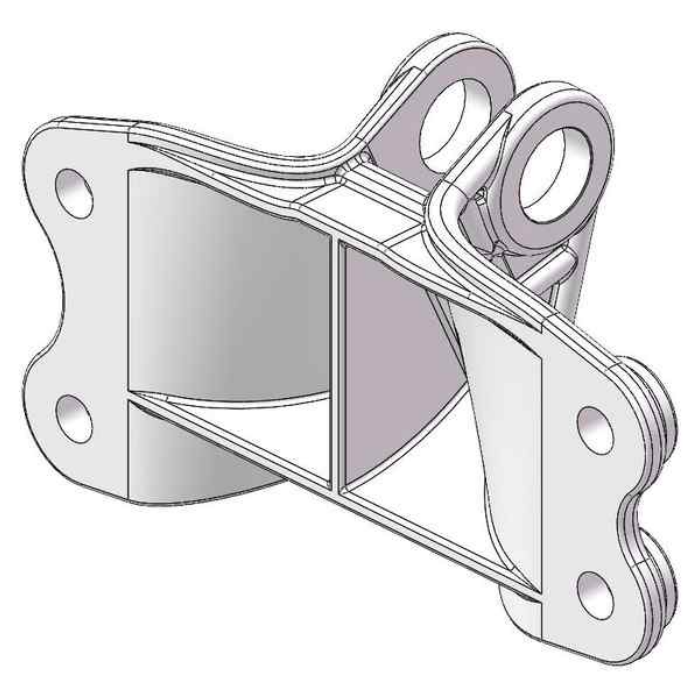 | 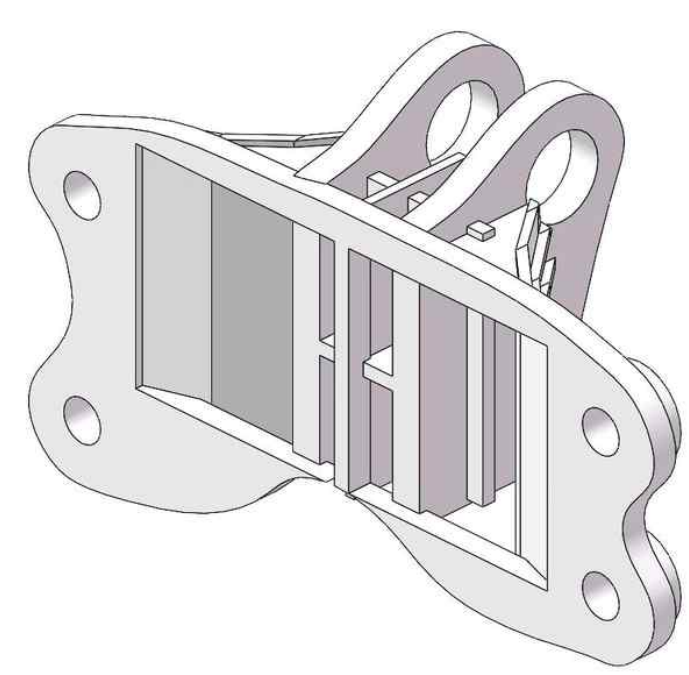 |
|---|---|
| **Reference CAD model** | **S1 generated model** |

Figure 6-27 Reference and generated models for I2C-05 with three input views

With three input views, the asymmetric mounting lugs and holes through the paired lugs are in approximately the correct positions. However, the front transverse ribs are replaced by several straight vertical ribs, and the wing surfaces are segmented, as shown in the close-up below. HD is 14.098 mm, above the main experiment's 9.959 mm, and the overall dimensions remain too small. The complex freeform surfaces at the bottom of the reference model are approximated by a collection of simple surfaces.

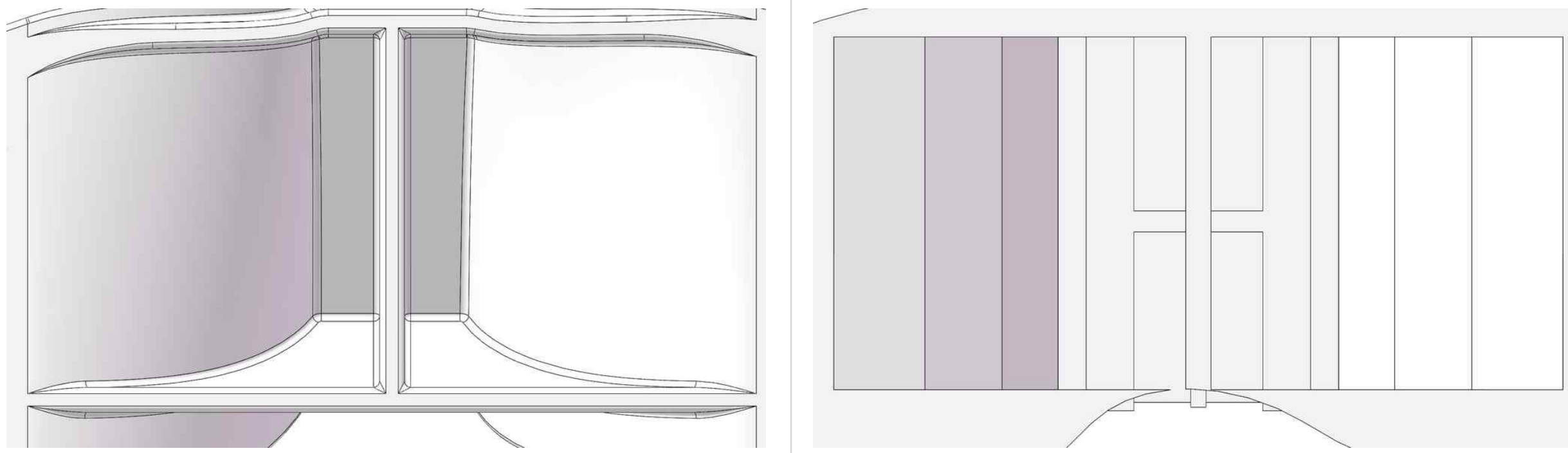

| Reference model: close-up | S1 generated model: close-up |
| --- | --- |

Figure 6-28  Rib directions in the three-view result for I2C-05

### I2C-06 Hammerhead robot chassis (complex)

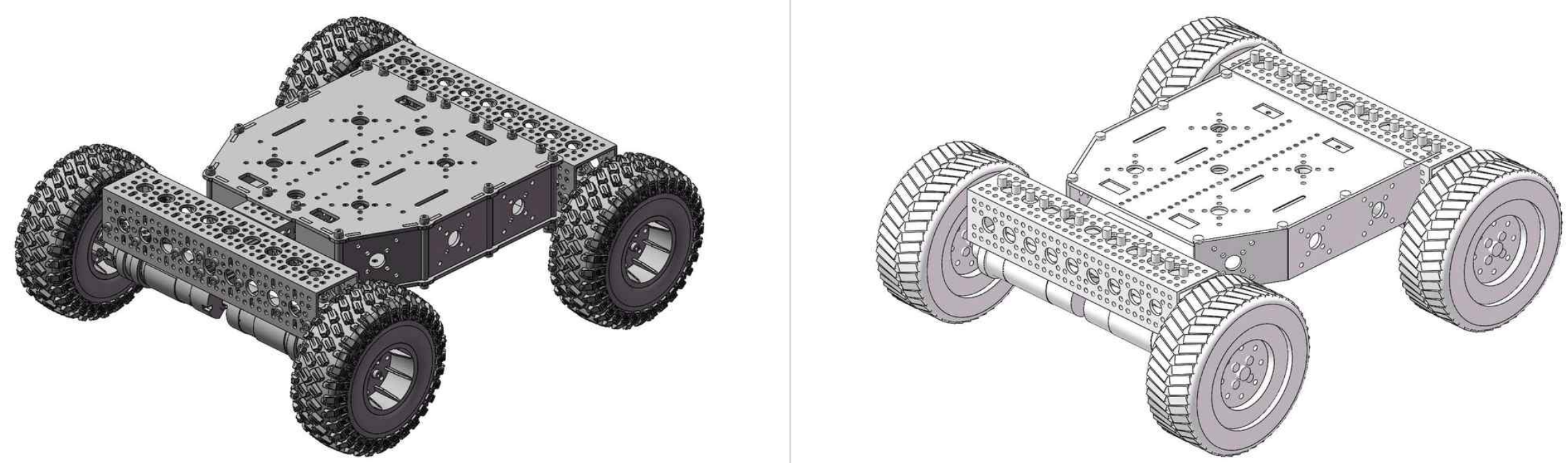

| Reference CAD model | S1 generated model |
| --- | --- |

Figure 6-29  Reference and generated models for I2C-06 with three input views

The three-view result retains the overall arrangement of the chassis plates, crossmembers, and four wheels. It contains 480 valid solids, compared with 721 in the main experiment. This difference cannot simply be interpreted as 241 missing parts: it also reflects whether details such as tire treads are modeled as separate solids or combined. HD is 78.811 mm, close to the main experiment's 80.297 mm, and the overall scale remains substantially too small. Hole patterns, wheel hubs, and connection structures have not been checked for correspondence individually.

#### 6.3.2.2 S2: Axonometric view removed

This group uses six orthographic views—front, rear, left, right, top, and bottom—with the axonometric view removed. Web search and the prompt are unchanged from the main experiment. Table 6-9 summarizes the results.

Table 6-9  Results with six input views (S2)

| Case ID | Validity of each solid | Mean IoU (six views) | HD (mm) | IoU by iteration | Main differences from the reference |
|---|---|---|---|---|---|
| I2C-01 | Pass | 0.9896 | 17.246 | 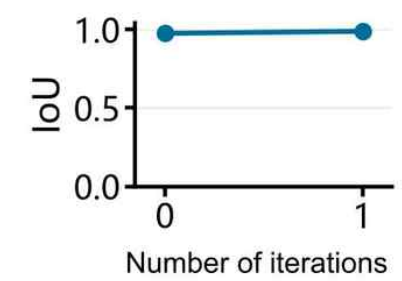 | Clips are simplified to straight slot-like shapes; hole-boss details and scale still differ. |
| I2C-02 | Pass | 0.9947 | 36.380 | 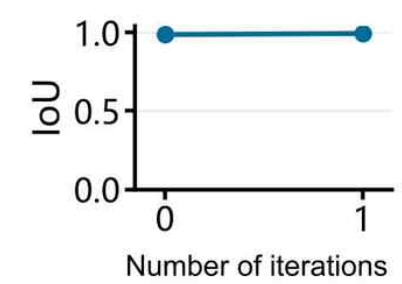 | The lower edge of the opening is incorrectly closed; dimensions fall between the three-view and main-experiment results. |
| I2C-03 | Pass | 0.9467 | 63.106 | 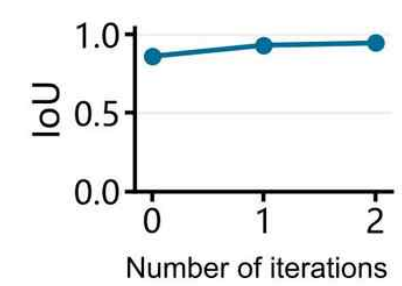 | One solid is formed, at a scale close to the main experiment; transitions between the lug and rib walls remain simplified. |
| I2C-04 | Pass | 0.9806 | 20.867 | 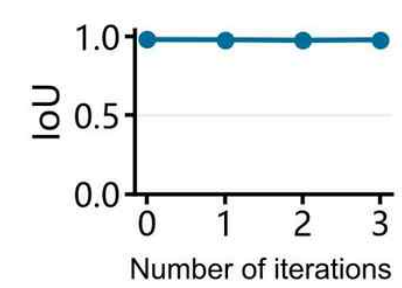 | Overall dimensions are close to the reference; hole openings and fillets at connections remain simplified. |
| I2C-05 | Pass | 0.9549 | 14.616 | 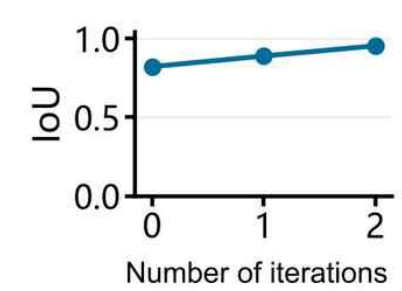 | Straight vertical front ribs differ from the reference layout; surface transitions are not accurately reconstructed. |
| I2C-06 | Pass | 0.9291 | 72.731 | 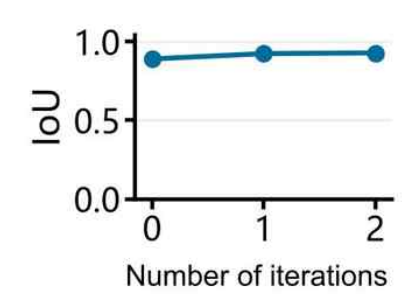 | Overall layout is retained, with dimensions closer to the reference than in the main experiment; connection details still differ. |

Table 6-10  Time taken for each case with six input views (S2)

| Case ID | I2C-01 | I2C-02 | I2C-03 | I2C-04 | I2C-05 | I2C-06 |
|---|---|---|---|---|---|---|
| Time | 9 min | 11 min | 17 min | 17 min | 20 min | 17 min |

With the axonometric view removed, all six cases pass the basic solid checks and largely retain their overall layouts. HD for I2C-03 falls from 112.981 mm in the three-view group to 63.106 mm, close to the seven-view main experiment's 62.401 mm. I2C-05 still fails to reproduce the front rib layout correctly. I2C-02 and I2C-04 have lower HD values than in the main experiment, largely accompanied by overall dimensions closer to the reference.

### I2C-01 TrackerOne backplate (simple)

| 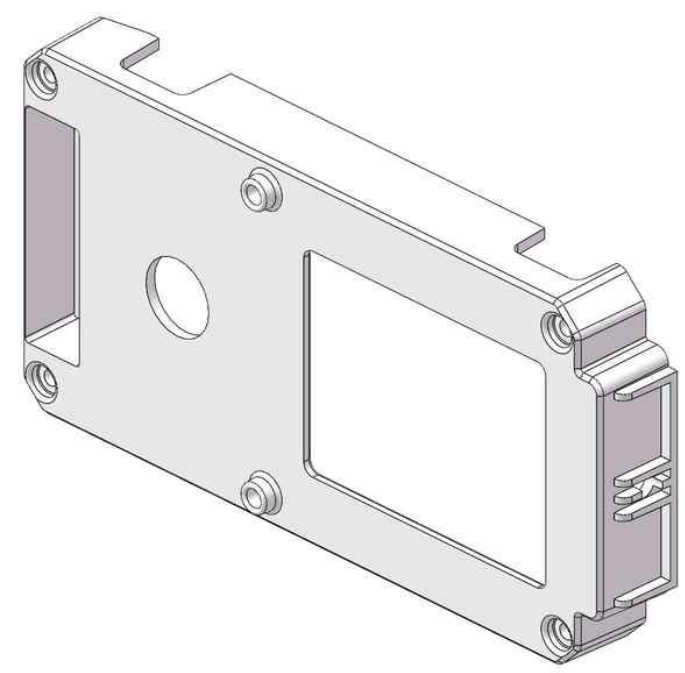 | 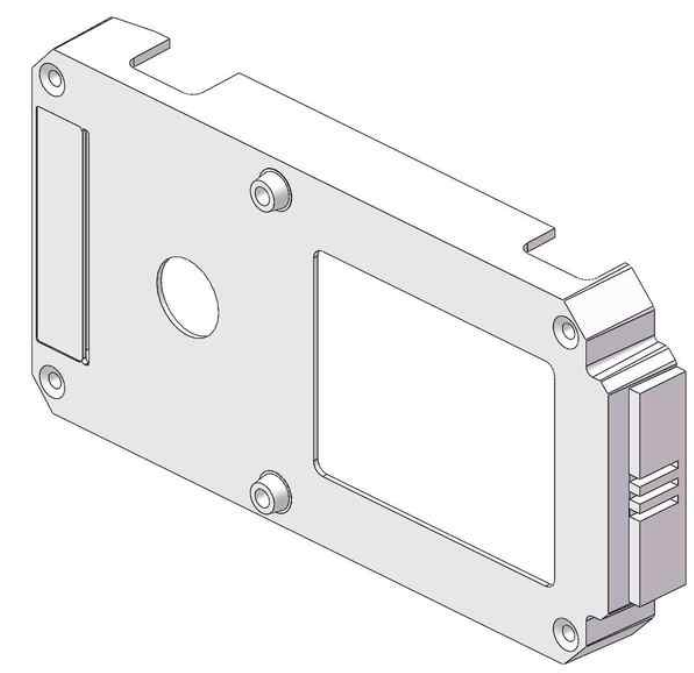 |
| --- | --- |
| **Reference CAD model** | **S2 generated model** |

Figure 6-30 Reference and generated models for I2C-01 with six input views

The six-view result retains the overall arrangement of the window and holes. Side clips are still simplified to straight slot-like shapes, and annular details at the corner hole bosses differ from the reference. HD is 17.246 mm, slightly above both the three-view result and the main experiment. The generated model is approximately 159.90 mm long, compared with 132.11 mm for the reference, and the local clip shapes have not been reconstructed accurately.

### I2C-02 KnowCO2 sensor housing (simple)

| 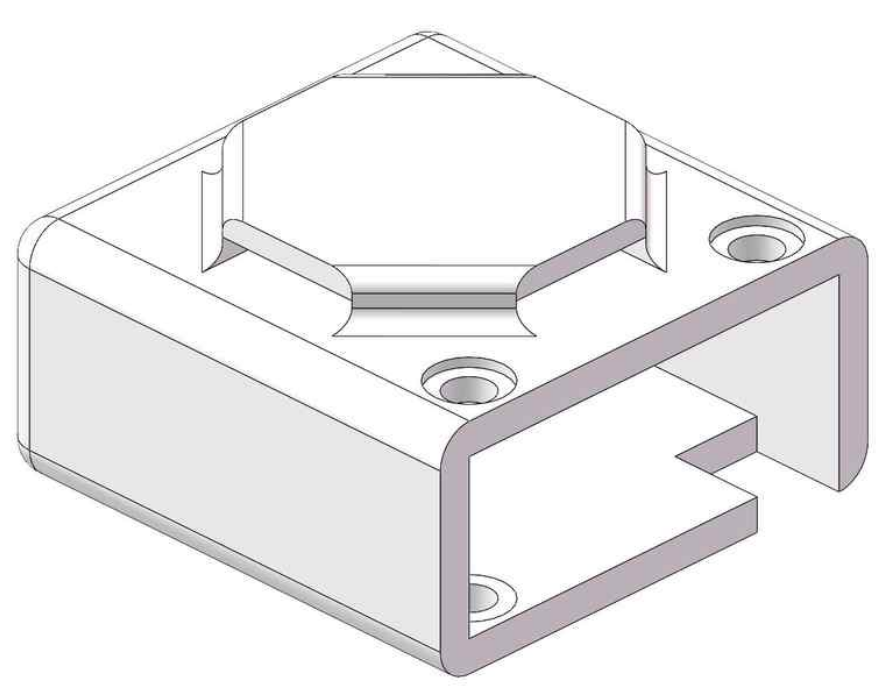 | 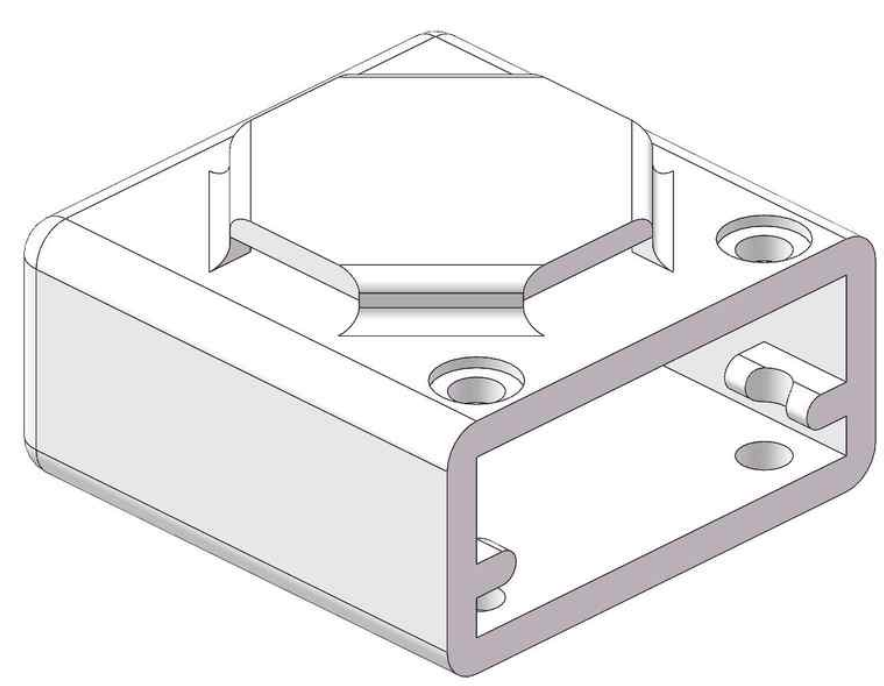 |
| --- | --- |
| **Reference CAD model** | **S2 generated model** |

Figure 6-31 Reference and generated models for I2C-02 with six input views

With six input views, the top platform and its root transitions are fairly close to the reference. However, a continuous lower edge remains across the side opening, reproducing the structural error seen in the three-view group. The generated model measures 75.00 × 73.00 × 42.00 mm, between the three-view result and the main-experiment result. HD is 36.380 mm, and silhouette IoU is 0.9947. Despite this high overlap, the model is substantially oversized, and the continuous frame along the bottom of the opening changes the original clearance space. These differences affect how mating parts are located and inserted and require further verification.

### I2C-03 C5 Corvette lower control arm (moderate)

| | |
|---|---|
| 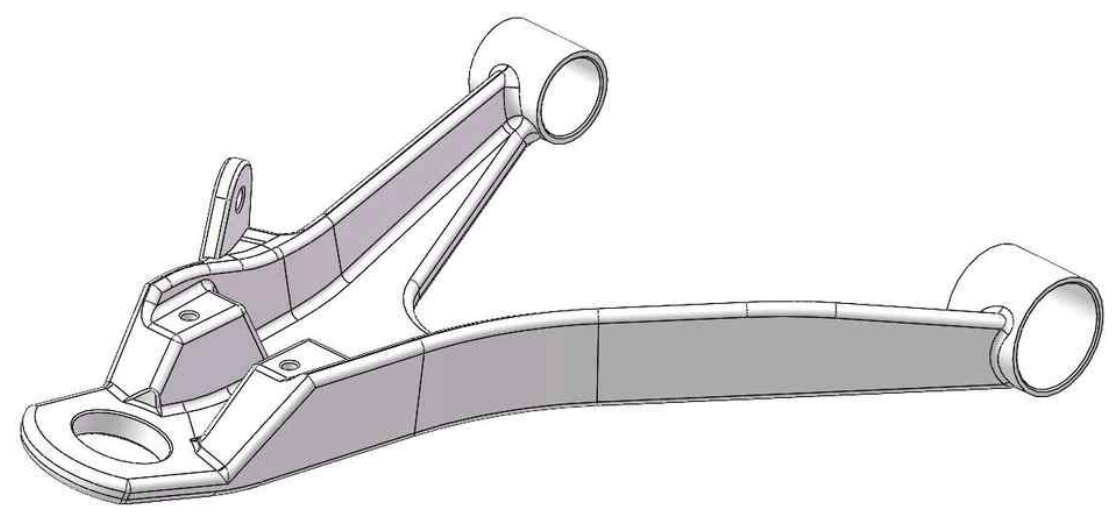 | 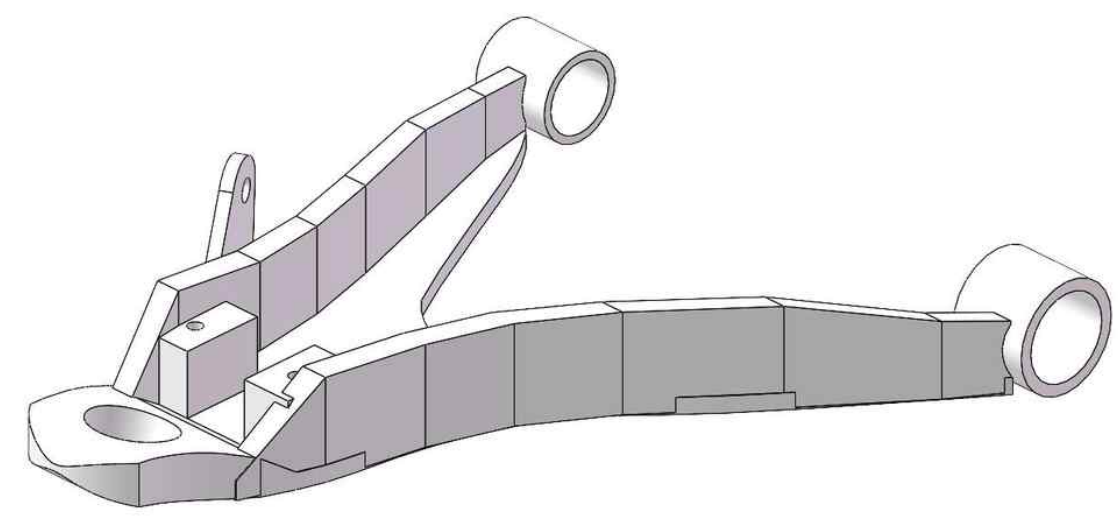 |
| **Reference CAD model** | **S2 generated model** |

Figure 6-32 Reference and generated models for I2C-03 with six input views

The six-view result forms one valid solid, with more complete connections between the forked arms, bushings, and front lug than in the three-view result. Its maximum extent increases to 394.50 mm, and HD falls from 112.981 mm in the three-view group to 63.106 mm, close to the main-experiment result. However, planar and segmented surfaces at the lug root and hole opening still approximate the original freeform surfaces and fillets. Combining the geometry into a single solid has not recovered the reference part's local connection sections and transition shapes.

### I2C-04 C5 Corvette steering knuckle (moderate)

| | |
|---|---|
| 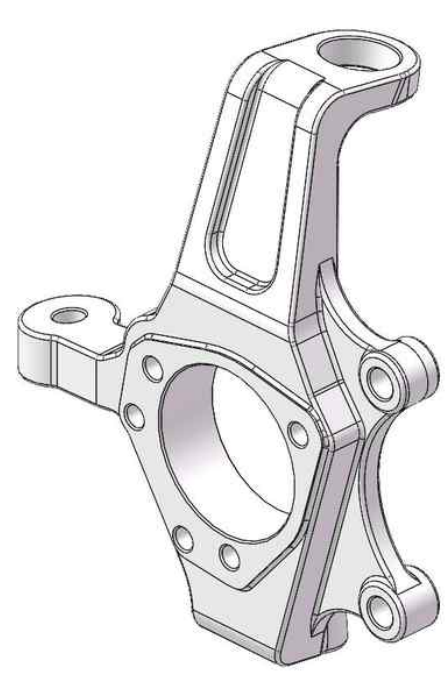 | 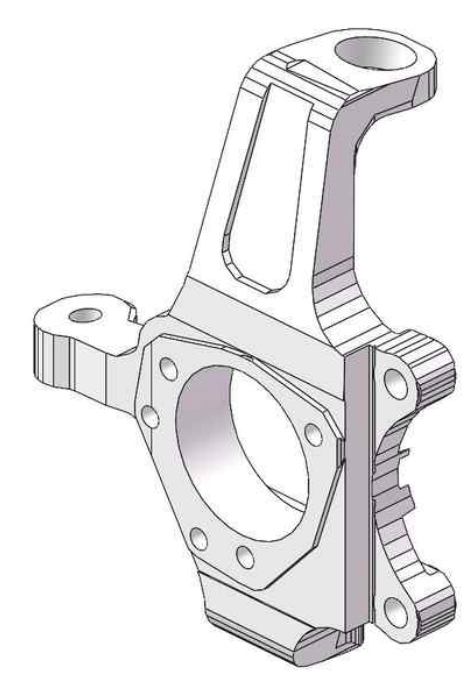 |
| **Reference CAD model** | **S2 generated model** |

Figure 6-33 Reference and generated models for I2C-04 with six input views

The six-view result retains the main holes, curved arm, and mounting lugs. Its maximum extent is 316.00 mm, close to the reference's 317.22 mm. HD is 20.867 mm, similar to the three-view result and lower than the main experiment. Curved transitions at the mounting lugs and connecting walls are still approximated by planar facets, and some hole steps and fillets are missing. The remaining differences involve local contact surfaces, wall thicknesses, and connection transitions, which matter for lug fits and local loading.

### I2C-05 GE jet engine bracket IVO (complex)

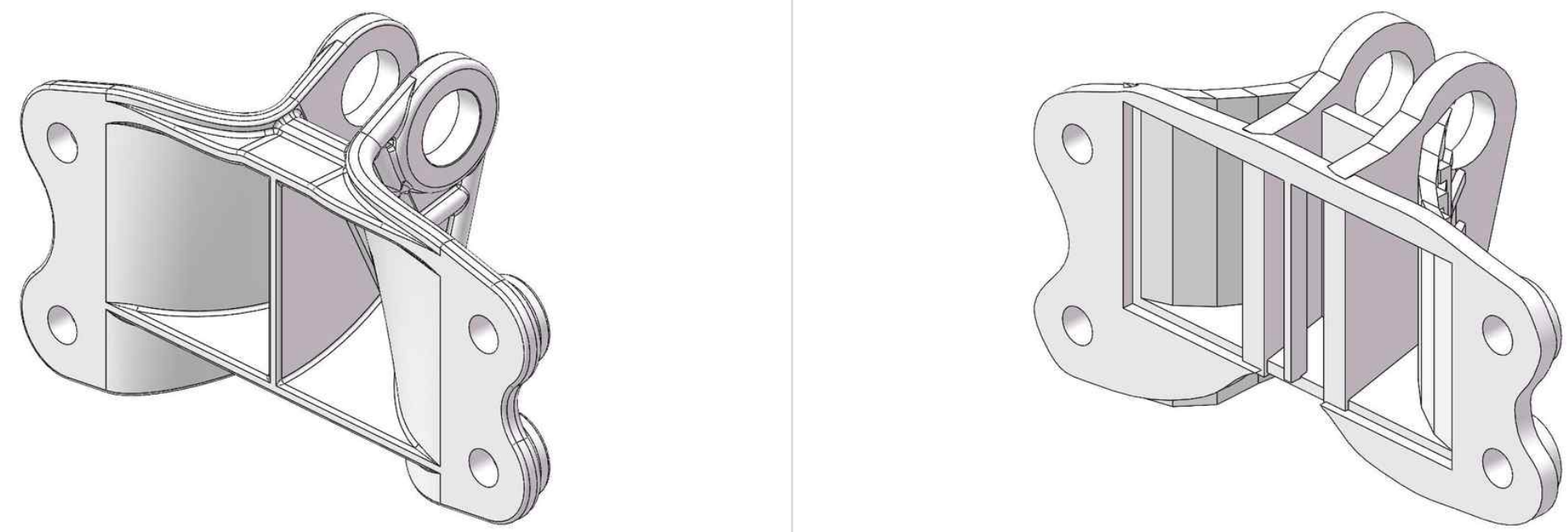

| **Reference CAD model** | **S2 generated model** |
|---|---|

Figure 6-34 Reference and generated models for I2C-05 with six input views

The six-view result retains the mounting holes at both ends of the wing and the holes through the upper paired lugs. However, several straight vertical ribs remain on the front, unlike the reference's transverse ribs. HD is 14.616 mm, close to the three-view result and above the main experiment's 9.959 mm. Overall dimensions are similar to those in the main experiment, but the number, direction, and connections of the ribs differ markedly. The local supports for the thin wing have not been reconstructed as in the reference.

### I2C-06 Hammerhead robot chassis (complex)

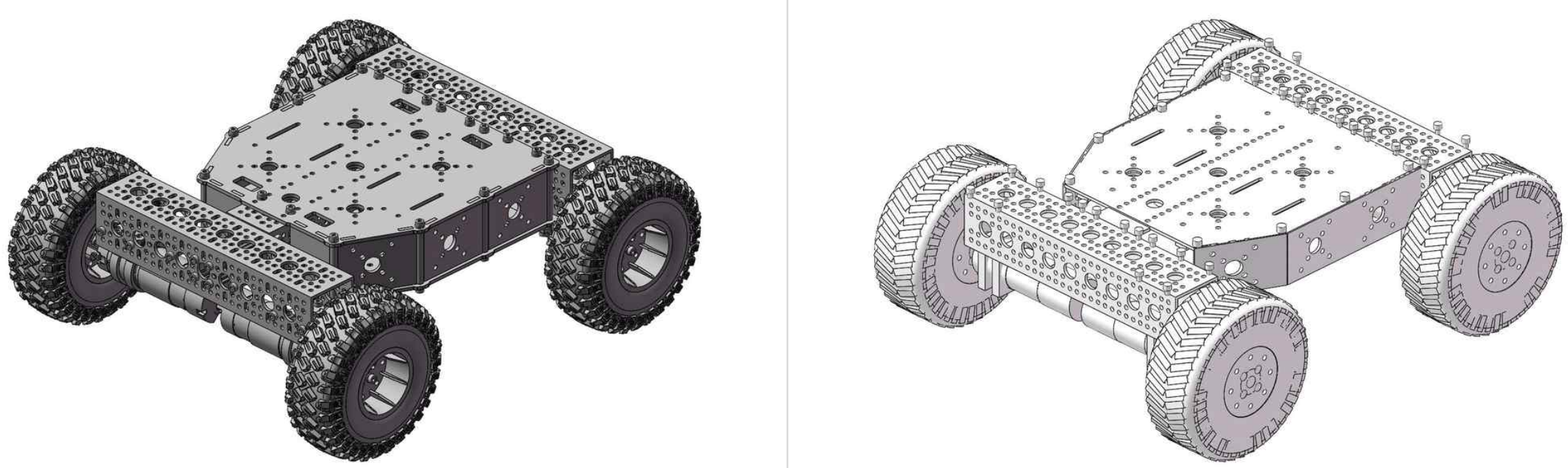

| **Reference CAD model** | **S2 generated model** |
|---|---|

Figure 6-35 Reference and generated models for I2C-06 with six input views

The six-view result largely reconstructs the overall arrangement of plates, crossmembers, and four wheels and contains 799 valid solids. The longest side of its bounding box is 317.33 mm, closer to the reference's 429.45 mm than either the three-view result or the main experiment. HD decreases to 72.731 mm, but the overall dimensions remain too small. Wheel hubs, hole patterns, and internal connection features still differ. Solid validity checks cannot replace checks of interface dimensions and assembly fits.

#### 6.3.2.3 S3: Web search disabled

This group tests I2C-05 and I2C-06 with external web search disabled. Inputs remain the six orthographic views plus an axonometric view, and the main-experiment prompt is retained. Table 6-11 summarizes the results.

Table 6-11 Results with web search disabled (S3)

| Case ID | Validity of each solid | Mean IoU (seven views) | HD (mm) | IoU by iteration | Main differences from the reference |
|---|---|---|---|---|---|
| I2C-05 | Pass | 0.9690 | 19.905 | 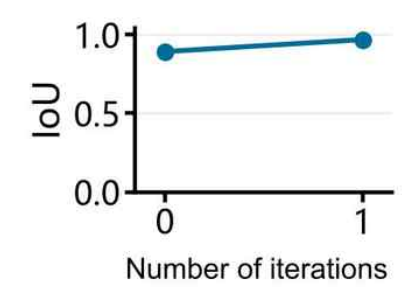 | A continuous plate approximates the front; the wing and main ribs are incompletely reconstructed. |
| I2C-06 | Pass | 0.9109 | 146.769 | 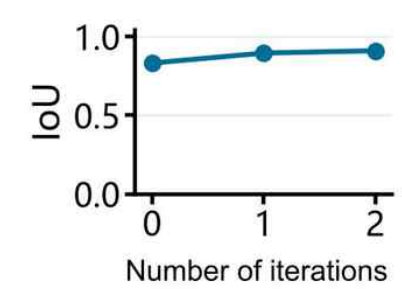 | Tires, wheel hubs, and dense hole patterns are substantially simplified; overall scale is too small. |

Table 6-12 Time taken for each case with web search disabled (S3)

| Case ID | I2C-05 | I2C-06 |
|---|---|---|
| Time | 24 min | 20 min |

Both cases produce models that pass the solid validity checks. After one revision, I2C-05's mean IoU across the seven views increases from 0.8927 to 0.9690. Its HD is 19.905 mm, however, above the main experiment's 9.959 mm with web search enabled. After two revisions, I2C-06's mean IoU increases from 0.8325 to 0.9109, but its HD is 146.769 mm, also above the main experiment's 80.297 mm. Both results retain the overall shape, while substantially simplifying the bracket's curved ribs and the chassis's tire and hole-pattern details.

### I2C-05 GE jet engine bracket IVO (complex)

| | |
|---|---|
| 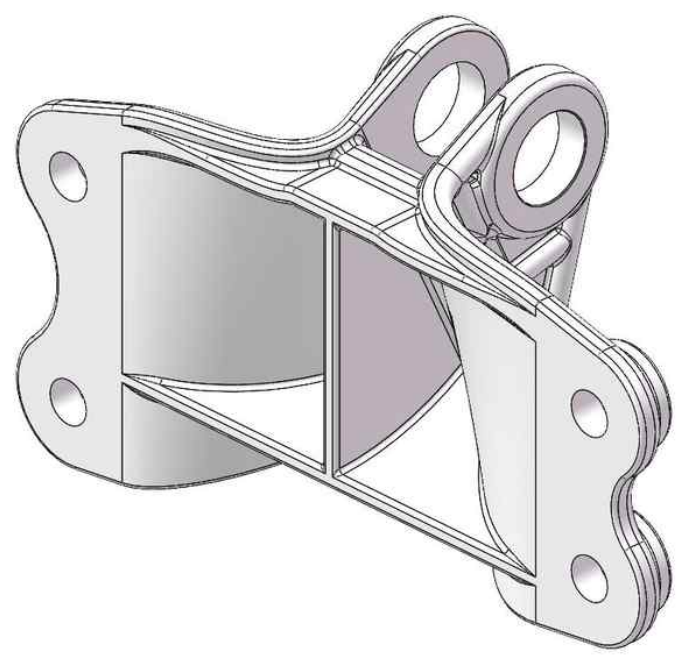 | 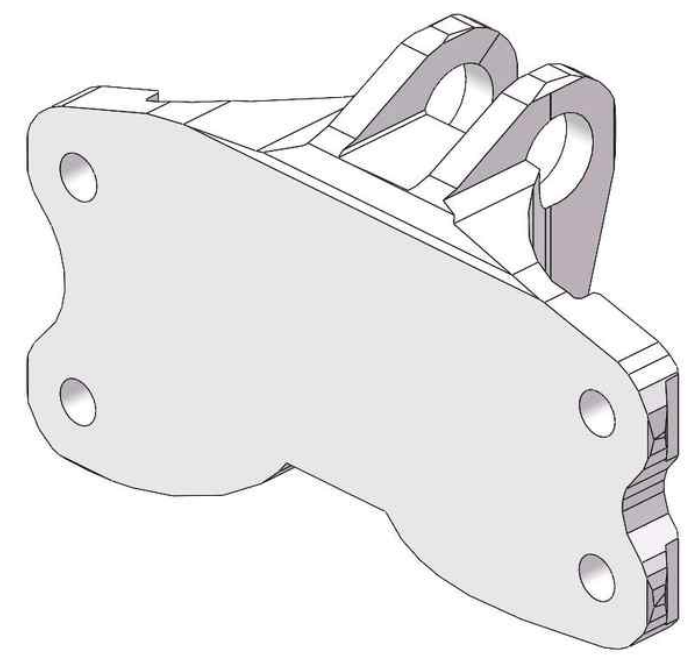 |
| **Reference CAD model** | **S3 generated model** |

Figure 6-36 Reference and generated models for I2C-05 with web search disabled

The generated bracket retains the mounting holes on both sides and the holes through the upper paired lugs. A continuous flat plate replaces the original freeform surfaces and rib structure at the front, leaving the curved wing, transverse ribs, and local fillets incompletely reconstructed. Overall dimensions are close to those in the main experiment with web search enabled, yet HD increases from 9.959 to 19.905 mm; IoU is 0.9690. The flat-plate approximation changes the wing's curvature, local sections, and internal open spaces, producing support geometry different from the reference. High silhouette overlap is insufficient to establish that the original structural function has been preserved.

### I2C-06 Hammerhead robot chassis (complex)

| | |
|---|---|
| 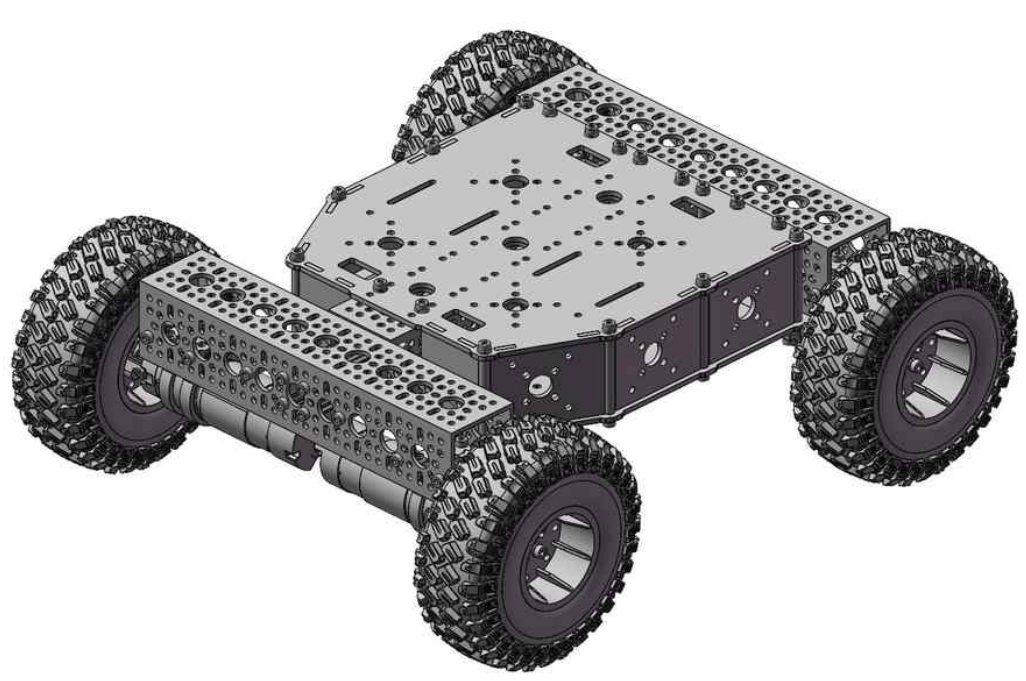 | 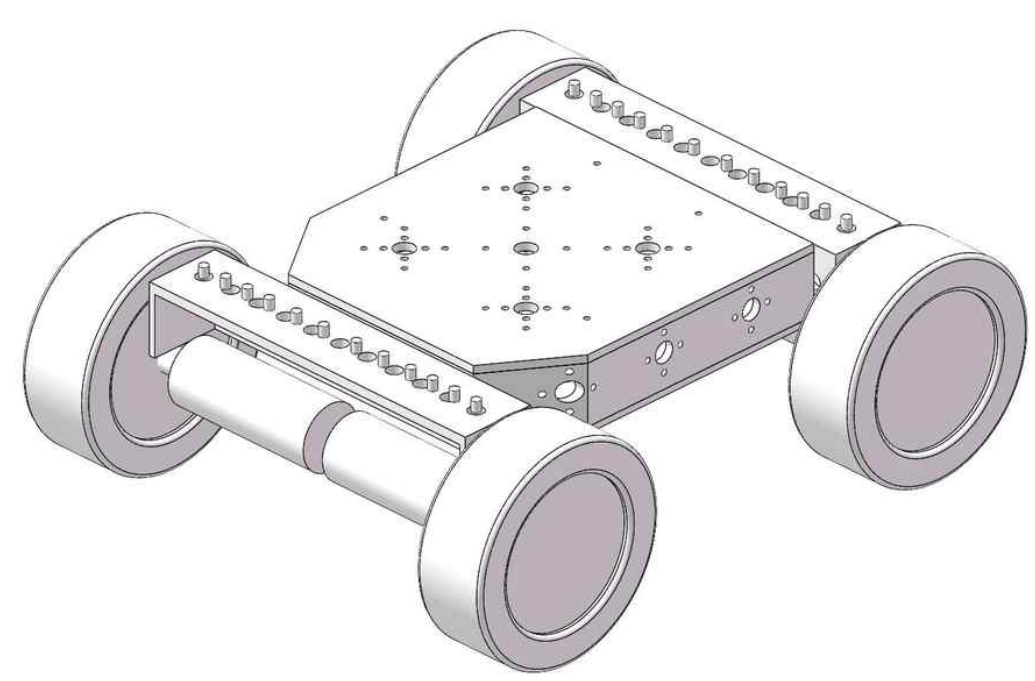 |
| **Reference CAD model** | **S3 generated model** |

Figure 6-37 Reference and generated models for I2C-06 with web search disabled

The generated chassis retains the overall arrangement of the bottom and top plates, crossmembers, and four wheels and contains 45 valid solids. Tire treads are absent, wheel hubs are simplified to discs, and many of the densely arranged holes and slots in the crossmembers and plates are missing. The maximum extent is approximately 201.58 mm, below both the reference's 429.45 mm and the main experiment's 308.47 mm with web search enabled. HD increases to 146.769 mm. In addition to the undersized overall geometry, simplified hubs and mounting holes and slots alter connection geometry and component insertion positions. Assembly and motion relationships require further verification.

| 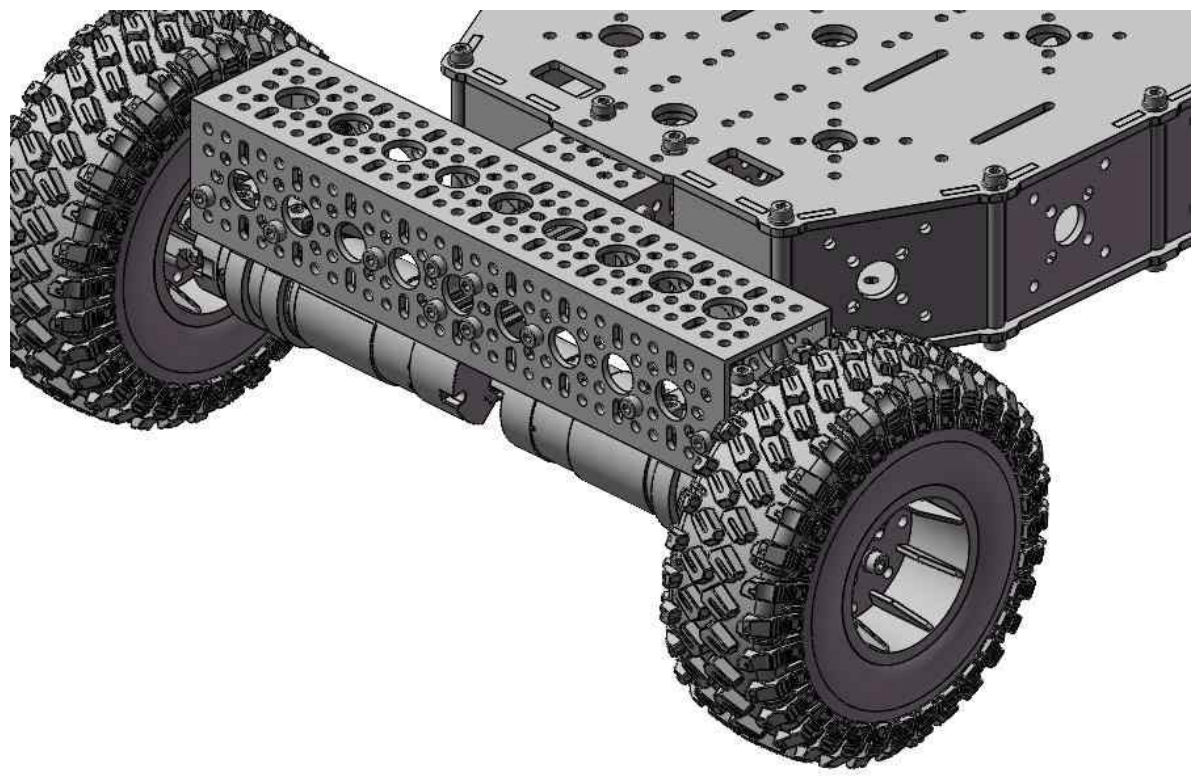 | 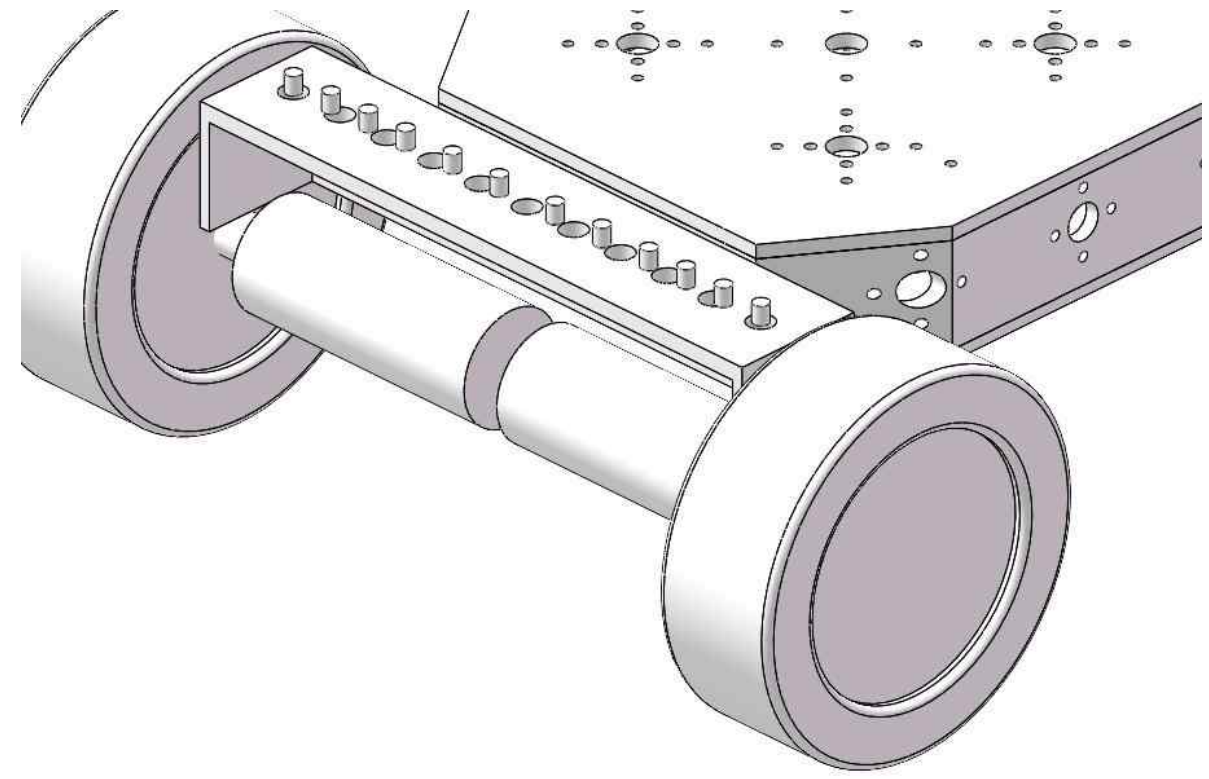 |
| --- | --- |
| **Reference model: close-up** | **S3 generated model: close-up** |

Figure 6-38  Wheel and crossmember features in I2C-06 with web search disabled

#### 6.3.2.4 S4: Stress test

The stress test uses I2C-07, a bell housing with multiple flange levels, hole bosses, side windows, and dense ribbing. Six orthographic views and an axonometric view are supplied, with web search enabled. This test further examines GPT-6's ability to reconstruct a complex model. Table 6-13 presents the results.

Table 6-13  Bell-housing stress-test results (S4)

| Case ID | Complexity | Validity of each solid | Mean IoU (seven views) | HD (mm) | IoU by iteration | Time |
| --- | --- | --- | --- | --- | --- | --- |
| I2C-07 | Complex | Pass | 0.9013 | 78.484 | 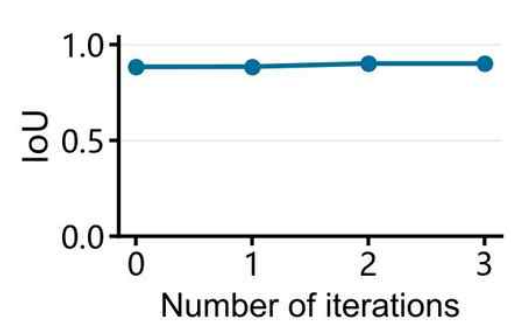  | 20 min |

## I2C-07 Bell housing (complex)

### Input renderings

| | |
|---|---|
| 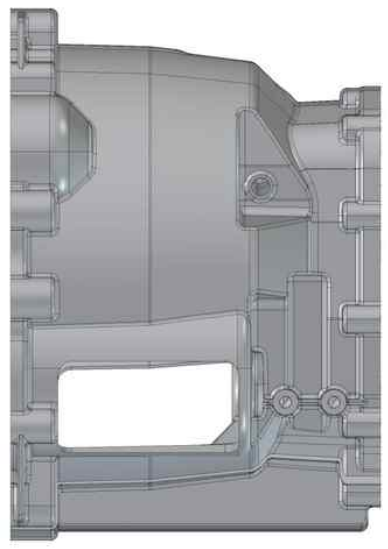 | 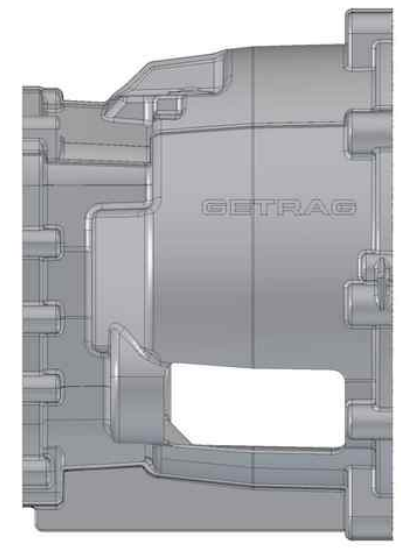 |
| **Front view** | **Rear view** |
| 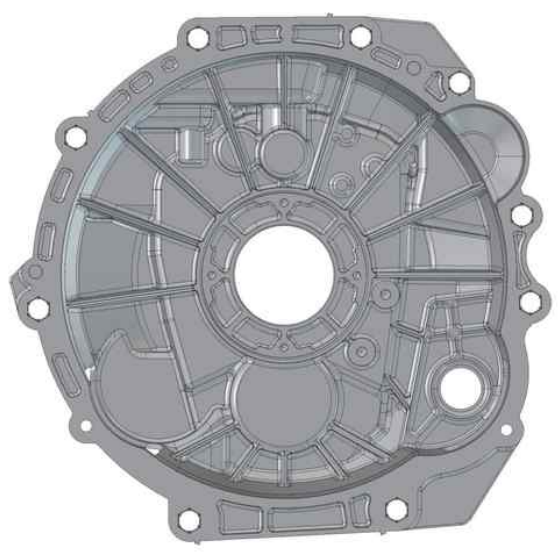 | 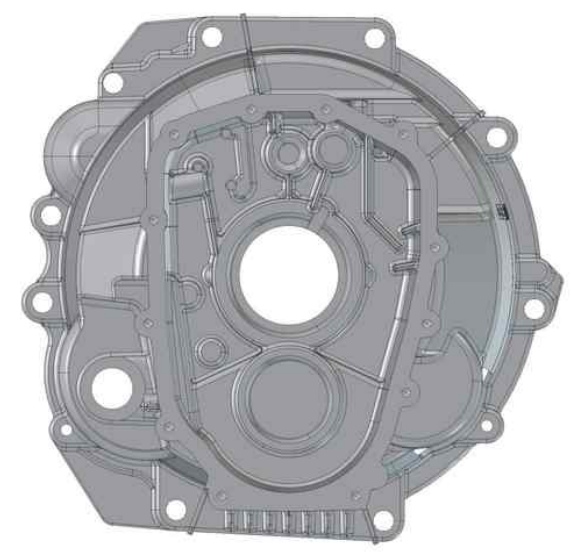 |
| **Left view** | **Right view** |
| 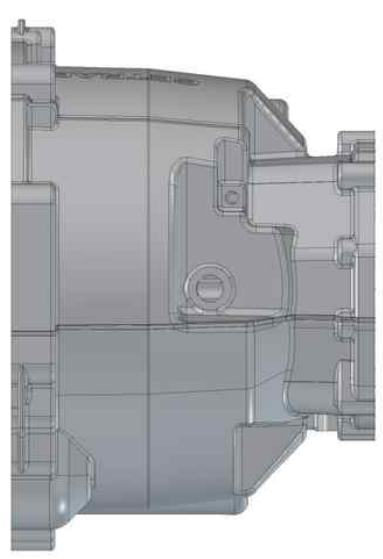 | 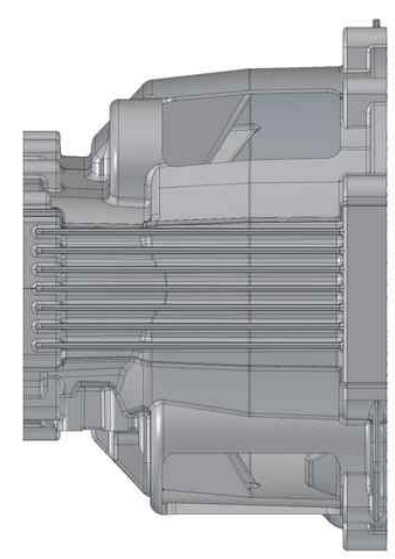 |
| **Top view** | **Bottom view** |
| 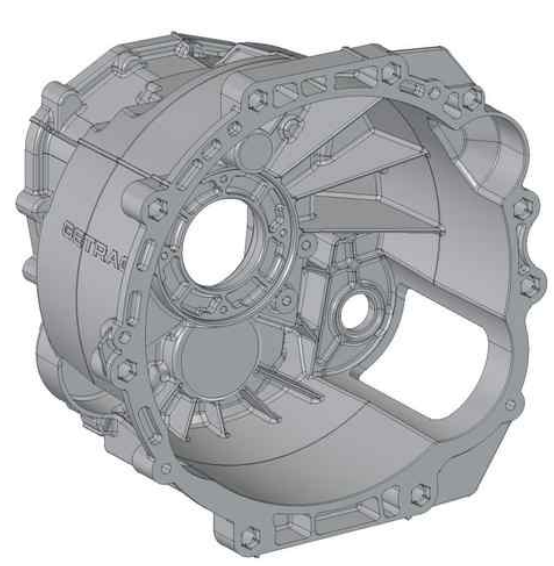 | |
| **Axonometric view** | |

Figure 6-39 Seven input renderings for I2C-07

## Results

| | |
|---|---|
| 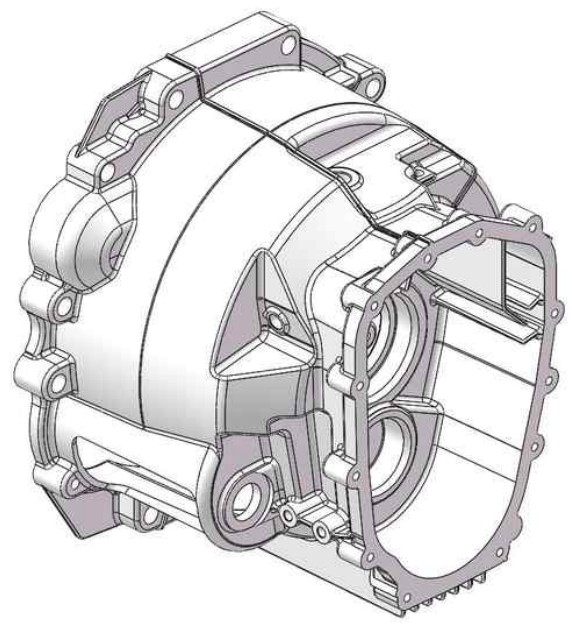 | 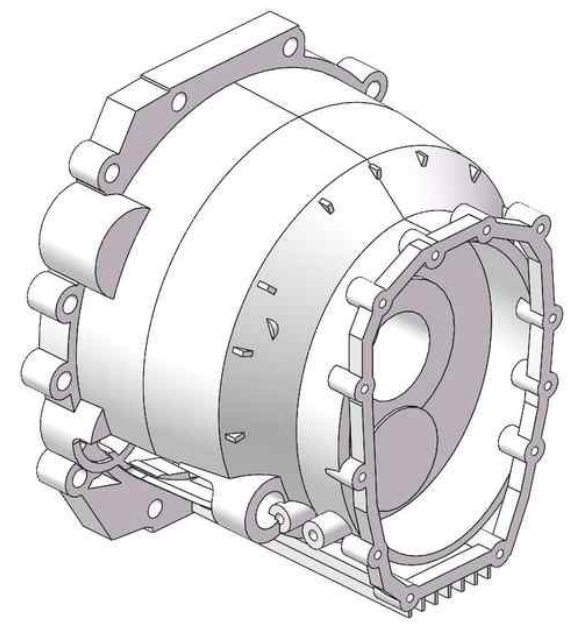 |
| **Reference model (axonometric view)** | **Generated model (axonometric view)** |
| 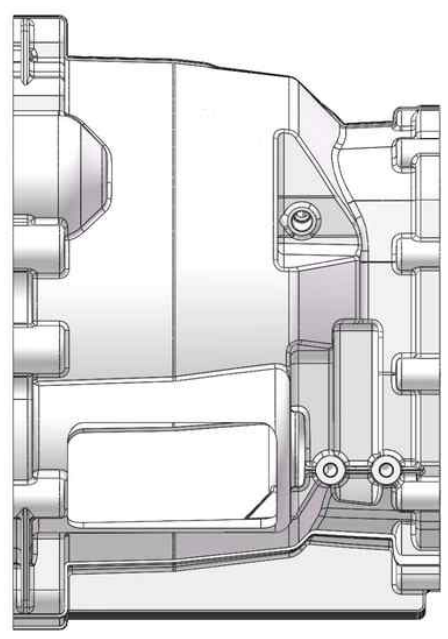 | 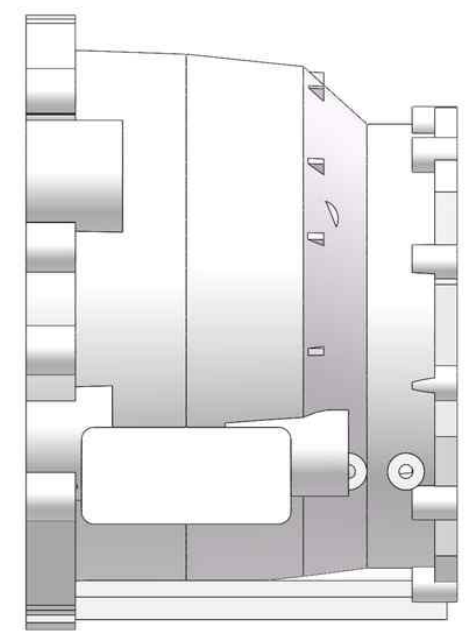 |
| **Reference model (front view)** | **Generated model (front view)** |
| 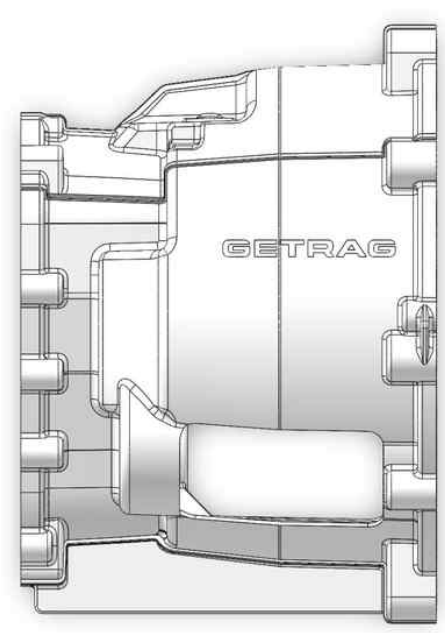 | 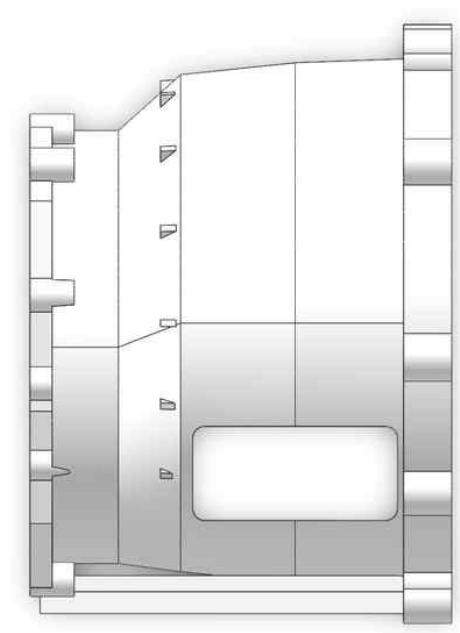 |
| **Reference model (rear view)** | **Generated model (rear view)** |
| 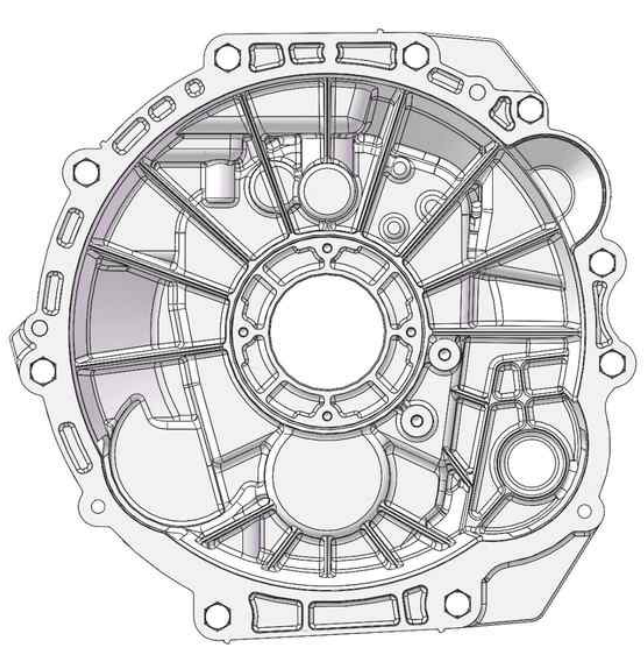 | 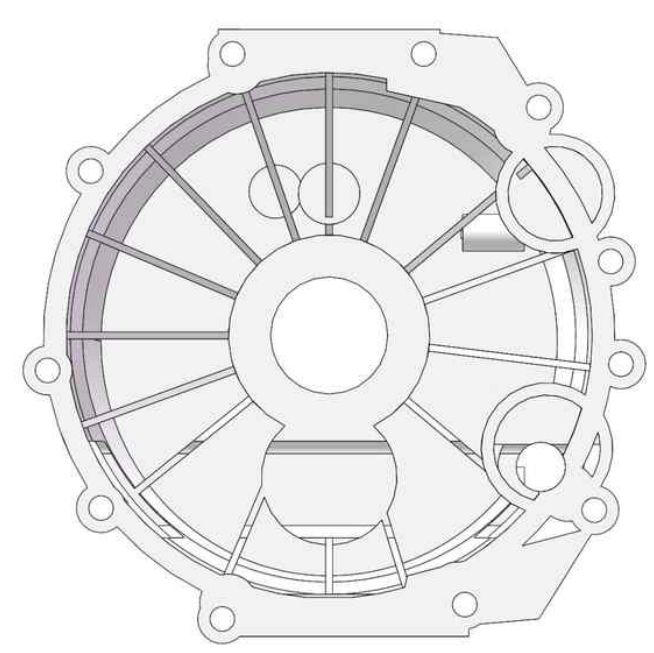 |
| **Reference model (left view)** | **Generated model (left view)** |
| 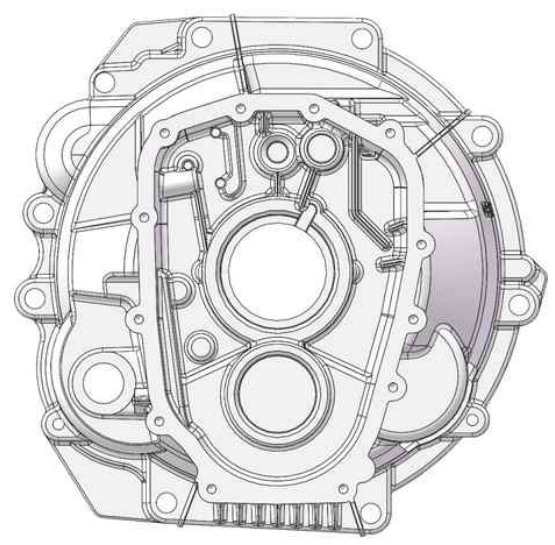 | 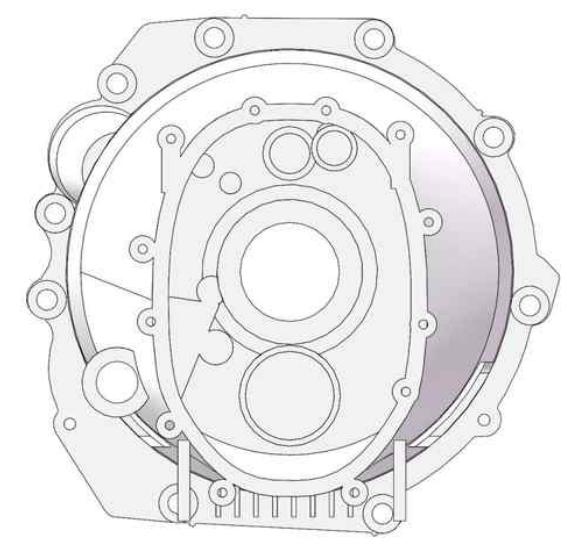 |
| **Reference model (right view)** | **Generated model (right view)** |
| 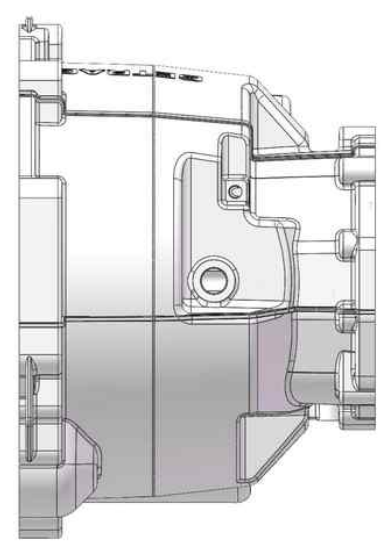 | 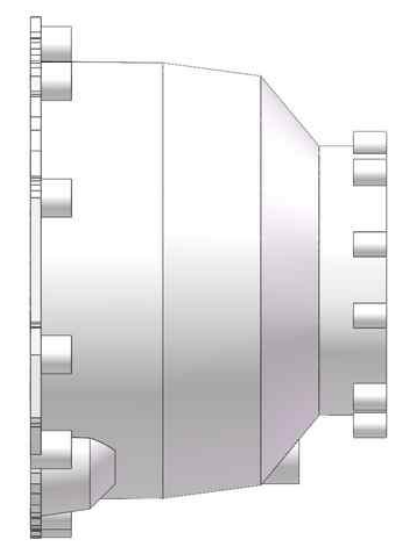 |
| **Reference model (top view)** | **Generated model (top view)** |
| 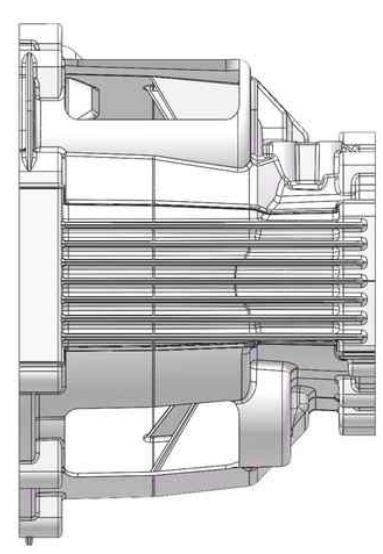 | 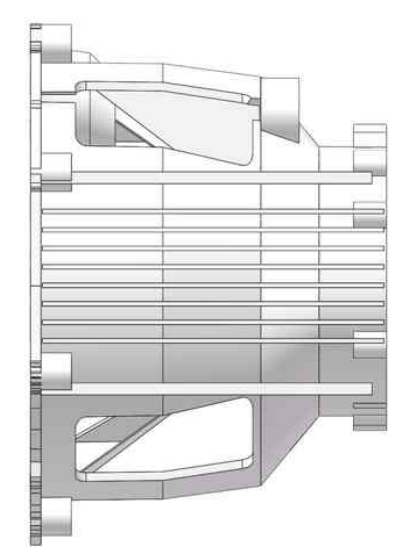 |
| **Reference model (bottom view)** | **Generated model (bottom view)** |

Figure 6-40 Reference and generated bell-housing models for I2C-07

The generated bell housing contains one valid solid and includes the main features: the large-end flange, main shaft hole, side windows, and radial ribs. Connections between the outer wall and flanges are substantially simplified. Casting transitions around hole bosses, fine ribs, and local recesses are incomplete; regular hole bosses and straight ribs approximate the local freeform surfaces and connection transitions. Mean IoU across the seven views is 0.9013, and HD is 78.484 mm. The generated model's maximum extent is 302.99 mm, compared with 392.17 mm for the reference, so it is substantially undersized.

Figure 6-41 plots silhouette IoU for each view against iteration number.

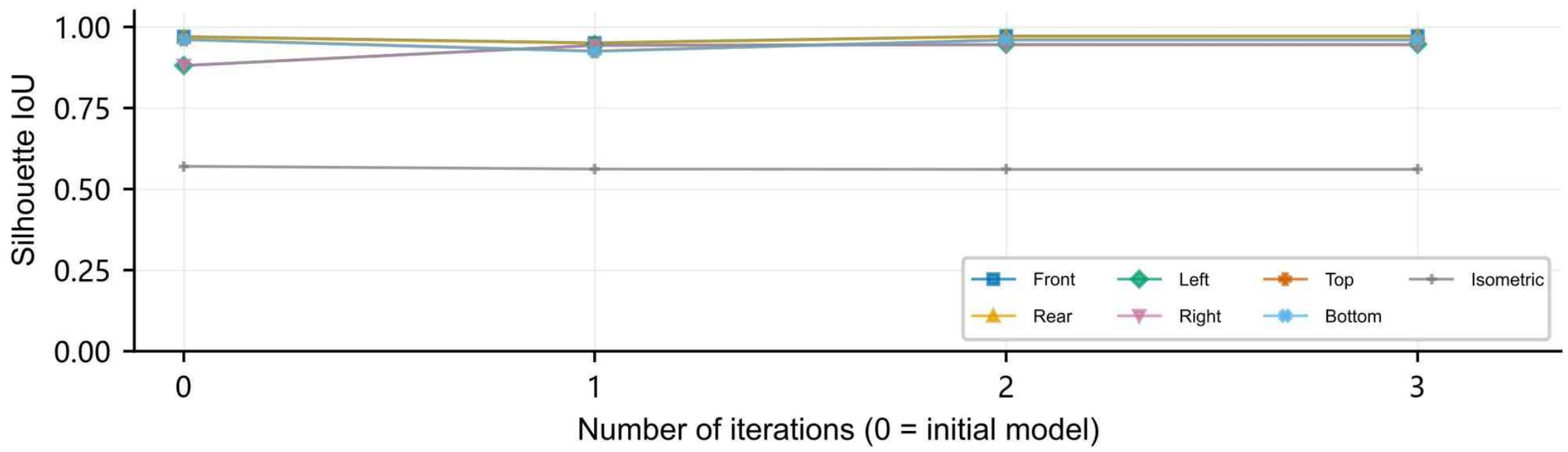

Figure 6-41 Silhouette IoU by iteration for I2C-07

## 6.4 Analysis of GPT-6's approach

The recorded outputs and tool operations suggest the workflow illustrated below. It describes observable modeling steps and does not establish the model's internal reasoning.

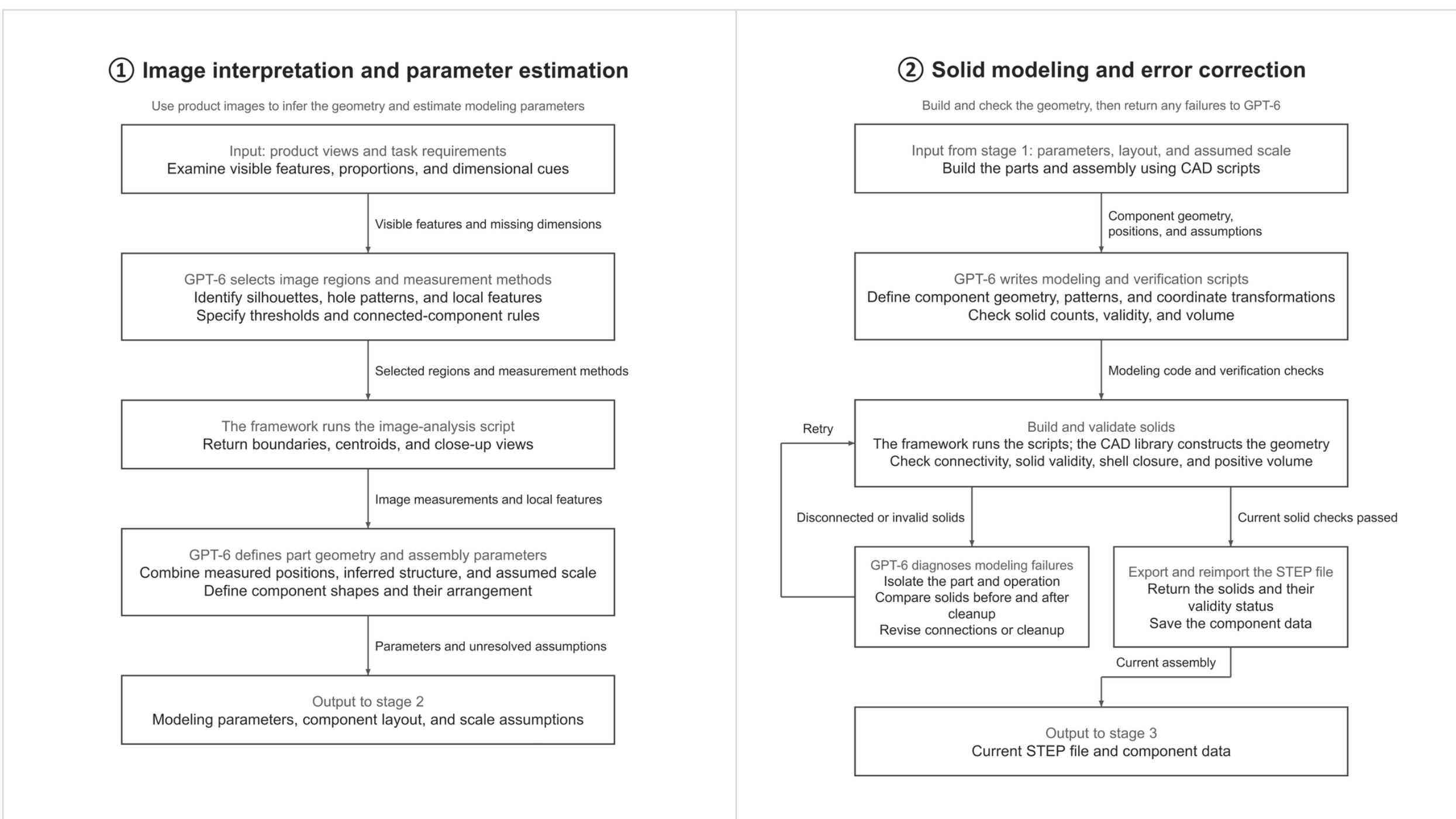

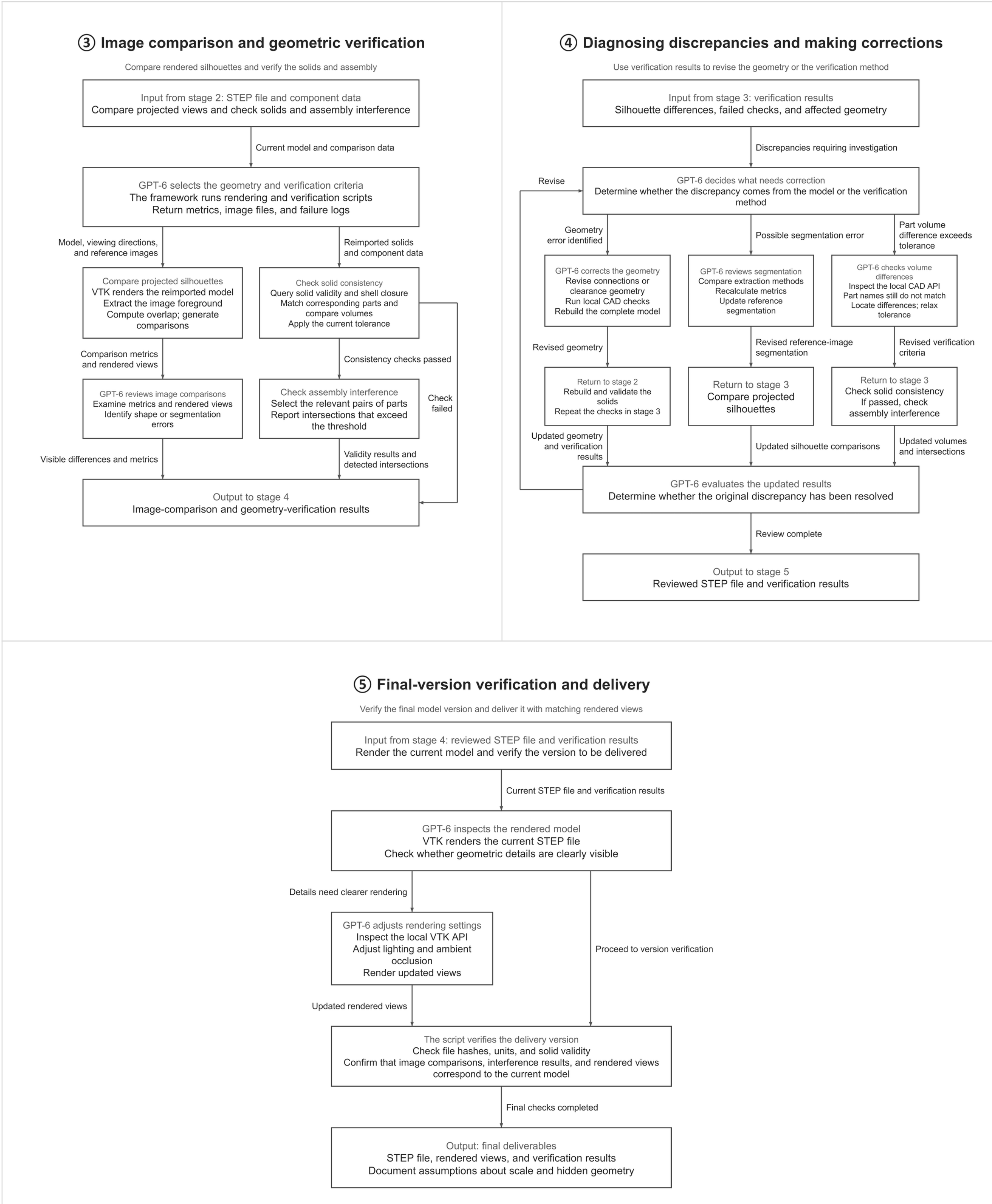

Figure 6-42  Inferred workflow for reconstruction from images

## 6.5 Capabilities and limitations

In CAD reconstruction from images, GPT-6 can combine silhouettes, holes and slots, and relative positions across multiple views to reconstruct the main structure of a part and produce a STEP model. In these tests, the overall shapes of regular plates and housings were usually recovered. Complex brackets, the chassis, and the bell housing also retained their main components and overall layouts.

The main limitations concern freeform surfaces, connections between solids, and dimensions that affect function. Complex freeform surfaces and transitions at rib intersections are often approximated by planes, ruled surfaces, or regular ribs, leaving local curvature, cross-sections, and fillets inaccurate. Some nominally single parts consist of separate solids that overlap at connections without a Boolean union. Validity of each solid does not establish that the overall connections are correct. Differences in holes, slots, clips, guides, and assembly interfaces may affect positioning, installation, or load bearing; assumed scales in the absence of dimensional cues may also affect fit. In the two cases tested with web search disabled, models were still generated, but local structures were more simplified than in their corresponding main-experiment results. These comparisons do not establish a general effect of web search.

# Chapter 7 Parametric reconstruction from B-rep models

## 7.1 Task objectives and evaluation criteria

This task evaluates whether GPT-6 can infer a plausible set of features, sketches, reference geometry, modeling history, and dependencies from the geometry and topology of a B-rep model, then reconstruct a corresponding parametric CAD model. It tests the combined ability to work backward from final geometry to modeling logic, plan a parametric modeling sequence, and execute it.

A B-rep model can result from several different, equally reasonable parametric modeling histories. An exact node-by-node match to the original feature tree is therefore not required for success. The evaluation separately considers the correctness of the final geometry, the suitability of the parametric structure, reliability under editing, and consistency between the reconstructed modeling history and the input B-rep model. Table 7-1 summarizes the task.

Table 7-1 Objectives and evaluation criteria for parametric reconstruction from B-rep models

| Test element | Description |
|---|---|
| Evaluation question | Can GPT-6 infer appropriate features, sketches, reference geometry, a modeling sequence, and dependencies from a B-rep model, then reconstruct a geometrically consistent, editable parametric CAD model? |
| Input | A B-rep model of a mechanical part. |
| Output | An editable parametric CAD model that matches the input B-rep geometry within the specified tolerance, with an appropriate feature tree, sketches, reference geometry, dimensional parameters, and feature dependencies. Explain the modeling approach and any uncertainty where necessary. |
| Evaluation criteria | First assess whether a parametric model is produced and how closely its final geometry matches the input B-rep. Then assess the suitability of its parametric structure and reliability under editing. |

## 7.2 Test design

### 7.2.1 Test case selection

The cases include brackets, connecting rods, pump housings, forked shafts, housing covers, engine blocks, pulley support assemblies, three-jaw chucks, and gearbox assemblies.

Complexity is classified primarily by the final geometry and structural relationships. The criteria consider the complexity of surfaces, holes, and slots; the numbers of solids and parts; interactions between local features; and the difficulty of organizing an assembly (Table 7-2).

Table 7-2 Complexity criteria for parametric reconstruction from B-rep models

| Complexity | Selection criteria | Typical characteristics |
|---|---|---|
| Simple | The main modeling sequence is straightforward, with clear extents for basic features. Even with many repeated holes or cuts, feature interactions are predominantly local and relatively independent. | Few dependencies between modeling steps; basic extrusions, revolves, and cuts; local hole features, patterns, and mirrors. |
| Moderate | Several feature types, multiple sketches, or additional reference geometry. Some sequencing decisions or local feature interactions are involved, and several reconstruction strategies may be reasonable. Assemblies are included. | Multiple sketches and reference elements; fillets and chamfers; patterns and hole features. Some steps depend on earlier sketches or features. |
| Complex | Many closely interdependent modeling steps, with extensive reference geometry and feature interactions. Later operations may overwrite or remove boundaries of earlier features. Assemblies are included. | Numerous revolved cuts, hole features, or finishing features; many later steps depend on earlier features, with strong interactions. |

Using these criteria, five simple, three moderate, and three complex cases are selected. Table 7-3 lists the cases.

Table 7-3 Test cases for parametric reconstruction from B-rep models

| Case ID | Rendered CAD model | Faces | Main features | Complexity | Basis for classification |
|---|---|---|---|---|---|
| B2P-01 | 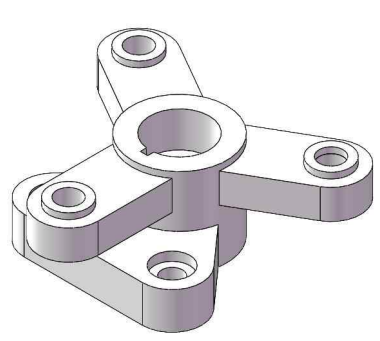 | 60 | Bracket, holes, stepped cross-sections | Simple | Regular layered structure, with clear local profiles and relationships between axial thicknesses. |
| B2P-02 | 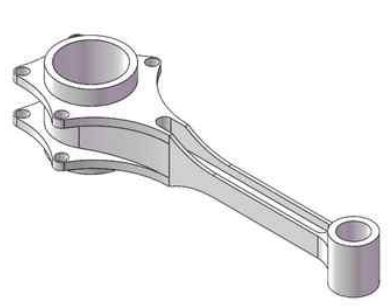 | 83 | Connecting rod, holes at both ends, layered profiles | Simple | Regular rod body and end holes, with clear relationships between the main dimensions. |
| B2P-03 | 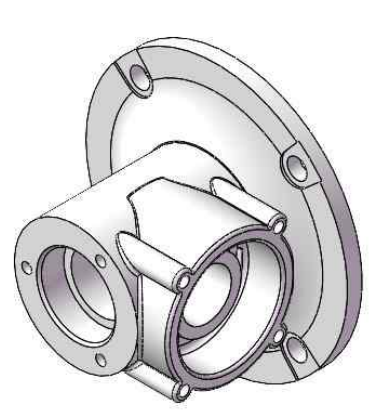 | 178 | Cylindrical sections in two directions, base flange, counterbores, mounting holes | Simple | Predominantly regular cylindrical sections and local holes, with clear axial dimensions and hole-feature extents. |

| Case ID | Rendered CAD model | Faces | Main features | Complexity | Basis for classification |
|---|---|---|---|---|---|
| B2P-04 | 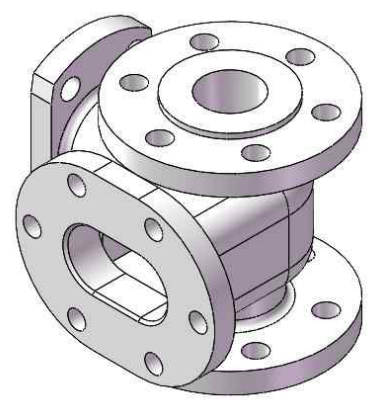 | 153 | Flow passages in several directions, flanges, locating bosses, hole patterns | Simple | Regular flow passages, flanges, and repeated mounting holes. Local dimensions and their dependent features can be checked separately. |
| B2P-05 | 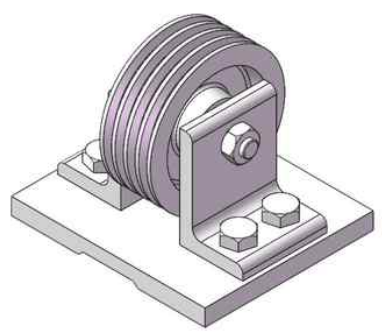 | 348 | Pulley, shaft, left and right supports, fasteners | Simple | 16 solids with clear component and positional relationships, including simple assembly constraints. |
| B2P-06 | 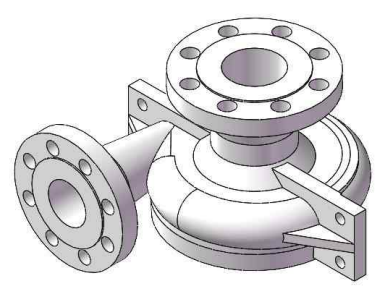 | 155 | Curved pump chamber, flow passages, flanges | Moderate | The pump chamber intersects the flanges; curved surfaces and internal passages make reconstruction more difficult. |
| B2P-07 | 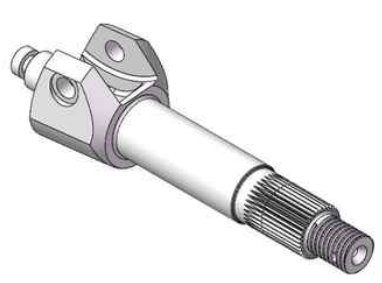 | 380 | Forked head, shaft section, slots and holes | Moderate | A junction between the forked head and shaft, with numerous local slots, holes, and profile changes. |
| B2P-08 | 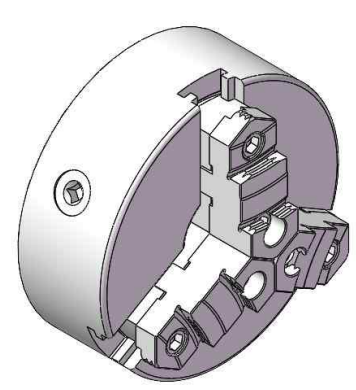 | 1628 | Chuck body, jaws, repeated components | Moderate | 24 solids with many repeated instances; parts must be organized and their relative positions retained. |
| B2P-09 | 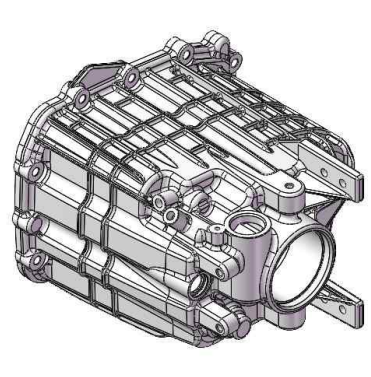 | 5639 | Ribs, holes and cavities, complex surfaces | Complex | 5,639 faces, with intersecting complex surfaces, ribs, holes, and cavities and complicated local connections. |
| B2P-10 | 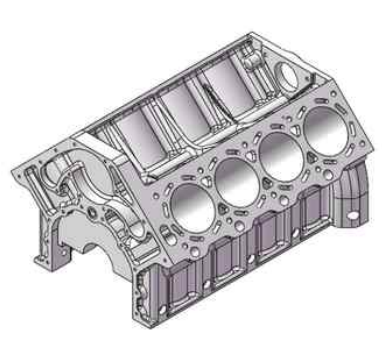 | 3422 | Cylinder bores, flow passages, dense arrangements of holes | Complex | Intersecting cylinder bores and internal passages, with intricate radial profiles and hole details. |

| Case ID | Rendered CAD model | Faces | Main features | Complexity | Basis for classification |
|---|---|---|---|---|---|
| B2P-11 | 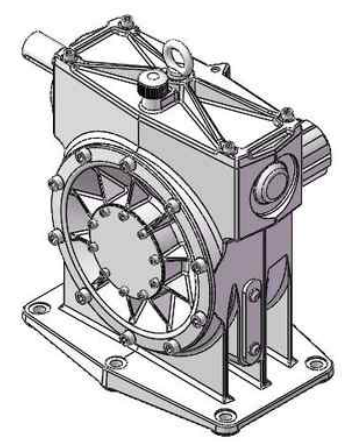 | 10391 | Housing, gears, shafts, bearing assemblies | Complex | 463 solids with densely arranged part instances, complex geometric details, and a complex assembly hierarchy. |

### 7.2.2 Experimental conditions

#### 7.2.2.1 Main experiment

All cases use the same prompt, with a B-rep model supplied as a STEP file. Only the input case changes; the prompt remains fixed so that results can be compared across cases and complexity levels. The prompt reads: "Use the SolidWorks modeling API to reconstruct an editable parametric model from the supplied STEP model. Requirements: 1. Reproduce the model's geometry and establish appropriate sketches, features, and parameter dependencies. For an assembly, retain separate parts and their assembly relationships. 2. If reconstruction involves multiple iterations, use a consistent evaluation method to calculate and record the geometric error at each iteration, and provide an error curve. State clearly where a calculation cannot be performed. 3. I will manually change parameters approximately 10 times for each model to check whether the changes take effect, dependent features update correctly, and any problems occur. Explain where the main parameters are located and what they control."

*The quoted prompt is an English translation of the Chinese prompt used in the experiment.*

#### 7.2.2.2 Additional experiments

Representative complex cases are selected for additional experiments on the effect of web-search access. These tests examine GPT-6's ability to analyze geometry, infer a modeling history, and reconstruct a parametric model from the input B-rep when external information is unavailable through web search. The results are recorded separately and are not pooled with the main experiment.

## 7.3 Results

The main experiment evaluates reconstruction under a uniform prompt, examining solid validity, geometric agreement with the reference CAD model, and editability across complexity levels. Additional experiments disable web search for representative complex cases to examine dependence on external information. The two sets of results are presented and evaluated separately.

### 7.3.1 Main experiment results

GPT-6 uses the SOLIDWORKS API to produce native project files containing sketches, dimensional parameters, and modeling operations. This section first summarizes all 11 cases, then compares reference and reconstructed models from the same viewpoint. Geometric measurements and parameter-editing tests provide the basis for the case analyses.

#### 7.3.1.1 Overview

All 11 cases use the prompt specified in Section 7.2.2. Geometric agreement is measured by bidirectional root-mean-square (RMS) surface distance. The reference and reconstructed models are first triangulated. Points are sampled uniformly over each mesh's surface, allocating samples in proportion to triangle area. The nearest distance from each point to the other mesh is then calculated. A lower RMS value indicates closer overall agreement between the surfaces.

Reconstructed models are first checked for solid validity. Models that fail this check are excluded from parameter-editing tests and are reported as "Not tested (invalid solids)." This designation records failure of the validity prerequisite, not an observed editing failure. Each model that passes undergoes 10 single-parameter edits in SOLIDWORKS. In each test, only one dimensional parameter is actively changed; all other input parameters remain fixed, while dependent features update through the existing relationships. The original value is restored after each test. An edit passes only if it takes effect, the model rebuilds successfully, the geometry changes as intended, dependent structures update correctly, the edited model passes the solid validity checks, and the original state can be restored. Hole closure, damage to the geometry, or misalignment of components constitutes a failure. Table 7-4 presents the results.

Table 7-4 Overall results for parametric reconstruction from B-rep models

| Case ID | Complexity | Solid validity | RMS surface distance (mm) | Editing-test results |
|---|---|---|---|---|
| B2P-01 | Simple | 1/1 passed | 0.000698 | 10/10 passed |
| B2P-02 | Simple | 1/1 passed | 0.000158 | 10/10 passed |
| B2P-03 | Simple | 1/1 passed | 0.05565 | 4/10 passed |
| B2P-04 | Simple | 1/1 passed | 0.04671 | 8/10 passed |
| B2P-05 | Simple | 16/16 passed | 0.00175 | 10/10 passed |
| B2P-06 | Moderate | 2/3 passed | 2.9687 | Not tested<br>(invalid solids) |
| B2P-07 | Moderate | 1/1 passed | 0.5304 | 10/10 passed |
| B2P-08 | Moderate | 23/24 passed | 0.4193 | Not tested<br>(invalid solids) |

| Case ID | Complexity | Solid validity | RMS surface distance (mm) | Editing-test results |
|---|---|---|---|---|
| B2P-09 | Complex | 20/23 passed | 9.9593 | Not tested<br>(invalid solids) |
| B2P-10 | Complex | 0/1 passed | 2.6181 | Not tested<br>(invalid solids) |
| B2P-11 | Complex | 262/278 passed | 2.2087 | Not tested<br>(invalid solids) |

Table 7-5 reports the time GPT-6 takes to reconstruct each case.

Table 7-5  Reconstruction time for each case

| Case ID | B2P-01 | B2P-02 | B2P-03 | B2P-04 | B2P-05 | B2P-06 |
|---|---|---|---|---|---|---|
| Time | 1 min | 1 min | 9 min | 16 min | 11 min | 13 min |
| **Case ID** | **B2P-07** | **B2P-08** | **B2P-09** | **B2P-10** | **B2P-11** | |
| Time | 6 min | 21 min | 33 min | 5 min | 1 h 52 min | |

### 7.3.1.2 Case analyses

### B2P-01 Three-arm flanged shaft (simple)

| Reference CAD model | GPT-6 reconstruction |
|---|---|
| 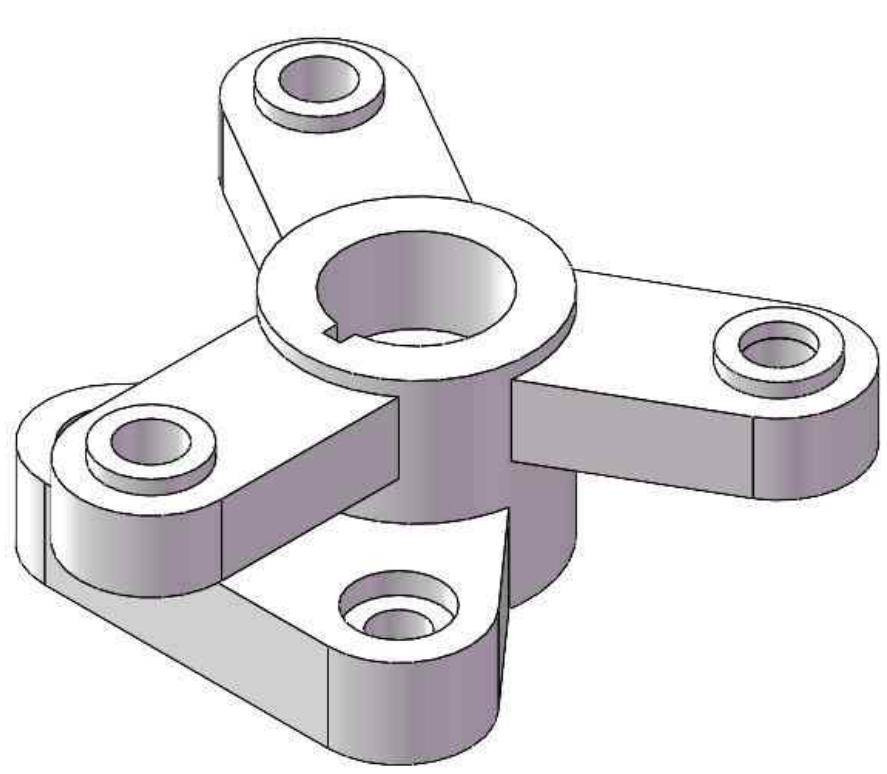 | 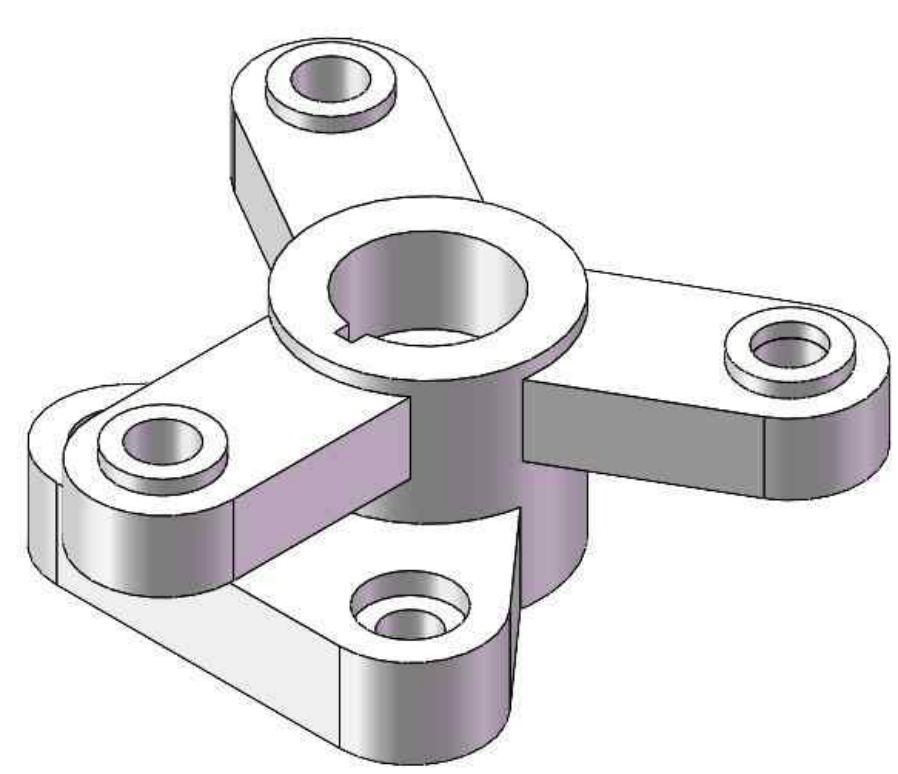 |
| Faces: 60 | Faces: 61 |

Figure 7-1  Reference and reconstructed models for B2P-01

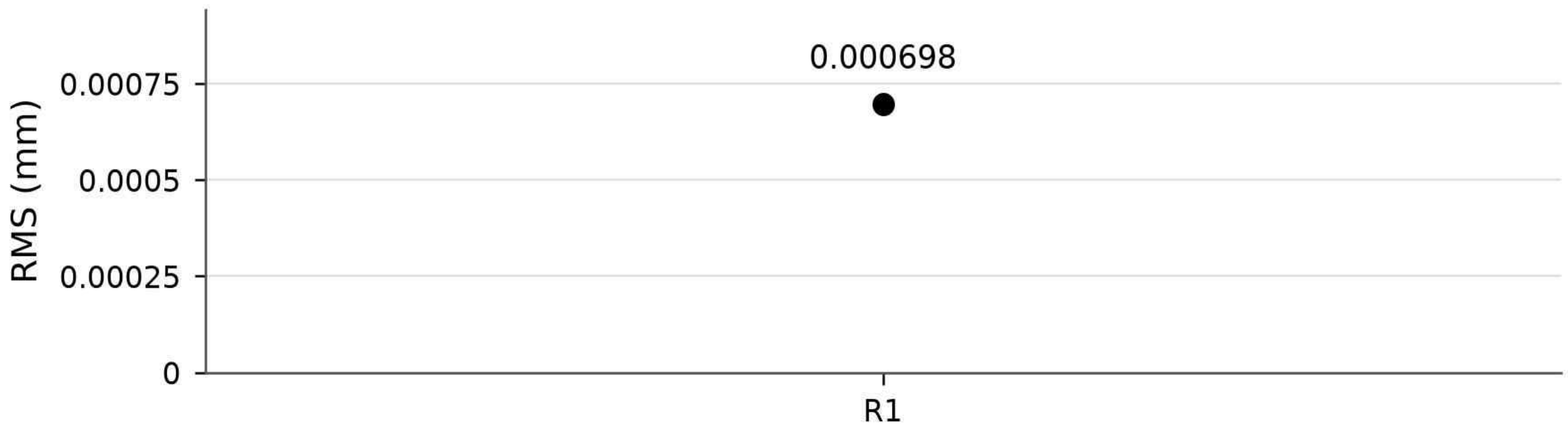

Figure 7-2  B2P-01 RMS surface distance by iteration

| **Before the edit** | **After the edit** |
| --- | --- |
| 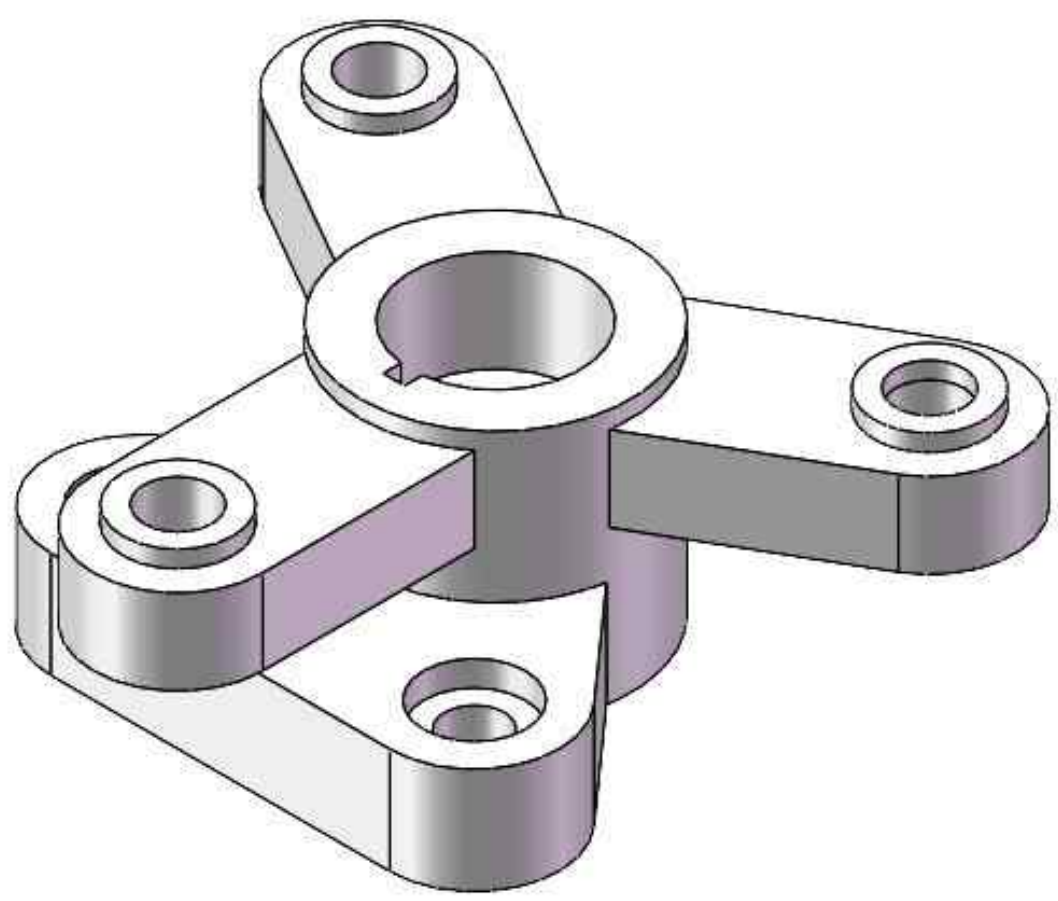 | 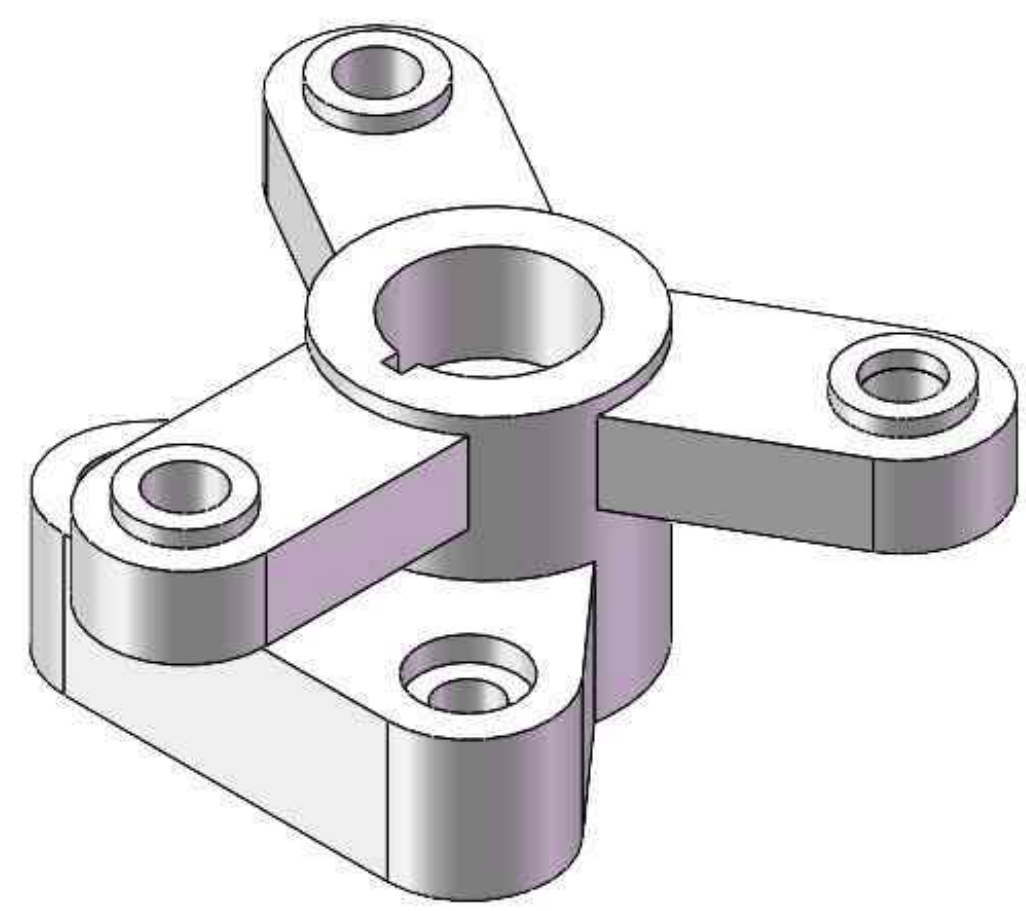 |

Figure 7-3  B2P-01 parameter edit: BasePlate, 19 → 28.5 mm; passed

The reconstructed model is a valid solid. GPT-6 completes only one iteration, recovering the main structure of the three-arm flanged shaft with geometry very close to the reference. All 10 parameter edits pass: the geometry updates correctly and can be restored after each edit.

## B2P-02 Connecting rod (simple)

| **Reference CAD model** | **GPT-6 reconstruction** |
|---|---|
| 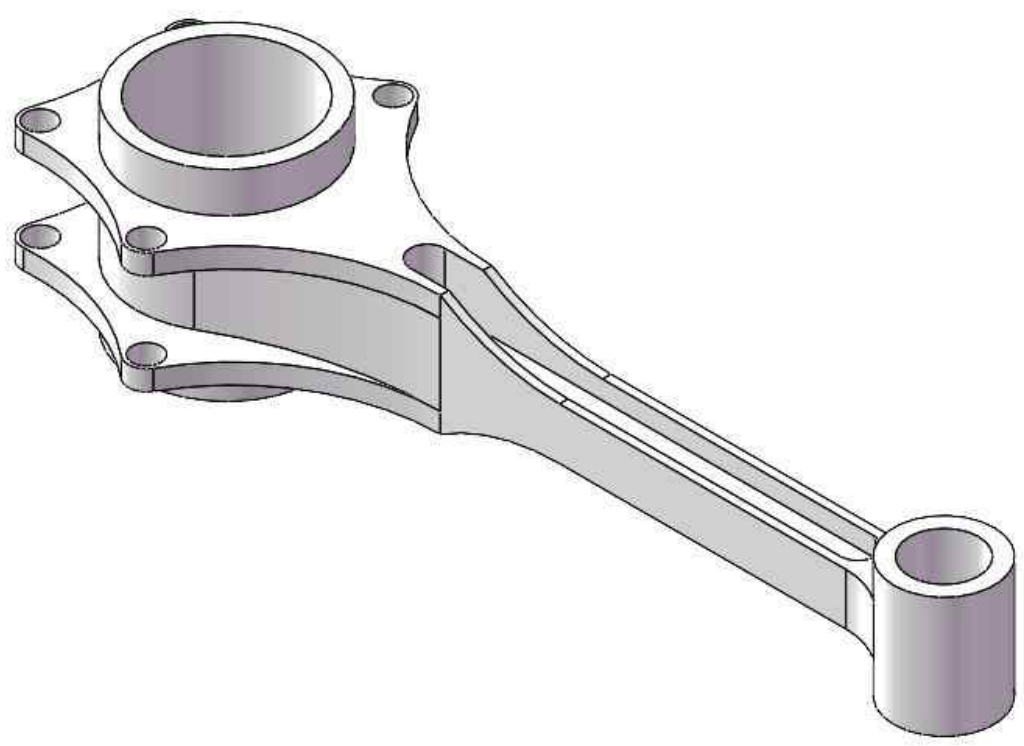 | 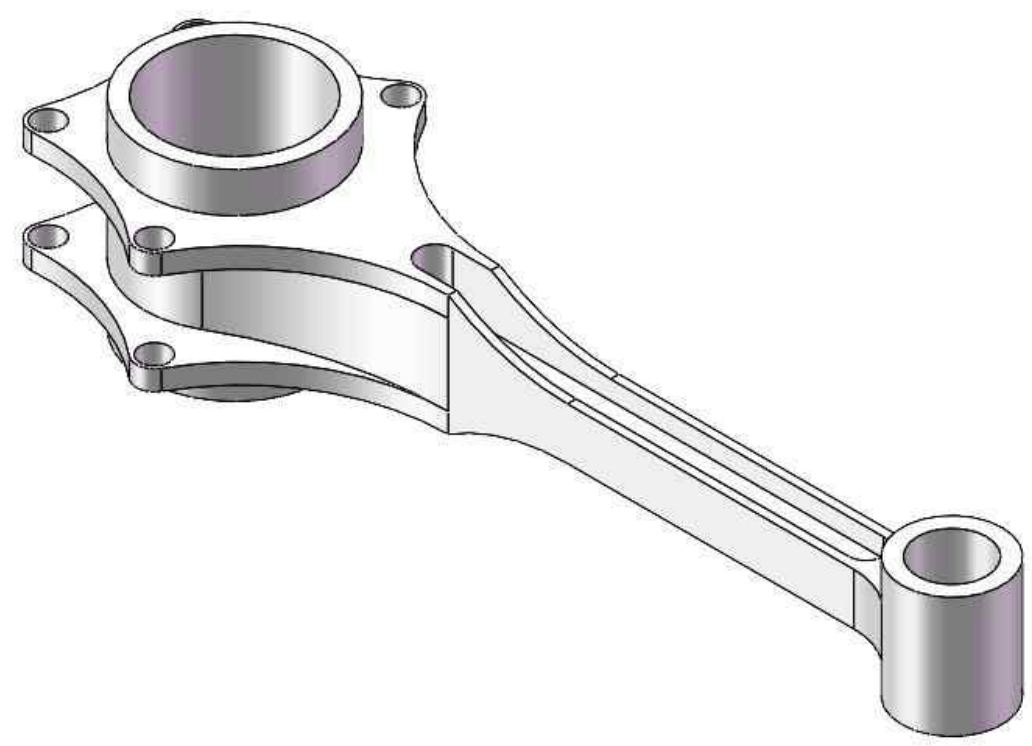 |
| Faces: 83 | Faces: 83 |

Figure 7-4  Reference and reconstructed models for B2P-02

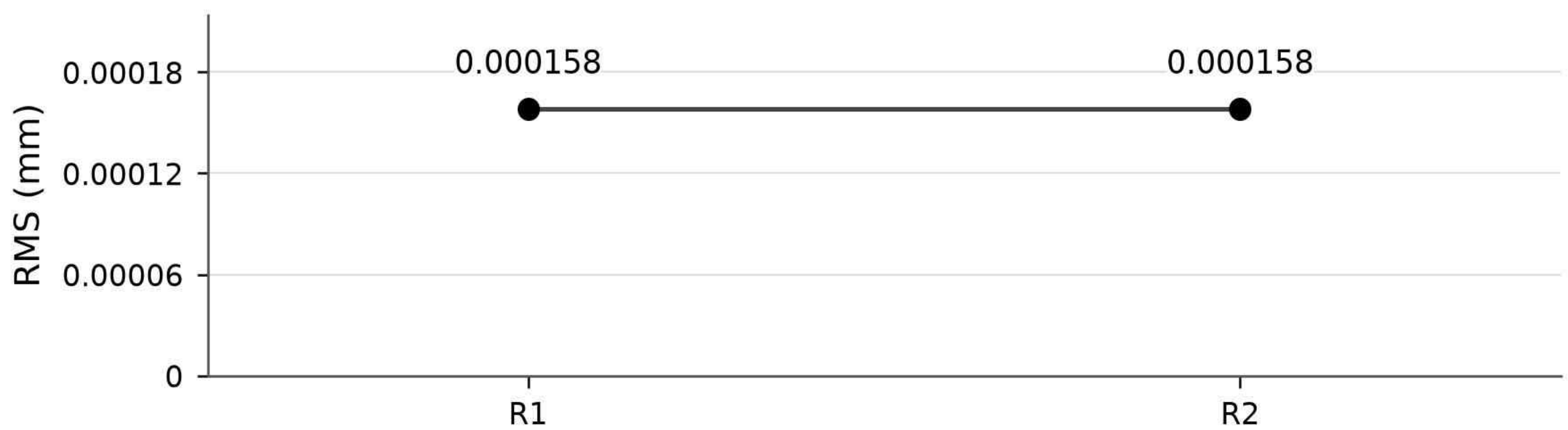

Figure 7-5  B2P-02 RMS surface distance by iteration

| **Before the edit** | **After the edit** |
|---|---|
| 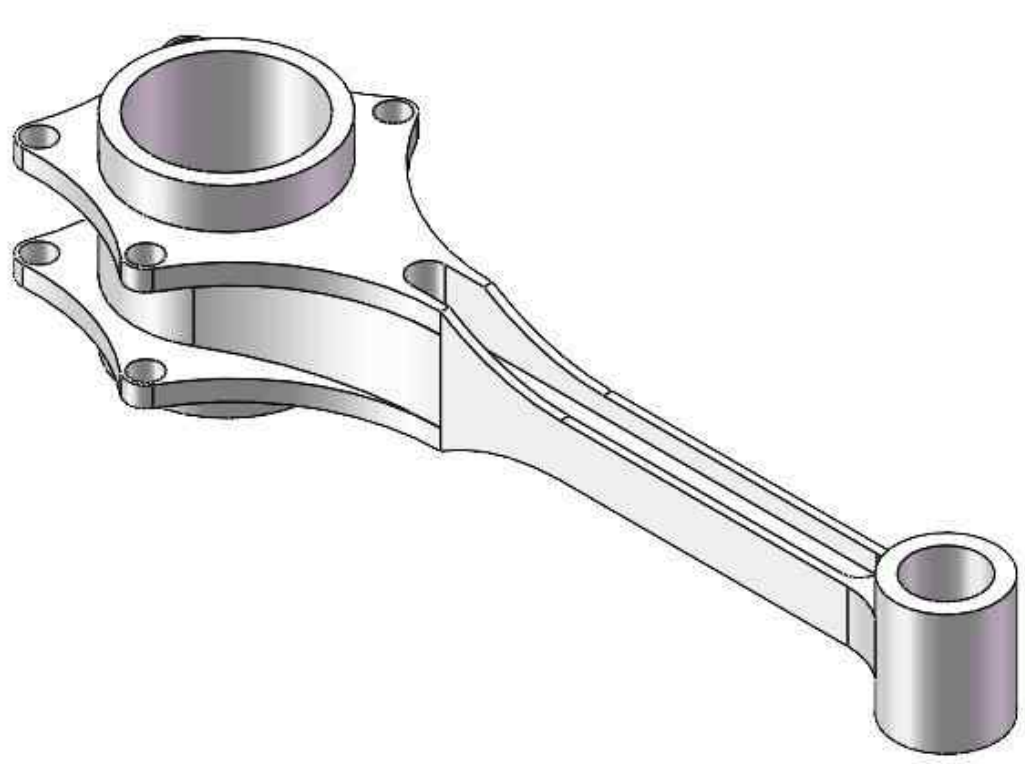 | 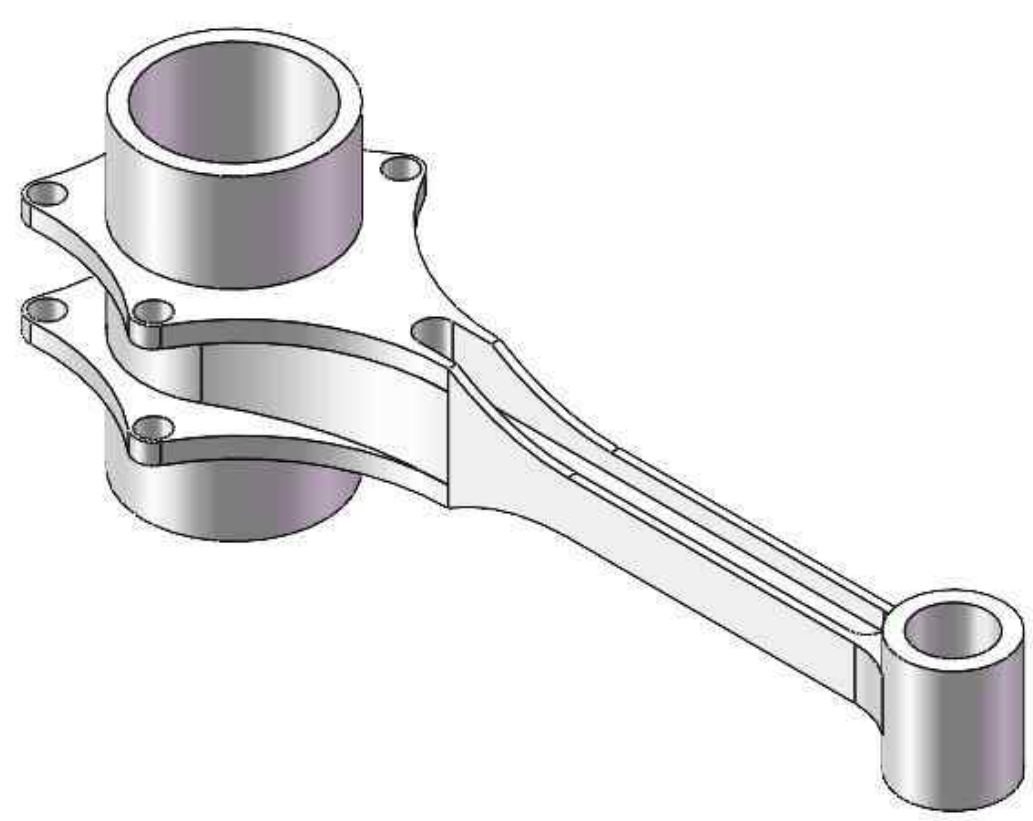 |

Figure 7-6  B2P-02 parameter edit: BigEndWidth, 70 → 109.9 mm; passed

The reconstructed model is a valid solid. RMS is 0.000158 mm in both iterations, with no evident geometric improvement between the saved versions. The overall geometry of the rod body and the holes at both ends is close to the reference. All 10 parameter edits pass, with the geometry updating correctly and returning to its original state after each test.

### B2P-03 Flanged housing (simple)

| Reference CAD model | GPT-6 reconstruction |
|---|---|
| 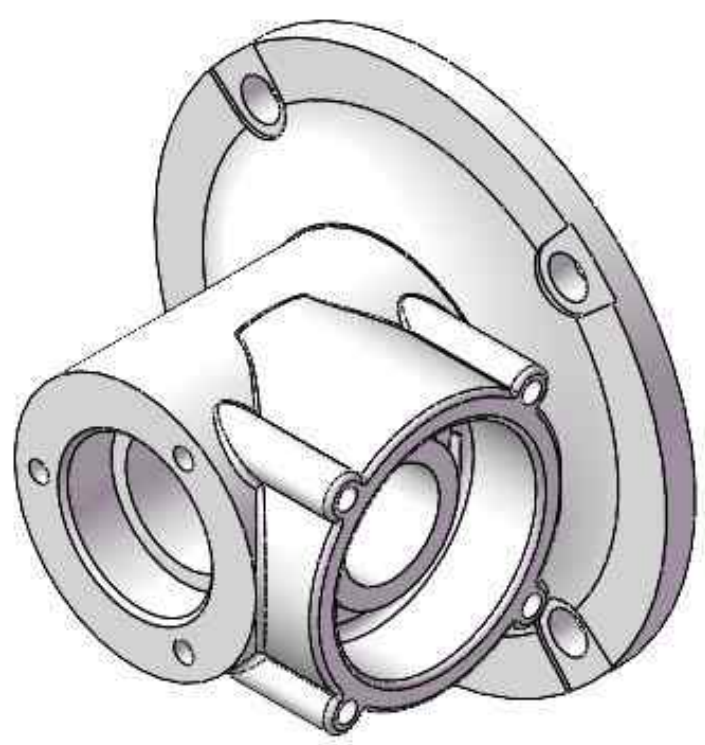 | 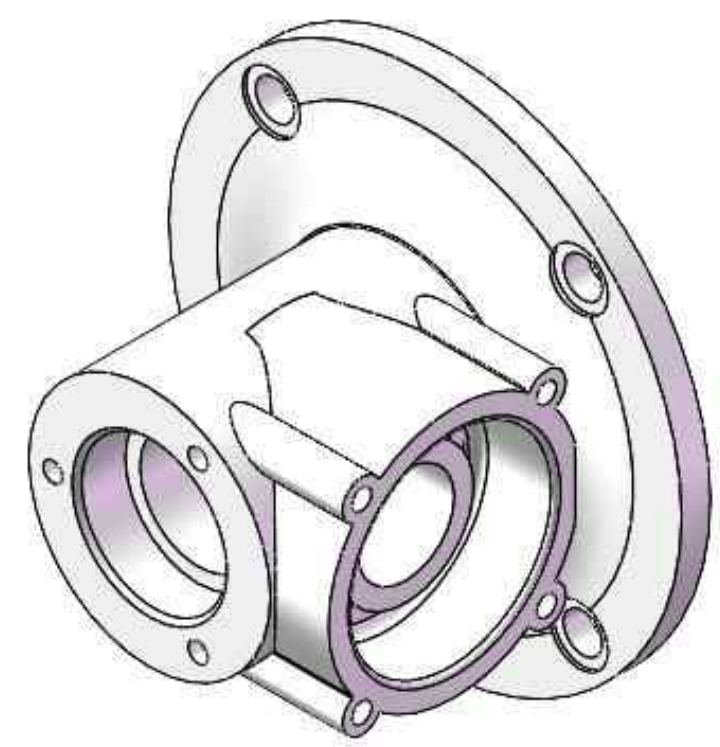 |
| Faces: 178 | Faces: 114 |

Figure 7-7 Reference and reconstructed models for B2P-03

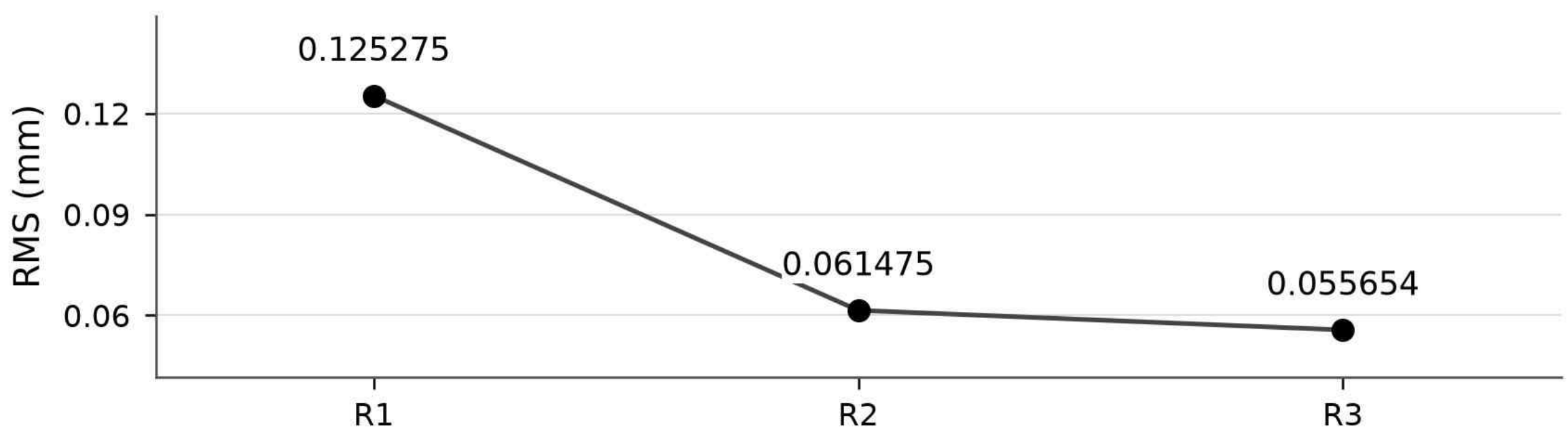

Figure 7-8 B2P-03 RMS surface distance by iteration

| Before the edit | After the edit |
|---|---|
| 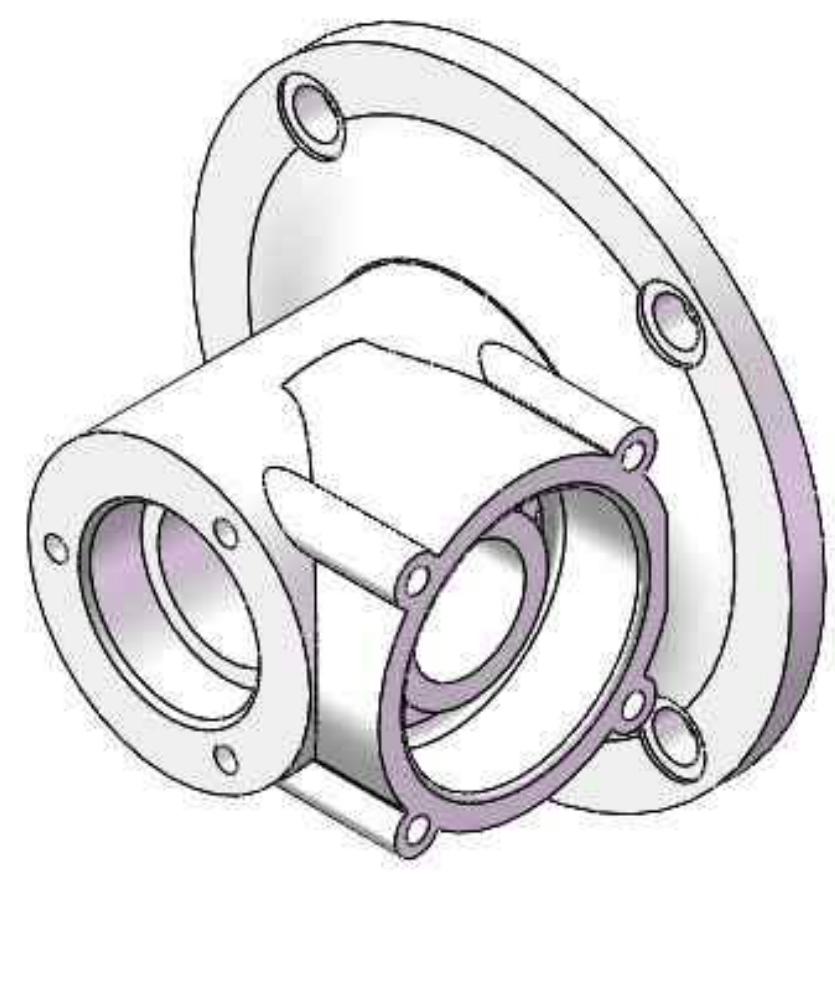 | 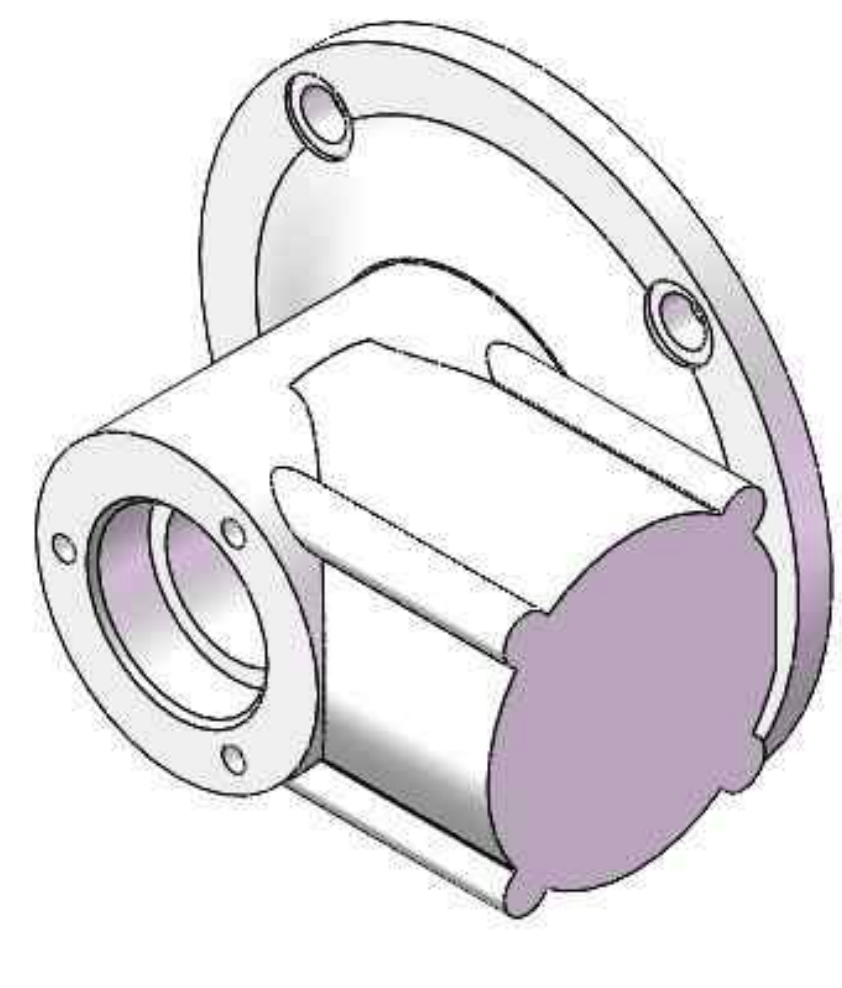 |

Figure 7-9 B2P-03 parameter edit: SideHousing_Boss depth, 36 → 54 mm; failed

The reconstructed model is a valid solid. Over three iterations, RMS decreases from 0.1252 to 0.05565 mm. The model closely reproduces the cylindrical sections in two directions, the base flange, and the main holes. Of the 10 parameter edits, four pass and six fail. In the illustrated test, increasing the side-housing depth from 36 to 54 mm closes the original hole opening. Other failures include truncation of the side structure, disappearance of mounting holes or counterbore shoulders, and errors when rebuilding dependent features.

### B2P-04 Multi-passage flanged housing (simple)

| Reference CAD model | GPT-6 reconstruction |
|---|---|
| 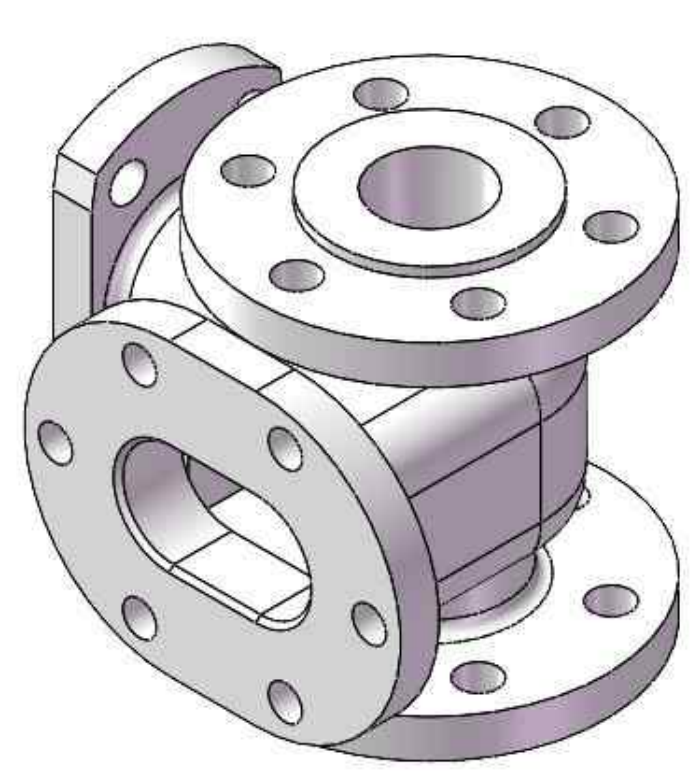 | 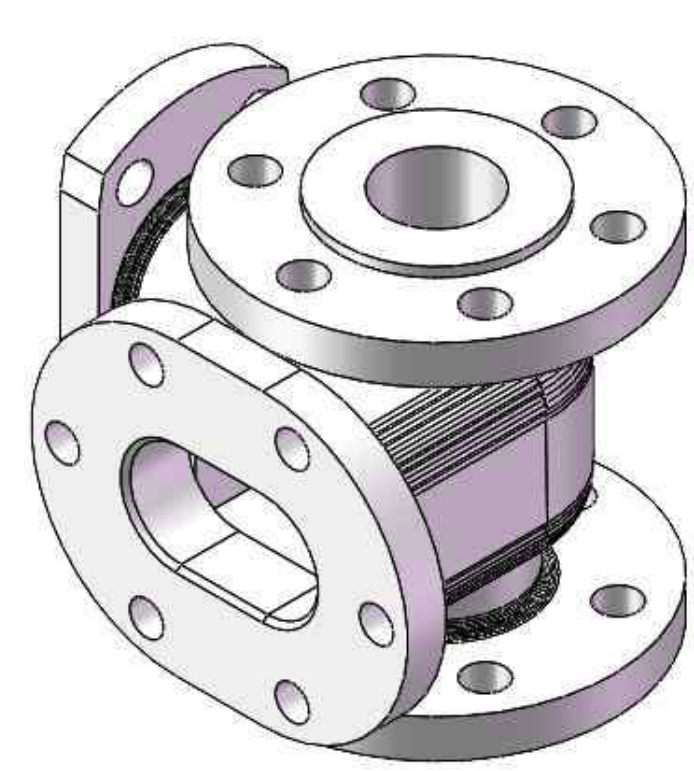 |
| Faces: 153 | Faces: 1,401 |

Figure 7-10 Reference and reconstructed models for B2P-04

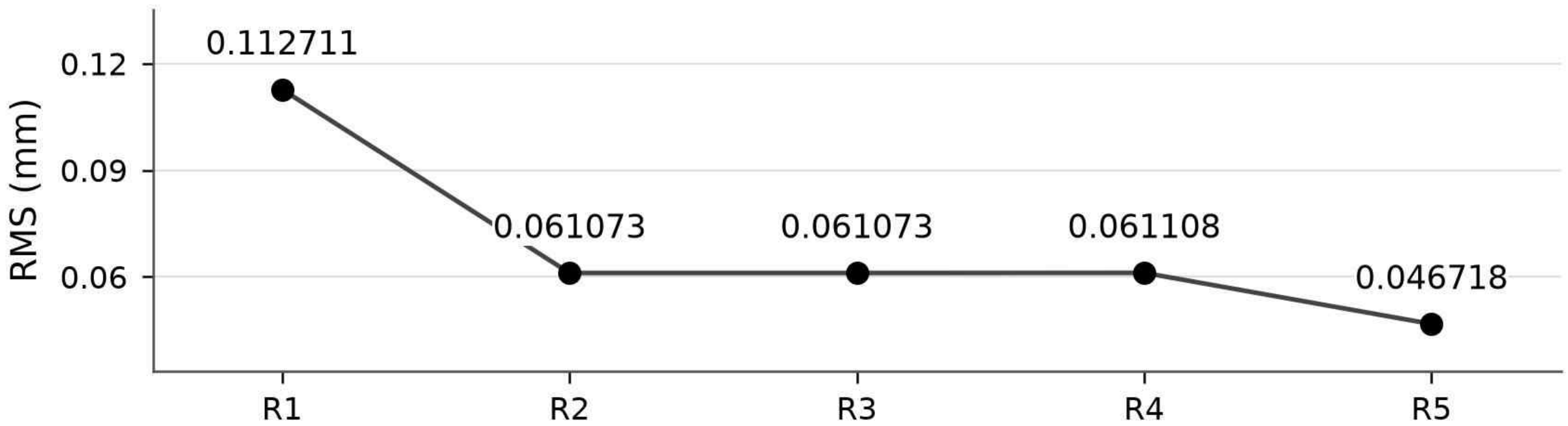

Figure 7-11 B2P-04 RMS surface distance by iteration

| **Before the edit** | **After the edit** |
| --- | --- |
| 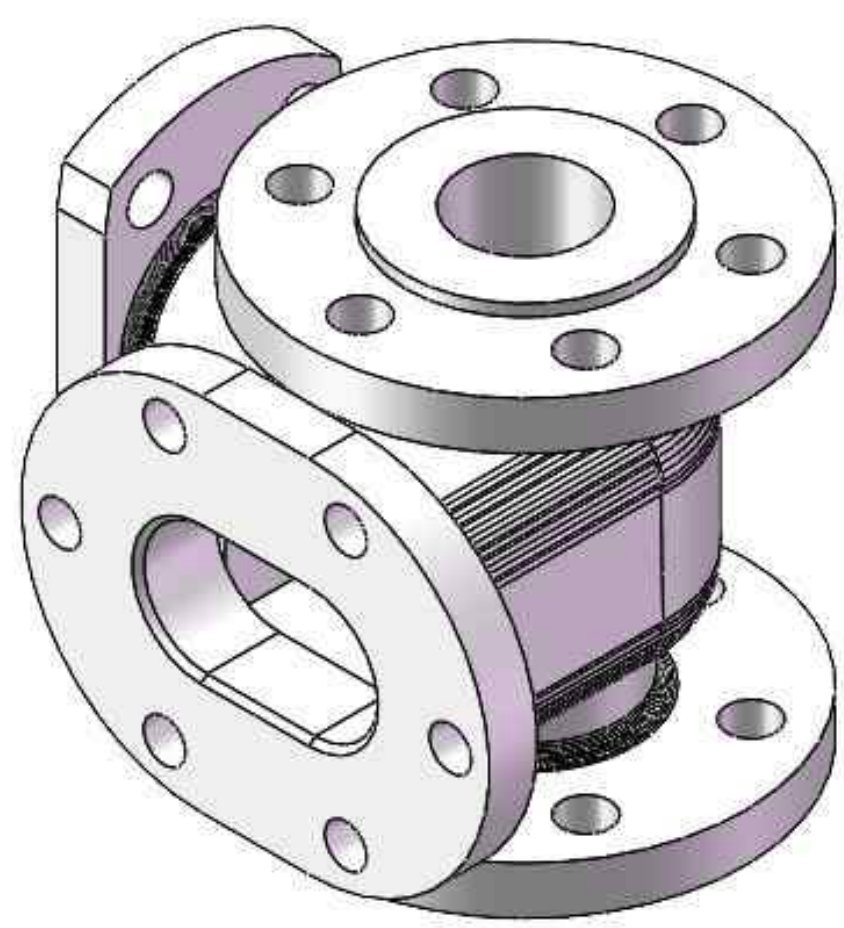 | 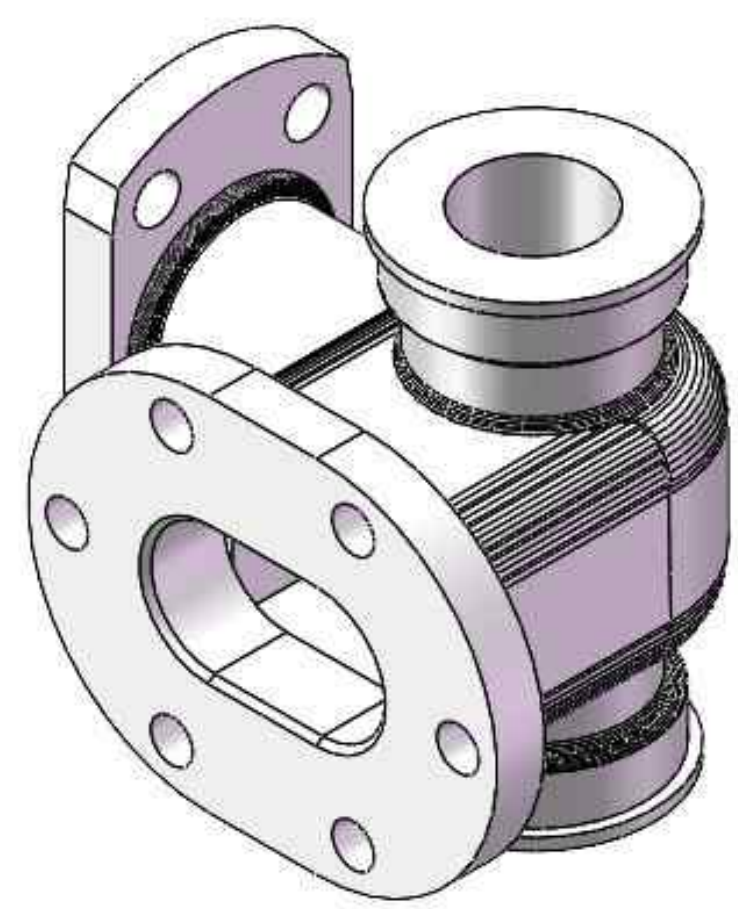 |

Figure 7-12 B2P-04 parameter edit: MainFlangeOD, 146 → 73 mm; failed

The reconstructed model is a valid solid. RMS decreases from 0.1127 to 0.04671 mm over five iterations. Eight of the 10 parameter edits pass and two fail. In the illustrated test, the main flange's outside diameter is reduced from 146 to 73 mm, but the mounting-hole positions do not update with it. The holes fall outside the smaller flange and disappear.

## B2P-05 V-belt pulley and bearing support assembly (simple)

| Reference CAD model | GPT-6 reconstruction |
|---|---|
| 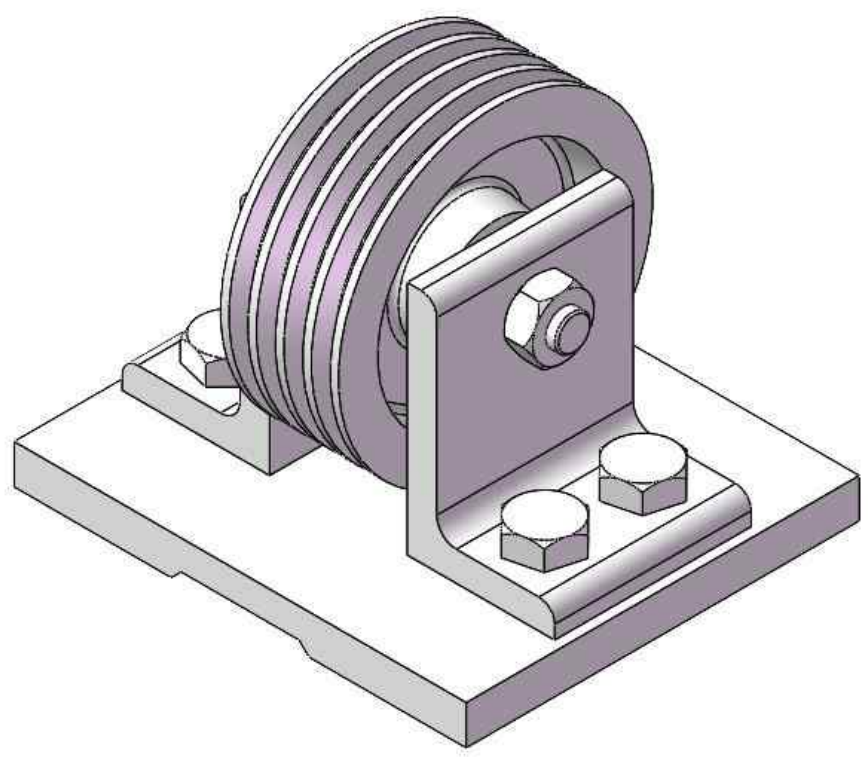 | 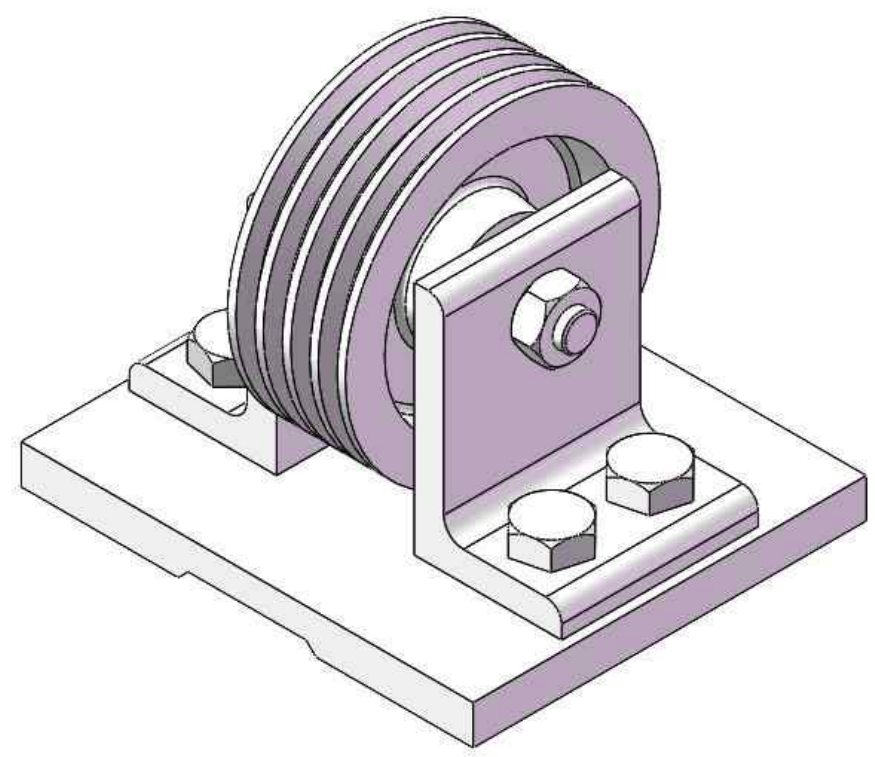 |
| Faces: 348; solids: 16 | Faces: 351; solids: 16 |

Figure 7-13 Reference and reconstructed models for B2P-05

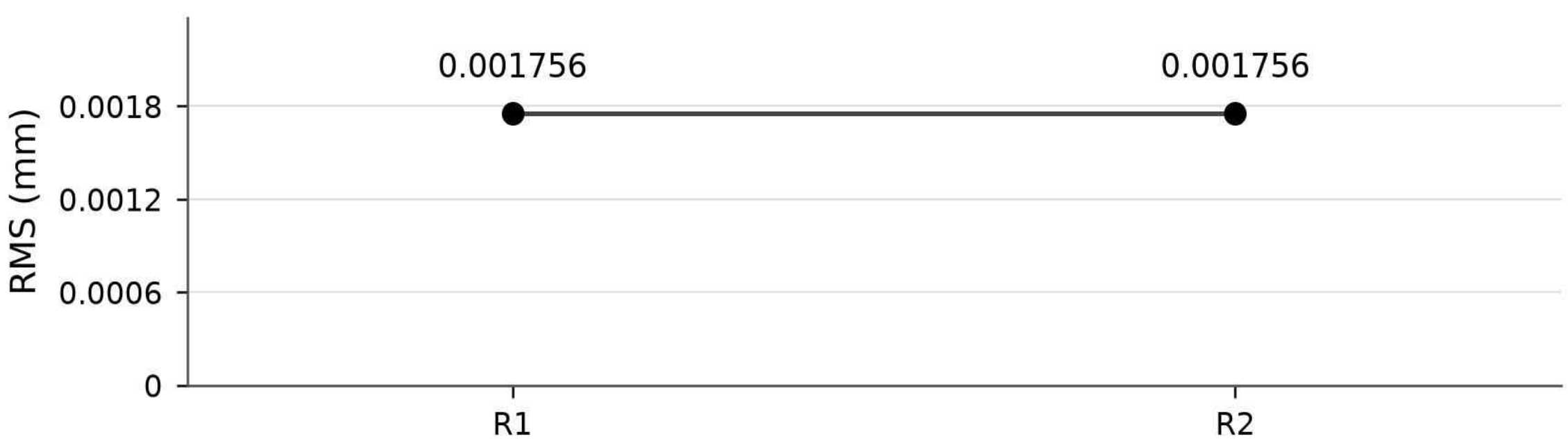

Figure 7-14 B2P-05 RMS surface distance by iteration

| Before the edit | After the edit |
|---|---|
| 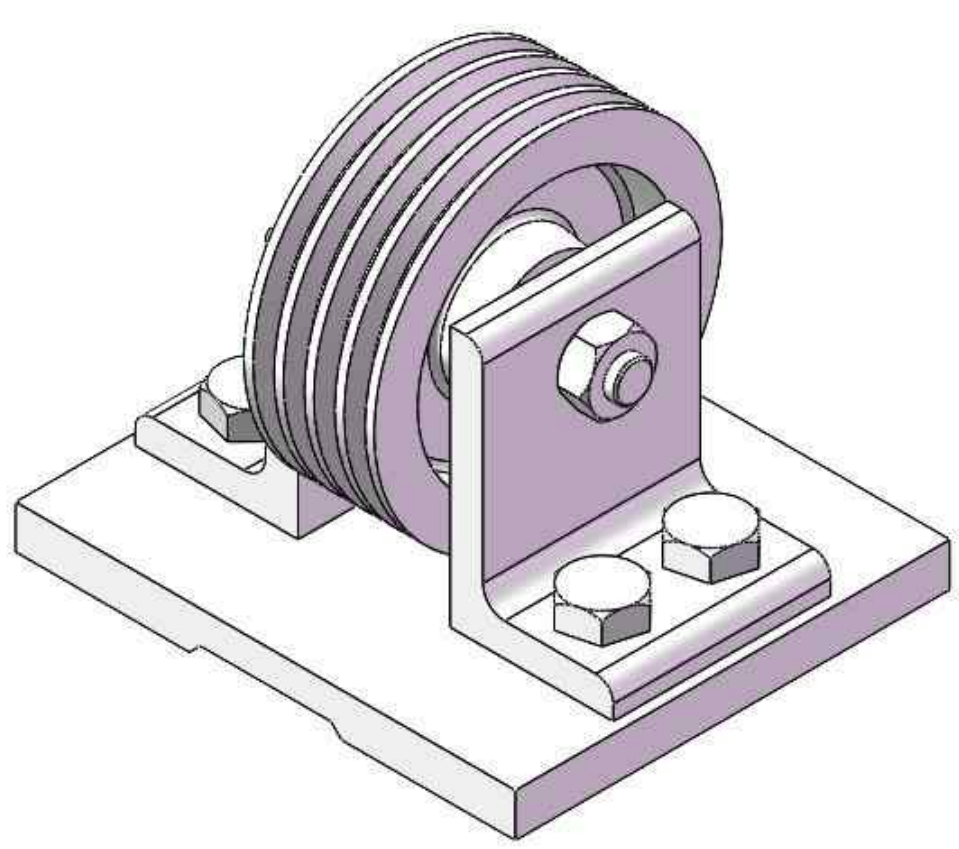 | 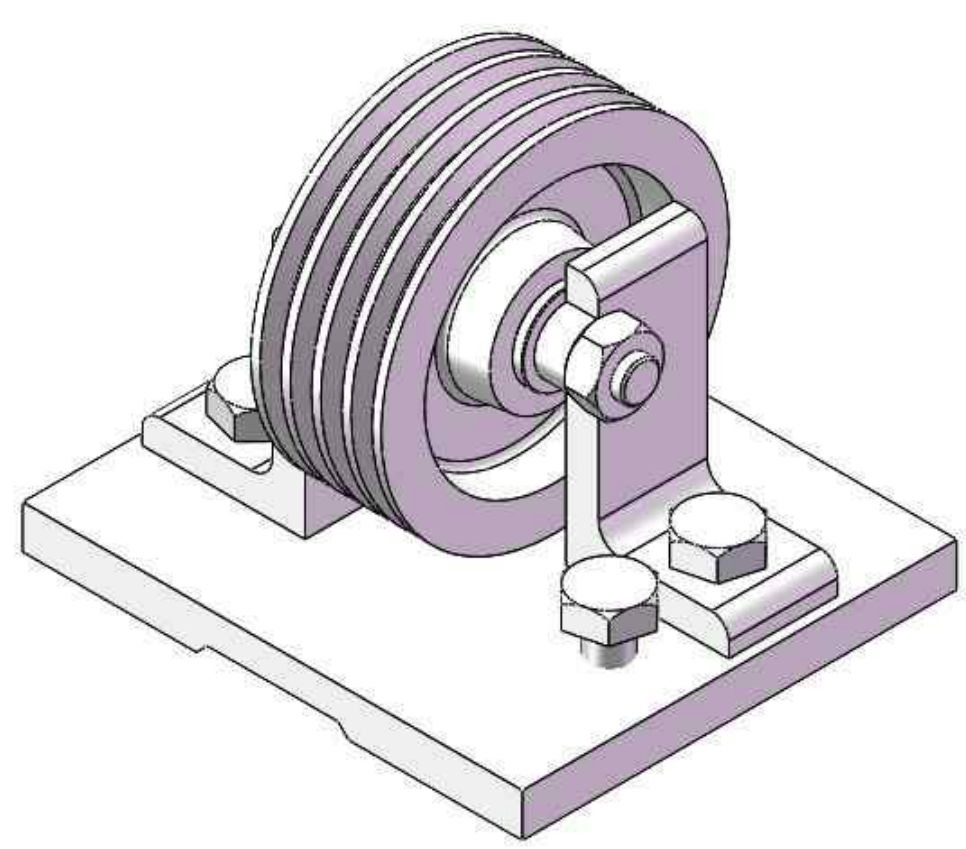 |

Figure 7-15 B2P-05 parameter edit: 15_Bearing_Right / BracketWidth, 100 → 50 mm; passed

All 16 parts in the reconstructed assembly are valid solids. RMS is approximately 0.0018 mm in both iterations. The reference and reconstructed STEP assemblies each contain 16 solids, and measurements indicate close agreement in their main structures. All 10 parameter edits pass, with the geometry updating correctly and returning to its original state. In the illustrated test, the right support's width is reduced from 100 to 50 mm, while the other components remain intact and behave normally.

**B2P-06 Centrifugal pump housing (moderate)**

| Reference CAD model | GPT-6 reconstruction |
|---|---|
| 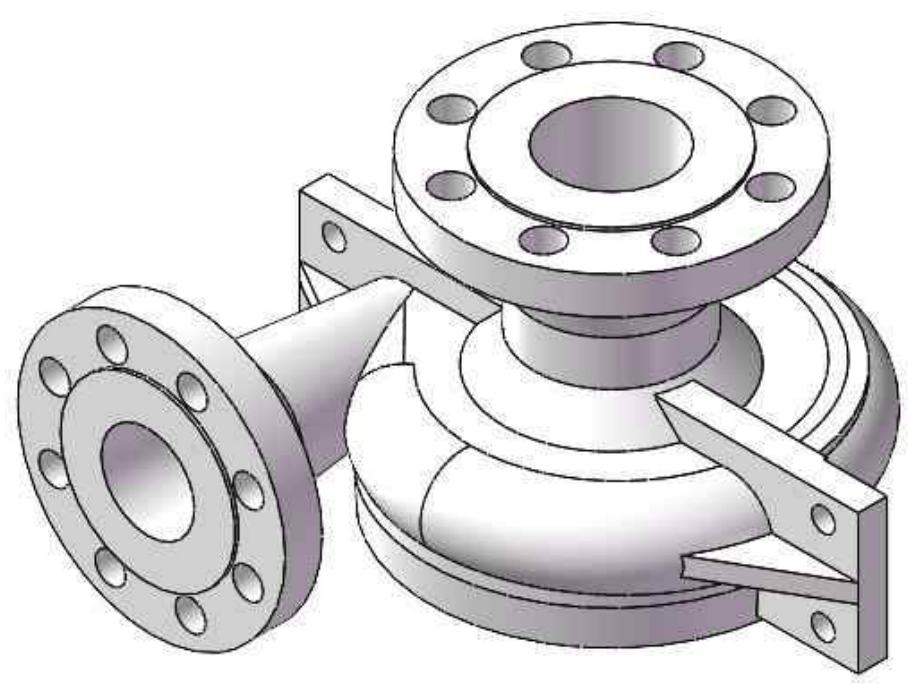 | 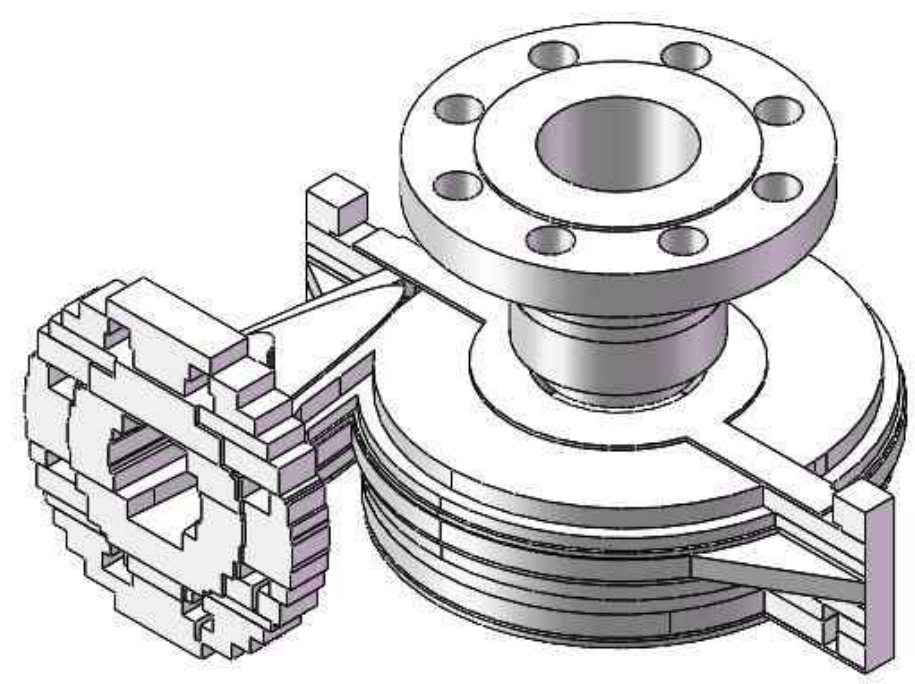 |
| Faces: 155 | Faces: 1,081 |

Figure 7-16  Reference and reconstructed models for B2P-06

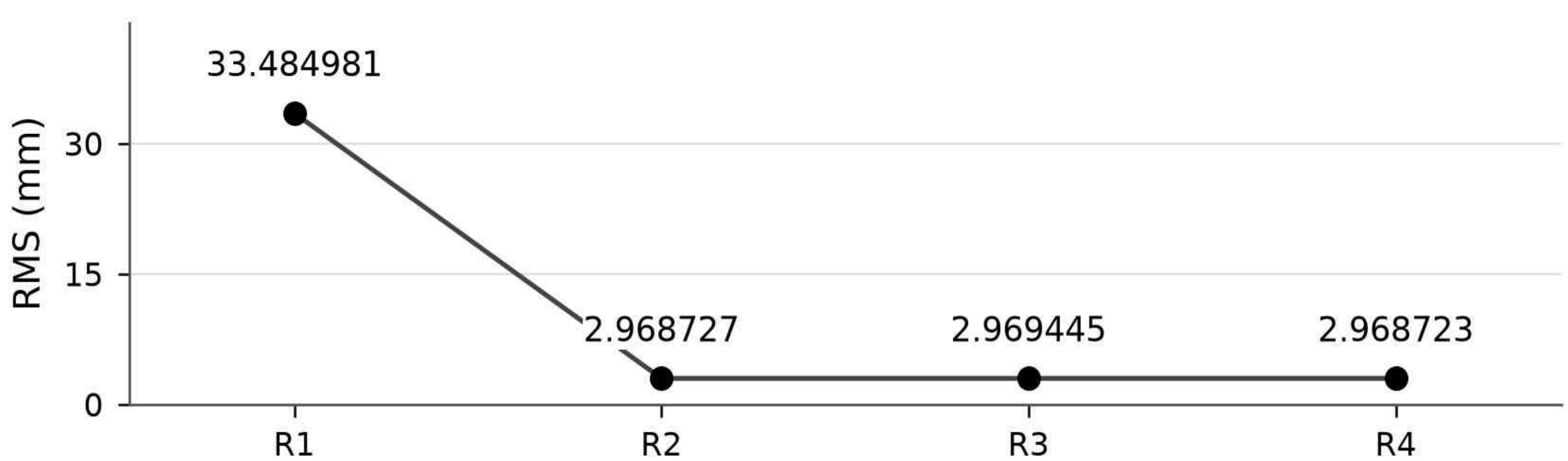

Figure 7-17  B2P-06 RMS surface distance by iteration

The reference model contains one solid; the reconstruction contains three, of which two are valid and one is invalid. Across four iterations, RMS falls from 33.485 to approximately 2.9687 mm and then levels off. The pump chamber, flow passages, and their connections still differ from the reference. Because one solid fails the validity check, the model is excluded from parameter-editing tests.

### B2P-07 Control rod shaft (moderate)

| Reference CAD model | GPT-6 reconstruction |
|---|---|
| 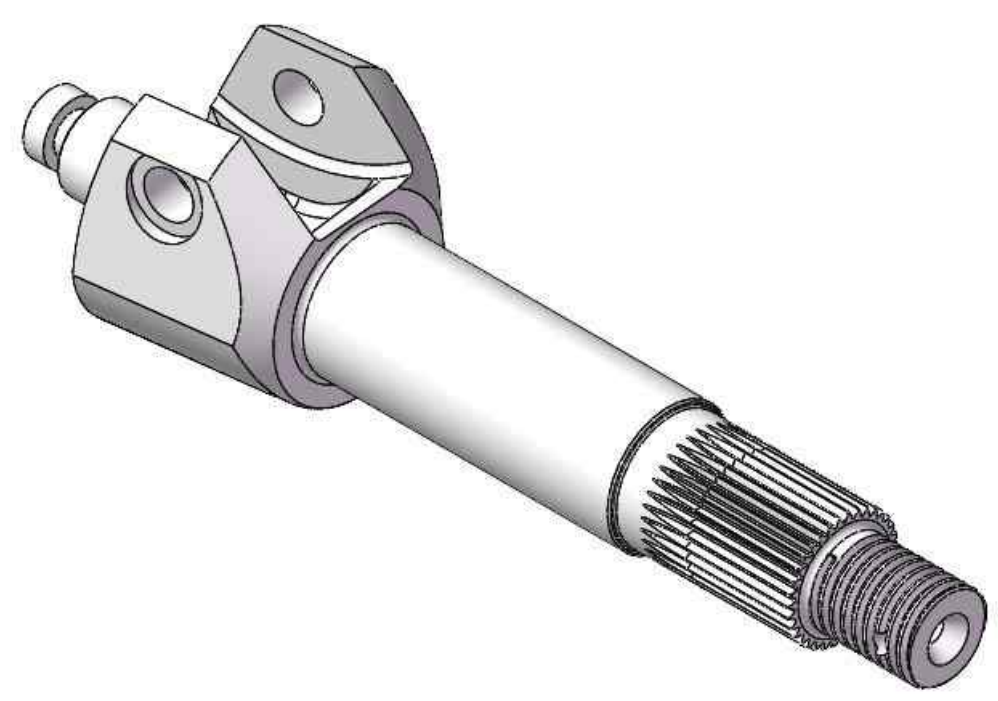 | 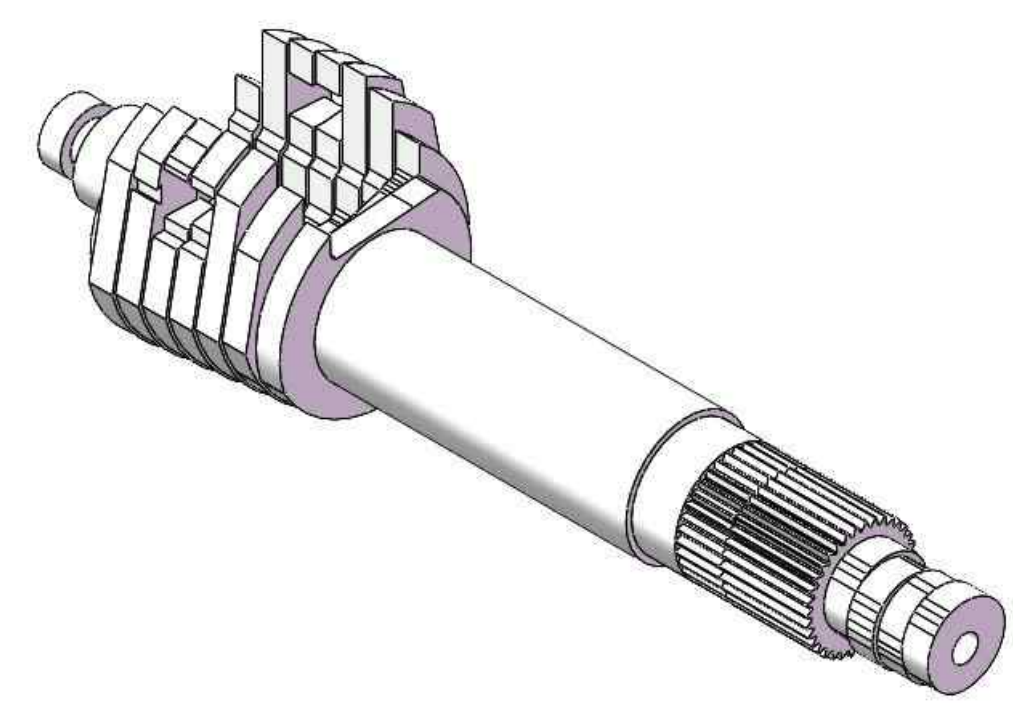 |
| Faces: 380 | Faces: 825 |

Figure 7-18  Reference and reconstructed models for B2P-07

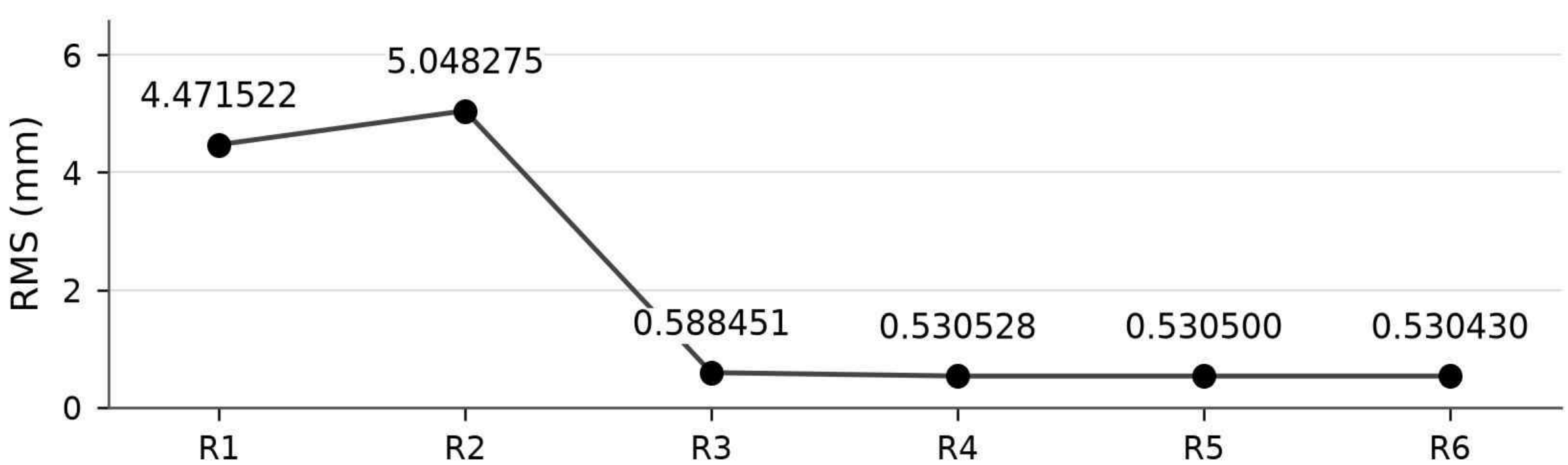

Figure 7-19  B2P-07 RMS surface distance by iteration

| Before the edit | After the edit |
|---|---|
| 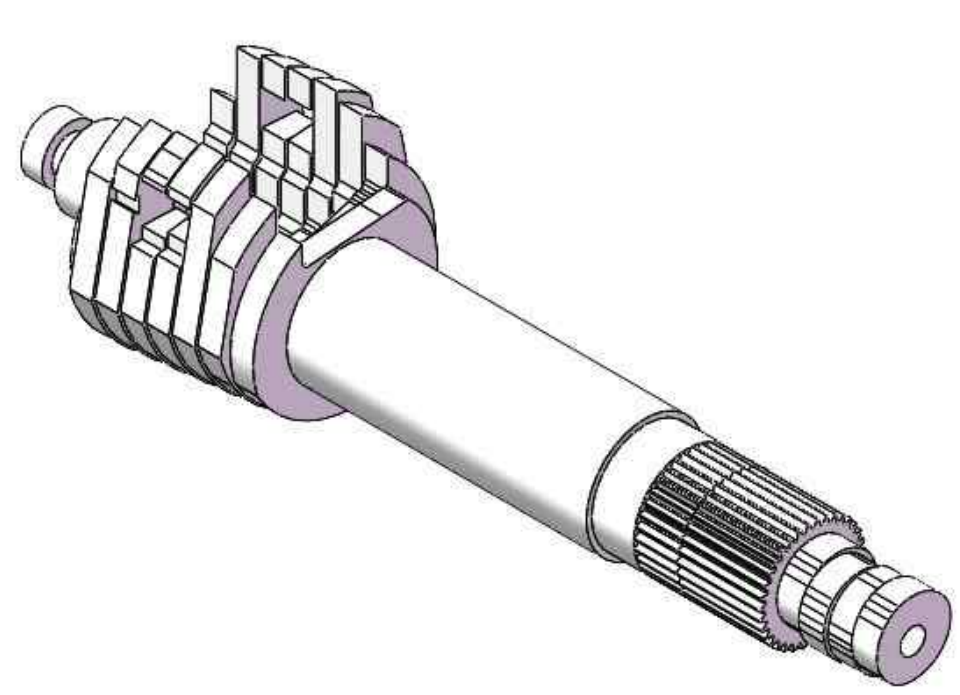 | 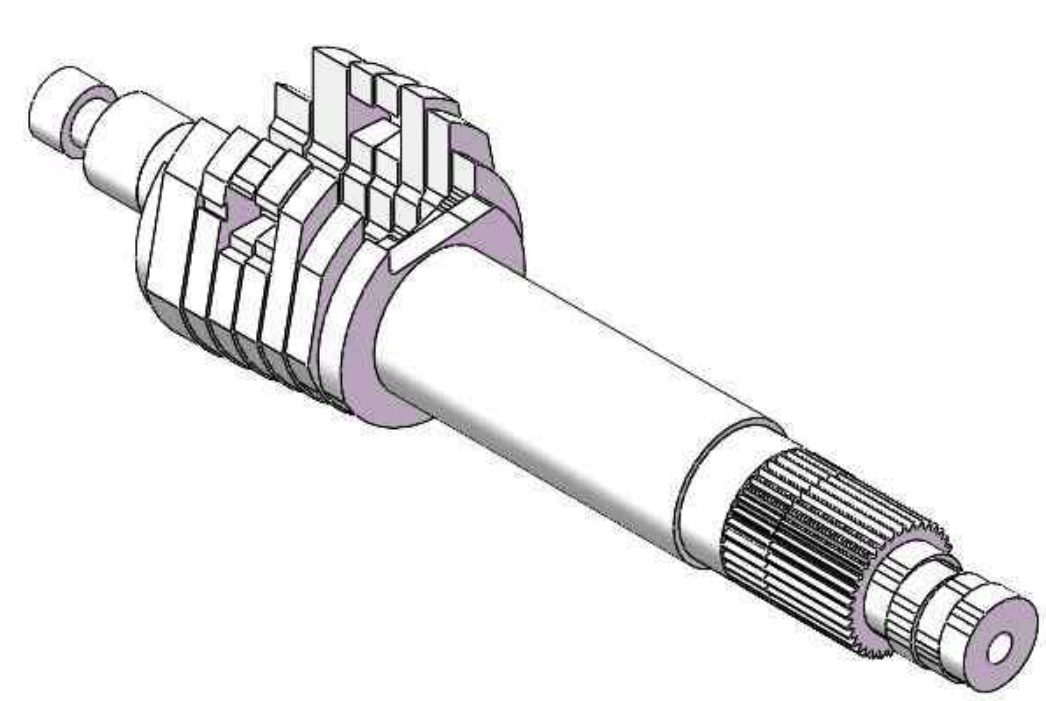 |

Figure 7-20  B2P-07 parameter edit: Zone1Length, 44.75 → 67.125 mm; passed

The reconstructed model is a valid solid. Across six iterations, RMS initially rises from 4.4715 to 5.0483 mm, then falls to approximately 0.5304 mm. Local geometry, including the junction between the forked head and shaft, still differs; the model is only visually similar to the reference. All 10 parameter edits pass, with the geometry updating correctly and returning to its original state.

**B2P-08 Lathe chuck (moderate)**

| Reference CAD model | GPT-6 reconstruction |
| --- | --- |
| 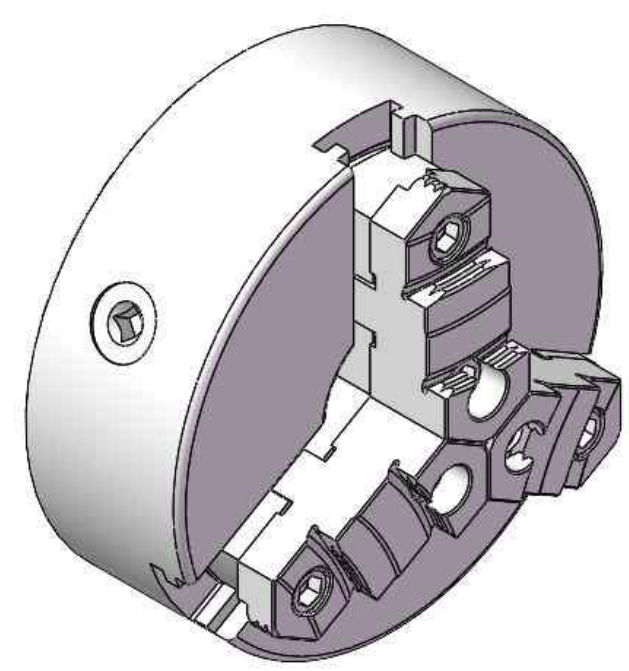 | 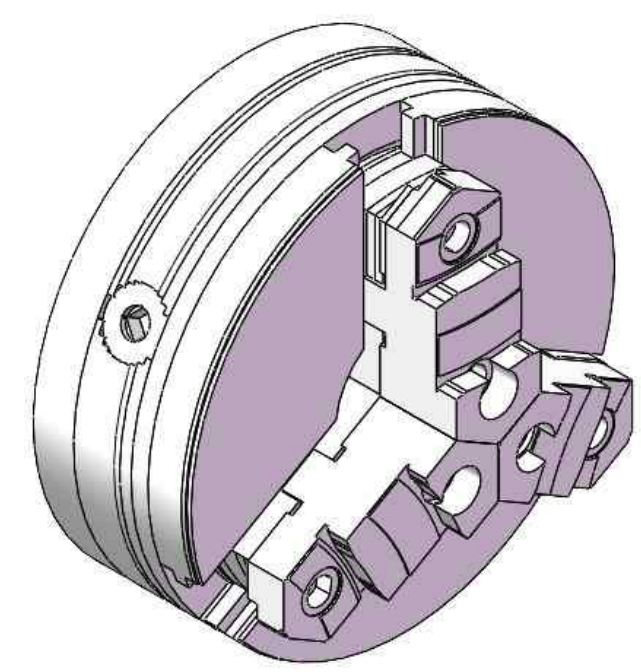 |
| Faces: 1,628; solids: 24 | Faces: 6,544; solids: 24 |

Figure 7-21 Reference and reconstructed models for B2P-08

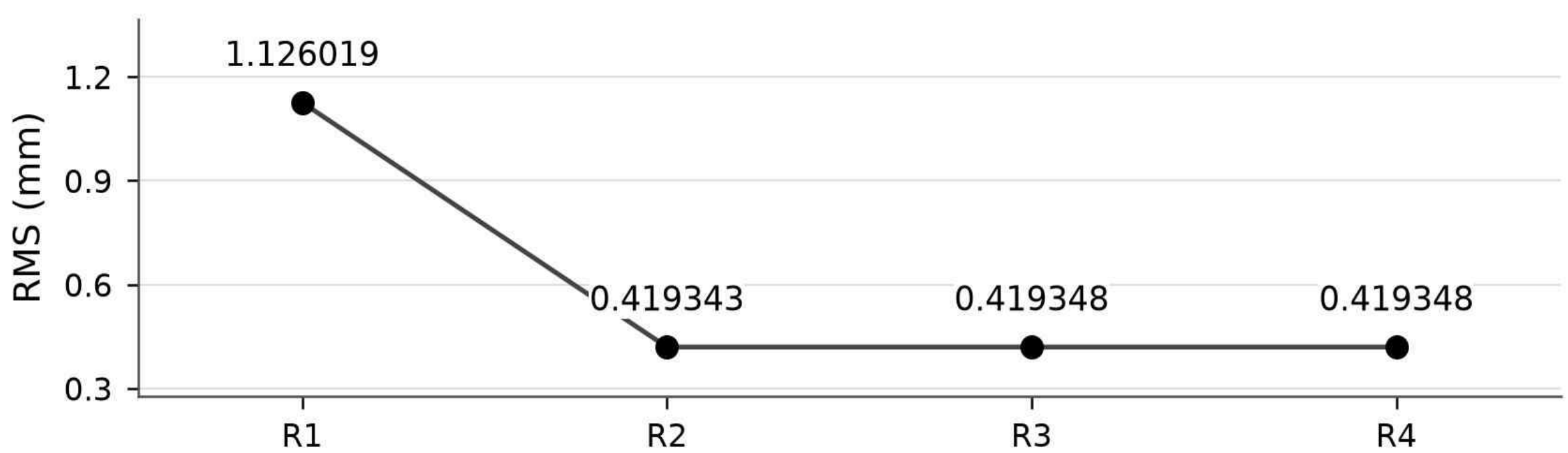

Figure 7-22 B2P-08 RMS surface distance by iteration

The reconstruction contains 23 valid solids and one invalid solid, as well as 371 faces not belonging to any solid. Across four iterations, RMS decreases from 1.1260 to approximately 0.4193 mm and then remains essentially unchanged. Because one solid fails the validity check, the model is excluded from parameter-editing tests.

### B2P-09 Gearbox housing cover (complex)

| Reference CAD model | GPT-6 reconstruction |
| --- | --- |
| 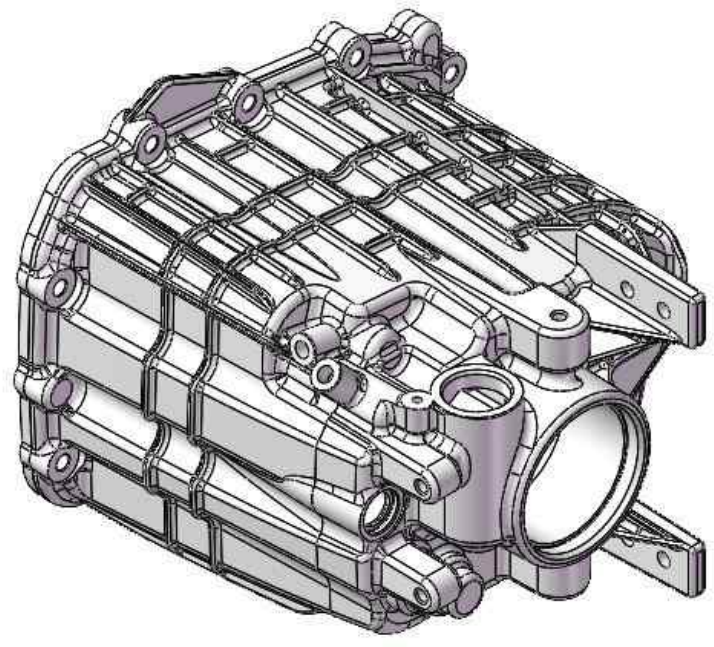 | 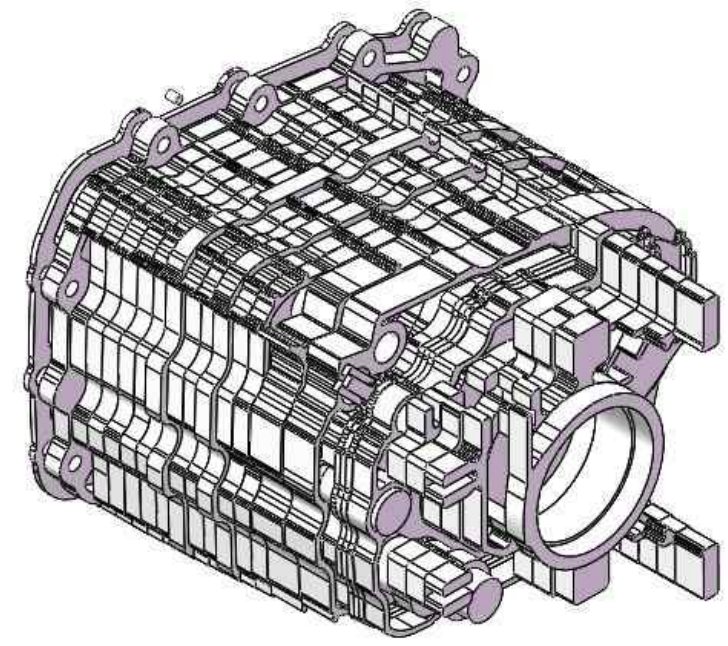 |
| Faces: 5,639 | Faces: 8,466 |

Figure 7-23  Reference and reconstructed models for B2P-09

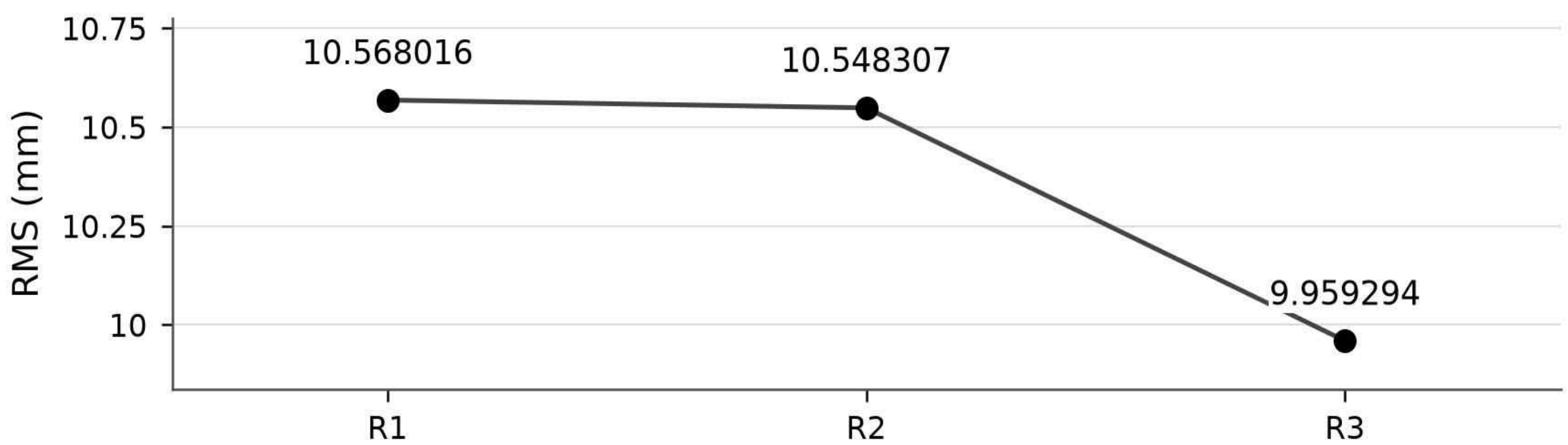

Figure 7-24  B2P-09 RMS surface distance by iteration

The reference contains only three solids, whereas the reconstruction contains 23: 20 valid and three invalid. RMS decreases from 10.5680 to 9.9593 mm over three iterations, leaving a large overall deviation. A series of extrusions produces simple surfaces that approximate the cover’s complex surfaces, with dense segmentation visible on the exterior. Because some solids fail the validity check, the model is excluded from parameter-editing tests.

### B2P-10 V8 engine block (complex)

| Reference CAD model | GPT-6 reconstruction |
| --- | --- |
| 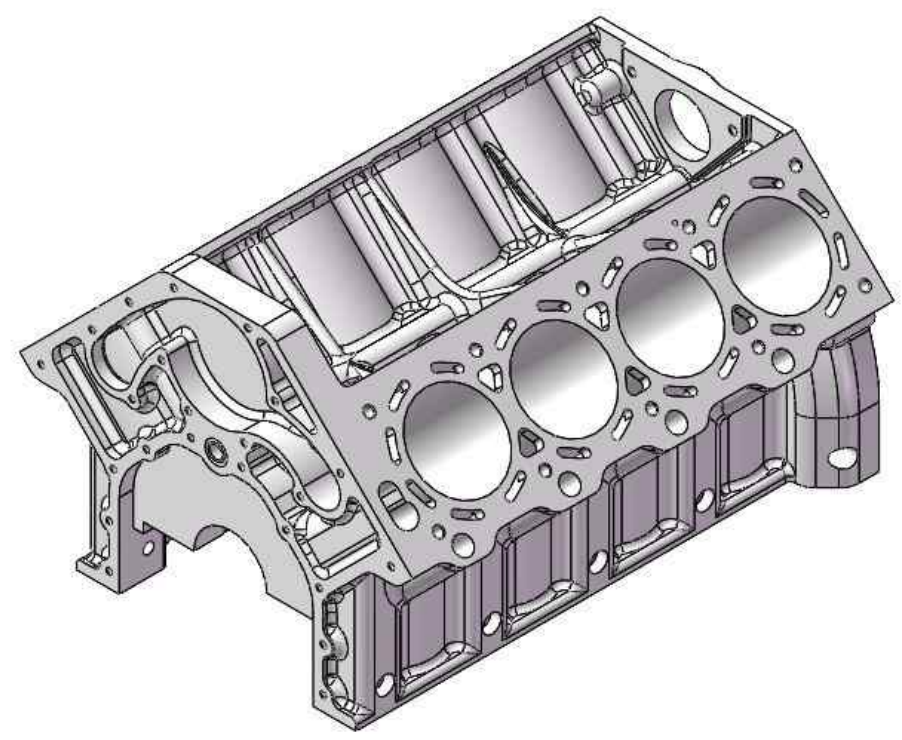 | 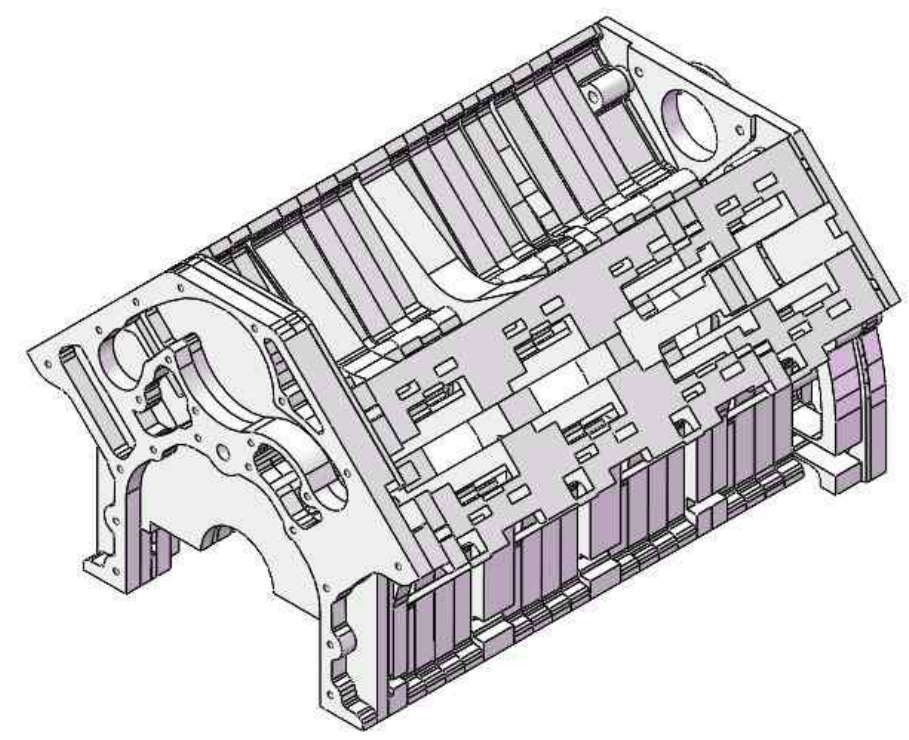 |
| Faces: 3,422 | Faces: 4,830 |

Figure 7-25  Reference and reconstructed models for B2P-10

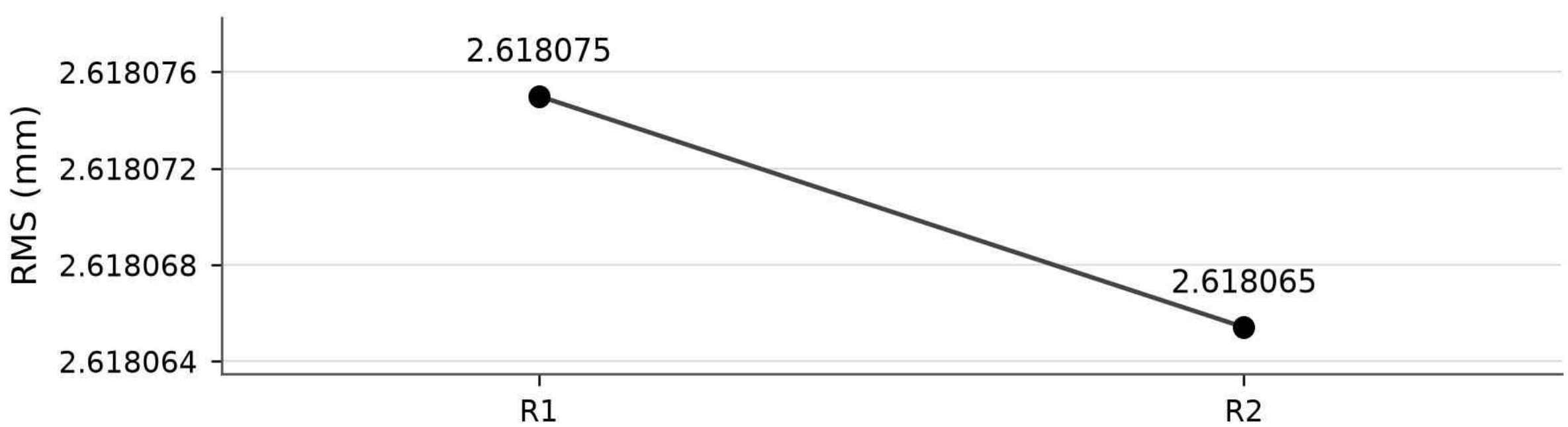

Figure 7-26  B2P-10 RMS surface distance by iteration

The reconstructed model is an invalid solid. RMS is approximately 2.6181 mm in both iterations, indicating little improvement, and the overall structure differs substantially from the reference. The reconstruction extrudes separate cross-sections, stacking simple surfaces to approximate the block’s curved surfaces and passage profiles. The face count increases from 3,422 to 4,830, while cylinder bores and flow passages remain incomplete. Having failed the solid validity check, the model is excluded from parameter-editing tests.

**B2P-11 Worm and planetary gearbox (complex)**

| Reference CAD model | GPT-6 reconstruction |
|---|---|
| 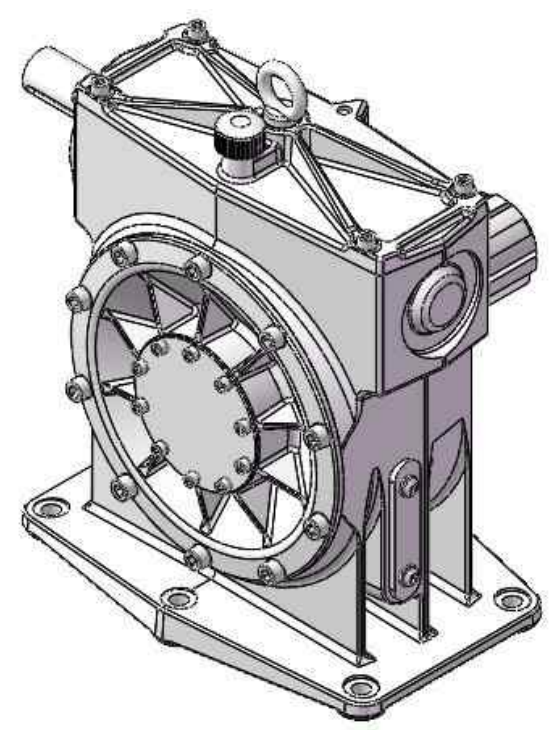 | 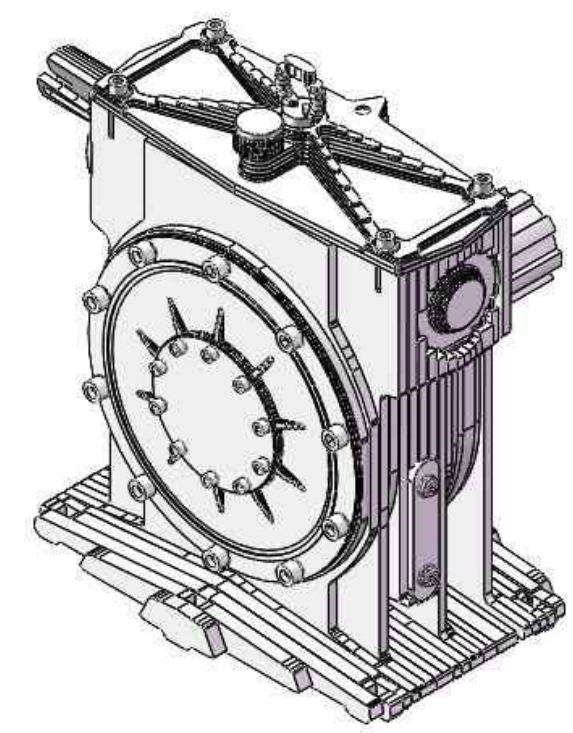 |
| Faces: 10,391; solids: 463 | Faces: 55,640; solids: 278 |

Figure 7-27 Reference and reconstructed models for B2P-11

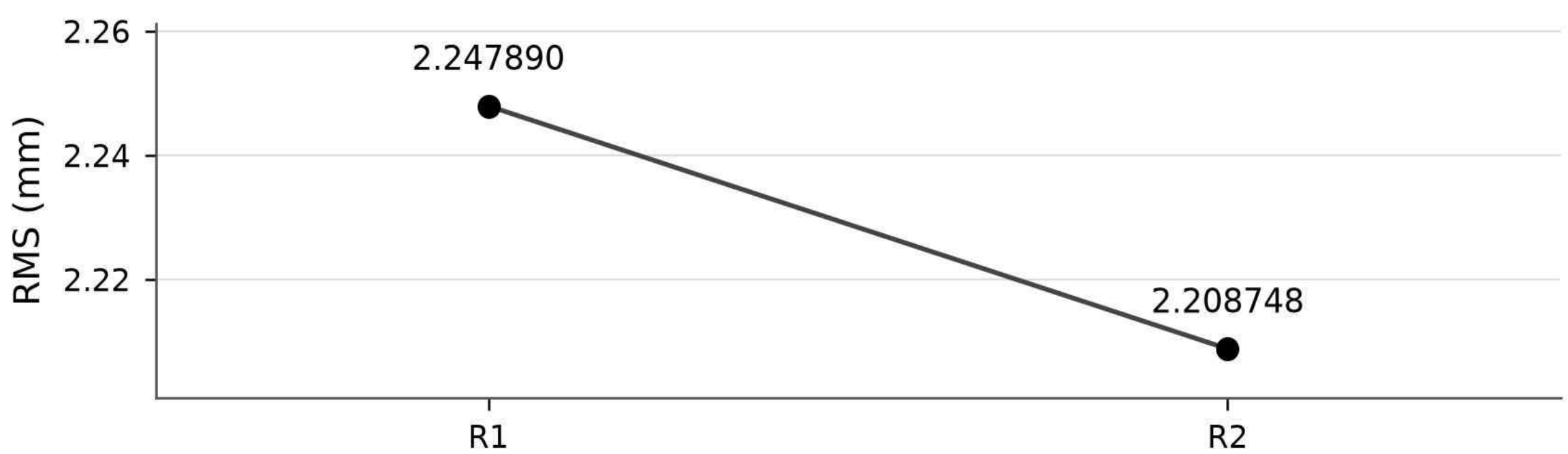

Figure 7-28 B2P-11 RMS surface distance by iteration

The reference contains 463 solids, whereas the reconstruction contains only 278: 262 valid and 16 invalid. RMS decreases only slightly, from 2.2479 to 2.2087 mm, over two iterations. The reconstructed parts use extrusions of separate cross-sections, approximating the curved surfaces of the housing and other components with simple surfaces. The face count increases from 10,391 to 55,640. Because some solids fail the validity check, the model is excluded from parameter-editing tests.

### 7.3.2 Additional experiment results

Web search is disabled for the representative complex cases selected in Section 7.2.2, and each result is compared with its main-experiment counterpart. The input model, prompt, model configuration, and output requirements remain the same; only web-search access changes. The comparison examines whether disabling search produces evident changes in the reconstruction process or results (Table 7-6).

Table 7-6  Results with web search enabled and disabled

| Case ID | Test condition | Solid validity | RMS (mm) | Editing-test results |
|---|---|---|---|---|
| B2P-09 | Main experiment | 20/23 passed | 9.9593 | Not tested (invalid solids) |
| B2P-09 | Web search disabled | 1/3 passed | 0.6074 | Not tested (invalid solids) |
| B2P-10 | Main experiment | 0/1 passed | 2.6181 | Not tested (invalid solids) |
| B2P-10 | Web search disabled | 0/2 passed | 1.6184 | Not tested (invalid solids) |
| B2P-11 | Main experiment | 262/278 passed | 2.2087 | Not tested (invalid solids) |
| B2P-11 | Web search disabled | 262/278 passed | 2.2067 | Not tested (invalid solids) |

Table 7-7 reports the time GPT-6 takes to reconstruct each case.

Table 7-7  Reconstruction time with web search enabled and disabled

| Case ID | B2P-09 Main experiment | B2P-09 Web search disabled | B2P-10 Main experiment | B2P-10 Web search disabled | B2P-11 Main experiment | B2P-11 Web search disabled |
|---|---|---|---|---|---|---|
| Time | 33 min | 1 h 0 min | 5 min | 48 min | 1 h 52 min | 1 h 15 min |

The additional-experiment cases are analyzed below.

### B2P-09 Gearbox housing cover (web search disabled)

| Reference CAD model | Main-experiment reconstruction | Reconstruction with web search disabled |
|---|---|---|
| 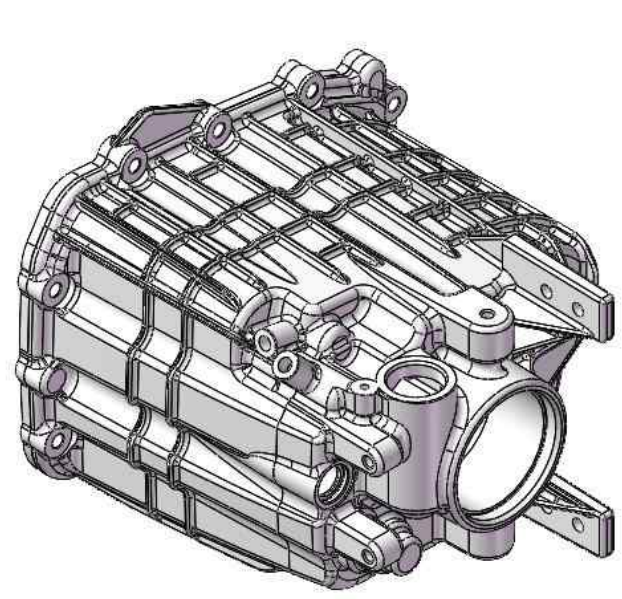 | 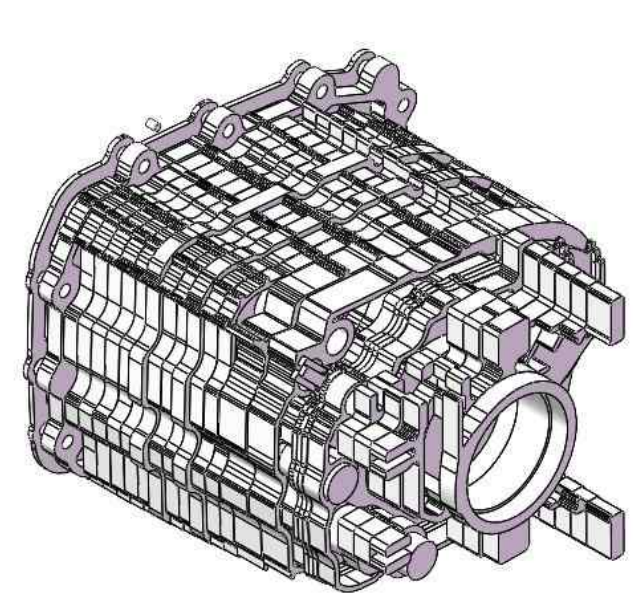 | 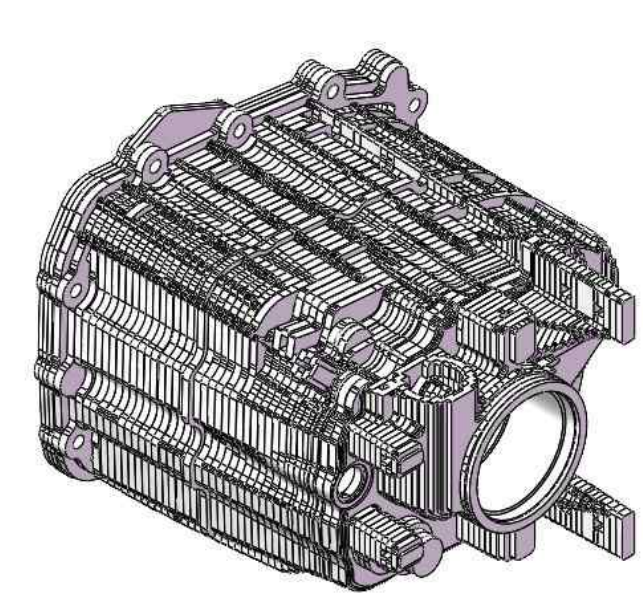 |
| Faces: 5,639; solids: 3 | Faces: 8,466; solids: 23 | Faces: 21,567; solids: 3 |

Figure 7-29  Reference and reconstructed models for B2P-09 with web search disabled

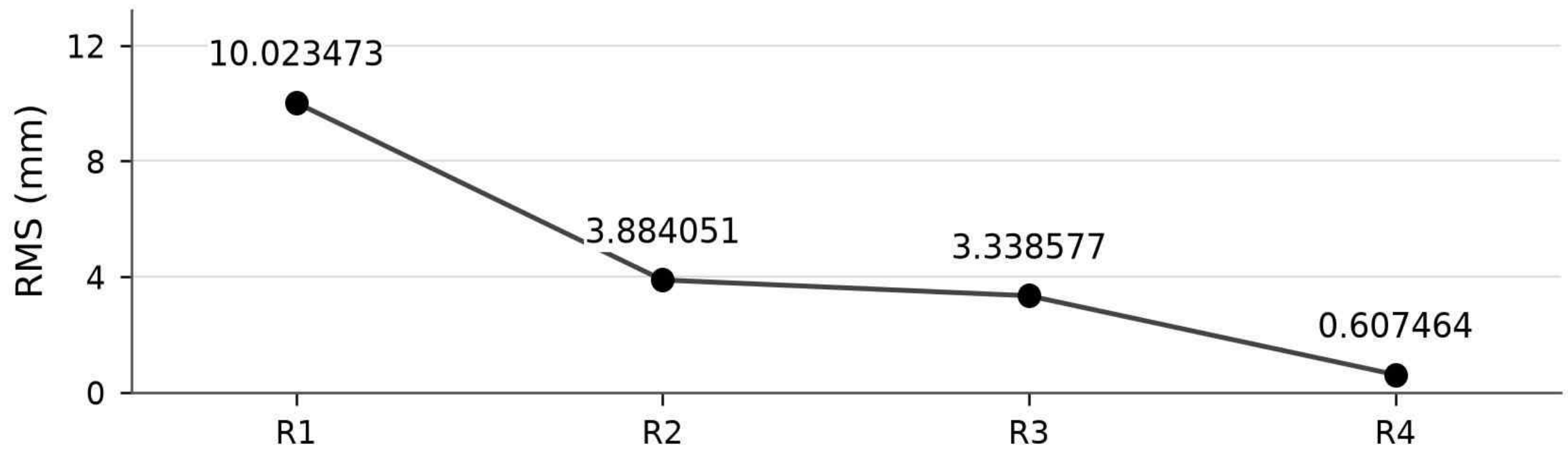

Figure 7-30  B2P-09 RMS surface distance by iteration with web search disabled

The solid count matches the reference, but only one solid passes the validity check and two fail. The model is therefore excluded from parameter-editing tests because it fails the solid validity prerequisite. RMS decreases from 10.0234 to 0.6074 mm over four iterations. Complex surfaces are still approximated by layers of simple surfaces, with the face count rising to 21,567 and even denser segmentation. Geometric error is lower than in the main experiment, but not all solids pass the validity check.

### B2P-10 V8 engine block (web search disabled)

| Reference CAD model | Main-experiment reconstruction | Reconstruction with web search disabled |
|---|---|---|
| 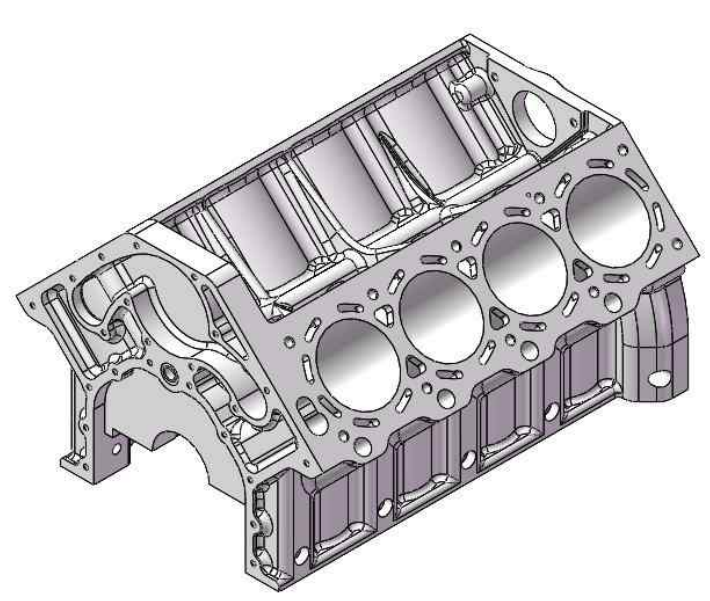 | 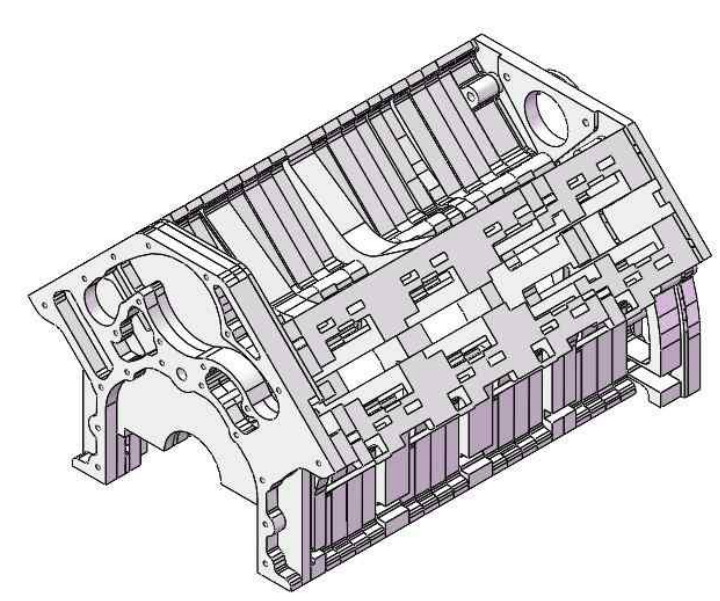 | 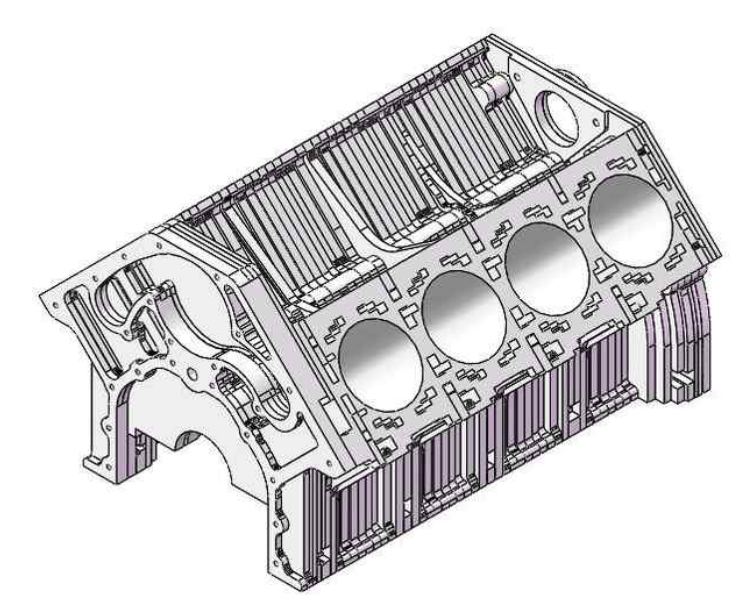 |
| Faces: 3,422; solids: 1 | Faces: 4,830; solids: 1 | Faces: 11,746; solids: 2 |

Figure 7-31  Reference and reconstructed models for B2P-10 with web search disabled

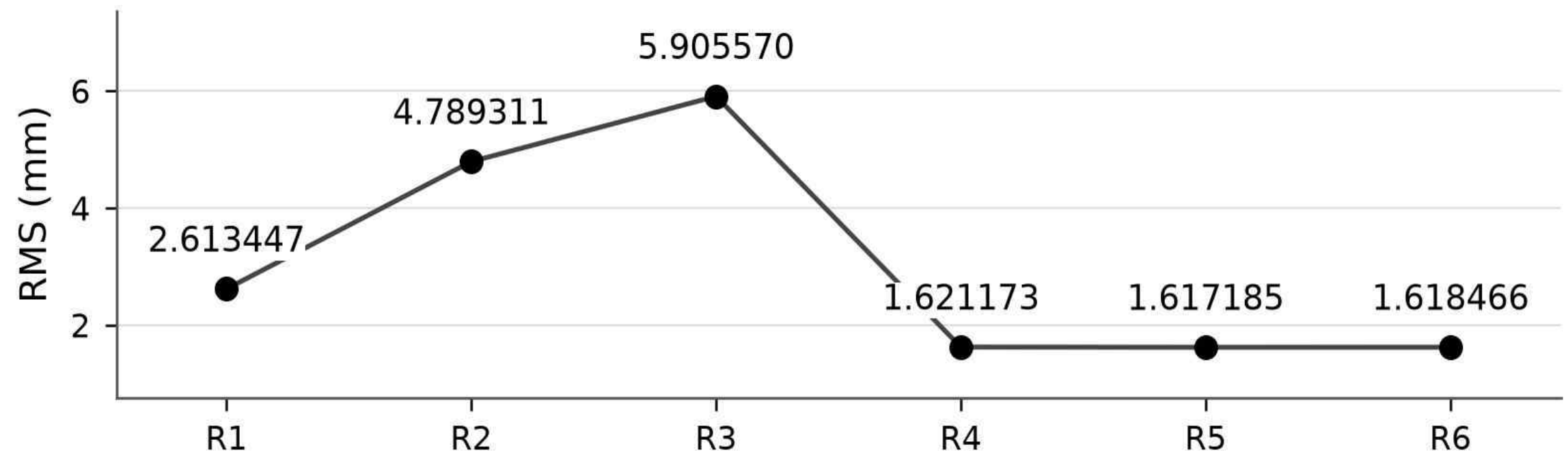

Figure 7-32 B2P-10 RMS surface distance by iteration with web search disabled

Neither of the two reconstructed solids passes the validity check. The model is therefore excluded from parameter-editing tests because it fails the solid validity prerequisite. Across six iterations, RMS initially rises from 2.6134 to 5.9055 mm, then falls to 1.6184 mm. The block's curved surfaces are still approximated by stacked simple surfaces.

### B2P-11 Worm and planetary gearbox (web search disabled)

| Reference CAD model | Main-experiment reconstruction | Reconstruction with web search disabled |
|---|---|---|
| 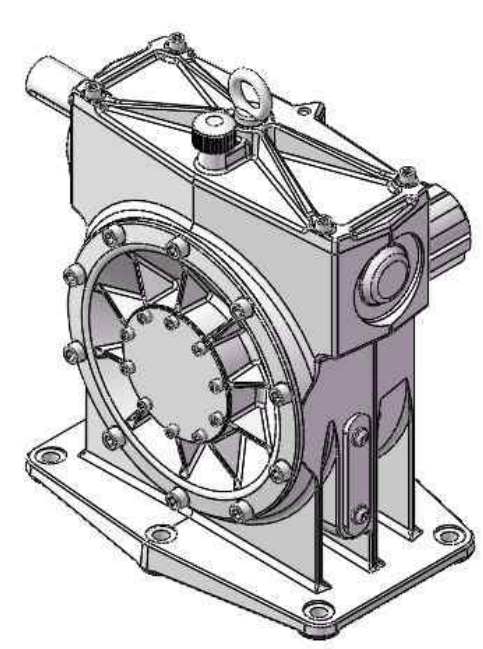 | 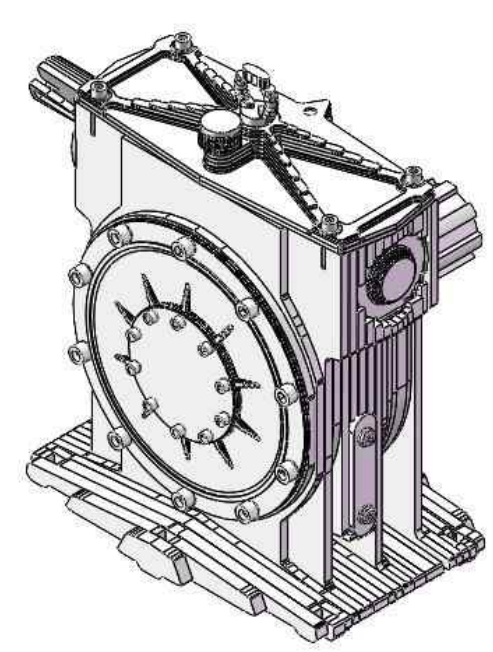 | 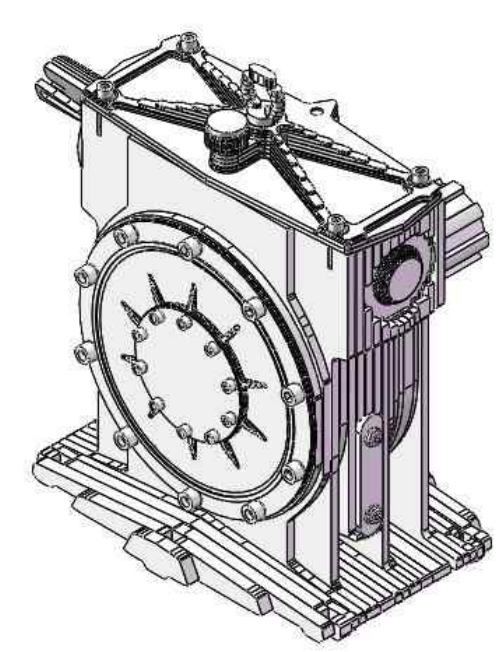 |
| Faces: 10,391; solids: 463 | Faces: 55,640; solids: 278 | Faces: 55,637; solids: 278 |

Figure 7-33 Reference and reconstructed models for B2P-11 with web search disabled

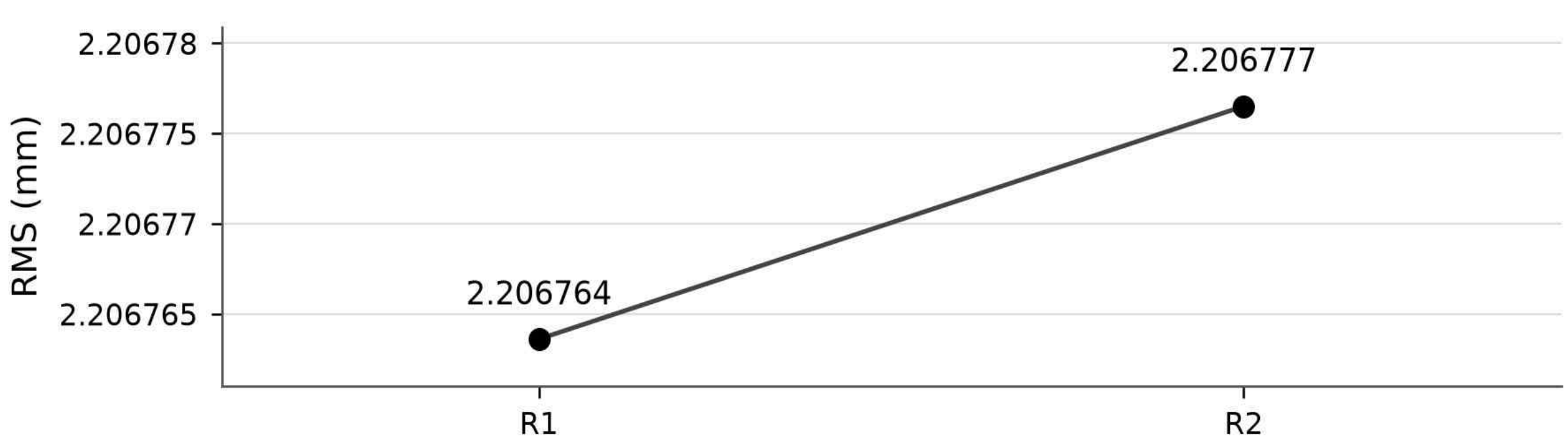

Figure 7-34 B2P-11 RMS surface distance by iteration with web search disabled

The validity check passes 262 solids and fails 16. The model is therefore excluded from parameter-editing tests because it fails the solid validity prerequisite. RMS is 2.2067 mm in each of the two iterations, remaining essentially unchanged. Simple surfaces still approximate the curved geometry, and overall geometric error is close to that in the main experiment.

GPT-6 can still produce native parametric models with web search disabled. Compared with the main experiment, there is no clear, consistent improvement or deterioration in model quality. Substantial limitations remain in solid validity and geometric agreement, and none of these outputs qualifies for the editing tests. These three comparisons are insufficient to establish a general effect of web search on parametric reconstruction.

## 7.4 Analysis of GPT-6's approach

The recorded outputs and tool operations suggest the following workflow for parametric reconstruction from B-rep models:

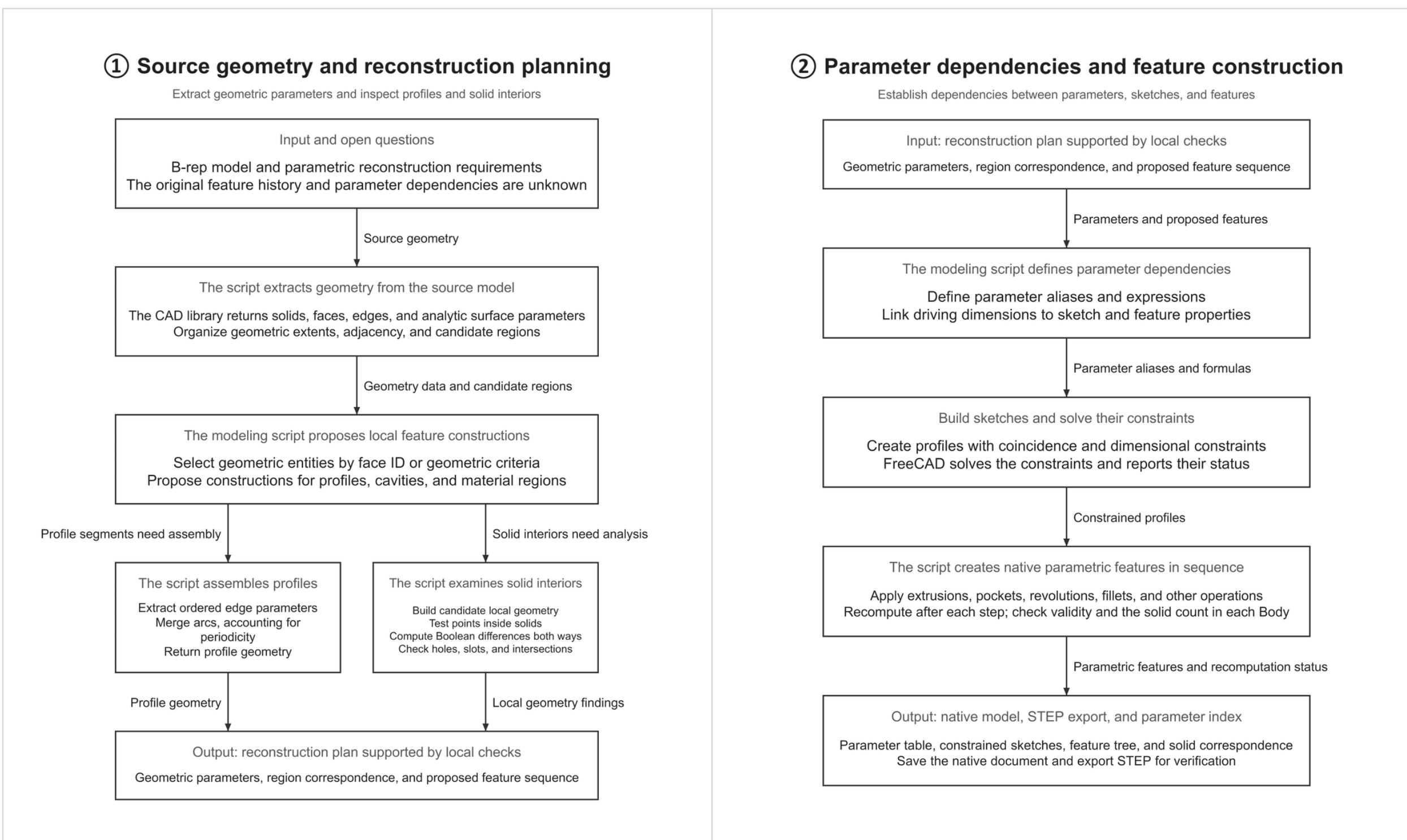

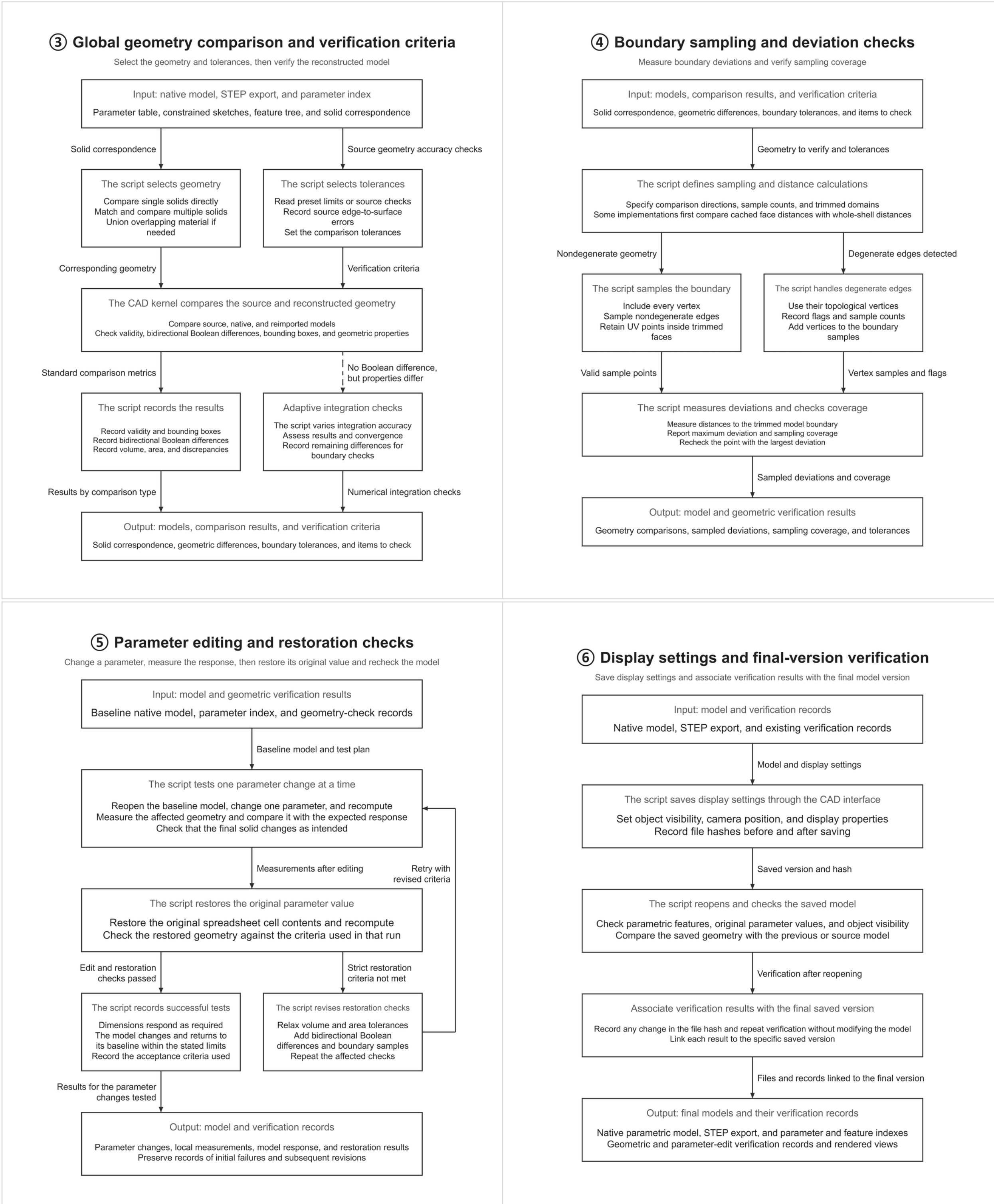

Figure 7-35 Inferred workflow for parametric reconstruction from B-rep models

## 7.5 Capabilities and limitations

In these tests, GPT-6 can read the geometry and topology of an input B-rep model, create sketches and dimensional parameters, build features such as extrusions, revolves, and cuts, and produce editable parts and assemblies. It recovers the main shapes of the three-arm flanged shaft, connecting rod, and pulley support assembly. Iterative modeling also reduces overall geometric error for the pump housing, forked shaft, and complex housings. However, reconstruction of complex surfaces and establishment of parameter dependencies remain incomplete. Some models approximate continuous surfaces by dividing them into cross-sections and stacking simple surfaces. The gearbox cover consequently has a densely segmented exterior, and the engine block's passage profiles are not fully recovered. Parameter edits also expose missing dependencies between features. Only six of the 11 models pass the solid validity check and qualify for editing tests: five simple and one moderate. Each undergoes 10 single-parameter edits, of which 52 pass in total. Only four models pass all 10 edits—three simple and one moderate—representing 36.4% of all 11 cases. Increasing the depth of the B2P-03 flanged housing closes a hole opening. In another test, reducing a flange diameter causes mounting holes to disappear. These failures show that dependent features do not update correctly when the main dimensions change.

GPT-6 can construct native parametric models from B-rep inputs and edit some dimensions, but it cannot yet reliably reconstruct both complex surfaces and assembly relationships correctly. Some reconstructed solids fail validity checks, and the chuck also contains faces not belonging to any solid. With web search disabled, complex surfaces are still approximated by layers of simple surfaces. The three comparison cases are insufficient to establish an effect of web search on reconstruction quality. These tests primarily cover single-parameter edits; combinations of parameter changes, feature addition or deletion, and full assembly functionality have not been evaluated.

# Chapter 8 Conclusions

The preceding chapters evaluate GPT-6 across six kinds of CAD tasks, from part classification and feature recognition to model generation and parametric reconstruction. This chapter draws together the evidence on the capabilities demonstrated, the requirements that remain difficult to satisfy, and the contribution and limitations of iterative, tool-assisted workflows.

The classification and feature-recognition results show that GPT-6 can combine specific geometric information with mechanical-design knowledge to interpret CAD models. For example, it identifies a four-throw crankshaft from the connections between its offset shaft sections and crank webs, and a double torsion spring from the arrangement of its coils and arms (PC-11 and PC-12). In the bearing case, it identifies the broad class as a roller bearing, but also notes that the main roller surfaces are cylindrical and uses this observation to assess whether the model represents a standard tapered roller bearing (PC-09). GPT-6 can also distinguish local features and identify relationships between them: in FR-04, it correctly identifies nested pockets at multiple levels, intersecting passages, and mounting holes interrupted by those passages. These results show that GPT-6 can combine geometry, mechanical-design knowledge, and retrieved information to classify parts, recognize features, and explain the basis for its judgments.

Correct recognition of local geometry does not, however, ensure a correct interpretation of a part's overall purpose. In the PC-10 test with neither candidate-class hints nor web search, GPT-6 recognizes the spring and valve stem but classifies the pipe coupling as a check valve, mistaking the function of its internal valve mechanism for the purpose of the entire coupling. Feature grouping can fail in a similar way. In FR-06, two spatially separate sections of hole wall are grouped as one hole, while an external cylindrical face is incorrectly identified as a hole. GPT-6 can use geometry to reason about a model, but it still makes mistakes about which faces belong to one feature and what role a local feature plays in the whole part.

Successful generation cases include solids with complex internal passages and assemblies containing several parts. Given detailed geometric requirements, GPT-6 produces TC-05 as a single valid V8 engine-block solid, with a connected water jacket isolated from the cylinder bores. In TC-06, it models and assembles the main drive and valve gear of a horizontal single-cylinder steam engine. The components and relative positions of the piston rod, crosshead, crank and connecting rod, and eccentric valve gear meet the requirements. These results show that GPT-6 can construct complex internal geometry and position and assemble multiple parts when geometric requirements are specified.

Modeling from functional requirements further demonstrates GPT-6's ability to develop a design from a set of requirements. In TRC-03, it designs a pump casing around a specified impeller, creating connected flow passages, pipe interfaces, and a rear opening for disassembly. It provides clearance for fastening tools and rotor removal and calculates head loss and pressure-bearing capacity for the specified operating conditions. For the pneumatic two-finger gripper in TRC-04, GPT-6 selects a parallel-gripping design driven by a pneumatic cylinder, synchronized by a rack-and-pinion mechanism, and guided by shafts that constrain finger motion. Although this differs from the reference model's linkage mechanism, it meets the functional requirements listed in the test. These tasks involve mechanism selection, interface design, spatial layout, and engineering calculations, showing that GPT-6 can make concrete design choices and implement them in CAD for some tasks.

Accurate generation and reconstruction of complex surfaces, thin walls, and local transitions remain recurring difficulties across tasks. D2C-A03 does not accurately recover complex surfaces and thin-walled regions from engineering drawings. In reconstruction from images, I2C-03 approximates the control-arm sidewalls with ruled lofts through multiple sections, leaving the smooth surfaces and root fillets incomplete. In parametric reconstruction from B-rep models, B2P-09 and B2P-10 use extrusions of separate cross-sections to approximate housing surfaces and passage profiles, with substantial geometric deviations or invalid solids. Although the main shape and some holes and slots can be generated, complex surfaces and connection transitions remain difficult to reproduce accurately.

Even when the overall shape is present, local passages, the way solids are connected, and assembly relationships may fail to meet the requirements. D2C-P05 produces a valid engine-block solid but turns some through holes into blind holes. The control-arm body, bushings, and lug are all present in the three-view I2C-03 result, but the reconstruction consists of separate solids that overlap locally and does not reproduce the reference part as one connected solid. The gripper in TC-04 includes a motor, gears, and the associated mechanism parts, yet a radial clearance remains between the motor shaft and the pinion's central bore, with no set screw or other fastening feature to secure them. The connection therefore cannot reliably transmit torque. Similar problems occur in the helicopter modeled from functional requirements: in TRC-07, the pitch links are not connected to the swashplate, and the gimbal motor is not secured to its support plate. These small geometric errors directly affect function. GPT-6 can thus create quite complex geometry and component layouts while still overlooking local requirements that determine whether the design works.

Insufficient input information cannot explain all these failures. Missing scale information in images and internal structures not fully shown in three orthographic views do limit geometric recovery. Yet the STEP files supplied for parametric reconstruction already contain the complete geometry, and the reconstructed models still show substantial approximation. B2P-09 and B2P-10, for example, approximate housing surfaces and passage profiles with extrusions of separate cross-sections, leaving large geometric deviations or multiple invalid solids. Five of the 11 parametric reconstructions fail the solid validity check. Complete geometric information must still be extracted, analyzed, and translated into suitable modeling features and operations to produce an accurate reconstruction. These tests show that, even with complete target geometry, GPT-6 cannot yet make full use of that information to reconstruct complex models accurately.

Parametric reconstruction exposes further limitations when a design is modified. The initial B2P-03 flanged housing is a valid solid, with a bidirectional RMS surface distance of approximately 0.05565 mm. Parameter edits nevertheless cause failures: increasing the side-housing depth from 36 to 54 mm closes an existing hole opening, while other edits remove mounting holes or counterbore shoulders or cause dependent features to fail during rebuilding. A B-rep describes the model's shape but does not uniquely determine the feature history or which dimensions and parts should update together when an edit is made. The tests show that GPT-6 cannot generally ensure the intended behavior under subsequent parameter edits.

These capabilities and limitations also need to be understood in terms of how GPT-6 performs CAD tasks. In most cases in this report, it uses an agent workflow that repeatedly generates a model, checks the result, and makes corrections based on feedback. In TC-05, for example, the water jacket initially opens incorrectly into the central cavity and longitudinal passages are closed. Iterative revisions are needed to establish the required connectivity and separation. In TRC-02, adjustments to thrust washers, stops, and jaw-fastening features eliminate local interference between moving parts. These revisions demonstrate that GPT-6 can analyze and correct specific problems in an existing model during iteration.

Correcting some problems does not mean that all requirements are met by the end of the process. TC-07 reaches 100% satisfaction of the listed geometric requirements, yet the final model still contains non-manifold connections and self-intersecting surfaces and fails the overall solid validity check. In B2P-08, RMS decreases from 1.1260 to approximately 0.4193 mm, but one solid is invalid and 371 faces belong to no solid, so the model is excluded from parameter-editing tests. The effectiveness of iterative improvement must therefore be assessed against each requirement separately; improvement in one metric is insufficient to establish that the task is complete.

Some error curves also fluctuate during iteration, showing that edits do not always yield the intended improvement. This suggests that AI models still struggle to identify the source of an error, decide what needs changing, and select the appropriate operation. Moreover, CAD features and parameters are interdependent: one edit can affect several design requirements and introduce new deviations while correcting the original problem. Effective iteration therefore requires accurate identification of both the geometry to change and the scope of the operation, together with an assessment of the consequences for dependent features and design requirements.

Iterative correction also incurs substantial time costs. Many modeling and reconstruction tasks take tens of minutes. Complex text-based modeling cases average 1 h 37 min 30 s, with one case taking 3 h 17 min. In reconstruction from images, I2C-06 takes 3 h 41 min, the longest recorded run in the report. Execution records show that some tasks require repeated construction, checking, and correction before the requirements are progressively met. Even where modeling succeeds, long completion times make it difficult to support the rapid feedback, repeated edits, and comparison of alternatives needed during design. Current systems have demonstrated the ability to complete complex tasks, but the time required remains a substantial limit on practical usability.

Table 8-1 GPT-6's current capabilities and main limitations in CAD tasks

| Task | Current capabilities | Main limitations |
|---|---|---|
| **Part classification** | GPT-6 uses geometric information and engineering knowledge to identify what kind of part a model represents. | GPT-6 may recognize the general type of part but still misidentify the specific type. It may also mistake the function of one structure within a part for the purpose of the whole part. |
| **Feature recognition** | GPT-6 identifies individual holes, slots, pockets, and other machining features in a model. | GPT-6 does not always distinguish individual features or identify their types correctly, especially where features are nested or intersect. |
| **CAD reconstruction from engineering drawings** | GPT-6 combines multiple views, section views, and dimensions to recover the main shape and internal structure. In some cases, the generated parts can be assembled as intended. | GPT-6 can misread how complex structures are arranged and get critical dimensions wrong, leaving out local features or creating interference between assembled components. |
| **CAD generation from text** | GPT-6 builds models from geometric descriptions, selects designs to meet functional requirements, and carries out modeling and engineering calculations. | GPT-6 still makes frequent errors in complex features, passage connectivity, local transitions, and connections that secure parts or transmit motion. The models do not consistently meet functional requirements or pass solid validity checks. |
| **CAD reconstruction from images** | GPT-6 recovers the main shape, main features, and component layout from images showing multiple views. | GPT-6 often simplifies complex surfaces and local transitions. Connections between solids and assembly interfaces may also be incorrect. Without a dimensional reference, absolute scale cannot be determined uniquely from the images. |
| **Parametric reconstruction from B-rep models** | GPT-6 recovers the main geometry of simple and moderately complex models, creates parametric models, and supports some dimensional changes. | Reconstructed complex surfaces and internal passages can differ substantially from the originals. GPT-6 often fails to establish the dependencies needed for related features to update correctly after an edit. |
| **Agentic iteration across tasks** | GPT-6 uses feedback to add missing features, adjust dimensions and positions, reduce geometric error, and meet more of the requirements. | GPT-6 still struggles to decide what to change and how far those changes should extend. Even after several iterations, models may contain invalid solids, lack required functions, or fail when a parameter is edited. Long iterations limit practical use. |

The evidence is limited to the selected cases, the stated system configuration, and the checks performed. Complexity is defined separately for each task, so the levels are not a common difficulty scale across chapters. The additional comparisons involve few cases and do not quantify variability across repeated runs. Publicly available source models and embedded STEP metadata may also supply prior information; the tests do not isolate those effects. Feature and geometry checks, calculation-based functional assessments, and single-parameter editing tests establish different aspects of performance and should not be combined into a single overall success rate.

These results support the use of general-purpose AI as an assistant for producing initial CAD models and exploring design alternatives in the demonstrated task settings. Reliable engineering use still requires verification of geometry, solid validity, functional connections, assembly relationships, and parameter dependencies. The central challenge is to make iterative correction consistent across these requirements while reducing the time needed to reach an acceptable model.